# Raumzeitkonzeptionen

# in der

# Quantengravitation

## Reiner Hedrich

Institut für Philosophie und Politikwissenschaft
Fakultät Humanwissenschaften und Theologie
Technische Universität Dortmund
Emil-Figge-Strasse 50
44227 Dortmund
reiner.hedrich@udo.edu

Zentrum für Philosophie und Grundlagen der Wissenschaft
Justus-Liebig-Universität Giessen
Otto-Behaghel-Strasse 10 C II
35394 Giessen
Reiner.Hedrich@phil.uni-giessen.de

*"I am convinced [...] of the utility of the dialog between physics and philosophy. [...] I think that most physicists underestimate the effect of their own epistemological prejudices on their research. [...] today foundational problems are back on the table, as they were at the time of Newton, Faraday, Heisenberg and Einstein.[...] I wish contemporary philosophers concerned with science would be more interested in the ardent lava of the fundamental problems science is facing today."*
(Rovelli (2004) 305)

*"Note that if our existing views on spacetime and/or quantum theory are* not *adequate, the question then arises of the extent to which they are* relevant *to research in quantum gravity. [...] how* iconoclastic *do our research programmes need to be?"* (Isham (1997) 2)

*"[...] nach einer, wie manche sagen, wahren Begebenheit, die in dem von ihr ausgeschwemmten Ozean der Mutmassungen, je nach charakterlicher Disposition, labyrinthisch aufschimmert oder schlichtweg unergründlich erscheinen mag [...]"* (Artrinaux (2009) vii)

# Inhalt

# 1.    Einleitung

## 1.1.    Motivationen für die Quantengravitation

### Die konzeptionelle Inkompatibilität von Allgemeiner Relativitätstheorie und Quantenmechanik bzw. Quantenfeldtheorien

Eine der entscheidendsten Motivationen für die Entwicklung einer Theorie der Quantengravitation besteht nach weit verbreiteter Auffassung in der erst einmal offensichtlichen wechselseitigen konzeptionellen Unverträglichkeit der Allgemeinen Relativitätstheorie mit der Quantenmechanik bzw. den Quantenfeldtheorien.

> *"General relativity and quantum theory are among the greatest intellectual achievements of the 20th century. [...] And yet, they offer us strikingly different pictures of physical reality. [...] Everything in our past experience in physics tells us that the two pictures we currently use must be approximations, special cases that arise as appropriate limits of a single, universal theory."* (Ashtekar (2005) 2) – *"[...] [Quantum Mechanics] and [General Relativity] have destroyed the coherent picture of the world provided by prerelativistic classical physics: each is formulated in terms of assumptions contradicted by the other theory."* (Rovelli (2004) 3)

Dass hier eine Inkompatibilität im eigentlichen Sinne vorliegt, setzt allerdings, wie im weiteren Verlauf aufzuzeigen sein wird, zwei meist stillschweigend unterstellte Annahmen voraus:[1]

(a)        Die Gravitation ist eine fundamentale Wechselwirkung.[2]
(b)        Die Quantenmechanik ist (soweit es ihre grundsätzlichen Einsichten und die fundamentale Ebene physikalischen Geschehens betrifft) universell gültig.[3]

Das Ziel der Entwicklung einer Theorie der Quantengravitation besteht vor diesem Hintergrund also entweder (i) in der Aufhebung bzw. Überwindung einer echten Unverträglichkeit zwischen den etablierten Theoriekomplexen unter Voraussetzung der Richtigkeit der beiden Annahmen (a) und (b) – oder eben (ii) darin, unter Einsatz der zu entwickelnden, neuen Theorie aufzuzeigen, dass es sich bei den Konflikten zwischen den etablierten Theoriekomplexen nur um eine scheinbare Unverträg-lichkeit handelt, deren Auftreten darin begründet liegt, dass mindestens eine der impliziten Annahmen (a) und (b) nicht zutrifft.

---

[1] Die gegenteilige Vermutung, dass die offensichtlichen Konflikte zwischen Allgemeiner Relativitätstheorie und Quantenmechanik bzw. Quantenfeldtheorien nur scheinbar, nicht aber tatsächlich mit einer konzeptionellen Unverträglichkeit und mit Widersprüchen zwischen den Theoriekonstrukten einhergeht, liesse sich also nur insofern begründen, dass die konkreten Instanzen zu benennen wären, die zur Ungültigkeit mindestens einer der beiden nachfolgenden Voraussetzungen der Inkompatibilitätsthese führen.

[2] Die Option, dass diese Annahme falsch ist, und die daraus erwachsenden Implikationen und Perspektiven für die Entwicklung einer Theorie der Quantengravitation werden explizit in Kap. 3.3. zu behandeln sein. Konkrete Theorieansätze, die von einer emergenten Gravitation auf der Grundlage eines prägeometrischen Substrats ausgehen, folgen in Kap. 4.6.

[3] Die Option, dass diese Annahme falsch sein könnte, wird nur am Rande einfliessen. Siehe wiederum Kap. 3.3. und Kap. 4.6.

Im Fall (i) kommt die Intuition zum Einsatz, dass es sich, obwohl die Annahmen (a) und (b) zutreffen, bei den beiden fundamentalsten, aber miteinander unverträglichen Theoriekomplexen, über welche die Physik zur Zeit verfügt, mindestens in einem Falle um eine Näherung zu einer noch fundamentaleren Beschreibung handelt, innerhalb derer die auftretenden konzeptionellen Unvereinbarkeiten auflösbar wären oder überhaupt nicht mehr in Erscheinung treten würden; diese fundamentalere Beschreibung wäre die gesuchte Theorie der Quantengravitation. Ob eine solche Theorie der Quantengravitation nur die Allgemeine Relativitätstheorie als klassische Theorie der Gravitation und der Raumzeit ablöst, oder ob sie auch gleich noch die Beschreibung der übrigen Wechselwirkungen im Rahmen der Quantenfeldtheorien korrigiert oder ersetzt, bleibt erst einmal offen.

Im Fall (ii) hätte eine Theorie der Quantengravitation vor allem zu erklären, unter welchen spezifischen Bedingungen sich mindestens eine der oben genannten Annahmen[4] als falsch erweist und inwiefern sich dadurch der konzeptionelle Konflikt als ein nur scheinbarer verstehen lässt, der etwa auf ein Missverständnis hinsichtlich des Geltungsbereichs mindestens einer der beteiligten Theorien zurückzuführen ist.

In beiden Fällen wäre also (mindestens) eine neue Theorie erforderlich. Im Fall (i) ist dies eine Theorie, mittels derer sich die dann tatsächliche Unverträglichkeit der Vorgängertheorien überwinden lässt. Im Fall (ii) benötigt man entweder, wenn die Gravitation keine fundamentale Wechselwirkung sein sollte, eine fundamentalere Theorie als die Allgemeine Relativitätstheorie, die diese als Grenzfall enthält, oder, wenn die Quantenmechanik nicht universell gültig sein sollte, eine Theorie, die diese begrenzte Gültigkeit konkretisiert, vielleicht sogar die Quantenmechanik gänzlich ablöst, sie aber dennoch als Grenzfall enthält und schließlich insbesondere ihren immensen Erfolg erklärt. Für den Fall, dass beide Annahmen (a) und (b) falsch sein sollten, kann man nur hoffen, dass man mit einer Theorie auskommt, die sowohl die Allgemeine Relativitätstheorie und die Quantenmechanik als Grenzfall enthält, und es nicht erneut zu widerstreitenden Konstrukten kommt, die das bestehende Problem nur auf eine tiefere Ebene verschieben.

In beiden Fällen ist zudem sehr wahrscheinlich damit zu rechnen, dass die Allgemeine Relativitätstheorie durch eine fundamentalere Beschreibung abgelöst wird: Im Fall (i) liegt der Grund dafür in der hier vorausgesetzten universellen Gültigkeit der Quantenmechanik, im Fall (ii) erfolgt diese Theorieablösung mindestens dann, wenn es sich bei der Gravitation um keine fundamentale Wechselwirkung handelt, man also eine fundamentalere Beschreibung benötigt, um die Gravitation als residuales Phänomen verstehen zu können. In beiden Fällen besteht zudem die Möglichkeit, dass das Standardmodell der Quantenfeldtheorien oder gar die gesamte Theoriestruktur und Methodologie der Quantenfeldtheorien eine Korrektur erfährt oder gänzlich abgelöst wird. Zwingend ist dies jedoch nur, wenn die Quantenmechanik keine universelle Gültigkeit besitzt. Und nur in diesem spezifischen Fall ist auch mit einer fundamentaleren Theorie als der Quantenmechanik zu rechnen. Das mit der Entwicklung einer Theorie der Quantengravitation einhergehende oder zumindest zu erwartende Gefährdungspotential für die etablierten Theorien nimmt also, wenn man die verschiedenen Möglichkeiten durchspielt, von der Allgemeinen Relativitätstheorie über die Quantenfeldtheorien bis zur Quantenmechanik ab. – In allen Fällen lohnt es sich jedoch, die konstatierte Unverträg-lichkeit der Allgemeinen Relativitätstheorie mit der Quantenmechanik bzw. den Quantenfeldtheorien genauer in Augenschein zu nehmen, unabhängig davon, ob sie echt oder nur scheinbar ist.

---

[4] Gemeinhin ist es die Annahme (a), die im Fokus der Versuche steht, die Unverträglichkeit als eine nur scheinbare zu erweisen.

*

Die Allgemeine Relativitätstheorie stellt heute gleichermassen unsere allgemeinste Theorie der Raumzeit wie auch jeglichen gravitativen Verhaltens dar.[5] Beide Aspekte werden im Rahmen der Theorie gleichgesetzt, indem die Gravitation als Ausdruck der dynamischen Raumzeit und ihrer differentialgeometrischen Eigenschaften verstanden wird. Die dynamischen Eigenschaften der Gravitation werden im Rahmen der Theorie durch die dynamischen Eigenschaften der Raumzeit repräsentiert. Von der Gravitation bleibt in dieser Darstellung nichts übrig als dynamische Raumzeit. Die Beschreibung der Gravitation wird zur Geometrodynamik.

Die Allgemeine Relativitätstheorie als Theorie der Gravitation und der Raumzeit ist jedoch das, was man in der Physik eine klassische Theorie nennt. D.h. sie ist keine Quantentheorie. Alle anderen Grundkräfte der Natur – die elektromagnetische, die schwache und die starke Wechselwirkung – werden, im Gegensatz zur Gravitation, mittels der Quantenfeldtheorien in zumindest konzeptionell einheitlicher Form beschrieben. Die Quantenfeldtheorien als mittelbare Folge des Bestrebens, die quantenmechanische Beschreibung der Wechselwirkungen (bzw. Wechselwirkungsfelder) mit der speziellen Relativitätstheorie in Einklang zu bringen, werden zur Zeit als die allgemeinsten Theorien des dynamischen Verhaltens der Materie und ihrer Wechselwirkungen angesehen. Nur die Gravitation wurde in ihrem konzeptionellen Rahmen bisher nicht erfasst.

Einer der wesentlichsten Gründe dafür besteht nach weit verbreiteter Auffassung eben gerade darin, dass die Quantenmechanik und die Quantenfeldtheorie konzeptionell nicht mit der Allgemeinen Relativitätstheorie vereinbar sind. Die Art und Weise, wie die heutige Physik die Gravitation beschreibt, erscheint als mit der zur Zeit verwendeten physikalischen Beschreibung aller anderen Wechselwirkungen letztlich nicht verträglich. Unsere grundlegendsten heutigen physikalischen Theorien entwerfen jeweils eigenständige Bilder der Wirklichkeit, die nicht nur von gänzlich unterschiedlichen und miteinander inkompatiblen konzeptionellen Voraussetzungen ausgehen, sondern spätestens für den Bereich der Überschneidung ihrer Gegenstandsfelder zu klaren Konflikten führen. Diese lassen sich nicht etwa auf einen singulären Konfliktpunkt reduzieren, sondern erweisen sich als mehr oder weniger vielgestaltig. Als paradigmatisch für die wechselseitigen Konflikte und die konzeptionelle Unverträglichkeit zwischen der Allgemeinen Relativitätstheorie und der Quantenmechanik bzw. den Quantenfeldtheorien lassen sich dennoch wohl die folgenden drei Problempunkte ansehen:

(1) Das Gravitationsfeld wird von der Allgemeinen Relativitätstheorie als klassisches dynamisches Feld behandelt, repräsentiert durch die (pseudo-)Riemannsche Metrik der Raumzeit. Dynamische Felder besitzen jedoch gemäss der Quantenmechanik Quanteneigenschaften. Setzt man die universelle Gültigkeit der Quantenmechanik voraus, so erwächst aus dieser Problemlage eine unmittelbare Motivation für eine (direkte oder indirekte) Quantisierung des Gravitationsfeldes. Ein wesentlich entschiedeneres Argument für eine Quantisierung des Gravitationsfeldes erwächst jedoch – die universelle Gültigkeit der Quantenmechanik wiederum vorausgesetzt – aus dem spezifischen Konflikt zwischen Allgemeiner Relativitätstheorie und Quantenmechanik, zu dem es dadurch kommt, dass erstere auch die Materie und die Energie, die gemeinsam über den Energie-Spannungs-Tensor in die Einsteinschen Feldgleichungen einfliessen, genauso klassisch behandelt wie die Gravitation.

---

[5] Zur Raumzeitkonzeption, den ontologischen Implikationen und den Interpretationsproblemen der Allgemeinen Relativitätstheorie sowie zu den damit einhergehenden Implikationen für die unterschiedlichen Strategien zur Entwicklung einer Theorie der Quantengravitation, siehe Kap. 2.1.

> *"The fact that general relativity treats matter classically, and gravity as curvature, while our best theories of matter are quantum theories using a flat metric, is enough to show that some sort of reconciliation is needed."* (Butterfield / Isham (1999) 128)

Gegen eine Lösung im Rahmen eines theoretischen Hybridkonstruktes, welches die Gravitation klassisch und alles andere, also die Materie und alle nicht-gravitativen Felder, insofern quantenmechanisch behandelt, dass man statt klassischer Grössen quantenmechanische Erwartungswerte in den Energie-Spannungs-Tensor einfügt, gibt es gewichtige Argumente.[6] Solche semi-klassischen Theorien führen nicht nur zu konzeptionellen Unverträglichkeiten, sondern zu direkten Widersprüchen.[7]

> *"The right-hand side of the field equations [of general relativity] describes matter sources, the behaviour of which is governed by quantum theory. The left-hand side of the field equations describes gravitation as a classical field. If the right-hand side represents quantized matter then the field equations as they stand are* inconsistent.*"* (Riggs (1996) 2)

(2) In der Allgemeinen Relativitätstheorie wird die Dynamik des Gravitationsfeldes durch die Dynamik der raumzeitlichen Metrik repräsentiert, so dass eine Quantisierung des Gravitationsfeldes einer Quantisierung der dynamischen Metrik entspräche. Die Quantenfeldtheorien, die das Verhalten dynamischer Quantenfelder beschreiben, arbeiten jedoch mit einem festen, nicht-dynamischen Hintergrundraum mit fester, statischer (im einfachsten Falle: Minkowskischer) Metrik. Sie kommen also in ihrer bisherigen Form nicht für die Erfassung einer dynamischen 'Quantengeometrie' in Frage.

> *"The whole framework of ordinary quantum field theory breaks down once we make the gravitational field (and the differential manifold) dynamical, once there is no background metric any longer!"*
> (Thiemann (2007) 5)

---

[6] Siehe etwa Kiefer (1994, 2004 [Kap. 1.2.], 2005), Peres / Terno (2001), Terno (2006); siehe auch Callender / Huggett (2001a, 2001b).

[7] Hier kommen jedoch die oben schon erwähnten, meist stillschweigend unterstellten Annahmen ins Spiel. Es kommt nur dann zum Problem mit semi-klassischen Hybridkonstrukten, wenn es sich bei der Gravitation um eine fundamentale Wechselwirkung handelt. Auch die Motivation für eine Quantisierung der Gravitation auf der Grundlage der Argumente gegen semi-klassische Theorien, wird nur dann wirksam, wenn die Gravitation eine fundamentale Wechselwirkung ist. Sollte die Gravitation hingegen ein emergentes, intrinsisch klassisches Phänomen sein, welches von fundamentalen (quantisierten) Wechselwirkungen induziert wird bzw. aus diesen resultiert, so ergibt sich aus den Argumenten gegen semi-klassische Theorien keine Motivation für eine Quantisierung der Gravitation, da die Gravitation als emergentes Phänomen dann eben nicht zu den fundamentalen Wechselwirkungen auf der Ebene des quantenmechanischen Substrats zählt. Sie kommt erst als aus dieser Substratdynamik resultierendes oder von ihr induziertes, emergentes Phänomen zustande. Sollte dies der Fall sein, dann gibt es keine semi-klassische Hybriddynamik, die zu Inkonsistenzen in der Quantentheorie führen könnte. Auf der fundamentalen Ebene gäbe es dann überhaupt keine Gravitation, die diese Inkonsistenzen hervorrufen könnte. Eine Quantisierung der Gravitation verliert unter diesen Umständen nicht nur eine ihrer wichtigsten Motivation, sondern erscheint vielmehr – als Quantisierung kollektiver, emergenter, nicht-fundamentaler Freiheitsgrade – völlig abwegig. Siehe Kap. 3.3.

Es ist jedoch nicht nur der dynamische Charakter des Gravitationsfeldes bzw. des metrischen Feldes, der hier zu Problemen führt. Wie noch aufzuzeigen sein wird, ist vor allem die substantiell interpretierbare Diffeomorphismusinvarianz[8] der Allgemeinen Relativitätstheorie – der substantielle, physikalisch relevante Aspekt der allgemeinen Kovarianz, die zu den essentiellen Grundvoraussetzung der Theorie zählt – ohnehin mit keinem festen Hintergrundraum vereinbar.[9] Die quantisierte Erfassung der metrischen Eigenschaften einer dynamischen Raumzeit lässt sich nicht wiederum auf einem Hintergrundraum beschreiben – und schon gar nicht einem statischen Hintergrundraum mit schon festgelegter Metrik. Für eine Quantentheorie des Gravitationsfeldes kommt auch aus diesem Grund kaum eine Quantenfeldtheorie im herkömmlichen Sinne in Frage.[10]

(3) In der Allgemeinen Relativitätstheorie ist die Zeit eine Komponente der dynamischen Raumzeit. Sie ist dynamisch in die Wirkung der Materie und Energie auf die Metrik und die Wirkung der Metrik auf die Materie eingebunden. Sie lässt sich nur lokal und intern definieren; es gibt keine globale Zeit, keinen physikalisch wirksamen globalen Zeitparameter.[11] – In der Quantenmechanik hingegen wird die Zeit als globaler Hintergrundparameter behandelt, nicht einmal als physikalische Observable, der ein quantenmechanischer Operator zugeordnet werden könnte. Vertraut man der Allgemeinen Relativitätstheorie als klassischer Raumzeittheorie hinsichtlich ihrer Implikationen für den Zeitbegriff mehr als der Quantentheorie, welche die Problematik der Zeit in der Art, in der sie mit ihr umgeht, gar nicht behandelt, so deutet sich hier die Notwendigkeit einer konzeptionellen oder modelltheoretischen Modifikation oder Erweiterung des formalen Instrumentariums der Quantenmechanik an.

*

Neben den auf diese Weise erst einmal rudimentär beleuchteten konzeptionellen Motivationen gibt es jedoch auch konkretere physikalische Problemlagen, die eine Theorie der Quantengravitation für wünschenswert oder vielleicht sogar unabdinglich erscheinen lassen:

## Ungelöste physikalische Problemstellungen

Obwohl im Kontext der etablierten Theorien allgemein angenommen wird, dass es sich bei der Gravitation um eine fundamentale Wechselwirkung handelt und dass die Quantenmechanik universell gültig ist, lässt sich für viele mikroskopische ebenso wie für viele makroskopische Systeme über die unter diesen Voraussetzungen vorliegende Unvereinbarkeit unserer fundamentalen physikalischen Theorien hinwegsehen: Für Elektronen und ihr dynamisches Verhalten etwa spielt die Gravitation im allgemeinen nur eine sehr untergeordnete Rolle. Für Berechnungen im Rahmen der

---

[8] Siehe Kap. 2.1. sowie Earman (2006, 2006a).

[9] Siehe Kap. 2.1. sowie Earman (1986, 1989, 2002, 2006, 2006a), Earman / Norton (1987), Norton (1988, 1993, 2004).

[10] Sollte sich eine Theorie der Quantengravitation, welche die Allgemeine Relativitätstheorie als fundamentalere Beschreibung ablöst, als notwendigerweise hintergrundunabhängig herausstellen, bliebe das konzeptionelle Spannungsgefüge zu den herkömmlichen Quantenfeldtheorien, mittels derer im Rahmen des Standardmodells die übrigen Wechselwirkungen erfasst werden, bestehen, was für die Quantenfeldtheorien mittelfristig eine modelltheoretische Modifikationen, eine gravierende konzeptionelle Korrektur oder gar, was am wahrscheinlichsten ist, eine vollständige Ablösung erforderlich machen würde.

[11] Die unter (2) schon erwähnte, aktive Diffeomorphismusinvarianz der Allgemeinen Relativitätstheorie führt nicht zuletzt zum *Problem der Zeit*. Siehe Kap. 2.1. und 4.4. sowie Belot / Earman (1999), Earman (2002), Pons / Salisbury (2005), Rickles (2005a), Rovelli (2001, 2006), Unruh / Wald (1989).

Himmelsmechanik kann man andererseits wahrscheinlich getrost Quanteneffekte ignorieren.[12] Und auch für die GPS-Positionierung lässt sich zwischen der raumzeitlichen Einordnung, welche die Allgemeine Relativitätstheorie voraussetzt, und dem Funktionieren der Atomuhren, für dessen Verständnis die Quantenmechanik erforderlich ist, gut trennen.

Die wechselseitige Unverträglichkeit von Allgemeiner Relativitätstheorie und Quantenmechanik (bzw. die Unzulänglichkeit mindestens einer dieser Theorien als fundamentale Beschreibung) wird jedoch vor allem dort zum Problem, wo aus physikalischen Gründen die widerstreitenden Partner gleichermassen Berücksichtigung finden müssten: etwa bei schwarzen Löchern (*Hawking-Strahlung*[13], *Bekenstein-Hawking-Entropie*[14]) oder dem hochverdichteten Zustand, der für den Beginn unseres Universums angenommen wird ('Urknall'; Physik des frühen Universums; Quantenkosmologie). Hier überlappen die Gegenstandsbereiche der beiden miteinander unverträglichen theoretischen Konstrukte, die das zur Zeit fundamentalste Instrumentarium der Physik darstellen. Gleichermassen gravitative wie Quanteneffekte spielen für diesen Bereich, wie man heute vermutet, eine entscheidende Rolle. Weder die Allgemeine Relativitätstheorie noch die Quantenfeldtheorien sind, einzeln betrachtet und alleine, hinreichend für die Beschreibung der physikalischen Probleme in diesem Überlappungsbereich ihrer Zuständigkeit. Und ihre konzeptionelle Unverträglichkeit (bzw. Unzulänglichkeit) ist spätestens hier kaum hinnehmbar.

Insbesondere sagt die Allgemeine Relativitätstheorie in beiden genannten, paradigmatischen Fällen Singularitäten voraus: Punkte, für die das Äquivalenzprinzip nicht mehr gilt, so dass die Theorie ihre Gültigkeit verliert. Dies zeigt, dass sie keine umfassende Theorie der Raumzeit liefert.

> *"Under very general conditions, it follows from general relativity that singularities are unavoidable. GR, therefore, predicts its own breakdown."* (Kiefer (2005) 1) – *"In a truly fundamental theory, there is no room for such breakdowns and it is suspected by many that the theory cures itself upon quantisation [...]."* (Thiemann (2007) 6)

Ohne eine Theorie der Quantengravitation, welche die Beschreibung der gravitativen Wechselwirkung mit der Quantenmechanik zumindest konzeptionell vereinbar und widerspruchsfrei macht, kommt man hier kaum weiter. Eine solche Theorie sollte nach heutiger Ansicht in der Lage sein, die möglichen Quanteneigenschaften der Gravitation (bzw. der dynamischen Raumzeit) zu erfassen – bzw. zu erklären, wie die Gravitation als vielleicht nur emergentes, intrinsisch klassisches Phänomen ohne eigene Quanteneigenschaften mit den Quanteneigenschaften der übrigen Wechselwirkungen und der Materie vereinbar ist.

Zudem sollte sie unter anderem auch erklären können, welche Mikrozustände zur *Bekenstein-Hawking-Entropie* führen,[15] ob und – wenn ja – wie es zur *Hawking-Strahlung* schwarzer Löcher kommt, ob diese in Durchbrechung der Unitarität der Quantenmechanik zu einem Informationsparadoxon[16] führt und was bei einer eventuellen vollständigen Zerstrahlung schwarzer Löcher passiert. In allen diesen Problempunkten ist eine Theorie erforderlich, die hinsichtlich der Beschrei-bung des Ereignishorizontes eines schwarzen Loches über klassische und quasi-klassische Ansätze hinausgeht.

---

[12] Für die Apollo-Missionen zum Mond und die dafür erforderlichen Berechnungen war sogar die Newtonsche Mechanik völlig ausreichend.

[13] Siehe Hawking (1974, 1975), Bardeen / Carter / Hawking (1973), Wald (1994, 2001).

[14] Siehe Kap. 3.1. sowie Bekenstein (1973, 1974, 1981, 2000, 2001, 2003), Wald (1994, 2001), Bousso (2002).

[15] Siehe Kap. 3.1.

[16] Siehe etwa Hawking (1976, 1982, 2005), Belot / Earman / Ruetsche (1999).

## *Nomologische Vereinigung*

Obgleich die konzeptionellen Konflikte zwischen den etablierten Theorien sowie die physikalischen Probleme, die eng mit diesen Konflikten verbunden sind, durchaus eine ausreichende Motivation für die Entwicklung einer Theorie der Quantengravitation liefern, wird manchmal – vor allem von den Vertretern der Stringtheorie[17] – auch die nomologische Vereinigung aller Wechselwirkungen als wesentliche Motivation für die Entwicklung einer solchen Theorie ins Feld geführt. Eine solche Motivation besitzt allerdings, im Vergleich zur ersteren, einen völlig anderen Status:

Die Vermeidung von Widersprüchen zwischen etablierten Theoriegebäuden ist – spätestens, wenn diese Überschneidungen in ihren Gegenstandsbereichen aufweisen – nicht nur ein wünschenswertes Ziel der Wissenschaft, sondern für die Erfassung dieser Gegenstandsbereiche unabdingbar. Solche Widersprüche sind also entweder durch eine Theorieablösung aufzuheben – oder aber im Rahmen einer theoretischen Weiterentwicklung als nur scheinbare Widersprüche verstehbar zu machen, die aus einem Missverständnis hinsichtlich der Geltungsbereiche der Vorgängertheorien resultieren.

Die Zielsetzung einer nomologischen Vereinigung aller Wechselwirkungen hat nicht diese zwingende Unmittelbarkeit. Sie beruht auf einer metaphysischen Annahme – der Idee der nomologischen Einheit der Natur –, die entweder zutreffen kann oder eben auch nicht. Sie ist keine zwingende Anforderung an physikalische Theorien. Vielmehr kann sie sich nur indirekt durch den Erfolg nomologisch vereinheitlichter Theorien als angemessen erweisen.

Eine *konzeptionelle Vereinheitlichung* unserer Theorien ist, soweit sie zur Ausräumung von tatsächlichen Widersprüchen dient, also unabdingbar. Eine *nomologische Vereinigung* aller Wechselwirkungen ist hingegen nur eine der möglichen Optionen für die Theorieentwicklung. Nur wenn sich zeigen sollte, dass sich eine konzeptionelle Vereinheitlichung unserer Theorien im Sinne der Beseitigung von Widersprüchen nur um den Preis einer nomologischen Vereinigung erreichen lässt, wäre diese nomologische Vereinigung Teil der unabdingbaren Voraussetzungen für die Entwicklung einer Theorie der Quantengravitation. Solange sich aber keine guten Gründe dafür finden lassen, dass die nomologische Vereinigung der Kräfte für die Beseitigung oder Aufhebung der Inkompatibilitäten zwischen Allgemeiner Relativitätstheorie und Quantenmechanik bzw. Quantenfeldtheorie im entstehenden Kontext einer Theorie der Quantengravitation eine entscheidende Rolle spielt oder gar unabdingbar ist, bleibt das nomologische Vereinheitlichungsprogramm nichts mehr als das Ergebnis einer metaphysischen Vorannahme.

Die bisherigen Erfolge des Vereinheitlichungskonzeptes in der Physikgeschichte ändern daran erst einmal wenig, zumal diese Erfolge auch im Bereich ausserhalb der Gravitation nur im konzeptionellen, modelltheoretischen, nicht aber im nomologischen Sinne umfassend sind: Bisher liessen sich nur die elektromagnetische und die schwache Wechselwirkung tatsächlich im Rahmen eines nomologisch einheitlichen Ansatzes (und ausschliesslich im Rahmen eines solchen Ansatzes) erfassen. Die starke Wechselwirkung bleibt mit dem bisherigen Misserfolg der *Grand Unified Theories* hier erst einmal aussen vor. Das Standardmodell der Quantenfeldtheorien ist also zwar konzeptionell einheitlich, aber nicht nomologisch vereinheitlicht. Dafür, und für die Tatsache, dass sich die Gravitation überhaupt nicht in diesem Rahmen erfassen lässt, könnte es gute Gründe geben. Die Natur könnte sich schlichtweg einer vollständigen nomologischen Vereinigung verschliessen. Über diese entscheiden jedenfalls nicht konzeptionelle Prädispositionen, sondern einzig und allein der empirische Erfolg oder Misserfolg nomologisch vereinheitlichter Theorien.[18]

---

[17] Siehe Kap. 4.2.

## 1.2.  Konzeptionelle Voraussetzungen

Dass eine physikalische Theorie, wie jede Theorie überhaupt, konzeptionell kohärent und widerspruchsfrei sein sollte, ist ebenso unstrittig wie die Tatsache, dass jede empirisch-wissenschaftliche Theorie empirisch adäquat sein muss, sich also in keinem klar etablierbaren Konflikt mit den verfügbaren empirischen Daten befinden darf. Gleichermassen unstrittig ist es wohl ebenso, dass eine neue Theorie, die dazu antritt, etablierte Theorien, aus welchen Gründen auch immer, abzulösen, über diese basalsten Anforderungen hinaus spezifische neue quantitative Vorhersagen machen muss, die sich im Rahmen der etablierten Theorien nicht erreichen lassen.[19] Diese neuen spezifischen Vorhersagen stellen die Basis für differentielle empirische Testmöglichkeiten dar, in deren Rahmen sich die neue Theorie bewähren muss; sie muss sich auch in diesem erweiterten Kontext als empirisch adäquat erweisen und damit ihre Vorteile gegenüber den etablierten Theorien unter Beweis stellen.

### *Keine empirischen Daten !*

Das in dieser Hinsicht immer noch virulente Problem aller Ansätze zu einer Theorie der Quantengravitation besteht nun darin, dass es einerseits keine empirischen Daten gibt, die zweifelsfrei über den Kontext der etablierten Theorien hinausweisen. Es gibt kein einziges unumstrittenes empirisches Faktum, welches mit den etablierten Theorien und ihren Vorhersagen und Implikationen grundsätzlich unverträglich wäre und als Anhaltspunkt für eine theoretische Weiterentwicklung dienen könnte.[20] Bislang können sich die Ansätze zur Quantengravitation also in ihrer Entwicklung ausschliesslich auf Indizien aus dem Kontext der etablierten, empirisch gestützten Vorläufertheorien berufen. – Andererseits war es bisher keinem der Ansätze zu einer Theorie der Quantengravi-tation möglich, spezifische quantitative Vorhersagen zustandezubringen, die über den Kontext der etablierten Theorien und ihrer Vorhersagen hinausgehen und insofern als Basis für einen differentiellen empirischen Test dienen könnten.

---

[18] Nicht unerwähnt bleiben sollte an dieser Stelle jedoch die Tatsache, dass es – von den in der Tradition und Erfolgsgeschichte einer nomologischen Vereinigung der Kräfte verankerten Theorieansätzen und Argumenten völlig unabhängig – innerhalb der prägeometrischen Ansätze zur Quantengravitation, wie noch zu erläutern sein wird, Argumente dafür gibt, dass die Entstehung von Raumzeit, Materie und allen Wechselwirkungen am besten aus einem gemeinsamen prägeometrischen Ursprung heraus verstanden werden kann. Siehe Kap. 4.6. Diese Argumente legen – für den Fall einer emergenten Raumzeit – für jede fundamentale Theorie eine vereinheitlichte Beschreibung von Raumzeit und Materie nahe. Die tatsächliche Relevanz dieser Argumente hängt allerdings davon ab, ob die Raumzeit tatsächlich eine emergente Grösse ist. Und dies lässt sich am besten am langfristigen konzeptionellen und schliesslich vor allem empirischen Erfolg bzw. Misserfolg der prägeometrischen Theorieansätze ablesen.

[19] Unstrittig ist dies wohl zumindest unter allen denen, die sich nicht entschlossen haben, die Stringtheorie ausschliesslich aufgrund von Konsistenzargumenten zur einzig akzeptablen fundamentalen Theorie zu küren.

[20] Die intensive Suche nach relevanten empirischen Daten, die über den Rahmen der etablierten Theorien hinausweisen, ist natürlich im Gange. Siehe etwa Giulini / Kiefer / Lämmerzahl (2003), Amelino-Camelia (2008), Amelino-Camelia / Smolin (2009).

Vor diesem Hintergrund bleibt die erst einmal augenscheinliche konzeptionelle Inkompatibilität zwischen Allgemeiner Relativitätstheorie und Quantenmechanik bzw. Quantenfeldtheorien, gemeinsam mit den aus ihr erwachsenden konkreten und bisher ungeklärten physikalischen Problemlagen, die einzige zwingende Motivation für eine Theorieentwicklung im Bereich der Quantengravitation. Die Ausgangsbasis für die Theorieentwicklung erschöpft sich in den etablierten Vorgängertheorien und den empirischen Daten, auf die diese sich schon stützen. Ob dies als Ausgangsbasis ausreicht, ist immerhin nicht unumstritten:

> *"Some critics have argued that the [quantum gravity] search is futile, because anything might happen in [quantum gravity] regimes, at scales far removed from our experience. Maybe the search is impossible because the space of possible theories is too large. / At present, this worry is probably unjustified. If this were the problem, we would have plenty of complete, predictive and coherent theories of [quantum gravity], and the problem would be the choice among them. Instead, the situation is the opposite: we haven't any. The fact is that we have plenty of information about [quantum gravity], because we have [quantum mechanics] and we have [general relativity]. Consistency with [quantum mechanics] and [general relativity], plus internal consistency, form an extremely strict set of constraints."* (Rovelli (2007) 1303)

Angesichts der konzeptionellen Konfliktpunkte innerhalb des Spektrums der etablierten Theorien bleibt aber kaum die Option, die Augen zu verschliessen und alle auf eine Theorie der Quantengravitation abzielenden Ambitionen vorerst ruhen zu lassen, bis konkrete empirische Befunde vielleicht irgendwann zu einer besseren Ausgangsbasis führen. Und diese Optionslosigkeit ist sicherlich erst einmal unabhängig davon, ob man die folgende Hoffnung teilt oder nicht:

> *"[...] hope that the knowledge of the world coded into [general relativity] and [quantum mechanics] can be a good guide for finding a theory capable of describing physical regimes that we have not yet explored."* (Rovelli (2007) 1305)

Die Anforderung der empirischen Adäquatheit reduziert sich jedenfalls bei dieser restriktiven Ausgangsbasis erst einmal darauf, dass die Theorieansätze zur Quantengravitation den phänomenalen Gehalt der etablierten Theorien zumindest soweit reproduzieren, dass keine Konflikte mit den empirischen Daten entstehen, auf die sich die etablierten Theorien schon stützen. Dies lässt sich im Kern wiederum erst einmal auf die Anforderung reduzieren, dass die neuen Theorieansätze diese etablierten Theorien zumindest als Grenzfall oder Näherung enthalten müssen. Darüber hinaus sollten sie zumindest mittelfristig zu spezifischen quantitativen Vorhersagen führen, die sich von denen der etablierten Theorien unterscheiden, nicht aber mit schon bestehender Empirie in Konflikt geraten. Dies ist, wie gesagt, bisher noch nicht gelungen.

Die gesamte Diskussion um eine Theorie der Quantengravitation bleibt, bis sich dies ändert, erst einmal im ausschliesslich konzeptionellen Rahmen gefangen. Dieser konzeptionelle Rahmen wiederum wird vorrangig durch die konzeptionellen Grundannahmen und Implikationen der etablierten Theorien abgesteckt. Und die möglichen Antworten auf die Frage, welche konzeptionellen Anforderungen tatsächlich an eine Theorie der Quantengravitation zu stellen sind, betreffen dann, soweit sie über die elementaren Forderungen konzeptioneller Kohärenz und Widerspruchsfreiheit sowie empirischer Adäquatheit hinausgehen, vor allem die jeweilige Einschätzung, welche der konzeptionellen oder modelltheoretischen Implikationen der Vorläufertheorien und welche der basalen physikalischen Grundüberzeugungen, die in ihnen zum Ausdruck kommen, für den Bereich der Quantengravitation weiterhin als relevant oder gar essentiell erachtet werden.

Infolge der konzeptionellen Unverträglichkeiten innerhalb des Spektrums der etablierten Theorien kann diese Relevanzfrage immer nur zu einer Auswahl konzeptioneller Komponenten führen; in ihrer Gesamtheit führen sie eben gerade zu den konstatierten Unverträglichkeiten. Für die Auswahl aber gibt es keine apriori gültigen objektiven Kriterien. Hier kommen in nicht unerheblicher Weise Komponenten ins Spiel, die mit den unterschiedlichen Vorerfahrungen der an den jeweiligen Theorieansätzen arbeitenden Forscher zusammenhängen. Dies führt dazu, dass es hinsichtlich der Relevanzfrage keinen durchgängigen Konsens zwischen den verschiedenen Theorieansätzen gibt. Wenngleich einige der zentralen konzeptionellen Implikation der Vorläufertheorien innerhalb eines grösseren Spektrums der Theorieansätze zur Quantengravitation weitgehend unstrittig sind, bleibt dennoch ein erheblicher Freiraum.

Dieser Freiraum wird in einem nicht unerheblichen Ausmass mit zum Teil durchaus einleuchtenden, zum Teil aber auch durchaus problematischen metaphysischen Hintergrundüberzeugungen angereichert, die sich oft nur sehr unzureichend auf stringente physikalische Motivation oder gar empirische Belege stützen lassen. Die schon erwähnte Problematik der nomologischen Vereinigung der Wechselwirkungen spielt hier gleichermassen eine Rolle wie etwa die jeweiligen Positionen hinsichtlich der um den Status der Raumzeit geführten Substantialismus-Relationalismus-Debatte. Es geht hier nicht zuletzt um die folgenden Fragen: Ist die Hintergrundunabhängigkeit der Allgemeinen Relativitätstheorie notwendigerweise konstitutiver Teil der konzeptionellen Basis für die Entwicklung einer Theorie der Quantengravitation? Ist die Raumzeit eine fundamentale Entität oder ein emergentes Phänomen? Ist sie eine Substanz oder ein relationales Konstrukt? Wenn sie eine Substanz sein sollte: Verfügt sie über Quanteneigenschaften? Wenn sie emergent sein sollte: Beruht sie auf einem Quantensubstrat oder etwas völlig anderem? Muss eine Theorie der 'Quantengravitation' notwendigerweise eine Quantentheorie sein? Muss es sich notwendigerweise um eine nomologisch oder ontologisch vereinheitlichte Theorie handeln? – Problematisch werden diese Fragestellungen spätestens dann, wenn sie für die einzelnen Beteiligten schon im Vorhinein entschieden sind und die entsprechenden Entscheidungen gar ohne explizite Thematisierung in die Theorieentwicklung eingehen.

Trotz des damit einhergehenden Gefahrenpotentials, das leicht zu einer radikalen Verunsicherung führen kann, und trotz der Berechtigtheit dieser Verunsicherung soll hier nun dennoch – im Sinne einer ersten vorsichtigen Annäherung, die im weiteren Verlauf dann zu differenzieren sein wird[21] – der Frage nach den möglichen konzeptionellen Voraussetzungen für die Theorieentwicklung im Bereich der Quantengravitation nachgegangen werden: Welche Elemente bestehender physikalischer Theorien lassen sich mit einer gewissen Berechtigung tatsächlich als relevant für die Entwicklung und Formulierung einer Theorie der Quantengravitation ansehen und sollten insofern erst einmal in die Theorieentwicklung Eingang finden, zumindest so lange, bis sich eventuell explizite Gründe gegen sie auftun? Von welchem konzeptionellen oder modelltheoretischen Ausgangspunkt könnte die Theorieentwicklung starten? Welche physikalischen Ideen, Konzepte, Prinzipien, An-sätze etc. könnten sich als heuristisch relevant oder zumindest vorteilhaft erweisen? Welche könnten sich gar als essentiell herausstellen? Und um die Problematik auf den Punkt zu bringen: Welche Elemente der Vorgängertheorien stellen vermutlich echte physikalische Einsichten dar, die vielleicht auch den nächsten Schritt in der Entwicklung physikalischer Theorien unbeschadet überstehen könnten?

---

[21] Eine detailliertere Behandlung dieser Problematik wird in Kap. 2. folgen.

Ohne dass alle diese Fragen nun schon beantwortet werden sollten oder könnten, ist eine partielle und vor allem vorläufige Festlegung zumindest hinsichtlich einiger der angerissenen Fragestellungen sicherlich unumgänglich, um überhaupt zu einer Ausgangsbasis für die Theorieentwicklung im Bereich der Quantengravitation zu gelangen. Eine solche Festlegung kann im Rahmen eines ausschliesslich konzeptionellen Diskurses immer nur mit Vorbehalt erfolgen – einerseits mit der Grundüberzeugung, dass konkretere konzeptionelle Randbedingungen für die Theorieentwicklung unabdinglich sind, und andererseits mit der Hoffnung, dass die damit erreichte Konkretisierung, wenn sie in die falsche Richtung führen sollte, immerhin die Gelegenheit bietet, aus den entsprechenden Fehleinschätzungen etwas zu lernen. Erst dies führt zu einer ausreichenden Motivationslage für eine Modifikation hinsichtlich der konzeptionellen Grundüberzeugungen. Ohne das konkrete Austesten der jeweiligen Implikationen wird dies nur schwerlich möglich sein.

Eine unter diesen Vorbehalten formulierte, erste, vorsichtige, konzeptionell breit angelegte Annäherung an eine mögliche konzeptionelle Ausgangsbasis für die Theorienbildung soll im folgenden nicht zuletzt dem Ziel dienen, zu einer vorläufigen Minimaldefinition dessen zu gelangen, was unter einer 'Theorie der Quantengravitation' (alles) zu verstehen sein könnte. Und sie sollte daher (soweit absehbar) alle wesentlichen Alternativen und Optionen offen lassen:

## *Konzeptionelle Extrapolationen*

Die Allgemeine Relativitätstheorie[22] ist, soweit es den Kontext der empirisch bestens bestätigten Theorien betrifft, die fundamentalste Theorie der Gravitation und der Raumzeit, über die wir zur Zeit verfügen. Solange es keine zwingenden Gründe gibt, im Rahmen einer fundamentaleren Theorie hinter die zentralen Einsichten der Allgemeinen Relativitätstheorie zurückzufallen, sollten diese bei der Entwicklung einer Theorie der Quantengravitation sicherlich erst einmal Berücksichtigung finden.

Eine der diesbezüglich wichtigsten Implikationen der Allgemeinen Relativitätstheorie besteht in der Identifizierung des Gravitationsfeldes mit der Metrik der Raumzeit. Mit dieser Geometrisierung der Gravitation geht die Einsicht einher, dass das Gravitationsfeld, im Gegensatz zu allen anderen Wechselwirkungs- und Materiefeldern, keine Entität in bzw. auf der Raumzeit ist, sondern vielmehr eine Manifestation der hier nun dynamisierten Raumzeit selbst. Es lässt sich folglich auch nicht auf einer schon vorgegebenen Raumzeit mit festgelegter Metrik – auf einem vorgegeben raumzeitlichen Hintergrund – erfassen und beschreiben. Diese Implikation der Geometrisierung der Gravitation liesse sich als ein erster, basaler Aspekt der *Hintergrundunabhängigkeit* der Allgemeinen Relativitätstheorie bezeichnen.

---

[22] Für eine über diese erste Annäherung hinausgehende, detailliertere Behandlung der aus der Allgemeinen Relativitätstheorie ableitbaren Implikationen für eine Theorie der Quantengravitation, siehe Kap. 2.1.

Ein weiterer zentraler Aspekt der mit der Allgemeinen Relativitätstheorie einhergehenden Hintergrundunabhängigkeit resultiert aus ihrer aktiven Diffeomorphismusinvarianz als physikalisch wirksamer, substantieller Komponente der allgemeinen Kovarianz, die wiederum zu den grundlegendsten Voraussetzungen der Theorie und ihrer Entwicklung zählt. Die Diffeomorphismusinvarianz der Theorie erschöpft sich nicht in einer simplen Koordinatentransformationsinvarianz, sondern schliesst die Invarianz gegenüber Punkttransformationen innerhalb der raumzeitlichen Mannigfaltigkeit ein. Dies lässt sich unter bestimmten Bedingungen hin auf eine relationalistische Auffassung der Raumzeit interpretieren.[23] Wenn diese Charakteristika beim Übergang von der Allgemeinen Relativitätstheorie zu einer Theorie der Quantengravitation erhalten bleiben sollten, ist auch für diese möglicherweise nicht nur von einer dynamischen, sondern vielleicht sogar von einer relationalen Raumzeitkonzeption auszugehen. Beide Eigenschaften sind ohne weiteres mit der Möglichkeit verträglich, dass es sich bei der Raumzeit um ein emergentes Phänomen handelt, welches seine Dynamik aus einer prägeometrischen Substratdynamik bezieht und seine ausschliesslich relationale Struktur durch die Konstituenten der Substratdynamik oder die einer intermediären Schicht erhält, welche die phänomenologische Raumzeit als solche erzeugt.[24]

Die Quantenmechanik[25] stellt die fundamentalste Theorie des dynamischen Verhaltens von Materie und Feldern dar, über die wir zur Zeit verfügen. Sie konnte sich bisher in allen bekannten empirischen Tests unumstritten behaupten. Dies lässt die Annahme ihrer universellen Gültigkeit zumindest als nicht völlig unvernünftig erscheinen.[26] Und unter diesen Bedingungen erscheint es auf den ersten Blick ebensowenig unvernünftig anzunehmen, dass auch das Gravitationsfeld, wie alle anderen Felder, über Quanteneigenschaften verfügt, die im Rahmen der Allgemeinen Relativitätstheorie als klassischer Theorie der Gravitation eben noch nicht erfasst werden.

Deutlich stringenter als diese Intuition legen allerdings die zuvor schon erwähnten Argumente gegen semi-klassische Theorien der Gravitation eine Quantisierung des Gravitationsfeldes nahe: Erweiterungen der Allgemeinen Relativitätstheorie, die im Sinne eines Hybridkonstruktes etwa quantenmechanische Erwartungswerte für Materie- und Wechselwirkungsfelder in den Energie-Spannungstensor einfliessen lassen, die Gravitation aber klassisch behandeln, sind nicht nur konzeptionell unschön, sondern sie führen zu direkten Widersprüchen.[27] Die Argumente gegen semi-klassische Theorien der Gravitation schliessen ein fundamentales klassische Gravitationsfeld in einer ansonsten umfassenden Quantenwelt aus.

Aber diese Argumente schliessen – auch unter der Annahme der universellen Gültigkeit der Quantenmechanik und im Kontrast zu der auf dieser Annahme beruhenden anfänglichen Intuition, dass auch das Gravitationsfeld, wie alle Wechselwirkungsfelder, Quanteneigenschaften haben sollte – nicht per se die Möglichkeit aus, dass es sich bei der Gravitation um ein intrinsisch klassisches Phänomen handelt. Sie schliessen nur aus, dass es sich, wenn die Quantenmechanik universell gültig sein sollte, bei der Gravitation um eine intrinsisch klassische und gleichzeitig fundamentale Wechselwirkung handelt. Sollte die Gravitation ein intrinsisch klassisches Phänomen sein, so müsste sie eine residuale Wechselwirkung sein: ein emergentes Phänomen. Auf einer fundamentaleren Ebene würde sie gar nicht vorkommen. Sie wäre nicht Teil des umfassenden Quantensubstrats, sondern würde von diesem in klassischer, makroskopischer Näherung als residuales Phänomen hervorgebracht. Sie wäre eine mehr oder weniger indirekte Folge von fundamentaleren Wechselwirkungen[28] und ihren jeweiligen Freiheitsgraden. Somit gäbe es auch keine Widersprüche mit einer eventuellen Hybriddynamik. Das Quantensubstrat selbst würde über gar keine gravitativen Freiheitsgrade verfügen.

---

[23] In der Allgemeinen Relativitätstheorie kann die raumzeitliche Mannigfaltigkeit aufgrund der substantiell interpretierbaren Diffeomorphismusinvarianz nicht (oder nur um den hohen Preis unmotivierter metaphysischer Annahmen) als substantielle Entität angesehen werden. Die damit einhergehende komplexe Problemlage wird in Kap. 2.1. ausführlich zu diskutieren sein. Siehe auch Earman (1986, 1989, 2002, 2006, 2006a), Earman / Norton (1987), Norton (1988, 1993, 2004).

[24] Siehe hierzu insbesondere Kap. 4.6.

[25] Für eine über diese erste Annäherung hinausgehende, detailliertere Behandlung der aus der Quantenmechanik ableitbaren Implikationen für eine Theorie der Quantengravitation, siehe Kap. 2.2.

[26] Weiteren Aufwind bekommt diese Annahme nicht zuletzt dadurch, dass inzwischen eine ganze Reihe makroskopischer Quantenphänomene und –systeme bekannt sind und experimentell erforscht wurden.

[27] Siehe etwa Kiefer (1994, 2004 [Kap. 1.2.], 2005), Peres / Terno (2001), Terno (2006), Callender / Huggett (2001a, 2001b).

Sollte die mit der Allgemeinen Relativitätstheorie einhergehende Geometrisierung der Gravitation dennoch einen physikalisch wesentlichen Sachverhalt und vielleicht sogar in indirekter Weise einen Charakterzug der Substratebene erfassen, so könnte dies bedeuten, dass es auf der Ebene des Quantensubstrats nicht einmal raumzeitliche Freiheitsgrade gibt, so dass auch die Raumzeit als makroskopisches, intrinsisch klassisches, emergentes Phänomen zu verstehen wäre, welches im Rahmen der gleichen Näherung bzw. des gleichen klassischen Grenzfalles zustandekäme, der auch die Phänomenologie der Gravitation hervorbringt.

Unter den Bedingungen einer emergenten Gravitation wäre eine Theorie, welche die Gravitation als ausschliesslich klassisches Phänomen beschreibt, als effektive Theorie anzusehen, die sich mit der intrinsisch klassischen Dynamik von kollektiven Freiheitsgraden beschäftigt. Das gleiche gilt für eine Theorie, welche Gravitation und Raumzeit als miteinander verbundene klassische Phänomene beschreibt, für den Fall, dass beide emergente Phänomene sein sollten, die auf der Ebene des Quantensubstrats gar nicht vorkommen. Die Allgemeine Relativitätstheorie wäre im Kontext der Emergenzszenarien eine solche Theorie. Sie müsste sich unter diesen Bedingungen aus einer fundamentaleren Theorie, welche die Dynamik auf der Substratebene beschreibt, als Näherung oder Grenzfall ableiten lassen.

Sollte die Gravitation jedoch, wie gemeinhin stillschweigend angenommen wird, eine fundamentale Wechselwirkung sein und sollte die Quantenmechanik universell gültig sein, so wäre davon auszugehen, dass das Gravitationsfeld tatsächlich über Quanteneigenschaften verfügt, die bisher, im Rahmen der Allgemeinen Relativitätstheorie, noch keine Berücksichtigung erfahren haben. Das Gravitationsfeld wäre zu 'quantisieren'. Was das im einzelnen bedeuten könnte, wäre zu klären: insbesondere, ob eine 'Quantisierung' des Gravitationsfeldes schon notwendigerweise mit einer direkten Quantisierung der Allgemeinen Relativitätstheorie gleichzusetzen wäre, und, wenn ja, welche Quantisierungsmethode dem Problem angemessen wäre. Eine Quantisierung analog zu der des elektromagnetischen Feldes im Kontext der Quantenelektrodynamik kommt – wie sich leicht einsehen lässt und noch explizit aufzuzeigen sein wird – aufgrund des Konfliktes zwischen fundamentalen Einsichten der Allgemeinen Relativitätstheorie und dem modelltheoretischen Instrumentarium der herkömmlichen Quantenfeldtheorien wohl kaum in Frage; die Problematik der Hintergrundunabhängigkeit des Gravitationsfeldes und der damit unverträglichen Hintergrundabhängigkeit des quantenfeldtheoretischen Instrumentariums wurde im Vorausgehenden schon angedeutet.

---

[28] Das heisst noch nicht, dass es sich bei diesen fundamentaleren Wechselwirkungen um die bekannten nicht-gravitativen Wechselwirkungen handeln muss.

Sollte die Allgemeine Relativitätstheorie einen für eine Quantisierung der Gravitation geeigneten Ausgangspunkt liefern – sollte sie also zumindest die relevanten klassischen Eigenschaften des Gravitationsfeldes angemessen erfassen –, so würde die mit ihr einhergehende Identifizierung des Gravitationsfeldes mit Eigenschaften einer dynamischen Raumzeit bedeuten, dass eine Quantisierung des Gravitationsfeldes einer Quantisierung dieser dynamischen Raumzeit bzw. ihrer Metrik entspräche.

> *"[...] general relativity is not just a theory of gravity – in an appropriate sense, it is also a theory of spacetime itself; and hence a theory of quantum gravity must have something to say about the quantum nature of space and time."* (Butterfield / Isham (2001) 34)

Die Quantisierung der Gravitation würde zu einer 'Quantengeometrie' führen. – Was dies bedeuten könnte, wäre wiederum im einzelnen zu klären: Würde die so entstehende Theorie, um konzeptionell und schliesslich auch empirisch erfolgreich zu sein, etwa die Metrik der Raumzeit als Erwartungswert einer Quantenvariablen beschreiben? Hätte man es mit Unschärfen der Raumzeit oder mit Superpositionen von Raumzeiten zu tun, mit Quantenfluktuationen der Metrik oder der gesamten raumzeitlichen Geometrie, vielleicht sogar der Topologie der Raumzeit?

Wie dem auch immer sein sollte: Eine solche Strategie für die Entwicklung einer Theorie der Quantengravitation, die ihren Ausgangspunkt von einer direkten Quantisierung der Allgemeinen Relativitätstheorie nimmt – mit der Absicht, die Quanteneigenschaften der Gravitation und der Raumzeit zu erschliessen und zu beschreiben –, kann überhaupt nur dann erfolgreich sein, wenn die Gravitation kein intrinsisch klassisches, emergentes Phänomen ist, sondern zu den fundamentalen Wechselwirkungen zählt und tatsächlich über bisher noch nicht erschlossene Quanteneigenschaften verfügt. Aber auch in diesem Fall garantiert die Strategie einer direkten Quantisierung der Allgemeinen Relativitätstheorie noch nicht selbstredend den Erfolg. Beschreitbare Wege zur Erschliessung möglicher Quanteneigenschaften der Gravitation könnten unter Umständen gänzlich andere Strategien erfordern.

Sollte aber die Gravitation emergent sein, so entspräche eine Quantisierung der Allgemeinen Relativitätstheorie, auch wenn diese die Gravitation als intrinsisch klassisches Phänomen adäquat beschreiben sollte, letztlich immer nur einer Quantisierung kollektiver, makroskopischer Freiheitsgrade, die auf der Grundlage eines gänzlich anders gearteten Substrats und grundsätzlich anders gearteter fundamentaler Freiheitsgrade zustandekommen. Es wären schlicht die falschen Freiheitsgrade, die hier quantisiert würden. Eine angemessene Theorie der Quantengravitation wäre nicht über die Quantisierung der Allgemeinen Relativitätstheorie zu erreichen. Das Ziel einer Theorie, die den Motivationen gerecht wird, die den Ansätzen zur Quantengravitation zugrundeliegen, müsste vielmehr die Identifizierung des Quantensubstrats, seiner Freiheitsgrade und seiner Dynamik sein. Das Zustandekommen der Gravitation und ihrer empirisch nachweislichen Dynamik als emergentes, intrinsisch klassisches Phänomen müsste auf dieser Grundlage erklärt werden. – Dies wäre gleichzeitig eine der fundamentalsten konzeptionellen Anforderungen an eine solche Substrattheorie.

Ob eine solche Theorie noch unter der Bezeichnung 'Quantengravitation' formieren sollte, ist dann letztlich nichts anderes als eine Frage der Konvention. Hier soll nun im folgenden – im Sinne der schon angekündigten *Minimaldefinition* – die Bezeichnung 'Theorie der Quantengravitation' für jede Theorie Anwendung finden, die sich mit ausreichenden Erfolgsaussichten anschickt, die kon-zeptionelle Unverträglichkeit zwischen Allgemeiner Relativitätstheorie und Quantenmechanik bzw. Quantenfeldtheorie zu überwinden, unabhängig davon, ob es sich um eine tatsächliche oder nur um eine scheinbare Unverträglichkeit handelt, und ebenso unabhängig davon, ob die Gravitation nun über Quanteneigenschaften verfügt oder ob sie ein intrinsisch klassisches Phänomen darstellt. Dies schliesst pro forma letztlich sogar eine Nicht-Quantentheorie ein, die sich diesem Ziel mit Aussicht auf Erfolg widmet.[29]

<div align="center">*</div>

Unter Berücksichtigung aller soweit erfassten Möglichkeiten wäre eine *Theorie der Quantengravitation'* also eine Theorie, welche die Dynamik (und möglicherweise die Emergenz) der Gravitation (und möglicherweise der Raumzeit) auf eine Art und Weise erklärt, die es möglich macht, die (tatsächliche oder nur scheinbare) wechselseitige konzeptionelle Unverträglichkeit zwischen Allgemeiner Relativitätstheorie und Quantenmechanik bzw. Quantenfeldtheorie zu überwinden oder aufzuheben. In jedem Falle wäre es eine Theorie, welche die der Gravitation (und möglicherweise auch der Raumzeit) zugrundeliegende Substratdynamik beschreiben würde, unabhängig davon, ob auf der Substratebene gravitative und/oder raumzeitliche Freiheitsgrade existieren, und ebenso unabhängig davon, ob es sich um ein Quantensubstrat oder etwas gänzlich anders Geartetes handeln sollte. Alle diese Optionen bleiben erst einmal offen. Die Gravitation könnte eine fundamentale Wechselwirkung oder eben auch ein emergentes Phänomen sein. Sie könnte über noch unerschlossene Quanteneigenschaften verfügen oder eben auch nicht. Die Raumzeit könnte eine substantielle Entität oder ein relationales Konstrukt sein; sie könnte schon auf der Substratebene in der einen oder anderen Form (vielleicht als Quantengeometrie) eine Rolle spielen oder eben auch ein emergentes und vielleicht intrinsisch klassisches Phänomen ohne quantengeometrische Substruktur sein. Eine sich schliesslich auch als empirisch adäquat erweisende Theorie der Substratdynamik könnte eine Quantentheorie im weitestgehend herkömmlichen Sinne sein oder eben auch etwas gänzlich anderes.

Eine solche Theorie müsste aber in jedem Falle die Bedingungen, unter denen sich die etablierten Beschreibungen als angemessen erwiesen haben, als (makroskopische oder niederenergetische) Näherung oder als (klassischen oder statistischen) Grenzfall enthalten, sie müsste den empirischen Gehalt dieser etablierten, empirisch gut bestätigten Beschreibungen reproduzieren, und sie müsste erklären, auf Grund welcher konzeptionellen Erweiterung oder auf Grund welcher Modifikation sich deren wechselseitige Unverträglichkeit aufheben lässt, etwa indem sie Grenzen der jeweiligen Geltungsbereiche aufzeigt. Darüberhinausgehend müsste sie schliesslich zu spezifischen, neuen, quantitativen, empirisch überprüfbaren Vorhersagen gelangen, die erst eine unabhängige, über die Vorgängertheorien und ihren empirischen Gehalt hinausgehende empirische Bestätigung möglich machen. – Eine solche Minimaldefinition lässt ein grosses Spektrum an Strategien und Ausgangspunkten für eine mögliche 'Theorie der Quantengravitation' zu.[30] Die verschiedenen Theorieansätze unterscheiden sich dabei, wie schon angedeutet, vor allem hinsichtlich der spezifischen Komponenten aus den etablierten Vorgängertheorien, die jeweils für den Bereich der 'Quantengravitation' als weiterhin relevant, essentiell oder gar unabdingbar angesehen werden und schliesslich in die Theoriebildung einfliessen.

---

[29] Beispiele für solche Ansätze werden zumindest kurz in den Kap. 3.3. und 4.6. zu erörtern sein. Es ist sicherlich Geschmackssache, ob man einen nicht-quantenphysikalischen Ansatz noch unter dem Überbegriff der 'Quantengravitation' fassen möchte. Die Alternative bestände darin, die oben angegebene Minimaldefinition mit Ausnahmeklauseln zu erweitern. Wenn es weniger um Bezeichnungen als um die Sache selbst gehen soll, erscheint dies jedoch wenig motiviert.

[30] Dieses Spektrum wird als solches in Kap. 1.3. zu systematisieren und in Kap. 4. hinsichtlich seiner konkreten Instantiierungen zu erörtern sein. In Kap. 4. stehen dabei die jeweiligen raumzeitlichen Konzeptionen im Vordergrund.

## *Elemente des Übergangs*

Im Vorausgehenden wurde behauptet, dass die etablierten, aber augenscheinlich erst einmal wechselseitig unverträglichen Vorgängertheorien die einzigen konkreten konzeptionellen Anhaltspunkte für die Entwicklung einer Theorie der Quantengravitation liefern. Dies ist bei genauerer Betrachtung nicht ganz richtig. Hinzu kommen nämlich, als potentielle Komponenten eines Ausgangspunktes für die weitere Theorieentwicklung, Ideen und Konzeptionen, die zwar aus dem Kontext der etablierten Theorien heraus motivierbar sind, aber eigentlich schon nicht mehr ganz zu diesem Kontext gehören; man könnte sie als so etwas wie 'Elemente des Übergangs' bezeichnen.[31] Was diese für die in Entwicklung befindlichen Theorien der Quantengravitation – mindestens in heuristischer Hinsicht, wenn nicht gar als Komponenten, die unmittelbar in die Theorieentwicklung einfliessen – interessant macht, ist vor allem Tatsache, dass diese 'Elemente des Übergangs' schon so etwas wie das Ergebnis eines Brückenschlages über die Grenzen der einzelnen, miteinander konzeptionell unverträglichen Theoriengebäude hinweg sind: Implikationen der Allgemeinen Relativitätstheorie finden gemeinsam mit solchen der Quantenmechanik bzw. der Quantenfeldtheorien Berücksichtigung. Weitere Komponenten, die weder dem einen noch dem anderen dieser widerstreitenden Theorienkomplexe entstammen, kommen meist noch hinzu.

Insbesondere die Thermodynamik schwarzer Löcher liefert in paradigmatischer Weise solche Elemente des Übergangs, die über den Kontext der etablierten Theorien hinausweisen. Von zentraler Bedeutung ist dabei die *Bekenstein-Hawking-Entropie* schwarzer Löcher, die sich aus einer gemeinsamen Berücksichtigung von Implikationen der Allgemeinen Relativitätstheorie und der Quantenmechanik bzw. der Quantenfeldtheorien im Verbund mit Überlegungen aus den Bereichen der Thermodynamik und der Informationstheorie ergibt.[32] Gemeinsam mit der ebenso im Rahmen der Thermodynamik schwarzer Löcher motivierbaren, aber in ihrer Bedeutung über diesen Kontext hinausreichenden *Holographischen* bzw. *Kovarianten Entropiegrenze*,[33] weist sie direkt auf eine diskrete Struktur auf der *Planck-Ebene*[34] hin. Dies lässt sich entweder – wenn es sich bei der Raumzeit um eine fundamentale Grösse handeln sollte – als Hinweis auf eine diskrete Raumzeitstruktur deuten, oder – wenn die Raumzeit nicht fundamental sein sollte – als Hinweis auf eine diskrete Substratstruktur, aus der heraus die Raumzeit als emergentes makroskopisches Phänomen zustandekommt. In jedem Fall deuten die sich im Rahmen der Thermodynamik schwarzer Löcher abzeichnenden, aber letztlich über diesen Kontext hinausweisenden Einsichten auf ein Substrat hin, welches nur über eine finite Zahl von physikalisch wirksamen Freiheitsgraden pro finiter raum-

---

[31] Für eine über diese erste Annäherung hinausgehende, detailliertere Behandlung, siehe Kap. 3.

[32] Siehe Kap. 3.1.

[33] Siehe Kap. 3.1.

[34] Die Planck-Ebene – die raumzeitliche Grössenordnung und der Energiebereich, für den die Gravitation etwa die gleiche Stärke wie die übrigen Wechselwirkungen aufweisen sollte – wird von nahezu allen diesbezüglichen Theorieansätzen als der originäre Bereich der Quantengravitation angesehen. Die Planck-Grössen ergeben sich als Kombination grundlegender Naturkonstanten:

Planck-Länge:  $l_P = (\hbar \cdot G / c^3)^{1/2} = 1{,}62 \cdot 10^{-35}$ m

Planck-Zeit:  $t_P = (\hbar \cdot G / c^5)^{1/2} = 5{,}40 \cdot 10^{-44}$ sec

Planck-Masse bzw. -Energie:  $m_P = (\hbar \cdot c / G)^{1/2} = 2{,}17 \cdot 10^{-5}$ g $= 1{,}22 \cdot 10^{19}$ GeV$/c^2$

($\hbar$: Plancksches Wirkungsquantum, G: Gravitationskonstante, c: Lichtgeschwindigkeit)

zeitlicher Region verfügt, unabhängig davon, ob diese raumzeitliche Region schon auf der Substrat-
ebene existiert oder, von dieser induziert, auf einer höheren Emergenzebene hervorgebracht wird,
also unabhängig davon, ob auf der Substratebene quantengeometrische oder ausschliesslich prä-
geometrische Freiheitsgrade bestimmend sind.

Für eine solche diskrete Substratstruktur sprechen vielleicht auch schon die in der Allgemeinen
Relativitätstheorie auftretenden (und deren modelltheoretischen, differentialgeometrischen Apparat
transzendierenden) Singularitäten, ebenso wie die Divergenzen, die für kleine Abstände bzw. hohe
Energien in den Quantenfeldtheorien auftreten; beide liessen sich möglicherweise als artifizielle
Konsequenzen der Kontinuumsannahme (bzw. der Annahme einer infiniten Zahl physikalisch rele-
vanter Freiheitsgrade pro raumzeitlicher Region) deuten, die hier an ihre Grenzen stösst.[35]

Obwohl die Allgemeine Relativitätstheorie ebenso wie die Quantenfeldtheorien von einem raum-
zeitlichen Kontinuum und einer infiniten Zahl von physikalisch relevanten Freiheitsgraden pro
raumzeitlicher Region ausgehen,[36] deutet sich interessanterweise bisher in nahezu allen Theorie-
ansätzen zur Quantengravitation eine diskrete Substratstruktur an.[37] Dies gilt nicht nur die radikale-
ren Ansätze, sondern insbesondere auch für die äusserst konservativ angelegten, in die in vielfälti-
ger Weise konzeptionelle Elemente aus den kontinuumsgestützten Vorgängertheorien Eingang fin-
den. Als paradigmatisches Beispiel für einen solchen konservativen Ansatz, der die grundsätzlichen
Einsichten der Allgemeinen Relativitätstheorie wie der Quantenmechanik als konstitutive Elemente
für die Theorieentwicklung im Bereich der Quantengravitation ansieht, lässt sich die *Loop Quantum
Gravity* anführen.[38] Sie ist das Ergebnis einer direkten, nicht-perturbativen Quantisierung der All-
gemeinen Relativitätstheorie; und sie führt dennoch zu einer diskreten Spinnetzstruktur, die in die-
sem Kontext als Quantensubstruktur der klassischen makroskopischen Raumzeit verstanden wird.

## *Extrapolationsgefahren und die Notwendigkeit eines Pluralismus der Theorieansätze*

Angesichts des rein konzeptionellen Rahmens, in dem sich die Entwicklung möglicher Theorien der
Quantengravitation infolge des Fehlens konkreter empirischer Anhaltspunkte immer noch bewegt,
soll hier schliesslich noch einmal explizit auf eine mögliche Gefahrenlage und ihre Konsequenzen
hingewiesen werden. Bei einer Theorieentwicklung, die sich ausschliesslich auf die Anforderungen
der Kohärenz und Widerspruchsfreiheit sowie auf die Notwendigkeit stützten kann, den empiri-
schen Gehalt der etablierten Theorien zu reproduzieren, und die dabei, um überhaupt zu konkreten

---

[35] Siehe auch Crane (2007). Dafür, dass sich Indizien für eine diskrete Raumzeit aus einer gemeinsamen Berücksichti-
gung quantenmechanischer Unschärfe und des im Rahmen der Allgemeinen Relativitätstheorie möglichen Gravita-
tionskollapses ableiten lassen, plädieren Calmet / Graeser / Hsu (2005). Als weitere mögliche Konsequenz einer über-
beanspruchten Kontinuumsannahme liesse sich die Nichtrenormierbarkeit der kovarianten Quantisierung der Allgemei-
nen Relativitätstheorie nennen. Siehe Kap. 4.1.

[36] Die in der Allgemeinen Relativitätstheorie verwendete Differentialgeometrie setzt notwendigerweise ein Raumzeit-
kontinuum voraus; Quantenfelder, die selbst wiederum eine infinite Zahl von Freiheitsgraden repräsentieren, setzen
einen kontinuierlichen raumzeitlichen Hintergrund voraus, auf dem sie definiert werden.

[37] Bei den Ansätzen, die von einer fundamentalen Raumzeit ausgehen, deren Quanteneigenschaften im Rahmen einer
Theorie erschlossen werden sollen, die gerade in diesem Punkt über die Allgemeine Relativitätstheorie hinausgeht, ist
dies eine diskrete Raumzeitstruktur. In anderen Fällen handelt es sich um ein prä-raumzeitliches, 'prägeometrisches'
Substrat.

[38] Siehe Kap. 4.4.

Aussagen zu gelangen, in vermutlich unvermeidlicher und gleichzeitig nicht vollständig motivier- und kontrollierbarer Weise konzeptionelle Komponenten aus den Vorgängertheorien einbeziehen muss, ist durchaus Vorsicht geboten. Eine übermässige Einbeziehung konzeptioneller Elemente aus den etablierten Theorien könnte, insbesondere wenn diese unhinterfragt im Kontext einer entstehenden Theorie der Quantengravitation Verwendung finden, durchaus in konzeptionelle Sackgassen führen.

Es ist also vermutlich nicht sonderlich ratsam, sich *allein* der Strategie einer Fortschreibung der etablierten Theorien unter Ausmerzung ihrer gegenseitigen Unverträglichkeiten zu verschreiben, wie dies insbesondere im Versuch einer direkten Quantisierung der Allgemeinen Relativitätstheorie geschieht. Es könnte eben ohne weiteres sein, dass das Gravitationsfeld zwar ein dynamisches Feld ist, nicht jedoch ein fundamentales Wechselwirkungsfeld, was eine Quantisierung des Gravitationsfeldes nicht nur zu einer sehr fragwürdigen Strategie werden lässt, sondern zu theoretischen Artefakten und damit in Sackgassen für die Theorienbildung führt. Ebenso ist schliesslich auch nicht gänzlich auszuschliessen, dass sich die physikalische, substantielle Deutung der allgemeinen Kovarianz der Allgemeinen Relativitätstheorie in letzter Instanz als ein theoretisches Artefakt erweist.[39] – Man sollte sich also vermutlich vor forschungskonservativen Überhöhungen der etablierten Theorien hüten, wie sie etwa im folgenden Zitat anklingen:

> *"[...] figuring out where the true insights are and finding the way of making them work together is the work of fundamental physics. This work is grounded on the* confidence *in the old theories, not on random search for new ones."* (Rovelli (2004) 306)

Vielmehr sollten unter den gegebenen Bedingungen und Unsicherheiten *auch* Alternativen erwogen werden, die auf den ersten Blick exzentrisch erscheinen mögen und nicht von den Vorgängertheorien und ihren konzeptionellen Implikationen nahegelegt werden, diese aber dennoch als Näherungen, Grenzfälle oder Niederenergieimplikationen enthalten könnten.

> *"Note that if our existing views on spacetime and/or quantum theory are* not *adequate, the question then arises of the extent to which they are* relevant *to research in quantum gravity. [...] how* iconoclastic *do our research programmes need to be?"* (Isham (1997) 2)

Da sowohl die konservativen als auch die spekulativeren (und insbesondere die hochspekulativen) Ansätze ihre jeweils eigenen Risiken bergen, stellt eine pluralistisch angelegte Herangehensweise in der Entwicklung einer Theorie der Quantengravitation wohl die einzige angemessene Strategie dar. Alle konzeptionell kohärenten Alternativen sollten in Betracht gezogen werden. Solange keine empirischen Daten vorliegen, bleiben die Ansätze zur Quantengravitation ohnehin allesamt spekulativ. Erst einmal handelt es sich unter diesen Bedingungen um keine Physik im strengen Sinne, sondern um eine mit ausschliesslich mathematisch-modelltheoretischen Mitteln und unter spezifischen konzeptionellen Bedingungen fortgesponnene Metaphysik der Natur, die dennoch insofern hochinteressant ist, als sie jederzeit, mit konkreten quantitativen Vorhersagen und entsprechenden empirischen Bestätigungen, wieder zur Physik mutieren könnte.

---

[39] Siehe Kap. 2.1. und Kap. 3.3.

## 1.3.  Alternativen für die Theorienbildung

Im folgenden soll nun versucht werden, das Spektrum möglicher Strategien für die Entwicklung einer Theorie der Quantengravitation unter Berücksichtigung der jeweiligen Voraussetzungen der einzelnen Strategien zu systematisieren. Die sich auf diese Weise ergebende Systematisierung ist sicherlich nicht die einzig mögliche; das sich ergebende Spektrum der strategischen Möglichkeiten ist nicht notwendigerweise umfassend und die Systematisierung ist nicht notwendigerweise in jeder Hinsicht eindeutig; einige Hybridkonstrukte lassen durchaus unterschiedliche Zuordnungen zu. Die Systematisierung erfolgt vor allem im Hinblick auf die leichtere Einschätzbarkeit und Zuordenbarkeit in den später nachfolgenden Einzeldarstellungen[40] einzelner Ansätze zu einer Theorie der Quantengravitation und ihrer jeweiligen Implikationen für die Raumzeitproblematik.

### *Umfassende Geometrisierung*

Die inzwischen nicht mehr als aktuelle Option für den Bereich der Quantengravitation angesehene Idee einer umfassenden Geometrisierung als Weg zu einer vereinheitlichten Fundamentaltheorie von Materie, Gravitation und nicht-gravitativen Wechselwirkungen geht in ihren Ursprüngen auf die Zeit zurück, in der die Geometrisierung der Gravitation im Rahmen der Allgemeinen Relativitätstheorie schon erfolgreich gelungen war, andererseits aber neben der Gravitation nur der Elektromagnetismus als fundamentale Wechselwirkung im Spiel war. Einstein selbst versuchte sich über mehrere Jahrzehnte hinweg bis zu seinem Tod erfolglos an einer umfassenden Geometrisierung im Rahmen seiner *Einheitlichen Feldtheorie*. Unabhängig davon wurde schon in den zwanziger Jahren von Kaluza und Klein[41] eine (konzeptionell heterogene[42]) Einbeziehung des Elektromagnetismus in das Geometrisierungsprogramm der Allgemeinen Relativitätstheorie um den Preis einer höheren Dimensionalität der Raumzeit vorgeschlagen: eine fünfte, eingerollte (kompaktifizierte) Dimension kam hinzu.[43]

Den einzigen weitergehenden Versuch einer umfassenden Geometrisierung von Materie und allen fundamentalen Kräften – mittlerweile waren neben dem Elektromagnetismus noch die starke und die schwache Kraft hinzugekommen – unternahm dann John A. Wheeler in den fünfziger und sechziger Jahren.[44] Seine *Geometrodynamik* hatte einen umfassenden Raumzeitmonismus zum Ziel: eine gekrümmte, dynamische Raumzeit mit möglicherweise komplexer (und dynamischer) Topologie (Einstein-Rosen-Brücken, Geone, etc.) als einzige Substanz. Alle Eigenschaften von Materie und Wechselwirkungen sollten als dynamische, metrische und topologische Eigenschaften der Raumzeit erfasst werden.[45] Letztendlich ist Wheelers monistische Geometrodynamik – im Versuch ihrer Er­weiterung auf eine ebenso monistische *Quantengeometrodynamik* – an den Quanteneigenschaften von Materie und Wechselwirkungsfeldern gescheitert, insbesondere an den für einen solchen An­satz zu erwartenden Quantenfluktuationen der raumzeitlichen Topologie (Wheelers *Spacetime foam*), die mit dem differentialgeometrischen Apparat, den der Ansatz von der Allgemeinen Relati­vitätstheorie übernommen hat, unvereinbar sind.

---

[40] Siehe vor allem Kap. 4.

[41] Siehe Kaluza (1921) und Klein (1926) sowie Wesson (1998).

[42] Die zusätzliche fünfte Dimension des Kaluza-Klein-Ansatzes ist keine Komponente einer dynamischen Raumzeit, wie sie der Gravitation entspricht, sondern vielmehr eine relativ schlichte formale Implementierung zusätzlicher elektromagnetischer Freiheitsgrade.

[43] Zwischenzeitlich gab es modifizierte und um quantenfeldtheoretische Elemente erweiterte Versuche einer Neuauflage von Konzepten aus der Kaluza-Klein-Theorie, allerdings ohne den Anspruch, damit zu einer umfassenden Geometrisierung gelangen zu wollen. Vielmehr sollten Materie- und Wechselwirkungsfelder einige ihrer Eigenschaften dadurch erhalten, dass ihre Dynamik in einem Kontext stattfindet, der zusätzliche (aufgerollte oder 'interne') Dimensionen der Raumzeit umfasst. Die Grundidee dieser quantenfeldtheoretischen Neuauflage der Kaluza-Klein-Theorie kommt schliesslich wieder im Stringansatz zum Einsatz. Siehe weiter unten in diesem Teilkapitel sowie vor allem Kap. 4.2.

[44] Siehe Wheeler (1957, 1962), Misner / Wheeler (1957).

Zu einer echten Neuauflage des Geometrisierungsgedankens – wenn auch auf kuriose Weise um eine Stufe tiefer gelegt – kommt es heute im Kontext prägeometrischer Theorieansätze.[46] Wie in Wheelers Geometrisierungsprogramm werden hier zur Erfassung physikalischer Grössen topologi­sche Freiheitsgrade ausgenutzt, dies allerdings ohne eine vorausgesetzte raumzeitliche Grundlage. Die *Topologisierung* von Elementen, die auf der Ebene einer emergenten Raumzeit in Form nicht­geometrischer Freiheitsgrade (Quantenzahlen von Materieteilchen und Wechselwirkungsquanten) in Erscheinung treten, erfolgt auf prä-raumzeitlicher Ebene. Die Konflikte einer dynamischen To­pologie mit raumzeitlichen Kontinuumsmodellen, an denen Wheelers monistische Quantengeome­trodynamik letztendlich scheiterte, bleiben hier aus; es wird kein Kontinuum vorausgesetzt.

## *Quantengravitation durch Quantisierung der Allgemeinen Relativitätstheorie*

Die Strategie, zu einer Theorie der Quantengravitation durch eine (direkte) Quantisierung der All­gemeinen Relativitätstheorie zu gelangen, wird heute von vielen Physikern als die naheliegendste Option angesehen.[47] – Eine solche Strategie muss aber noch nicht notwendigerweise zum ge­wünschten Resultat führen. Eine der essentiellsten Anforderungen an eine Theorie der Quantengra­vitation besteht in der Ableitbarkeit der Allgemeinen Relativitätstheorie als Implikation oder Grenz­fall – oder zumindest in der Reproduktion ihres empirischen bzw. phänomenalen Gehalts. Das Er­gebnis einer Quantisierung der Allgemeinen Relativitätstheorie erfüllt noch nicht notwendigerweise und per se diese Anforderung – und eine Theorie, welche diese Anforderung erfüllt, muss nicht notwendigerweise aus der direkten Quantisierung der Allgemeinen Relativitätstheorie resultieren.

Versucht man dennoch die Allgemeine Relativitätstheorie direkt zu quantisieren, so ist erst einmal zu entscheiden, welche Quantisierungsmethode verwendet wird. Es gibt diverse methodische Mög­lichkeiten für eine solche Quantisierung. Auch wenn die klassische Ausgangstheorie bekannt und klar definiert ist, ist ihre Quantisierung eine in keiner Weise eindeutig festgelegte Prozedur.

> *"Given a classical theory, one cannot derive a unique 'quantum theory' from it. The only possibility is to 'guess' such a theory and to test it by experiment."* (Kiefer (2004 [²2007]) 133)

Nicht zuletzt stellt sich die Frage, welche physikalischen Grössen quantisiert werden: etwa die Me­trik oder Grössen, die auf der Ebene der Metrik anzusiedeln sind, oder die Topologie, die kausale Struktur etc. Zudem ist zu klären, inwiefern die Hintergrundunabhängigkeit der Allgemeinen Rela­tivitätstheorie bei ihrer Quantisierung berücksichtigt wird. – Alle bestehenden Ansätze beginnen mit einer Quantisierung der Metrik oder von Grössen, die auf der strukturellen Ebene der Metrik definiert sind (Konnektionen, Holonomien). Entscheidende Unterschiede bestehen jedoch im Hinblick auf die Berücksichtigung der Hintergrundunabhängigkeit:

---

[45] Dass ein solcher substantialistischer Raumzeitmonismus nicht unbedingt mit den Deutungsmöglichkeiten hinsichtlich des Status der Raumzeit verträglich sein muss, wie sie sich im Kontext der Allgemeinen Relativitätstheorie abzeichnen, sollte im Rahmen des Kap. 2.1. deutlich werden.

[46] Siehe Kap. 4.6.

[47] Die Idee einer nomologischen Vereinigung aller Wechselwirkungen spielt dabei erst einmal keine Rolle.

## (1)  *Kovariante Quantisierung*[48]:

Ziel der *Kovarianten Quantisierung* ist die Formulierung einer Quantenfeldtheorie der Gravitation analog zur Quantenelektrodynamik. Der hier verwendete herkömmliche quantenfeldtheoretische Formalismus benötigt jedoch grundsätzlich einen Hintergrundraum mit fester Metrik, auf dem die Operatorfelder definiert werden. Quantisiert werden im kovarianten Ansatz konsequenterweise nur Fluktuationen der Metrik auf einem festen Hintergrundraum, gemeinhin einem Minkowski-Raum. Ein solcher störungstheoretischer Ansatz, der die Dynamik von Feldquanten der Gravitation (*Gravitonen*) auf einer festen Raumzeit beschreibt, ist notwendigerweise hintergrundabhängig. Dies entspricht einer Verletzung der aktiven Diffeomorphismusinvarianz der Allgemeinen Relativitätstheorie[49] in ihrer Quantisierung. Die *Kovariante Quantisierung* der Allgemeinen Relativitätstheorie startet also von vornherein mit einem konzeptionellen Widerspruch: Sie entspricht einer hintergrundabhängigen Quantisierung einer hintergrundunabhängigen Theorie. – Konsequenterweise erweist sich der Ansatz als nicht-renormierbar und kommt somit als fundamentale Theorie nicht in Frage.

## (2)  *Kanonische Quantisierung*[50]:

Die *Kanonische Quantisierung* umgeht diese Probleme von vornherein. Ihr Ziel ist die direkte, vollständige, nicht-perturbative, hintergrundunabhängige Quantisierung der Allgemeinen Relativitätstheorie, ausgehend von ihrer Hamiltonschen Formulierung. Sie setzt zwar, zumindest was ihren modelltheoretischen Ausgangspunkt betrifft, eine raumzeitliche Mannigfaltigkeit voraus, nicht jedoch eine schon vorgegebene Hintergrundmetrik. Die sich in ihrem Rahmen ergebende *Wheeler-DeWitt-Gleichung*, die quantisierte Form der Hamiltonschen Zusatzbedingung, mit der im klassischen Fall, in der Hamiltonschen Formulierung der Allgemeinen Relativitätstheorie,[51] der zeitliche Aspekt der Diffeomorphismusinvarianz Berücksichtigung findet, führt jedoch in der ursprünglichen *geometrodynamischen*[52] Form des kanonischen Quantisierungsansatzes zu gravierenden (und mutmasslich unlösbaren) mathematischen und konzeptionellen Problemen.

Besser sieht es diesbezüglich in einer neueren Variante des kanonischen Quantisierungsansatzes aus: der *Loop Quantum Gravity*[53]. Hier erfolgt die *Kanonische Quantisierung* der Allgemeinen Relativitätstheorie zwar weiterhin ausgehend von ihrer Hamiltonschen Formulierung, jedoch nicht mehr auf der Grundlage der geometrischen Variablen (Metrik und Krümmung) des älteren Ansatzes, sondern auf der der sogenannten Ashtekar-Variablen[54] bzw. ihrer konzeptionellen Weiterent-wicklung. Dennoch bestehen insbesondere immer noch ungelöste Probleme hinsichtlich der Ableitbarkeit der Niederenergiephysik und des klassischen Grenzfalles. Die basale Anforderung an eine Theorie der Quantengravitation, die Reproduktion der bekannten Phänomenologie der Gravitation bzw. der Einsteinschen Feldgleichungen als Näherung bzw. klassischem Grenzfall, konnte bisher nicht erfüllt werden. Auch wenn die *Loop Quantum Gravity* von einer Quantisierung der Allgemeinen Relativitätstheorie ihren Ausgang nimmt, führt sie nicht so ohne weiteres zu ihr zurück. Wie alle anderen Ansätze zur Quantengravitation, macht die *Loop Quantum Gravity* zudem keine empirisch direkt überprüfbaren Vorhersagen.

---

[48] Siehe Kap. 4.1.
[49] Siehe Kap. 2.1.
[50] Siehe Kap. 4.3. und 4.4.
[51] Siehe Kap. 2.1.
[52] Siehe Kap. 4.3.
[53] Siehe Kap. 4.4.
[54] Siehe Ashtekar (1986, 1987).

Darüberhinausgehend gibt es konzeptionelle Probleme, die vor allem die Vieldeutigkeit in der Formulierung des Hamilton-Operators sowie in der Festlegung der Repräsentation der Operator-Algebra betreffen. Hinzu kommen radikale Konsequenzen, verglichen mit den etablierten Theorien, wie etwa das virulente *Problem der Zeit*[55]: Die Theorie beschreibt eine eingefrorene Dynamik; die Zeit ist eine unphysikalische Eichvariable; es gibt keine zeitabhängigen Observablen. Alle Observablen erweisen sich zudem als nichtlokal und die beschriebene Dynamik ist nicht unitär.

(3) *Quantisierung einer diskretisierten Version der Allgemeinen Relativitätstheorie*[56]:

Es gibt eine ganze Reihe von Ansätzen, die (aus z.T. unterschiedlichen Gründen) anstatt einer vollständigen und direkten Quantisierung der Allgemeinen Relativitätstheorie die Strategie einschlagen, diese in einer diskretisierten Form zu quantisieren. Zu diesen Ansätzen gehören insbesondere die *Konsistente Diskretisierung*, der *Regge Calculus* sowie die verschiedene (Euklidischen und Lorentzschen) Varianten der *Dynamischen Triangulation*. Diese Ansätze unterscheiden sich in der Diskretisierungsmethode, der jeweils schon vorausgesetzten Signatur der Raumzeit sowie insbesondere dem Erfolg in den Quantisierungsbemühungen. Sie lassen sich jedoch vermutlich auch im Falle ihres konzeptionellen wie empirischen Erfolgs bestenfalls als effektive Theorien, vergleichbar den Gittereichtheorien, ansehen.

(4) *Quantisierung einer erweiterten Version der Allgemeinen Relativitätstheorie bzw. Erweiterung einer Quantisierung der Allgemeinen Relativitätstheorie*:

Eine weitere Möglichkeit, die Direktheit der Quantisierung der Allgemeinen Relativitätstheorie aufzuweichen, um mit bestimmten konzeptionellen Problemen besser fertig werden zu können, besteht darin zu versuchen, entweder eine erweiterte oder modifizierte Version der Allgemeinen Relativitätstheorie zu quantisieren oder die Quantisierung der Allgemeinen Relativitätstheorie nachträglich zu erweitern oder zu modifizieren, etwa mit dem Ziel die in der *Kovarianten Quantisierung* auftretende Nichtrenormierbarkeit zu beheben. Im letzteren Fall sind vor allem die zugrundegelegten Symmetrien das vorrangige Ziel der Erweiterung. In diesem Kontext fällt etwa die *Supergravity*[57], ein um die Supersymmetrie angereicherter und mit zusätzlichen Raumdimensionen ausgestatteter

---

[55] Siehe Kap. 2.1. und 4.4.
[56] Siehe Kap. 4.5.
[57] Siehe Cremmer / Julia / Scherk (1978).

Ansatz, der in den siebziger und achtziger Jahren zeitweise heftig diskutiert wurde, aber aufgrund konzeptioneller Probleme heute allgemein nicht mehr als aktuell angesehen wird.[58]

> *"Another response to the non-renormalizability was to consider 'corrections' to the theory in the form of additional particles with quantum loop amplitudes that serve to cancel out the divergences associated with the gravitons. This is the way of 'supergravity' theories."* (Rickles / French (2006) 17)

## (5) *Bedingungen, unter denen eine Quantisierung der Allgemeinen Relativitätstheorie als Weg zu einer Theorie der Quantengravitation unangemessen wäre*:

Angesichts der Bestrebungen, die Allgemeine Relativitätstheorie zu quantisieren, sollte daran erinnert werden, dass eine solche Strategie mit der Ziel der Entwicklung einer Theorie der Quantengravitation nicht voraussetzungslos ist und nur unter bestimmten Bedingungen überhaupt als sinnvoll angesehen werden kann. Sollte die Gravitation keine fundamentale Wechselwirkung, sondern ein residuales bzw. induziertes Phänomen sein, oder sollte die Raumzeit nicht fundamental sein, sondern Ausdruck anderer nicht-raumzeitlicher Freiheitsgrade, so wäre, wie schon vorausgehend erläutert, eine Quantisierung der Allgemeinen Relativitätstheorie konzeptionell unsinnig. Wenn die Gravitation bzw. die Raumzeit emergente Phänomene sind und die Allgemeine Relativitätstheorie entsprechend nur als effektive Theorie zur Erfassung dieser emergenten Grössen und ihrer Dynamik angesehen werden kann, so führt eine Quantisierung dieser effektiven Theorie ganz sicher nicht zu einer angemessenen Theorie der Quantengravitation – genausowenig wie etwa eine Quantisierung der Navier-Stokes-Gleichung zu einer realistisch interpretierbaren Quantenhydrodynamik führen würde; man würde schlichtweg die falschen Freiheitsgrade quantisieren, nämlich kollektive, makroskopische Grössen, denen eine völlig andere Mikrodynamik zugrundeliegt.

Sollten Raumzeit und/oder Gravitation emergent sein, so wäre also vielmehr zu versuchen, diese aus einer fundamentaleren (mikroskopischen), möglicherweise prä-raumzeitlichen Theorie bzw. Dynamik als (makroskopische) Näherungen, etwa nach Einführung und Berücksichtigung von Ordnungsparametern oder kollektiven Anregungszuständen, abzuleiten. Es wäre aber umgekehrt völlig unsinnig, diese nicht-fundamentalen Kollektivparameter zu quantisieren.

Im Falle einer emergenten, intrinsisch klassischen Raumzeit ginge es in einer Theorie der 'Quantengravitation' nicht mehr um Quantenkorrekturen zu einer klassischen Raumzeitauffassung, etwa Quantenfluktuationen der Metrik oder quantenmechanische Unschärfen der Raumzeit. Eine emergente, klassische Raumzeit hätte keine solchen Quanteneigenschaften. Sie käme vielmehr auf der Grundlage eines gänzlich anders gearteten Substrats zustande. Sollten Raumzeit und/oder Gravitation emergent und intrinsisch klassisch sein, so ginge es in einer Theorie der 'Quantengravitation' nicht um die Quanteneigenschaften von Gravitation oder Raumzeit – diese gäbe es dann eben gar nicht –, sondern vielmehr um die Quanteneigenschaften des Substrats, auf dessen Grundlage sich Gravitation und klassische Raumzeit ergeben. Die konzeptionellen Voraussetzungen, die sich für eine Theorie der Quantengravitation vielleicht aus den etablierten Theorien, insbesondere der Allgemeinen Relativitätstheorie, ableiten liessen, würden sich unter diesen Umständen nur bedingt als relevant und verlässlich erweisen. Vielleicht würden sie sogar in eine völlig falsche Richtung weisen.

---

[58] Die *Supergravity* wird zumindest nicht mehr als fundamentale Beschreibung ernstgenommen. Allerdings spielt sie seit einiger Zeit im Rahmen des Superstring-Ansatzes wieder eine Rolle: als (zusätzlicher) perturbativer Grenzfall einer angezielten nicht-perturbativen Verallgemeinerung der Stringtheorien.

Relevant werden diese Überlegungen vor allem vor dem Hintergrund der Tatsache, dass die Möglichkeit, dass Raumzeit und/oder Gravitation emergente Phänomene sein könnten, inzwischen mehr als nur eine abstrakte Idee ist. Es gibt mittlerweile eine Vielzahl von mehr oder weniger konkreten theoretischen Szenarien, die zu erklären versuchen, wie es zu einer emergenten Raumzeit bzw. zu einer emergenten Gravitation kommen könnte. Da diese später im einzelnen zu erörtern sein werden,[59] soll hier nur eine kurze stichpunktartige Auflistung erfolgen:

(a) *Raumzeit als Ausdruck eines prägeometrischen quantenmechanischen Zustandsspektrums*:[60]
Dimensionalität und Topologie der Raumzeit werden als Ergebnis der Herausbildung von Ordnungsparametern innerhalb des Zustandsspektrums relativ einfacher Quantensysteme dynamisch erzeugt. Fermionische Freiheitsgrade führen zu einer flachen Raumzeit, bosonische zu einer aufgerollten Raumzeit. Es kommt unter bestimmten Bedingungen zu Phasenübergängen zwischen Raumzeiten unterschiedlicher Dimensionalität.

(b) *Raumzeit als emergentes thermodynamisches bzw. statistisches Phänomen*:[61]
Die Einsteinschen Feldgleichungen lassen sich aus einer Verallgemeinerung der Proportionalität von Entropie und Horizontfläche bei schwarzen Löchern (*Bekenstein-Hawking-Entropie*) ableiten. Die Grundlage dafür liefert die thermodynamische Beziehung zwischen Wärme, Temperatur und Entropie, gemeinsam mit der Deutung der Temperatur als Unruh-Temperatur eines beschleunigten Beobachters innerhalb eines lokalen Rindler-Horizontes. Wärme wird als Energiefluss durch einen kausalen (Vergangenheits-)Horizont gedeutet. Dieser Energiefluss tritt als Krümmung der Raumzeit und mithin als Gravitationsfeld in Erscheinung.

(c) *Raumzeit als emergentes hydrodynamisches Phänomen*:[62]
Die Raumzeit ergibt sich als kollektiver Quantenzustand vieler Mikrokonstituenten mit einer makroskopischen Quantenkohärenz, vergleichbar einem Bose-Einstein-Kondensat.

(d) *Raumzeit als emergentes festkörperphysikalisches Phänomen*:[63]
Gravitation und Raumzeit kommen als emergente Phänomene auf der Grundlage von masselosen Spin-2-Anregungszuständen eines Fermionen-Systems mit Fermi-Punkt zustande.

(e) *Raumzeit als phänomenologisches Ergebnis eines computationalen Prozesses*:
Hierzu gibt es eine Vielzahl von Ansätzen mit den unterschiedlichsten Voraussetzungen.[64] Manche gehen sogar von einem Nicht-Quanten-Substrat und der Emergenz der Quanteneigenschaften mikroskopischer Systeme aus.

---

[59] Siehe Kap. 3.3. Soweit diese konzeptionellen Szenarien noch nicht den Status konkreter Theorieansätze haben, werden sie im Kontext der 'Elemente des Übergangs' (Kap. 3.) und nicht in dem der konkreteren Theorieansätze zur Quantengravitation (Kap. 4.) behandelt. Siehe auch Teil (2) im nachfolgenden Abschnitt ('Quantengravitation ohne Quantisierung der Allgemeinen Relativitätstheorie', des vorliegenden Kapitels 1.3.

[60] Siehe Kap. 3.3 sowie Kaplunovsky / Weinstein (1985); vgl. auch Dreyer (2004) sowie Kober (2009).

[61] Siehe Kap. 3.3. sowie Jacobson (1995, 1999), Eling / Guedens / Jacobson (2006), Jacobson / Parentani (2006), Padmanabhan (2002a, 2004, 2007a).

[62] Siehe Kap. 3.3. sowie Hu (2005, 2009), Hu / Verdaguer (2003, 2004), Oriti (2006), Sakharov (2000), Visser (2002), Barcelo / Liberati / Visser (2005), Weinfurtner (2007).

[63] Siehe Kap. 3.3. sowie Volovik (2000, 2001, 2003, 2006, 2007), Zhang (2002), Tahim et al. (2007), Hu (2009).

## *Quantengravitation ohne Quantisierung der Allgemeinen Relativitätstheorie*

Es gibt ausser dem Versuch der direkten Quantisierung der Allgemeinen Relativitätstheorie im wesentlichen zwei Strategien um zu einer Theorie der Quantengravitation (im Sinne einer Quantentheorie der Gravitation) zu gelangen: durch die Quantisierung einer anderen klassischen Theorie oder durch die Formulierung einer Quantentheorie, die nicht aus der Quantisierung einer klassischen Theorie resultiert.[65]

(1)    *Entwicklung einer Theorie der Quantengravitation durch (direkte) Quantisierung einer anderen klassischen Theorie*:

Als auf den ersten Blick überraschende Instantiierung dieser Strategie hat sich der perturbative Stringansatz erwiesen.[66] Ausgehend von einer Entwicklung, die in den sechziger Jahren in der Hadronenphysik ihren Anfang nahm, hat sich post hoc herausgestellt, dass die Quantisierung der speziell-relativistischen Dynamik eines eindimensional ausgedehnten, schwingungsfähigen Objekts (*String*) unter bestimmten Bedingungen zumindest formal die Dynamik von *Gravitonen* auf einer Minkowski-Hintergrundraumzeit beschreibt und näherungsweise die Phänomenologie der Gravitation bzw. die Einsteinschen Feldgleichungen (mit Stringkorrekturen) reproduziert.

Neben den Spin-2-Zuständen, die als *Gravitonen* interpretiert werden können, gehören zum Schwingungsspektrum des *String* Skalarteilchen, Spin-1-Eichbosonen und fermionische Materiebestandteile. Formal lässt sich der perturbative Stringansatz damit als Instantiierung einer nomologischen Vereinheitlichung aller Wechselwirkungen inklusive der Gravitation ansehen.

Zu den spezifischen Bedingungen, unter denen dies alles jedoch erst gilt, zählt, dass (i) die Hintergrundraumzeit zehn Dimensionen umfasst – und insofern erklärt werden muss, wie es zu unserer vierdimensionalen phänomenologischen Raumzeit kommt –, und dass (ii) die Dynamik des *String* die Supersymmetrie erfüllt, die für jedes bekannte Fermion einen unbekannten bosonischen Partner – und umgekehrt – fordert, so dass erklärt werden muss, wieso wir diese Teilchen noch nicht beobachtet haben und wieso – und auf welche Weise – die Supersymmetrie, wenn sie überhaupt vorliegt, gebrochen ist.

Abgesehen von seiner Hintergrundabhängigkeit, die der Stringansatz aufgrund seiner modelltheoretischen Basis von den Quantenfeldtheorien (und der *Kovarianten Quantisierung*) geerbt hat und die für eine Theorie, welche die Einsteinschen Feldgleichungen der Allgemeinen Relativitätstheorie zumindest formal als klassischen Grenzfall reproduziert, erst einmal unangemessen ist, solange dafür keine guten Gründe angeführt werden können, weist der Stringansatz eine Vielzahl weiterer Probleme auf: So ist kein physikalisch motivierbares, fundamentales Prinzip bekannt, aus dem heraus sich der Stringansatz motivieren oder gar entwickeln liesse. Es gibt keine analytische, nicht-perturbative Formulierung. Stattdessen gibt es eine zu grosse Zahl perturbativer Szenarien: die fünf bekannten perturbativen zehndimensionalen Stringtheorien (vielleicht schon vier zu viel) führen zu einem riesigen Spektrum von (etwa $10^{500}$) vierdimensionalen Modellen mit unterschiedlicher Niederenergiephysik – der sogenannten *String-Landscape*. Es ist aber nicht ein einziges Modell bekannt, welches zu einer Phänomenologie entsprechend dem quantenfeldtheoretischen Standardmodell führen würde. Und es lassen sich keine quantitativen, experimentell überprüfbaren Vorhersagen ableiten – wie bei allen anderen Ansätzen zur Quantengravitation.

---

[64] Siehe Kap. 3.3. und 4.6. sowie Lloyd (1999, 2005, 2005a, 2007), Cahill (2005), Cahill / Klinger (1997, 2005), Hsu (2007), Livine / Terno (2007), Zizzi (2001, 2004, 2005), Hardy (2007).

[65] Beide Strategien sind grundsätzlich mit und ohne nomologische Vereinheitlichung der Gravitation mit den nichtgravitativen Wechselwirkungen denkbar. Eine nomologische Vereinheitlichung ist, wie zuvor schon erörtert, als Ziel erst einmal unabhängig von der grundsätzlichen Motivation zur Entwicklung einer Theorie der Quantengravitation, nämlich die wechselseitigen Unvereinbarkeiten zwischen Allgemeiner Relativitätstheorie und Quantenmechanik bzw. Quantenfeldtheorien zu überwinden. Erst im Bereich der prägeometrischen Theorien, die von einer emergenten Raumzeit ausgehen, gibt es schliesslich Gründe, von einer gemeinsamen Materie- und Geometrogenese auszugehen. Siehe Kap. 1.4. sowie 4.6.

[66] Siehe Kap. 4.2.

(2)     *Entwicklung einer Theorie der Quantengravitation in Form einer Quantentheorie, die nicht durch die Quantisierung einer klassischen Theorie zustandekommt*:

Angesichts der Probleme sowohl des Stringansatzes (dem immer noch populärsten und personalstärksten Ansatz im Bereich der Quantengravitation) als auch der *Loop Quantum Gravity* (dem vielversprechendsten Vertreter des direkten kanonischen Quantisierungsansatzes) und vor dem Hintergrund der Option, dass die Gravitation und/oder die Raumzeit emergente Phänomene sein könnten, lohnt es sich, nach Alternativen Ausschau zu halten. Die direkte Quantisierung einer klassischen Theorie – sei dies die Allgemeine Relativitätstheorie oder eine andere klassische Dynamik – ist nicht der einzige Weg zu einer Theorie der Quantengravitation. Das Spektrum der Alternativen, die ohne eine direkte Quantisierung auskommen, und ihrer jeweiligen konzeptionellen Hintergründe und Motivationen ist inzwischen schon recht vielfältig. Und es bietet nicht zuletzt Aussichten, die jeweiligen Probleme, die sich für die direkten Quantisierungsansätze abzeichnen, zu vermeiden.

Insbesondere formieren diverse, mehr oder weniger konkret ausformulierte, diskrete prägeometrische Ansätze in diesem Spektrum: *Causal Sets*, computationale Ansätze, prägeometrische *Quantum Causal Histories* etc.[67] Diese Ansätze lassen sich – über ihre jeweiligen Unterschiede hinweg – nicht zuletzt aus den zahlreichen Indizien für ein diskretes Substrat der Raumzeit heraus motivieren, wie sie sich schon im (erweiterten) Kontext der etablierten Theorien abzeichnen. Zu nennen wären hier insbesondere die *Bekenstein-Hawking-Entropie* als Ergebnis des Zusammenspiels von Argumenten aus Allgemeiner Relativitätstheorie, Quantenfeldtheorien und Thermodynamik, als auch darüber hinausgehend die Ableitung diskreter Spinnetz-Strukturen in der *Loop Quantum Gravity*, die Anzeichen für eine minimale Länge im Stringansatz sowie die Reproduktion der *Bekenstein-Hawking-Entropie* im Rahmen beider Ansätze.

### Fundamentaltheoretische Neukonzeption

Die Frage bleibt jedoch, ob die angesprochenen Ansätze zu einer Theorie der Quantengravitation hinreichend und insbesondere radikal genug sind, um einerseits die konzeptionellen Widersprüche zwischen Allgemeiner Relativitätstheorie und Quantenmechanik bzw. Quantenfeldtheorien in der einen oder der anderen Weise aufzuheben und andererseits tatsächlich zu einer durchweg empirisch angemessenen Beschreibung der Natur zu gelangen. – Ist insbesondere die Quantenmechanik auf der fundamentalsten Ebene immer noch gültig, wie dies nahezu alle Theorieansätze zur Quantengravitation annehmen? Oder müsste ein noch radikalerer Weg beschritten werden? Könnte es möglicherweise erforderlich sein, zur Ausräumung der konzeptionellen Unvereinbarkeiten zwischen Allgemeiner Relativitätstheorie und Quantenmechanik eine 'Theorie der Quantengravitation' zu formulieren, die selbst keine Quantentheorie im engeren Sinne mehr ist?

---

[67] Siehe Kap. 3.3. und 4.6.

> *"[...] this type of approach [...] is often motivated by the view that the basic ideas behind general relativity and quantum theory are so fundamentally incompatible that any complete reconciliation will necessitate a total rethinking of the central categories of space, time, and matter."* (Butterfield / Isham (2001) 60)

Bisher gibt es nur wenige, die der Auffassung sind, dass die Quantenmechanik keine universelle Gültigkeit besitzt und somit nicht Teil einer fundamentalen Beschreibung sein kann, sondern vielmehr einer emergenten Systemebene angehört. Das prominenteste Beispiel ist Gerard 't Hooft, der eine nicht-quantenmechanische, prägeometrische, deterministische Basaldynamik postuliert, aber bisher nicht ausformuliert hat.[68]

## *Anomologisches Substrat*

Eine weitere Idee geht in die Richtung, nicht speziell die universelle Gültigkeit der Quantenmechanik, sondern vielmehr die Annahme einer grundlegenden Gesetzesartigkeit der Natur überhaupt in Frage zu stellen: Vielleicht besitzt die Nomologie, die wir der Natur im Rahmen unserer empirisch-wissenschaftlichen Erschliessung zusprechen, nur approximative Gültigkeit für den makroskopischen und den niederenergetischen Bereich. Vielleicht ist die Natur auf ihrer fundamentalsten Ebene anomologisch, in gewisser Hinsicht chaotisch. Vielleicht ist die scheinbare Nomologie der Natur nicht zuletzt das Ergebnis unserer wissenschaftlichen Methodologie und der mit ihr verbundenen Suche nach Regularitäten und gesetzesartigen Strukturen.

John Wheeler hat diese Idee als *Law-without-law*-Physik[69] ins Spiel gebracht. Weiter ausgearbeitet wurde sie von Holger Nielsen im Rahmen seiner *Random Dynamics*[70], in deren Kontext es sogar gelang, einige der gemeinhin der Natur zugeschriebenen Symmetrien und Regularitäten als approximatives Ergebnis einer statistischen Mittelung aus einem anomologischen, chaotischen Substrat abzuleiten.

---

[68] Siehe 't Hooft (1999, 2000a, 2001, 2001a, 2007) sowie Suarez (2007). Weitere Beispiel für solche prä-quantenmechanischen Ansätze sind etwa Adlers *Matrixtheorie* (Adler (2002)), Requardts *zelluläre Netze* (Requardt (1995, 1996, 1996a, 1996b, 2000) sowie Requardt / Roy (2001)) sowie Cahills *Prozessphysik* (Cahill (2002, 2005), Cahill / Klinger (1996, 1997, 1998, 2005)); siehe auch Elze (2009).
[69] Siehe Wheeler (1979, 1983).
[70] Siehe Nielsen (1983), Frogatt / Nielsen (1991), Nielsen / Rugh (1994), Nielsen / Rugh / Surlykke (1994), Bennett / Brene / Nielsen (1987); vgl. auch Donoghue / Pais (2009). Man könnte durchaus das Landscape-Fiasko des *Stringansatzes* (siehe Kap. 4.2.) als Widerlegung der Grundthese der *Random Dynamics* ansehen. Unsere bekannte Niederenergiephysik setzt offensichtlich sehr spezifische Bedingungen voraus. Sie lässt sich nicht aus beliebigen Voraussetzungen ableiten.

*"The ambition of the Random Dynamics project is a sort of derivation of the laws of physics as we know them from an almost empty set of assumptions. [...] we postulate and then attempt to show that any set of sufficiently general assumptions made at some fundamental scale will ultimately lead to the same description. [...] In essence we consider natural laws to be the result of a kind of selection taking place in the transitional region between a very chaotic stage at some high energy level (Planck scale?) and our level. Whether or not there exists order at some even higher energy level is only of philosophical interest."* (Bennett / Brene / Nielsen (1987) 158)

### Flickwerk-Physik

Sollten alle auch noch so radikalen Versuche einer Beseitigung der konzeptionellen Inkompatibilitäten zwischen den etablierten physikalischen Theorien auf lange Sicht erfolglos bleiben, bliebe als letzte Option die Auffassung, dass eine einheitliche, konzeptionell kohärente physikalische Beschreibung der Natur vielleicht nicht erreichbar ist. Vielleicht sind physikalische Theorien ausschliesslich als Instrumentarien mit begrenzter explikatorischer Reichweite zu sehen. Vielleicht sind die Bemühungen um eine fundamentale physikalische Weltbeschreibung, die auf die Erfassung einer einheitlichen Substratdynamik abzielt, als unangemessene Extrapolation nomologischer Ambitionen zu sehen. Vielleicht ist also insbesondere die Annahme einer Einheit der Welt letztendlich falsch. Vielleicht haben wir es mit einer *'gesprenkelten Welt'*[71] zu tun, die schlichtweg disparate Phänomenbereiche aufweist, welche jeweils eine völlig eigenständige wissenschaftliche Herangehensweise erfordern. Vielleicht gibt es sogar Bereiche der Welt, die sich jeder wissenschaftlichen Erschliessung versperren. – Im Rahmen eines solchen Verständnisses physikalischer Theorien und ihres Bezugs zur Realität gäbe es kaum Aussichten für eine Theorie der Quantengravitation im Sinne der zuvor ausgeführten Minimaldefinition, die sich vorrangig auf die Zielsetzung der Beseitigung bzw. Aufhebung konzeptioneller Widersprüche innerhalb unseres Theoriengebäudes beruft. Bevor man sich jedoch damit zufrieden gibt, sollte man alles versucht haben und keine der Alternativen – seien sie auch noch so radikal – ausser Acht gelassen haben. Und wann könnte man das schon sagen!

## 1.4.  Alternativen hinsichtlich möglicher Raumzeitkonzeptionen

Da die Ansätze zu einer Theorie der Quantengravitation nicht nur hinsichtlich ihrer jeweiligen physikalischen wie konzeptionellen Voraussetzungen und ihrer konzeptionellen Kohärenz, sondern vor allen Dingen auch hinsichtlich ihrer jeweiligen Implikationen für unser Verständnis der Raumzeit[72],

---

[71] Siehe etwa Cartwright (1983, 1989, 1994, 1999). Vgl. auch Morrison (2000).

[72] An dieser Stelle ist nun spätestens ein klärendes Wort zur thematischen Abgrenzung erforderlich: Es soll im Folgenden um die Raumzeit (als Nachfolgekonzept zum klassischen Raumbegriff) und um ihr tatsächliches und mögliches Schicksal im Übergang von der Allgemeinen Relativitätstheorie zu den in Entwicklung befindlichen Ansätzen zu einer Theorie der Quantengravitation gehen.

Dabei soll mit der Begrifflichkeit der 'Raumzeit' und ihrer thematischen Kopplung an die Problematik der Gravitation bzw. der Quantengravitation jedoch nicht schon gesagt sein, dass die strukturelle bzw. nomologische Verbindung zwischen Raum und Zeit, wie sie seit der Speziellen Relativitätstheorie besteht, oder die Geometrisierung der Gravitation, wie sie mit der Allgemeinen Relativitätstheorie vorliegt, als unerschütterliche Gegebenheiten betrachtet werden sollen. Ganz im Gegenteil sollen gerade auch diese konzeptionellen Elemente als solche verstanden werden, die bei entspre-ihrer physikalischen Eigenschaften und ihres ontologischen Status zu untersuchen sein werden, soll im folgenden erst einmal ein vorläufiges Spektrum der diesbezüglich grundsätzlich in Frage kommenden Optionen entwickelt werden. Dieses Spektrum betrifft einerseits die Raumzeit, ihren ontologischen Status und ihre strukturellen und physikalischen Eigenschaften, andererseits das Verhältnis der Raumzeit zu den weiteren konstitutiven Komponenten des physikalischen Geschehens: der Materie und den nicht-gravitativen Wechselwirkungsfeldern – und dies sowohl für den Fall, dass diese fundamentalen Status besitzen, als auch für den, dass sie als Resultat eines Emergenzprozesses zu sehen sind. Die Klärung der tatsächlich in der Natur vorliegenden Bedingungen für die Existenz und Beschaffenheit der Raumzeit und der weiteren Konstituenten physikalischen Geschehens – sowie nicht zuletzt des Verhältnisses dieser zueinander – gehört, soweit diese sich im Rahmen eines empirisch-wissenschaftlichen Zugangs erschliessen lassen, zu den wesentlichsten Desiderata im Rahmen der Suche nach einer konzeptionell und empirisch adäquaten Theorie der Quantengravitation.

## *Die Natur von Raum und Zeit*

Hinsichtlich der Natur (und Struktur) von Raum und Zeit kommen zwei entscheidende Dichotomien zum Tragen: die zwischen kontinuierlicher und diskreter Raumzeit und die zwischen fundamentaler und emergenter Raumzeit.

### (1)    *Kontinuierliche vs. diskrete Raumzeit*

Die Allgemeine Relativitätstheorie geht ebenso wie die Quantenfeldtheorien von einem raumzeitlichen Kontinuum aus.[73] Es gibt allerdings schon im Kontext dieser etablierten Theorien Anzeichen,

---

chender Notwendigkeit im Kontext neuer Theorieentwicklungen durchaus problematisiert und modifiziert werden können.

Andererseits soll mit der Begrifflichkeit der 'Raumzeit' – und ihrer Ankopplung an die Probleme unseres Verständnisses der Gravitation, an die Allgemeine Relativitätstheorie als unsere bisher allgemeinste Theorie von Gravitation und Raumzeit, und schliesslich an die Perspektiven, die sich mit einer und für eine Theorie der Quantengravitation ergeben könnten – durchaus das Spektrum der vorrangig zu behandelnden Thematik abgesteckt werden: Dieses Spektrum schliesst Fragen des wechselseitigen Verhältnisses von Raum, Zeit und Materie ebenso ein wie etwa spezielle Probleme der Zeit und des Zeitbegriffs, soweit sie sich im Kontext spezifischer Ansätze zur Quantengravitation (etwa der *Loop Quantum Gravity*) ergeben.

Es schliesst hingegen erst einmal alle anderen Problemstellungen und Thematisierungen, die diesbezüglich in den Sinn kommen könnten, insbesondere etwa hinsichtlich einer Physik, Wissenschaftsphilosophie oder Metaphysik der Zeit, explizit aus. Dieser Ausschluss betrifft – solange sich keine spezifische Verbindung zu Ansätzen aus dem Kontext der Quantengravitation ausmachen lässt – die ansonsten vieldiskutierten Aspekte der Präsentismus-Endurantismus-Debatte, die ebenso weit verbreitete Diskussion um die Philosophie des Werdens und um die Struktur von Zeitreihen, und nicht zuletzt alle mit der Gerichtetheit der Zeit bzw. dem 'Fluss der Zeit' zusammenhängenden Themen, insbesondere die unterschiedlichen Ansätze zur Erklärung des Zeitpfeiles (oder der Zeitpfeile), wie sie etwa im Kontext der statistischen Physik oder der Quantenkosmologie diskutiert werden. Vor allem die Problemstellungen der Quantenkosmologie, die eine gänzlich andere Motivationslage und Zielrichtung aufweist als die Ansätze zu einer Theorie der Quantengravitation, werden im Folgenden grundsätzlich aussen vor bleiben.

[73] Die in der Allgemeinen Relativitätstheorie verwendete Differentialgeometrie setzt notwendigerweise ein Raumzeitkontinuum voraus; Quantenfelder setzen gemeinhin ein raumzeitliches Kontinuum voraus, auf dem sie definiert werden.

die sich hin auf eine letztendliche Unangemessenheit des Kontinuumskonzepts deuten lassen.[74] Insbesondere ist es aber die Thermodynamik schwarzer Löcher, die Hinweise auf eine diskrete Substratstruktur liefert.[75] Diese lassen sich entweder als Anzeichen für eine diskrete Raumzeitstruktur deuten, oder aber als Hinweis auf eine diskrete Substratstruktur, auf deren Grundlage sich die Raumzeit als emergentes Phänomen ergibt. In jedem Fall deuten sie auf ein Substrat hin, welches nur über eine finite Zahl von Freiheitsgraden pro finiter raumzeitlicher Region verfügt, unabhängig davon, ob diese raumzeitliche Region schon auf der Substratebene existiert oder von dieser induziert bzw. auf einer höheren Emergenzebene hervorgebracht wird.

Und interessanterweise finden sich bisher in nahezu allen Ansätzen zu einer Theorie der Quantengravitation Hinweise auf eine diskrete Substratstruktur[76] – auch in den Ansätzen, die sich konzeptionell sehr eng an die durchweg kontinuumsorientierten Vorgängertheorien anlehnen. Eine definitive Theorie der Quantengravitation wird diesen Sachverhalt klären müssen.

## (2)  *Fundamentale vs. emergente Raumzeit*

Die Annahme, dass die Raumzeit fundamental ist, ist gleichbedeutend damit, dass die Raumzeit auf der fundamentalsten Ebene der Natur schon eine Rolle spielt, oder, dass sie, wenn es keine solche fundamentalste Ebene geben sollte, sondern vielmehr einen infiniten Regress von strukturellen Schichten, auf allen Ebenen der Realität für das physikalische Geschehen konstitutiv ist. Die Frage nach der Fundamentalität der Raumzeit ist dabei erst einmal völlig unabhängig von der Kontinuums-Diskretheit-Dichotomie. Eine fundamentale Raumzeit könnte letztlich durchaus eine diskrete Struktur aufweisen. Sie könnte etwa über Quanteneigenschaften verfügen, die vielleicht gerade zu dieser diskreten Struktur führen.

Die Annahme, dass es sich bei der Raumzeit um ein emergentes Phänomen handelt, bedeutet hingegen, dass es eine fundamentalere Ebene unter der von der Allgemeinen Relativitätstheorie beschriebenen gibt, die über keine raumzeitlichen Freiheitsgrade (also solche im Sinne der makroskopischen Raumzeit) verfügt. Die Raumzeit würde dann von einer nicht-raumzeitlichen Substratebene hervorgebracht, vielleicht als intrinsisch klassisches, makroskopisches Phänomen. Allerdings schliesst die Emergenz der Raumzeit noch nicht per se die Möglichkeit aus, dass diese auf einer intermediären, ebenfalls schon emergenten Ebene nicht vielleicht doch Quanteneigenschaften aufweist. Diese intermediäre Ebene wäre dann jedoch nicht mit der Substratdynamik zu verwechseln, die selbst keine raumzeitlichen Eigenschaften oder Freiheitsgrade aufweist, auch keine solchen, die Quanteneigenschaften der Raumzeit entsprächen.

Grundsätzlich denkbar sind auch hybride Varianten, welche die fundamental-emergent-Dichotomie in der Hinsicht durchbrechen, dass sie den Raum als emergentes Phänomen behandeln, der Zeit jedoch fundamentalen Status zusprechen. Theorien, die mit einem solchen Szenario arbeiten, müssten jedoch erklären, wie es dann zu der in der Allgemeinen Relativitätstheorie vorliegenden und empirisch erfolgreichen strukturellen und dynamischen Kopplung von Raum, Zeit und Materie kommt.[77] Dies macht solche Szenarien nicht gerade zur naheliegendsten Option; die Hybridisierung hinsichtlich des Status von Raum und Zeit führt vielmehr zu Problemen, die erst einmal, solange für sie keine wirklich guten Motivationen vorgebracht werden, als Argument gegen die entsprechenden Ansätze wirksam werden.[78]

---

[74] Wie schon in Abschnitt 1.2. erwähnt, liessen sich hier etwa die für kleine Abstände bzw. hohe Energien in den Quantenfeldtheorien auftretenden Divergenzen anführen, oder eben auch die in der Allgemeinen Relativitätstheorie auftretenden Singularitäten. Beide lassen sich als mögliche Artefakte deuten, die aus der Kontinuumsannahme bzw. der Annahme einer infiniten Zahl physikalisch relevanter Freiheitsgrade pro raumzeitlicher Region resultieren.
[75] Siehe Kap. 3.1.
[76] Siehe Kap. 4.

Eine erfolgreiche Theorie der Quantengravitation wird jedenfalls die Frage klären müssen, ob es sich bei der Raumzeit um ein fundamentales oder ein emergentes Phänomen handelt. In beiden Fällen wird sie insbesondere klären müssen, welche Eigenschaften und welche Struktur die Raumzeit aufweist, im zweiten Fall kommt die Frage hinzu, von welcher nicht-raumzeitlichen Substratstruktur und -dynamik die Raumzeit auf welche Art und Weise und unter welchen Bedingungen hervorgebracht wird.

<p style="text-align:center">*</p>

Durchaus zu unterscheiden von der Fragestellung, ob die Raumzeit fundamental oder emergent ist, ist die, ob es sich bei der Raumzeit um eine Substanz, d.h. eine ontologisch eigenständige Entität, oder vielmehr um ein relationales Gefüge handelt, welches sich in seinen Eigenschaften, seiner Struktur und seiner Dynamik auf andersgeartete Entitäten, ihre relationalen Eigenschaften und die ihnen zugrundeliegende Dynamik reduzieren lässt. Schon innerhalb der Allgemeinen Relativitätstheorie ist diese Entscheidung zwischen einer substantiellen und einer relationalen Sichtweise bezüglich der Raumzeit nicht unproblematisch; wie noch im einzelnen zu erörtern sein wird, ist sie an hochgradig verästelte Interpretationskonstrukte gekoppelt, die gemeinhin der Substantialismus-Relationalismus-Debatte subsumiert werden.[79]

Die Annahme der Substantialität der Raumzeit setzt zwar durchaus voraus, dass diese fundamental ist. (Und sollte die Gravitation tatsächlich im Sinne der Allgemeinen Relativitätstheorie Ausdruck von spezifischen Eigenschaften der Raumzeit sein, so wäre auch diese Teil des fundamentalen Geschehens.) – Nicht ganz so einfach ist es jedoch im Falle der Relationalität der Raumzeit. Sollten die Entitäten, durch die die Raumzeit als relationales Gefüge aufgespannt wird, selbst der fundamentalen Strukturebene angehören, sollten diese also nicht emergent sein, wird man auch einer relational verstandenen Raumzeit, die sich dann letztlich auf diese fundamentalen Entitäten und ihre relationalen Eigenschaften reduzieren lässt, kaum Emergenzstatus zusprechen wollen, auch wenn die Raumzeit dann in ihrem relationalen Zustandekommen nicht selbst als fundamental anzusehen ist. Sind die Entitäten, welche die Raumzeit aufspannen, hingegen selbst schon emergent, so lässt sich zwar grundsätzlich nichts dagegen einwenden, dann auch von einer emergenten Raumzeit zu sprechen; allerdings geht bei dieser Sprechweise die entscheidende Pointe verloren; man sollte in einem solchen Fall vielleicht genauer von einer relationalen Raumzeit sprechen, die noch dazu von emergenten Entitäten aufgespannt wird. Dann ist vor allem zu klären, auf welcher Grundlage und unter welchen Bedingungen es zur Emergenz der Entitäten kommt, welche die Raumzeit aufspannen, und schliesslich, unter welchen Bedingungen diese emergenten Entitäten die Raumzeit aufspannen. – In allen Fällen sind also grundsätzlich zwei Typen von Fragen zu klären:

---

[77] Solche Szenarien stehen natürlich auch schon im Konflikt mit der Speziellen Relativitätstheorie.
[78] Siehe Kap. 3.3.
[79] Siehe Kap. 2.1.

– Welchen Status hat die Raumzeit?
– In welchem Verhältnis steht die Raumzeit zu weiteren Komponenten des Naturgeschehens?

Unglücklicherweise sind diese beiden Fragen nicht unabhängig voneinander.

## Das Verhältnis von Raum und Zeit zur Materie

Sollten sich gute Gründe dafür finden lassen, dass es sich bei der Raumzeit um eine eigenständige Substanz handelt, ist alles noch am einfachsten. Es wäre dann zu klären, welche Eigenschaften und Strukturen diese Substanz aufweist – und, ob es noch weitere von ihr unabhängige Substanzen gibt. Hinsichtlich der letzten Frage ergeben sich grundsätzlich zwei verschiedene Szenarien:

### Monistischer Raumzeitsubstantialismus:

Es gibt nur eine Substanz: die Raumzeit. Alle Phänomene und vermeintlich andersgearteten Entitäten sind Ausdruck der Struktur und der Dynamik der Raumzeit. Diese hat vermutlich Quanteneigenschaften und zeichnet sich, über ihre metrische Struktur hinaus, durch komplexere topologische Strukturen aus; nur unter diesen Umständen besteht überhaupt eine Aussicht auf eine umfassendere Erklärung des Zustandekommens und der Dynamik der vermeintlich nicht-raumzeitlichen Entitäten (z.B. Materieteilchen als stabile topologische Konstellationen). – Diese Idee eines monistischen Raumzeitsubstantialismus stellt seit dem Scheitern der Versuche einer umfassenden Geometrisierung von Materie und Wechselwirkungen[80] (inklusive ihrer Quanteneigenschaften) wohl bis auf weiteres keine ernstzunehmende Option mehr dar.

### Nicht-monistischer Raumzeitsubstantialismus:

Es gibt mehr als eine Substanz. Eine davon ist die Raumzeit.[81] Die Gravitation ist im Sinne der Allgemeinen Relativitätstheorie Ausdruck der dynamischen Struktur der Raumzeit. Da die Raumzeit als Substanz fundamentalen Status besitzt, ist unter der Annahme, dass die Quantenmechanik universell gültig ist, davon auszugehen, dass die Raumzeit über Quanteneigenschaften verfügt, die von der Allgemeinen Relativitätstheorie nicht erfasst werden. Sie müssten im Rahmen einer Theorie der Quantengravitation erschlossen und in ihrer Dynamik beschrieben werden. Eine solche Theorie müsste den empirischen Gehalt der Allgemeinen Relativitätstheorie als Implikation eines klassischen Grenzfalles reproduzieren.

Das Zusammenspiel zwischen der Raumzeit und den weiteren Substanzen, die ihren Ausdruck in der Materie und den nicht-gravitativen Wechselwirkungen finden, wäre ebenso im Kontext einer zu entwickelnden Theorie der Quantengravitation (oder einer über diese noch einmal hinausgehenden Theorie) zu erfassen.

---

[80] Siehe Kap. 1.3.
[81] Unglücklicherweise ist diese Auffassung schon innerhalb der Allgemeinen Relativitätstheorie durchaus problematisch. Siehe Kap. 2.1.

*

Sollten sich hingegen gute Gründe dafür finden lassen, dass die Raumzeit keine eigenständige Substanz, sondern ein relationales Konstrukt darstellt, wird es ein wenig komplizierter. Es wäre zu klären, von welchen Entitäten dieses Konstrukt aufgespannt wird, welchen Status diese Entitäten selbst wiederum haben, unter welchen Bedingungen dieses Konstrukt zustandekommt, und nicht zuletzt, wann diese Bedingungen erfüllt sind. Erst dann könnte geklärt werden, welche Eigenschaften und Strukturen die Raumzeit aufweist und auf welchen Ebenen des Naturgeschehens sie eine Rolle spielt.

Dem klassischen Leibnizschen Relationalismus zufolge wird der dreidimensionale Raum durch die Materie aufgespannt; er ist nichts mehr als der Ausdruck der relationalen Verhältnisse von Materiekonstellationen. Für die vierdimensionale Raumzeit der Allgemeinen Relativitätstheorie ist eine solche Auffassung schon aus strukturellen Gründen nicht mehr angemessen. Als basale Elemente, welche die Raumzeit im Sinne eines relationalen Szenarios aufspannen, kommen hier nur noch elementare Ereignisse in Frage. Zu klären ist dann, welche Art von Ereignissen dies sind, ob diese Ereignisse schon die elementarste Schicht des Geschehens darstellen, ob sie bestimmte Entitäten mit bestimmten Eigenschaften (etwa Materieteilchen) voraussetzen, ob diese Entitäten schliesslich selbst wiederum fundamentalen Status besitzen oder vielleicht von einer gänzlich anders gearteten Substratebene hervorgebracht werden.

Im Bereich der (radikaleren) Theorieansätze zur Quantengravitation finden sich inzwischen Szenarien,[82] in denen die Raumzeit zwar als Ausdruck des Verhaltens von Materiekonstituenten bzw. als sich aus diesem Verhalten ergebendes relationales Gefüge verstanden wird; jedoch gehören die Materiekonstituenten, deren Verhalten die Raumzeit hervorbringt, hier noch nicht der fundamentalen Ebene des Geschehens an (auf der aber immerhin noch die Quantenmechanik gilt), vielmehr sind sie selbst wiederum emergent. Raumzeit ist hier eine dynamische, relationale Implikation des Verhaltens von Entitäten, die selbst wiederum emergent sind. Raumzeit und Materie entstehen gemeinsam; Raumzeit ist dabei aber nichts anderes als der Ausdruck des dynamischen Verhaltens der Materie. Auf der Substratebene gibt es keine Materie und keine Raumzeit, sondern nur elementarste kausale Relationen bzw. Informationsflüsse.

Durchaus unabhängig von diesem konkreten Ansatz wie von seiner spezifischen Substratkonzeption, zeichnen sich inzwischen gute Gründe dafür ab, dass für den Fall einer emergenten Raumzeit von einem gemeinsamen Entstehen von Raumzeit und Materie aus einer gemeinsamen Basis auszugehen ist: also von einer gemeinsamen Materie- und Geometrogenese. Zwei getrennte Basen für die Materie- und die Geometrogenese würden aufgrund ihrer dynamischen Kopplung schlichtweg eine Zusammenführung auf einer wiederum fundamentaleren Ebene nahelegen.[83] Eine Chance für eine ontologische Unabhängigkeit von Materie und Raumzeit besteht folglich eigentlich nur, wenn die

---

[82] Siehe Kap. 4.6.: Abschnitt 'Quantum Causal Histories'.

[83] Die Kopplung zwischen Materie- und Geometrogenese liefert schliesslich auch die überzeugendsten Gründe für eine nomologische Vereinigung aller Wechselwirkungen, wenn auch vielleicht nicht im Sinne der von den Quantenfeldtheorien und vom Stringansatzes vorgeschlagenen Mechanismen. Dies gibt zumindest Anlass zur Hoffnung, dass sich die Frage, ob eine nomologische Vereinigung tatsächlich vorliegt bzw. ob die Annahme ihrer Richtigkeit für eine konsistente und empirisch erfolgreiche Beschreibung der Natur tatsächlich erforderlich ist, im Kontext einer erfolgreichen Theorie der Quantengravitation beantworten lassen sollte.

Raumzeit eine fundamentale, substantielle Entität sein sollte – eine Annahme, die schon im Rahmen der Allgemeinen Relativitätstheorie problematisch wird.

# 2. Raumzeit im Kontext der etablierten Theorien

## 2.1. Raumzeit in der Allgemeinen Relativitätstheorie

Die Allgemeine Relativitätstheorie stellt heute gleichermassen unsere allgemeinste Theorie der Raumzeit wie auch jeglichen gravitativen Verhaltens dar. Solange es keine guten Gründe gibt, die dagegen sprechen, sollten die Einsichten, die mit der Allgemeinen Relativitätstheorie hinsichtlich der Raumzeit gewonnen wurden, als Teil eines möglichen Ausgangspunktes in der Entwicklung einer Theorie der Quantengravitation Berücksichtigung finden. Diese Einsichten sollten jedoch nicht unreflektiert und jenseits jeder Revisionsmöglichkeit einbezogen werden. Eine unreflektierte Festschreibung der mit der Allgemeinen Relativitätstheorie einhergehenden Raumzeitkonzeption könnte nämlich für den Bereich der Quantengravitation durchaus in konzeptionelle Sackgassen führen. Es sollten also grundsätzlich auch Alternativen erwogen werden, die vielleicht nicht auf den ersten Blick durch die Raumzeitkonzeption der Allgemeinen Relativitätstheorie nahegelegt werden, die aber dennoch den phänomenologischen Gehalt der Allgemeinen Relativitätstheorie zu reproduzieren in der Lage sind.

Unter diesen Vorbehalten steht nun einer genaueren Erörterung der in der Allgemeinen Relativitätstheorie vorliegenden Raumzeitkonzeption nichts mehr im Wege. Denn, bevor sich mögliche Implikationen für die Theorieansätze im Bereich der Quantengravitation überhaupt sondieren lassen, ist natürlich erst einmal zu klären, welche Einsichten hinsichtlich der Raumzeit denn tatsächlich mit der Allgemeinen Relativitätstheorie einhergehen. – Wie sich herausstellen wird, ist dies eine Frage, die in ihren möglichen Antworten in mancher Hinsicht weit von einem Konsens entfernt ist.

*

Die Allgemeine Relativitätstheorie lässt sich hinsichtlich der Festlegung physikalisch möglicher Raumzeiten und ihrer mathematischen Erfassung in Form eines zweistufigen Prozesses rekonstruieren:

> "First, one carefully specifies the mathematical structure to be employed. This is usually achieved by specifying the form of the theories' models. General relativity, for example, will have models of the form $< M, g_{ab}, T_{ab} >$, where $M$ is a differentiable manifold and $g_{ab}$ and $T_{ab}$ tensors of appropriate type. Second, one formulates the conditions specifying the set of models allowed by the theory. These conditions are roughly the 'laws' of the theory. They contain, for example, in the case of general relativity, the field equation $G_{ab} = \kappa\, T_{ab}$." (Norton (1989) 1220)

Der erste Schritt besteht also in der Charakterisierung der im Rahmen der Theorie verwendeten mathematischen Modelle der Raumzeit und der auf ihr definierten Grössen. Das Raumzeitkontinuum, welches die Theorie beschreibt, wird durch eine vierdimensionale, kontinuierlich differenzierbare (pseudo-Riemannsche) Punktmannigfaltigkeit $M$ repräsentiert: eine Ansammlung von Punkten mit definierter Topologie und Dimensionalität. Auf der gesamten Mannigfaltigkeit ist ein metrischer Feldtensor $g_{ab}$ definiert; dieser repräsentiert die geometrischen und chronometrischen

Beziehungen zwischen den Punkten der Mannigfaltigkeit bzw. den Ereignissen, die in der Raumzeit stattfinden.[84] Hinzu kommt der Energie-Spannungs-Tensor $T_{ab}$, der ebenso auf der gesamten Mannigfaltigkeit definiert ist; er repräsentiert die Materie und alle nicht-gravitativen Felder.

Der zweite Schritt besteht dann in der Auswahl gerade der mathematischen Modelle einer Raumzeit, die gemäss der Allgemeinen Relativitätstheorie physikalisch mögliche, reale Entsprechungen haben können. Diese Auswahl erfolgt über die Einsteinschen Feldgleichungen. Diese entsprechen dem physikalischen Gesetz, welches das Spektrum mathematischer Modelle (wie sie im ersten Schritt definiert wurden) gerade auf solche einschränkt, denen im Rahmen der Theorie physikalische Möglichkeit zugesprochen wird, die also physikalische Situationen repräsentieren.

> *"The set of models of the theory are exactly those which satisfy a set of conditions L, called the laws of the theory."* (Norton (1989) 1223)

Die Einsteinschen Feldgleichungen beschreiben die Beziehung zwischen dem metrischen Tensor und dem Energie-Spannungs-Tensor, der die Materie und alle nicht-gravitativen Felder repräsentiert. Sie erfassen die Dynamik zwischen den raumzeitlichen Strukturen und der in der Raumzeit vorfindlichen Materie (zuzüglich der nicht-gravitativen Felder und ihrer Energiedichten).

## Die Identität von Gravitation und Raumzeitmetrik

Die Allgemeine Relativitätstheorie ist jedoch nicht nur eine Theorie der Raumzeit, sondern vor allem auch eine Gravitationstheorie. Die Gravitation wird jedoch nicht durch ein weiteres auf der Raumzeit definiertes Feld repräsentiert, sondern entspricht vielmehr Eigenschaften der Raumzeit: Die von der Materie und den Energiedichten der nicht-gravitativen Felder hervorgerufenen Gravitationswirkungen werden von der Theorie in Form von Eigenschaften einer dynamischen Raumzeit erfasst. Wollte man weiterhin am Gravitationsfeld als selbständigem Begriff festhalten, so liesse sich sagen, dass dieses – durch die Gravitationswirkung von Materie und anderen Feldern hervorgebracht – die Metrik (bzw. die chronogeometrischen Eigenschaften) der Raumzeit bestimmt bzw. in diese eingeschrieben wird. In der Allgemeinen Relativitätstheorie ist das Gravitationsfeld jedoch keine eigenständige Grösse. Vielmehr wird es schlichtweg durch die Raumzeitmetrik repräsentiert.[85] Es ist eine Manifestation der Raumzeitmetrik. Die Raumzeitmetrik entspricht dem Gravitationsfeld und das Gravitationsfeld entspricht der Raumzeitmetrik.

> *"[...] the metric plays a dual role in general relativity: it serves to generate both the gravitational field structures and the chronometric, spatio-temporal structures."* (Rickles / French (2006) 1) – *"[...] a mathematical model of [general relativity] is specified by a four-dimensional mathematical manifold $M_4$ and by the metrical tensor field g, where the latter dually represents* both *the chrono-geometrical structure of spacetime* and *the potential for the inertial-gravitational field. Non-gravitational physical fields, when they are present, are also described by dynamical tensor fields, which appear as sources of the Einstein equations."* (Dorato / Pauri (2006) 126)

---

[84] Der metrische Tensor repräsentiert gleichzeitig das Gravitationsfeld, welches mit den chronogeometrischen Eigenschaften der Raumzeit identifiziert wird. Siehe weiter unten.

[85] Die Theorie lässt sich auch so umformulieren, dass andere chronogeometrische Grössen als Repräsentationen des Gravitationsfeldes in Frage kommen. Es sind jedoch immer Grössen, die auf der Beschreibungsebene des metrischen Tensors operieren.

Die in der Allgemeinen Relativitätstheorie vorliegende Identität (oder Identifizierung) von Gravitation und chronogeometrischen Eigenschaften der Raumzeit lässt grundsätzlich zwei Interpretationsweisen zu, deren eine gerade schon angeklungen ist: Man kann diese Identität so lesen, dass es eigentlich gar keine Gravitation gibt, und dass das, was wir als Gravitation ansehen, eigentlich nur Eigenschaften der Raumzeit sind. Man kann aber die Identifizierung der Metrik mit dem Gravitationsfeld auch, spätestens wenn man dieses Gravitationsfeld zu den Inhalten der Raumzeit rechnet, dahingehend deuten, dass es in der Allgemeinen Relativitätstheorie keine substantiellen Anteile der Raumzeit mehr gibt:

> *"Einstein's major discovery is that spacetime and gravitational field are the same object. A common reading of this discovery is that there is no gravitational field: just a dynamical spacetime. In view of quantum theory, it is more illuminating and more useful to say that there is no spacetime, just the gravitational field."* (Rovelli (2004) 266) – *"The gravitational field is not located in spacetime: it is with respect to* it *that things are localized."* (Rovelli (1998) 19)

Diese beiden Interpretationsweisen der Identifizierung der Gravitation mit chronogeometrischen Eigenschaften der Raumzeit entsprechen im wesentlichen Fokussierungen, die sich den beiden Polen der Substantialismus-Relationalismus-Debatte zuordnen lassen.[86] Völlig unabhängig von dieser Interpretationsfrage ist jedoch das, was sich aus der Identifizierung der Gravitation mit den chronometrischen Eigenschaften der Raumzeit folgern und als *erster Aspekt* der in der Allgemeinen Relativitätstheorie vorliegenden *Hintergrundunabhängigkeit* bezeichnen liesse: *Da das Gravitationsfeld kein Feld auf der Raumzeit ist, sondern als Manifestation der Raumzeit(metrik) letztlich (Aspekten) der Raumzeit entspricht, lässt es sich konsequenterweise nicht wiederum auf einer Raumzeit (die schon über eine Metrik verfügt[87]) beschreiben.*

Eine Theorie der Quantengravitation müsste, wenn sie diese Einsicht der Allgemeinen Relativitätstheorie zu ihren konzeptionellen Voraussetzungen zählt – was sicherlich zumindest für die Ansätze zu einer direkten Quantisierung der Allgemeinen Relativitätstheorie gelten sollte –, die Dynamik der Gravitation (und damit, infolge der Identität von Gravitation und raumzeitlicher Metrik, konsequenterweise auch die Dynamik der chronogeometrischen Komponenten der Raumzeit) auf eine Weise beschreiben, bei der nicht schon eine Raumzeit mit chronogeometrischen Eigenschaften vorausgesetzt wird.

> *"[...] gravity is geometry whence, in a fundamental theory, there should be no background metric."*
> (Ashtekar (2005) 12)

Infolgedessen ist es konzeptionell völlig unsinnig, die Gravitation auf einer schon vorgegebenen Raumzeit zu quantisieren.

---

[86] Zur relationalen Auffassung hinsichtlich der Raumzeit gleich mehr. Zur Substantialismus-Relationalismus-Debatte siehe weiter unten.

[87] Wie weiter unten zu motivieren sein wird, kann die raumzeitliche Mannigfaltigkeit aufgrund der aktiven Diffeomorphismusinvarianz der Allgemeinen Relativitätstheorie nicht oder nur um einen sehr hohen Preis als substantielle Entität angesehen werden.

## *Dynamizität der Raumzeit*

Die Allgemeine Relativitätstheorie beschreibt eine dynamische Raumzeit. – Damit kann natürlich kaum gemeint sein, dass die Raumzeit einer zeitlichen Veränderung unterworfen ist; sie enthält nämlich schon alle zeitlichen Veränderungen.

> *"Considering the fact that in [general relativity] temporal relations themselves are specified by the metric tensor, to say that the metric tensor is dynamical in the sense of temporal change would be apparently nonsensical. Thus being dynamical for the metric tensor should be primarily understood as meaning being subject to causal interactions with matter or other physical entities [...]."* (Cao (2001) 189)

Wenn also hinsichtlich der Raumzeit von Dynamizität die Rede ist, kann damit nur der Bezug auf einen dynamischen, wechselseitigen Wirkungszusammenhang gemeint sein, in den diese eingebunden ist: Eine Entität (die Raumzeit) wirkt auf andere Entitäten (Materie und nicht-gravitative Felder) ein und erfährt ebenso Einwirkungen von diesen anderen Entitäten. Dieser wechselseitige Wirkungszusammenhang lässt sich, wenn man das Gravitationsfeld als Hilfsbegriff verwendet, folgendermassen beschreiben: Die Materie (zuzüglich der Energie, die den nicht-gravitativen Feldern zukommt) erzeugt ein Gravitationsfeld, das einer Raumzeitkrümmung entspricht, also im metrischen Tensor erfasst wird. Diese Raumkrümmung, die das Gravitationsfeld repräsentiert, bestimmt wiederum die Bewegung der Materie. Materie, Energie und Raumzeitmetrik sind auf diese Weise dynamisch verbunden.

Das heisst aber noch nicht, dass es im Falle der Abwesenheit von Materie und nicht-gravitativen Feldern keine Raumkrümmung gibt und dann notwendigerweise eine flache Minkowskische Raumzeit vorliegt. Komplexere raumzeitliche Strukturen setzen, wie die Vakuumlösungen der Einsteinschen Feldgleichungen deutlich machen, nicht notwendigerweise Materie oder nicht-gravitative Felder voraus. Es gibt – wiederum unter Verwendung der semi-autonomen Begrifflichkeit des Gravitationsfeldes als einer Entität, die in dieser Sprechweise noch dazu eine zeitliche Entwicklung haben kann – so etwas wie eine Eigendynamik des Gravitationsfeldes (und seiner intrinsischen Energiedichte). Letztlich ist diese Sprechweise aber nichts weiter als eine Umschreibung einer materiefreien und dennoch nicht-flachen, komplexeren Raumzeit. In allen Fällen repräsentiert die Dynamik der Raumzeitmetrik die Dynamik des Gravitationsfeldes. Oder, um noch einmal die schon verwendete Sprechweise zu bemühen: Die Dynamik der Raumzeitmetrik wird durch die Dynamik des Gravitationsfeldes, die einerseits eine durch Massen und Energien induzierte Dynamik ist, andererseits aber auch eine Eigendynamik sein kann, vollständig bestimmt.

Die Dynamizität der Raumzeit führt – im Verbund mit der oben beschriebenen Identifizierung der Gravitation mit den chronogeometrischen Eigenschaften der Raumzeit – insbesondere zu einer Einsicht, die man als *zweiten Aspekt* (und gleichzeitig als Spezifizierung) der *Hintergrundunabhängigkeit* der Allgemeinen Relativitätstheorie bezeichnen könnte, und dies wiederum mit Auswirkungen auf den Bereich der Quantengravitation: *Die Gravitation lässt sich nicht* – weder klassisch noch im Rahmen einer Quantengravitationstheorie, welche die Einsichten der klassischen Allgemeinen Relativitätstheorie zu ihren konzeptionellen Grundlagen zählt – *auf einer vorgegebenen Raumzeit mit schon vorgegebener nicht-dynamischer (statischer) Metrik beschreiben.*

Es soll aber an dieser Stelle nicht verschwiegen werden, dass die Allgemeine Relativitätstheorie keine vollständig hintergrundunabhängige Theorie ist. Sie verfügt vielmehr über einen rudimentären raumzeitlichen Hintergrund, dessen Komponenten nicht in die Dynamik einbezogen und festgeschrieben sind. Die Komponenten dieses Hintergrundes sind:

– die differenzierbare (pseudo-Riemannsche) Punktmannigfaltigkeit[88]
– die Dimensionalität der Raumzeit
– die Signatur der Raumzeit
– die (zumindest für eine spezifische Lösung unveränderliche) Topologie bzw. Zusammenhangsform der Raumzeit.

Ob eine Quantengravitationstheorie an diesem Residuum an Hintergrundabhängigkeit, wie sie die Allgemeine Relativitätstheorie immer noch aufweist, festhalten sollte (wie dies etwa die kanonischen Quantisierungsansätze tun) ist zumindest fraglich. Vielmehr sollte sich im Rahmen einer Quantengravitationstheorie klären lassen, ob diese residuale Hintergrundabhängigkeit ein theoretisches Artefakt der Allgemeinen Relativitätstheorie oder eine notwendige physikalische Implikation im Bereich der Beschreibung makroskopischer Raumzeiten darstellt.

Unzweifelhaft ist jedenfalls, dass die Dynamizität der Raumzeit und die Identifizierung des Gravitationsfeldes mit den chronogeometrischen Eigenschaften der Raumzeit nicht mit einer Beschreibung des Gravitationsfeldes auf einem Hintergrundraum vereinbar ist, insbesondere wenn dessen Metrik schon fest vorgegeben und statisch ist. – Quantisiert man also (indem man versucht, die Allgemeine Relativitätstheorie zu quantisieren) die Gravitation auf einer statischen flachen Hintergrundraumzeit, so begeht man gleich mehrere Fehler. Wenn Gravitation und dynamische Raumzeit(metrik) letztlich identisch sind, so kann man nicht die Gravitation wiederum auf einer Raumzeit beschreiben wollen – und schon gar nicht auf einer statischen Raumzeit. Dies heisst insbesondere, dass die herkömmlichen Methoden, die innerhalb der Quantenfeldtheorien Verwendung finden, für eine Quantisierung der Allgemeinen Relativitätstheorie (und mithin der Gravitation) vollkommen ungeeignet sind. Theorieansätze im Kontext der Quantengravitation werden sich also kaum auf die Methoden stützen können, die gewöhnlich im Rahmen der Quantenfeldtheorien verwendet werden.

*"The whole formalism of ordinary [quantum field theory] relies heavily on this background structure and collapses to nothing when it is missing."* (Thiemann (2007) 9)

## Relationalität der Raumzeit ?

Die weitgehende Hintergrundunabhängigkeit der Allgemeinen Relativitätstheorie wird oft als Indiz für die Relationalität der von ihr beschriebenen Raumzeit gedeutet oder gar als ausreichende Grundlage für einen entsprechenden Relationalismus hinsichtlich der Raumzeit angesehen.

*"This is the source of the claim that general relativity, and background independent theories, are* relational*: it simply means that the states and observables of the theory do not make reference to background structures. "* (Rickles (2008a) 6)

---

[88] Es gibt gute Argumente dafür, dass dieser Punktmannigfaltigkeit selbst kein ontischer Status zukommt. Siehe hierzu die Diskussion um die relationale Auffassung hinsichtlich der Raumzeit und die Erörterungen zur Substantialismus-Relationalismus-Debatte weiter unten.

Die Annahme der Relationalität der Raumzeit entspricht der Auffassung, dass es sich bei der Raumzeit nicht um eine eigenständige substantielle Entität handelt, sondern um nichts anderes als den Ausdruck der relationalen, raumzeitlich deutbaren Bezüge zwischen den eigentlichen physikalischen Systemen bzw. Ereignissen. Im Hinblick auf die Allgemeine Relativitätstheorie heisst dies, dass die Raumzeit nichts mehr und nichts anderes ist als der Ausdruck der relationalen Bezüge zwischen der Materie (bzw. den Ereignissen, in die diese eingebunden ist) und den vorhandenen physikalischen Feldern – inklusive des (im raumzeitlichen Bild durch die Metrik repräsentierten) Gravitationsfeldes. Spätestens wenn man die quantenfeldtheoretische Materieauffassung noch mit ins Spiel bringt, gibt es dann nur noch dynamische Felder und ihre relationalen Beziehungen:

> *"The clean way of expressing Einstein's discovery is to say that there are no space and time: there are only dynamical objects. The world is made by dynamical fields. These do not live in, or on, spacetime: they form and exhaust reality."* (Rovelli (2006) 27) – *"[...] with general relativity, we have understood that the world is* not *a non-dynamical metric manifold with dynamical fields living over it. Rather, it is a collection of dynamical fields living, so to say, on top of each other."* (Gaul / Rovelli (2000) 4)

Es gibt natürlich Stimmen, die demgegenüber geltend machen, dass dem Gravitationsfeld ein besonderer Status zukommt, der sich dann auf die chronometrischen Eigenschaften der Raumzeit übertragen lässt und der sich schliesslich vielleicht sogar auf die Raumzeit selbst und ihren ontologischen Status hin interpretieren liesse:

> *"Rovelli is lumping all of the dynamical fields together, as being ontologically 'on all fours'; but this is a mistake: we can remove all fields with the exception of the gravitational field and still have a dynamically possible world – i.e. there are* vacuum *solutions to Einstein's field equations. But we cannot remove the gravitational field in the same way, leaving the other fields intact. This is an indication that there is something* special *about the gravitational field: it cannot be switched off; it is not just one field among many. / Hence, the substantivalist would be perfectly within her rights to claim ownership. But, so would the relationalist since there is this ambiguity over the ontological nature of the field: spacetime or material object?"* (Rickles (2008a) 3f)

Unter diesen Voraussetzungen erscheint eine genauere Sichtung der Sachlage und ihrer möglichen Interpretation unumgänglich: Das primäre Argument für die Relationalität der Raumzeit im Kontext der Allgemeinen Relativitätstheorie ergibt sich aus dem, was sich als *substantielle Form der allgemeinen Kovarianz* bezeichnen liesse.[89] Die allgemeine Kovarianz ist eines der elementarsten Grundprinzipien der Allgemeinen Relativitätstheorie. Ihre Elementarität wird nur noch durch die Problemgeladenheit ihrer möglichen Deutungen sowie die der Deutung ihrer Relevanz übertroffen. Insbesondere ist erst einmal zwischen zwei sehr unterschiedlichen Ausprägungen der allgemeinen Kovarianz zu unterscheiden, worauf vor allem John Earman eindringlich hingewiesen hat:

> *"The first point that needs to be emphasized is that the existing philosophical literature does not show an appreciation for the fact that the requirement of general covariance comes in two versions, weak and strong. The weak version requires that the laws of a theory be written in a form that is valid in all coordinate systems or, equivalently that the laws retain their form under an arbitrary coordinate*

---

[89] Siehe insbesondere Earman (2002, 2006, 2006a) sowie Norton (1989).

*transformation. The strong version requires that the spacetime diffeomorphism group be a gauge group of the theory."* (Earman (2002) 16)[90]

Die *formale (bzw. schwache) allgemeine Kovarianz* ist nichts mehr als eine *Invarianz unter Koordinatentransformation*. Sie entspricht der vollständigen *Koordinatenunabhängigkeit* einer Theorie und ihres physikalischen Gehalts. Sie ist für sehr viele physikalische Theorien erfüllt bzw. zumindest erfüllbar, etwa auch für die Newtonsche Mechanik oder die Spezielle Relativitätstheorie.

> *"The weak version is so called because practically any theory can, with suitable ingenuity, be massaged so as to fulfill this requirement; for example, using the tensor calculus Newtonian gravitational theory and special relativistic theory of motion can be written in generally covariant form."* (Earman (2002) 16)

Die *substantielle (bzw. starke) Form der allgemeinen Kovarianz* ist dagegen etwas sehr spezielles und sehr restriktives. Sie ergibt sich auf der Grundlage des Unterschieds zwischen einer simplen Koordinatentransformation und einer echten Punkttransformation; sie entspricht der Invarianz unter Punkttransformationen:

> *"To get a substantive requirement, think of the coordinate transformation $x^\mu \rightarrow x'^\mu (x^\nu)$ not as a mere relabeling of the points of **M** (where a point p initially assigned the coordinates $x^\mu(p)$ is assigned the new coordinate $x'^\mu (p)$) but as indicating a (local) diffeomorphism d of **M** (i.e. a point transformation $p \rightarrow p'$ where the new point p' is related to the old point p by $x'^\mu(p) \rightarrow x^\mu(p')$. Associated with d are new object fields $d^*O_1, d^*O_2, ... , d^*O_N$ on **M** that, in general, will be different from the original object fields, i.e., there will be points $p \in M$ such that $d^*O_i (p) \neq O_i(p)$ for some i."* (Earman (2006) 447)

Die Allgemeine Relativitätstheorie ist nun insbesondere nicht nur unter beliebigen Koordinatentransformationen, sondern eben auch unter echten Punkttransformationen invariant.[91] Grundlage (bzw. formal-mathematisches Pendant) dieser *Invarianz unter Punkttransformationen*, die nichts anderes als das entscheidende Element der substantiellen Form der allgemeinen Kovarianz ist, ist die *aktive Diffeomorphismusinvarianz* der Allgemeinen Relativitätstheorie.

Ein Diffeomorphismus ist eine differenzierbare, umkehrbare Abbildung der Punktmenge einer Mannigfaltigkeit, im vorliegenden Falle auf sich selbst. Ein solcher Diffeomorphismus lässt nun grundsätzlich wiederum zwei verschiedene Interpretationen zu: Der *passiven Deutung* zufolge entspricht der Diffeomorphismus einer *Koordinatentransformation*; in der *aktiven Deutung* entspricht er einer *Punkttransformation*. Die in der Allgemeinen Relativitätstheorie vorliegende aktive Diffeomorphismusinvarianz entspricht gerade der Invarianz der Theorie unter echten Punkttransformationen – und eben nicht nur unter Koordinatentransformationen. Aktive Diffeomorphismen, also echte Punkttransformationen, sind das entscheidende Kennzeichen der substantiellen Form der allgemeinen Kovarianz, wie sie in der Allgemeinen Relativitätstheorie vorliegt. Solche aktiven Diffeomorphismen entsprechen einer Verschiebung aller physikalischen Felder innerhalb der gleichen raumzeitlichen Punktmannigfaltigkeit. Solche Punkttransformationen haben nichts mit einer simplen Umbenennung der Koordinaten zu tun. Die Allgemeinen Relativitätstheorie erschöpft sich nicht in einer trivialen Koordinatensysteminvarianz. Sie ist nicht nur unter Koordinatentransformationen, sondern auch unter echten Punkttransformationen invariant.

---

[90] Zur Interpretation der Diffeomorphismusinvarianz als Eichinvarianz, siehe weiter unten.
[91] Siehe auch Norton (1987, 1989, 1992a).

*"[...] active diffeomorphisms are* dynamical symmetries *of the Einstein's tensor equations, i.e.,* they map solutions into solutions." (Lusanna / Pauri (2005) 7)

Die Invarianz der Allgemeinen Relativitätstheorie unter aktiven Diffeomorphismen impliziert insbesondere, dass die in der Theorie formulierbaren physikalischen Beschreibungen der Raumzeit, die sich mittels eines Diffeomorphismus ineinander überführen lassen, hinsichtlich ihrer beobachtbaren Konsequenzen gleich sind. Die durch einen aktiven Diffeomorphismus vollzogene Verschiebung aller physikalischen Felder auf der Punktmannigfaltigkeit führt zu keinen beobachtbaren physikalischen Unterschieden. Die durch einen aktiven Diffeomorphismus ineinander überführbaren Beschreibungen der Raumzeit unterscheiden sich in keiner Weise hinsichtlich ihrer beobachtbaren Konsequenzen. Sie sind *Leibniz-äquivalent*.

Lässt man die *Leibniz-Äquivalenz* als Kennzeichen von Identität gelten, so betreffen diese Beschreibungen der Raumzeit die gleiche physikalische Situation. Grundlage für das hier zur Anwendung kommende Kriterium physikalischer Identität ist Leibniz' *Prinzip der Identität des Ununterscheidbaren (Principium identitatis indiscernibilium).*[92] Die durch Verschiebung aller physikalischen Felder hervorgerufene Situation wird, da keine beobachtbaren Unterschiede zwischen ihnen vorliegen, sie also *Leibniz-äquivalent* sind, mit der Situation ohne Verschiebung physikalisch gleichgesetzt. Es wird hier also, im Sinne des Leibnizschen Identitätskriteriums, nach der Punkttransformation keine andere Raumzeit beschrieben, sondern nur eine weitere Beschreibung der gleichen Raumzeit hinzugewonnen.

*"Associated with d are new object fields* $d^*O_1, d^*O_2, ... , d^*O_N$ *on* **M** *that, in general, will be different from the original object fields, i.e., there will be points* $p \in M$ *such that* $d^*O_i(p) \neq O_i(p)$ *for some i. It is consistent with formal general covariance that (***M***,* $d^*O_1, d^*O_2, ... , d^*O_N$*) and (***M***,* $O_1, O_2, ... , O_N$*) correspond to different physical situations. Substantive general covariance denies that this is so and maintains that, once again, different but equivalent descriptions of the same physical situation are being produced."* (Earman (2006) 447)

Diese Identitätsausweisung aufgrund der *Leibniz-Äquivalenz* ist neben dem Unterschied zwischen simpler Koordinatentransformation und echter Punkttransformation die zweite entscheidende Komponente der substantiellen Deutung der allgemeinen Kovarianz der Allgemeinen Relativitätstheorie. Die *Leibniz-Äquivalenz* wird als Kennzeichen für physikalische Identität verwendet. – Damit entspricht aber die Transformation durch einen aktiven Diffeomorphismus letztlich einer reinen *Eichtransformation*, der selbst kein physikalischer Gehalt zukommt. Die Identifizierbarkeit der durch einen aktiven Diffeomorphismus ineinander überführbaren Beschreibungen der Raumzeit aufgrund ihrer *Leibniz-Äquivalenz* führt zur Deutung der aktiven Diffeomorphismusinvarianz als *Eichinvarianz*. Und die Allgemeine Relativitätstheorie lässt sich als *Eichtheorie* deuten. – Dies gilt zumindest, wenn man eine so allgemeine Begriffsfassung von Eichtheorien verwendet, wie die folgende:[93]

---

[92] Welche Konsequenzen es hat, wenn man die *Leibniz-Äquivalenz* als Kriterium für physikalische Identität zurückweist, wird weiter unten zu erörtern sein. Eine der prominentesten Formen der Ablehnung der *Leibniz-Äquivalenz* als Kriterium für Identität findet sich etwa im Kontext des Haecceitismus. Hier wird einer Entität eine primitive Dasheit ('*haecceitas*') zugesprochen, jenseits jeder beobachtbaren Eigenschaft. Diese *haecceitas* individuiert die entsprechende Entität, ohne dass auf eine weitere Eigenschaft, etwa eine beobachtbare, zurückgegriffen werden müsste.

[93] Dass es auch andere Verwendungsweisen des Begriffs 'Eichtheorie' gibt, bleibt davon unbenommen:

*"These are theories in which the physical system being dealt with is described by more variables than there are physically independent degrees of freedom. The physically meaningful degrees of freedom then reemerge as being those invariant under a transformation connecting the variables (gauge transformation). Thus, one introduces extra variables to make the description more transparent and brings in at the same time a gauge symmetry to extract the physically relevant content."* (Henneaux / Teitelboim (1992) Preface)

Eichtheorien arbeiten demzufolge, entweder um eine transparentere oder praktikablere Darstellung zu erreichen oder um überhaupt eine physikalische Beschreibung zu ermöglichen, mit unphysikalischen Redundanzen in der Theoriestruktur: mit einer Überschussstruktur (der Eichfreiheit), der selbst kein physikalischer Gehalt zukommt.[94] Der eigentliche physikalische Gehalt der Theorie ergibt sich erst bei Eliminierung bzw. bei Sondierung der Redundanz, also bei Reduzierung um die verwendete Überschussstruktur bzw. bei ihrer entsprechenden Berücksichtigung in der Deutung der Theorie.

Für viele Eichtheorien wäre eine redundanzfreie Beschreibung sehr unpraktikabel. Man verwendet die Eichfreiheitsgrade, um (wenngleich physikalisch gegenstands- und wirkungslose) Symmetrien, und damit bestimmte mathematische Instrumentarien, in der Beschreibung physikalischer Systeme ausnutzen zu können. Im Rahmen der Allgemeinen Relativitätstheorie geht die Motivation für die als Eichinvarianz deutbare substantielle Form der Kovarianz jedoch über das Praktikable hinaus. Wie noch zu verdeutlichen sein wird, ist bisher nicht einmal im Ansatz klar, wie eine redundanzfreie Beschreibung aussehen könnte.

Solange man aber innerhalb der Allgemeinen Relativitätstheorie mit der redundanzbehafteten eichtheoretischen Erfassung physikalisch möglicher Situationen arbeiten muss, bleibt die *Leibniz-Äquivalenz* als Kriterium für physikalische Identität das entscheidende Instrument für die Sondierung der vorliegenden Überschussstruktur. Der als reine Eichtransformation interpretierte aktive Diffeomorphismus überführt der eichinvarianten Deutung zufolge eine Beschreibung der entsprechenden physikalischen Situation in eine andere Beschreibung der gleichen physikalischen Situation. Die Punkttransformation entspricht also einer Transformation innerhalb einer Äquivalenzklasse von Beschreibungen der gleichen physikalischen Situation.[95] Dass es solche verschiedenen Beschreibungen der gleichen physikalischen Situation gibt, ist gerade die Folge der Redundanz bzw. Über-schussstruktur innerhalb des theoretischen Instrumentariums, wie es die Allgemeine Relativitäts­theorie liefert. Die Theorie verwendet zur Beschreibung eines physikalischen Systems mehr unab­hängige dynamische Variablen als es echte physikalische Freiheitsgrade gibt. Der aktive Diffeo­morphismus vermittelt nur zwischen verschiedenen Beschreibungen der gleichen physikalischen Situation. Der Transformation selbst kommt kein physikalischer Gehalt zu. Die aktive Diffeo­morphismusinvarianz reflektiert und erfasst gerade die Redundanz innerhalb des theoretischen Instru­mentariums der Allgemeinen Relativitätstheorie. Sie ist genau in dem Sinne als Eichinvarianz zu verstehen, in dem dieser Begriff für die Erfassung und Ausweisung unphysikalischer Redundanzen und Überschussstrukturen innerhalb einer Theorie steht.

---

*"Now some writers want to reserve the label 'gauge theory' for Yang-Mills theories. This seems to me to be a merely terminological matter – if you do not wish to call [the general theory of relativity] a gauge theory because it is not Yang-Mills, that is fine with me; but please be aware that the constrained Hamiltonian formalism provides a perfectly respectable sense in which the standard textbook formulations of [the general theory of relativity] using tensor fields on differentiable manifolds does contain gauge freedom."* (Earman (2002b) 8)
Zum '*constrained Hamiltonian formalism*' siehe weiter unten den Abschnitt *Exkurs: Zur Hamiltonschen Formulierung der Allgemeinen Relativitätstheorie* bzw. Earman (2002b).

[94] Welche spezifischen unphysikalischen Freiheitsgrade in der Allgemeinen Relativitätstheorie zur Anwendung kommen, lässt sich, ebenso wie ihr Status als Eichtheorie, genauer im Kontext der Hamiltonschen Formulierung der Theorie einschätzen. Siehe weiter unten.

[95] Zur Problematik der Eichinvarianz interpretierten Diffeomorphismusinvarianz sowie schliesslich zum *Lochargument*, in dem die Diffeomorphismusinvarianz der Allgemeinen Relativitätstheorie als Motivation für die These der Hintergrundunabhängigkeit der Theorie eine weitere argumentative Komponente erhält, siehe weiter unten die Erörterung der Substantialismus-Relationalismus-Debatte sowie die Ausführungen zur Hamiltonschen Darstellung der Allgemeinen Relativitätstheorie.

Die *substantielle (starke) Form der allgemeinen Kovarianz* umschliesst somit hinsichtlich ihrer konzeptionellen Abgrenzung zur schwachen Allerweltskovarianz die folgenden *Komponenten*:

–   den *Unterschied* zwischen einer simplen *Koordinatentransformation* (passiver
    Diffeomorphismus) und einer echten *Punkttransformation* (aktiver Diffeomorphismus)
    – und die Identifizierung der letzteren als Invarianz der Theorie,

–   die *Leibniz-Äquivalenz* der mittels Punkttransformation (aktive Diffeomorphismen)
    ineinander überführbaren Modelle (Beschreibungen) der Raumzeit als Kriterium für die
    Identität der physikalischen Situation, auf die sich diese Beschreibungen beziehen,

–   die Deutung der aktiven Diffeomorphismusinvarianz als *Eichinvarianz* aufgrund der
    *Leibniz-Äquivalenz* der durch Diffeomorphismen ineinander überführbaren
    Beschreibungen.

                                    *

Das Besondere an der substantiellen Form der allgemeinen Kovarianz besteht nicht zuletzt darin, dass sie im Gegensatz zur rein formalen Variante, der Koordinatenunabhängigkeit einer Theorie, gerade nicht für nahezu jede Theorie erfüllt ist oder zumindest erreicht werden kann.

> *"[...] the substantive requirement of general covariance lies in the demand that diffeomorphism inva­riance is a gauge symmetry of the theory at issue. This requirement is termed 'substantive' because it is* not *automatically satisfied by a theory that is formally generally covariant, i.e., a theory whose equations of motion / field equations are written in generally covariant coordinate notation or, even better, in coordinate-free notation."* (Earman (2006) 444)

Vielmehr findet sich die substantielle Form der allgemeinen Kovarianz (soweit es die etablierten Theorien und nicht etwa den Bereich der Ansätze zur Quantengravitation betrifft) ausschliesslich in der Allgemeinen Relativitätstheorie. Darüberhinausgehend gilt sie nur noch für einige exotische Varianten bzw. Vorläufer der Allgemeinen Relativitätstheorie, wie etwa die skalare Gravitations­theorie von Nordström.

Die aktive Diffeomorphismusinvarianz ist eine sehr spezifische Eigenschaft der Dynamik, wie sie von der Allgemeinen Relativitätstheorie bzw. von den Einsteinschen Feldgleichungen beschrieben wird: Wenn ein Modell (eine Beschreibung) einer Raumzeit eine Lösung der Einsteinschen Feld­gleichungen ist, so führt ein auf dieses Modell angewandter aktiver Diffeomorphismus immer zu einem weiteren Modell, welches ebenfalls eine Lösung der Einsteinschen Feldgleichungen ist. Solche Modelle von Raumzeiten, die über aktive Diffeomorphismen ineinander überführbar sind, sind grundsätzlich physikalisch (hinsichtlich beobachtbarer Grössen) ununterscheidbar. Sie erfüllen gerade die *Leibniz-Äquivalenz*. Genau dies macht die aktive Diffeomorphismusinvarianz aus, durch die sich die Allgemeine Relativitätstheorie vor anderen Theorien auszeichnet.

> *"This is a nontrivial property of a system of differential equations, and should not be confused with the mathematical possibility of formulating a theory using tensors – a property with little physical relevance [...]."* (Westman / Sonego (2007) 1) – *"General relativity is distinguished from other dynamical field theories by its invariance under* active *diffeomorphisms. Any theory can be made invariant under* passive *diffeomorphisms. Passive diffeomorphism invariance is a property of the* formulation *of a dynamical theory, while active diffeomorphism invariance is a property of the dynamical theory* itself. *Invariance under smooth displacements of the* dynamical *fields holds only in general relativity and in any generally relativistic theory. It does not hold in [...] any [...] theory on a fixed (flat or curved) background."* (Gaul / Rovelli (2000) 30)

Die Allgemeine Relativitätstheorie hebt sich insbesondere auch in der Weise von anderen physikalischen Theorien ab, dass sie im Gegensatz zu diesen gar nicht anders als allgemein-kovariant formulierbar ist. Sie ist im Gegensatz zu diesen anderen Theorien mit einem festen raumzeitlichen Hintergrund prinzipiell unvereinbar.

> *"Das Auftreten der Kovarianzgruppe als Symmetriegruppe der [Allgemeinen Relativitätstheorie] ist aber wohl zu unterscheiden von ihrem Auftreten im Rahmen der trivialen Koordinatenkovarianz jeglicher Raumzeit-Theorien! Demnach besitzt das Kovarianzprinzip also sowohl eine triviale als auch eine nicht-triviale Komponente. Letztere besteht in der Aussage, dass die [Allgemeine Relativitätstheorie] keine absoluten Objekte kennt – oder um es noch pointierter auszudrücken: Zwar müssen sich alle grundlegenden physikalischen Theorien allgemein kovariant formulieren lassen, das ist trivial, die [Allgemeine Relativitätstheorie] ist aber dadurch charakterisiert, dass sie* gar keine andere *als eine kovariante Formulierung gestattet!"* (Lyre (2004) 132)

Die substantielle Form der allgemeinen Kovarianz der Allgemeinen Relativitätstheorie bringt damit letztlich nichts anderes als die schon konstatierte Hintergrundunabhängigkeit der Theorie zum Ausdruck, die sich ebenso in der Dynamizität der Raumzeit und der Identifizierung von Gravitation und Raumzeitmetrik widerspiegelt. Auch diese Hintergrundunabhängigkeit kommt unter den etablierten Theorien nur der Allgemeinen Relativitätstheorie zu.

> *"Das Prinzip der allgemeinen Kovarianz hat eine triviale und eine nicht-triviale Komponente und kann als Eichprinzip rekonstruiert werden. Sein nicht-trivialer physikalischer Gehalt kann auf mehrfache Weise zum Ausdruck gebracht werden: a) als das Fehlen absoluter Objekte in der [Allgemeinen Relativitätstheorie], b) als Verschmelzungseigenschaft des Faserbündels und c) als Universalität der gravitativen Wechselwirkung."* (Lyre (2004) 137)[96]

*

---

[96] Lyre verwendet den Faserbündelansatz, beruhend auf der Lagrange-Formulierung der Allgemeinen Relativitätstheorie, die er für durchsichtiger hält als die im folgenden noch zu bemühende und für die kanonische Quantisierung der Allgemeinen Relativitätstheorie bedeutsame Hamiltonsche Formulierung.

Dass es sich bei der aktiven Diffeomorphismusinvarianz der Allgemeinen Relativitätstheorie nicht einfach nur um ein theoretisches Artefakt handelt, sondern diese vielmehr integraler Bestandteil einer empirisch adäquaten Beschreibung ist, untermauern die vielfältigen empirischen Belege, die sich über die Jahrzehnte hinweg für die Allgemeine Relativitätstheorie ergeben haben. Im Besonderen gilt dies für empirische Daten, die differentiell im Hinblick auf die Implikationen der aktiven Diffeomorphismusinvarianz gedeutet werden können. Solche differentiell deutbaren Belege lassen sich etwa aus dem Verhalten von Pulsardoppelsternsystemen gewinnen; dieses lässt Rückschlüsse auf die tatsächliche Energieabstrahlung des jeweiligen Systems und mithin auf die Zahl der Freiheitsgrade von Gravitationswellen zu:[97]

> *"The observations that show that gravitational radiation carries energy away from binary pulsars in two degrees of freedom of radiation, exactly as predicted by Einstein's theory, may be considered the experimental death blow to the* absolute *point of view. The fact that two, and not five, degrees of freedom are observed means that the gauge invariance of the laws of nature includes spacetime diffeomorphism invariance. This means that the metric is a completely dynamical entity, and no component of the metric is fixed and non-dynamical. / As argued by Einstein and many others since, the diffeomorphism invariance is tied directly to the background independence of the theory. This is shown by the hole argument [...]. / Thus, classical general relativity is background independent. The arena for its dynamics is no spacetime, instead the arena is the configuration space of all the degrees of freedom of the gravitational field, which is the metric modulo diffeomorphisms."* (Smolin (2003) 12)[98]

Dies legt erst einmal nahe, die aktive Diffeomorphismusinvarianz der Allgemeinen Relativitätstheorie als wichtige Komponente bei der Entwicklung einer Theorie der Quantengravitation anzusehen – zumindest, solange sich keine entscheidenden Gegenargumente geltend machen lassen.

<center>*</center>

Eine der diesbezüglich möglicherweise entscheidenden Implikationen der substantiellen Form der allgemeinen Kovarianz der Allgemeinen Relativitätstheorie bzw. ihrer aktiven Diffeomorphismusinvarianz wurde jedoch im Vorausgehenden noch nicht ausreichend deutlich gemacht. Und es ist gerade diese Implikation, mittels derer sich die Rückbindung an die Behauptung erreichen lässt, dass mit der Allgemeinen Relativitätstheorie eine relationalistische Auffassung hinsichtlich der Raumzeit einhergeht:

Wenn man die *Leibniz-Äquivalenz* als Kriterium für physikalische Identität heranzieht und damit der aktiven Diffeomorphismusinvarianz der Allgemeinen Relativitätstheorie den Status einer Eichinvarianz zuspricht, hat dies – wie im Nachfolgenden noch weitergehend zu motivieren sein wird – zur Folge, dass die Annahme, die Raumzeitpunkte seien die Träger von physikalischen Eigenschaften, jede Grundlage verliert. Vielmehr kommen als Träger physikalischer Eigenschaften bestenfalls noch Äquivalenzklassen von raumzeitlichen Modellen, die durch einen Diffeomorphismus ineinander überführt werden können, in Frage. Die Punkte der Mannigfaltigkeit – und damit die raumzeitliche Mannigfaltigkeit selbst – verlieren jede physikalische Bedeutung. Sie sind, wenn das Kriterium der *Leibniz-Äquivalenz* erst einmal zur Anwendung gekommen ist, nur noch Teil der un-physikalischen Überschussstruktur innerhalb einer eichtheoretischen, redundanzbehafteten Erfassung der tatsächlichen physikalischen Gegebenheiten.

---

[97] Zum Zusammenhang zwischen aktiver Diffeomorphismusinvarianz und der für das Gravitationsfeld vorliegenden (und damit auch für Gravitationswellen zu erwartenden) Zahl physikalisch wirksamer Freiheitsgrade, siehe etwa Kiefer (2004), S. 114f.
[98] Zum *Lochargument* siehe weiter unten.

> *"[...] in the mathematics of classical [general relativity] we employ a background 'spacetime' manifold and describe the fields as living on the manifold. However, the diffeomorphism invariance of the theory demands that the localization on this manifold is pure gauge. That is, it is physically irrelevant. The manifold is just an artifice for describing a set of fields and other physical objects whose only 'localization' is with respect to one another. / A state of the universe does not correspond to one field configuration over the spacetime manifold M. It corresponds to an equivalence class under active diffeomorphisms of field configurations. In fact, M has no physical interpretation. It is a mathematical device without physical counterpart. [...] The only possibility of locating points is with respect to the dynamical fields and particles of the theory itself."* (Rovelli (2006) 31)

Die Interpretation der aktiven Diffeomorphismusinvarianz als Eichinvarianz wird insofern von einigen Autoren als entscheidendes Kennzeichen eines Relationalismus hinsichtlich der von der Allgemeinen Relativitätstheorie beschriebenen Raumzeit angesehen. Eine solche Einschätzung beruht vor allem darauf, dass eine substantialistische Deutung der Raumzeitpunkte der Mannigfaltigkeit (und mithin der Mannigfaltigkeit selbst) insbesondere infolge ihrer Unvereinbarkeit mit der *Leibniz-Äquivalenz* als Identitätskriterium (und Grundlage der eichinvarianten Deutung) sehr problematisch wird und nur um den Preis sehr kurioser Konsequenzen oder begrifflicher Konstrukte, von denen noch die Rede sein wird, aufrechterhalten werden kann.

Bringt man hingegen die eichinvariante Deutung der Allgemeinen Relativitätstheorie als naheliegendste und konsequenteste Sicht auf die Implikationen dieser immerhin empirisch bestens gestützten Theorie zur Anwendung, so ist die Raumzeit nichts mehr und nichts anderes als der Ausdruck der relationalen Bezüge zwischen den eigentlichen physikalischen Systemen bzw. Ereignissen. Sie ist nichts anderes als der Ausdruck der relationalen Bezüge zwischen den vorhandenen physikalischen Feldern bzw. den diese Felder konstituierenden Ereignissen.

> *"This substantive version of general covariance is satisfied in general relativity: arbitrary diffeomorphisms – applied actively – are gauge transformations and therefore do not change the physical situation. This is because there is no fixed spatiotemporal background structure against which the action of the diffeomorphisms can be set out. It is only the coincidence relations between the dynamical fields (including the metric field) that count."* (Dieks (2006) xii)

Auf diese Weise wird die eichinvariante Deutung zur entscheidenden Grundlage für eine relationalistische Auffassung bezüglich der Raumzeit. – Die relationalistischen Implikationen der aktiven Diffeomorphismusinvarianz der Allgemeinen Relativitätstheorie sollten sich nach Ansicht einiger Physiker dann aber auch unmittelbar in einer konzeptionell adäquaten Theorie der Quantengravitation niederschlagen. Ein besonderes Interesse haben in dieser Hinsicht die Vertreter der *Loop Quantum Gravity*, die eine direkte (kanonische) Quantisierung der Allgemeinen Relativitätstheorie anstreben und die schon im klassischen Fall gewonnenen Einsichten hinsichtlich der Relationalität der Raumzeit, soweit möglich, unmittelbar auf eine entsprechende Quantentheorie zu übertragen versuchen.[99]

---

[99] Im Kontext der *Loop Quantum Gravity* ist die aktive Diffeomorphismusinvarianz fest im Formalismus des Ansatzes verankert. Siehe Kap. 4.4.

> *"But the world is not formed by a fixed background over which things happen. The background itself is dynamical. [...] The absence of a fixed background in nature (or active diffeomorphism invariance) is the key general lessons we have learned from gravitational theories."* (Rovelli (1998) 5) – *"Therefore, what we need in quantum gravity is* a relational notion of a quantum spacetime.*"* (Rovelli (1998) 19)

So klar diese Sachlage vielleicht auf den ersten Blick erscheinen mag: Im Kontext der Wissenschaftsphilosophie der Physik ist die These der Relationalität der Allgemeinen Relativitätstheorie hinsichtlich ihrer Raumzeitkonzeption durchaus umstritten. Einerseits hängt einiges daran, ob man sich tatsächlich gezwungen sieht, die aktive Diffeomorphismusinvarianz der Allgemeinen Relativitätstheorie, unter Anerkennung der Forderung nach *Leibniz-Äquivalenz*, als Eichinvarianz zu interpretieren, oder, ob man dies – um den Preis der schon beschworenen recht kuriosen Konsequenzen – zu verneinen bereit ist. Eine wiederum andere Frage ist es, ob man die Deutung der aktiven Diffeomorphismusinvarianz als Eichinvarianz, wenn man diese akzeptieren sollte, schon als konstitutiv und hinreichend für den Relationalismus ansieht. Die Interpretationslandschaft hinsichtlich der Allgemeinen Relativitätstheorie ist vielgestaltig und hochgradig verästelt. Hinsichtlich der jeweils zugrundegelegten Begrifflichkeiten herrscht keineswegs Konsens. Die Frage der relationalistischen Implikationen der Theorie lässt sich somit nicht ohne weiteres entscheiden, solange man den Kontext dieser Frage nicht genauer sondiert und solange man vor allem nicht spezifiziert, was genau hinter dem Relationalismus bzw. seinem Gegenspieler, dem Substantialismus, an Begrifflichkeiten, Thesen und Implikationen steht. Diese Probleme und Fragestellungen sind Gegenstand der gleichermassen schillernden wie vielfach geschmähten, aber immer noch sehr lebendigen *Substantialismus-Relationalismus-Debatte*. Und in diesem Kontext ist es, wie sich zeigen wird, ganz und gar nicht in jedem Falle eindeutig und klar, ob eine bestimmte Einsicht und ihre Konsequenzen nun besser unter der Flagge des *Relationalismus* oder der des *Substantialismus* zu verorten wären.

Bei der Erörterung der Substantialismus-Relationalismus-Debatte sollte man sich – ob der Unschärfe und Mehrdeutigkeit der Begrifflichkeiten des 'Substantialismus' und des 'Relationalismus' – also von Zeit zu Zeit daran erinnern, dass diese Begrifflichkeiten letztlich vielleicht ohnehin zweitrangig sind. Entscheidend sind vielmehr, insbesondere für die Bemühungen um eine Theorie der Quantengravitation, die entsprechenden inhaltlichen Problemlagen, die zu den jeweiligen begrifflichen Einschätzungen führen.

## Substantialismus vs. Relationalismus

Der Substantialismus hinsichtlich der Raumzeit entspricht der These, dass diese gegenüber den in ihr befindlichen Dingen und den in ihr stattfindenden Prozessen einen eigenständigen ontologischen Status besitzt, dass sie eine eigenständige Substanz ist.

> *"[...] taking substantivalism to be the position that views spacetime to be an entity which exists* over and above *any material objects it might contain; or, in Earman's words, 'prior to the objects it contains' instead of being 'nothing but (might be constituted by, might be reducible to) the mutual relations among coexistent objects' ([Earman, 1989], p. 289)."* (Rickles (2008a) 3)

Diese These wird oft dahingehend erweitert, dass auch für die Teile der Raumzeit, die Raumzeitpunkte, ein eigenständiger ontologischer Status behauptet wird.[100]

> *"In debates concerning the nature of spacetime,* substantivalism *is simply the view that spacetime and its pointlike parts exist as fundamental, substantial entities."* (Pooley (2006a) 99)

Eines der oft angeführten Argumente für den Substantialismus stützt sich auf den Formalismus der Feldtheorien. Sowohl die klassischen Feldtheorien als auch die Quantenfeldtheorien schreiben ihre Feldwerte bzw. ihre Feldoperatoren jedem einzelnen Raumzeitpunkt zu. Im Kontext dieses modelltheoretischen Instrumentariums erscheinen Raumzeitpunkte als die basalen Entitäten, denen spezifische Eigenschaften, etwa Feldwerte, zugeschrieben werden können. Dem ersten Anschein nach wird auch das metrische Feld in der Allgemeinen Relativitätstheorie, im Sinne einer klassischen feldtheoretischen Auffassung, erst einmal den Punkten der Raumzeit zugeschrieben. Erst die Berücksichtigung der Implikationen der aktiven Diffeomorphismusinvarianz der Theorie führt zu der Einsicht, dass diese Auffassung nicht unproblematisch ist. – Auch hinsichtlich der anderen klassischen Feldtheorien und der Quantenfeldtheorien könnte es sich schliesslich herausstellen, dass die mit ihrem modelltheoretischen Apparat einhergehende Auffassung, Raumzeitpunkte als basale Entitäten zu sehen, denen Eigenschaften zugeschrieben werden, vielleicht nichts mehr als ein modelltheoretisches Artefakt des feldtheoretischen Formalismus ist.

> *"[...] we should be wary of taking as the basic objects of our ontology (according to some theory) those items that are postulated as the initial elements in a mathematical presentation of a theory. For it might be just happenstance of our formulation of the theory that these objects 'come first': a happenstance avoided by another formulation that can be agreed, or at least argued, to be better."* (Butterfield (2002) 295)

Ein weiteres, oft angeführtes Argument für den Substantialismus setzt bei den Lösungen der Allgemeinen Relativitätstheorie mit unterschiedlicher Metrik bei gleicher Materieverteilung an. So wird etwa die Möglichkeit des Auftretens von Gravitationswellen als Indiz für den eigenständigen ontologischen Status der Raumzeit angeführt. Spätestens die mit der Allgemeinen Relativitätstheorie kompatiblen Vakuumlösungen zeigen nach Auffassung der Substantialisten, dass Raumzeit als eigenständige Substanz völlig ohne das Vorhandensein von Materie auskommen kann. – Hier liesse sich jedoch ohne weiteres einwenden, dass die vermeintlich leere Raumzeit der Vakuumlösungen nicht wirklich leer ist; vielmehr enthält sie mindestens das metrische Feld, welches dem Gravitationsfeld entspricht: also eine dynamische, energie- und impulstragende Entität.

> *"[...] energy and momentum are carried by the metric in a way that forces its classification as part of the contents of spacetime."* (Earman / Norton (1987) 519)

---

[100] Manche Autoren setzen den Substantialismus mit einem Realismus hinsichtlich dieser raumzeitlichen Substanz bzw. ihrer Konstituenten gleich.
> *"The modern* substantivalist *position is a statement of spatio-temporal realism: its adherents claim that the individual* points *of the manifold, for a given solution of the Einstein equations, represent* directly *the physical points of space-time, as they would occur in the actual or in some possible world."* (Pauri / Vallisneri (2002) 7)

Diese Gleichsetzung ist jedoch insofern nicht unproblematisch, als auch dezidiert anti-substantialistische Auffassungen noch nicht notwendigerweise der Raumzeit den Realstatus absprechen, sondern vielmehr nur ihre eigenständige Existenzweise.

Und auch Gravitationswellen sind genau solche dynamischen, energie- und impulstragenden Entitäten. – Alles läuft schliesslich auf die Frage hinaus, was nun zur substantiell verstandenen Raumzeit und was zu ihren Inhalten gezählt wird.

> *"[...] in the general theory of relativity the metrical field does become dynamical, so that within this theory the state of the universe may be considered as completely specified by the coincidence relations between physical systems. The plausibility of this viewpoint obviously depends on whether one is prepared to go along with accepting the metrical field as a physical system that is on a par with the matter fields. If one does, general relativity appears as the vindication of relationalism. If one does not, general relativity appears as not amiable to relationalism after all: the theory allows possible universes in which there are no matter fields, so that in those universes there is only empty spacetime."*
> (Dieks (2006) xi)

Die These des Substantialismus, dass die Raumzeit gegenüber den in ihr befindlichen Dingen und den in ihr stattfindenden Prozessen einen eigenständigen ontologischen Status besitzt, lässt sich also eigentlich erst verhandeln, wenn geklärt ist, was zu diesen Dingen gezählt werden soll und was für die Raumzeit bleibt.

## *Varianten des Substantialismus*[101]

Die im Rahmen der Allgemeinen Relativitätstheorie verwendeten mathematischen Modelle der Raumzeit und der auf ihr definierten Grössen enthalten drei Komponenten:

– die vierdimensionale, kontinuierlich differenzierbare (pseudo-Riemannsche) Punktmannigfaltigkeit: eine Ansammlung von Punkten mit definierter Topologie und Dimensionalität,

– den auf der gesamten Mannigfaltigkeit definierten metrischen Tensor, der die geometrischen und chronometrischen Beziehungen zwischen den Punkten der Mannigfaltigkeit bzw. den Ereignissen, die in der Raumzeit stattfinden, – und gleichzeitig das Gravitationsfeld – repräsentiert,

– den Energie-Spannungs-Tensor, der ebenso auf der gesamten Mannigfaltigkeit definiert ist und der die Materie und die Energiedichten aller nicht-gravitativen Felder repräsentiert.

Der Energie-Spannungs-Tensor gehört unzweifelhaft zum Inhalt der Raumzeit. Wie ist es aber um die Metrik bestellt? – Insbesondere hinsichtlich dieser Frage unterscheiden sich die verschiedenen Varianten des Substantialismus.

*

Die erste Möglichkeit, die sich anbietet, besteht darin, die Metrik, da sie das Gravitationsfeld – als dynamische, energie- und impulstragende Entität – repräsentiert, zu den Inhalten der Raumzeit zu zählen und die Raumzeit mit der raumzeitlichen Mannigfaltigkeit gleichzusetzen. Alle Komponenten ausser der raumzeitlichen Mannigfaltigkeit selbst, also nicht zuletzt auch das metrische Feld, das die chronogeometrische Struktur der Raumzeit erfasst, sind dynamische Entitäten, die in ihrer Dynamik von partiellen Differentialgleichungen beschrieben werden. Diese Entitäten lassen sich prima facie eigentlich nur als Inhalte der Raumzeit deuten. So bleibt für die Raumzeit selbst, von ihren Inhalten getrennt, nur die raumzeitliche Mannigfaltigkeit.

---

[101] Ein Überblick über die Varianten des Substantialismus findet sich etwa in Hoefer (1996).

> *"The advent of general relativity has made most compelling the identification of the bare manifold with spacetime. For in that theory geometric structures, such as the metric tensor, are clearly physical fields in spacetime."* (Earman / Norton (1987) 519)

Ein entsprechender *Substantialismus bezüglich der raumzeitlichen Mannigfaltigkeit* spricht dann dieser Mannigfaltigkeit und den individuierten Raumzeitpunkten dieser Mannigfaltigkeit eigenständigen ontologischen Status zu. – Dass ein Basal-Substantialismus dieser Form für den Substantialisten nicht gerade die erste Wahl ist, liegt an den unmittelbaren Problemen, in die er gerät. Entsprechend wird die Auffassung, dass die Raumzeit mit der raumzeitlichen Mannigfaltigkeit gleichzusetzen ist, vor allem von denen vertreten, die zeigen wollen, dass ein Substantialismus hinsichtlich der Raumzeit nicht haltbar ist.

Die Probleme des Substantialismus bezüglich der raumzeitlichen Mannigfaltigkeit bestehen vor allem darin, dass dieser, wie zuvor schon deutlich geworden sein könnte, nicht mit der aktiven Diffeomorphismusinvarianz bzw. der substantiellen Form der allgemeinen Kovarianz verträglich ist, wie sie in der Allgemeinen Relativitätstheorie realisiert sind, – zumindest dann nicht, wenn man nicht bereit ist, den Bezug auf unbeobachtbare Entitäten und einen ziemlich unmotivierten Indeterminismus für die Allgemeine Relativitätstheorie mit ebenso unbeobachtbaren Konsequenzen hinzunehmen. Letzteres bedarf nun einer weiteren Erläuterung:

Die aktive Diffeomorphismusinvarianz der Allgemeinen Relativitätstheorie entspricht, wie zuvor erörtert, einer Invarianz unter echten Punkttransformationen, die einer Verschiebung von physikalischen Grössen, Feldern und Ereignissen von einem Raumzeitpunkt auf einen anderen entsprechen. Durch diese aktiven Diffeomorphismen werden keine beobachtbaren Grössen verändert. Dies hat aber zur Folge, dass der Bezug auf individuierte Raumzeitpunkte einem Bezug auf unbeobachtbare Entitäten entspricht. Er entspricht damit einer Zurückweisung des auf der *Leibniz-Äquivalenz* beruhenden Identitätskriteriums. Für den Substantialismus bezüglich der raumzeitlichen Mannigfaltigkeit sind zwei Raumzeiten (bzw. Modelle der Raumzeit), die sich durch einen Diffeomorphismus ineinander überführen lassen, die also insbesondere die physikalischen Felder im allgemeinen unterschiedlichen Raumzeitpunkten zuordnen, als unterschiedliche physikalische Raumzeiten zu verstehen, wenngleich diese sich in keiner beobachtbaren Weise unterscheiden und sich in Bezug auf die Dynamik völlig gleich verhalten. Der Substantialismus bezüglich der raumzeitlichen Mannigfaltigkeit impliziert auf diese Weise einen Bezug auf grundsätzlich unbeobachtbare Grössen: auf Unterschiede, die grundsätzlich keine beobachtbaren Folgen nach sich ziehen.

Dass er darüberhinausgehend einen letztlich völlig unmotivierten (oder nur auf der Grundlage seiner Raumzeitmetaphysik motivierten) Indeterminismus hinsichtlich der Allgemeinen Relativitätstheorie impliziert, wird schliesslich im *Lochargument* deutlich:[102] Das Lochargument ist wiederum eine direkte Implikation der aktiven Diffeomorphismusinvarianz der Allgemeinen Relativitätstheorie: Man konstruiere einen Diffeomorphismus, der (mit einem stetigen Anschluss an den Bereich ausserhalb) nur innerhalb eines umgrenzten Raumzeitbereichs (dem 'Loch'), in dem es nur das metrische bzw. gravitative Feld gibt, zu einer Verschiebung dieses metrischen Feldes in Bezug auf die jeweiligen Raumzeitpunkte führt:

> "Let us assume that $M_4$ contains a hole H: that is, an open region where all the non-gravitational fields vanish. On $M_4$ we can define an active diffeomorphism $D_A$: $x' = D_A\,x$ that re-maps the points inside H, but blends smoothly into the identity map outside H and on the boundary. If we consider the transformed tensor fields $g' \equiv D^*_A\,g$ (where $D^*_A$ denotes the action of a diffeomorphism on tensor fields) then, by construction, for any point $x \in H$ we have (in the abstract tensor notation) $g'(D_A\,x) = g(x)$, but of course $g'(x) \neq g(x)$ (in the same notation). The crucial fact to keep in mind at this point is that the Einstein equations are generally covariant: this means that if g is one of their solutions, so is the drag-along field g'." (Dorato / Pauri (2006) 127)

---

[102] Siehe Earman (1986, 1989), Earman / Norton (1987), Norton (1987, 1988, 2004).

Wählt man den Diffeomorphismus entsprechend einer solchen Lochtransformation, so können aus gleichen Ausgangszuständen – gemessen am Unterscheidungskriterium des *Substantialismus bezüglich der raumzeitlichen Mannigfaltigkeit* – unterschiedliche zeitliche Entwicklungen resultieren, deren Unterschiedlichkeit jedoch grundsätzlich unbeobachtbar ist. Es ist innerhalb der Allgemeinen Relativitätstheorie auf der Grundlage ihrer Dynamik nicht vollständig festgelegt, an welchen Punkten der raumzeitlichen Mannigfaltigkeit zukünftige Ereignisse stattfinden werden. Unter Verwendung der Unterscheidungskriterien des *Substantialismus bezüglich der raumzeitlichen Mannigfaltigkeit* wird die Beschreibung, welche die Allgemeine Relativitätstheorie von der Raumzeit liefert, also indeterministisch:

> "Now, if we think of the points of H as intrinsically individuated physical events, where 'intrinsic' means that their identity is independent of the metric – a claim that is associated with manifold substantivalism – then g and g' must be regarded as physically distinct solutions of the Einstein equations (after all, $g'(x) \neq g(x)$ at the same point x). This is a devastating conclusion for the causality (or, in other words, the determinism) of the theory, because it implies that, even after we completely specify a physical solution for the gravitational and non-gravitational fields outside the hole – in particular, on a Cauchy surface for the initial value problem – we are still unable to predict uniquely the physical solution within the hole." (Dorato / Pauri (2006) 127)

Dieser Indeterminismus führt jedoch eben gerade zu keinen beobachtbaren Unterschieden. – Warum sollte man also von der unabhängigen Existenz der raumzeitlichen Mannigfaltigkeit ausgehen, wenn dies zu solch unnötigen (und noch dazu unbeobachtbaren) radikalen Konsequenzen führt? Für die Auffassung, dass die Allgemeine Relativitätstheorie eine indeterministische Theorie ist, sollte es – so die Kritiker des Substantialismus – schon bessere Gründe geben als die Einführung eines unproduktiven metaphysischen Konstrukts.

> "Determinism may fail, but if it fails it should fail for a reason of physics, not because of a commitment to substantial properties which can be eradicated without affecting the empirical consequences of the theory." (Earman / Norton (1987) 524)

Das Lochargument zeigt somit, dass die These der Individuierbarkeit der Raumzeitpunkte der Mannigfaltigkeit nur mit erheblichen Schwierigkeiten motiviert und aufrechterhalten werden kann. Dass die Allgemeine Relativitätstheorie mit ihrem feldtheoretischen Formalismus und ihrer modelltheo­retischen Verwendung einer differenzierbaren Punktmannigfaltigkeit auf den ersten Blick den An­schein erweckt, die Existenz individuierter Punkte nahezulegen, sollte man also vermutlich nicht zu wörtlich nehmen. Der Preis dafür wäre vermutlich deutlich zu hoch.

> *"[...] the hole argument shows that some limits must be placed on our literal reading of a successful theory. [...] The hole argument shows us that we might want to admit that there is something a little less really there than the literal reading, lest we be forced to posit physically real properties that transcend both observation and the determining power of our theory."* (Norton (2004) 14)

Der Substantialismus bezüglich der raumzeitlichen Mannigfaltigkeit ist ohnehin sehr artifiziell: Die raumzeitliche Mannigfaltigkeit alleine macht kaum die Raumzeit aus. Sie enthält nur Topologie und Dimensionalität, aber keine Lichtkegelstruktur, keine Unterscheidung in Vergangenheit und Zu­kunft und keine Festlegung hinsichtlich der raumzeitlichen Distanzen. Und in der Allgemeinen Re­lativitätstheorie gibt es ohnehin keine raumzeitliche Mannigfaltigkeit ohne metrisches Feld; besten­falls gibt es eine solche mit materiefreier Metrik, im einfachsten Fall der Minkowski-Metrik. Lässt man die Metrik gänzlich aussen vor, so gibt es nur noch die Raumzeitpunkte selbst, die infolge der aktiven Diffeomorphismusinvarianz der Allgemeinen Relativitätstheorie nicht nur physikalisch be­deutungslos, sondern vor allem unbeobachtbar sind. Warum sollte man der Mannigfaltigkeit und den sie konstituierenden Raumzeitpunkten also einen eigenständigen Existenzstatus zubilligen?

<p style="text-align:center">*</p>

Konsequenterweise versuchen die Substantialisten den soeben erörterten Problemen dadurch zu entgehen, dass sie die Begrifflichkeit der Raumzeit um weitere, über die raumzeitliche Punktman­nigfaltigkeit hinausgehende Komponenten anreichern. – Ein Versuch, dies in einer Weise zu tun, die hinsichtlich der hinzukommenden Komponenten möglichst nur solche wählt, die nicht sofort im Verdacht stehen, unzweifelhaft zu den Inhalten der Raumzeit gezählt werden zu müssen, findet sich im *Substantialismus bezüglich der Mannigfaltigkeit plus basaler (Minkowski-) Metrik*.[103] Dieser beruht offensichtlich auf der durchaus vernünftigen Auffassung, dass die eigentliche raumzeitliche Struktur von der Metrik repräsentiert wird. Von dieser Auffassung ausgehend nimmt er dann jedoch eine höchst kuriose Wendung, die letztlich zur Ursache aller seiner Probleme wird. Die Idee ist hier nämlich die, dass die Raumzeit in basalster Form (etwa als Minkowski-Hintergrundmetrik) gemein­sam mit der raumzeitlichen Mannigfaltigkeit Gegenstand einer substantialistischen These sein könnte. Die naheliegende Motivation für diese kuriose Strategie lässt sich in der Auffassung ver­muten, dass einerseits die volle Metrik wohl doch zu sehr im Verdacht steht, zu den Inhalten der Raumzeit zu zählen, und andererseits die Mannigfaltigkeit ganz ohne Metrik zu wenig ist und be­kanntermassen dem Lochargument zum Opfer fällt.

Das fundamentale Problem, auf das diese Auffassung stösst, besteht nun darin, dass eine feste Hin­tergrundmetrik letztlich mit der Dynamizität der Raumzeit in der Allgemeinen Relativitätstheorie wie mit ihrer aktiven Diffeomorphismusinvarianz völlig unverträglich ist. – Hinzu kommt, dass eine verallgemeinerte Form des Locharguments unter Umständen immer noch greift, und zwar für den Fall, dass die Metrik auch nur irgendeine Form von Symmetrie aufweist.[104]

---

[103] Siehe Maudlin (1988).
[104] Siehe Norton (1988).

*"[...] the presence of these symmetries represents a failure of the further structure to individuate fully the points of the manifold."* (Norton (1988) 60)

*

Wieso also sollte man nicht gleich alles wagen und einen umfassenden *Substantialismus bezüglich der Metrik* bzw. des metrischen Feldes vertreten?[105] Zugrunde liegt dieser Variante wiederum – und diesmal ohne Einschränkung und Kompromisse – die durchaus vernünftige Auffassung, dass die Metrik am ehesten dem entspricht, was innerhalb der Allgemeinen Relativitätstheorie die Raumzeit ausmacht.

*"The substantivalist's natural response to the hole dilemma is to insist that space-time is represented not by the bare manifold but by the manifold plus metric, by the metric space."* (Maudlin (1990) 546)

Die Tatsache, dass dem metrischen Feld Energie und Impuls zukommen, die mit anderen Feldern, etwa auch Materiefeldern, ausgetauscht werden, wird dann gar als Positivum für den Substantialismus hinsichtlich der mit dem metrischen Feld gleichgesetzten Raumzeit gedeutet:

*"If this is an energy-bearing structure, all the better for substantivalism."* (Maudlin (1990) 548)[106]

Gemeinhin werden Energie und Impuls jedoch als charakteristisch für die Inhalte der Raumzeit angesehen. Wenn man das metrische Feld aber aufgrund seiner Eigenschaften und seiner Gemeinsamkeiten mit allen anderen physikalischen Feldern zu den Inhalten der Raumzeit zählt, bleibt kein Behältnis zurück – zumindest, wenn man die raumzeitliche Mannigfaltigkeit nicht mehr als substantielle Entität ernstzunehmen bereit ist, da dies hiesse, sich auf einen unmotivierten Bezug auf unbeobachtbare Grössen und eine ebenso unmotivierte Indeterminismusthese bezüglich der Allgemeinen Relativitätstheorie einzulassen. Der Substantialismus hinsichtlich einer mit der Metrik gleichgesetzten Raumzeit wird daher in den Augen seiner Kritiker zu einer (relativ schlecht getarnten) Form von Relationalismus. – Für den Fall, dass man eine umfassende Geometrisierung im Sinne von Wheeler oder dem späten Einstein als strategisches Fernziel ins Auge fassen sollte, würde sie ihn gar, wie John Earman und John Norton betonen, zu einer ziemlich trivialen Position werden lassen.

*"[...] classifying the metric as part of the container spacetime leads to a trivialisation of the substantivalist view in unified field theories of the type developed by Einstein, in which all matter is represented by a generalised metric tensor. For there would no longer be anything contained in spacetime, so that the substantivalist view would in essence just assert the independent existence of the entire universe."* (Earman / Norton (1987) 519)

Aber auch dieser extreme und gleichzeitig hypothetische Fall einer gelingenden umfassenden Geometrisierung lässt sich vom Substantialisten ohne weiteres als Triumph seiner Behauptung umdeuten:

---

[105] Siehe etwa Maudlin (1990, 1992) sowie Bartels (1994, 1996). Manche der Ausprägungen eines Substantialismus bezüglich der Metrik formieren unter der Bezeichnung *'Metrischer Essentialismus'*.

[106] Vgl. Pooley (2006a), der aber eine im Vergleich zu Maudlin andersgeartete substantialistische Position vertritt (siehe weiter unten) und entsprechend aus gänzlichen anderen Motivationen heraus argumentiert:
*"And the substantivalist will, of course, see this as making his realism about spacetime all the more plausible: as Rovelli says, spacetime now obeys the action-reaction principle."* (Pooley (2006a) 100)

*"[...] that space-time is a substance does not become trivialized if it could be argued that space-time is the only substance."* (Maudlin (1990) 548)

Dem Substantialisten, der die Raumzeit mit der Metrik gleichsetzt, bleibt letztlich aber vor allem die Option, auch unabhängig von den Extremvisionen einer umfassenden Geometrisierung, die Unterscheidung in Raumzeit und Raumzeitinhalte überhaupt in Frage stellen:

*"[...] the metaphor of container and contained is hopelessly obscure when applied to spacetime."* (Maudlin (1990) 547)

Dass aber auch eine solche Strategie nicht unbedingt zu einer Stützung der substantialistischen Sichtweise beitragen muss, zeigen nicht zuletzt die Einwände von Robert Rynasiewicz.[107] Dieser hält die Unklarheit darüber (bzw. das Desinteresse daran), was nun zur Raumzeit und was zu ihren Inhalten zählt, für ein klares Zeichen, dass die Substantialismus-Relationalismus-Debatte (seit Newton und Leibniz) völlig ihren Gegenstand eingebüsst hat.[108]

*"There simply is not a fact of the matter here as to what should count as 'space' and what should count as a 'physical object' in the sense of the original 'space-matter' distinction that spawned the debate."* (Rynasiewicz (1996) 301) – *"The distinction between 'space' and 'matter', or alternatively, between 'container' and 'contents', necessary for the debate, though immediate to common sense and sustainable to the microlevel for the earlier participants of the debate, ultimately evaporates in the course of the developments of the physics of the aether."* (Rynasiewicz (1996) 306)

Eine neuere Strategie, mit dieser Problemlage innerhalb der Substantialismus-Relationalismus-Debatte umzugehen, besteht in dem Vorschlag, diese Debatte zugunsten einer strukturenrealistischen Sichtweise aufgeben:[109]

*"I think this is evidence in favour of the view that the time has come to forget about the 'debate' between substantivalism and relationalism, and focus on an alternative. Here I argue that structuralism offers a suitable alternative. The position involves the idea that physical systems (which I take to be characterized by the values for their observables) are exhausted by extrinsic or relational properties: they have no intrinsic properties at all! This is a consequence of background independence coupled with gauge invariance. This leads to a rather odd picture in which objects and structure are deeply entangled in the sense that, inasmuch as there are objects, any properties they possess are structurally conferred: they have no reality outside the correlations. What this means is that the objects don't* ground *the structure; they are nothing independently of the structure, which takes the form of a (gauge-invariant) correlation between (gauge variant) field values. [...] This admittedly rather wild-sounding metaphysics can be made more precise through the use of Rovelli's framework of partial and complete observables."* (Rickles (2008a) 16)[110] – *"[...] we have an interpretative underdetermination: both substantivalists and relationalists can lay claim to this setup. [...] Our response is to evade the underdetermination by adopting a structuralist metaphysics: forget points and forget individual mate-*

---

[107] Siehe Rynasiewicz (1996, 1994).

[108] Vgl. Rovelli (2007), Kap. 4.1.

[109] Siehe etwa Rickles (2008, 2008a), Dorato (2000), Rickles / French (2006), Cao (2006), vgl. Pauri / Vallisneri (2002), Lusanna / Pauri (2005), Dorato / Pauri (2006).

[110] Zu Rovellis partiellen und vollständigen Observablen siehe den Abschnitt zum *Problem der Zeit* in Kap. 4.4.

*rial fields, the structure as characterized by the equivalence class of metrics is where our ontological commitments should lie. / [...] what is called relationalism can be understood as a form of structuralism."* (Rickles / French (2006) 24)

Man kann leicht den Eindruck gewinnen, dass hier der Relationalismus tatsächlich nur einen neuen Namen bekommt. Was diese Auffassung unterstützt, ist die Tatsache, dass für die strukturenrealistische Sichtweise die Deutung der aktiven Diffeomorphismusinvarianz als Eichinvarianz konstitutiv ist. Diese Deutung wird jedoch, wie weiter unten zu erörtern sein wird, aus nachvollziehbaren Gründen gemeinhin als konstitutives Element einer relationalistischen Auffassung hinsichtlich der Raumzeit aufgefasst. Man benötigt dafür nicht unbedingt eine Umdeutung in Richtung auf einen Strukturenrealismus.

Unabhängig davon, ob man auf der Grundlage der Infragestellung einer sinnvollen Unterscheidung zwischen der Raumzeit und ihren Inhalten nun für einen Strukturenrealismus, für den Relationalismus oder für den Niedergang oder gar die gänzliche Irrelevanz der Substantialismus-Relationalismus-Debatte argumentieren möchte, ist eines klar: Für einen Substantialismus hinsichtlich einer mit dem metrischen Feld identifizierten Raumzeit nutzt diese Unklarheit in Bezug auf die Unterscheidung zwischen Raumzeit und ihren Inhalten jedenfalls gar nichts. – Sinnvoller wäre da wohl, wie es etwa Mauro Dorato und Massimo Pauri tun, diese Unterscheidung wiederzubeleben und für die Priorität des metrischen Feldes gegenüber den anderen physikalischen Feldern zu argumentieren:

*"[...] although the metric tensor field, qua physical field, cannot be regarded as the traditional empty container of other physical fields, we believe that it has ontological priority over all other fields. This pre-eminence has various reasons, but the most important is that the metric field tells all other fields how to move causally. In agreement with the general-relativistic practice of not counting the gravitational energy induced by the metric as a component of the total energy, we believe that physical spacetime should be identified with the manifold endowed with its metric, thereby leaving the task of representing matter to the stress-energy tensor."* (Dorato / Pauri (2006) 129)

Hier wird dann jedoch wiederum sehr schnell deutlich, dass die Entscheidung, ob man, aufgrund der in der Allgemeinen Relativitätstheorie vorliegenden Sachlage, eine jeweilige Interpretation der von ihr beschriebene Raumzeit dem Substantialismus oder dem Relationalismus subsumiert, vor allem von den begrifflichen Vorentscheidungen und den an die begrifflichen Alternativen geknüpften Erwartungen abhängt. Und zu diesen begrifflichen Vorentscheidungen zählt letztlich eben immer noch insbesondere die Antwort auf die Frage, was als Raumzeit und was als Inhalt der Raumzeit zählen soll.

*"[...] which is the best candidate to interpret the role of space and time in [general relativity], the manifold or the (manifold plus) the metric? Those opting for the bare manifold $M_4$ (like Earman and Norton) correctly point out that g cannot be understood as interpreting the role of the 'empty spacetime' of the traditional debate: by embodying the potential of the gravitational field, g is to be regarded as a (special) type of 'physical field'. Those opting – much more reasonably, in our opinion – for the 'manifold plus the metric field' (Maudlin 1990, Stachel 1993) also correctly point out that the metric provides the chrono-geometrical structure as well as, most significantly, the causal structure of spacetime."* (Dorato / Pauri (2006) 127)

Konsequenterweise kommt die Kritik am vermeintlichen strukturenrealistischen Ausweg aus der Substantialismus-Relationalismus-Debatte auch vor allem von Seiten eines *subtileren Substantialismus*, der gelernt hat, den Problemen eines Substantialismus hinsichtlich der Punktmannigfaltigkeit ebenso wie denen eines solchen hinsichtlich der mit dem metrischen Feld identifizierten Raumzeit zu entgehen.[111] Die hierbei herangezogen argumentativen Mittel basieren etwa auf unterschiedlichen Definitionen des Identitätsbegriffs sowie auf z.T. recht kuriosen Argumentationen, die das der analytischen Philosophie mittlerweile liebgewonnene Instrumentarium des Rekurses auf 'mögliche Welten' bemühen.

> *"[...] Butterfield portraits diffeomorphic models as different possible worlds and invokes counterpart theory to argue that at most one can represent an actual space-time."* (Pauri / Vallisneri (2002) 7)

Eine besondere Rolle spielen im Kontext dieser subtileren Formen des Substantialismus zudem haecceitistische Argumente. Der *Haecceitismus* besteht in der Zuschreibung einer primitiven (metaphysischen) Dasheit ('*haecceitas*') jenseits jeder beobachtbaren (und damit physikalisch erfassbaren) Eigenschaft. Der *Haecceitismus* weist also insbesondere die *Leibniz-Äquivalenz* als Kriterium für Identität zurück. – Die subtileren Formen des Substantialismus sollen hier aber deshalb nicht weiterverfolgt werden, weil sie sich hinsichtlich der Deutung der aktiven Diffeomorphismusinvarianz der Allgemeinen Relativitätstheorie – und insbesondere hinsichtlich ihrer physikalischen Implikationen – nicht von dem unterscheiden, was im folgenden unter der Flagge des 'Relationalismus' formieren wird. Es ist nämlich gerade die aktive Diffeomorphismusinvarianz, die in ihrer Deutung als Eichinvarianz hier als konstitutiv für den 'Relationalismus' behandelt werden wird – eine begriffliche Festlegung, die zwar nicht zwingend ist, aber infolge der argumentativen Motivationen, die sich für den Relationalismus im Kontext der Allgemeinen Relativitätstheorie anführen lassen, durchaus begründet und nachvollziehbar ist.

Im folgenden geht es jedoch nur in zweiter Linie um den Relationalismus selbst; im Fokus stehen vielmehr die Konsequenzen der aktiven Diffeomorphismusinvarianz der Allgemeinen Relativitätstheorie und ihre Deutung als Eichinvarianz. Der Grund für diese Fokussierung auf die als Eichinvarianz gedeutete aktive Diffeomorphismusinvarianz als eigentlichen Gegenstand des Interesses – jenseits ihrer (nicht unumstrittenen) konstitutiven Rolle für den Relationalismus – ist insbesondere in der Tatsache zu sehen, dass diese unmittelbar in die direkte, *Kanonische Quantisierung* der Allgemeinen Relativitätstheorie Eingang findet – und schliesslich nicht zuletzt auch Motivationen liefert, nach Lösungen für das Problem der Quantengravitation jenseits einer solchen direkten Quantisierung zu suchen.[112]

Werfen wir aber dennoch zunächst einen genaueren Blick auf das, was im Rahmen der Allgemeinen Relativitätstheorie unter 'Relationalismus' bezüglich der Raumzeit verstanden werden kann:

---

[111] Siehe etwa Butterfield (1989) und Pooley (2006a).

[112] Die Alternativen zu einer direkten Quantisierung transzendieren in einigen Fällen, wie sich zeigen wird, in grundlegender Weise die Problematik einer substantialistischen oder relationalistischen Deutung der Raumzeit bzw. die Begrifflichkeiten des Substantialismus und Relationalismus selbst. Sie liefern, wenn man die alten Begrifflichkeiten über ihren ursprünglichen Kontext hinaus überhaupt noch bemühen möchte, z.T. bestenfalls noch Argumente für eine aus einer fundamentaleren Substratdynamik heraus motivierbare relationalistische Deutung der Raumzeit. Die Raumzeit wird in einigen dieser Szenarien nicht mehr von Materie und Feldern aufgespannt, vielmehr kommt sie aus einer basaleren, prägeometrischen Ebene heraus zustande, besitzt also selbst bestimmt keinen substantiellen Charakter. – Siehe insb. Kap. 4.6.

## Relationalismus

Der Relationalismus bezüglich der Raumzeit entspricht der Auffassung, dass die Raumzeit und raumzeitliche Verhältnisse nichts anderes sind als Ausdruck der relationalen Bezüge zwischen den eigentlichen physikalischen Systemen bzw. Ereignissen. Raumzeit und raumzeitliche Verhältnisse supervenieren also dieser Auffassung zufolge auf den physikalischen Systemen bzw. Ereignissen. Sie haben keinen eigenständigen ontologischen Status. Oder um es in einer paradoxen Weise aus-zudrücken: Raumzeit ist eine relationale Implikation ihrer Inhalte.

Dabei spricht der Relationalismus der Raumzeit noch nicht notwendigerweise ihre Existenz ab,[113] sondern erst einmal nur eine unabhängige Existenz, die einen eigenständigen ontologische Status begründen könnte. Sie besitzt also bestenfalls einen ontologischen Status, der sich von dem anderer Entitäten herleitet. Nur diese anderen, nicht-raumzeitlichen Entitäten besitzen einen eigenständigen ontologischen Status. Es geht dem Relationalismus also nicht primär um die Existenz der Raumzeit, sondern um die Zurückweisung einer Auffassung, welche die Raumzeit als eigenständige Entität, als eigenständige Substanz, versteht oder gar elementare Konstituenten der Raumzeit – die Raum-zeitpunkte – postuliert, denen dann ein eigenständiger ontologischer Status zugesprochen wird.

> *"Substantivalists and relationalists will [...] agree that the world has some given geometrical struc-*
> *ture. Substantivalists understand the existence of spacetime in terms of the existence of its pointlike*
> *parts, and gloss spatio-temporal relations between material events in terms of the spatio-temporal re-*
> *lations between points at which the events occur. Relationalists will deny that spacetime points enjoy*
> *this robust sort of existence, and will accept spatio-temporal relations between events as primitive."*
> (Belot / Earman (2001) 227)

Um eine relationalistische Position als tragfähig zu erweisen, wäre also vor allem zu zeigen, dass sich alle als fundamental angesehenen physikalischen Theorien grundsätzlich in vollständig relatio-nalistischer Form ohne Bezug auf eine vorgegebene eigenständige Raumzeit formulieren oder re-konstruieren lassen. Die Raumzeit selbst müsste in diesen Theorien in Form einer relationalen Struktur von Beziehungen zwischen anderen Entitäten, die keinen vorgegebenen Raum und keine vorgegebene Zeit voraussetzen, rekonstruiert werden können. Ebenso müsste sich der Feldbegriff, der in der Physik eine zentrale Rolle spielt, auf eine Weise rekonstruieren lassen, die ohne den Be-zug auf einen vorgegebenen Raum und eine vorgegebene Zeit auskommt.

Bei der traditionellen Leibnizschen bzw. Machschen Form des Relationalismus handelte es sich um einen solchen bezüglich des dreidimensionalen Raumes. Diesem zufolge gibt es keinen substantiel-len Raum, sondern nur räumliche Beziehungen zwischen materiellen Körpern. Der Raum wird als relationales Gefüge verstanden, welches von materiellen Körpern aufgespannt wird. Später waren die Entitäten, auf denen räumliche Relationen beruhen sollten (bzw. auf denen die Raumzeit super-venieren sollte), dann Punktteilchen und ausgedehnte Felder – wobei letztere sich wiederum rela-tionalistisch deuten lassen, wenn man annimmt, dass sie aus Teilen bestehen, die zueinander in einem räumlichen Verhältnis stehen.

---

[113] Es gibt allerdings durchaus radikalere Varianten des Relationalismus, wie etwa die weiter oben schon angeführte Position von Rovelli (2007), für die dies anders ist:
> *"The clean way of expressing Einstein's discovery is to say that there are no space and time: there are only dy-*
> *namical objects. The world is made by dynamical fields. These do not live in, or on, spacetime: they form and*
> *exhaust reality."* (Rovelli (2006) 27)

Mit dem Aufkommen der Allgemeinen Relativitätstheorie als Theorie, die eine vierdimensionale Raumzeit beschreibt, für die räumliche und zeitliche Aspekte dynamisch gekoppelt sind, war dieser Relationalismus bezüglich des dreidimensionalen Raumes nicht mehr aufrechtzuerhalten. Wollte man weiterhin eine relationalistische Position beziehen, so musste dies angesichts der Weiterentwicklung der Physik nun ein Relationalismus bezüglich der vierdimensionalen Raumzeit sein.

Im Gegensatz zum dreidimensionalen Raum des traditionellen Relationalismus, der von Punktteilchen und Feldern aufgespannt wird, kommen für die vierdimensionale Raumzeit als konstitutive Elemente, auf denen diese supervenient bzw. von denen diese als relationales Gefüge aufgespannt wird, nur elementare Ereignisse in Frage. Wenn eine physikalische Theorie der vierdimensionalen Raumzeit in konsistenter Weise relationalistisch formuliert werden soll, müssen also vor allem die nach heutiger physikalischer Auffassung omnipräsenten Quantenfelder der Materie und der nichtgravitativen Wechselwirkungen, in Form solcher elementaren Ereignisse, die keine Raumzeit voraussetzen, sondern erst zustandebringen bzw. definieren, rekonstruierbar sein.

Im Formalismus der Quantenfeldtheorien werden diese Quantenfelder als Operatorfelder den Raumzeitpunkten zugeordnet. Es sieht hier also erst einmal so aus, als wären sie so etwas wie Eigenschaften der Raumzeitpunkte. Das modelltheoretische Instrumentarium der Quantenfeldtheorien leistet somit einem Basal-Substantialismus bezüglich der Mannigfaltigkeit und der sie konstituierenden Raumzeitpunkte erst einmal formalen Vorschub – in der gleichen Weise, in der dies schon die klassischen Feldtheorien mit ihrer Zuschreibung von Feldwerten als Eigenschaften von Raumzeitpunkten getan haben. Eine relationalistische Rekonstruktion müsste also an erster Stelle plausibel machen, dass diese formale Implikation des modelltheoretischen Ansatzes der Quantenfeldtheorien ein theoretisches Artefakt darstellt und sich Quantenfelder zumindest im Prinzip ohne Rekurs auf eine schon bestehende Raumzeit und ihre Punkte rekonstruieren lassen. Der elementarste Schritt hierzu besteht vermutlich darin, Felder als ausgedehnte Entitäten aufzufassen, die über elementare Teile verfügen, die wiederum zueinander in räumlichen bzw. raumzeitlichen Relationen stehen, ohne schon eine Raumzeit vorauszusetzen. Die entsprechenden elementaren Teile von Feldern wären im relationalistischen Verständnis als elementare Ereignisse anzusehen.

> *"In the case of field theories, the relationist has to assume that elementary field-parts possess spatial relations with respect to each other, and that there are coincidence relations between the parts of different fields."* (Dieks (2006) xi)

Eine entscheidende Rolle bei der relationalistischen Rekonstruktion der Raumzeit kommt mutmasslich – wenn sich die diesbezüglich mit der Allgemeinen Relativitätstheorie abzeichnenden Einsichten bewahrheiten sollten – dem Gravitationsfeld (bzw. dem mit diesem zu identifizierenden metrischen Feld zu), welches mit allen anderen Feldern (und sich selbst) wechselwirkt.[114]

> *"There are only interacting fields (including the gravitational field) and particles: the only notion of localization which is present in the theory is relative: dynamical objects (fields and particles) are localized only with respect to one another."* (Rovelli (2007) 1313)

---

[114] Das metrische Feld wird hierbei – im Gegensatz etwa zur Auffassung des metrischen Essentialismus – zu den 'Inhalten der Raumzeit' gezählt, die erst die Raumzeit aufspannen.

Die Etablierung eines umfassenden Relationalismus hinsichtlich der Raumzeit müsste jedoch, über diese ersten Schritte hinausgehend, sehr wahrscheinlich mit einer grundlegenden Modifikation physikalischer Begrifflichkeiten und der zur Anwendung kommenden modelltheoretischen Prozeduren einhergehen. Mit einer solchen Etablierung eines umfassenden Relationalismus und der dafür erforderlichen grundlegenden Modifikation unserer theoretischen Instrumentarien dürften jedoch nicht unerhebliche Probleme für die Formulierung fundamentaler physikalischer Theorien verbunden sein:

> *"[...] a physics where space and time are absolute can be developed one particle at a time, while a relational view requires that the properties of any one particle are determined self-consistently by the whole universe."* (Smolin (2006c) 202)

Das Verhältnis zwischen

- – einer Theorie, die mit einem im wesentlichen dem Relationalismus nicht angemessenen formalen und modelltheoretischen Instrumentarium eine letztlich – so zumindest die Vermutung – relationalistisch zu verstehende raumzeitliche Dynamik beschreibt, und
- – der angestrebten vollständig relationalistischen Beschreibung dieser raumzeitlichen Dynamik

könnte womöglich in einiger Hinsicht einem Verhältnis entsprechen, welches etwas konkretere Intuitionen und Assoziationen weckt: dem zwischen

- – einer Eichtheorie, die mit unphysikalischen Überschussstrukturen und Redundanzen ausgestattet ihren Gegenstandsbereich erfasst, und
- – einer um diese unphysikalischen Überschussstrukturen und Redundanzen erleichterten Beschreibung des gleichen Gegenstandsbereichs – einer Beschreibung, die sich ausschliesslich auf die echten physikalischen Freiheitsgrade bezieht.

Die Eichtheorien sind insbesondere deshalb so erfolgreich, weil es oft nicht so einfach und manchmal sogar gänzlich unmöglich ist, eine redundanzfreie Beschreibung zu erreichen. Oft ermöglicht erst die Einführung unphysikalischer Symmetrien – Eichsymmetrien bzw. Eichinvarianzen eben – eine physikalisch tragfähige und modelltheoretisch handhabbare Beschreibung. Für vollständig relationalistische Theorien liegt vielleicht eine in mancher Hinsicht analoge, wenngleich wohl gravierendere Problemlage vor.

*

Solange jedoch eine vollständig relationalistische Rekonstruktion der bestehenden physikalischen Theorien noch nicht gelungen ist und diese Theorien, zumindest was die formale, modelltheoretische Seite betrifft, immer noch vielfältige nicht-relationalistische Implikationen in sich tragen, kann der Relationalismus nur auf eine (meist mühsame) Sondierung der Tiefenstruktur dieser bestehenden Theorien setzen, in der Hoffnung, dass sich dabei hinreichende Argumente finden lassen, die eine relationalistische Sichtweise hinsichtlich der Raumzeit in direkter oder indirekter Weise stützen. Der aussichtsreichste Kandidat für diese Strategie ist die Allgemeine Relativitätstheorie, aus der sich, wie im Vorausgehenden schon erörtert wurde, in der Tat gute Argumente für den Relationalismus ableiten lassen.

Wie wir gesehen haben, beruht das zentrale Argument, das sich im Kontext einer Sondierung der Tiefenstruktur der Allgemeinen Relativitätstheorie für den Relationalismus gewinnen lässt, auf der in dieser Theorie vorliegenden substantiellen Form der allgemeinen Kovarianz: ihrer aktiven Diffeomorphismusinvarianz.[115] Diese führt – im Verbund mit einer ihrer direkten Implikationen: dem *Lochargument* – erst einmal zu einem Argument gegen den *Substantialismus bezüglich der raumzeitlichen Mannigfaltigkeit*. Dieser lässt sich nur um den Preis eines Bezuges auf unbeobachtbare Entitäten und einen ebenso unbeobachtbaren, ausschliesslich metaphysisch motivierbaren Indeterminismus, welcher der Allgemeinen Relativitätstheorie zuzuschreiben wäre, aufrechterhalten. – Nun liesse sich jedoch einwenden, dass ein Argument gegen eine spezifische Form des Substantialismus, nämlich den bezüglich der raumzeitlichen Mannigfaltigkeit, noch nicht notwendigerweise mit einem Argument für eine relationalistische Sichtweise zu verwechseln ist:

> *"Now,* if *relationalism in [general relativity] were entailed by the claim that diffeomorphically related mathematical models don't represent physically distinct solutions, most physicists would count themselves as relationalists. After all, the assumption that* an entire equivalence class of diffeomorphically related mathematical solutions represents only one physical solution *is regarded as the most common technical way out of the strictures of the hole argument (in the philosophical literature such an assumption is known, after Earman and Norton (1987), as* Leibniz equivalence*). However, we believe that it is not at all clear whether Leibniz equivalence grinds corn for the relationalist's mill, since the spacetime substantivalist can always ask: (1) why on earth should we identify* physical *spacetime with the bare manifold* deprived of the metric field*? (2) Why should we assume that the points of the mathematical manifold have an intrinsic* physical *identity independently of the metric field?."* (Dorato / Pauri (2006) 128)

Hierauf sind zwei naheliegende Reaktionen denkbar: Zum einen liesse sich argumentieren, dass die Argumente gegen einen Substantialismus bezüglich der raumzeitlichen Mannigfaltigkeit dem Relationalismus tatsächlich schon in ausreichender Weise Vorschub leisten – und dies insbesondere, da alle anderen Formen des Substantialismus einer solchen Auffassung zufolge keine wirklichen Alternativen zum Relationalismus bieten: Sie akzeptieren in der einen oder anderen Form genau das, was der Relationalismus als entscheidend ansieht: entweder, indem sie Entitäten, die auch im substantialistischen Bild eher als Inhalte der Raumzeit zu deuten wären, mit dieser identifizieren, was letztlich nur zu substantialistischen Sprechweisen führt, die nach einer entsprechenden Sprachregulierung besser im Rahmen des Relationalismus zu verorten wären, oder indem sie sich in eine mehr oder weniger ausgeprägte strukturalistische Sichtweise retten, die ebenfalls besser unter der Flagge des Relationalismus segeln sollte. Die einzige Form des Substantialismus, für die dies vielleicht nicht geltend gemacht werden kann, findet sich in den Varianten, die auf dem *Haecceitismus* beruhen und die *Leibniz-Äquivalenz* als Identitätskriterium zurückweisen. Mit diesen Ausreissern, die auf eine metaphysische Identifizierung jenseits jeder physikalischen Beobachtbarkeit setzen, kann der an rein physikalischen Belangen orientierte und zu allem entschlossene Relationalismus leben, ohne sie wirklich ernst zu nehmen.

---

[115] Schon hier sollte zumindest nicht unerwähnt bleiben, dass eine der unabdingbaren Voraussetzungen für die Diffeomorphismusinvarianz in der Annahme einer differenzierbaren Mannigfaltigkeit besteht, in der Kontinuumsannahme also. Aus ihr lassen sich also nur im Kontinuumsfall Argumente für einen Relationalismus ableiten. Im Falle diskreter Strukturen müssten andersgeartete Argumente gefunden werden bzw. diskrete Pendants zur Diffeomorphismusinvarianz formuliert werden.

Zum anderen liesse sich aber auch – und dies ist vielleicht die angemessenere Alternative – argumentieren, dass es nicht wirklich um die Begrifflichkeiten des 'Substantialismus' bzw. des 'Relationalismus' geht, sondern vielmehr um die konkreteren physikalischen Implikationen der aktiven Diffeomorphismusinvarianz der Allgemeinen Relativitätstheorie und um die Möglichkeit, auf ihrer Grundlage etwas über die fundamentale Struktur des Weltgeschehens und die ihm zugrundeliegenden Entitäten zu erfahren. Ob diese Einsichten dann schon als konstitutiv für eine Sichtweise angesehen werden können, die das Etikett 'Relationalismus' verdient, bleibt vor diesem Hintergrund eine Frage der sprachlichen Konvention. Eine solche Konvention könnte eben auch gerade darin bestehen einen Relationalismus bezüglich der Allgemeinen Relativitätstheorie schlichtweg per definitionem mit ihrer eichinvarianten Deutung gleichzusetzen, die sich in ihren entscheidenden Motivationen gerade auf die aktive Diffeomorphismusinvarianz der Theorie stützt. Wenn dies dem Relationalisten schon genügt, wer sollte ihm seine Zufriedenheit dann schon absprechen. – Im Weiteren soll es, dieser Strategie der Unterordnung oberbegrifflicher Zuweisung unter inhaltliche Fragen und Konsequenzen mit direkter physikalischer Relevanz folgend, vorrangig um die unmittelbaren und die mittelbaren Implikationen der aktiven Diffeomorphismusinvarianz gehen.

<p style="text-align:center">*</p>

Wenn man die *Leibniz-Äquivalenz* als im Kontext der Physik unabweisliches Kriterium ernst nimmt und damit Ununterscheidbarkeit mit physikalischer Identität gleichsetzt, führt die aktive Diffeomorphismusinvarianz der Allgemeinen Relativitätstheorie zu der Einsicht, dass es nicht etwa Raumzeitpunkte sind, die als Träger von physikalischen Eigenschaften betrachtet werden können, sondern bestenfalls Äquivalenzklassen raumzeitlicher Zuordnungen, die durch einen Diffeomorphismus ineinander überführt werden können. Wie gesagt: Ob man das schon 'relationalistisch' nennen sollte, ist eine Entscheidungssache. Unstrittig ist aber, dass eine Zurückweisung der *Leibniz-Äquivalenz* als Identitätskriterium einem Bezug auf unbeobachtbare Grössen entspricht und zudem die Allgemeine Relativitätstheorie ohne weitere Motivationen zu einer indeterministischen Theorie macht, ohne dass dies wiederum beobachtbare Folgen hätte. Mit einer solchen Zurückweisung überschreitet man die Grenzen der Physik hin auf das Feld der Naturmetaphysik.

Also: Unter Berücksichtigung der aktiven Diffeomorphismusinvarianz der Allgemeinen Relativitätstheorie und unter Verwendung der *Leibniz-Äquivalenz* als Kriterium für physikalische Identität entsprechen physikalische Raumzeiten – dieser Theorie zufolge – Äquivalenzklassen von raumzeitlichen Modellen, die sich aufgrund der Diffeomorphismusinvarianz ergeben. Alle Modelle innerhalb einer solchen Äquivalenzklasse sind dann Beschreibungen der gleichen Raumzeit. Diese Beschreibungen unterscheiden sich lediglich hinsichtlich der jeweiligen Darstellung.

> *"Relationalists [...] will maintain that all instantiations of a given four-dimensional geometry are numerically identical – x(t) and x'(t) correspond to the same physical possibility."* (Belot / Earman (1999) 176)

Ereignisse lassen sich dann aber nicht mehr den Punkten auf der Mannigfaltigkeit zuordnen; vielmehr lassen sich nur über diffeomorphismusinvariante Grössen identifizieren. Nur diese diffeomorphismusinvarianten Grössen entsprechen den eigentlichen intrinsischen geometrischen und dynami-schen Eigenschaften, denen physikalische Relevanz zukommt. – Wenn man die Konsequenzen dieser Einsicht schon als Relationalismus bezeichnen möchte, dann entspricht der Relationalismus damit einer Deutung der Allgemeinen Relativitätstheorie, welche deren aktive Diffeomorphismusinvarianz als im physikalischen Sinne identitätsstiftend interpretiert. Die Hauptmotivation für die Auffassung, dass nur allgemein-kovariante, diffeomorphismusinvariante Grössen über einen eigenständigen ontologischen Status verfügen und nur diese Grössen den 'echten' Freiheitsgraden der Theorie entsprechen können, beruht jedenfalls auf der Tatsache, dass Grössen, die nicht diffeomorphismusinvariant sind, weder direkt, noch hinsichtlich ihrer Konsequenzen beobachtbar oder messbar sind. Sie haben grundsätzlich metaphysischen Status. Behandelt man sie dennoch als ernstzunehmende physikalische Grössen, so führt dies, wie das Lochargument zeigt, zudem zu einem wiederum ausschliesslich metaphysisch motivierten Kryptoindeterminismus.

Der nächste gedankliche Schritt, der sich in der Auslotung der Konsequenzen der aktiven Diffeomorphismusinvarianz der Allgemeinen Relativitätstheorie ergibt, besteht dann in der folgenden Einsicht: Wenn man Ununterscheidbarkeit (im Sinne der *Leibniz-Äquivalenz*) als Kriterium für physikalische Identität einfordert, und wenn man daher physikalische Raumzeiten mit Äquivalenzklassen von raumzeitlichen Modellen gleichsetzt, die sich durch einen aktiven Diffeomorphismus ineinander überführen lassen, so behandelt man damit aktive Diffeomorphismen als Eichtransformationen und die aktive Diffeomorphismusinvarianz als Eichinvarianz.

> *"[...] either (i) we interpret the theory as an indeterministic theory, where the future is not determined by the past. Or (ii) we interpret active diffeomorphisms as a 'gauge invariance': that is, we postulate that the complete observables of the theory are only given by quantities that are invariant under this transformation."* (Rovelli (2007) 1309)

Damit wird die Allgemeine Relativitätstheorie als Eichtheorie verstanden.

> *"[...] the lesson of the hole argument is [...] that the spacetime diffeomorphism group is a gauge group of [general relativity]."* (Earman (2002a) 22)

Wird die Allgemeine Relativitätstheorie jedoch als Eichtheorie verstanden und werden aktive Diffeomorphismen als Eichtransformationen behandelt, so lassen sich die Äquivalenzklassen von raumzeitlichen Modellen, die durch einen aktiven Diffeomorphismus ineinander überführt werden können, (erst einmal metaphorisch) als Eichorbits verstehen.[116] Alle 'Punkte' (Modelle einer Raumzeit) innerhalb eines Eichorbits entsprechen Beschreibungen der gleichen Raumzeit. Aussagen der Form

– Grössen, die nicht diffeomorphismusinvariant sind, sind nicht beobachtbar oder messbar.
– Nur diffeomorphismusinvariante Grössen haben ontologischen Status.

lassen sich dann übersetzen in die entsprechenden eichtheoretischen Pendants:

– Grössen, die nicht eichinvariant sind, sind nicht beobachtbar oder messbar.
– Nur eichinvariante Grössen haben ontologischen Status.

---

[116] Im konkreteren Sinne zu Eichorbits werden diese Äquivalenzklassen in der Hamiltonschen Darstellung der Allgemeinen Relativitätstheorie. Siehe weiter unten.

> *"It seems far more natural to insist that the* only *physically real quantities are gauge-invariant quantities (call this strategy the adoption of a* gauge-invariant *interpretation)."* (Belot / Earman (2001) 221)

Und wenn man eine Deutung der Allgemeinen Relativitätstheorie, welche die aktive Diffeomorphismusinvarianz gemeinsam mit der *Leibniz-Äquivalenz* als konstitutiv für die Ausweisung physikalisch realer Grössen ansieht, schon als Relationalismus bezeichnen möchte, dann wird hier schliesslich der Relationalismus mit der eichinvarianten Interpretation der Allgemeinen Relativitätstheorie gleichgesetzt:[117]

> *"[...] relationalism is a gauge invariant interpretation of general relativity."* (Belot / Earman (1999) 176)

Viel entscheidender als die Frage, ob diese tatsächlich schon das Etikett 'Relationalismus' verdient, ist aber, dass eine solche eichinvariante Interpretation unumgänglich ist, wenn man die aktive Diffeomorphismusinvarianz der Allgemeinen Relativitätstheorie im Verbund mit der *Leibniz-Äquivalenz* als Kriterium zur Identifizierung physikalisch realer Grössen verwendet. – So unumgänglich wie eine eichinvariante Interpretation der Allgemeinen Relativitätstheorie unter diesen Bedingungen erscheinen mag, so radikal sind jedoch ihre Folgen.

## *Zur Hamiltonschen Formulierung der Allgemeinen Relativitätstheorie*

Die zum Teil äusserst radikalen Implikationen der eichinvarianten Interpretation der Allgemeinen Relativitätstheorie sollen nun im Kontext ihrer Hamiltonschen Formulierung verdeutlicht werden. Die ansonsten eher metaphorische Redeweise von Äquivalenzklassen von raumzeitlichen Modellen als Eichorbits wird im Übergang zur Hamiltonschen Darstellung konkret. In diesem Kontext wird die Struktur der Allgemeinen Relativitätstheorie als Eichtheorie deutlicher. Eine Alternative zur Hamiltonschen Darstellung hinsichtlich der durchsichtigeren Erfassung der Allgemeinen Relativitätstheorie als Eichtheorie würde sich mit ihrer auf dem Lagrange-Formalismus beruhenden Faserbündeldarstellung ergeben. Die Wahl der Hamiltonschen Darstellung als Ausgangspunkt zur Erörterung der Implikationen der eichinvarianten Interpretation der Allgemeinen Relativitätstheorie hat jedoch gute Gründe: Sie spielt nicht zuletzt für die Ansätze zur Quantengravitation eine entscheidende Rolle; die Hamiltonsche Formulierung der Allgemeinen Relativitätstheorie stellt insbesondere den Ausgangspunkt der direkten *Kanonischen Quantisierung* dar – in ihrer geometrodynamischen Variante[118] wie als *Loop Quantum Gravity*[119].

In der Hamiltonschen Darstellung der Allgemeinen Relativitätstheorie wird die Raumzeit in dreidimensionale raumartige Hyperflächen und einen Zeitparameter zerlegt. Die formale Auszeichnung des Zeitparameters ist insbesondere zur Definition des kanonischen Impulses erforderlich, der nicht zuletzt für die *Kanonische Quantisierung* der Allgemeinen Relativitätstheorie unabdinglich ist.

---

[117] Es gibt hinsichtlich dieser sprachlichen Festlegung jedoch durchaus abweichende Ansichten. So widerspricht etwa Rickles (2008) der Auffassung, dass die Deutung der aktiven Diffeomorphismusinvarianz als Eichinvarianz schon für den Relationalismus hinreichend ist. Und es gibt viele weitere Versuche, dem Relationalisten abzusprechen, dass er tatsächlich ein Relationalist ist. Aber, wie gesagt, es geht nicht vorrangig um den Relationalismus und seine begrifflichen Voraussetzungen, sondern um die Konsequenzen, die aus der aktiven Diffeomorphismusinvarianz erwachsen.
[118] Siehe Kap. 4.3.
[119] Siehe Kap. 4.4.

*"The [...] Hamiltonian approach starts with a slicing of the four-dimensional manifold $M_4$ into constant-time hypersurfaces $\Sigma_\tau$, indexed by the* parameter time $\tau$ *[...]."* (Dorato / Pauri (2006) 131)

Die Diffeomorphismusinvarianz (bzw. Eichinvarianz) der Allgemeinen Relativitätstheorie wird in der Hamiltonschen Darstellung in Form der sogenannten Zusatzbedingungen[120] erfasst.[121] Formal betrachtet sind die (primären) Zusatzbedingungen eine direkte Folge der Legendre-Transformation beim Übergang von der Lagrange- zur Hamilton-Darstellung.

*"Since the original Einstein equations are not hyperbolic, it turns out that the canonical momenta are not all functionally independent, but satisfy four conditions known as* primary *constraints. Four other* secondary *constraints arise when we require that the primary constraints be preserved through evolution [...]."* (Dorato / Pauri (2006) 132)

Die (primären) Zusatzbedingungen sind erforderlich, da die in den dynamischen Grundgleichungen des Hamilton-Formalismus verwendeten kanonischen Variablen noch nicht die Diffeomorphismusinvarianz der Theorie berücksichtigten und daher mit einer noch ungekennzeichneten Redundanz in der Darstellung einhergehen. Zu solchen Zusatzbedingungen kommt es grundsätzlich in der Hamiltonschen Darstellung für Theorien mit einer unphysikalischen Überschussstruktur, wie sie etwa in Form einer Eichfreiheit vorliegen kann. So wie die Kennzeichnung von Eichinvarianzen innerhalb einer Eichtheorie zur Erfassung unphysikalischer Redundanzen dient, werden diese Redundanzen in der Hamiltonschen Darstellung durch die Zusatzbedingungen erfasst. Die Zusatzbedingungen lassen sich in dieser Hinsicht als Generatoren von Eichtransformationen verstehen.

Nach der Zerlegung der Raumzeit in dreidimensionale raumartige Hyperflächen und einen Zeitparameter ergeben sich für die Allgemeine Relativitätstheorie vier Zusatzbedingungen: die *Hamiltonsche Zusatzbedingung* und drei *Impuls-* bzw. *Diffeomorphismus-Zusatzbedingungen*. Die Hamiltonsche Zusatzbedingung spiegelt die zeitliche Komponente der Diffeomorphismusinvarianz in der Hamiltonschen Darstellung der Allgemeinen Relativitätstheorie wider. Sie betrifft also nicht zuletzt die Dynamik des Systems. Die Impuls- bzw. Diffeomorphismus-Zusatzbedingungen erfassen die räumlichen Komponenten der Diffeomorphismusinvarianz.

In der Hamiltonschen Formulierung der Allgemeinen Relativitätstheorie wird mittels dieser Zusatzbedingungen eine Festschreibung der aktiven Diffeomorphismusinvarianz in der formalen Darstellung der Theorie vollzogen. Die Zusatzbedingungen erfassen die aktive Diffeomorphismusinvarianz der Allgemeinen Relativitätstheorie im Kontext ihrer Hamiltonschen Darstellung. Sie entsprechen ihrer jeweiligen Einforderung im Rahmen einer spezifischen Wahl kanonischer Variablen. – Wie diese Implementierung der aktiven Diffeomorphismusinvarianz der Theorie durch die Zusatzbedingungen im einzelnen funktioniert, lässt sich am besten im (gedanklichen) Übergang vom vollen (kinematischen) zum (stufenweise) reduzierten (und schliesslich physikalischen) Phasenraum in der Hamiltonschen Darstellung verstehen:

Die Darstellung des Hamiltonschen Systems mittels der jeweils gewählten kanonischen Variablen genügt, wie schon angedeutet, noch nicht ohne weiteres der Diffeomorphismusinvarianz der Theorie. Der durch die kanonischen Variablen definierte volle *kinematische Phasenraum* des Hamilton-schen Systems entspricht – ohne Berücksichtigung der Zusatzbedingungen – noch nicht dem tatsächlichen physikalischen Zustandsraum des Systems; vielmehr enthält er physikalisch bzw. dynamisch unmögliche Zustände und er beschreibt die physikalisch bzw. dynamisch möglichen Zustände in einer vieldeutigen, redundanten Weise. Die Zusatzbedingungen erfassen gerade diese Redundanz der Theorie, die sich auf der Grundlage der Wahl der kanonischen Variablen ergibt.

---

[120] In der englischsprachigen Literatur formieren diese als *'Constraints'*.
[121] Siehe Belot / Earman (1999, 2001), Henneaux / Teitelboim (1992), Govaerts (2002).

Mit der Berücksichtigung der Zusatzbedingungen[122] reduziert sich der volle kinematische Phasenraum des Hamiltonschen Systems dann erst einmal auf *die durch die Zusatzbedingungen definierte Oberfläche*[123] im Phasenraum. Die auf dieser Oberfläche liegenden Phasenraumpunkte beschreiben die entsprechend der Theorie (bzw. der ihr zugrundeliegenden Dynamik) konsistent möglichen physikalischen Zustände des Systems; die ausserhalb dieser Oberfläche liegenden Phasenraumpunkte entsprechen physikalisch und dynamisch unmöglichen Zuständen.

> *"The existence of constraints implies that not all the points of phase space represent physically meaningful states: rather, we are restricted to the* constraint surface *where all the constraints are satisfied. The dimensionality of the constraint surface is given by the number of the original canonical variables, minus the number of functionally independent constraints."* (Pauri / Vallisneri (2002) 12)

Aber die auf der durch die Zusatzbedingungen definierten Oberfläche liegenden Phasenraumpunkte beschreiben die konsistent möglichen physikalischen Zustände des Systems noch nicht in eindeutiger, sondern in vieldeutiger Weise. Die Punkte dieser Oberfläche, die eine präsymplektische Struktur aufweist, gehören vielmehr zu Äquivalenzklassen von Zuständen, die untereinander physikalisch nicht unterscheidbar sind.

> *"The first change to note in the shift from a Hamiltonian system to a constrained Hamiltonian system is that the symplectic form is replaced by a* presymplectic form *[...]. The presymplectic form induces a partitioning of the phase space into subspaces (not necessarily manifolds) known as* gauge orbits.*"* (Rickles (2006) 156)

Wenn man die Diffeomorphismusinvarianz, die innerhalb der Hamiltonschen Darstellung in Form der Zusatzbedingungen eingefordert wird, als Eichinvarianz deutet, lassen sich diese Äquivalenzklassen nun im Phasenraumbild tatsächlich als Eichorbits verstehen. Die Zusatzbedingungen erzeugen gerade Eichtransformationen, die (infinitesimalen) Diffeomorphismen entsprechen.

---

[122] Hier und im folgenden geht es erst einmal nur um die Zusatzbedingungen erster Klasse, zu denen alle oben genannten gehören:

> *"Another important distinction [...] between constraints is that holding between* first class *and* second class *constraints. A constraint [...] is said to be first class if its Poisson bracket with any other constraint is given as a linear combination of the constraints [...] Any constraint not satisfying these relations is second class. Our sole concern is with the first class constraints. The appearance of such constraints in a theory implies that the dynamics is restricted to a submanifold [...] of the full phase space [...]; i.e. the* constraint surface.*"* (Rickles (2006) 156)

Die Unterscheidung im primäre (Resultat der Legendre-Transformation) und sekundäre Zusatzbedingungen ist kategorial verschieden von der Unterscheidung in Zusatzbedingungen erster Klasse (solche, die mit allen anderen Zusatzbedingungen kommutieren) und solche zweiter Klasse.

[123] In der englischsprachigen Literatur formiert diese als '*Constraint Surface*'.

*"When used as generators of canonical transformations, some constraints, known as* first class*, will map points on the constraint surface to points on the same surface; these transformations are known as* gauge transformations.*"* (Pauri / Vallisneri (2002) 12)

Alle Phasenraumpunkte innerhalb eines Eichorbits entsprechen (hinsichtlich beobachtbarer Grössen ununterscheidbaren) Beschreibungen der gleichen physikalischen Situation: d.h. der gleichen Raumzeit. Diese Beschreibungen unterscheiden sich lediglich formal, etwa hinsichtlich der jeweiligen Koordinatenwahl auf den Hyperflächen. – Als 'Observablen' kommen, so man die Vorhersagekraft der Theorie erhalten will und deshalb nur auf beobachtbare Grössen Bezug nimmt, nur solche Grössen in Frage, die eichinvariant sind.

*"[...] finding a constraint for a Hamiltonian system often [...] indicates that we are dealing with a gauge theory: there is some symmetry transformation between states in the phase space that leaves all dynamical parameters unchanged. The usual understanding is that since any physical quantities must 'make a difference' dynamically, all observables (physically real quantities) must be gauge invariant."* (Callender / Huggett (2001a) 18)

Diese Forderung nach Eichinvarianz der Observablen entspricht im Rahmen der Hamiltonschen Darstellung formal der Forderung, dass alle Observablen Grössen sind, deren Poisson-Klammern mit allen Zusatzbedingungen (erster Klasse[124]) verschwinden.[125] – Die nicht unbedingt wünschenswerten Alternativen hierzu sind wiederum die gleichen wie beim Ignorieren der aktiven Diffeomorphismusinvarianz: Einerseits entspricht der Bezug auf Grössen, die nicht eichinvariant sind, einer Zurückweisung der *Leibniz-Äquivalenz* als Identitätskriterium und damit einer Bezugnahme auf grundsätzlich unbeobachtbare Unterschiede, denen physikalischer Realstatus zugesprochen wird. Andererseits führt die Annahme, dass Beschreibungen, die sich durch eine Eichtransformation ineinander überführen lassen, als Beschreibungen unterschiedlicher physikalischer Zustände betrachtet werden können, gerade zum Indeterminismus der Theorie. Dieser Indeterminismus hat jedoch, wie wir schon wissen, keine beobachtbaren Konsequenzen. Alles spricht dafür, dass er ein Artefakt einer unzureichenden Darstellung des entsprechenden physikalischen Systems ist. Dies ist gerade die Problematik, um die es im Lochargument ging.

Will man diese Probleme gänzlich vermeiden, so könnte man nun versuchen, eine – vielleicht physikalisch angemessenere – Beschreibung zu finden, in der alle Uneindeutigkeiten und Redundanzen vollständig eliminiert werden. Eine Strategie, zu einer solchen Beschreibung zu gelangen, besteht darin zu versuchen, jedem Eichorbit einen Punkt in einem abstrakten Phasenraum zuzuordnen, dem *reduzierten Phasenraum*, der dann in einer Eins-zu-eins-Zuordnung nur noch Darstellungen von physikalisch unterscheidbaren Zuständen enthalten sollte.[126][127] Spricht man ausschliesslich eichinva-

---

[124] Zusatzbedingungen erster Klasse sind, wie schon erwähnt, solche, deren Poisson-Klammern mit allen weiteren Zusatzbedingungen verschwinden.

[125] In einer Quantentheorie, die durch die Quantisierung einer klassischen Theorie in der Hamiltonschen Formulierung mit Zusatzbedingungen gewonnen wird, entspricht dies dann gerade der Forderung, dass Quantenobservablen mit allen Quanten-Zusatzbedingungen kommutieren müssen.

[126] Fischer / Moncrief (1996) versuchen dies zumindest unter spezifischen Einschränkungen der Allgemeinheit.

[127] Bei einer solchen Strategie können aber so einige Probleme auftreten:

"*[...] the reduced phase space might not have the structure of a manifold, and so will not be able to play the role of a phase space [...].*" (Rickles (2006) 161)

Zur Beseitigung der Mehrdeutigkeit der Beschreibung, wie sie im Rahmen der Darstellung der Zustände des Systems auf der durch die Zusatzbedingungen definierten Oberfläche des kinematischen Phasenraums mit ihrer präsymplekti-rianten (diffeomorphismusinvarianten) Grössen physikalische Relevanz bzw. einen eigenständigen ontologischen Status zu, so repräsentiert dieser reduzierte Phasenraum gerade die physikalisch realen Zustände des Systems.

> *"By taking the set of gauge orbits as the points of a new phase space, and endowing this set with a symplectic structure, one can construct a phase space for a Hamiltonian system – this new space is known as the* reduced phase space *[...]. In effect, the structure of the reduced space* encodes *all the gauge-invariant information [...] even though no gauge symmetry remains [...]."* (Rickles (2006) 161)

Wenn diese Strategie gelingen sollte, entsprächen die Punkte dieses reduzierten Phasenraums dann den tatsächlich physikalisch und dynamisch möglichen raumzeitlichen Modellen der Allgemeinen Relativitätstheorie, die hier nun in eindeutiger und eben nicht mehr in mehrdeutiger Weise erfasst würden. Alle Beschreibungen einer bestimmten Raumzeit fielen in diesem reduzierten Phasenraum auf den gleichen Punkt. In einer solchen Darstellung gäbe es keine Redundanzen in der Beschreibung mehr. Es gäbe konsequenterweise keine Eichinvarianzen mehr, die zur Erfassung solcher Redundanzen zu berücksichtigen wären. Eine solche Eins-zu-eins-Beschreibung enthielte nur noch echte physikalische Grössen.

Die Zusatzbedingungen in der Hamiltonschen Darstellung der Allgemeinen Relativitätstheorie repräsentieren in dieser Hinsicht gerade die Anforderungen, unter denen der volle kinematische Phasenraum auf den Phasenraum der eichinvarianten Grössen bzw. der 'echten' Freiheitsgrade des Systems – den Phasenraum der echten physikalischen Zustände – reduziert wird. Der Übergang auf den reduzierten (physikalischen) Phasenraum – und damit die Eliminierung der Eichinvarianz und der mit ihr einhergehenden Redundanz in der Beschreibung – wird daher als 'Lösung' der Zusatzbedingungen bezeichnet. Durch die Zusatzbedingungen (die auf der durch sie definierten Oberfläche im kinematischen Phasenraum des Hamiltonschen Systems eine präsymplektische Struktur induzieren, die sich als Eichorbitstruktur verstehen lässt) und schliesslich durch die Lösung dieser Zusatzbedingungen werden gerade die eich- bzw. diffeomorphismusinvarianten Grössen und mithin die physikalisch realen Zustände des Systems in der Hamiltonschen Darstellung ausgezeichnet.

Wie sehen diese eichinvarianten Grössen bzw. diese physikalisch realen Zustände des Systems aber nun aus? – Hier ist es nun unabwendlich, langsam einen Blick auf die Konsequenzen zu werfen, die sich auf der Grundlage der eichinvarianten Deutung der Allgemeinen Relativitätstheorie und ihrer Darstellung im Rahmen des Hamiltonschen Formalismus ergeben:

---

schen Eichorbitstruktur vorliegt, wird von manchen Autoren – neben dem Übergang auf einen reduzierten Phasenraum – als Alternative eine *Eichfixierung* vorgeschlagen, die in Form eines Schnittes durch die Eichorbits jeweils gerade einen repräsentativen Phasenraumpunkt auswählt, der dann als Darstellung für den physikalischen Zustand des Systems dient. Siehe etwa Pauri / Vallisneri (2002), Lusanna / Pauri (2005), Dorato / Pauri (2006). Auch dabei können jedoch erhebliche Probleme auftreten, die nicht geringer sind als die beim Übergang auf einen reduzierten Phasenraum:

> *"However in the case of general relativity [...] the geometrical structure of the constraint surface and the gauge orbits can prohibit the implementation of gauge conditions, so that some gauge slices will intersect some gauge orbits more than once, or not at all. If the former occurs then some states will be multiply represented (i.e. surplus remains); if the latter occurs, some genuine possibilities will not be represented in the phase space and, therefore, will not be deemed possible."* (Rickles (2006) 166)

## *Die eichinvariante Interpretation und die Konsequenzen*

Wie im Vorausgehenden deutlich geworden sein dürfte, gibt es zur eichinvarianten Interpretation der Allgemeinen Relativitätstheorie keine allzu überzeugenden Alternativen, wenn man sich nicht auf eine Naturmetaphysik einlassen möchte. Grössen, die nicht diffeomorphismusinvariant sind, sind weder direkt, noch hinsichtlich ihrer Konsequenzen beobachtbar oder messbar. Sie haben grundsätzlich metaphysischen Status. Ihre Interpretation als ernstzunehmende physikalische Grössen führt zudem, wie das Lochargument zeigt, zu einem Kryptoindeterminismus. Wenn man dies alles nicht will, bleibt nicht viel mehr, als die *Leibniz-Äquivalenz* als physikalisch relevantes Identitätskriterium zu akzeptieren, die aktive Diffeomorphismusinvarianz der Allgemeinen Relativitätstheorie als Eichinvarianz zu verstehen und schliesslich die Theorie als Eichtheorie zu deuten.

Mit dieser Deutung geht dann insbesondere die Auffassung einher, dass nur die diffeomorphismusinvarianten und damit eichinvarianten Grössen physikalisch ernst zu nehmen sind. Sie entsprechen den 'echten' Freiheitsgraden der Theorie. Die Punktmannigfaltigkeit der Allgemeinen Relativitätstheorie hat im Rahmen dieser Deutung der aktiven Diffeomorphismusinvarianz als Eichinvarianz keinen ontologischen Status. – Dies wird dann, was allerdings letztendlich eine Frage der begrifflichen Konventionen bleibt, von einigen Autoren mit dem Relationalismus bezüglich der von der Allgemeinen Relativitätstheorie beschriebenen Raumzeit gleichgesetzt. Die Allgemeine Relativitätstheorie ist dieser Auffassung zufolge schon eine relationalistische Theorie, die ihren Relationalismus jedoch teilweise hinter einer redundanten, unphysikalischen Überschussstruktur verbirgt.

Was hier jedoch entscheidender ist: Die so plausible eichinvariante Interpretation der Allgemeinen Relativitätstheorie hat einige äusserst radikale Konsequenzen:

> *"For a theory that satisfies the proposed version of requirement of substantive general covariance, the 'observables', or quantities that the theory counts as genuine physical magnitudes as opposed to mere mathematical artifice, will be diffeomorphic invariants [...]. Some philosophers pay lip service to this consequence without realizing just how radical it is. To illustrate it suffices to focus on the vacuum Einstein field equations [...]. For generic solutions of these equations no local object field constructed from $g_{ab}$ and its derivatives – not even 'scalar invariants' such as the Ricci curvature scalar R that appears in the Einstein equations – is a diffeomorphic invariant."* (Earman (2006) 448)

Das Problem besteht einerseits darin, dass so gut wie keine eichinvarianten Grössen in der Allgemeinen Relativitätstheorie bekannt sind:

> *"Very few gauge invariant quantities are known. [...] Thus, it is not at all trivial to formulate a gauge invariant interpretation of general relativity. As long as the existence of a sufficient number of suitable gauge invariant quantities of general relativity remains an open question, a dark cloud hangs over the programme of giving a gauge invariant interpretation of the theory."* (Belot / Earman (1999) 177)

Insbesondere gehören das metrische Feld und die Materiefelder nicht zu den eichinvarianten Grössen innerhalb der Allgemeinen Relativitätstheorie:

> *"In [general relativity], for example, the metric and matter fields [...] are not observables [...]."* (Earman (2006a) 13)

Sollten diesbezüglich noch Zweifel bestanden haben, so zeigt sich spätestens hier, dass die sich an der Metrik als konstitutivem Element der Raumzeit orientierenden substantialistischen Deutungen der Allgemeinen Relativitätstheorie, wie etwa der metrische Essentialismus, unvereinbar sind mit der eichinvarianten Deutung der Theorie; sie sind letztendlich unverträglich mit der *Leibniz-Äquivalenz* als physikalischem Identitätskriterium.

Es gibt – und dies ist die andere, vielleicht noch gravierendere Seite des Problems – überhaupt keine lokalen (oder auch nur quasi-lokalen) eichinvarianten Grössen in der Allgemeinen Relativitätstheorie.[128]

> *"[...] it is known that there are* no *local gauge-invariant quantities."* (Belot / Earman (2001) 229)

Das ist letztlich nicht sonderlich erstaunlich; es hat vielmehr einen ziemlich leicht nachvollziehbaren, durchaus anschaulichen Grund:

> *"[...] a diffeomorphism maps one spacetime point to another, and therefore one obvious way of constructing a diffeomorphism-invariant object is to take a scalar function of spacetime fields and integrate it over the whole of spacetime, which gives something that is very* non-*local. The idea that 'physical observables' are naturally non-local is an important ingredient in some approaches to quantum gravity."* (Butterfield / Isham (2001) 61)

Es bleibt, auch wenn man dies schon in Betracht zieht, erst einmal weitgehend unklar, welche Observablen für die Allgemeine Relativitätstheorie überhaupt eine Rolle spielen könnten.[129]

> *"[...] one can wonder whether there exists a full set of gauge-invariant quantities [...] of general relativity."* (Belot / Earman (2001) 229)

Übersetzt in die Begrifflichkeit der Darstellung der eichinvarianten Grössen im (um die Eichsymmetrien) reduzierten (physikalischen) Phasenraum, dessen Punkte den Eichorbits bzw. den Äquivalenzklassen der durch Diffeomorphismen verbundenen Lösungen der Allgemeinen Relativitätstheorie entsprechen, stellt sich das gleiche Problem dann folgendermassen dar:

> *"Relatively little is presently known about the structure of the reduced phase space of general relativity. It is known that this space has singularities corresponding to models of general relativity with symmetries, and is smooth elsewhere [...]."* (Belot / Earman (2001) 229)

---

[128] Siehe etwa Torre (1993).

[129] Lyre (2004) zeigt am Beispiel der Maxwellschen Elektrodynamik sehr überzeugend, dass es im Kontext von Eichtheorien, wenn man sich unter Berücksichtigung der Forderung nach *Leibniz-Äquivalenz* auf eichinvariante Grössen als ernstzunehmende physikalische Grössen beschränkt, wohl vor allem nichtlokale Holonomien und Wilson-Loops sind, die als Repräsentanten der echten, physikalischen Freiheitsgrade in Frage kommen. Solche Holonomien und Wilson-Loops stellen dann auch schliesslich den Ausgangspunkt für die Quantisierung der Allgemeinen Relativitätstheorie im Rahmen der *Loop Quantum Gravity* dar (siehe Kap. 4.4.). Das Problem der nichtlokalen echten Freiheitsgrade tritt, wie das Beispiel der Elektrodynamik zeigt, wohl nicht erst im Kontext der Allgemeinen Relativitätstheorie auf, wird aber dort durch die aktive Diffeomorphismusinvarianz, die hier die Rolle der Eichinvarianz spielt und dabei einen unproblematischen Bezug auf Raumzeitpunkte (und damit auf Lokalisierungen) unmöglich macht, brisanter und gleichzeitig in gewisser Weise anschaulicher nachvollziehbar.

Und hier lässt sich dann munter über den reduzierten Phasenraum der echten Freiheitsgrade speku-
lieren: seine absolute Fremdartigkeit, seine Abstraktheit und Unanschaulichkeit – gemessen an den
bisherigen Erfahrungen mit physikalischen Theorien.

> *"[...] would it be possible to build a fundamental theory that is grounded in the reduced phase space
> parametrized by the Dirac observables? This would be an abstract and highly non-local theory of
> classical gravitation but, transparency aside, it would lack all the epistemic machinery (the gauge
> freedom) which is indispensable for the application of the theory."* (Dorato / Pauri (2006) 146)[130] –
> *"Indeed, would it be possible to build a fundamental theory that is grounded in the reduced phase
> space parametrized by the Dirac observables? This would be an abstract and highly nonlocal theory
> of gravitation that would admit an infinity of gauge-related, spatio-temporally local realizations.
> From the mathematical point of view, however, this theory would be just an especially perspicuous in-
> stantiation of the relation between canonical structure and locality that pervades contemporary theo-
> retical physics nearly everywhere."* (Pauri / Vallisneri (2002) 27)

Die Nichtlokalität der 'echten' physikalischen Freiheitsgrade, die sich hinter der Allgemeinen Rela-
tivitätstheorie verbergen, und die daraus resultierende Konsequenz, für eine Beschreibung auf der
Grundlage der 'echten' physikalischen Freiheitsgrade ohne eine raumzeitliche Mannigfaltigkeit aus-
kommen zu müssen, hat nicht zuletzt auch die Notwendigkeit des Verzichts auf das gewohnte
mathematisch-modelltheoretische Instrumentarium der Physik zur Folge. Dies schliesst insbeson-
dere Differentialgleichungen, Differentialgeometrie etc. ein. Dieses gewohnte mathematische In-
strumentarium setzt eine differenzierbare Mannigfaltigkeit voraus, die für eine Beschreibung auf
der Grundlage der 'echten' physikalischen Freiheitsgrade ganz sicher nicht mehr vorliegen wird. Es
könnte also sein, dass sich dieser reduzierte Phasenraum der echten physikalischen Freiheitsgrade in
seiner Fremdartigkeit und Abstraktheit mit dem methodologischen Instrumentarium der heutigen
Physik gar nicht entwickeln oder gar nicht handhaben lässt.

Bei der eichtheoretischen Beschreibung der Raumzeit und ihrer 'Dynamik' – einer Beschreibung,
die offensichtlich mit vielfältigen Redundanzen arbeitet und nur sehr indirekte Hinweise auf die
echten, physikalisch wirksamen Freiheitsgrade enthält – handelt es sich also, wie spätestens vor
diesem Hintergrund deutlich wird, keinesfalls um das leicht vermeidbare Ergebnis einer nur be-
quemen, sich an den gewohnten Instrumentarien der Physik orientierenden, symmetriegestützten
Herangehensweise. Vielmehr könnte es sich durchaus herausstellen, dass diese redundanzbehaftete
eichtheoretische Erfassung tatsächlich die einzige Möglichkeit für eine (klassische) Beschreibung
der Raumzeit darstellt, die uns zur Zeit (und auf der Grundlage der zur Zeit verfügbaren mathemati-
schen und modelltheoretischen Instrumentarien) zu Gebote steht.

Andererseits könnte es aber auch durchaus sein, dass eine Herangehensweise, die mit einem völlig
anderen mathematischen Instrumentarium arbeitet, das gerade zu einer physikalisch angemessene-
ren, redundanzfrei(er)en Beschreibung führen würde, eigentlich schon mit den heute verfügbaren
Mitteln realisiert werden könnte, dass aber der Blick auf diese Alternative bisher noch durch die
methodischen Traditionen der Physik verstellt wird. Eine solche Alternative, die vielleicht zu einer
vollständig relationalen, originär nichtlokalen Beschreibung führen würde, könnte sich vielleicht gerade im Kontext der Bemühungen um eine Theorie der Quantengravitation aufzeigen lassen. Dabei könnte sich die phänomenologische Raumzeit ohne weiteres als Artefakt einer redundanzbehafteten (eichtheoretischen) Beschreibung erweisen.[131] Hierauf deuten immerhin schon die wenigen bekannten Eigenschaften (insbesondere die Nichtlokalität) der noch unbekannten echten Freiheitsgrade hin, die der Allgemeinen Relativitätstheorie offenbar zugrundeliegen. – Doch kehren wir nach diesen Spekulationen erst einmal zurück zu den etwas konkreteren Problemlagen, die sich mit der eichinvarianten Deutung der Allgemeinen Relativitätstheorie ergeben:

---

[130] Dorato / Pauri (2006) versuchen, die echten Freiheitsgrade der Allgemeinen Relativitätstheorie mittels einer intrinsi-
schen Eichfixierung zu erschliessen. Siehe ebenso Pauri / Vallisneri (2002), Lusanna / Pauri (2005). Diese Eichfixie-
rung führt zu einer Aufspaltung des metrischen Tensors, mittels derer die echten Freiheitsgrade der Allgemeinen Rela-
tivitätstheorie erschlossen werden sollen.

## Das Problem der Zeit

Ein gleichermassen plakatives wie virulentes Problem, das aus der eichtheoretischen Deutung und der entsprechenden Forderung nach Eichinvarianz für die echten physikalischen Grössen der Allgemeinen Relativitätstheorie resultiert, wurde im Vorausgehenden noch nicht explizit benannt. Es betrifft die Zeit und wie sie sich im Rahmen der Allgemeinen Relativitätstheorie darstellt:

Einerseits gibt es in der Allgemeinen Relativitätstheorie keinen externen Zeitparameter. Die Koordinatenzeit (als Entwicklungsparameter der Feldgleichungen) ist als Eichvariable nicht beobachtbar oder messbar. Andererseits ist die von Uhren gemessene physikalische Grösse (die Eigenzeit) eine nicht-triviale Funktion des Gravitationsfeldes. Damit gibt es keine übergreifend wirksame, global verfügbare und zudem messbare Zeit in der Allgemeinen Relativitätstheorie – schon gar keine, die für alle ihre Lösungen gleichermassen anwendbar wäre.

Dass die Zeit in der Allgemeinen Relativitätstheorie keine eichinvariante Grösse ist, lässt sich in der Hamiltonschen Darstellung leicht nachvollziehen: Die Hamiltonsche (skalare) Zusatzbedingung lässt sich, wie wir gesehen haben, als Generator einer Eichtransformation verstehen. Da die Hamiltonsche Zusatzbedingung aber auch (zumindest für den generischen Fall eines kompakten Raumes) die zeitliche Entwicklung von einer Hyperfläche zur nächsten erfasst, entspricht diese nun einer Eichtransformation.

> "The Hamiltonian constraint of the theory is a first class constraint and should therefore be viewed as a gauge transformation. However, since the constraint is responsible for generating the time evolution of the data from one hypersurface to the next it looks as if time evolution corresponds to the unfolding of a gauge transformation. Gauge transformations are, of course, unphysical." (Rickles (2005a) 11f)

Wenn die zeitliche Entwicklung jedoch einer Eichtransformation entspricht, so gibt es in der Allgemeinen Relativitätstheorie, zumindest im Kontext ihrer eichinvarianten Deutung, keine physikalisch wirksame Zeit und keine zeitlichen Entwicklungen, die sich als physikalisch real deuten liessen.

> "[...] prima facie, gauge invariant interpretations of general relativity imply that time and change are illusions." (Belot / Earman (1999) 177)

---

[131] Dies hiesse aber andererseits, dass gerade die Eichinvarianzen eine entscheidende Rolle spielen könnten, wenn es um die Rekonstruktion des herkömmlichen raumzeitlichen Bildes geht, das dann aus einer fundamentaleren Beschreibung gewonnen werden müsste, die nur mit den zugrundeliegenden echten Freiheitsgraden arbeitet:

> "In other words the gauge structure could be seen as playing a crucial role in the re-construction of the concrete spatiotemporal continuum representation from a non-local structure." (Lusanna / Pauri (2005) 25)

Das Problem betrifft nicht etwa nur die Suche nach einem physikalisch ernstzunehmenden Zeit-parameter innerhalb der Allgemeinen Relativitätstheorie, sondern den Versuch der Identifizierung von physikalisch ernstzunehmenden zeitlichen Entwicklungen überhaupt: Wenn die zeitliche Ent-wicklung einer Eichtransformation entspricht, sind alle Observablen des Systems notwendigerweise zeitunabhängig. Zeitabhängige Grössen kommen, da sie eichabhängig und somit nicht beobachtbar und nicht messbar sind, als Observablen nicht in Frage. Wenn die Hamiltonsche Zusatzbedingung, welche die zeitliche Entwicklung des Systems beschreibt, als Generator einer Eichtransformation verstanden wird, so entsprechen alle Verlaufsschritte einer zeitlichen Entwicklung Punkten inner-halb des gleichen Eichorbits im kinematischen Phasenraum des Hamiltonschen Systems (bzw. auf der durch die Zusatzbedingungen definierten Oberfläche). Im reduzierten Phasenraum der eichinva-rianten, echten physikalischen Freiheitsgrade, dessen Punkte jeweils vollen Eichorbits entsprechen, fallen dann alle Verlaufsschritte einer zeitlichen Entwicklung auf den gleichen Punkt zusammen. In der Darstellung des Systems im reduzierten Phasenraum kommen zeitliche Entwicklungen also überhaupt nicht mehr vor. Die Darstellung im reduzierten Phasenraum löscht jegliche zeitliche Entwicklung aus.

> *"However, since one of the constraints (the Hamiltonian constraint) was associated with time evolu-tion, in factoring its action out the dynamics is eliminated, since time evolution unfolded along a gauge orbit (i.e. instants of time correspond to the points 'parametrizing' a gauge orbit)."* (Rickles (2006) 166)

Die Allgemeine Relativitätstheorie beschreibt also der eichinvarianten Interpretation zufolge nicht etwa ein System, das sich als 'sich in der Zeit entwickelnd' verstehen lässt, zumindest, wenn man damit eine Zeit meint, die der Beobachtung zugänglich ist und die unabhängig ist von der in ihr ablaufenden Entwicklung.

> *"[General relativity] teaches us that we must abandon the idea that the flow of time is an ultimate aspect of reality. [...] The dynamics of [general relativity] itself cannot be cleanly described in terms of evolution in time."* (Rovelli (2006) 34)

Aber welche Alternative gibt es schliesslich zur eichinvarianten Interpretation? Die einer zeitlichen Entwicklung (im absoluten Sinne) entsprechenden Zustände sind der Allgemeinen Relativitätstheo-rie zufolge – unabhängig von der Interpretationsfrage – empirisch nicht unterscheidbar. Sollte die Theorie in irgendeiner nicht offensichtlichen, verborgenen Weise dann doch noch die Beschreibung einer zeitlichen Entwicklung enthalten, so ist dies zumindest nicht so ohne weiteres aus ihrer formalen Struktur heraus erkennbar.[132]

---

[132] Dies lässt sich, insbesondere in Bezug auf die relationale Einbindung der Zeit in das von der Allgemeinen Relativi-tätstheorie beschriebene Gesamtgeschehen, auch anders ausdrücken:

> *"One important consequence is that one cannot define the physical observables of the theory without solving the dynamics. In other words, as Stachel emphasizes,* there is no kinematics without dynamics. *This is because all observables are relational, in that they describe relations between physical degrees of freedom."* (Smolin (2006c) 207)

Insbesondere Rovelli versucht einen relationalen Zeitbegriff durch die Auszeichnung interner zeitlicher Grössen zu etablieren:

> *"[General relativity] does not describe evolution in time: it describes the relative evolution of many variables with respect to each other. [...] This is the temporal aspect of [general relativity]'s relationalism."* (Rovelli (2001) 112)

*"If there is a true time hiding in the formalism of [the general theory of relativity], it is well hidden indeed."* (Earman (2002) 19)

## Implikationen für die Quantengravitation

Die zum Teil äusserst radikalen Konsequenzen einer eichinvarianten Interpretation der Allgemeinen Relativitätstheorie übertragen sich insbesondere dann auf die Ansätze zu einer Theorie der Quantengravitation, wenn diese versuchen, die Allgemeine Relativitätstheorie unter Berücksichtigung ihrer grundlegenden konzeptionellen Implikationen direkt zu quantisieren.

*"Indeed, in order to formulate a gauge-invariant quantum theory, one would like to be able to find [...] a full set of gauge-invariant quantities [...]. This would amount to isolating the true (i.e. gauge-invariant) degrees of freedom of the theory."* (Belot / Earman (2001) 229)

Die mit der eichinvarianten Deutung der klassischen Ausgangstheorie verbundene Problematik der Suche nach den echten, physikalisch wirksamen Freiheitsgraden betrifft insofern vor allem den Ansatz der *Kanonischen Quantisierung*, also der direkten nicht-perturbativen Quantisierung der Allgemeinen Relativitätstheorie in ihrer Hamiltonschen Darstellung. Sowohl in ihrer geometrodynamischen Variante[133] als auch in der *Loop Quantum Gravity*[134] spielt die Idee der Identifizierung der eichinvarianten, echten Freiheitsgrade eine entscheidende Rolle bei der Auszeichnung der Quantenobservablen, für die eben auch wieder Eichinvarianz zu fordern ist.

*"What to take as the basic gauge invariant quantities or observables of classical [general theory of relativity] is a much discussed issue, especially for theorists who pursue the canonical approach to quantum gravity since, presumably, the observables of the canonical quantum gravity (i.e. the self-adjoined operators of the relevant Hilbert space) will correspond to gauge invariant quantities of classical general relativity."* (Earman (2006) 448)

Die Auszeichnung dieser echten Freiheitsgrade lässt sich im Rahmen der *Kanonischen Quantisierung* in mancher Hinsicht sogar als noch unabwendbarer ansehen als in der klassischen Theorie, wie bei der Erörterung der *Loop Quantum Gravity* zu verdeutlichen sein wird.[135] Dort wird nämlich die eichinvariante Deutung der Diffeomorphismusinvarianz fest im Formalismus der Theorie verankert; sie lässt sich dann nicht mehr als Komponente ansehen, deren Lesart vielleicht noch im Kontext der Klärung interpretatorischer Fragen zu verhandeln wäre. Es gibt im Rahmen der *Loop Quantum Gravity* einfach nicht mehr diesen interpretatorischen Spielraum, den es in Bezug auf die Allgemeine Relativitätstheorie vielleicht grundsätzlich noch gibt, auch wenn letztlich schon hier, für den klassischen Fall, die Alternativen zur eichinvarianten Deutung nicht sehr verlockend erscheinen.

---

Mögliche Reaktionen auf bzw. Lösungsansätze zum *Problem der Zeit* werden, da dieses im Rahmen der *Kanonischen Quantisierung* der Allgemeinen Relativitätstheorie wieder – und diesmal noch hartnäckiger – auftritt, im Kontext der *Loop Quantum Gravity* zu diskutieren sein. Siehe Kap. 4.4.

[133] Siehe Kap. 4.3.

[134] Siehe Kap. 4.4.

[135] Siehe Kap. 4.4.

Insbesondere werden in der *Kanonischen Quantisierung* die radikalen Implikationen und Probleme der eichinvarianten Interpretation, die sich schon in der klassischen Theorie aufzeigen lassen, allesamt wieder virulent.

> *"Accepting a gauge-invariant interpretation of general relativity, and thus treating the general covariance of general relativity as analogous to the gauge invariance of electromagnetism, leads to nasty technical and interpretive problems when one attempts to quantize the theory."* (Belot / Earman (2001) 230)

Und die Probleme, die sich mit der eichinvarianten Deutung der Allgemeinen Relativitätstheorie abzeichnen, werden im Rahmen ihrer Quantisierung teilweise sogar noch gravierender. Einerseits kommen manche der pragmatischen Auswege, die im Kontext der klassischen Theorie noch zur Verfügung stehen – wenn es etwa um konkrete Fragen geht, die sich ausschliesslich auf spezifische Lösungen beschränken –, hier im Kontext ihrer Quantisierung, wo man es nun mit quantenmechanischen Superpositionen von Raumzeiten zu tun hat, einfach nicht mehr in Betracht. Andererseits führt auch das *Problem der Zeit* hier zu noch grundlegenderen Schwierigkeiten als im klassischen Fall. Es hat für eine Quantentheorie insbesondere zur Folge, dass diese ohne Unitarität auskommen muss. Es gibt keine unitäre Entwicklung mehr, da es überhaupt keine zeitliche Entwicklung im ursprünglichen quantenmechanischen Sinne mehr gibt. Und die Tatsache, dass im Rahmen der herkömmlichen Quantenmechanik erst einmal ein Zeitparameter vorausgesetzt werden muss, führt letztlich zur Notwendigkeit einer grundlegenden Modifikation des quantenmechanischen Formalismus.[136] Hinzu kommen die Probleme, die sich aus der unabwendlichen Nichtlokalität der eichinvarianten Observablen ergeben.

Diese Implikationen der eichinvarianten Deutung werden zwar im kanonischen Quantisierungsansatz, der auf dem Hamilton-Formalismus beruht, besonders deutlich; aber letztlich betreffen sie jegliche Theorie der Quantengravitation, welche die aktive Diffeomorphismusinvarianz der Allgemeinen Relativitätstheorie und ihre Deutung als Eichinvarianz in ihren Formalismus überführt. Sie betreffen letztlich jede Theorie, die sich in ernsthafter Weise noch als Ergebnis einer direkten Quantisierung der Allgemeinen Relativitätstheorie ansehen lässt. – Oder wie Karel Kuchar diesbezüglich konstatiert:

> *"The profound message of general relativity is that spacetime does not have any fixed structure which is not dynamical but governs dynamics from outside as an unmoved mover. The problem of time is only one facet of the missing unmoved mover, and canonical quantization only a handy tool for laying bare the consequences. [...] If so, the problem of time in quantum geometrodynamics may be only a wind that precedes the storm."* (Kuchar (1999) 193)

Der Erfolg der Ansätze zur Quantengravitation, welche die Allgemeine Relativitätstheorie als entscheidenden konzeptionellen Ausgangspunkt der Theorieentwicklung ansehen, hängt damit offensichtlich nicht unwesentlich davon ab,

– ob die Identifizierung der echten Freiheitsgrade gelingt,
– ob dies zu einer konsistenten Quantisierung und damit schliesslich zu einer konsistenten Theorie der Quantengravitation führt,

---

[136] Siehe Kap. 4.4. sowie Rovelli (1991, 1991a, 1991c, 1996, 2002).

– ob diese dann – die Grundanforderung an jede Theorie der Quantengravitation – die Allgemeine Relativitätstheorie als klassischen Grenzfall enthält bzw. ihre Phänomenologie reproduziert, und

– ob sie schliesslich – die Grundanforderung an jede physikalische Theorie – eine (differentielle) empirische Überprüfung ermöglicht (und dann auch besteht).

Für die Modalitäten der Identifizierung der echten, physikalischen Freiheitsgrade, die von einer solchen Theorie der Quantengravitation erfasst werden sollen, gibt es letztlich zwei Möglichkeiten: Diese Identifizierung kann einerseits schon im Rahmen der klassischen Theorie, also der Allgemeinen Relativitätstheorie, angestrebt werden, etwa im Kontext der abstrakten Darstellung auf der Grundlage des reduzierten Phasenraums. Dies hätte den Vorteil, dass die echten Freiheitsgrade der klassischen Theorie als Ausgangspunkt für eine Quantisierung genutzt werden können. Andererseits könnte man jedoch auch versuchen, nach einer Quantisierung der redundanzbehafteten kinematischen Formulierung der klassischen Theorie quasi post-hoc die echten Freiheitsgrade der Quantengravitation zu identifizieren. Da aber eine solche, hier nun auf die Quantentheorie verschobene Suche nach den echten Freiheitsgraden überhaupt nur dann als sinnvoll erscheint, wenn die entsprechende Theorie der Quantengravitation die Einsichten der eichinvarianten Deutung der Allgemeinen Relativitätstheorie perpetuiert – insbesondere auch die entsprechenden Implikationen hinsichtlich der Raumzeit –, muss auch eine solche Quantisierung der immer noch redundanzbehafteten klassischen Theorie im Rahmen eines hintergrundunabhängigen Ansatzes erfolgen.

Welche dieser beiden Strategien man auch verfolgt: Die Identifizierung der echten, physikalisch wirksamen Freiheitsgrade ist sicherlich – sowohl im klassischen Fall, als auch im Kontext einer entsprechenden Quantentheorie, wenn diese die diesbezüglichen Implikationen der klassischen Theorie fortschreibt – alles andere als ein triviales Problem:

> *"Although we sometimes use Einstein's equations as if they were a machine for generating solutions, within which we then study the motion of particles or fields, this way of seeing the theory is inadequate as soon as we want to ask questions about the gravitational degrees of freedom themselves. Once we ask about the actual local dynamics of the gravitational field, we have to adopt the viewpoint which understands general relativity to be a background-independent theory within which the geometry is completely dynamical, on an equal footing with the other degrees of freedom. The correct arena for this physics is not a particular spacetime, or even the linearized perturbations of a particular spacetime. It is the infinite dimensional space of gravitational degrees of freedom. From this viewpoint, individual spacetimes are just trajectories in the infinite dimensional phase or configuration space; they can play no more of a role in a quantization of spacetime than a particular classical orbit can play in the quantization of an electron."* (Smolin (2006c) 218)

*

Vor dem Hintergrund der Probleme, die den Ansätzen zu einer Theorie der Quantengravitation aus der eichinvarianten Interpretation der Allgemeinen Relativitätstheorie erwachsen, erscheint es also durchaus sinnvoll, die Frage nach der Unabdinglichkeit einer solchen Ausgangslage für die Quantengravitation und entsprechend die nach möglichen Alternativen zu stellen. Letztlich heisst dies aber nicht anderes, als die Frage nach der Angemessenheit der eichinvarianten Deutung der Allgemeinen Relativitätstheorie erneut aufzurollen, nun vor dem Hintergrund der möglichen Perspekti-ven, die sich mit ihr für eine Quantisierung ergeben. – So bestreitet Weinstein[137] angesichts der daraus resultierenden Probleme etwa, dass Diffeomorphismen tatsächlich als Eichtransformationen gedeutet werden können, und dass die Allgemeine Relativitätstheorie aufgrund ihrer Diffeomorphismusinvarianz als Eichtheorie verstanden werden kann.

> *"[...] the diffeomorphism group is unlike an ordinary gauge group in that it is not a 'local' (in the sense of 'at the same manifold point') transformation."* (Weinstein (1999) S152)

Nach Weinsteins Auffassung ist es keinesfalls klar, wie die Allgemeine Relativitätstheorie zu interpretieren ist, und vor allem unter Vorgabe welcher konzeptionellen Randbedingungen sie zu quantisieren wäre.

> *"[...] it is not at all clear what it means to quantize a diffeomorphism-invariant theory."* (Weinstein (1999) S154)

Aber wie, wenn nicht analog zu einer Eichinvarianz, sollte sich die aktive Diffeomorphismusinvarianz deuten lassen, wenn die Modelle der Raumzeit, die sich durch Diffeomorphismen ineinander überführen lassen, nicht physikalisch unterscheidbar sind, insofern also einer Äquivalenzklasse von physikalisch ununterscheidbaren Beschreibungen entsprechen, die sich, wenn man keinen Bezug auf unbeobachtbare Grössen und keinen Kryptoindeterminismus akzeptieren möchte, nur auf ein und den selben physikalischen Zustand beziehen können? Welche Alternativen gibt es also zur eichinvarianten Interpretation der Allgemeinen Relativitätstheorie? Anders gefragt: Um welchen Preis wären diese Alternativen zu haben? Möchte man an der Allgemeinen Relativitätstheorie als Ausgangsbasis festhalten, so wäre dieser für die Alternativen zu zahlende Preis vermutlich zu hoch – oder die Alternativen zu unmotiviert – und es bliebe letztlich nur die eichinvariante Interpretation. Zu ihr gibt es in letzter Konsequenz nur sehr kuriose Alternativen, die, wenn man sich auf die prinzipielle Beobachtbarkeit physikalischer Grössen als realitätsstiftendes Kriterium beruft und metaphysische Konstruktionen in dieser Hinsicht ablehnt, inakzeptabel sind.

Erhellend im Hinblick auf die von Weinstein aufgeworfene Problematik sind die Ausführungen von Dorato und Pauri, die auf den entscheidenden Unterschied der Eichinvarianz der Allgemeinen Relativitätstheorie im Vergleich mit den Eichinvarianzen anderer Theorien hinweisen, aber dennoch weiterhin an einer eichinvarianten Deutung der Allgemeinen Relativitätstheorie festhalten und ihre Konsequenzen hervorheben:

> *"[...] while from the point of view of the constrained Hamiltonian mathematical formalism general relativity is a gauge theory like any other (e.g. electromagnetism and Yang-Mills theories), from the physical point of view it is radically different. For, in addition to creating the distinction between what is observable and what is not, the gauge freedom of [general relativity] is unavoidably entangled with the constitution of the very stage, spacetime, where the play of physics is enacted [...]. In other words, the gauge mechanism has the dual role of making the dynamics unique (as in all gauge theories), and of fixing the spatio-temporal, dynamical background [...]."* (Dorato / Pauri (2006) 144)[138]

---

[137] Siehe Weinstein (1999). Vgl. auch Maudlin (2002) und Kuchar (1986, 1991, 1992).

[138] In der Faserbündeldarstellung lässt sich diese Besonderheit der Eichinvarianz der Allgemeinen Relativitätstheorie in der Verschmelzung des Basis- und des Faserraums zu einem Verschmelzungsbündel erfassen. Siehe Lyre (2004).

Es gibt jedoch auch die überraschende Einschätzung, die Deutungsfrage hinsichtlich der Diffeomorphismusinvarianz der Allgemeinen Relativitätstheorie als nicht im engeren Kontext dieser Theorie zu entscheidendes Problem anzusehen. So ist etwa Norton der Auffassung, dass die Frage, ob es sich bei der Allgemeinen Relativitätstheorie um eine Eichtheorie handelt, nicht im Rahmen des mathematischen Formalismus, sondern im Kontext 'physikalischer Überlegungen' zu klären ist – wobei nicht ganz klar wird, ob mit letzteren vielleicht gar empirisch überprüfbare Konsequenzen gemeint sein könnten:

> *"The decision as to whether a transformation is a gauge transformation cannot merely be decided by the mathematics; it is a physical issue that must be settled by physical considerations."* (Norton (2004) 14)

Aber letztlich steht nicht die Frage der eichinvarianten Deutung der Allgemeinen Relativitätstheorie der Empirie gegenüber. Letztlich steht hier vielmehr die Frage zur Debatte, ob die Allgemeine Relativitätstheorie überhaupt – zumindest im klassischen, makroskopischen Fall – als empirisch adäquate Theorie angesehen werden kann: mit allen ihren radikalen Implikationen.

Ein weiterer überraschender Vorschlag, der wohl vor allem die Unsicherheit in dieser Fragestellung deutlich macht, stammt vor Earman. Dieser sieht erstaunlicherweise Chancen dafür, dass mögliche Bestätigungsinstanzen für eine eichinvariante Deutung der Allgemeinen Relativitätstheorie auf dem Umweg über eine entsprechende Quantisierung gewonnen werden können.

> *"If the loop formulation of quantum gravity continues to make theoretical progress and, eventually, passes experimental checks, then I would take these successes to be confirmation of the gauge interpretation of [general relativity] dictated by the Dirac constraint formalism."* (Earman (2002a) 21)

Führt man sich Earmans überzeugende Argumente für die eichinvariante Deutung der Allgemeinen Relativitätstheorie vor Augen, so erscheint diese Einschätzung recht kurios. Wieso sollte der Umweg über die Quantisierung zu einer Neueinschätzung hinsichtlich der Deutung der Allgemeinen Relativitätstheorie führen? Zwar würde ein Erfolg der direkten *Kanonischen Quantisierung* der Allgemeinen Relativitätstheorie ihren klassischen Ausgangspunkt – inklusive seiner eichinvarianten Interpretation – als adäquat erweisen. Aber die eichinvariante Deutung der Allgemeinen Relativitätstheorie würde über die schon bestehenden Argumente hinaus keinen weiteren Aufwind erhalten – und auch nicht benötigen. Andererseits würden gravierende und sich als unlösbar herausstellende Probleme einer direkten Quantisierung der Allgemeinen Relativitätstheorie umgekehrt nicht ihre eichinvariante Deutung, sondern entweder die Strategie einer direkten Quantisierung oder aber ihren klassischen Ausgangspunkt als unangemessen erscheinen lassen.

Sollte die gesuchte Quantengravitationstheorie also aus der Quantisierung der Allgemeinen Relativitätstheorie heraus gewonnen werden können und (auch empirisch) erfolgreich sein, und sollten die für unser Verständnis der Raumzeit massgeblichen Implikationen der Allgemeinen Relativitätstheorie sich in ihr perpetuieren, so würde dies (zumindest indirekt) die Allgemeine Relativitätstheorie für den klassischen, makroskopischen Bereich bestätigen – inklusive ihrer ohnehin schon im klassischen Fall ausreichend motivierten eichinvarianten Interpretation.

Umgekehrt verhält es sich aber folgendermassen: Sollten die Implikationen und Probleme der eichinvarianten Interpretation der Allgemeinen Relativitätstheorie im Kontext ihrer direkten Quan-tisierung zu unüberwindlichen Problemen führen, so wäre dies nicht etwa ein Zeichen für die Unangemessenheit ihrer eichinvarianten Interpretation, sondern entweder für die mangelnde Tragfähigkeit der Strategie einer direkten Quantisierung oder für die zumindest partielle Unangemessenheit der Allgemeinen Relativitätstheorie selbst als konzeptionellem Ausgangspunkt für eine Theorie der Quantengravitation. Es würde zumindest deutlich machen, dass die Allgemeine Relativitätstheorie als allumfassende Leitlinie für die Theoriebildung im Bereich der Quantengravitation nicht unproblematisch ist. Ihre eichinvariante Deutung selbst würde dabei jedoch nicht separat zur Disposition stehen.

Die entscheidende Frage ist und bleibt letztendlich, ob die Allgemeine Relativitätstheorie eine in ihrem Gegenstandsbereich empirisch adäquate Theorie ist oder durch eine bessere Theorie, etwa eine Quantengravitationstheorie, zu ersetzen ist, die nicht nur <u>nicht</u> die Allgemeine Relativitätstheorie als klassischen Ausgangspunkt hat, sondern vielleicht sogar zu Implikationen führt, die auch im klassischen, makroskopischen Bereich – dem ureigenen Gegenstandsbereich der Allgemeinen Relativitätstheorie – von dieser abweichen, und die vielleicht einige ihrer radikalen Konsequenzen aufhebt.

Ohnehin liesse sich argumentieren, dass eine Theorie, die bei einer adäquaten Interpretation unabwendbar (so dies der Fall sein sollte, was noch lange nicht klar ist[139]) zum *Problem der Zeit* führt, letztendlich nicht mit den methodologischen Voraussetzungen empirischer Wissenschaft vereinbar ist – und letztlich ihre eigenen empirischen Bestätigungsinstanzen ad absurdum führt:

*"[...] at least in the typical case, a physical theory is confirmed by testing its* predictions *– statements made* at an earlier time *in ignorance of their truth-value and then checked by making observations* at a later time. *Both the formulation of a prediction and the performance of a subsequent observation to test it are* acts *– events of a particular kind involving different intentional states that the observer is in at different times. It follows that the testing of a prediction presupposes the possibility of* change *– in the mental state of an observer, if not also in the physical state of the world that he or she is observing. / All those points are blindingly obvious. But note what follows from them. There can be no reason whatever to accept any theory of gravity – quantum or classical – which entails that there can be no observers, or that observers can have no experiences, some occurring later than others, or that there can be no change in the mental states of observers, or that observers cannot perform different acts at different times. It follows that there can be no reason to accept any theory of gravity – quantum or classical – which entails that there is no time, or that there is no change. Now it is important to note that it does not follow that no such theory can be* true. *But any such theory would have the peculiar feature that, if true, there could be no reason to accept it. [...] In the case of a quantum theory of gravity, this negative conclusion may not come as a surprise. [...] But classical general relativity is a different matter: we take ourselves to have considerable evidence supporting this theory [...] But if general relativity, correctly interpreted, implies the nonexistence of time, or of change, then we must be wrong to take this evidence to support the theory after all. For the correct interpretation of this supposed evidence must be undercut its epistemic credentials. Put bluntly, a radically timeless interpretation of general relativity entails the impossibility of performing any of the experiments and observations, the performance of which we ordinarily take to provide our reasons to believe that theory. Such an interpretation makes the theory empirically self-refuting.* (Healey (2002) 300f)

---

[139] Siehe Kap. 4.4.

Der einzige Ausweg aus diesem Problem stützt sich dann tatsächlich auf die Hoffnung, dass es – auch im Rahmen einer eichinvarianten Deutung der Allgemeinen Relativitätstheorie – durchaus Auswege aus der vermeintlichen Zeitlosigkeit geben könnte, etwa in Form eines internen, relationalen Zeitkonzeptes, mittels dessen sich dieser Konflikt mit den methodologischen Voraussetzungen empirischer Wissenschaft verhindern lässt.[140]

*

Angesichts der radikalen Konsequenzen der eichinvarianten Interpretation der Allgemeinen Relativitätstheorie und angesichts der ausschliesslich kuriosen Alternativen, die für eine Interpretation der Allgemeinen Relativitätstheorie ansonsten in Frage kämen, sollte man dennoch nicht unbedingt warten, bis die direkt und unumschränkt auf ihr aufbauenden Ansätze zu einer Theorie der Quantengravitation vielleicht tatsächlich unzweifelhaft in unüberwindliche Probleme geraten. Wie auch immer könnte man letztgültig feststellen, ob es sich um unüberwindliche Probleme handelt? – Man sollte vielmehr schon zuvor parallel laufende Strategien verfolgen. Im Rahmen solcher Alternativstrategien ist es sicherlich nicht unsinnig, sich die grundsätzliche Frage zu stellen, ob die Allgemeine Relativitätstheorie tatsächlich mit allen ihren konzeptionellen Implikationen als angemessene Grundlage für eine zukünftige Theorie der Quantengravitation angesehen werden muss. Die unkritische Übernahme der mit ihr gewonnenen Einsichten könnte nämlich, wenn diese theoretische Artefakte enthalten sollten, ohne weiteres zu Erkenntnishemmnissen beitragen.

Eine mögliche Alternative zu den sich anbahnenden Problemen einer direkten Quantisierung der Allgemeinen Relativitätstheorie besteht also gerade darin, letztere nicht in allen ihren fundamentalen konzeptionellen Implikationen als unzweifelhaft relevant und konstitutiv für die Theorienbildung im Bereich der Quantengravitation anzusehen. Explizit könnte eine solche Alternative etwa die Annahme beinhalten, dass die Diffeomorphismusinvarianz vielleicht ein theoretisches Artefakt ist, das für den intendierten Bereich der Quantengravitation nicht mehr (umfassend) gilt. Die Diffeomorphismusinvarianz, die in ihrer eichinvarianten Interpretation als wesentliche Quelle für die geschilderten Probleme angesehen werden muss, spielt für die Quantengravitation ohnehin nur dann eine unmittelbare Rolle, wenn eine entsprechende Theorie von einem Raumzeitkontinuum ausgeht, oder zumindest von einer differenzierbaren Mannigfaltigkeit als rudimentärem Hintergrund.[141] Mit den noch zu erörternden Argumenten für eine diskrete Raumzeit- bzw. Substratstruktur[142] erscheint es jedoch sehr plausibel, dass dieses Raumzeitkontinuum auf fundamentaler Ebene ohnehin nicht existiert und dass damit die Diffeomorphismusinvarianz auf fundamentaler Ebene irrelevant ist.

> *"The concept of a smooth spacetime should not have any meaning in a quantum theory of the gravitational field where probing distances beyond the Planck length must result in black hole creation which then evaporate in Planck time, that is, spacetime should be fundamentally discrete. But clearly smooth diffeomorphisms have no room in such a discrete spacetime. The fundamental symmetry is probably something else, maybe a combinatorial one, that looks like a diffeomorphism group at large scales.."*
> (Thiemann (2001) 117)

---

[140] Siehe Kap. 4.4. sowie Rovelli (1991, 1991a, 1991c, 2001, 2002, 2004, 2007, 2009).

[141] Die *Loop Quantum Gravity* setzt beispielsweise eine solche differenzierbare Mannigfaltigkeit (eine kontinuierlich differenzierbare Pseudo-Riemannsche Punktmannigfaltigkeit mit vorgegebener Topologie, vorgegebener Dimensionalität und vorgegebener Signatur der Raumzeit) als rudimentären Hintergrund voraus, auch wenn sie diese dann schliesslich wegdiskutieren möchte. Siehe Kap. 4.4.

[142] Siehe Kap. 3.1..

Wenn es kein raumzeitliches Kontinuum mehr gibt, verliert also die Diffeomorphismusinvarianz, zumindest in der Form, in der sie innerhalb der Allgemeinen Relativitätstheorie in Erscheinung tritt, ihre formale Relevanz. Für Ansätze, die ohne ein Raumzeitkontinuum auskommen, jedoch weiterhin an der von der Allgemeinen Relativitätstheorie nahegelegten Hintergrundunabhängigkeit festhalten, wären eventuelle Pendants zur Diffeomorphismusinvarianz bzw. entsprechende Realisierungsbedingungen zur substantiellen Form der allgemeinen Kovarianz erst einmal hinsichtlich ihrer möglichen Relevanz und ihrer jeweiligen Konsequenzen auszuloten.

Eine weitere durchaus interessante Option ist die folgende: Die aktive Diffeomorphismusinvarianz könnte eine Eigenschaft einer makroskopischen Dynamik sein, zu der es auf der Grundlage eines völlig anders gearteten Substrats kommt.[143]

> *"Many symmetries we are familiar with associated with the macroscopic spacetime will become meaningless at the microscopic level (quantum gravity)."* (Hu (2009) 7)

Und die makroskopische Raumzeit, für die sich die aktive Diffeomorphismusinvarianz (dann vielleicht sogar in umfassender Form) geltend machen lässt, könnte vom Substrat dynamisch vollständig entkoppelt sein.[144]

---

[143] Girelli / Liberati / Sindoni (2008, 2009) (siehe auch Liberati / Girelli / Sindoni (2009)) beschreiben ein Szenario, für das sich die aktive Diffeomorphismusinvarianz, wie sie für die Allgemeine Relativitätstheorie vorliegt, (nicht jedoch die Theorie selbst) als emergentes Ergebnis einer Dynamik einstellt, für die diese Diffeomorphismusinvarianz noch nicht vorliegt. Es handelt sich um eines der ganz wenigen emergenten Szenarien, für die dies gelingt.

> *"[...] a typical drawback of analogue gravity models is related to the fact that they show only the emergence of a background Lorentzian geometry while they are unable to reproduce a geometrodynamics of any sort. [...] our model overcomes this drawback and indeed is able to describe the emergence of a full gravitational, diffeomorphism invariant, theory for scalar gravity. This theory will come out to be the only known other theory of gravitation, apart from General Relativity, which satisfy the strong equivalence principle, i.e. Nordstrom gravity."*
> (Girelli / Liberati / Sindoni (2008) 4f)

Ob sich ein solches Modell in einer Weise ausweiten liesse, dass es in der Lage wäre, die Allgemeine Relativitätstheorie selbst zu reproduzieren ist bisher noch unklar.

> *"[...] Nordström gravity is only a scalar gravity theory, which has been falsified by experiments [...]. In order to obtain a more physical theory we should look for a more complicated emergent Lagrangian."* (Girelli / Liberati / Sindoni (2008) 8)

Immerhin hat das beschriebene Modell empirisch nachweisbare Signaturen, die, wie man hofft, auch für mögliche realistischer Erweiterungen erhalten bleiben:

> *"In agreement with the fact that diffeomorphism invariance is emergent in our system, it can be noted that the cubic contribution [...] ends up breaking it at the same level it breaks Lorentz invariance. [...] in our framework one would predict strong deviations from the weak field limit of the theory whenever the gravitational field becomes very large."* (Girelli / Liberati / Sindoni (2008) 7)

Eine entscheidende Bedingung dafür, dass es überhaupt zur Emergenz der aktiven Diffeomorphismusinvarianz kommt, ist jedoch, dass Raumzeit und Materie im gleichen Schritt als emergente Phänomene aus der Substratdynamik hervorgehen.

> *"[...] gravity and matter are both emergent at the same level."* (Girelli / Liberati / Sindoni (2008) 7)

Auch darin unterscheidet sich das Modell von vielen der bekannten (und noch vorzustellenden) Emergenz-Szenarien (vgl. Kap. 3.3.). Gemeinsam ist ihm diese Implikation jedoch etwa mit den prägeometrischen *Quantum Causal Histories* (vgl. Kap. 4.6.).

Zur Diffeomorphismusinvarianz als Eigenschaft einer emergenten raumzeitlichen Ebene, die sich auf der Grundlage eines gänzlich anders beschaffenen Substrats ergibt, für das diese nicht gilt, siehe auch Jannes (2008).

[144] Vgl. insbesondere die prägeometrischen *Quantum Causal Histories* in Kap. 4.6.

Der residuale Sinngehalt der makroskopischen Diffeomorphismusinvarianz reduziert sich bei einem Übergang zu einer solchen völlig anders gearteten fundamental(er)en Ebene womöglich sehr schnell auf die relationalistischen Intuitionen, die sich mit ihr im Kontext der Allgemeinen Relativitätstheorie ergeben. Gerade diese relationalistischen Intuitionen könnten es sein, die sich im Kontext der Ansätze zu einer Theorie der Quantengravitation weiterhin als relevant erweisen: als so etwas wie die Essenz der mit der Allgemeinen Relativitätstheorie gewonnenen Einsichten für den zukünftigen Bereich der Quantengravitation. Da die Diffeomorphismusinvarianz selbst für diesen Bereich dann aber keine Gültigkeit mehr besitzt, müssten sowohl die Motivationen für einen solchen Relationalismus als auch die möglichen Instantiierungen der Hintergrundunabhängigkeit gänzlich anders formuliert werden. Die Hoffnungen eines solchen Relationalismus liefen unter diesen Randbedingungen – was im weiteren noch zu motivieren sein wird – vor allem auf eine präraumzeitliche Theorie hinaus: eine 'prägeometrische' Theorie der Quantengravitation.[145]

> *"The relationist claim then, put generally, is that there is some different theory to be had, one that duplicates at least all the successes of current theory, and which does not appeal to substantial space(time) in its formulation, making do instead with some set of primitive relations that hold between actual material things."* (Hoefer (1998) 463)

Im Rahmen eines solchen prägeometrischen Szenarios kommt die Raumzeit als emergentes Phänomen auf der Grundlage eines nicht-raumzeitlichen Substrats zustande. In welcher Hinsicht und in welchem Ausmass eine solche Theorie dann tatsächlich hintergrundunabhängig ist und die entsprechenden relationalistischen Intuitionen umsetzt, wäre dann unter wesentlich weiter gesteckten Voraussetzungen zu klären, als sie in Bezug auf die Allgemeine Relativitätstheorie Anwendung finden. Die Hintergrundunabhängigkeit im klassischen, makroskopischen Bereich könnte sich durchaus, infolge einer vollständigen dynamischen Entkopplung, aus einer Substratebene heraus ergeben, für die diese Hintergrundunabhängigkeit nicht gilt.

Und hier lohnt sich vielleicht schon jetzt ein Seitenblick auf andersgeartete, nicht in jedem Falle als prägeometrisch einzustufende Emergenzszenarien, die diese Hintergrundunabhängigkeit in Frage stellen könnten:[146] Diese Ansätze, die es in diversen Ausprägungen gibt, gehen erst einmal von einer Emergenz der Gravitation aus, was eine Emergenz der Raumzeit einschliessen kann, aber nicht muss. Die gemeinsame Grundidee dieser Ansätze ist, dass das Gravitationsfeld zwar ein dynamisches Feld ist, aber nicht notwendigerweise ein fundamentales Wechselwirkungsfeld.[147] Dies würde einerseits eine Quantisierung des Gravitationsfeldes zu einer sehr fragwürdigen Sache machen, die sehr wahrscheinlich zu theoretischen Artefakten und Sackgassen für die Theorienbildung führt. Andererseits könnte es zusätzlich aber auch bedeuten, dass nicht nur die allgemeine Kovarianz bzw. die aktive Diffeomorphismusinvarianz der Allgemeinen Relativitätstheorie in ihrer umfassenden Form letztlich theoretische Artefakte sind, sondern auch die Idee der Geometrisierbarkeit der Gravitation keine umfassende Gültigkeit besitzt. Die Idee einer geometrischen Erfassbarkeit der Gravitation würde zwar im makroskopischen Kontext unter bestimmten Bedingungen zu reliablen Vorhersagen führen. Eine Theorie der Quantengravitation, welche die Geometrisierbarkeit der Gravitation in Form einer Forderung nach Hintergrundunabhängigkeit als Ausgangspunkt wählt, hätte dann jedoch kaum Aussicht auf Erfolg. – Um in dieser Weise die Forderung nach Hintergrundunabhän-gigkeit aufzugeben, müssen allerdings gute Gründe in Form diverser, sich ergänzender konzeptioneller Motivationen, besser noch in Form empirischer Indizien vorliegen.

---

[145] Argumente für eine solche Theorie finden sich in den Kap. 3. und 4.6.
[146] Siehe Kap. 3.3.
[147] Vgl. auch Kap. 1.

Die orthodoxeren prägeometrischen Ansätze halten hingegen erst einmal an der in der Allgemeinen Relativitätstheorie vorliegenden Kopplung von Raumzeit und Gravitation fest, nur dass sie beide als emergente, in ihrem Zustandekommen aneinander gekoppelte bzw. miteinander identifizierbare Phänomene behandeln. Diese prägeometrischen Ansätze sind hintergrundunabhängig, und dies in zum Teil noch deutlich ausgeprägterem Masze als die Allgemeine Relativitätstheorie selbst.[148] Das heisst, sie transzendieren die auch in der Allgemeinen Relativitätstheorie immer noch vorliegende residuale Hintergrundabhängigkeit, die eine vorgegebene Dimensionalität, eine vorgegebene Signatur und (zumindest für die jeweiligen Lösungen) eine vorgegebene Topologie der Raumzeit einschliesst. – Ohnehin könnte man sich fragen, ob eine solche radikalere Form von Hintergrundunabhängigkeit, die noch deutlich über die in der Allgemeinen Relativitätstheorie realisierte hinausgeht, und vielleicht sogar zu einem Verzicht auf jegliche Form von Hintergrund abzielt, nicht vielleicht am ehesten dem Geist der schon in der Allgemeinen Relativitätstheorie ausgemachten relationalistischen Tendenzen entspricht.

Zielt eine Theorie der Quantengravitation, wie dies in den prägeometrischen Ansätzen der Fall ist, auf die Beschreibung einer Substratstruktur ab, die Raumzeit als emergentes Phänomen verstehbar werden lassen soll, so stellt sich – wie zuletzt auch im Hinblick auf eine durchaus umfassendere, möglichst vollständige Hintergrundunabhängigkeit, auf die die relationalistischen Intuitionen letztendlich abzielen – vor allem die Frage, von welchen Entitäten und welchen Relationen eine solche prägeometrische, prä-raumzeitliche Theorie handelt und wie auf der Grundlage des von ihr beschriebenen Substrats Raumzeit als makroskopisches, quasi-kontinuierliches Phänomen zustandekommt.

> *"Crucial to the clarification of relationism, then, is an explanation of what sorts of 'material things' are supposed to exist and what kinds of primitive relations are to be posited between these things."*
> (Hoefer (1998) 463)

Unabhängig von der jeweiligen spezifischen Antwort auf diese Frage nach der Beschaffenheit des Substrats, ist eine emergente Raumzeit ganz sicher nicht mehr in einem fundamentaleren Sinne substantialistisch deutbar, was eben nichts anderem als der Umsetzung der relationalistischen Intuitionen durch die Emergenzszenarien entspricht. Die Frage der Substantialität wäre im Kontext prägeometrischer Theorien bestenfalls für die Entitäten zu stellen, die das Substrat konstituieren.[149]

---

[148] Vgl. Kap. 4.6.

[149] Die traditionelle Substantialismus-Relationalismus-Debatte dreht sich ausschliesslich um die Frage der Substantialität des Raumes bzw. der Raumzeit und die Ableitbarkeit dieser aus einem anderen substantiell-materialen Gefüge. Der Relationalismus bezüglich der Raumzeit ist dabei in keiner Weise gezwungen, das Konzept der Substantialität selbst in Frage zu stellen, sondern nur die Anwendbarkeit dieses Konzeptes auf die Raumzeit. Eine Infragestellung des Konzeptes der Substantialität selbst ergibt sich erst im Zusammenspiel mit ganz anderen Kontexten:

> *"Both sides in the absolute/substantivalist vs. relational debate accept the traditional subject-predicate parsing of spacetime ontology and ideology. For the relationalist, the subjects are material bodies and/or events in their histories, and the spatiotemporal predicates are relational properties of these subjects. For the absolute / substantivalist, the subjects are points or regions of space or spacetime, and the spatiotemporal predicates are relational and non-relational properties of these subjects. Twentieth-century physics has been unkind to both sides. First, classical field theory elevated fields to coequal status with particles, and subsequently quantum field theory (arguably) demoted particles to a second class if not epiphenomenal status. This ascendency of fields*

Damit relativiert sich die konzeptionelle Zielrichtung, die der Substantialismus-Relationalismus-Debatte zugrundeliegt, im Kontext dieser prägeometrischen Ansätze zur Quantengravitation noch einmal grundlegend – wahrscheinlich grundlegender als mit dem Aufkommen der Feldtheorien, mit dem der Allgemeinen Relativitätstheorie und mit dem der Quantenfeldtheorien: Diese Debatte ist dann eben vor allem ganz sicher nicht mehr in Bezug auf die Raumzeit zu führen, sondern, wenn überhaupt, bestenfalls noch in Bezug auf das Substrat.

Von grösserer Relevanz als diese transponierte Substantialismus-Relationalismus-Debatte wird sich im Kontext der prägeometrischen Ansätze zur Quantengravitation dann aber vermutlich die grundsätzlichere Frage des Verhältnisses von emergenter Raumzeit zu ihren 'Inhalten' (etwa Materie und Feldern) erweisen: Sollten die emergente makroskopische Raumzeit und ihre 'Inhalte', wie man vielleicht erst einmal (unüberlegt) annehmen könnte, auf unterschiedliche Ursprünge zurückgehen, so würde dies, wenn sich das prägeometrische Substrat, auf dessen Grundlage die Raumzeit dann – isoliert von ihren 'Inhalten' – zustandekäme, selbst wieder substantialistisch deuten liesse, vielleicht sogar mit einem Wiederaufleben eines (allerdings nicht fundamental zu verstehenden) Quasi-Substantialismus hinsichtlich der Raumzeit einhergehen. Es wäre jedoch in einem solchen Szenario mit unterschiedlichen Ursprüngen für die Raumzeit und ihre 'Inhalte' ebenso denkbar, dass die spezifische Geartetheit des Substrats schon für dieses eine substantialistische Deutung ausschliesst. Ein rein computational-informationstheoretisches Substrat etwa liesse höchstens eine Form von quasi-relationalistischer oder strukturalistischer Deutung zu. Und auch allen aus diesem Substrat hervorgehenden emergenten Phänomenen könnte dann kaum so etwas wie Substantialität zugesprochen werden.

Aber letztendlich ist ein solches Szenario mit einer prägeometrischen Basis, aus der die makroskopische Raumzeit hervorgeht, und einem davon getrennten Ursprung für die Materie und die nichtgravitativen Wechselwirkungen nicht nur sehr künstlich, sondern völlig absurd: Die Interaktion zwischen beiden Komponenten macht deutlich, dass es eine umfassende, gemeinsame Basis für das Zustandekommen von Raumzeit und ihren 'Inhalten' geben muss.[150] Im Kontext der prägeometrischen Szenarien ist notwendigerweise von einem gemeinsamen prägeometrischen Ursprung für die Raumzeit und ihre 'Inhalte' auszugehen. Die dann in einem solchen Szenario hinsichtlich des ontologischen Status der Raumzeit zu stellenden Frage transzendieren in jedem Falle den Kontext der traditionellen Substantialismus-Relationalismus-Debatte. Ein Festhalten an den mit ihr verbundenen Begrifflichkeiten in Bezug auf die Raumzeit ist in diesem veränderten Kontext wohl kaum noch motivierbar. Hinsichtlich des Substrats hingegen stehen die mit diesen Begrifflichkeiten verbundenen Optionen grundsätzlich weiterhin offen: Ein mögliches Wiederaufleben eines Substantialismus hinsichtlich des prägeometrischen Substrats ist bei gegebener substantieller Deutbarkeit ebenso denkbar wie ein mögliches Wiederaufleben eines Relationalismus für den Fall einer ausschliesslich strukturalistischen oder computational-informationstheoretischen Deutbarkeit des Substrats und seiner Dynamik. – Vielleicht ist also die explizit in Bezug auf die Allgemeine Relativitätstheorie geführte traditionelle Substantialismus-Relationalismus-Debatte hinsichtlich unserer Einschätzung des ontologischen Status der Raumzeit letztendlich gar nicht einmal so sonderlich interessant, weil die Allgemeine Relativitätstheorie noch kein adäquates und vollständiges Bild der Raumzeit und ihrer Existenzbedingungen liefert.[151] Es ist also durchaus möglich, dass der traditionellen Substantialismus-Relationalismus-Debatte nur der Verdienst zukommt, für die nächste Theoriengeneration, die Theorien der Quantengravitation, Inspirationen zu liefern.[152]

---

*seems at first to be a god-send for the substantivalist since it seems to lend itself to the view that spacetime is the only basic substance qua object of predication. But diffeomorphism invariance as a gauge symmetry seems to wipe away spacetime points as objects of predication."* (Earman (2006a) 16)

[150] Hier sei auch an das in Girelli / Liberati / Sindoni (2008, 2009) beschriebene Szenario erinnert, für das sich die aktive Diffeomorphismusinvarianz der Raumzeit als emergentes Ergebnis auf der Grundlage einer Dynamik einstellt, für die diese Diffeomorphismusinvarianz noch nicht vorliegt. Die entscheidende Bedingungen dafür, dass es zur Emergenz der Diffeomorphismusinvarianz kommt, ist hierbei jedoch gerade, dass Raumzeit und Materie im gleichen Schritt als emergente Phänomene aus der Substratdynamik hervorgehen. Siehe auch Kap. 4.6.

## 2.2. Raumzeit in der Quantenmechanik und in den Quantenfeldtheorien

Es ist nicht zuletzt die wechselseitige konzeptionelle Unverträglichkeit der Quantenmechanik und der Quantenfeldtheorien mit der Allgemeinen Relativitätstheorie, die – unabhängig von der spezifischen Deutung des Konflikts – eine entscheidende Motivation für eine theoretische Weiterentwicklung jenseits dieser etablierten Theorien liefert. Andererseits legt sie aber auch nahe, dass diese etablierten Theorien, die unser zur Zeit fundamentalstes Instrumentarium zur Erfassung physikalischer Gegebenheiten darstellen, erst einmal von Bedeutung für diese Entwicklung einer Theorie der Quantengravitation sein könnten; die etablierten fundamentalen Theorien, über die wir zur Zeit verfügen, sollten also bei der Entwicklung einer Theorie, mittels derer gerade ihre wechselseitigen Konflikte überwunden werden sollen, allesamt Berücksichtigung finden. Dies gilt jedoch nicht in der gleichen Weise für alle Aspekte dieser etablierten Theorien. Insbesondere gilt es nicht für die Aspekte, die sich auf unser Verständnis der Raumzeit beziehen. Quantenmechanik und Quantenfeldtheorien sind im Gegensatz zur Allgemeinen Relativitätstheorie keine Theorien, denen es spezifisch und vorrangig um die Raumzeit ginge. Einerseits stehen sie mit ihrer, im Vergleich, unkritischen und unreflektierten, geradezu naiven Behandlung von Raum und Zeit in z.T. krassem Widerspruch zur dezidierten, theoretisch motivierten und empirisch gestützten Raumzeitauffassung der Allgemeinen Relativitätstheorie. Andererseits liefern sie vielleicht dennoch – zumindest für den Fall, dass die Gravitation eine fundamentale Wechselwirkung sein sollte und sich zudem, im Sinne der Allgemeinen Relativitätstheorie, geometrisieren lassen sollte – Einsichten hinsichtlich der dann zu erwartenden, aber in der Allgemeinen Relativitätstheorie als klassischer Theorie nicht erfassten Quanteneigenschaften der Raumzeit.

### *Quantenmechanik und Quantenfeldtheorien: Konzeptionelle Voraussetzungen*

Es gibt naheliegenderweise in der Quantenmechanik und in den Quantenfeldtheorien nicht die geringsten Anzeichen für die im Rahmen der Allgemeinen Relativitätstheorie deutlich werdende Form der Dynamizität der Raumzeit, noch viel weniger für eine eventuelle Relationalität der Raumzeit.

---

[151] Ein Indiz für die letztendliche Unangemessenheit der Allgemeinen Relativitätstheorie als realistisch deutbarer Beschreibung der Natur liefern nicht zuletzt die Singularitäten in ihren Lösungen:
*"That [general relativity] cannot be true at the most fundamental level is clear from the singularity theorems [...]."* (Kiefer (2004) 2)

[152] Vgl. Hoefer (1998):
*"[...] substantivalism and relationism today must be understood in part as bets, bets about how a successful quantum gravity theory – if one proves possible – will change or fail to change our current [general relativity]-based understanding of spacetime."* (Hoefer (1998) 462)

Vielmehr gehen Quantenmechanik und Quantenfeldtheorie von einer grundsätzlich undynamischen Hintergrundraumzeit aus und stehen damit in deutlichem Konflikt zur Allgemeinen Relativitäts-theorie.

Die nicht-relativistische Quantenmechanik setzt einen statischen Newtonschen Hintergrundraum voraus und behandelt die Zeit schlicht als Hintergrundparameter. Die räumliche Lokalisierung der von ihr mittels der Wellenfunktion bzw. des Zustandsvektors beschriebenen Systeme erfolgt über den Ortsoperator. Die Kommutatorregeln zwischen den quantenmechanischen Operatoren setzen den Hintergrundraum ebenso voraus wie die Heisenbergschen Unschärfen. Die Zeit wird in der Quantenmechanik nicht einmal durch einen Operator repräsentiert; sie läuft im Sinne einer Newton-schen absoluten Zeit einfach als Parameter mit. In der relativistischen Quantenmechanik wird der Newtonsche absolute Raum als Hintergrundraum dann durch eine feste Hintergrundraumzeit mit flacher Minkowski-Metrik ersetzt; die Zeit wird aber weiterhin formal als für jeden Raumpunkt kontinuierlich laufender Parameter behandelt. Und auch in den Quantenfeldtheorien ändert sich diesbezüglich wenig. Es bleibt bei einer festen Hintergrundraumzeit mit festgelegter (meist flacher, Minkowskischer) Metrik, die notfalls durch eine gekrümmte, aber eben wiederum statische Hinter-grundraumzeit ersetzt werden kann. Die Feldoperatoren der Quantenfeldtheorie werden den Raum-zeitpunkten dieser Hintergrundraumzeit zugeordnet. Physikalische Grössen treten, sobald man die quantenfeldtheoretischen Zustände raumzeitlich interpretiert, immer als Eigenschaften von Raum-zeitpunkten in Erscheinung.[153]

Wenn also für eine zukünftige Theorie der Quantengravitation etwas Essentielles aus dem Kontext der Quantenmechanik oder dem der Quantenfeldtheorien übernommen oder wenigstens gelernt werden kann, so betrifft dies sicherlich nicht vorrangig und unmittelbar unser Verständnis der

---

[153] Es gibt jedoch Stimmen, die einer solchen Sicht auf die Quantenfeldtheorien widersprechen. So argumentiert Dieks (2001a) etwa dafür, dass die Hintergrundraumzeit der Quantenfeldtheorien nicht nur nicht substantiell gedeutet werden kann, sondern nicht einmal mehr tatsächlich als raumzeitliche Arena angesehen werden muss. Die Hintergrundraumzeit der Quantenfeldtheorien hat seiner Auffassung nach vielmehr nur noch den Status eines Systems von Ordnungspara-metern:

> *"In relativistic quantum field theory the physical meaning of the spacetime manifold seems to become even more indirect. Not only does the manifold lose its status as a substantival container, but also its function as a repre-sentation of spacetime positions inherent in physical objects becomes problematic. 'Space and time' become ordering parameters in the web of properties of physical subsystems. They seem to regain their traditional meaning only in the non-relativistic limit in which the classical particle concept becomes applicable."* (Dieks (2001a) 241)

Von einer solchen Sichtweise ist es vielleicht nur noch ein kleiner Schritt zur nicht-kommutativen Geometrie. Siehe etwa Connes (1998, 2000), Connes / Marcolli (2006), Muller-Hoissen (2008), Yang (2007).

> *"Non-commutative geometry takes as its starting point the fact that the geometric properties of any differentiable manifold can be expressed fully in terms of the (commutative) algebra of functions of the manifold."* (Monk (1997) 17)

Die nicht-kommutative Geometrie entspricht einer Erweiterung auf eine nicht-kommutative, algebraische Struktur der Raumzeit. Ihre formale Grundlage besteht in der Tatsache, dass sich jede differenzierbare Mannigfaltigkeit durch die algebraische Struktur eines kommutativen Rings differenzierbarer Funktionen bestimmen lässt. Und diese algebraische Struktur lässt sich eben auch auf nicht-kommutative Algebren erweitern. Das sich dabei einstellende Ergebnis:

> *"What emerges is a structure which is neither continuous nor discrete; rather it resembles two continuous (4-dimensional) space-time manifolds, 'separated' by a distance ~ $10^{-16}$ cm."* (Monk (1997) 17)

Die Tatsache, dass die (in solcher Weise erweiterte) Raumzeit durch eine (nicht-kommutative) Operatoralgebra darge-stellt wird, wie sie in der Quantenmechanik und auch in anderen Kontexten vorkommt, legt nach Ansicht mancher Autoren nicht zuletzt die Möglichkeit der Verwendbarkeit dieses Ansatzes im Kontext der Quantengravitation nahe: als Instrumentarium zur Erfassung und Beschreibung der Quanteneigenschaften der Raumzeit.

Raumzeit. Bei den sich im Rahmen der Quantenmechanik abzeichnenden Einsichten geht es vielmehr erst einmal um die Modalitäten und modelltheoretischen Mechanismen der Zuschreibung von Quanteneigenschaften zu physikalischen Entitäten, die in dynamische Prozesse eingebunden sind. Dies betrifft insbesondere erst einmal die möglichen Quanteneigenschaften, die dem Gravitationsfeld zugeschrieben werden müssten, so es ein fundamentales Feld sein sollte und so die Quantenmechanik universell gültig sein sollte, ebenso wie die Quanteneigenschaften, die in diesem Fall der Raumzeit zugeschrieben werden müssten, wenn die Annahme der Geometrisierbarkeit der Gravitation, wie sie der Allgemeinen Relativitätstheorie zugrundeliegt, zutreffen sollte.

## Was heisst 'Quantisierung' der Gravitation bzw. der Raumzeit ?

Gemäss der Quantenmechanik kommen dynamischen Entitäten (etwa Feldern) Quanteneigenschaften zu. Sollte die Quantenmechanik universell gültig sein und das Gravitationsfeld Ausdruck einer fundamentalen Wechselwirkung sein, so wären diesem also Quanteneigenschaften zuzuschreiben, analog zu den Quanteneigenschaften der Materie und der anderen für fundamental angesehenen Wechselwirkungsfelder. Sollte die Gravitation zudem im Sinne der Allgemeinen Relativitätstheorie geometrisierbar sein, sich die Gravitation also in Form einer dynamischen Geometrie und ihrer Eigenschaften erfassen lassen, liesse sich diese Einsicht auf die Raumzeit übertragen. Auch dieser kämen Quanteneigenschaften zu. Man hätte es letztendlich mit so etwas wie einer 'Quantengeometrie' zu tun.

Wie lassen sich diese möglichen Quanteneigenschaften der Gravitation und der Raumzeit jedoch unter den gegebenen Voraussetzungen erschliessen? Durch eine Quantisierung der Gravitation und damit der Raumzeit? – Letztendlich herrscht ohne spezifische Vorgaben erst einmal eine erhebliche Unklarheit darüber, welche Entität überhaupt quantisiert werden soll:

> "Quantization of gravity means quantization of geometry. But which structures should be quantized, that is, to which structures should one apply the superposition principle? [...] Structures that are not quantized remain as absolute (non-dynamical) entities in the formalism. One would expect that in a fundamental theory no absolute structure remains. This is referred to as background independence of the theory." (Kiefer (2005) 2)

Hält man sich hinsichtlich der Quantisierung der Gravitation an eine direkte Quantisierung der Allgemeinen Relativitätstheorie, so ist immerhin noch klar, welche Entität dynamisch ist, und damit wohl die erste Wahl als Ansatzpunkt für eine Quantisierung darstellt: die Metrik (bzw. in alternativen Darstellungen: Grössen auf der Ebene der Metrik, wie etwa Konnektionen oder Holonomien).[154] Es sind jedoch auch indirektere Formen einer Quantisierung der Gravitation und der Raumzeit denkbar, die gerade nicht von einer direkten Quantisierung der Allgemeinen Relativitätstheorie ihren Ausgang nehmen und dennoch zu einer Quantentheorie führen, die der Gravitation bzw. der Raumzeit Quanteneigenschaften zuspricht, die also eine fundamentale Raumzeit mit zusätzlichen Quanteneigenschaften beschreibt.

---

[154] Es ist aber immerhin fraglich, ob man sich diesbezüglich uneingeschränkt auf die Allgemeine Relativitätstheorie verlassen sollte. Welche Entität zu quantisieren wäre, lässt sich, wenn überhaupt, erst im Kontext konkreter Ansätze zur Theorienbildung und ihrer jeweiligen Erfolge und Misserfolge beurteilen. (Siehe Kap. 4.) Es gibt dafür, auch unter den Voraussetzungen, die überhaupt für eine Quantisierung der Gravitation sprechen, keine apriori-Argumente. Andererseits gibt es aber Argumente, die unter bestimmten Voraussetzungen gänzlich gegen die Quantisierung der Allgemeinen Relativitätstheorie sprechen, und einen völlig anderen Weg zu einer Theorie der Quantengravitation nahelegen. Eine solche Theorie müsste dann aber immer noch die Allgemeine Relativitätstheorie als klassischen Grenzfall, als Implikation oder als Näherung enthalten. (Siehe Kap. 3.3. und Kap. 4.)

Bestenfalls in einem metaphorischen Sinne hingegen liesse sich von einer 'Quantisierung der Gravitation' für den Fall reden, dass die Gravitation keine fundamentale Wechselwirkung sein sollte. Für diesen Fall wäre, unter Voraussetzung der universellen Gültigkeit der Quantenmechanik, das Ziel, eine Quantentheorie zu konstruieren, die das Zustandekommen der Gravitation und vielleicht auch der Raumzeit verstehbar macht, ohne dass diesen notwendigerweise Quanteneigenschaften zugeschrieben werden müssten. Im Falle einer intrinsisch klassischen, makroskopischen, emergenten Gravitation bzw. Raumzeit, die auf einem gänzlich anders gearteten Quantensubstrat ohne gravitative Freiheitsgrade – oder gar einem prä-raumzeitlichen Quantensubstrat – beruhen, gäbe es eben gerade keine solche Quanteneigenschaften.

## *Hat die Raumzeit Quanteneigenschaften ?*

Welche Quanteneigenschaften wären jedoch dem Gravitationsfeld zuzuschreiben, wenn es sich bei diesem tatsächlich um ein fundamentales Wechselwirkungsfeld handeln sollte? Und welche Quanteneigenschaften wären für die Raumzeit zu erwarten, wenn zudem die Annahme der Geometrisierbarkeit der Gravitation zuträfe? – Wäre etwa damit zu rechnen, dass es zu Superpositionen von Raumzeiten kommt? (Zu solchen Superpositionen kommt es gemeinhin in einer Quantentheorie für alle quantisierten Grössen.) Wäre mit quantenmechanischen Unschärfen der Raumzeit zu rechnen? Würde sich die Metrik der Quantengeometrie als Erwartungswert einer Quantenvariablen ergeben? (Wenn die Metrik der Raumzeit die zu quantisierende Grösse darstellt, so wäre dies zumindest im Rahmen eines naiven Schlusses aus den bisherigen Erfahrungen mit der Quantisierung klassischer Grössen zu erwarten.)

Wie steht es um Quantenfluktuationen der Raumzeit, die ebenso aus den bisherigen Erfahrungen mit der Quantisierung klassischer Theorien zu erwarten wären?[155] Betreffen sie nur die Metrik oder betreffen diese Fluktuationen auch die Topologie der Raumzeit, wie dies etwa Wheeler mit seiner Idee des *Spacetime foam* einmal nahegelegt hat?[156] – Letzteres würde massive konzeptionelle Probleme nach sich ziehen (oder zumindest eine grundlegende Modifikation hinsichtlich der verwendeten modelltheoretischen Instrumentarien erforderlich machen): Quantenfluktuationen der raumzeitlichen Topologie sind nämlich nicht nur mit dem differentialgeometrischen Instrumentarium der Allgemeinen Relativitätstheorie unverträglich. Sie sind ebenso unverträglich mit dem Instrumentarium sowohl der kanonischen als auch der kovarianten Ansätze zu einer Quantisierung der Allgemeinen Relativitätstheorie; beide setzen eine differenzierbare Mannigfaltigkeit mit fester Topologie voraus.[157]

Aber soweit muss man gar nicht gehen, um zu sehen, dass sich eine direkte Quantisierung der Allgemeinen Relativitätstheorie durchaus als äusserst problematischer und fragwürdiger Weg hin zu einer Theorie der Quantengravitation herausstellen könnte, wenn sich die bisherigen Erfahrungen mit der Quantisierung klassischer Grössen hier in elementarster Weise bestätigen sollten. Um in Probleme zu geraten, müssen die zu erwartenden Quantenfluktuationen der Raumzeit gar nicht erst die Topologie der Raumzeit betreffen. Quantenfluktuationen der Metrik reichen hier völlig aus:

> *"In fact, the possibility of referring directly to 'the quantum structure of spacetime' faces at the very least a serious conceptual difficulty, concerning the localization of the gravitational field: what does it mean to talk about the values of the gravitational field at a point if the metric field itself is subject to quantum fluctuations? [...] In this case, we could no longer tell whether the separation between two points is spacelike, null, or timelike since quantum fluctuations of the metric could exchange past and future."* (Dorato / Pauri (2006) 142)

---

[155] Dabei sollte man sich vor Augen führen, dass 'Quantenfluktuationen der Raumzeit' keine zeitlichen Fluktuationen im üblichen quantenmechanischen Sinne sein können.

[156] Vgl. Kap. 1.3. Siehe auch Callender / Weingard (2000), Anderson / DeWitt (1986).

[157] Siehe Kap. 4.

Quantenfluktuationen der Raumzeit führen, da sie Fluktuationen der basalen Kausalstruktur implizieren, zu wahrscheinlich unüberwindlichen Problemen bei der Quantisierung des Gravitationsfeldes – zumindest, wenn man hierbei die bekannten Quantisierungsverfahren zur Anwendung bringt:

> *"Assuming that the metric fluctuations do affect the causal structure, one would expect that the commutation relations themselves should reflect this by also undergoing fluctuations of some sort. However, it is not at all clear what this means, or how it might be represented."* (Weinstein (2001) 97)

Die bekannten Quantisierungsverfahren setzen schlichtweg für ihre Anwendbarkeit eine stabile Kausalstruktur voraus:

> *"[...] fluctuations in the gravitational field imply fluctuations in the spatiotemporal, and hence causal, structure of the world. But it is hard to see how one can make sense of canonical commutation relations and hence quantize anything in the absence of a stable causal structure."* (Callender / Huggett (2001a) 22)

Unabhängig von diesen Problemen könnte es durchaus sein, dass es tatsächlich Fluktuationen der kausalen Struktur gibt. Vielleicht ist die Kausalstruktur ja nur ein approximatives oder emergentes Konstrukt. – Methodisch stellen Fluktuationen der Kausalstruktur aber in jedem Fall ein erhebliches Problem für eine entsprechende Quantentheorie der Raumzeit dar:

> *"[...] once we embark on constructing a quantum theory of gravity, we expect some sort of quantum fluctuations in the metric, and so also in the causal structure. But in that case, how are we to formulate a quantum theory with a fluctuating causal structure?"* (Butterfield / Isham (2001) 64)

Sie führen zu Implikationen, die eindeutig über den Kontext einer direkten Quantisierung der Allgemeinen Relativitätstheorie mittels der bekannten Quantisierungsverfahren, wie sie etwa in den Quantenfeldtheorien (insbesondere in der Quantenelektrodynamik) verwendet werden, hinausweisen.

> *"Recall that in a quantum theory of gravity the spacetime metric will be an operator. Yet the metric field is responsible for chronogeometrical structure in addition to gravitational field structure, which implies that it is responsible for the causal structure too (microcausal structure induced). In other words, since the causal structure is dependent on the metric and the causal structure determines whether two events are spacelike or not, and given that the metric is prone to quantum fluctuations, it follows that the notion of spacelikeness, and therefore microcausality itself, becomes subject to quantum* *fluctuations: one of the fundamental axioms of quantum field theory is thus rendered meaningless. We need, then, a new conception of spacetime that goes beyond the conceptions that we find in quantum field theory and general relativity, and that will take something special indeed: prima facie, we need to either reject background-independence (scrap general relativity) or else find a way to set up a background independent quantum field theory (scrap quantum field theory as it is understood at present)."* (Rickles / French (2006) 14)

Sollte eine Quantisierung der Gravitation (bzw. der Allgemeinen Relativitätstheorie) also zu Quantenfluktuationen der Raumzeit führen, wie es zumindest eine naive Übertragung der Erfahrungen mit der Quantisierung anderer klassischer Theorien nahelegen könnte und wie sie bei Anwendung der bekannten Quantisierungsverfahren (unter Voraussetzung der Gültigkeit der in der Allgemeinen Relativitätstheorie vorliegenden Beziehung zwischen Gravitation und Raumzeit) auch tatsächlich zu erwarten sind, so wäre damit eigentlich schon klar, dass auf dieser Grundlage eine direkte Quantisierung der Gravitation durch eine Quantisierung der Allgemeinen Relativitätstheorie – was dann einer Quantisierung der Raumzeit selbst (bzw. ihrer Metrik) entspräche – kein gangbarer Weg hin zu einer Theorie der Quantengravitation ist. Denn gerade unter den Bedingungen, für die eine direkte Quantisierung der Gravitation als gangbarer Weg erscheinen könnte – also im Szenario: fundamentale gravitative Wechselwirkung mit zusätzlichen Quanteneigenschaften –, ergeben sich Konsequenzen, unter denen sich diese direkte Quantisierung der Gravitation (zumindest unter Anwendung der bekannten Quantisierungsverfahren) sehr schnell als sehr problematisch erweist.[158]

Vielleicht sind die Allgemeine Relativitätstheorie und die Quantenmechanik also viel zu verschieden, um eine simple Amalgamierung zuzulassen. Vielleicht muss eine Theorie der Quantengravitation, auch für den Fall, dass die Gravitation eine fundamentale Wechselwirkung sein sollte, der (in der Allgemeinen Relativitätstheorie noch nicht erfasste) Quanteneigenschaften zuzuschreiben sind, einen völlig anderen Weg beschreiten.[159] – Wenn die Gravitation und vielleicht sogar die Raumzeit hingegen emergente Grössen sein sollten, erscheint eine direkte Quantisierung der Allgemeinen Relativitätstheorie ohnehin völlig unsinnig. Und dann sind es ohnehin nicht Quanteneigenschaften der Gravitation bzw. der Raumzeit, die einen Ansatzpunkt für eine Theorie der 'Quantengravitation' liefern. Eine solche Theorie hätte sich in diesem Falle vielmehr mit den Quanteneigenschaften des Substrats zu beschäftigen.

## Nichtlokalitäten

Es gibt in der Quantenmechanik vielfältige Hinweise auf Phänomene, die das klassische Lokalitätsprinzip überschreiten. Dies betrifft insbesondere die Nichtseparierbarkeit verschränkter Systeme sowie die Implikationen, die sich aus der Bellschen Ungleichung und den entsprechenden empirischen Bestätigungen der Quantenmechanik ergeben; letztere machen insbesondere deutlich, dass es nicht möglich ist, die Dynamik quantenmechanischer Systeme auf der Grundlage einer strikt lokalen Substruktur zu reproduzieren.

---

[158] Zu weiteren konzeptionellen Problemen einer direkten Quantisierung der Gravitation, siehe Weinstein (2001).

[159] Ein Mittelweg, der darin besteht, an einer möglichst konservativen Form der Amalgamierung von Allgemeiner Relativitätstheorie und Quantenmechanik festzuhalten, dies aber unter Einsatz von deutlich subtileren Quantisierungsverfahren, die über die traditionellen Ansätze und ihre naiverweise zu erwartenden Implikationen hinsichtlich der Quanteneigenschaften von Gravitation und Raumzeit entscheidend hinausgehen, liegt in der *Loop Quantum Gravity* vor. Ob dieser Ansatz zum Erfolg führt, ist jedoch immer noch unklar. Siehe Kap. 4.4.

Die quantenmechanische Nichtseparierbarkeit beschränkt sich dabei nicht notwendigerweise auf die Mikrowelt. Es liegen inzwischen experimentelle Belege für makroskopische verschränkte Systeme und damit für die Relevanz nichtlokaler quantenmechanischer Zustände im Makrobereich vor. Es wäre also nicht zuletzt damit zu rechnen, dass eine Quantisierung der Gravitation, die nach weit verbreiteter Auffassung mit einer Quantisierung der Raumzeit einhergeht, ebensolche makroskopischen Ausprägungen quantenmechanischer Nichtseparierbarkeit nach sich zieht.

> *"It must be emphasized that if gravity is quantized, the kinematical non-separability of quantum theory demands that the whole universe must be described in quantum terms."* (Kiefer (2004) 3)

Quantenmechanische Nichtseparierbarkeiten transzendieren jedoch den konzeptionellen Kontext der Speziellen wie der Allgemeinen Relativitätstheorie. Insbesondere sind sie mit der raumzeitlich gebundenen Kausalität, wie sie diesen klassischen Theorien zugrundeliegt, unverträglich. Daran ändert sich auch erst einmal wenig angesichts der Tatsache, dass die echten physikalischen Freiheitsgrade, die entsprechend ihrer eichinvarianten Interpretation[160] aus der Tiefenstruktur der Allgemeinen Relativitätstheorie aufscheinen, ohnehin nichtlokalen Charakter besitzen. Diese im Kontext ihrer eichinvarianten Deutung unvermeidlichen nichtlokalen Eigenschaften der klassischen Theorie, die das Konzept einer raumzeitlichen (und damit auch kausalen) Geordnetheit letztendlich vollständig transzendieren und als nur phänomenologisch wirksame Oberflächenstruktur erscheinen lassen, haben erst einmal nicht das geringste mit den quantenmechanischen Formen von Nichtlokalität bzw. Nichtseparierbarkeit zu tun, die diese raumzeitliche Geordnetheit, um sie in sehr spezifischer und selektiver Weise zu durchbrechen, gerade voraussetzen müssen. Es müsste sich schon – was einer nicht geringen Überraschung gleichkäme – eine konzeptionell gut gestützte Verbindung zwischen den nichtlokalen klassischen Freiheitsgraden der Allgemeinen Relativitätstheorie und den quantenmechanischen Nichtseparierbarkeiten herstellen lassen, etwa im Rahmen einer geeigneten Theorie der Quantengravitation, um hier nicht nur eine konzeptionelle Vereinbarkeit, sondern vielleicht darüber hinaus sogar eine physikalisch interpretierbare gemeinsame Grundlage zu etablieren. Man kann bestenfalls hoffen, dass sich dieses Problem im Rahmen der Entwicklung einer Theorie der Quantengravitation, die vielleicht von ganz anderen Ausgangspunkten und Problemstellungen ausgeht, lösen lässt – oder sich (im günstigsten Fall) von selbst löst: etwa, indem es sich als gänzlich irrelevant herausstellt, da sich zeigt, dass die Raumzeit als emergentes Phänomen ohnehin nur als klassische Näherung für den makroskopischen Bereich eine Rolle spielt. Dann käme es zu keinem Konflikt mit den nichtlokalen Zügen der Quantenmechanik; die raumzeitliche Näherung und das Lokalitätsprinzip würden sich schlichtweg als nur begrenzt gültig und verlässlich herausstellen.

Fazit: Eine spezifische Perspektive, die sich auf der Grundlage der quantenmechanischen Nichtseparierbarkeiten für eine Theorie der Quantengravitation und ihre Entwicklung ergeben könnte, zeichnet sich nicht ab – ausser man ist mutig genug, gleich zu behaupten, dass diese Form der Nichtlokalität ein indirekter Hinweis auf eine emergente Raumzeit und mithin eine emergente Gravitation darstellt.[161]

---

[160] Siehe Kap. 2.1.
[161] Siehe hierzu Kap. 3.3.

# 3. Elemente des Übergangs und spekulative Ideen

## 3.1. Bekenstein-Hawking-Entropie, holographische Grenze und finite Informationsdichten

### Die Bekenstein-Hawking-Entropie

Die *Bekenstein-Hawking-Entropie*[162] gehört zu einem Komplex von Einsichten und Argumenten, die sich aus der Verbindung von Ideen und konzeptionellen Überlegungen aus dem hier nun gemeinsamen Kontext der Allgemeinen Relativitätstheorie, der Quantenmechanik, den Quantenfeldtheorien, der Thermodynamik und der Informationstheorie ergeben.

Die intertheoretische Verbindung nimmt ihren Anfang, als sich Jacob Bekenstein in den frühen siebziger Jahre die Frage stellt, wie schwarze Löcher mit der Thermodynamik vereinbar sein könnten. Sollten nämlich die wenigen Eigenschaften (Masse, Drehimpuls, Ladung), die man im Rahmen der Allgemeinen Relativitätstheorie schwarzen Löchern zuschreibt, deren umfassende Beschreibung liefern, so hiesse dies, dass Entropie 'vernichtet' wird, wenn Materie in ein schwarzes Loch hineinfällt. Dies steht jedoch im Widerspruch zum Entropiesatz der Thermodynamik. Bekensteins Schluss:[163] Schwarze Löcher müssen eine Entropie besitzen, wenn der Entropiesatz der Thermodynamik nicht grundlegend falsch sein sollte. Hätten sie keine Entropie, so würde die Entropie insgesamt abnehmen, wenn Materie in ein schwarzes Loch fällt.[164]

Unabhängig davon kommen etwa zur selben Zeit James Bardeen, Brandon Carter und Stephen Hawking[165] mit Überlegungen, die sich vollständig im Rahmen der klassischen Allgemeinen Relativitätstheorie bewegen, zu dem Schluss, dass schwarze Löcher basalen Gesetzen gehorchen, die formal denen der gewöhnlichen Thermodynamik entsprechen, wenn man – jeweils bis auf einen Proportionalitätsfaktor – ihre Oberflächenschwerkraft mit der Temperatur und die Fläche ihres Ereignishorizontes mit der Entropie identifiziert. – Im Zusammenfluss dieser Ergebnisse, ergibt sich dann die Hypothese, dass schwarze Löcher tatsächlich über eine Entropie verfügen, die *Bekenstein-Hawking-Entropie*, und dass diese proportional zur Fläche ihres Ereignishorizonts ist.[166]

Diese Hypothese gewinnt dann schliesslich weitere Plausibilität, als Hawking[167] in der Folge Überlegungen zum Verhalten des quantenfeldtheoretischen Vakuums am Ereignishorizont schwarzer Löcher anstellt. Er kommt im Rahmen einer semi-klassischen Näherung zu dem Schluss, dass diese offenbar tatsächlich eine Schwarzkörperstrahlung aussenden: in Übereinstimmung mit der Tatsache, dass ihnen, erst einmal formal, eine Temperatur zugesprochen werden kann, die proportional zu ihrer Oberflächenschwerkraft ist.[168] – Aus dieser unabhängigen Herleitung der Schwarzkörperstrahlung schwarzer Löcher lässt sich dann schliesslich auch der Proportionalitätsfaktor zwischen der Oberflächengravitation des schwarzen Lochs und seiner Temperatur ableiten. Mit den Ergebnissen aus der Thermodynamik schwarzer Löcher führt dies dann zum Rückschluss auf den Proportionalitätsfaktor zwischen der Entropie und der Fläche des Ereignishorizontes: *Die Entropie schwarzer Löcher entspricht demnach einem Viertel der Planck-Einheiten auf der Fläche des Ereignishorizontes.* Jeweils vier Planck-Quadrate auf dieser Oberfläche entsprechen genau einem Freiheitsgrad. Dieses Resultat ergibt sich mithin aus einer Kombination von Komponenten, die sich über die Allgemeine Relativitätstheorie, die Quantenmechanik und die Thermodynamik erstrecken. Und es ermöglicht schliesslich die Formulierung eines verallgemeinerten Entropiesatzes der Thermodynamik, der die Entropie schwarzer Löcher mit einschliesst.

---

[162] Siehe Bekenstein (1973, 1974, 1981, 2001, 2003), Wald (1994, 2001), Bousso (2002), Carlip (2008a).

[163] Siehe Bekenstein (1973, 1974).

[164] Eine Alternative hierzu bestände darin anzunehmen, dass schwarze Löcher imSinne der Allgemeinen Relativitätstheorie gar nicht existieren, sondern ein theoretisches Artefakt darstellen.

[165] Siehe Bardeen / Carter / Hawking (1973).

[166] Es gibt jedoch bis heute immer noch keine allgemein gültige Auffassung hinsichtlich des Entropiegehaltes des Gravitationsfeldes, wie es von der Allgemeinen Relativitätstheorie beschrieben wird.

[167] Siehe Hawking (1974, 1975) sowie Wald (1994, 2001).

## Die 'holographische Grenze'

Man kann sich nun angesichts der *Bekenstein-Hawking-Entropie* schwarzer Löcher zwei grundsätzliche Fragen stellen. Die erste lautet: *Wie kommt die Bekenstein-Hawking-Entropie schwarzer Löcher zustande?* Der Hintergrund ist dabei der folgende: Gemeinhin geht die Entropie eines physikalischen Systems auf das statistische Verhalten seiner mikroskopischen Konstituenten zurück.

> *"In a typical thermodynamic system, thermal properties are the macroscopic echoes of microscopic physics."* (Carlip (2008a) 2)

Worin bestehen nun aber diese mikroskopischen Konstituenten im Falle eines schwarzen Loches?

> *"What is the microscopic, statistical origin of black hole entropy? We have learned that a black hole, viewed from the outside, is unique classically. The Bekenstein-Hawking formula, however, suggests that it is compatible with $e^{S_{BH}}$ independent quantum states. The nature of these quantum states remains largely mysterious."* (Bousso (2002) 7)

Die Allgemeine Relativitätstheorie gibt nicht die geringste Auskunft über mögliche mikroskopische Konstituenten eines schwarzen Loches. Sie beschreibt schwarze Löcher ausschliesslich durch die Parameter Masse, Drehimpuls und elektrische Ladung. Diese wenigen klassischen Parameter können jedoch kaum für die (gemeinhin riesige) Entropie eines (makroskopischen) schwarzen Loches verantwortlich sein. Es müssten vielmehr diskrete mikroskopische Freiheitsgrade existieren, die von der klassischen Theorie nicht erfasst werden, mit einer (sehr grossen, aber) finiten Informations-

---

[168] Diese Strahlung schwarzer Löcher lässt sich auch als Konsequenz des Unruh-Effekts verstehen – eine weitere noch in mancher Hinsicht unausgelotete Verbindung zwischen Gravitationseinflüssen, Quantenmechanik und Thermodynamik:

> *"[...] an influential paper by Bill Unruh convincingly argued that an observer that accelerates in the vacuum state of a conventional [quantum field theory] interacts with the quantum field as if this was in a thermal bath. This shed light on black hole radiation, because an observer that remains at a fixed distance from a black hole is in constant acceleration (in order not to freely fall), and therefore black hole radiation can be interpreted simply as an Unruh effect. But at the same time this result appears to suggest that there is a deep general relation, which we do not yet understand, tying together gravity, thermodynamics and quantum theory."* (Rovelli (2007) 1295) menge auf der Ebene der Mikrokonstituenten einhergehen und gerade zur *Bekenstein-Hawking-Entropie* schwarzer Löcher führen.

> *"What is the statistical mechanical origin of the black hole entropy $S_{BH} = a_{hor}/4 \, l_{Pl}^2$? What are the microscopic degrees of freedom that account for this entropy? [...] Where do all these states reside? To answer these questions, in the early nineties John Wheeler suggested the following heuristic picture, which he christened 'It from Bit'. Divide the black hole horizon into elementary cells, each with one Planck unit, $l_{Pl}^2$, of area and assign to each cell two microstates, or one 'bit'."* (Ashtekar (2002) 11)

Auch wenn Wheelers *It-from-bit*-Ansatz hier vielleicht eine Veranschaulichung der mikroskopischen Partitionierung des Ereignishorizontes liefert, so gibt er dennoch letztendlich keine Antwort auf die eigentliche Frage, welche Mikrokonstituenten und welche Mikrozustände diese Partitionierung hervorrufen. – Eine solche Antwort erhofft man sich nicht zuletzt von einer erfolgreichen Theorie der Quantengravitation.[169]

> *"Die ultimative Informationskapazität eines Stücks Materie – mit anderen Worten, seine wahre thermodynamische Entropie – kann nicht berechnet werden, solange das Wesen der ultimativen Bestandteile der Materie oder der tiefsten Strukturebene nicht bekannt ist."* (Bekenstein (2003) 36)

Die zweite der grundsätzlichen Fragen hinsichtlich der Entropie schwarzer Löcher klingt dann zwar wesentlich unspektakulärer, erweist sich aber schliesslich als die perspektivenreichere: *Welche grundsätzlichen Einsichten gehen mit der Bekenstein-Hawking-Entropie einher, solange die Mikrokonstituenten, auf denen sie beruht, noch nicht bekannt sind?* – Die Antwort: Die *Bekenstein-Hawking-Entropie* liefert, durchaus unabhängig von der Klärung der Frage nach den spezifischen, für sie verantwortlichen Mikrokonstituenten, Rückschlüsse auf den Informationsgehalt bzw. die Informationsdichte auf der fundamentalsten Ebene der Natur.

> *"How much complexity [...] lies at the deepest level of nature? How much information is required to specify any physical configuration completely, as long as it is contained in a prescribed region."* (Bousso (2002) 12)

Die Antwort der gesamten Kontinuumsphysik – der Quantenfeldtheorien wie der klassischen Feldtheorien, insbesondere auch der Allgemeinen Relativitätstheorie – auf die *Frage nach dem Informationsgehalt in einer eingegrenzten, finiten räumlichen Region* lautet: unendlich.

> *"A quantum field theory consists of one or more oscillators at every point in space. Even a single harmonic oscillator has an infinite-dimensional Hilbert space. Moreover, there are infinitely many points in any volume of space, no matter how small."* (Bousso (2002) 12)

Die diesbezügliche Antwort, die sich nun mittels der Thermodynamik schwarzer Löcher ableiten lässt, liefert hier einen deutlichen Kontrast zur Kontinuumsphysik. Sie lautet: *Es gibt eine definitive,*

---

[169] Es gibt mehr als genug Lösungsvorschläge hierfür, etwa auch einen solchen von Seiten der *Loop Quantum Gravity*. Siehe Kap. 4.4. Von einer definitiven Lösung sind aber alle Ansätze noch weit entfernt.

*finite, 'holographische Grenze'*[170] *für den Informationsgehalt, der maximal einer eingegrenzten, finiten Raumregion zukommen kann.* Denn die Entropie schwarzer Löcher ist die maximale Entropie, die für ein Raumvolumen erreichbar ist. Und diese maximale finite Entropie entspricht einem maximalen finiten Informationsgehalt auf der Ebene des zugrundeliegenden Substrats. Die sich auf diese Weise ergebende finite Informationsdichtegrenze lässt sich anschaulich motivieren:[171]

> *"Die holographische Grenze gibt an, wie viel Information in einem gegebenen Raumgebiet enthalten sein kann. Man betrachte eine Materieansammlung, die in eine Kugel der Oberfläche A passt und zu einem Schwarzen Loch kollabiert. Die Horizont-Oberfläche des Schwarzen Loches muss kleiner als A sein, seine Entropie somit kleiner als A/4. Da die Entropie nicht abnehmen kann, muss auch die ursprüngliche Materieverteilung weniger als A/4 Entropie- oder Informationseinheiten enthalten haben. Das bedeutet, dass der maximale Informationsgehalt eines Raumgebietes durch dessen Oberfläche festgelegt wird – entgegen der nahe liegenden Annahme, die Speicherkapazität eines Gebiets hänge von seinem Volumen ab."* (Bekenstein (2003) 38)

Übersetzt in den Kontext der Quantenmechanik, welche die Informationserhaltung in Form der Forderung nach Unitarität erfasst, lautet die Motivation für die holographische Grenze etwa folgendermassen:

> *"A more compelling consideration is based on unitarity. Quantum-mechanical evolution preserves information; it takes pure state to a pure state. But suppose a region was described by a Hilbert space of dimension $e^V$, and suppose that region was converted to a black hole. According to the Bekenstein entropy of a black hole, the region is now described by a Hilbert space of dimension $e^{A/4}$. The number of states would have decreased, and it would be impossible to recover the initial state from the final state. Thus, unitarity would be violated. Hence, the Hilbert space must have had dimension $e^{A/4}$ to start with."* (Bousso (2002) 14)

Eine stringentere Herleitung der holographischen Entropiegrenze unter noch wesentlich verallgemeinerten Bedingungen liegt schliesslich mit Raphael Boussos *'kovarianter Entropiegrenze'* vor.[172]

## Implikationen der holographischen Grenze

Sollte der Entropiesatz der Thermodynamik tatsächlich universelle Gültigkeit für sich beanspruchen können und auch auf schwarze Löcher anwendbar sein, wie es die zuvor geschilderten Überlegun-gen nahelegen, so gibt es also offensichtlich eine maximale, finite Entropie, die innerhalb eines Raumvolumens erreichbar ist. Diese maximale finite Entropie entspricht einer maximalen, finiten Zahl von Freiheitsgraden, die in einem Raumvolumen physikalisch wirksam werden und entsprechend für dieses zu berücksichtigen sind.

---

[170] Der Begriff der *'holographischen* Grenze' geht auf die Tatsache zurück, dass der maximale Informationsgehalt, der einer finiten räumlichen Region zukommen kann, proportional zur Oberfläche dieser Region ist, was wiederum die Grundlage für das *'holographische Prinzip'* darstellt. Siehe Kap. 3.2.

[171] Die Voraussetzungen für eine formale Ableitung sind einerseits der *verallgemeinerte Entropiesatz* der Thermodynamik, der eben gerade auch die *Bekenstein-Hawking-Entropie* mitberücksichtigt, andererseits die Annahme, dass nur schwache Gravitationswirkungen vorliegen. – Die Voraussetzungen für die Ableitung werden von den sogenannten *'Monstern'* unterlaufen: semi-klassischen Lösungen, die sich per se nicht auf entsprechende Mikrozustände zurückführen (und somit auch nicht in den Kontext einer Theorie der Quantengravitation überführen) lassen, die also in mancher Hinsicht (insbesondere was die Modalitäten ihres Zustandekommen betrifft) als unphysikalisch gelten können; siehe Hsu / Reeb (2007, 2009a).

[172] Siehe Bousso (2002). Hier stellen nun auch die o.g. *'Monster'* kein Problem bzw. Gegenbeispiel mehr dar; siehe Hsu / Reeb (2009a). Bisher sind nur sehr wenige, höchst exotische und physikalisch mutmasslich völlig unrealistische Szenarien bekannt, für die diese kovariante Entropiegrenze durchbrochen wird.

> *"Since it involves no assumptions about the microscopic properties of matter, it places a fundamental limit on the number of degrees of freedom in nature."* (Bousso (2002) 36)

Die finite Entropie, die einem Raumvolumen maximal zukommen kann, entspricht damit gleichzeitig einem maximal erreichbaren Informationsgehalt, der auf physikalischem Wege tatsächlich innerhalb dieses Raumvolumens implementiert bzw. kodiert werden kann. – Jede darüber hinausgehende Information kann, was die tatsächlichen physikalischen Verhältnisse betrifft, nur als theoretisches (bzw. modelltheoretisches) Artefakt betrachtet werden, dem kein physikalisches Gegenstück entspricht. Jede Beschreibung, die ein Mehr an Information impliziert, als es die holographische Grenze zulässt, enthält solche unphysikalischen Artefakte.

Und die maximale, finite Entropie, die innerhalb eines Raumvolumens erreichbar ist, ist erstaunlicherweise proportional zur begrenzenden Oberfläche dieses Raumvolumens. Sie entspricht wiederum einer maximalen, finiten Zahl von Freiheitsgraden für dieses Raumvolumen, die sich – in einer direkten, proportionalen Zuordnung –letztlich schon der einschliessenden Oberfläche zuschreiben liessen. Die einem Raumvolumen zukommende maximale, finite Informationsmenge lässt sich also offensichtlich vollständig auf der einschliessenden Oberfläche implementieren bzw. kodieren. Es gibt keine Information, die das gesamte Raumvolumen betrifft, die über die grundsätzlich auf der Oberfläche kodierbare finite Information hinausginge. – Dies ist auf den ersten Blick (und vor allem vor dem Hintergrund unserer Erfahrungen mit der Kontinuumsphysik der Feldtheorien und Quantenfeldtheorien) nichts weniger als kontraintuitiv. Für den maximalen Entropiegehalt, die maximale Zahl der Freiheitsgrade, den maximalen Informationsgehalt innerhalb eines Raumvolumen, würde man – wenn diese Grössen schon in einer grundsätzlichen Weise begrenzt sind – erst einmal eine Proportionalität zum jeweiligen Volumen vermuten, nicht aber zu dessen eingrenzender Oberfläche.

> *"Naively one would expect the maximal entropy to grow with the volume of spatial regions. Instead, it is set by the area of surfaces."* (Bousso (2002) 36)

Die Proportionalität zur Oberfläche ist dann schliesslich der Ausgangspunkt von Spekulationen, die unter der Bezeichnung *'Holographisches Prinzip'* diskutiert werden und auf die weiter unten einzugehen sein wird.[173] Durchaus unabhängig vom 'Holographischen Prinzip' liefert die holographische bzw. kovariante Entropiegrenze mit ihrer Festschreibung finiter maximaler Informationsdichten (bzw. einer finiten maximalen Zahl physikalisch relevanter Freiheitsgrade pro Raumvolumen) jedoch schon deutliche Anzeichen für eine tiefste Strukturebene, die wiederum Gegenstand der in Entwicklung befindlichen Theorien der Quantengravitation sein sollte bzw. in diesen ihren Niederschlag finden sollte.

> *"We conclude that* the fundamental theory responsible for the [holographic] bound unifies matter, gravity, and quantum mechanics.*"* (Bousso (2002) 37)

---

[173] Siehe Kap. 3.2.

Im Kontext einer erfolgreichen Theorie der Quantengravitation wäre also nicht zuletzt das konkrete Zustandekommen der holographischen bzw. kovarianten Entropiegrenze zu klären. Dies wird wahrscheinlich nicht möglich sein, ohne die tatsächlich vorliegenden mikroskopischen Freiheitsgrade zu bestimmen, die für diese Grenze verantwortlich sind und die auf der makroskopischen Ebene etwa zur *Bekenstein-Hawking-Entropie* schwarzer Löcher führen.

Und eine solche Theorie der Quantengravitation wird – wenn der holographischen bzw. kovarianten Entropiegrenze universelle Gültigkeit zukommt – gleichermassen die raumzeitlichen Freiheitsgrade *und* die nicht-raumzeitlichen Materiefreiheitsgrade in ihrem Zustandekommen erfassen müssen. Beide Komponenten sind dieser Grenze schliesslich gemeinsam unterworfen, unabhängig davon, ob sie nun schon auf der Substratebene vorliegen oder erst von dieser ausgehend auf einer emergenten Ebene hervorgebracht werden:

> ʺThe covariant entropy bound is a law of physics which must be manifest in an underlying theory. This theory must be a unified quantum theory of matter and space-time. From it, Lorentzian geometries and their matter content must emerge in such a way that the number of independent quantum states describing the light-sheets of any surface B is manifestly bounded by the exponential of the surface area.ʺ (Bousso (2002) 37)

Bei der Entwicklung einer solchen Theorie der Quantengravitation könnten sich die bestehenden Einsichten aus dem Kontext der Thermodynamik schwarzer Löcher, deren mikroskopisches Pendant dann im Rahmen dieser Theorie in spezifischer Weise geklärt werden müsste, immerhin schon als – zumindest in heuristischer Hinsicht – hochgradig relevant erweisen. Wenn es eine maximale Informationsdichte gibt, so deutet diese auf eine grundlegende diskrete Struktur hin, auch wenn erst einmal unklar bleibt, ob es sich hierbei um eine raumzeitliche oder eine prä-raumzeitliche Struktur handelt. Die maximal möglichen, finiten Informationsdichten können jedoch nur eine Folge einer diskreten Substratstruktur sein.

Das heisst aber gleichzeitig, dass die Idee eines Raumzeitkontinuums, wie es der Allgemeinen Relativitätstheorie, der Quantenmechanik und den Quantenfeldtheorien, aber auch etwa dem Stringansatz zugrundeliegt, letztendlich unangemessen ist. Man könnte sich ohnehin fragen – bzw. schon vor dem Aufkommen der Thermodynamik schwarzer Löcher und der holographischen Entropiegrenze gefragt haben: Was berechtigt uns überhaupt zu der Annahme, dass es sich bei der Raumzeit um ein Kontinuum handelt? Man hätte also nicht unbedingt auf die Einsichten warten müssen, die sich im Kontext der Thermodynamik schwarzer Löcher ergeben haben. Es wäre schon vorher vielleicht nicht allzu fernliegend gewesen, die Singularitäten, welche die Allgemeine Relativitätstheorie vorhersagt, ohne sie zu beschreiben,[174] die Divergenzen, zu denen es in den Quantenfeldtheorien kommt, oder die Nichtrenormierbarkeit der Ansätze zu einer *Kovarianten Quantisierung* der Allgemeinen Relativitätstheorie als Anzeichen für die letztendliche Unangemessenheit der Kontinuumsannahme und mithin als solche für eine diskrete Raumzeit (oder eine emergente Raumzeit auf der Basis eines diskreten Substrats) zu deuten. – Mit der Thermodynamik schwarzer Löcher wird diese Einschätzung nun manifest.

---

[174] Vgl. auch Crane (2007).

"Thermal nature of horizons cannot arise without the spacetime having a microstructure." (Padmanabhan (2007a) 2)

Mit den Indizien für eine diskrete, fundamentale Struktur, wie sie sich im Rahmen der Thermodynamik schwarzer Löcher ergeben, erweist sich ein modelltheoretisches Instrumentarium, welches etwa auf eine differenzierbare Mannigfaltigkeit setzt, als unangemessen für eine Beschreibung der Substratebene. Nun gehen aber die Allgemeine Relativitätstheorie ebenso wie alle Ansätze zu ihrer direkten Quantisierung, insbesondere auch die ansonsten in ihrer Vorgehensweise durchaus subtile *Loop Quantum Gravity*, von einer solchen differenzierbaren Mannigfaltigkeit aus. Schon die Formulierbarkeit der Diffeomorphismusinvarianz setzt diese voraus. Auch der Substantialismus bezüglich der Raumzeit setzt sie voraus, entweder indem er von Raumzeitpunkten ausgeht, denen Substantialität zugesprochen wird, oder sobald er diese Substantialität am metrischen Feld festmacht, das auf dem Kontinuum definiert ist.

Das Problem der letztendlichen Unangemessenheit der Kontinuumsannahme vor dem Hintergrund der Implikationen der Thermodynamik schwarzer Löcher betrifft nahezu die gesamte Physik, insbesondere aber die Allgemeine Relativitätstheorie und die Quantenfeldtheorien. Spätestens vor dem Hintergrund dieser Unangemessenheit wird die Anwendung herkömmlicher quantenmechanischer und quantenfeldtheoretischer Verfahren im Bereich der Quantengravitation hochgradig fragwürdig. Mit den Argumenten für eine diskrete, fundamentale Struktur erweist sich das Zahlenkontinuum als modelltheoretisch nur begrenzt angemessenes Instrumentarium zur Beschreibung der physikalischen Realität, das endgültig unangemessen wird, wenn es um deren Substratebene geht. Der einzige offensichtliche Ausweg, der sich hier anbietet, besteht im Versuch, von vornherein mit diskreten Strukturen zu arbeiten, vielleicht sogar mit prägeometrischen diskreten Strukturen.[175]

Ein Hoffnungsschimmer – auch für die umwegsameren Ansätze, die (zumindest in modelltheoretischer Hinsicht) erst einmal am Kontinuum festhalten – ist darin zu sehen, dass sich auch hier indirekte Hinweise auf ein diskretes Substrat abzeichnen. Dies gilt ansatzweise vielleicht schon für die Quantenmechanik und die Quantenfeldtheorien, ausgeprägter immerhin für den Stringansatz[176] und schliesslich sehr deutlich für die *Loop Quantum Gravity*.[177] Für letztere ergibt sich so etwas wie eine Diskretisierung durch die Hintertür auf der Basis eines formal immer noch mit dem Kontinuum arbeitenden Ansatzes, der sich – trotz aller Subtilität im Vergleich mit diversen Konkurrenten – in dieser post-hoc-Relativierung seines formalen Ausgangspunktes als modelltheoretisch nur begrenzt angemessen erweist. Und man könnte sich angesichts dessen dann fragen: Geht das auch besser? Warum sollte man nicht, zumindest in einer parallel laufenden Alternativstrategie, versuchen, gleich mit einem diskreten Ansatz zu starten?[178]

---

[175] Siehe insbesondere Kap. 4.6.
[176] Siehe Kap. 4.2.
[177] Siehe Kap. 4.4.
[178] Siehe wiederum Kap. 4.4. für die Details solcher Argumentationen.

## 3.2. Das Holographische Prinzip

Wie wir gesehen haben, gibt es offensichtlich eine finite Zahl von physikalisch relevanten Freiheitsgraden, die einem Raumvolumen maximal zugeschrieben werden können, und damit einen maximal erreichbaren Informationsgehalt, der innerhalb eines Raumvolumens physikalisch implementiert bzw. kodiert werden kann. Und diese Grössen sind nicht etwa proportional zum Raumvolumen, dem sie zugeschrieben werden, sondern vielmehr proportional zur begrenzenden Oberfläche des Raumvolumens. Die maximale, finite Zahl von physikalisch relevanten Freiheitsgraden, die einem Raumvolumen zugeschrieben werden können, erschöpft sich also offensichtlich gerade in der Zahl der Freiheitsgrade, die sich in einer direkten, proportionalen Zuordnung grundsätzlich auch schon auf der einschliessenden Oberfläche definieren liessen. Die einem Raumvolumen zukommende maximale, finite Informationsmenge lässt sich also offenbar vollständig auf der einschliessenden Oberfläche implementieren bzw. kodieren. Es gibt keine Information, die das gesamte Raumvolumen betrifft, die über die grundsätzlich schon auf der Oberfläche implementierbare bzw. kodierbare finite Information hinausginge.

Diese Proportionalität zur volumenbegrenzenden Fläche ist gerade das Charakteristikum der holographischen bzw. kovarianten Entropiegrenze. Auf ihrer Grundlage formulierten Gerard 't Hooft und Leonard Susskind Anfang der neunziger Jahre das *'Holographische Prinzip'*:[179] Danach gilt, spätestens für die fundamentale Strukturebene physikalischen Geschehens, dass die fundamentalen physikalischen Freiheitsgrade, die der Dynamik auf der Oberfläche eines Raumvolumens zugesprochen werden können, grundsätzlich – und nicht nur im Hinblick auf den maximal einschliessbaren Informationsgehalt – die Freiheitsgrade im eingeschlossenen Raumvolumen vollständig bestimmen.

> *"The holographic principle [...] implies that the number of fundamental degrees of freedom is related to the area of surfaces in spacetime."* (Bousso (2002) 2)

Damit liessen sich in einer zukünftigen Theorie der Quantengravitation – wenn diese tatsächlich die fundamentalste Strukturebene in adäquater Weise erfassen sollte – die Freiheitsgrade, die einem Raumvolumen und dem in ihm (scheinbar) stattfindenden Geschehen zugesprochen werden können, als auf der Oberfläche dieses Raumvolumens vollständig definiert betrachten. Die Freiheitsgrade auf der umgebenden Oberfläche würden die Freiheitsgrade in einem Raumvolumen vollständig festlegen. Eine vollständige Oberflächen-Theorie würde notwendigerweise auch die Beschreibung des gesamten Geschehens im eingeschlossenen Raumsegment umfassen. Das physikalische Geschehen in einem Raumvolumen liesse sich als eine holographische Projektion des (echten) physikalischen Geschehens, das auf seiner umgrenzenden Oberfläche stattfindet, ansehen. Eine vollständige fundamentale Theorie müsste also nur die Freiheitsgrade auf der umgebenden Oberfläche beschreiben, um alles zu beschreiben, was es zu beschreiben gibt. – Dies entspricht allerdings einer radikalen Transzendierung der Lokalitätsannahme.[180]

---

[179] Siehe 't Hooft (1993), Susskind (1995), Bousso (2002), Bekenstein (2003).

[180] Man könnte hier durchaus spekulieren, ob die gesuchten 'echten', physikalischen Freiheitsgrade der Allgemeinen Relativitätstheorie (vgl. Kap. 2.1.) vielleicht das Ergebnis des Vorliegens einer holographischen Festlegung der physikalischen Freiheitsgrade auf der Ebene eines Quantensubstrats sind, auf dessen Grundlage die klassische Raumzeit erst zustandekommt.

*"[...] the holographic principle calls into question [...] the very notion of locality."* (Bousso (2002) 2)

Wie könnte es aber auf der fundamentalen Ebene zu einer solchen holographischen Determinierung des (scheinbaren) physikalischen Geschehens in einem Raumvolumen durch das (tatsächliche) physikalische Geschehen auf seiner eingrenzenden Oberfläche kommen? Lässt sich das Holographische Prinzip vielleicht schon als Hinweis auf die Emergenz der phänomenologischen Raumzeit deuten? Lässt es sich darüberhinausgehend womöglich sogar als Indiz für ein computationales Universum lesen: für eine körnige, ausschliesslich auf (mit finiten Mitteln) berechenbaren Zahlen beruhende fundamentale Basis des Weltgeschehens, die uns ein Mehr an Dimensionen und dynamisch wirksamen Freiheitsgraden vorgaukelt als tatsächlich berechnet werden? Dies hätte zur Folge, dass mit dem Übergang von der (phänomenologischen) Beschreibung des vermeintlichen, zumindest z.T. vorgegaukelten raumzeitlichen Geschehens zu einer auf den echten physikalisch wirksamen Freiheitsgraden beruhenden fundamentalen Beschreibung eine signifikante Datenreduktion oder – kompression einherginge.[181] Ist also nicht nur das raumzeitliche Kontinuum eine Fiktion, sondern darüberhinausgehend die meisten seiner vermeintlich physikalisch wirksamen Freiheitsgrade, mindestens aber eine seiner Dimensionen?

Die Vermutung, dass es eine finite computationale Basis gibt, die dem gesamten physikalischen Geschehen zugrundeliegt, erhält zuletzt zusätzlichen Aufwind durch die sich mittlerweile einstellenden Hinweise auf die Möglichkeit einer über das Holographische Prinzip noch deutlich hinausgehenden Universalität von grundlegend unterschiedlichen dynamischen Theorien (etwa Gravitations- und Eichtheorien), die jeweils zur gleichen niederenergetischen Phänomenologie führen.[182] Solche Hinweise zeichnen sich schon mit der AdS/CFT-Korrespondenzvermutung im Stringansatz[183] ab, aber insbesondere auch mit ihren möglichen, über diesen engeren Kontext hinausweisenden Verallgemeinerungen:

*"[...] there are already examples of large equivalence classes amongst gauge theories. [...] These include conjectures of dualities between theories with different gauge groups and, as in the case of the AdS/CFT conjectures, different numbers of dimensions. Others do not require supersymmetry but involve dualities among non-commutative and matrix formulations of gauge theories. It is then natural to ask if all of these conjectured dualities, supersymmetric and not, may be several tips of a single iceberg involving a much wider class of dualities. If so, the question is what principle underlies all these dualities."* (Smolin (2008) 4)

Die Universalität von grundlegend verschiedenen dynamischen Theorien, die zur gleichen niederenergetischen Phänomenologie führen, wird vor allem dadurch brisant, dass die Theorien innerhalb einer Universalitätsklasse im allgemeinen von gänzlich verschiedenen basalen Freiheitsgraden ausgehen:

*"Suppose that a large class of theories, which induced the standard model as well as a large set of plausible alternatives, were actually equivalent to each other, in the sense that there were transformations that mapped the degrees of freedom and solutions of any two of these theories into each other."* (Smolin (2008) 3)

---

[181] Diese Annahme und ihre Motivation gehen über das auf der Grundlage der Diskretheit der Raumzeitstruktur formulierbare Argument für ein computationales Universum hinaus.

[182] Siehe Smolin (2008).

[183] Siehe Kap. 4.2.

Insbesondere gehören Theorien mit unterschiedlicher Dimensionalität[184] und unterschiedlichen Eichgruppen zur gleichen Universalitätsklasse, d.h. sie führen zur gleichen niederenergetischen Phänomenologie. Hinzu kommen deutliche Hinweise auf Transformationsbezüge zwischen dynamischen Theorien innerhalb einer Universitätsklasse, die raumzeitliche mit internen Freiheitsgraden bzw. Symmetrien mischen und austauschen:[185]

> "[...] whether two degrees of freedom are related by a translation in space or by an internal symmetry transformation will not be absolute." (Smolin (2008) 5)

Eine solche Austauschbarkeit von raumzeitlichen und internen Freiheitsgraden liesse sich, wenn diesen Theorien eine unmittelbare physikalische Relevanz zugesprochen werden kann, als direktes Indiz für eine emergente Raumzeit lesen.[186] – Die mit einem solchen Szenario einhergehenden Implikationen, die deutlich über die des Holographischen Prinzips hinausgehen, führen insbesondere auch zu einer zusätzlichen epistemischen Problemlage: Lassen sich nämlich die von grundlegend verschiedenen Theorien innerhalb einer Universalitätsklasse jeweils verwendeten, unterschiedlichen (Gesamtheiten von) basalen Freiheitsgrade(n) durch einen Theoriewechsel, der die niederenergetischen Implikationen unverändert lässt, gegeneinander austauschen, so lässt dies einigen Zweifel daran zu, dass die basalen Freiheitsgrade einer spezifischen (empirisch adäquaten) Beschreibung der vorliegenden Phänomenologie – also etwa die einer bestimmten empirisch erfolgreichen Gravitationstheorie – tatsächlich für das vorliegende physikalische Geschehen verantwortlich sind, das hiermit gerade beschrieben werden soll. Sie lassen sich vielmehr als das Ergebnis einer kontingenten Auswahl aus dem Pool aller der Theorien ansehen, die zur Erfassung der entsprechenden niederenergetischen Phänomenologie geeignet sind und diesbezüglich eben gerade eine Universalitätsklasse bilden.

> "Were this true, the holographic principle might be a special case of a wider class of equivalences amongst theories." (Smolin (2008) 4)

Es wäre dann vielleicht nach der elementarsten Beschreibung innerhalb einer Universalitätsklasse zu suchen: einer Beschreibung mit so wenig Freiheitsgraden wie möglich. Die Lösung dieses Problems wiederum könnte sich infolge einer möglichen Vieldeutigkeit gleichwertig elementarer Beschreibungen als deutlich schwieriger herausstellen als das analoge Problem im Falle des Holographischen Prinzips; hier ist die Sachlage einfacher: die hinsichtlich der Zahl der verwendeten Freiheitsgrade elementarste, redundanzfreie Beschreibung ist zweifelsohne die auf der Grundlage der Oberflächenfreiheitsgrade; es liesse sich also argumentieren, dass diese gerade die echten, dem Geschehen zugrundeliegenden Freiheitsgrade erfasst. – Die Implikationen im Hinblick auf die Emergenz der Raumzeit und auf eine mögliche computationale Basis des physikalischen Geschehens sind in beiden Szenarien identisch, nur dass sich das Holographische Prinzip in seiner Berufung auf die kovariante Entropiegrenze als deutlich besser gestützt ansehen lässt.

---

[184] Dies gilt etwa schon für die AdS/CFT-Vermutung im Stringansatz. Siehe Kap. 4.2.

[185] Damit ist die entsprechende Universalitätsklasse als Referenzgrösse der Beschreibung notwendigerweise hintergrundunabhängig.

[186] Es finden sich in Smolin (2008) sogar Hinweise auf eine mögliche Emergenz der Voraussetzungen, die für eine quantenmechanische Beschreibung erforderlich sind.

## 3.3.   Gravitation und/oder Raumzeit als emergente Phänomene

Angesichts der schon lange während Bemühungen um die Entwicklung einer Theorie der Quantengravitation, die schon in den dreissiger Jahren des zwanzigsten Jahrhunderts begonnen haben, der grossen Probleme, die dabei immer wieder aufgetreten sind, und schliesslich der immer noch sehr begrenzten Erfolge, auf die sich diese Ansätze berufen können,[187] drängen sich – und dies erst einmal völlig unabhängig von den vorausgehenden Überlegungen zur Thermodynamik schwarzer Löcher und zum Holographischen Prinzip – die folgenden Fragen geradezu auf: Warum stellt die Gravitation ein solches Problem für die entsprechenden Quantisierungsbestrebungen dar? Was ist in dieser Hinsicht das Besondere an der Gravitation?

> *"What is so special about the gravitational force that it has persisted in its quantisation for about 70 years already?"* (Thiemann (2007) 9)

Von da ist es dann kein grosser Schritt zu der Vermutung, dass vielleicht einfach eine der grundlegenden Annahmen, von denen alle traditionellen Ansätze zur Entwicklung einer Theorie der Quantengravitation – insbesondere die, die auf einer direkten Quantisierung beruhen – ausgegangen sind, schlichtweg falsch sein könnte. Vielleicht ist die Gravitation einfach keine fundamentale Wechselwirkung, der dann im Übergang zu einer Theorie der Quantengravitation Quanteneigenschaften zugesprochen werden müssten:

> *"[...] gravity could all in all be an intrinsically classic / large scale phenomenon [...]."* (Girelli / Liberati / Sindoni (2008) 1)

Die Argumente gegen semi-klassische Theorien der Gravitation[188] schliessen zwar für den Fall, dass die Quantenmechanik universell gültig sein sollte, ein fundamentales klassisches Gravitationsfeld in einer ansonsten umfassenden Quantenwelt aus. Aber sie schliessen – auch unter der Annahme der universellen Gültigkeit der Quantenmechanik – nicht die Möglichkeit aus, dass es sich bei der Gravitation um ein intrinsisch klassisches Phänomen handelt. Sie schliessen für diesen Fall nur aus, dass es sich bei der Gravitation gleichzeitig um eine fundamentale Wechselwirkung handelt.

### Emergente Gravitation und/oder emergente Raumzeit

Sollte also die Gravitation ein intrinsisch klassisches Phänomen sein, so müsste sie eine residuale bzw. induzierte Wechselwirkung sein: ein emergentes Phänomen. Auf einer fundamentaleren Ebene würde sie gar nicht vorkommen. Sie wäre nicht Teil des Quantensubstrats, sondern würde von diesem in klassischer, makroskopischer Näherung als residuales Phänomen hervorgebracht. Sie wäre eine mehr oder weniger indirekte Folge fundamentalerer Wechselwirkungen und der entsprechenden fundamentaleren Freiheitsgrade des Substrats. Vielleicht wäre sie in ihrem Zustandekommen in irgendeiner Weise vergleichbar mit den residualen Van-der-Waals-Kräften in der Elektrodynamik, vielleicht käme sie aber auch auf eine ganz andere Weise zustande. Jedenfalls gäbe es für eine solche emergente Gravitation keine konzeptionellen Widersprüche mit der universellen Gültigkeit der

---

[187] Siehe Kap. 4.
[188] Vgl. Kap. 1.1. und 1.2.

Quantenmechanik. Es gäbe insbesondere keine semi-klassischen Hybriddynamik. Das Quantensubstrat selbst würde über gar keine gravitativen Freiheitsgrade verfügen.

Sollte die mit der Allgemeinen Relativitätstheorie einhergehende (und mithin nicht zuletzt auch empirisch bewährte) Geometrisierung der Gravitation dennoch einen wesentlichen Charakterzug der nunmehr emergenten, intrinsisch klassischen Gravitation erfassen, so könnte man dann schliesslich vermuten, dass auch die Raumzeit ein makroskopisches, intrinsisch klassisches, emergentes Phänomen ist, dass es also auf der Ebene des Quantensubstrats keine raumzeitlichen Freiheitsgrade gibt.

> *"Within the [emergent gravity framework] the very concepts of geometry and gravitational interaction are not seen as elementary aspects of Nature but rather as collective phenomena associated to the dynamics of more fundamental objects."* (Girelli / Liberati / Sindoni (2008) 1)

Und es wäre unter diesen Bedingungen plausibel anzunehmen, dass die Raumzeit im Rahmen der gleichen Näherung bzw. des gleichen klassischen Grenzfalles hervorgebracht wird, die bzw. der auch die Phänomenologie der Gravitation hervorbringt. – Der aus diesen Überlegungen resultierende Ansatzpunkt für eine zu entwickelnde Theorie der Quantengravitation bestände dann gerade in der Hoffnung, mittels dieser Theorie im Rahmen der Erfassung einer Substratdynamik, für die diese emergenten Grössen noch keine Rolle spielen, eine Erklärung für ihr Zustandekommen zu erreichen, etwa im Sinne einer zumindest näherungsweisen Ableitbarkeit.

Aber Vorsicht: Szenarien, welche die Frage nach der Emergenz der Gravitation von der nach der Emergenz der Raumzeit getrennt behandeln, sind nicht nur grundsätzlich denkbar, sondern wurden explizit vorgeschlagen.[189] Diesen zufolge könnte es durchaus auf der Grundlage einer schon vorhandenen Raumzeit zur Emergenz der Gravitation kommen. Im Rahmen solcher Hybridszenarien wären dann jedoch hinreichende Argumente anzuführen, die erklärbar werden lassen, wie die Allgemeine Relativitätstheorie mit ihrer Geometrisierung der Gravitation – auch in empirischer Hinsicht – erfolgreich sein konnte. Das plausiblere Szenario ist also erst einmal ein solches, das mit einer emergenten Gravitation auch eine emergente Raumzeit verbindet.[190]

Inzwischen gibt es diverse, mehr oder weniger ausformulierte und mehr oder weniger überzeugende Ideenansätze, die von einer emergenten Gravitation und/oder einer emergenten Raumzeit ausgehen. – Die grundsätzliche Frage, die im Rahmen solcher Ansätze zu klären ist, lautet: Wenn die Raumzeit eine abgeleitete, emergente Grösse ist, woraus leitet sie sich dann ab? Woraus besteht das Sub-strat und welcher Dynamik gehorcht es? Hierauf gibt es, wie die folgenden Beispiele zeigen sollten, die unterschiedlichsten Antworten:[191]

---

[189] Siehe weiter unten.

[190] Und hier wäre nun daran zu erinnern, dass es, wie wir schon gesehen haben, immerhin einige Argumente gibt, die in direkter oder indirekter Weise für eine emergente Raumzeit sprechen. Die *Bekenstein-Hawking-Entropie* schwarzer Löcher liefert zwar erst einmal nur ein Indiz für diskrete mikroskopische Freiheitsgrade, die sich nur unter spezifischen Bedingungen hin auf eine emergente Raumzeit deuten lassen. Es gibt hier immer noch die Möglichkeit, dass die Raumzeit fundamental ist, aber über eine diskrete Mikrostruktur verfügt. Ein solches Bild zeichnet etwa die *Loop Quantum Gravity*. Siehe Kap. 4.4. – Das holographische Prinzip jedoch lässt sich nicht nur als deutliches Indiz, sondern geradezu als strukturelle Vorstufe einer theoretischen Etablierung der Idee einer emergenten Raumzeit verstehen. In dieser Vorstufe wird eine vierdimensionale Raumzeit (mit Gravitation) durch eine Dynamik hervorgebracht, die, was die echten physikalischen Freiheitsgrade betrifft, auf einer dreidimensionalen Raumzeit vollständig definiert ist (und, wie sich zeigt, ohne Gravitation auskommt). Zur Verdeutlichung der (im vorausgehenden Satz nur angedeuteten) Einbeziehung der Problematik einer emergenten Gravitation in das sich mit dem holographischen Prinzip abzeichnende Szenario, vgl. insbesondere auch die AdS/CFT-Korrespondenz im Stringansatz. Siehe Kap. 4.2.

## *Raumzeit als thermodynamisches Phänomen*

Auf einen Vorschlag von Ted Jacobson[192] geht die Idee zurück, dass sich die Raumzeit durchaus als thermodynamische Konsequenz einer *grundsätzlich unbekannten* fundamentalen (mikroskopischen) Dynamik ansehen liesse. Das diesbezügliche Argument beruht auf der Umkehrung der Schlussweise, die von der allgemein-relativistischen Beschreibung schwarzer Löcher im Verbund mit einer semi-klassischen Behandlung von Quantenfeldern an ihrem Ereignishorizont zur Thermodynamik schwarzer Löcher und der *Bekenstein-Hawking-Entropie* führt.

> *"It is difficult to resist concluding [...] that the horizon entropy density proportional to area is a more primitive concept than the classical Einstein equation, which appears as a thermodynamic consequence of the interplay of entropy and causality."* (Jacobson / Parentani (2006) 337)[193]

Die Einsteinschen Feldgleichungen lassen sich nämlich, wie Jacobson zeigt, aus einer Verallgemeinerung der Proportionalität von Entropie und Horizontfläche bei schwarzen Löchern (*Bekenstein-Hawking-Entropie*) ableiten. Zu den Voraussetzungen für diese abgeleitete Ableitung zählen die thermodynamischen Beziehungen zwischen Wärme, Temperatur und Entropie. Die Temperatur wird dabei als Unruh-Temperatur eines beschleunigten Beobachters innerhalb eines lokalen Rindler-Horizontes interpretiert, Wärme als Energiefluss durch einen kausalen (Vergangenheits-) Horizont. Dieser Energiefluss tritt als Krümmung der Raumzeit und mithin als Gravitationsfeld in Erscheinung.

Die Allgemeine Relativitätstheorie wäre dann nichts anderes als eine effektive Theorie: ein thermodynamisches Näherungsresultat, das für seine Gültigkeit Gleichgewichtsbedingungen voraussetzt.[194]

---

[191] Alle hier im folgenden zu findenden Ansätze haben gemein, dass sie zwar versuchen, eine Antwort auf die Frage zu geben, welcher Typus von Substrat (bzw. welcher grundsätzliche Mechanismus) der Emergenz der Gravitation bzw. der Emergenz der Raumzeit zugrundeliegt, nicht aber auf die Frage nach der konkreten Substratdynamik und den konkreten relationalen Beziehungen zwischen den jeweils angenommenen Substratkonstituenten. In einigen der zu besprechenden Ansätzen lässt sich die letztere Frage aus prinzipiellen konzeptionellen Gründen nicht beantworten. Die Tatsache, dass es sich bei diesen Ansätzen eher um konzeptionelle Ideen und noch nicht um tatsächliche Theorieentwürfe zur Quantengravitation handelt, lässt ihre Besprechung im Kontext dieses Kapitels, in dem es um spekulative Ideen im Übergang zwischen den etablierten Theorien und einer zukünftigen Theorie der Quantengravitation geht, angemessen erscheinen – und nicht etwa in Kap. 4., das konkrete Theorieansätze zur Quantengravitation behandelt. Vgl. insbesondere auch Kap. 4.6., in dem einige der hier zu besprechenden Ideenansätze in neuem Gewand wieder auftauchen und dann vielleicht auch – nach weiteren Einsichten in das Spektrum der alternativen Möglichkeiten – besser beurteilt werden können.
Also, ohne dass dies immer eindeutig zu entscheiden wäre: Ansätze mit einer konkreten prägeometrischen Substratkonstruktion finden sich in Kap. 4.6., Ansätze zu einer emergenten Raumzeit und/oder Gravitation ohne eine solche konkrete Bestimmung des Substrats finden sich hier im Kap. 3.3.
[192] Siehe Jacobson (1995, 1999), Eling / Guedens / Jacobson (2006), Jacobson / Parentani (2006), Padmanabhan (2002a, 2004, 2007a).
[193] Vgl. die auf der gleichen Annahme beruhende Idee der *Holographischen Schirme*. Siehe weiter unten in diesem Teilkapitel.
[194] Eine Quantisierung der Allgemeinen Relativitätstheorie wäre, wie Jacobson deutlich macht, unter diesen Bedingungen kaum der richtige Weg zu einer physikalisch angemessenen Theorie der Quantengravitation:

> *"[...] one might expect that sufficiently high frequency or large amplitude disturbances of the gravitational field would no longer be described by the Einstein equation, not because some quantum operator nature of the metric would become relevant, but because the local equilibrium conditions would fail. It is my hope that [...] we shall eventually reach an understanding of the nature of 'non-equilibrium spacetime'."* (Jacobson (1995) 7)

Das energieflusserzeugende System hinter dem Horizont ist allerdings in Jacobsons Szenario grundsätzlich unbeobachtbar und bleibt daher in seinen Eigenschaften unbekannt. Der Weg zu einer empirisch direkt überprüfbaren Substrattheorie ist damit im Rahmen eines solchen Ansatzes versperrt. Dieser führt vielmehr zur Unmöglichkeit der Formulierung einer konkreten Theorie der Quantengravitation, die je über den Status reiner Spekulation hinausgelangen könnte. Unsere Erschliessungsmöglichkeiten beschränken sich unter diesen Voraussetzungen letztendlich auf effektive, phänomenologische Theorien.

### Gravitation als hydrodynamisches oder festkörperphysikalisches Phänomen

Die durchaus sehr unterschiedlichen Szenarien, welche die Gravitation als emergentes hydrodynamisches oder festkörperphysikalisches Phänomen zu rekonstruieren versuchen,[195] gehen zumindest zum Teil historisch auf Andrej Sakharovs Idee einer *Induzierten Gravitation* aus den sechziger Jahren zurück:[196] Gravitation wird von Sakharov als residualer Effekt der elektromagnetischen Wechselwirkung gedeutet, zu dem es infolge von Vakuumfluktuationen kommt. Die Gravitation könnte sich nach seiner Auffassung auf ähnliche Weise auf der Grundlage der Quantenelektrodynamik ergeben wie die Hydrodynamik auf der Grundlage der Molekülphysik. Die Einstein-Hilbert-Wirkung der Allgemeinen Relativitätstheorie wäre Teil der effektiven Wirkung einer Quantenfeldtheorie.

Dies setzt letztlich allerdings eine feste (Minkowski-)Hintergrundraumzeit voraus, die mit der Allgemeinen Relativitätstheorie konzeptionell unvereinbar ist, so dass sich diese per se nicht in vollem Umfang und mit allen ihren konzeptionellen Implikationen reproduzieren lässt. – Dies ist nicht nur in Sakharovs Ansatz ein zentrales Problem, sondern ebenso für viele der weiteren hydrodynamischen[197] und festkörperphysikalischen Szenarien. Manche dieser Szenarien liefern jedoch immerhin

---

> *"This perspective suggests that it may be no more appropriate to quantize the Einstein equation than it would be to quantize the wave equation for sound in air."* (Jacobson (1995) 2)

[195] Siehe insbesondere Hu (2005, 2009), Hu / Verdaguer (2003, 2004, 2008), Oriti (2006), Volovik (2000, 2001, 2003, 2006, 2007, 2008), Zhang (2002), Tahim et al. (2007), Padmanabhan (2004), Eling (2008), Jannes (2008, 2009), Liberati / Girelli / Sindoni (2009).

[196] Siehe Sakharov (2000), Visser (2002), Barcelo / Liberati / Visser (2005), Weinfurtner (2007).

[197] Ein expliziter neuerer Versuch, die Gravitation als emergentes hydrodynamisches Phänomen und die makroskopische Raumzeit als Kondensat (vergleichbar etwa mit einem Bose-Einstein-Kondensat) zu verstehen – als kollektiven Quantenzustand vieler Mikrokonstituenten mit einer makroskopischen Quantenkohärenz also –, geht auf B.L. Hu zurück. (Siehe Hu (2005, 2009) sowie Jannes (2008, 2009) und Liberati / Girelli / Sindoni (2009).)

> *"The metric or connection forms are hydrodynamic variables, and most macroscopic gravitational phenomena ca be explained as collective modes and their excitations (of the underlying micro-theory)."* (Hu (2009) 10)

Immerhin ergibt sich im Kontext eines solchen Modells nicht zuletzt auch ein natürlicher Erklärungsansatz für die kosmologische Konstante:

> *"One obvious phenomenon staring at our face is the vacuum energy of the spacetime condensate, because if spacetime is a quantum entity, vacuum energy density exists unabated for our present day late universe, whereas its origin is somewhat mysterious for a classical spacetime in the conventional view."* (Hu (2005) 4) nachvollziehbare Begründungen für ihre jeweiligen Einschränkungen hinsichtlich der konzeptionellen Voraussetzungen der Allgemeinen Relativitätstheorie. Diese Einschränkungen betreffen neben der Frage der Hintergrundunabhängigkeit insbesondere die der Geometrisierbarkeit der Gravitation.

*

Der avancierteste der festkörperphysikalischen Ansätze zur Erklärung des Zustandekommens der Gravitation geht auf Grigory E. Volovik zurück.[198] Als Ausgangspunkt dienen hier die im Rahmen der Festkörperphysik erforschten Charakteristika des Verhaltens von Fermi-Quantenflüssigkeiten. Diese zeigen unter spezifischen Voraussetzungen ein niederenergetisches Verhalten, das sowohl die grundsätzliche Phänomenologie des Standardmodells der Quantenfeldtheorien als auch die der Gravitation zu reproduzieren scheint. Dieses Verhalten kommt dadurch zustande, dass die Dynamik dieser Fermi-Quantenflüssigkeiten unter bestimmten Voraussetzungen niederenergetische fermionische Quasiteilchen hervorbringt, die sich als chirale Fermionen deuten lassen, ebenso wie kollektive bosonische Anregungszustände, die sich als Pendants zu den Wechselwirkungsbosonen sehen lassen.

> "The quasiparticles and collective bosons perceive the homogeneous ground state of condensed matter as an empty space – a vacuum – since they do not scatter on atoms comprising this vacuum state: quasiparticles move in a quantum liquid or in a crystal without friction just as particles move in empty space. The inhomogeneous deformations of this analog of the quantum vacuum is seen by the quasiparticles as the metric field of space in which they live. It is an analog of the gravitational field."
> (Volovik (2003) 3)

Die niederenergetische Phänomenologie dieser Fermi-Quantenflüssigkeiten zeichnet sich damit durch effektive Eichfelder und ein effektives Gravitationsfeld mit den entsprechenden Symmetrien (insbesondere der Lorentz-Invarianz) aus. Das effektive Gravitationsfeld lässt sich wiederum als Ausprägung einer effektiven Raumzeit deuten, die dann raumzeitliche Krümmungen, schwarze Löcher, Ereignishorizonte etc. aufweist. – Zu diesem Verhalten kommt es jedoch für Fermi-Quantenflüssigkeiten nur unter der Voraussetzung, dass diese einen topologischen Defekt im Impulsraum aufweisen: einen *Fermi-Punkt*.[199] [200]

---

Auch Hu macht deutlich, dass eine Quantisierung der Allgemeinen Relativitätstheorie kaum als adäquater Weg zu einer Theorie der Quantengravitation angesehen werden könnte, wenn die Grundidee des hydrodynamischen Szenarios zuträfe:

> "In our view general relativity is the hydrodynamical (the low energy, long wavelength) regime of a more fundamental microscopic theory of spacetime, and the metric and the connection forms are the collective variables derived from them. At shorter wavelength or higher energies, these collective variables will lose their meaning, much as phonon modes cease to exist at the atomic scale. This view marks a big divide on the meaning and practice of quantum gravity. In the traditional view, quantum gravity means quantizing general relativity, and in practice, most programs under this banner focus on quantizing the metric of the connection functions. Even though the stated goals of finding a microstructure of spacetime is the same, the real meaning and actual practice between these two views are fundamentally different. If we view [general relativity] as hydrodynamics and the metric or connection forms as hydrodynamic variables, quantizing them will only give us a theory for the quantized modes of collective excitations, such as phonons in a crystal, but not a theory of atoms or QED. [...] we find it more useful to find the micro-variables than to quantize macroscopic variables." (Hu (2005) 2)

[198] Siehe Volovik (2000, 2001, 2003, 2006, 2007, 2008) sowie Jannes (2008) und Hu (2009).

[199] Als Paradigma für eine Fermi-Quantenflüssigkeit dient Volovik erst einmal superfluides ³He-A. Die Allgemeine Relativitätstheorie wird für dieses Beispiel aber erst einmal nur näherungsweise reproduziert.

*"The universality classes are determined by the momentum space topology of fermion zero modes, in other words by the topological defects in momentum space [...]."* (Volovik (2003) 5)

Angesichts der grundsätzlichen festkörperphysikalischen Reproduzierbarkeit der Phänomenologie gravitativen Verhaltens schliesst Volovik dann, dass die Gravitation und – unter Berücksichtigung ihrer Geometrisierbarkeit – auch die (effektive) Raumzeit ohne weiteres emergente Phänomene sein könnten, die auf der Grundlage einer festkörperphysikalischen Mikrodynamik zustandekommen.[201] [202] Sollte es sich bei dem Substrat, auf dem diese emergenten Phänomene beruhen, etwa um eine Quantenflüssigkeit mit Fermi-Punkt handeln, zu deren kollektiven Anregungszuständen masselose Spin-2-Anregungszustände zählen, so würden sich etwa die folgenden Implikationen ergeben:

---

*"This metric theory is a caricature of the Einstein gravity, since the metric field does not obey the Einstein equations."* (Volovik (2003) 4)

Volovik vermutet, dass sich dies durch eine weitere Spezifizierung der festkörperphysikalischen Voraussetzungen beheben lässt.

*"The realization of a quantum liquid with the completely covariant effective theory at low energy requires some effort. We need such a 'perfect' quantum liquid, where in the low-energy corner the symmetries become 'exact' to a very high precision, as we observe today in our Universe. The natural question is: are there any guiding principles to obtain the perfect quantum vacuum?"* (Volovik (2003) 463) – *"From the model it follows that to bring the effective gravity closer to Einstein theory [...] one must suppress all the effects of the broken symmetry: the 'Planck matter' must be in the non-superfluid disordered phase without Goldstone bosons, but with Fermi points in momentum space."* (Volovik (2003) 8)

[200] Zu den Voraussetzungen zählt ebenso die Annahme, dass das fundamentale festkörperphysikalische System den Gesetzen der Quantenmechanik gehorcht.

*"We started with a system of many atoms which obeys the conventional quantum mechanics and is described by the many-body Schrödinger wave function. The creation and annihilation of atoms is strongly forbidden at low energy scale, and thus there is no quantum field theory for the original atoms. The quantum field theory emerges for excitations – fermionic and bosonic quasi-particles – which can be nucleated from the vacuum. In the systems with Fermi points, this emergent quantum field theory becomes relativistic."* (Volovik (2003) 467)

Die Gültigkeit der Quantenmechanik muss zumindest bis zu einer Energieebene angenommen werden, die signifikant oberhalb der Emergenzebene der Raumzeit und des quantenfeldtheoretischen Standardmodells liegt.

*"[...] if quantum mechanics is not fundamental, the scale at which it emerges should be far above the Planck scale.."* (Volovik (2008) 15)

[201] Auch hier ist wiederum klar: Sollte das Emergenzszenario zutreffen, so würde eine Quantisierung der Allgemeinen Relativitätstheorie zu keinen Einsichten in die fundamentale Dynamik, die ihrer Emergenz zugrundeliegt, führen:

*"Gravity cannot be quantized. [...] it is the low-energy macroscopic classical output of the high energy microscopic quantum vacuum."* (Volovik (2007) 6)

[202] Die einzige echte fundamentale Konstante wäre dann das Plancksche Wirkungsquantum; alle anderen vermeintlichen Konstanten, einschliesslich der Lichtgeschwindigkeit, wären abgeleitete Grössen, die nur für die emergente Dynamik relevant wären und darüber hinaus keine Gültigkeit besässen. Das fundamentale festkörperphysikalische System (das fundamentale 'Quanten-Vakuum') wäre mit seiner Dynamik in einem sehr hohen Energiebereich angesiedelt: der Planck-Ebene oder sogar höher, wie Volovik vermutet:

*"To obtain the observed high precision of physical laws, the Lorentz symmetry must persist well above the Planck energy. [...] the Einstein equations emerge in the limit $E_{Lorentz} \gg E_P$. [...] All this implies that physics continues far beyond the Planck scale, and this opens new possibilities for construction of microscopic theories."* (Volovik (2008) 9)

Volovik spekuliert, dass Symmetrien für diesen Energiebereich des Substrats vielleicht gar keine Rolle spielen, dass vielleicht alle Symmetrien Niederenergie-Symmetrien sind. Vgl. Nielsens *Random Dynamics*. (Siehe Kap. 1.3.) Der Rekurs auf eine Trans-Planckebenen-Physik flösst aber – nebenbei gesagt – nicht unbedingt grösstes Vertrauen in das Fermi-Punkt-Szenario ein.

*"If gravity emerges from the Fermi point scenario, then: / Gravity emerges together with matter. [...] / Fermionic matter which emerges together with gravity consists of Weyl fermions. [...] / In the Fermi point scenario space-time is naturally 4-dimensional. This is the property of the Fermi-point topology [...] / The underlying physics must contain discrete symmetries. [...] As a side effect, in the low-energy corner discrete symmetries are transformed into gauge symmetries and give rise to gauge fields. "*
(Volovik (2007) 6)

Aus diesen Implikationen lassen sich dann aber auch umgekehrt, bei ihrem tatsächlichen Vorliegen in der Natur, Indizien für das Fermi-Punkt-Szenario ableiten. Plausibler wird dieses allerdings, wenn sich darüberhinausgehend auch unabhängige Anzeichen für das Vorliegen eines Fermi-Punktes aufzeigen liessen:

*"The existence of the Fermi point in the vacuum of our Universe is an experimental fact, since all our elementary particles, quarks and leptons, are chiral Weyl fermions. It is still not excluded that some of neutrinos are Majorana fermions, but this only changes the topological characteristic of the Fermi point."* (Volovik (2007) 5)

Wenngleich die Reproduktion einer notwendigerweise vierdimensionalen Raumzeit im Fermi-Punkt-Szenario als zwar interessant, aber letztlich immer noch vergleichsweise unspektakulär bezeichnet werden kann, so bietet die Reproduktion eines flachen Universums schon ein deutlich klareres, differentielles Indiz für dieses Szenario:

*"The universe is naturally flat. [...] In general relativity the flatness of the Universe requires either fine tuning or inflationary scenario [...] The observed flatness of our Universe is in favor of emergent gravity."* (Volovik (2007) 7)

Zudem liefert das festkörperphysikalische Szenario – im Gegensatz zu nahezu allen anderen Theorieansätzen – eine natürliche Grundlage für die Annahme einer kleinen Kosmologischen Konstante.[203]

*"Cosmological constant is naturally small or zero. In general relativity, the cosmological constant is arbitrary constant, and thus its smallness requires fine-tuning. Thus observations are in favor of emergent gravity."* (Volovik (2007) 7)

Ebenso ergeben sich im Rahmen des festkörperphysikalischen Szenarios Erklärungsansätze für die Proportionalität der Entropie, die einem Raumvolumen zugeschrieben werden kann, zu seiner begrenzenden Oberfläche, entsprechend der *Bekenstein-Hawking-Entropie*.

*"If the spacetime has microscopic degrees of freedom, then any bulk region will have an entropy and it has always been a surprise why the entropy scales as the area rather than volume. Our analysis shows that, the semiclassical limit, when Einstein's equations hold to the lowest order,* the entropy is contributed only by the boundary term and the system is holographic.*"* (Padmanabhan (2004) 4)

Insbesondere scheint aber das Fermi-Punkt-Szenario grundsätzlich in der Lage zu sein, die Phänomenologie der Allgemeinen Relativitätstheorie, und damit diese selbst als effektive Theorie, zu re-produzieren (ebenso wie die Phänomenologie dynamischer Strukturen, die denen des Standardmodells der Quantenfeldtheorien ähneln).

---

[203] Zu den Erklärungsansätzen für die kosmologische Konstante auf der Grundlage einer festkörperphysikalischen Sichtweise der Raumzeit siehe auch Padmanabhan (2004, 2007a, 2008).

Dennoch bleibt im Fermi-Punkt-Szenario – wie auch im Rahmen der weiteren hydrodynamischen und festkörperphysikalischen Ansätze – erst einmal weitgehend unklar, in welcher Weise immer noch ein raumzeitlicher Hintergrund (oder zumindest eine Zeit, die als externer Parameter mitläuft) vorausgesetzt wird. Volovik thematisiert diese Problematik immerhin noch in der Weise, dass er deutlich macht, dass etwa die Hintergrundunabhängigkeit (und mithin die Diffeomorphismusinvarianz bzw. die allgemeine Kovarianz) der Allgemeinen Relativitätstheorie in letzter Instanz vielleicht gerade ein theoretisches Artefakt sein könnte.

> *"The effective gravity may essentially differ from the fundamental gravity even in principle. Since in the effective gravity the general covariance is lost at high energy, the metrics which for the low-energy observers look as equivalent, since they can be transformed to each other by coordinate transformations, are not equivalent physically. As a result, in emergent gravity some metrics, which are natural in general relativity, are simply forbidden. [...] Some coordinate transformations in GR are not allowed in emergent gravity; [...] The non-equivalence of different metrics is especially important in the presence of the event horizon."* (Volovik (2007) 6)

Eine mögliche Hintergrundabhängigkeit ist für einen Ansatz, der von einer emergenten Gravitation ausgeht, zwar sicherlich nicht wünschenswert, aber immer noch eher hinnehmbar als für einen Ansatz, der von einer direkten Quantisierung der Allgemeinen Relativitätstheorie – einer hintergrundunabhängigen Theorie – ausgeht. Dies gilt allerdings nur dann, wenn ein auf der Emergenzannahme hinsichtlich der Gravitation beruhender Ansatz erklären kann, wie es zum Anschein einer umfassenden Gültigkeit der grundsätzlichen konzeptionellen Implikationen der Allgemeinen Relativitätstheorie – insbesondere ihrer Hintergrundunabhängigkeit und der Annahme einer vollständigen Geometrisierbarkeit der Gravitation – kommen konnte, inwiefern diese konzeptionellen Elemente in letzter Instanz als theoretische Artefakte anzusehen sind, unter welchen Bedingungen ihnen zumindest annähernde Gültigkeit zugesprochen werden kann, und mithin, worauf der Erfolg der Allgemeinen Relativitätstheorie, die in grundlegender Weise mit diesen Artefakten operiert, beruht . Nur dann ist es konzeptionell möglich, von einer Emergenz der Gravitation unter Voraussetzung eines raumzeitlichen Hintergrundes auszugehen. Insbesondere sind in diesem Rahmen dann auch Szenarien denkbar, die in subtilerer Weise gleichzeitig an der Geometrisierbarkeit der Gravitation festhalten und dennoch mit einem festen raumzeitlichen Hintergrund arbeiten, der dann jedoch nicht mit der phänomenologischen, effektiven, emergenten Raumzeit identisch ist; und es wäre diese emergente, phänomenologische Raumzeit, der die geometrisierbaren Eigenschaften der Gravitation eingeschrieben wären – und dies, ohne dass eine absolute Hintergrundunabhängigkeit der entsprechenden Theorie vorausgesetzt werden müsste.

Ziel bleibt bei alledem aber immer die Identifizierung des Substrats: die Beantwortung der Frage nach den mikroskopischen Konstituenten und ihrer Dynamik, auf deren Grundlage Gravitation und/oder Raumzeit zustandekommen.

> *"Our ultimate goal is to reveal the still unknown structure of the ether (the quantum vacuum) using our experience with quantum liquids."* (Volovik (2003) 462) – *"A first step towards the elusive theory of 'quantum gravity' would be to identify the microscopic constituents ('atoms') of space."* (Volovik (2008) 9)

Und genau hierin liegt das grundsätzliche – und letztendlich in seiner Konzeption begründete – Problem der festkörperphysikalischen Ansätze. Aus der niederenergetischen Phänomenologie lässt sich im Rahmen eines solchen Ansatzes nicht (oder zumindest nicht so ohne weiteres) auf das Substrat zurückschliessen.

> *"Compared to working out the macroscopic limits of a known microscopic theory, going the reverse way is always difficult because of the missing information."* (Hu (2009) 5)

Und dieses allgemeinere Problem hat hier noch einen wesentlich spezielleren, zusätzlichen Aspekt: Denn ganz verschiedene festkörperphysikalische Strukturen und Dynamiken führen zu den gleichen niederenergetischen Phänomenologien. Die effektive niederenergetische Phänomenologie eines festkörperphysikalischen Systems hängt in vieler Hinsicht nicht von der jeweils zugrundeliegenden mikroskopischen Dynamik ab.[204] Es gibt vielmehr Universalitätsklassen von gänzlich verschiedenen mikroskopischen, festkörperphysikalischen Systemen mit den gleichen niederenergetischen Implikationen.

> *"An effective theory in condensed matter does not depend on details of microscopic (atomic) structure of the substance. The type of effective theory is determined by the symmetry and topology of zero modes of the quantum vacuum, and the role of the underlying microscopic physics is only to choose among different universality classes of quantum vacua on the basis of the minimum energy consideration."* (Volovik (2003) 5)

Hinzu kommen dann auch noch grundsätzliche Unsicherheiten bei der Ableitung des niederenergetischen Verhaltens aus einer entsprechenden Substratdynamik.

> *"As is known from many examples of condensed matter physics, it is often a nontrivial task to deduce mesoscopic behavior from micro-dynamics: one usually encounters nonlinear or even nonlocal interactions in strongly correlated systems, and needs some ingenuity to identify collective variables at succesive levels of structure to make the analysis simpler."* (Hu (2009) 4)

Das eigentliche Ziel einer Theorie der Quantengravitation, wenn es eben nicht mehr um die Quantisierung einer klassischen Theorie geht, weil diese nichts mehr als eine effektive Theorie ist, bleibt also ungelöst. Die konkrete Identifizierung der Substratdynamik ist offenbar mit den Mitteln des festkörperphysikalischen Ansatzes grundsätzlich nicht einlösbar.

### Raumzeit als Ausdruck eines Ordnungsparameter- bzw. Zustandsspektrums

Könnte es sich bei der Raumzeit bzw. bei der Möglichkeit einer raumzeitlichen Lokalisierung um das phänomenologisch wirksame Ergebnis einer Auffächerung von quantenmechanischen Zuständen eines nicht-raumzeitlichen Quantensubstrats handeln: eine Auffächerung, die vielleicht durch spezifische Zustandsbesetzungs- oder Ausschlussregeln, etwa eine verallgemeinerte Form des Paulischen Ausschlussprinzips, erzwungen wird?

---

[204] Ganz besonders gilt dies für die thermodynamische Beschreibungsebene:
*"The most general physical laws which do not depend on the details of the underlying microscopic system are the laws of thermodynamics."* (Volovik (2008) 10)

Dann wären vermeintlich raumzeitliche Freiheitsgrade nichts anderes als 'interne' Freiheitsgrade, die gerade zu einer Phänomenologie führen, die wir als raumzeitlich interpretieren. Die Unterscheidung in raumzeitliche und interne Freiheitsgrade wäre nichts anderes als ein phänomenologisches Artefakt. Die Raumzeit wäre eine ausschliesslich phänomenologische Implikation eines originär prägeometrischen quantenmechanischen Zustandsspektrums. Prägeometrisch wäre dieses Zustandsspektrum in der Hinsicht, dass bestimmte Freiheitsgrade überhaupt erst auf der Grundlage ihrer Phänomenologie post hoc als raumzeitlich interpretiert werden. Grundsätzlich würden sie sich in nichts von den internen Freiheitsgraden unterscheiden. Letztlich gäbe es nichts anderes als interne Freiheitsgrade.

Eine Vorstufe für diese Verwischung der Unterscheidung in raumzeitliche und interne Freiheitsgrade wird schon im Holographischen Prinzip deutlich. Diese Vorstufe würde nun durch die umfassendere Einsicht ersetzt, dass alle Freiheitsgrade interne Freiheitsgrade sind. Die Raumzeit wäre nichts anderes als eine phänomenologisch wirksame Illusion.

Die vielleicht entscheidende Frage, die man bezüglich einer solchen Raumzeit dann zu stellen hätte, wäre, ob ihr ein kontinuierliches oder ein diskretes Zustandsspektrum zugrundeliegt. Die Thermodynamik schwarzer Löcher hätte diese Frage inzwischen wohl schon zugunsten der letzteren Option beantwortet.

*

Ein konkretes Szenario, welches in die Richtung einer solchen Auffassung der Raumzeit als phänomenologischem, letztlich illusionärem Ergebnis einer originär prägeometrischen quantenmechanischen Dynamik zielt, die nur interne Freiheitsgrade kennt, geht auf Vadim Kaplunovsky und Marvin Weinstein zurück, bei denen – es ist die Zeit vor der heissen Phase der Diskussion um die Quantengravitation – von der Gravitation an keiner Stelle in irgendeiner Weise die Rede ist:[205]

> "[...] the space-time continuum as an illusion of low-energy dynamics." (Kaplunovsky / Weinstein (1985) 1879)

Die Raumzeit – genauer: der Raum; denn die Zeit wird im Sinne der Quantenmechanik als Hintergrundparameter behandelt – stellt sich in diesem Szenario als das Ergebnis der Herausbildung von Ordnungsparametern dar, die von einem prägeometrischen quantenmechanischen System hervorgebracht werden, aber nur für den Niederenergiebereich Relevanz besitzen. Die Dimensionalität und die Topologie der Raumzeit werden dynamisch erzeugt.[206] Insbesondere gibt es Phasenübergänge zwischen Raumzeiten unterschiedlicher Dimensionalität:

> "[...] dimension can be thought of as an integer-valued order parameter which characterizes distinct phases of a single dynamical system." (Kaplunovsky / Weinstein (1985) 1895) – "The existence of these phases implies the possibility that finite-temperature effects can cause dimension-changing phase transitions." (Kaplunovsky / Weinstein (1985) 1893)

---

[205] Siehe Kaplunovsky / Weinstein (1985); vgl. auch Dreyer (2004, 2007) sowie Kober (2009).

[206] Fermionische Freiheitsgrade führen zu einer flachen, bosonische zu einer aufgerollten Raumzeit.
"[...] if the system was dominated by bosonic rather than fermionic fields then space-time would curl up instead of flattening." (Kaplunovsky / Weinstein (1985) 1896)

Und es gibt diesem Szenario zufolge nach Etablierung der Raumzeit eventuell noch übrigbleibende Eichfreiheitsgrade, die sich letztendlich wieder als Indiz für eine nur geringe Relevanz der Unterscheidung in raumzeitliche und interne Freiheitsgrade lesen lassen:

> *"[...] residual interactions among the low-energy degrees of freedom which have the structure of a gauge theory."* (Kaplunovsky / Weinstein (1985) 1896) – *"[...] one appears to trade off dimensions against SU(n) flavor symmetries in a manner reminiscent of Kaluza-Klein theories [...]."* (Kaplunovsky / Weinstein (1985) 1893)

Das Kaplunovsky-Weinstein-Szenario setzt jedoch, wie schon angedeutet, einen externen Zeitparameter voraus:

> *"There seem to be quantum systems which start out with a well-defined notion of time but no notion of space, and dynamically undergo a transition to a space-time phase [...]."* (Kaplunovsky / Weinstein (1985) 1879)

Aber diese Einschränkung – dass also nur der Raum durch die Herausbildung von Ordnungsparametern hervorgebracht wird – liesse sich, wie Florian Girelli, Stefano Liberati und Lorenzo Sindoni gezeigt haben,[207] grundsätzlich überwinden: Es ist nämlich durchaus möglich, eine zeitliche Entwicklung als emergentes Phänomen aus einer zeitlosen 'Dynamik' zu gewinnen.

<div align="center">*</div>

Gewisse Ähnlichkeiten, aber auch gravierende Unterschiede zum Kaplunovsky-Weinstein-Szenario weist der Ansatz von Manfred Requardt auf.[208] Die Gemeinsamkeit besteht vor allem darin, dass auch hier die Emergenz einer raumzeitlichen Mannigfaltigkeit als das Ergebnis der Herausbildung von Ordnungsparametern gesehen wird, diesmal auf der Grundlage zellulärer Netzwerke.[209]

> *"[...] emergence and reconstruction of a protoform of continuum space-time as, what we call, an order parameter manifold [...]."* (Requardt (2000) 1)

Der gravierendste Unterschied zum Kaplunovsky-Weinstein-Szenario besteht darin, dass Requardt für diese zellulären Netze noch nicht notwendigerweise von der Gültigkeit der Quantenmechanik ausgeht, dass vielmehr Quantenmechanik und Raumzeit gleichermassen als emergente Phänomene behandelt werden.[210] Requardt geht erst einmal von einem nichtlokalen stochastischen Substrat mit möglicherweise deterministischer Grunddynamik aus:

---

[207] Siehe Girelli / Liberati / Sindoni (2008, 2009) sowie Liberati / Girelli / Sindoni (2009). Eine solche 'Dynamik' hätte noch dazu empirisch überprüfbare Konsequenzen.
> *"[...] the invariance under Lorentz transformations is only an approximate property of the field equations [...].*
> *[...] our theory will show aether effects beyond second order."* (Girelli / Liberati / Sindoni (2008) 4)

[208] Siehe Requardt (1996, 1996a, 2000).

[209] Der Begriff des Ordnungsparameters wird dabei im Sinne der Synergetik verstanden; er impliziert insbesondere die 'Versklavung' mikroskopischer Freiheitsgrade durch die Dynamik der makroskopischen Ordnungsparameter.

[210] Es gibt einige weitere Ansätze, die von einer Emergenz der Quantenmechanik auf der Grundlage eines nicht-quantenmechanischen computationalen Substrats ausgehen. Siehe insbesondere 't Hooft (1999, 2000a, 2001, 2001a, 2007), Cahill (2002, 2005), Cahill / Klinger (1996, 1997, 1998, 2005). 't Hooft hält an einer notwendigerweise deterministischen Basis fest.

*"Beneath the surface there exists a more irregular and wildly fluctuating 'underworld' of a distinctly discrete and stochastic nature ([...]; at the very bottom the underlying dynamics may well be deterministic!). Its perhaps most characteristic feature is a peculiar* non-local *dynamical behavior [...]."* (Requardt (2000) 1)

Der Determinismus des Substrats gehört jedoch nicht notwendigerweise zu den konstitutiven Elementen des Ansatzes, da er für die emergenten Ebenen ohnehin wirkungslos bliebe:

*"Our networks are, among other things, complex dynamical systems. On the most fundamental level their dynamics is assumed to be* deterministic *(whereas this is not necessarily a crucial point due to the* shielding phenomenon *which decouples the various levels from each other to some extent). [...] on the higher levels the dynamics is no longer deterministic due to the 'integrating out' of degrees of freedom and corresponding loss of information."* (Requardt (2000) 26f)

Entscheidend ist vielmehr die Nichtlokalität des Substrats, die sich nach Requardts Auffassung in den nichtlokalen Eigenschaften der Quantenmechanik, insbesondere in ihrer Nichtseparierbarkeit, in Form eines emergenten Residuums niederschlägt:

*"We will argue in the following that this almost hidden non-local web of exchange of information, which arises quite naturally in our approach, plays a decisive role in the formation of quantum theory as an effective continuum theory incorporating certain non-local gross features of the depth structure of space-time [...]."* (Requardt (2000) 2)

Requardt geht von keiner regulären Struktur für das Substrat aus,[211] da sich eine solche reguläre Struktur nur schwerlich ohne eine wiederum tiefer liegende strukturelle und dynamische Ursache motivieren lässt; vielmehr geht er von einer völlig irregulären, dynamischen Substratstruktur aus, die alles dies nicht voraussetzt:

*"[...] we do not base our analysis on a rigid a priori fixed lattice structure but, instead of that, regard the geometric wiring diagram underlying the cellular network as a full fledged dynamical system of its own which interacts with the node states [...]. Thus, geometry, metrical properties, near- and far-orders all become dynamical* collective quantities *which are assumed to* coevolve. *The underlying philosophy is of course that ultimately both quantum theory* and *gravity emerge as two seemingly different but in fact related large scale aspects of one and the same underlying theory. "* (Requardt (2000) 8f)

Die gemeinsame Emergenz von Raumzeit und Quantenmechanik führen nach Requardts Auffassung insbesondere zu einer Kopplung von Geometro- und Materiegenese, wie sie auch für weiter fortgeschrittene prägeometrische Ansätze zu einer Theorie der Quantengravitation, wie etwa die *Quantum Causal Histories*, vorliegt.[212] Die entscheidende Randbedingung, die bei einer solchen gemeinsamen Geometro- und Materiegenese zu berücksichtigen ist, besteht dann nicht zuletzt darin, die wesentlichen Implikationen der Allgemeinen Relativitätstheorie zu reproduzieren; dies betrifft hier insbesondere die Wechselwirkung zwischen Raumzeit und Materie, wie sie von der Allgemeinen Relativitätstheorie beschrieben wird, und mithin die Geometrisierbarkeit der Gravitation.

---

[211] 't Hooft (1999, 2000a, 2001, 2001a, 2007) vermutet im Gegensatz dazu für das Substrat eine reguläre Struktur: etwa einen deterministisch arbeitenden zellulären Automaten.
[212] Siehe Kap. 4.6.

Hinsichtlich der konkreten Substratdynamik ist der Requardtsche Ansatz allerdings ebenso unentschieden wie die im Vorausgehenden schon beschriebenen Ansätze, nur dass er nicht durch eine prinzipielle konzeptionelle Unmöglichkeit an der Erschliessung des Substrats gehindert wird – und die Unentschiedenheit zum gegenwärtigen Stand der Entwicklung einer unnötigen und vermutlich falschen, vorauseilenden Festlegung vorzieht.

> *"[...] it is not reasonable (in the beginning) to concentrate too much on certain (possibly wrong or unimportant) microscopic details but better have the gross features right. That is, the real task may rather consist of extracting the possibly few crucial characteristics which the primordial theory must contain and which, in the end, survive the coarse graining limit."* (Requardt (2000) 9f)

Immerhin handelt es sich um keine konkrete Theorie zur Quantengravitation, sondern vielmehr um einen konzeptionellen Denkansatz, dessen grundsätzliche Möglichkeiten erst einmal ausgelotet werden müssen, bevor man sich hinsichtlich eines spezifischen Substrats festlegt – wenn dies dann überhaupt noch erforderlich sein sollte.

<p style="text-align:center">*</p>

Ein weiterer Denkansatz dieser Art stammt von David R. Finkelstein.[213] Dieser geht von einer basalen Prozessontologie als prägeometrischem Substrat aus, mit der Absicht, die Modalitäten einer Emergenz der Raumzeit zu rekonstruieren und dabei gleichzeitig eine nomologische Vereinheitlichung aller Wechselwirkungen zu erreichen. Wiederum ist die Kopplung von Geometro- und Materiegenese konstitutiv. Der Ansatz soll sowohl das von der Allgemeinen Relativitätstheorie gelieferte Bild der Raumzeit und der geometrisierten Gravitation reproduzieren, als auch das quantenfeldtheoretische Standardmodell der Elementarteilchenphysik. – Dabei setzt Finkelstein, im Gegensatz zu Requardt, die Quantenmechanik in ihrer Gültigkeit für das Substrat voraus. Für die emergente Raumzeit geht er von einer diskreten Struktur mit finiter Informationsdichte aus – entsprechend den schon erörterten Argumenten, wie sie sich im Kontext der Thermodynamik schwarzer Löcher abzeichnen:[214]

> *"Let us suppose that physical spacetime is a locally finite assemblage of discrete finite quantum elements."* (Finkelstein (1996) 478)

Das Besondere an Finkelsteins Ansatz ist, dass er in seiner konkreten Ausgestaltung eine Kombination der Grundgedanken nahezu aller auf den vorausgehenden Seiten schon diskutierten, als auch weiterer, noch zu erörternder Denkansätze darstellt. Der Ansatz ist in gewisser Weise eine Synthese (i) der Auffassung, dass die Raumzeit Ausdruck des Zustandsspektrums eines prägeometrischen quantenmechanischen Systems ist, (ii) der festkörperphysikalischen Modelle des Zustandekommens einer emergenten Raumzeit und schliesslich (iii) der noch zu diskutierenden Ansätze, welche die Raumzeit als Ergebnis eines computationalen Szenarios[215] zu rekonstruieren versuchen. In letzterer

---

[213] Siehe Finkelstein (1996, 2004), Finkelstein / Gibbs (1993).
[214] Siehe Kap. 3.1.
[215] Siehe weiter unten in diesem Teilkapitel.

Hinsicht finden sich vor allem Anklänge an das *Causal-Set*-Szenario.[216] – Die festkörperphysikali­schen Aspekte des Ansatzes, in die insbesondere auch die Argumente für eine diskrete, finite Basis­struktur einfliessen, werden etwa im Folgenden deutlich:

> *"When we model spacetime as a classical continuum, we encounter some of the same problems as when we treat (say) a diamond crystal as a continuum: We leave out important internal symmetries, bring in many arbitrary constants, and, if we take the theory seriously, run into mathematical non­sense and infinite physical results. Let us suppose that the spacetime continuum too is a smoothing approximation, omitting quantum spacetime structure, and seek a finer theory, consolidating the im­portant gains of both Einstein's and Dirac's relativities."* (Finkelstein (1996) 475)

Naheliegend (und von ausreichender Allgemeinheit) erscheint Finkelstein dann die Auffassung, dass die Raumzeit so etwas wie ein makroskopisches Quantenkondensat sein könnte, zu dem es auf der Grundlage eines irregulären Substrates kommt:

> *"In theories of spacetime microstructure, the POINCARÉ-invariant vacuum is supposed to be a macroscopic quantum condensation of a discrete structure that generally has no global symmetry. The POINCARÉ-invariance of the vacuum net is to be another instance of quantum self-organization, like the Heisenberg ferromagnet, Nambu's theories of the internal symmetries of the nucleus and the vacuum, the Landau theory of superfluid helium, and the Bardeen-Cooper-Schrieffer theory of super­conductivity."* (Finkelstein (1996) 482)

Insbesondere ein Superkondensat, wie es den Theorien der Superleitfähigkeit und der Suprafluidität zugrundeliegt, erscheint ihm als Modell der diskreten Struktur einer emergenten, makroskopischen Raumzeit geeignet, in die gleichermassen die (geometrisierte) Gravitation wie die Materie und die nicht-gravitativen Wechselwirkungen eingeschrieben sind:[217]

> *"[...] we count on macroscopic quantum self-organization, like that of a superconductor or superfluid, to propagate this Lorentz invariance from the local elements of spacetime to the global vacuum, which we therefore call a* supercrystal. *According to Newton's first law of mechanics, particles in the vacuum have infinite mobility [...]. Presumably this is a supermobility, and the physical elementary particles are manifestations of the vacuum Meissner effect, which concentrates net defects into stringlike tubes of disordered nets. If so, then the supercrystal was the first supercondensate to be discovered, followed some centuries later by the superfluid and the superconductor / Our first task, then, will be to derive the homogeneity of spacetime (translational symmetry), its isotropy (Lorentz invariance), and its supermobility from a macroscopic quantum condensation of quasibosons (pairs of fermions) with off-diagonal long-range order."* (Finkelstein (1996) 478)

---

[216] Siehe Kap. 4.6.

[217] Die methodologische Konsequenz, die sich für die möglichen Strategien zur Entwicklung einer Theorie der Quan­tengravitation ergibt, ist bei grundsätzlicher Adäquatheit des Szenarios, wie Finkelstein herausstreicht, die gleiche wie für alle anderen Emergenzszenarien:

> *"We should not apply canonical quantization to spacetime structure if spacetime is a supercondensate. Canoni­cal quantization is a reasonable way to reconstruct a quantum theory from its classical behavior at high quan­tum numbers, but it will not recover a quantum theory from the behavior of a low-temperature supercondensate. / For example, one could not discover the helium atom by canonically quantizing the macroscopic two-fluid field theory of superfluid helium. Nor could one discover the electron theory of solids by canonically quantizing the field theory of the Josephson potential of a superconductor."* (Finkelstein (1996) 482)

Gleichzeitig gibt es Anklänge an die Ansätze, die versuchen, die Raumzeit als das emergente, phä-nomenologische Ergebnis eines originär computationalen Prozesses zu rekonstruieren,[218] hier nun allerdings erweitert um die Aspekte, die eine Genese der nicht-gravitativen makroskopischen Frei-heitsgrade einschliessen. So heisst es etwa im Abstract von Finkelsteins "*CoSmputation*":

> *"We model the cosmos as a computation using a high-order Clifford algebra as quantum logic. The split between space-time and field variables arises from a condensation at the second level of analysis (bits of bits)."* (Finkelstein (o.J.))

In der konkreteren Ausformulierung dieser Idee einer Emergenz der Raumzeit, die eine Emergenz aller weiteren, nicht-raumzeitlichen, nicht-gravitativen Freiheitsgrade einschliesst und auf diesem Wege eine nomologische Vereinigung aller Wechselwirkungen ermöglichen soll, wird dann die Parallele zum *Causal-Set*-Szenario deutlich.[219]

> *"We propose a unified description of the known forces. We formulate a quantum relativistic spacetime as a (directed) graph of causal arrows with indefinite Hilbert metric, whose physical meaning is given. The simplest graph whose quantum relativity supports conservation of energy-momentum also supports a semidirect product of the cyclic groups 2 and 3 and the four-group $2^2$. We call these lattice degrees of freedom (permutational) twain, trine, and spin. Quantized $2^2$ becomes Lorentz Spin(4). Gauged, the energy-momentum and spin groups lead to gravity and torsion, and twain and trine lead to $SU_2$ and $SU_3$. We infer that color is actually trine, and the z component of isospin is twain."* (Finkel-stein / Gibbs (1993) 1801/Abstract)

Prägnanter formuliert:

> *"To put it simply, we conclude that spacetime is like a quantum relativistic Rubik cubic lattice with an indefinite Hilbert metric. [...] Gravity comes from lattice translations and torsion from spin."* (Finkel-stein / Gibbs (1993) 1813)

Finkelsteins Ansatz, in dem sich immerhin auf formaler Ebene eine nomologische Vereinigung aller Wechselwirkungen abzeichnet, ist jedoch vom Status einer vollständigen Theorie noch weit ent-fernt, insbesondere da die konkrete Substratdynamik bisher noch völlig im Unklaren bleibt:

> *"Since we do not have the dynamical action principle yet, there is not yet a unified theory of the known forces, but at least it is a unified description of them."* (Finkelstein / Gibbs (1993) 1812)

Es gibt zwar einen weitergehenden Vorschlag Finkelsteins[220] für ein dynamisches Wirkungsprinzip im Kontext eines festkörperphysikalischen Szenarios, das auf einem vierdimensionalen 'Hyperdia-manten' beruht. Einerseits sind aber die Konsequenzen und damit die Adäquatheit des Vorschlags noch weitgehend im Unklaren. Andererseits besteht das einzige Argument für diese Spezifizierung darin, dass nach Finkelsteins Auffassung der vierdimensionale 'Hyperdiamant' das einfachste Szenario liefert, das zu den phänomenologisch richtigen räumlichen Symmetrien sowie zu denen des teilchenphysikalischen Standardmodells führt. Grundsätzlich kämen, wenn man sich auf die Hypothese einliesse, dass diese festkörperphysikalischen Modelle eine angemessene Basis für die Emergenz der Raumzeit zuzüglich aller nicht-raumzeitlichen Freiheitsgrade liefern, neben dem 'Hyperdiamanten' eine grosse Zahl komplexerer vierdimensionaler 'Hyperkristalle' in Frage, deren entsprechende Konsequenzen noch nicht absehbar sind.

---

[218] Siehe weiter unten in diesem Teilkapitel sowie Lloyd (1999, 2005, 2005a, 2007), Hsu (2007), Livine / Terno (2007), Zizzi (2001, 2004, 2005), Hardy (2007). Diese Ansätze setzen allesamt die Quantenmechanik voraus. Das Substrat lässt sich dann als so etwas wie einen Quantencomputer deuten. Voraussetzung bleibt dabei erst einmal eine zumindest lokal wirksame, vielleicht relationale Zeit. Die dabei möglicherweise auftretenden Probleme und insbesondere ihre mögliche Lösung werden in Kap. 4.6. zu diskutieren sein.

[219] Siehe Kap. 4.6.

[220] Siehe Finkelstein (1996).

## Raumzeit als emergentes Ergebnis eines computationalen Prozesses

Schon in Wheelers Agenda[221] für eine zukünftige Theorie, mittels derer die Unvereinbarkeit von Quantenmechanik und Allgemeiner Relativitätstheorie überwunden werden soll, findet sich die Empfehlung:

> *"Translate the quantum versions of string theory and of Einstein's geometrodynamics from the language of continuum to the language of bits."* (Wheeler (1989) 362)

Dies ist einer der Kerngedanken von Wheelers *It-from-bit*-Konzeption.[222] Dieser zufolge geht die Raumzeit aus einem Substrat hervor, das von nicht-raumzeitlichen, nicht kausal gebundenen oder in irgendeiner Weise durch Gesetze bestimmten[223] Elementarereignissen konstituiert wird, die sich nach Wheelers Auffassung zahlen- bzw. informationstheoretisch, etwa auf der Grundlage von Borel-Mengen, erfassen lassen sollten.[224] Auf der Grundlage eines solchen stochastischen Substrats ergeben sich nach Wheelers Auffassung, allein infolge des 'Gesetzes der grossen Zahlen', (spätestens) auf der makroskopischen Ebene strukturelle Formen, Regularitäten, und Invarianzen – und damit nicht zuletzt auch die phänomenologische quasi-kontinuierliche Raumzeit der Allgemeinen Relativitätstheorie.

*

Eine schliesslich vor allem um einige der spezifischen Festlegungen hinsichtlich des Substrats – insbesondere seine 'Gesetzlosigkeit' bzw. seine Stochastizität – gereinigte Neuauflage von Wheelers *It-from-bit*-Konzeption findet sich inzwischen in Seth Lloyds *Computational-Universe*-Ansatz.[225] Hier wird aus Wheelers *It from bit* nun dezidiert ein *It from Qubit*.[226] Zentrales Ziel des Ansatzes ist es, die Emergenz der Raumzeit (einschliesslich aller makroskopisch relevanten nicht-raumzeitlichen, nicht-gravitativen Freiheitsgrade) auf der Grundlage einer vollständig hintergrundunabhängigen Quantencomputation verständlich zu machen: einem Quantencomputer. Dieser Quantencomputer wird als azyklische, gerichtete Vernetzungsstruktur modelliert, innerhalb derer Quanteninformationen fliessen:

> *"Each quantum computation corresponds to a directed, acyclic graph G, the 'wiring diagram' for the computation. The initial vertices of the graph correspond to input states. The directed edges of the graph correspond to quantum wires that move quantum information from place to place. The internal vertices of the computational graph represent quantum logic gates that describe interactions between qubits. The final vertices of the graph correspond to output states. Infinite computations correspond to graphs that need not have final states."* (Lloyd (2005a) 4)

---

[221] Siehe Wheeler (1989).

[222] Siehe Wheeler (1979, 1983, 1989).

[223] Dies ist die Grundannahme von Wheelers *Law-without-Law*-Konzeption. Siehe Kap. 1.3. sowie Stuckey (2001).

[224] Für dieses Substrat gelten nach Wheeler elementare Quantenprinzipien, insbesondere eine Form von quantenmechanischer Selbstreferenz. Wheeler bezeichnet diese quantenmechanische Selbstreferenz als *'Observer Participancy'*. Sie ist eine Implikation einer spezifischen Variante der epistemischen Deutung der quantenmechanischen Zustandsreduktion, und als solche weder unproblematisch, noch unumstritten.

[225] Siehe Lloyd (2005, 2005a). Vgl. auch Lloyd (1999, 2007), Cahill (2005), Cahill / Klinger (1997, 2005), Hsu (2007), Livine / Terno (2007), Zizzi (2001, 2004, 2005), Hardy (2007).

[226] Der Ansatz ist zudem einerseits entfernt verwandt mit dem *Causal-Set*-Ansatz (siehe Kap. 4.6.), andererseits mit den Versuchen einer hydrodynamischen Generierung einer emergenten Raumzeit. Ansätze, die im Gegensatz zu Lloyds Konzeption von einer Emergenz der Quantenmechanik auf der Grundlage eines nicht-quantenmechanischen computationalen Substrats ausgehen, finden sich nicht zuletzt in Requardt (1996, 1996a, 2000) (s.o.), Cahill (2002, 2005), Cahill / Klinger (1996, 1997, 1998, 2005).

Die Quantencomputationen selbst werden als quantenmechanische Superpositionen von *Computational Histories* verstanden. Der Übergang von einer solchen quantenmechanischen Superposition zu einer klassischen makroskopischen Raumzeit erfolgt dem Ansatz zufolge über Dekohärenzprozesse:

> *"The visible universe that we see around us presumably corresponds to one such decoherent history."* (Lloyd (2005a) 21)

Die emergente Raumzeit ist dabei notwendigerweise dynamisch, da sie überhaupt erst auf der Grundlage einer prägeometrischen Dynamik zustandekommt. Der Ansatz ist zudem insofern hintergrundunabhängig, als auf der Substratebene selbst keine Raumzeit vorausgesetzt wird.[227]

> *"Because distances are derived from dynamics, without reference to an underlying spacetime manifold, the resulting theory is intrinsically covariant and background independent."* (Lloyd (2005a) 2)

Lloyds zentrale These besteht nun in der Behauptung, dass sich unter diesen Voraussetzungen die Einsteinschen Feldgleichungen (zumindest in ihrer diskreten Form als Einstein-Regge-Gleichungen) notwendigerweise von selbst ergeben – einfach schon infolge der vollständigen Hintergrundunabhängigkeit der Quantencomputationen:

> *"Since general covariance [...] implies Einstein's equations, the geometry induced by the computational universe obeys Einstein's equations (in their discrete, Regge calculus form)."* (Lloyd (2005a) 7)

Lloyd behauptet dabei insbesondere, dass sich die Metrik der Raumzeit, die dann gerade den Einstein-Regge-Gleichungen gehorcht, unmittelbar aus den Quantencomputationen ergibt, also auf der Grundlage eines Substrates, das selbst über keine Metrik verfügt:

> *"The information that moves through the computation effectively 'measures' distances in spacetime in the same way that the signals passed between members of a set of GPS satellites measure spacetime."* (Lloyd (2005a) 7) – *"[...] distances are quantities that are derived from the underlying dynamics of*

---

[227] Die diskrete, lokale Zeitentwicklung innerhalb der Quantencomputation kann dabei naheliegenderweise nicht mit der beobachtbaren Zeit auf der emergenten raumzeitlichen Ebene identisch sein. Zur entsprechenden dynamischen Entkopplung, die hier anzunehmen wäre, siehe die prägeometrischen *Quantum Causal Histories* in Kap. 4.6.

*quantum systems."* (Lloyd (2005a) 2) – *"[...] those derived distances automatically conform to the laws of general relativity."* (Lloyd (2005a) 2)

Im Rahmen des *Computational-Universe*-Ansatzes sei damit ohne weiteres die Rückwirkung der emergenten Raumzeit (bzw. ihrer Metrik) auf die (ebenso von der quantencomputationalen Dynamik hervorgebrachte) Materie nachvollziehbar:

*"The computational universe model is intrinsically a theory of quantum matter coupled to gravity, and not a theory of either quantum matter or quantum gravity on its own."* (Lloyd (2005a) 13)

Dies alles gilt jedoch, wie er einschränkt, nur unter der Voraussetzung 'normaler' Materie und ausschliesslich positiver Krümmungen. – Ansonsten:

*"[...] space would have to curve more than it can curve."* (Lloyd (2005a) 12)

Lloyd behauptet, dass jede universelle Quantencomputation unter diesen Bedingungen eine emergente Raumzeit hervorbringt:

*"Any local quantum theory involving pairwise interactions allows the construction of a theory of quantum gravity."* (Lloyd (2005a) 20) – *"[...] any computation, for example, one that calculates the digits of π, corresponds to a class of spacetimes that obeys the Einstein-Regge equations."* (Lloyd (2005a) 41)

Dies gilt seiner Auffassung nach insbesondere auch für ein quantencomputationales Szenario, welches das quantenfeldtheoretische Standardmodell in Form einer Gittereichtheorie reproduziert.

*"For example, the quantum computation could reproduce the standard model as a lattice gauge theory, which would in turn give rise to quantum gravity by 'constructing' the distances between lattice points [...]."* (Lloyd (2005a) 20)

So liesse sich das Zustandekommen von Gravitation und gekrümmter Raumzeit ohne weiteres als residuales Phänomen einer quantenfeldtheoretischen Computation – aber auch jeder anderen – verstehen. Ob ein quantencomputationales Szenario, welches das quantenfeldtheoretische Standardmodell in Form einer Gittereichtheorie zum Ausgangspunkt nimmt, jedoch schon *unsere* phänomenologische Raumzeit reproduziert, ist natürlich sehr fraglich.

Das zentrale Problem, auf das der *Computational-Universe*-Ansatzes stösst, wenn man versucht, auf seiner Grundlage eine konkrete Theorie der Quantengravitation zu formulieren, verwundert angesichts der Erfahrungen mit den vorausgehend besprochenen Emergenzszenarien nicht sonderlich: Die konkrete Substratstruktur und –dynamik bleibt erst einmal wieder völlig im Unklaren. Übersetzt in die Terminologie dieses Denkansatzes heisst dies: Welche spezifische Quantencomputation unserem Universum zugrundeliegt, mithin also die spezifische Raumzeit dieses Universums hervorbringt, ist grundlegend unklar.

*"Every quantum computation corresponds to a family of metrics, each of which obeys the Einstein-Regge equations. But which computation corresponds to the universe we see around us? What is the 'mother' computation? We do not know."* (Lloyd (2005a) 23)

Es wären also erst einmal die Konsequenzen der unterschiedlichsten Computationen auszuloten und mit den entsprechenden empirischen Daten, die uns zur Verfügung stehen, abzugleichen:

> *"[...] we should investigate different candidates and compare the results of those investigations with observation."* (Lloyd (2005a) 42)

Oder man könnte, ohne eine entsprechende Auslotung der Details, für das Substrat schlichtweg in grösster Allgemeinheit des Ansatzes von einer Superposition aller möglichen Quantencomputationen ausgehen:

> *"An appealing choice of quantum computation is one which consists of a coherent superposition of all possible quantum computations [...]."* (Lloyd (2005a) 23)

Unabhängig von dieser umfassendsten aller Möglichkeiten sieht Lloyd in zellulären Quantenautomaten oder ihren randomisierten Varianten die vielleicht aussichtsreichsten Kandidaten.[228] Aber all dies ist nicht mehr als Spekulation ohne die geringste Ankopplung an die Empirie oder auch nur eine signifikante Einbeziehung von Einsichten aus den etablierten Theorien.

## *'Holographische Schirme': Von Quanteninformationsflüssen zur Raumzeit*

Spätestens angesichts der computationalen Szenarien muss man sich irgendwann einer entscheidenden Frage stellen: Wie kann es überhaupt zur Emergenz der Raumzeit auf der Grundlage eines rein computationalen Substrats, einer reinen Informationsdynamik, kommen? Wie kann Raumzeit aus etwas so völlig von der Raumzeit verschiedenem wie Quanteninformationen, Informationsflüssen oder elementaren kausalen Relationen hervorgehen? – Dies ist wahrscheinlich eine der fundamentalsten Fragen, die sich hinsichtlich einer emergenten Raumzeit stellen. In gewisser Weise stehen unsere Intuitionen hinsichtlich der Raumzeit erst einmal in deutlichem Konflikt mit der Grundannahme der prägeometrischen und insbesondere der rein informationstheoretischen, computationalen Emergenzszenarien – und dies erst einmal völlig unabhängig von ihrer konkreten Ausformulierung und ihrer konkreten Substratkonstruktion.

Dieses Spannungsgefüge zwischen unseren Intuitionen hinsichtlich der Raumzeit und ihrer möglicherweise rein computationalen Basis lässt sich jedoch auflösen: im Rahmen einer Idee, die auf Fotini Markopoulou und Lee Smolin zurückgeht und unter der Bezeichnung *Holographische Schirme* formiert.[229] Diese Idee liefert zumindest eine intuitiv nachvollziehbare Antwort auf die Frage, wie es zur Emergenz der Raumzeit auf der Grundlage eines originär prägeometrischen, nicht-raumzeitlichen Substrats kommen kann, wie also Raumzeit aus etwas völlig anderem, nicht-raumzeitlichen hervorgehen könnte.

Die entscheidende Voraussetzung, die hierfür benötigt wird, besteht in der Annahme, dass kausale Relationen auf der elementarsten Ebene schon existieren und durch den Fluss von Quanteninformation repräsentiert bzw. instantiiert werden. Die Quantenmechanik gilt also schon auf der Substratebene. Diese besteht dann im einfachsten Fall aus nichts anderem als einer dynamischen Struktur von ausschliesslich relational bestimmten Quantensystemen ohne raumzeitlichen Hintergrund: einem relationalen, sich ausschliesslich lokal verändernden, finiten Netzwerk von elementaren Quantensystemen.[230] Dieses hat in seiner einfachsten und allgemeinsten Form die Struktur eines finiten azyklischen Graphen, der aus gerichteten Relationen (Linien) zwischen elementaren Quantensystemen bzw. Quantenereignissen (Vertices) besteht. Die gerichteten Relationen werden durch Flüsse von Quanteninformation zwischen elementaren Quantensystemen instantiiert. Jede der Linien verfügt dabei über eine spezifische elementare Informationsdurchflusskapazität. Die Flüsse von Quanteninformation lassen sich somit als elementare kausale Relationen zwischen den elementaren Quantensystemen bzw. Quantenereignissen interpretieren. Da solche ausschliesslich relational bestimmten Quantensysteme über keine geometrischen Freiheitsgrade verfügen, sind sie – im Hinblick auf die makroskopische Raumzeit – notwendigerweise hintergrundunabhängig. Dynamische Prozesse innerhalb des Netzwerks finden in diskreten, ausschliesslich lokal wirksamen Schritten statt; sie bestehen gerade im Austausch von Quanteninformation zwischen den elementaren Quantensystemen des Netzwerks.

---

[228] Vgl. Wolfram (2002), Poundstone (1985).
[229] Siehe Markopoulou / Smolin (1999).

Nun definiere man (holographische) Schirme, die das relationale Netzwerk und damit einige seiner gerichteten Linien (Quanteninformationsflüsse) durchschneiden. Diesen Schirmen lässt sich eine spezifische Durchflusskapazität für Quanteninformationen zuordnen; sie ergibt sich einfach aus der Summe der Informationsdurchflusskapazitäten der durchschnittenen Linien.

Die nun zur Anwendung kommende zentrale Idee, die den *Holographischen Schirmen* schliesslich auch zu ihrem Namen verhilft, besteht in einer gedanklichen Umkehrung der Einsichten, die sich mit der *Bekenstein-Hawking-Entropie* und ihren Konsequenzen im Rahmen der Thermodynamik schwarzer Löcher abzeichnen.[231] Der *Bekenstein-Hawking-Entropie* zufolge ist die Entropie eines schwarzen Loches proportional zur Oberfläche seines Ereignishorizontes. Im Rahmen des *Holographischen Prinzips* legt diese Tatsache dann eine Sichtweise nahe, der zufolge der (aufgrund der holographischen bzw. kovarianten Entropiegrenze grundsätzlich finite) Informationsgehalt, der einem Raumvolumen zugeschrieben werden kann, vollständig durch den auf seiner Oberfläche kodierten bzw. kodierbaren Informationsgehalt bestimmt ist. Diese Auffassung wird, wenn sie zutreffen sollte, spätestens wirksam, wenn man sich der Substratebene zuwendet. In jedem Fall wird, gemäss der holographischen bzw. kovarianten Entropiegrenze, der maximale finite Informationsgehalt, der einem Raumvolumen zugeschrieben werden kann, durch dessen Oberfläche definiert; er ist proportional zu dieser Oberfläche.

Mit den *Holographischen Schirmen* wird diese Relation zwischen Fläche und Information nun gerade umgekehrt. Danach legt nicht die Fläche die maximal mögliche Information fest, sondern die ihr maximal zuschreibbare Information definiert überhaupt erst die Fläche. Genauer: Die Fläche, die einem holographischen Schirm zugeschrieben werden kann, wird durch dessen jeweilige Informationsdurchflusskapazität (also die Summe der Informationsdurchflusskapazitäten der durchschnittenen Quanteninformationsflusslinien) definiert; sie wird mit dieser gleichgesetzt.

> *"This leads us to suggest that the Bekenstein bound may be inverted and* area be defined to be a measure of the capacity of a screen for the transmission of quantum information." (Markopoulou / Smolin (1999) 3)

---

[230] Vgl. die prägeometrischen *Quantum Causal Histories* in Kap. 4.6.
[231] Siehe Kap. 3.1. und 3.2. Vgl. auch Jacobsons emergente Raumzeit auf thermodynamischer Grundlage (s.o.).

Hat man auf diese Weise die Fläche holographischer Schirme definiert, so lässt sich dann aus einem geeigneten Konstrukt aus holographischen Schirmen – etwa einem auf ihrer Grundlage zu definierenden sekundären Netzwerk – eine raumzeitliche Geometrie konstruieren. Oder man betrachtet, im Sinne des *Holographischen Prinzips*, die raumzeitliche Geometrie gleich als etwas, das jeweils schon von einem einzelnen holographischen Schirm hervorgebracht wird; die Freiheitsgrade, die in einer solchen raumzeitlichen Geometrie zum Ausdruck kommen, wären dann allesamt Freiheitsgrade der entsprechenden Fläche: des holographischen Schirms. Nur der holographische Schirm existierte; die Raumzeit wäre nicht mehr als eine Illusion, die sich auf seiner Grundlage einstellt.

Doch schon unabhängig von der Entscheidung zwischen diesen beiden Alternativen wird mit der Idee der *Holographischen Schirme* grundsätzlich nachvollziehbar, wie es zu einer raumzeitlichen Geometrie auf der Grundlage eines grundlegend nicht-raumzeitlichen Substrats von Informationsflüssen bzw. kausalen Relationen zwischen elementaren Quantensystemen kommen kann. Die Geometrie lässt sich grundsätzlich als das Ergebnis der relationalen Struktur und Dynamik eines vollständig prägeometrischen Substrats verstehen. Die Raumzeit wäre dann nichts anderes als ein Ausdruck der Relationalität von Informationsflüssen, die sich wiederum als elementare kausale Relationen deuten lassen. – Wheelers *It from bit*, wiederum transponiert zu einem *It from Qubit*, ist also gar nicht so unrealistisch wie es auf den ersten Blick erscheinen mag.

## *Implikationen der Emergenzszenarien*

Sollte die Raumzeit ein emergentes, intrinsisch klassisches Phänomen sein,[232] so ginge es in einer Theorie der 'Quantengravitation', wie im Vorausgehenden deutlich geworden sein sollte, nicht mehr einfach nur um Quantenkorrekturen zum klassischen Bild der Raumzeit. Es ginge nicht mehr um Quantenfluktuationen der Metrik,[233] nicht mehr um quantenmechanische Unschärfen der Raumzeit, nicht mehr um Superpositionen von Raumzeiten; es ginge um keine Gravitonenphysik[234] mehr. Eine emergente, intrinsisch klassische Raumzeit hat keine solchen Quanteneigenschaften. Vielmehr ginge es dann um die Quanteneigenschaften des Substrats, auf dessen Grundlage sich die klassische Raumzeit ergibt – so dieses Substrat überhaupt noch über Quanteneigenschaften verfügen sollte.

Die konzeptionellen Voraussetzungen, die sich für eine Theorie der Quantengravitation vielleicht aus den etablierten Theorien, insbesondere der Allgemeinen Relativitätstheorie, ableiten lassen, würden sich dann nur bedingt als relevant und verlässlich erweisen. Von der Dynamizität der Raumzeit in der Allgemeinen Relativitätstheorie liesse sich unter Umständen immer noch behaupten, dass sie vielleicht gerade auf die Ursprünge der Raumzeit in einer fundamentaleren prä-raumzeitlichen Dynamik hinweist. Die Diffeomorphismusinvarianz der Allgemeinen Relativitätstheorie jedoch könnte sich unter (dann genau zu klärenden) Umständen ohne weiteres als modelltheoretisches Artefakt erweisen. Dies gilt dann aber noch nicht unbedingt, soweit es die über diese hinausreichende Perspektive betrifft, für den im Rahmen der Allgemeinen Relativitätstheorie auf der Grundlage ihrer Diffeomorphismusinvarianz motivierbaren Relationalismus. Eine prägeometrische Theorie würde – hinsichtlich der somit emergenten Raumzeit – vermutlich ohnehin nur relationalistisch interpretierbar sein. Eine emergente Raumzeit wäre als solche jedenfalls kaum substantialistisch deutbar. Die ontologische Frage wäre dann aber ohnehin in Bezug auf die basalen Konstitu-enten des Geschehens zu stellen. Die Substantialismus-Relationalismus-Debatte bezüglich der Raumzeit wäre als Debatte über den ontologischen Status von Konzepten innerhalb einer effektiven Theorie letztendlich obsolet.

---

[232] Vgl. auch Kap. 4.6.
[233] Zu den Problemen, die sich aus der Annahme von Quantenfluktuationen der Metrik ergeben, siehe Kap. 2.2.
[234] Vgl. Kap. 4.1. und 4.2.

Schon im Kontext der Szenarien, die nur von einer emergenten Gravitation ausgehen, ohne dabei notwendigerweise auch gleich eine Emergenz der Raumzeit einzuschliessen, wäre die Allgemeine Relativitätstheorie nur noch als effektive Theorie anzusehen, die sich mit der intrinsisch klassischen Dynamik von kollektiven Freiheitsgraden beschäftigt; diese wären keinesfalls mit den gänzlich anders gearteten Freiheitsgraden zu verwechseln, welche die Dynamik auf der Substratebene bestimmen. – Das gleiche gilt erst recht für eine Theorie, welche Gravitation und Raumzeit als miteinander verbundene, intrinsisch klassische, emergente Phänomene beschreibt: Phänomene also, die auf der Ebene des Quantensubstrats gar nicht vorkommen. Die Allgemeine Relativitätstheorie sollte sich dann als effektive Theorie aus einer fundamentaleren Theorie, welche die Dynamik auf der Substratebene beschreibt, als Näherung oder Grenzfall ableiten lassen.

Eine solche fundamentalere Theorie liesse sich dann aber nicht über die Quantisierung ihres klassischen, makroskopischen Grenzfalles erreichen, der gerade die Dynamik von kollektiven Freiheitsgraden erfasst. Sollte die Raumzeit nicht fundamental sein, sondern Ausdruck anderer nicht-raumzeitlicher Freiheitsgrade, oder sollte die Gravitation keine fundamentale Wechselwirkung, sondern ein residuales bzw. induziertes Phänomen sein, so wäre eine Quantisierung der Allgemeinen Relativitätstheorie konzeptionell der falsche Weg.[235] Eine Quantisierung der Allgemeinen Relativitätstheorie würde dann als Quantisierung einer effektiven Theorie, die gerade die Dynamik kollektiver Freiheitsgrade erfasst, zu keinen Einsichten in die fundamentalere Dynamik, auf der diese kollektiven Freiheitsgrade beruhen, führen. Die Quantisierung der Allgemeinen Relativitätstheorie entspräche, auch wenn diese die Gravitation als intrinsisch klassisches Phänomen adäquat beschriebe, letztlich immer nur einer Quantisierung kollektiver, makroskopischer Freiheitsgrade, die auf der Grundlage eines gänzlich anders gearteten Substrats und grundsätzlich anders gearteter fundamentaler Freiheitsgrade hervorgebracht werden. Es wären schlichtweg die falschen Freiheitsgrade, die man hier quantisiert. Man würde damit nicht zu einer angemessenen Theorie der Quantengravitation gelangen – genausowenig wie etwa eine Quantisierung der Navier-Stokes-Gleichung zu einer realistisch interpretierbaren Quantenhydrodynamik führt.

Sollten Raumzeit und/oder Gravitation emergent sein, so wäre eine angemessene Theorie der Quantengravitation also nicht über die Quantisierung der Allgemeinen Relativitätstheorie zu erreichen. Ziel und Gegenstand einer Theorie, die den Motivationen gerecht wird, die den Ansätzen zur Quantengravitation zugrundeliegen, müsste vielmehr die Identifizierung des Substrats, seiner elementaren Konstituenten und der zwischen diesen bestehenden relationalen Struktur, seiner fundamentalen Freiheitsgrade und seiner Dynamik sein. Das Zustandekommen der Gravitation und ihrer empirisch nachweislichen Dynamik wäre auf dieser Grundlage als emergentes und vielleicht eben intrinsisch klassisches Phänomen zu erklären.

---

[235] Vgl. die Zitate von Jacobson, Volovik, Hu und Finkelstein in den vorausgehenden Fussnoten in diesem Teilkapitel. Vielleicht lassen sich aber noch bessere Argumente gegen eine Quantisierung der Allgemeinen Relativitätstheorie finden als die in der Existenz der geschilderten emergenten Szenarien gründenden. Vgl. Kap. 4.4. und 4.6.

# 4. Raumzeit in der Quantengravitation

Solange es noch keine spezifischen empirischen Daten gibt, die über die etablierten Vorgängertheorien hinausweisen, bewegen sich alle Ansätze zur Entwicklung einer Theorie der Quantengravitation im Kontext rein konzeptioneller Spekulationen. Die elementarste Anforderung an eine solche Theorie besteht unter diesen Bedingungen, neben ihrer logischen Widerspruchsfreiheit und ihrer konzeptionellen Kohärenz, vor allem darin, dass sie den empirischen Gehalt ihrer Vorgängertheorien reproduziert. Dies heisst insbesondere, dass sie die Allgemeine Relativitätstheorie als (klassischen) Grenzfall enthalten muss – vor allem, aber nicht nur, wenn sie mit der Strategie einer direkten Quantisierung dieser klassischen Vorgängertheorie arbeitet. Es wird sich im Folgenden zeigen, dass die Erfüllung dieser Anforderung keineswegs trivial ist, sondern durchaus auf hartnäckige Probleme stossen kann.

## 4.1. Kovariante Quantisierung der Allgemeinen Relativitätstheorie

Die *Kovariante Quantisierung*[236] ist eine Strategie zur Entwicklung einer Theorie der Quantengravitation, bei der versucht wird, eine (herkömmliche) Quantenfeldtheorie des Gravitationsfeldes zu formulieren, vergleichbar etwa der quantisierten Erfassung des elektromagnetischen Feldes in der Quantenelektrodynamik.

> *"The idea was [...] to do unto the gravitational field as was done to the electromagnetic field: quantize the gravitational field to a get a particle (the* graviton*) that mediates the interaction."* (Rickles / French (2006) 16)

Quantenfeldtheorien benötigen jedoch einen Hintergrundraum, auf dem erst die entsprechenden Operatorfelder definiert werden können. Da dieser Hintergrundraum schon über eine feste Metrik verfügen muss, andererseits aber das Gravitationsfeld (entsprechend der Allgemeinen Relativitätstheorie) selbst durch die Metrik repräsentiert wird, startet die quantenfeldtheoretische Erfassung des Gravitationsfeldes von vornherein mit einem perturbativen Ansatz, der metrische Fluktuationen auf einem Hintergrundraum mit fester Metrik beschreibt.

> *"The important assumption is the presence of an (approximate)* background *with respect to which standard perturbation theory (formulation of Feynman rules, etc.) can be applied."* (Kiefer (2005) 5)

Die *Kovariante Quantisierung* entspricht also einer Quantenfeldtheorie der Fluktuationen der Metrik[237] auf einem festen Hintergrundraum, für den im allgemeinen eine Minkowski-Metrik angenommen wird:

> *"Field-theoretic techniques are put at the forefront. The first step in this program is to split the spacetime metric $g_{\mu\nu}$ in two parts, $g_{\mu\nu} = \eta_{\mu\nu} + \sqrt(G\,h_{\mu\nu})$, where $\eta_{\mu\nu}$ is to be a background, kinematical metric,*

---

[236] Siehe DeWitt (1967a, 1967b).
[237] Zu den massiven konzeptionellen Problemen, die sich aus der Annahme von Quantenfluktuationen der Metrik ergeben, siehe Kap. 2.2.

*often chosen to be flat, G is Newton's constant, and $h_{\mu\nu}$, the deviation of the physical metric from the chosen background, the dynamical field. The two roles of the metric tensor are now split. The overall attitude is that this sacrifice of the fusion of gravity and geometry is a moderate price to pay for ushering-in the powerful machinery of perturbative quantum field theory. [...] it is only $h_{\mu\nu}$ that is quantized. Quanta of the field propagate on the classical background space-time with metric $\eta_{\mu\nu}$. If the background is in fact chosen to be flat, one can use the Casimir operators of the Poincaré group and show that the quanta have spin two and rest mass zero. [...] Thus, in this program, quantum general relativity was first reduced to a quantum field theory in Minkowski space."* (Ashtekar (2005) 5)

Die *Kovariante Quantisierung* beschreibt auf diese Weise ein Metrik-Operatorfeld auf einer festen Hintergrundraumzeit. Die Quantenaspekte der Fluktuationen des Gravitationsfeldes (bzw. der Metrik) werden in Form des Austausches von Wechselwirkungsbosonen erfasst, die sich auf dieser festen Hintergrundraumzeit bewegen. Dass diese *Gravitonen* masselose Spin-2-Teilchen sein müssen, ergibt sich auf der Grundlage der dynamischen Struktur der gravitativen Wechselwirkung: langreichweitig, ausschliesslich anziehend.

> *"The helicity states of the gravitational waves on the background become the quantum states of the graviton. Utilizing the representations of the Poincaré group, one is able to define the graviton as a spin-2-particle. We know that this particle must be massless because the gravitational interaction works long range, and the slightest mass would contradict results concerning the deflection of light."*
> (Rickles / French (2006) 16)

Auf dieser Grundlage lassen sich dann die für die quantisierten Spin-2-Felder formulierbaren Feynmanschen Gravitonengraphen nach Potenzen der Gravitationskonstanten entwickeln. Hierbei stellt sich schliesslich heraus, dass die Quantenfeldtheorie der Fluktuationen der Metrik nicht renormierbar ist.

> *"It is generally agreed that this non-renormalisability renders perturbatively quantised Einstein gravity meaningless as a fundamental theory because an infinite number of parameters would be required to make any physical prediction."* (Nicolai / Peeters / Zamaklar (2005) 3)

Als fundamentale Theorie ist die *Kovariante Quantisierung* damit ungeeignet – wenn eine definitiv nicht-renormierbare Theorie überhaupt zu irgend etwas Konkretem taugt, das über das Lernen aus konzeptionellen Fehlern hinausgeht. Das Instrumentarium der Quantenmechanik und der Quantenfeldtheorien ist offensichtlich mit dem der Allgemeinen Relativitätstheorie nicht auf diese simple Weise in Einklang zu bringen.

Dies ist nicht sonderlich überraschend. Eine Quantisierung der Gravitation im Sinne einer perturbativen Quantenfeldtheorie des Gravitationsfeldes scheitert letztendlich an der konzeptionellen Unverträglichkeit des modelltheoretischen Apparates der perturbativen Quantenfeldtheorien mit der Allgemeinen Relativitätstheorie. Eine der naheliegenden Ursachen für die Nichtrenormierbarkeit der *Kovarianten Quantisierung* hängt mit der Nichtlinearität der Gravitation zusammen, die schon in der Allgemeinen Relativitätstheorie zum Ausdruck kommt – und die sich im (wie sich zeigt: unangemessenen) perturbativ-quantenfeldtheoretischen Bild als Selbstwechselwirkung der Gravitonen darstellt. Die Gravitation koppelt an die Masse und, infolge der Masse-Energie-Äquivalenz, an die Energie an.[238] So ergeben sich im perturbativen Ansatz für kleiner werdende Abstände und steigende Energien immer höhere Selbstwechselwirkungsbeiträge zur Gravitation, da hier – mit zunehmender Energie der virtuellen Vakuumteilchen – die höheren Glieder der perturbativen Entwicklung eine immer wichtigere Rolle spielen. Dies führt für steigende Energien zu immer weiter steigenden, divergenten Beiträgen in der Reihenentwicklung. Und diese Divergenz ist, im Gegensatz zu den für die anderen Wechselwirkungen in ihrer perturbativen Erfassung auftretenden Divergenzen, infolge der Kopplung der Gravitation an die Energie nicht mehr kontrollierbar. Gerade dies ist die Ursache für die Nichtrenormierbarkeit der Gravitation in einem perturbativen quantenfeldtheoretischen Ansatz:

> *"[...] such non-renormalizable theories become pathological at short distances [...] – perhaps not too surprising a result for a theory which attempts in some sense to 'quantize distance'."* (Callender / Huggett (2001 a) 5)

Die Nichtrenormierbarkeit der *Kovarianten Quantisierung* und die in ihr zum Ausdruck kommende konzeptionelle Unvereinbarkeit der Allgemeinen Relativitätstheorie mit dem perturbativen quantenfeldtheoretischen Ansatz sind also alles andere als verwunderlich – insbesondere, wenn man sein Augenmerk auf das eigentliche konzeptionelle Problem der *Kovarianten Quantisierung* lenkt: ihre Hintergrundabhängigkeit.

> *"[...] perturbation theory, by its very definition, breaks background independence and diffeomorphism invariance at every finite order of perturbation theory. Thus one can restore background independence only by summing up the entire perturbation series, which is of course not easy. Not surprisingly, [...] applying this programme to Einstein's theory itself results in a mathematical disaster, a so-called non-renormalizable theory without any predictive power."* (Thiemann (2007) 10)

Geht man, wie die direkten Quantisierungsansätze dies tun, von der konzeptionellen Angemessenheit der Allgemeinen Relativitätstheorie im klassischen Bereich aus, und damit von der Identifizierbarkeit von Gravitation und Raumzeitmetrik, so lässt sich für eine quantenmechanische Behandlung der Gravitation – für eine Theorie der Quantengravitation – auf die Notwendigkeit einer Beschreibung ohne festen Hintergrundraum schliessen, zumindest solange sich keine sehr guten Gründe für eine Hintergrundabhängigkeit anführen lassen (zu denen simple modelltheoretische Gepflogenheiten ganz sicher nicht zählen).

> *"It appears that the project of quantising gravity can succeed only if the quantisation procedure does not* start *from a spacetime formalism in the usual sense, i.e. a framework using a differential manifold with given geometrical properties which can be interpreted in terms of distances between physical systems, intervals between events, and so on."* (Dieks (2001) 152)

Die Hintergrundabhängigkeit der *Kovarianten Quantisierung* entspricht einerseits einer Verletzung der allgemeinen Kovarianz der Allgemeinen Relativitätstheorie. Darüber hinaus, und im Gegensatz zu anderen hintergrundabhängigen Ansätzen zu einer Theorie der Quantengravitation, die aus mehr oder weniger überzeugenden konzeptionellen Gründen nicht von einer direkten Quantisierung der Allgemeinen Relativitätstheorie ihren Ausgang nehmen, stellt die *Kovariante Quantisierung* zudem noch ein konzeptionelles Unding dar: eine hintergrundabhängige Quantisierung einer hintergrund-unabhängigen Theorie bzw. eine Quantisierung der Raumzeit auf der Raumzeit.

---

[238] Alle anderen Wechselwirkungen koppeln nur an ihre entsprechenden 'Ladungen' an, nicht jedoch an die Energie.

## 4.2. Die Stringtheorien

Je nach eingenommener Perspektive lässt sich der Stringansatz[239] unterschiedlich charakterisieren (ohne dass sich diese unterschiedlichen Charakterisierungen gegenseitig ausschliessen würden): Einerseits lässt er sich in formaler Hinsicht als das Ergebnis der Quantisierung einer klassischen Theorie – der speziell-relativistischen Dynamik eindimensional ausgedehnter Objekte: der 'Strings'[240] – ansehen, einer Theorie also, die nicht mit der Allgemeinen Relativitätstheorie identisch ist; nach der Quantisierung lassen sich jedoch – wie zumindest plausibel gemacht werden kann – die Einsteinschen Feldgleichungen (unter gewissen Vorbehalten) als klassische, niederenergetische Implikation reproduzieren.

Was die modelltheoretischen Grundlagen des Stringansatzes betrifft, ist er andererseits als Erweiterung der *Kovarianten Quantisierung* der Allgemeinen Relativitätstheorie interpretierbar (wie sie zuvor schon einmal mit der zwischenzeitlich aufgegebenen *Supergravity*-Theorie[241] angestrebt wurde): ein Ansatz, der insbesondere, was den modelltheoretischen Ausgangspunkt, nicht aber die Konsequenzen betrifft, das quantenfeldtheoretische Instrumentarium nur moderat erweitert.

Die in einer solchen Perspektive auf den Stringansatz zur Geltung kommende Hoffnung ist, die Allgemeine Relativitätstheorie als Niederenergienäherung einer allgemeineren Quanten(feld)theorie zu reproduzieren. Und die aus dieser Idee resultierende Suche nach einer quantisierten Hochenergievariante der Allgemeinen Relativitätstheorie, die im Gegensatz zur *Kovarianten Quantisierung* renormierbar ist oder erst gar nicht renormiert werden muss, hat hier offensichtlich Erfolg: Im Ge-

---

[239] Siehe etwa Polchinski (2000, 2000a), Kaku (1999), Green / Schwarz / Witten (1987), Alvarez-Gaumé / Vázquez-Mozo (1995), Lerche (2000), Mohaupt (2003), Zwiebach (2004). Eine Übersicht über die zahlreiche Literatur gibt Marolf (2004).

[240] Der klassische String ist ein eindimensional ausgedehntes Objekt, dessen Dynamik auf einem d-dimensionalen Minkowski-Raum erfasst wird. (Formal ist dieser *target space* auf gekrümmte Mannigfaltigkeiten erweiterbar. Siehe Sanchez (2003).) Das zweidimensionale Weltblatt der klassischen Stringbewegung auf diesem d-dimensionalen Minkowski-Raum wird durch die von 0 bis $2\pi$ laufenden internen Stringkoordinaten $\sigma$ und die Eigenzeit $\tau$ beschrieben. Die Einbettung des Weltblatts in den d-dimensionalen Minkowski-Raum wird durch d Koordinaten-Funktionen $X^\mu(\sigma,\tau)$ erfasst. Unter Berücksichtigung der Stringspannung bzw. der Masse/Energie pro Längeneinheit lässt sich eine klassische Wirkungsfunktion für den String formulieren, die im wesentlichen der Fläche des Weltblattes des String entspricht. Nach einer geeigneten Transformation zeigt diese Wirkungsfunktion Symmetrieeigenschaften, die zu einer zweidimensionalen klassischen konformen Feldtheorie gehören. Die konforme Invarianz entspricht der Tatsache, dass eine Umbenennung der Koordinaten $\sigma$ auf dem Weltblatt des String die physikalischen Zustände nicht verändert.

Es gibt zwei verschiedene Stringtopologien: offene und geschlossene Strings. Für geschlossene Strings gilt: $X^\mu(\sigma + 2\pi,\tau) = X^\mu(\sigma,\tau)$. Kein Punkt des geschlossenen String ist topologisch ausgezeichnet. Daher gibt es auch keine Möglichkeit für fixierte Ladungen ohne Symmetriebrechung. Offene Strings hingegen können an ihren Enden Ladungen tragen, die als Quellen der entsprechenden Felder wirksam werden. Diese Ladungen sind hier erst einmal als Verallgemeinerung elektrischer Ladungen zu verstehen.

Ebenso gibt es zwei verschiedene Bewegungsformen für den klassischen String (sowohl für den offenen, als auch für den geschlossenen): die Translationsbewegung und die Oszillation. Für letztere sind rechtsorientierte (holomorphe) und linksorientierte (antiholomorphe) Schwingungen zu berücksichtigen.

[241] Siehe Kap. 1.3.

gensatz zur *Kovarianten Quantisierung* und zur *Supergravity* hat der Stringansatz den Vorzug, dass er offensichtlich nicht nur *nicht* nicht-renormierbar ist; er bedarf, so wie es aussieht, nicht einmal einer Renormierung; es kommt offensichtlich zu keinen Divergenzen; die perturbative Entwicklung bleibt von vornherein finit.

Dies hängt mit der wichtigsten konzeptionellen Besonderheit des Stringansatzes zusammen: Was diesen nämlich vor allem von der *Kovarianten Quantisierung* und der *Supergravity* unterscheidet, ist, dass mit ihm eine nomologische Vereinigung aller Wechselwirkungen inklusive der Gravitation angestrebt wird. Es ist offensichtlich diese, zumindest formal realisierte, nomologische Vereinigung, die zu seiner perturbativen Finitheit führt. Offensichtlich erfüllt sich im Stringansatz die Hoffnung, dass sich die Nichtrenormierbarkeit der *Kovarianten Quantisierung* vermeiden lässt, wenn man eine nomologische Vereinigung aller Wechselwirkungen anstrebt, die selbstredend nicht über eine direkte Quantisierung der Allgemeinen Relativitätstheorie erreicht werden kann.

Aber, wie wir sehen werden, wird auch für den Stringansatz die Hintergrundabhängigkeit zum Problem: kein Wunder bei der modelltheoretischen Basis, die er weitgehend mit der *Kovarianten Quantisierung* gemein hat.

*

Was seine faktische Entwicklungsgeschichte – die dritte der möglichen Perspektiven – betrifft, stellt sich der Stringansatz noch einmal anders dar. Hier deutet anfänglich nichts auf den Versuch oder eine Intention hin, eine Theorie der Quantengravitation, die den empirischen Gehalt der Allgemeinen Relativitätstheorie reproduziert, durch die Quantisierung einer geeigneten klassischen Ausgangstheorie zu erreichen, oder darauf, dass die *Kovariante Quantisierung* der Allgemeinen Relativitätstheorie in irgendeiner geeigneten Weise erweitert werden soll, um deren Probleme zu beheben. Der Stringansatz ist nicht das Ergebnis solcher strategischer Überlegungen, sondern vielmehr das einer Reihe von Zufällen, für die sich schliesslich post hoc ein Bezug auf mögliche Strategien zur Entwicklung einer Theorie der Quantengravitation hat herstellen lassen:

Die Geschichte des Stringansatzes beginnt im Kontext der Hadronenphysik der sechziger Jahre, wo er sich als nicht sonderlich erfolgreich herausstellt – oder zumindest als weniger erfolgreich als die *Quantenchromodynamik* und das *Quark-Modell*. Die Entdeckung schliesslich, dass die Quantisierung der klassischen (speziell-relativistischen) Dynamik eines eindimensional ausgedehnten, oszillierenden Strings zu einer perturbativen Quantentheorie führt, die Spin-2-Bosonen als Anregungszustände des Strings enthält, wurde dann zum Auslöser für die Reinkarnation des Stringansatzes als Anwärter auf den Status einer Theorie der Quantengravitation. In der Hadronenphysik kann man mit Spin-2-Bosonen nur wenig anfangen; in der Quantengravitation hingegen lassen sie sich entsprechend der *Kovarianten Quantisierung* naheliegenderweise als Gravitonen interpretieren. Der Übergang von der Hadronenphysik zur Quantengravitation impliziert dann nicht viel mehr als eine Verschiebung der anzusetzenden Stringspannung, die sich direkt in den Energieniveaus des oszillierenden Strings niederschlägt: von den Energien der Hadronenphysik zur Planck-Energie.

Damit – und unter Eliminierung diverser innertheoretischer Anomalien – ist es dann möglich, die Einsteinschen Feldgleichungen der Allgemeinen Relativitätstheorie, zumindest formal, zu reprodu-zieren.[242] Die Eliminierung der innertheoretischen Anomalien und die Reproduktion der Einsteinschen Feldgleichungen funktioniert jedoch, wie sich herausstellte, nur unter der Bedingung, dass man für die Dynamik des Strings (und damit ebenso für die als Gravitonen gedeuteten Stringanregungszustände) eine *zehndimensionale Hintergrundraumzeit* voraussetzt.[243]

## Die nomologische Vereinigung aller Wechselwirkungen

Die Entdeckung, dass die Quantisierung der speziell-relativistischen Dynamik eines oszillierenden Strings nicht nur zu Spin-2-Bosonen als Anregungszuständen des Strings, sondern – unter spezifischen Bedingungen – ebenso zu skalaren und fermionischen Anregungszuständen sowie schliesslich zu Spin-1-Eichbosonen führt, legte dann eine Interpretation des Ansatzes als nomologisch vereinheitlichte Beschreibung aller Wechselwirkungen inklusive der Gravitation nahe.[244]

Und genau dieses Vorliegen des ganzen Spektrums von Anregungszuständen des Strings ist es, das erst zur Finitheit der perturbativen Entwicklungen des Stringansatzes führt. Dies lässt sich dann als weiteres Argument für eine nomologische Vereinigung aller Wechselwirkungen auf der Grundlage der Quantisierung der Dynamik klassischer eindimensionaler Strings anführen. – Und hier kommen nun weitere, zum Teil sehr alte metaphysische Motive mit ins Spiel, die eine nomologische Vereinigung nahelegen: insbesondere die metaphysische Idee einer Einheit der Natur.[245] Aber es lässt sich ebenso auf die vielfältigen Erfolge von nomologischen Vereinheitlichungen im Laufe der Wissenschaftsgeschichte verweisen: von Newtons Vereinigung von irdischer und himmlischer Mechanik über Maxwells Vereinigung von Elektromagnetismus und Optik bis hin zur Quantenfeldtheorie der elektroschwachen Wechselwirkung (*Glashow-Salam-Weinberg-Theorie*). – Für metaphysische Zweifler gibt es zudem noch das rein physikalische Argument der (annähernden) Konvergenz der Wechselwirkungsstärken der bekannten fundamentalen Kräfte inklusive der Gravitation auf der Planck-Ebene, die sich als Indiz für eine nomologisch vereinigte Dynamik auf dieser Ebene anführen lässt.

Zu den notwendigen Voraussetzungen für ein Auftreten des gesamten Spektrums von String-Anregungszuständen – und mithin zu den Voraussetzungen für eine erfolgreiche nomologische Vereinigung aller Wechselwirkungen auf der Grundlage des Stringansatzes – zählt jedoch, ausser dem Vorliegen einer zehndimensionalen Hintergrundraumzeit, die *Supersymmetrie*:[246] eine Symmetrie zwischen fermionischen und bosonischen Zuständen. Interpretiert man das Spektrum dieser Zu-stände im Sinne des Stringansatzes als ein solches, das auch unsere bekannten Materieteilchen und Wechselwirkungsbosonen enthält, so heisst dies, dass es zu jedem bekannten Materieteilchen und zu jedem bekannten Wechselwirkungsboson – da diese sich infolge ihrer jeweiligen Eigenschaften in keine supersymmetrische Beziehung zueinander setzen lassen – einen bisher unbekannten supersymmetrischen Partner geben muss.

---

[242] Das heisst noch nicht notwendigerweise, dass der Stringansatz tatsächlich die Allgemeine Relativitätstheorie mit allen ihren konzeptionellen Implikationen oder phänomenologischen Konsequenzen reproduziert. Siehe weiter unten.

[243] Wollte man diese aus Gründen innertheoretischer Konsistenz anzunehmende höhere Dimensionalität der Raumzeit anschaulich motivieren, so liesse sich hier etwa darauf verweisen, dass schon der *Kaluza-Klein-Ansatz* zur Einbeziehung des Elektromagnetismus in den formalen Apparat der Allgemeinen Relativitätstheorie eine zusätzliche fünfte Dimension annehmen musste, so dass für die Einbeziehung weiterer Wechselwirkungen also mit weiteren zusätzlichen räumlichen Dimensionen zu rechnen war.

[244] Nach der Quantisierung eines klassischen, speziell-relativistischen Strings ohne interne Struktur auf einer flachen Raumzeit ergeben sich Schwingungszustände mit unterschiedlicher Masse und Spin. Neben (in erster Näherung) masselosen Zuständen, die alle schon genannten Zustände einschliessen, gibt es eine unendliche Serie von Zuständen mit Massen im Bereich von jeweils ganzzahligen Vielfachen der Planck-Masse. Alle diese Zustände sind als dynamische Erscheinungsformen des String anzusehen.

[245] Siehe Hedrich (2007).

[246] Siehe Zumino (1979).

Und die Supersymmetrie ist nicht nur für das Auftreten des gesamten Spektrums von String-Anregungszuständen unabdinglich. Vielmehr ist der perturbative Stringansatz nur dann mathematisch konsistent, wenn man die Supersymmetrie einbezieht. – Das einzige originär physikalische Argument dafür, die Supersymmetrie ernstzunehmen, besteht jedoch – solange keine supersymmetrischen Partner zu unseren bekannten Materieteilchen und Wechselwirkungsbosonen gefunden werden – darin, dass sich die Konvergenz der Wechselwirkungsstärken der bekannten fundamentalen Kräfte inklusive der Gravitation mit Einbeziehung der Supersymmetrie noch einmal verbessert.

## Erfolge ?

Als der grösste Erfolg des Stringansatzes wird von seinen Befürwortern gemeinhin die Tatsache angeführt, dass er die Einsteinschen Feldgleichungen als Niederenergienäherung reproduziert.

> *"Moreover, these theories have [...] the remarkable property of* predicting gravity *[...]."* (Witten (1996) 25)

Dass er dies in formaler Hinsicht tut, verdankt sich schlichtweg dem Umstand, dass Spin-2-Bosonen zu den Oszillationszuständen des Strings zählen. Und es ist unter diesen Umständen keinesfalls eine Überraschung; vielmehr lässt sich zeigen, dass das Vorliegen von Spin-2-Bosonen unter sehr allgemeinen Bedingungen notwendigerweise die Einsteinschen Feldgleichungen impliziert.[247]

Aber stellt nun die Reproduktion der Einsteinschen Feldgleichungen als Niederenergienäherung tatsächlich einen entscheidenden Erfolg des Stringansatzes dar?[248] – Einerseits ist es überhaupt erst die Tatsache, dass der Stringansatz infolge des Vorhandenseins von Spin-2-Anregungszuständen die Einsteinschen Feldgleichungen reproduziert, die diesen Ansatz als Anwärter auf eine Theorie der Quantengravitation ins Spiel gebracht haben. Es ist also ganz und gar nicht so, dass man eine Theorie der Quantengravitation auf der Grundlage der Quantisierung der klassischen, speziell-relativistischen Dynamik eines String entwickelt hätte, um daraufhin mit Genugtuung festzustellen, dass diese Theorie tatsächlich in der Weise ihren Zweck erfüllt, indem sie in der Lage ist, die Einsteinschen Feldgleichungen als Niederenergienäherung zu reproduzieren. Vielmehr hat der Stringansatz erst dadurch, dass sich gezeigt hat, dass er über diese Eigenschaft verfügt, seine Rolle als Theorieanwärter im Bereich der Quantengravitation gefunden, anstatt in der Rumpelkammer der Geschichte kurioser, aber erfolgloser Theorieansätze aus der frühen Hadronenphysik vorsichhinzudämmern. – Man könnte also mit einiger Berechtigung sagen, dass er den Bonus, die Einsteinschen Feldgleichungen als Niederenergienäherung zu reproduzieren, für den Wiedereintritt aus der Geschichte in die Aktualität physikalischer Theoriebildung aufgebraucht hat. Erst weitere Erfolge können dann die Angemessenheit dieser Reaktivierung erweisen.

---

[247] *"Now we can use a general result that goes back to Feynman: any theory of an interacting spin two massless particle must describe* gravity. *So string theory must reproduce gravitational physics."* (Giddings (2005) 6) – *"[...] with appropriate caveats, general relativity is necessarily recovered as the low-energy-limit of* any *interacting theory of massless spin-2 particles propagating on a Minkowski background, in which the energy and momentum are conserved (Boulware and Deser 1975)."* (Butterfield / Isham (2001) 59) – *"Weinberg (1995), in his discussion of covariant quantum gravity, showed that, in the vacuum case, one can* derive *the equivalence principle and general relativity from the Lorentz-invariance of the spin-2 quantum field theory of the graviton: the spin-2 theory is* equivalent *to general relativity and follows* from the quantum theory. The upshot of this is that any theory with gravitons is a theory that can accommodate general relativity (in some appropriate limit). This analysis forms the basis of string theory's claim that it is a candidate theory of quantum gravity: since there is a string vibration mode corresponding to a massless spin-2 particle, there is an account of general relativity [...]."* (Rickles (2005) 9)
Dass hier zwar formal die Einsteinschen Feldgleichungen, nicht aber die Allgemeine Relativitätstheorie im vollen Sinne reproduziert wird, wird sogleich zu erörtern sein.
[248] Immerhin gelingt eine solche Reproduktion nahezu keinem der anderen Ansätze zu einer Theorie der Quantengravitation, insbesondere auch nicht der *Loop Quantum Gravity*, die sich ansonsten als fortgeschrittenste, subtilste Variante einer direkten, nicht-perturbativen Quantisierung der Allgemeinen Relativitätstheorie ansehen lässt; siehe Kap. 4.4.

Andererseits sollte die Reproduktion der Einsteinschen Feldgleichungen als Niederenergienäherung vor dem Hintergrund des konzeptionellen Apparats und des modelltheoretischen Instrumentariums des Stringansatzes ohnehin nicht überbewertet werden. Denn sie bedeutet noch nicht, dass der Stringansatz tatsächlich die Allgemeine Relativitätstheorie reproduziert:

> *"It is sometimes asserted that string theory incorporates general relativity, because the Einstein equations (up to higher-order corrections) are a necessary condition that must be satisfied by a spacetime on which a string is to propagate consistently. This is true, but it is not the only necessary condition that must be satisfied. All string theories so far formulated explicitly in terms of the dynamics of the string worldsheets require supersymmetry or an equivalent constraint, to cancel the instability which manifests itself as the presence of a tachyon. In all known cases this requires that the spacetime have a timelike or null killing field. This reduces the possible cases to the measure zero subset of solutions to Einstein's equations in which the geometry is stationary and hence non-dynamical."* (Smolin (2006c) 198f)

Es bleibt ohnehin erst einmal völlig unklar, wie eine hintergrundabhängige Theorie – und alle perturbativen Stringtheorien sind aufgrund des hier verwendeten modelltheoretischen Instrumentariums notwendigerweise hintergrundabhängig[249] – überhaupt eine hintergrundunabhängige Theorie wie die Allgemeine Relativitätstheorie als Niederenergienäherung reproduzieren soll. Damit bleibt es ebenso unklar, welche Beziehung zwischen beiden Theorien überhaupt besteht.

Ein tatsächlicher Erfolg, den der Stringansatz seiner konzeptionellen Ausrichtung – nämlich der mit ihm einhergehenden nomologischen Vereinheitlichung – zuschreiben kann, ist jedoch die schon erwähnte Tatsache, dass sich die perturbative Stringtheorie offensichtlich (soweit dies bis heute tatsächlich überprüft ist) nicht nur als nicht nicht-renormierbar, sondern als finit herausstellen.[250] Zu verdanken ist diese Eigenschaft der Tatsache, dass der Stringansatz ein Spektrum von Oszillationszuständen beschreibt, das sich nicht nur auf natürliche Weise als Realisierung einer nomologischen Vereinheitlichung aller Wechselwirkungen verstehen lässt, sondern insbesondere auch zu einer ge-

---

[249] Siehe dazu weiter unten.
[250] Diese Finitheit ist für eine fundamentale Theorie letztlich sogar unabdinglich, da eine Renormierung immer empirisch zu bestimmende, freie Parameter ins Spiel bringt, die wiederum auf einer fundamentaleren Ebene zu erklären wären. Die Notwendigkeit der Renormierung zeigt also gerade, dass eine Theorie nicht fundamental ist. Nachweise für die Divergenzfreiheit der Stringtheorien gibt es jedoch bisher nur bis zur zweiten störungstheoretischen Ordnung. Es lässt sich aber immerhin plausibel machen, dass die Divergenzfreiheit auch für die höheren Ordnungen gilt.

genseitigen Auslöschung divergenter Komponenten in der perturbativen Entwicklung führt. – Es bleibt jedoch die Frage, welchen Wert dieser Erfolg vor dem Hintergrund der noch zu erörternden konzeptionellen Probleme des Stringansatzes besitzt. Dies gilt gleichermassen auch für die weiteren Erfolge, die der Stringansatz für sich geltend machen kann. Diese sind ohnehin eher von einer Art, die über den Stringansatz und sein modelltheoretisches Instrumentarium hinausweist und Ansätze auf einer anderen modelltheoretischen Grundlage nahelegt:

Obwohl der Stringansatz (genauer: die perturbativen Stringtheorien, in denen sich dieser im wesentlichen erschöpft) mit einer *kontinuierlichen* Hintergrundraumzeit arbeitet, führt er zu Indizien, die sich eindeutig hin auf eine *minimale Länge* interpretieren lassen.[251] Eine solche minimale Länge lässt sich einerseits schon als eine basale Implikation der Ausgedehntheit des String ansehen; sie ergibt sich andererseits aber in konkreter Weise insbesondere im Rahmen der *T-Dualität*.[252] Und sie spielt ebenso für die *D-Brane*[253] eine Rolle: dynamische Entitäten unterschiedlicher Dimensionalität,

---

[251] Siehe Amati et al. (1989).

[252] Dualitätsbeziehungen bestehen darin, dass die Zustandsspektren zweier zueinander dualer Modelle bzw. Theorien unter bestimmten, eindeutig zu spezifizierenden parametrischen Bedingungen identisch sind. Dies deutet auf die unter den jeweiligen spezifischen Bedingungen gegebene 'physikalische' Äquivalenz der zueinander dualen Beschreibungen hin. Dualitätsbeziehungen sind nach Auffassung einiger Theoretiker ein spezifisches Charakteristikum des Stringansatzes. (Siehe etwa Hull / Townsend (1995), Polchinski (1996).) Sie werden offensichtlich durch die Supersymmetrie begünstigt und lassen sich nach heutigem Ermessen vor allem auf die Tatsache zurückführen, dass es zu den für die herkömmlichen Quantenfeldtheorien typischen Divergenzen in supersymmetrischen Theorien infolge der gegenseitigen Auslöschung von bosonischen und fermionischen Anteilen oft gar nicht erst kommt. Die für den Stringansatz relevanten Dualitäten sind bisher jedoch nur für einige Spezialfälle nachgewiesen. Ein allgemeiner Beweis steht immer aus, so dass sie bestenfalls den Status plausibler Hypothesen haben. Dies gilt gleichermassen für alle auf ihnen aufbauenden Schlussfolgerungen.
Der einfachste Typus der im Stringansatz auftretenden Dualitätsbeziehungen ist die *T-Dualität* ('*target-space duality*'). (Siehe etwa Giveon / Porrati / Rabinovici (1994).) Sie ergibt sich für geschlossene Strings, die sich um eine kompakte zylindrische Dimension winden, in der Form, dass das Zustandsspektrum einer Stringtheorie A bei einer zylindrischen Kompaktifizierung mit dem Radius $R_A = R \, l_S$ ($l_S$ steht für die Stringlänge) identisch ist mit dem Zustandsspektrum einer Stringtheorie B bei einer zylindrischen Kompaktifizierung mit dem Radius $R_B = l_S/R$. Für $R_A \, R_B = l_S^2$ kommt es zu den gleichen Zustandsspektren für die beiden T-dualen Stringtheorien A und B. Anschaulich entspricht die T-Dualität dem Austausch von Windungs- und Oszillationszuständen beim Übergang von einer Beschreibung zur anderen. Dies führt zur Etablierung einer kleinsten Länge für die Stringtheorien: Verkleinert man für eine Stringtheorie A den Kompaktifizierungsradius unter die Stringlänge $l_S$, so entspricht das Resultat hinsichtlich der Stringzustände einer Beschreibung auf der Grundlage der zu A T-dualen Theorie B mit einem Kompaktifizierungsradius grösser als $l_S$. Die T-Dualität spiegelt insofern Prozesse, die sich scheinbar im Längenbereich unterhalb von $l_S$ abspielen, grundsätzlich auf Prozesse im Längenbereich oberhalb von $l_S$. Versucht man innerhalb des Stringansatzes Abstände zu beschreiben, die kleiner sind als die Stringlänge, so führt dies nur dazu, dass ein Alternativkonstrukt, das von Abständen handelt, die grösser sind als die Stringlänge, qua Äquivalenz die Beschreibung genauso gut übernehmen kann. Das heisst, dass unterhalb von $l_S$ keine Prozesse stattfinden, die nicht auch schon oberhalb von $l_S$ existieren. Und genau dies lässt sich nun so lesen, dass für die Stringtheorien mittels der T-Dualität die Stringlänge $l_S$ als kleinster Abstand festgeschrieben wird.

[253] Bei der tentativen Erschliessung des Übergangs vom perturbativen zum nicht-perturbativen Bereich ist im Stringansatz zunehmend deutlich geworden, dass die Stringdynamik höherdimensionale Entitäten, topologische Defekte und solitonische Zustände einschliessen kann und diese den nicht-perturbativen Bereich wahrscheinlich sogar dominieren. Es sind bei weitem noch nicht alle Entitäten bekannt, die im Rahmen der Stringtheorien und ihrer nicht-perturbativen Erweiterung auftreten können. Dies ist nicht zuletzt darauf zurückzuführen, dass der nicht-perturbative Bereich der Stringtheorien bisher nur ansatzweise erschlossen werden konnte und die zugrundeliegenden Prinzipien bisher völlig unbekannt sind.
*"[...] every indication is that the string description is useful only [...] [when] the string coupling becomes weak. In the center of the parameter space, not only do we not know the Hamiltonian but we do not know what degrees of freedom are supposed to appear in it. It is likely that they are not the one-dimensional objects that one usually*

die insbesondere für die zaghaften Ansätze einer nicht-perturbativen Erweiterung der perturbativen Stringtheorien von entscheidender Bedeutung sind.

> *"Moreover, there is a sense in which the spacetime coordinates for D-Branes are elevated from numbers to matrices [...]; only at low energy the matrices are diagonal and an ordinary spacetime picture holds."* (Polchinski (1996) 35)

Die darauf aufbauenden neueren, immer noch sehr spekulativen Entwicklungen innerhalb des Stringansatzes – insbesondere die *Matrix-Theorie*[254], von der manche vermuten, dass sie für bestimmte Aspekte der avisierten nicht-perturbativen Formulierung des Stringansatzes eine Rolle spielen könnte – liefern dann weitere Indizien für die Fehlerhaftigkeit der Annahme einer kontinuierlichen Raumzeit:

> *"[...] in the Matrix-theory nonperturbative formulation, the space-time coordinates of the string $x^i$ are replaced by the matrices $(X^i)_m{}^n$. This can perhaps be viewed as a new interpretation of the space-time structure. The continuous space-time manifold emerges only in the long distance region, where these*

---

*thinks of in string theory; it is more likely that they are the coordinate matrices of the D-Branes."* (Polchinski (1999) 21)

Zu den wichtigsten der mittlerweile bekannten Entitäten, die für die Stringtheorien beim Übergang in den nicht-perturbativen Bereich auftreten, gehören *p-* und *D-Branen*. *p-Branen* sind Lösungen des Stringansatzes, für welche die Energie auf eine p-dimensionale, räumliche Hyperfläche konzentriert ist. (Siehe etwa Townsend (1995a), Strominger (1996) und Witten (1996a).) Sie lassen sich nicht aus rein perturbativen Modellen ableiten, sondern erst im Rahmen einer zumindest ansatzweise nicht-perturbativen Beschreibung. p-Branen dehnen sich in p räumlichen Dimensionen (tangentiale Dimensionen) aus und lassen sich in allen anderen räumlichen Dimensionen (transversale Dimensionen) bewegen bzw. lokalisieren. Sie sind invariant unter Translation entlang der tangentialen räumlichen Dimensionen und der Zeit. Bei Entfernung von der Bran in transversaler Richtung nähern sich die Zustandswerte sehr schnell an die jeweilige Vakuumlösung an. Die zulässigen räumlichen Dimensionen p sind für die IIA-Theorie gerade und für die IIB-Theorie ungerade. (Zu den verschiedenen Typen von perturbativen Stringtheorien siehe weiter unten.) – Und manche p-Branen sind *Dirichlet-Branen*, kurz: *D-Branen*. Dies sind dynamische Raumzeit-Submannigfaltigkeiten, die sich dadurch auszeichnen, dass offene Strings auf ihnen enden. (Siehe etwa Douglas (1996, 1997), Bachas (1996, 1997), Polchinski (1995, 1996a), Polchinski / Chaudhuri / Johnson (1996), Johnson (2002), Aspinwall (2004).)

> *"D-branes are topological defects on which the ends of a string can be trapped."* (Polchinski (1996) 22)

Die Anbindung an die D-Bran ist formal dadurch gegeben, dass für die Stringendpunkte transversal zur Bran *Dirichlet-Randbedingungen* gelten. Diese entsprechen gerade einer Fixierung der Endpunkte der Strings auf der Bran. Tangential zur Bran können sich die Enden der Strings (entsprechend den hierfür geltenden *Neumann-Randbedingungen*) frei bewegen, aber eben (gemäss den transversal wirksamen Dirichlet-Randbedingungen) nicht von ihr weg. Die D-Bran definiert sich über die infolge der Dirichlet-Randbedingungen an ihr ankoppelnden offenen Strings. Die Integration der Dirichlet-Randbedingungen führt zur Spezifizierung der Raumzeitpunkte, an denen Strings ankoppeln. Sie führt zur Festlegung der D-Bran. Die Dynamik der D-Bran (Oszillationen, Translationen, Wechselwirkung mit anderen D-Branen und Strings) wird durch die Strings, die auf ihr enden, vollständig bestimmt. D-Branen kommen, wie vermutet wird, ausschliesslich infolge nicht-perturbativer Effekte zustande und treten in den Stringtheorien des Typs I, IIA und IIB als quasi-klassische, dynamisch stabile Lösungen in Erscheinung. Die zulässigen räumlichen Dimensionalitäten p von Dp-Branen (= p-dimensionale D-Branen) hängen vom Typus der entsprechenden Stringtheorie ab: Für den Typ IIA ist p gerade und kann die Werte 0, 2, 4, 6 und 8 annehmen. Für den Typ IIB ist p ungerade und kann die Werte -1, 1, 3, 5, 7 und 9 annehmen. Für den Typ I sind es die Werte 1, 5 und 9. Eine D9-Bran ist raumfüllend und entspricht somit durchgängigen Neumann-Randbedingungen; offene Strings können sich für diesen Fall frei bewegen. Unter Berücksichtigung der Dualitätsbeziehungen zwischen den Theorietypen, für die sich D-Branen ergeben, verkompliziert sich die Sachlage noch einmal. Insbesondere verwandelt die T-Dualität Neumann-Randbedingungen in Dirichlet-Randbedingungen um.

[254] Siehe Banks (1998, 2001a), Banks / Fischler / Shenker / Susskind (1997), Bigatti / Susskind (1997), Taylor (2001).

*matrices are diagonal and commute; while the space-time appears to have a noncommutative discretized structure in the short distance regime."* (Rovelli (1998) 4)

Eine solche Diskretisierung durch die Hintertür[255] auf der Basis eines formal mit dem Kontinuum arbeitenden Ansatzes weist deutlich über das modelltheoretische Instrumentarium des Stringansatzes hinaus. Und man könnte sich wiederum fragen: Wäre ein Ansatz, der auf einer anderen modelltheoretischen Grundlage arbeitet, die von Anfang an auf das Kontinuum verzichtet, für den Bereich der Quantengravitation nicht vielleicht angemessener?

Zu den weiteren erstaunlichen Erfolgen des Stringansatzes, die in dieselbe Richtung weisen, zählt schliesslich die Tatsache, dass sich auf seiner Grundlage – unter Voraussetzung der Gültigkeit der *S-Dualität*[256] – die *Bekenstein-Hawking-Entropie* reproduzieren lässt,[257] allerdings nur für physikalisch hochgradig unrealistische, extremale schwarze Löcher, die mit *BPS-Zuständen*[258] identifiziert werden; diese BPS-Zustände haben die Eigenschaft, dass sie die jeweils maximal mögliche (finite) stringspezifische Ladung tragen.

*"[...] no result has been obtained for a generic black hole."* (Kiefer (2005) 14)

---

[255] In der *Loop Quantum Gravity* fällt die entsprechende Diskretisierung durch die Hintertür noch etwas direkter aus. Siehe Kap. 4.4.

[256] Die *S-Dualität* ('strong-weak duality') ermöglicht es, Konsequenzen des Stringansatzes abzuleiten, die über den perturbativen Kontext hinausgehen. (Siehe etwa Polchinski (1996, 1999), Hull / Townsend (1995), Vafa (1997), Witten (1997).) Sie etabliert Äquivalenzen der Zustandsspektren zwischen Bereichen schwacher Kopplung bei einer Stringtheorie A und solchen starker Kopplung bei einer Theorie B. Die effektive Kopplungsstärke $g_s$ der Stringtheorien ist durch den Erwartungswert des *Dilaton-Feldes* gegeben: $g_s = e^{<\varphi>}$. Dieser Erwartungswert muss nicht notwendigerweise klein sein. Der perturbative Stringansatz funktioniert aber nur bei schwacher Kopplung (kleinem $g_s$). Ansonsten divergieren die störungstheoretischen Entwicklungen. Mit den perturbativen Stringtheorien lässt sich also für eine starke Kopplung (grosses $g_s$) keine Beschreibung der Stringdynamik erreichen. Hier bringt nun die S-Dualität zumindest partielle Abhilfe. Da sie Äquivalenzen der Zustandsspektren zwischen Bereichen starker Kopplung bei einer Stringtheorie A und solchen schwacher Kopplung bei einer Theorie B etabliert, ist es, wenn die perturbativen Prozeduren für eine Theorie A infolge starker Kopplung nicht mehr zum Einsatz kommen können, mit ihrer Hilfe möglich, diese durch die perturbativen Prozeduren einer zu ihr S-dualen Theorie B bei schwacher Kopplung zu ersetzen. Beide führen unter der Bedingungen $g_{sA} = 1/ g_{sB}$ zu identischen Resultaten hinsichtlich der Zustandsspektren. Man kann also den nicht-perturbativen Bereich der einen Theorie mit Hilfe der S-Dualität durch den perturbativen Bereich einer zweiten Theorie erfassen.

Die Gültigkeit der S-Dualität lässt sich jedoch nur indirekt nachweisen, da der perturbative Stringansatz eben nur bei schwacher Kopplung funktioniert. Entscheidend ist hierbei wiederum die Supersymmetrie. Für supersymmetrische Theorien lässt sich nämlich der Bereich der starken Kopplung zumindest punktuell über die nicht-perturbativen *BPS-* (*Bogomol'nyi-Prasad-Sommerfield*)-Zustände erschliessen. Die Energie ist für diese Zustände infolge der Supersymmetrie vollständig durch die Ladung festgelegt, also ohne den Einsatz perturbativer Prozeduren ermittelbar. Sie ist eine Folge der Symmetrie und nicht von der Dynamik abhängig. Für Theorien mit ungebrochener Supersymmetrie gilt: Jeder BPS-Zustand mit Ladung Null hat auch die Energie Null; alle BPS-Zustände mit Ladung Null sind degeneriert; aufgrund der Antikommutatorbeziehungen gibt es keinen Zustand mit geringerer Energie. BPS-Zustände mit Ladung und Energie Null sind notwendigerweise Grundzustände des Systems. Mit den BPS-Zuständen lässt sich der nicht-perturbative Bereich aber eben nur punktuell erschliessen. Tatsächlich ist die S-Dualität daher bisher eine nur für wenige Spezialfälle nachweisbare Vermutung.

[257] Siehe Strominger / Vafa (1996), Das / Mathur (2001), Peet (1998, 2001), Maldacena (1996), Bigatti / Susskind (2001a), David / Mandal / Wadia (2002), Carlip (2008a).

[258] Siehe die vorausgehende Anmerkung zur S-Dualität.

Auch dies spricht einerseits dafür, dass der Stringansatz auf irgendeine (wenn auch vielleicht modelltheoretisch unangemessene oder unvorteilhafte) Weise, die fundamentale Diskretheit auf der Substratebene erfasst. Andererseits lässt sich ihre Ableitbarkeit aus dem Stringansatz, wenn auch unter sehr unrealistischen Bedingungen, als weitere Motivation für die *Bekenstein-Hawking-Entropie* und damit für die Idee der Diskretheit von Raumzeit und physikalischen Prozessen ansehen.

Darüber hinaus kommt es im Stringansatz, wenn sich die diesbezüglich bestehenden Vermutungen erhärten sollten, sogar zu einer Instantiierung des *Holographischen Prinzips*: in Form der sogenannten *AdS/CFT-Dualität* ('*Maldacena Conjecture*').[259] Bisher ist die AdS/CFT-Dualität jedoch nicht mehr als eine gut gestützte, sehr plausible Vermutung.

> *"One cannot prove the AdS/CFT correspondence since we do not have an independent nonperturbative definition of string theory to compare it to."* (Horowitz (2005) 11)

Sollte diese Vermutung zutreffen, so hiesse dies, dass sich im Stringansatz eine direkte Korrespondenz aufzeigen lässt zwischen

(a) einer Stringtheorie, die auf einem *Anti-de-Sitter-Raum*[260] definiert ist, d.h. einer Quantentheorie, die gravitative Wechselwirkungen einschliesst, und

(b) einer *konform-invarianten Eichtheorie*[261], die auf der begrenzenden Oberfläche dieses Anti-de-Sitter-Raums definiert ist und keine gravitative Wechselwirkung enthält.

> *"The basic idea of AdS/CFT duality is to identify a conformally invariant field theory (CFT) on the n-dimensional boundary with a suitable quantum gravity theory in the (n+1)-dimensional AdS bulk."* (Schwarz (1998) 8f)

Die AdS/CFT-Dualität stellt also eine Korrespondenz (Identifizierbarkeit) her zwischen einer Gravitationstheorie, definiert auf einer n-dimensionalen Raumzeit, und einer gravitationsfreien Feldtheorie, definiert auf dem (n-1)-dimensionalen Rand dieser Raumzeit:

> *"The theory in AdS includes gravity, since any string theory includes gravity. So in the end we claim that there is an equivalence between a gravitational theory and a field theory. However, the mapping between the gravitational and field theory degrees of freedom is quite non-trivial since the field theory lives in a lower dimension. In some sense the field theory (or at least the set of local observables in the field theory) lives on the boundary of spacetime. One could argue that in general any quantum gravity theory in AdS defines a conformal field theory (CFT) 'on the boundary'."* (Aharony et al. (1999) 8)

---

[259] Siehe Maldacena (2003), Aharony et al. (1999), Klebanov (2001).

[260] Der *De-Sitter-Raum* (*dS*) ist eine maximal-symmetrische Lösung der Einsteinschen Feldgleichungen mit leerem Raum und positiver kosmologischer Konstante, die zu einer beschleunigten Ausdehnung des Raumes führt. Diese Lösung wurde von Willem de Sitter im Jahre 1917 gefunden. Der *Anti-de-Sitter-Raum* (*AdS*) ist entsprechend eine maximal-symmetrische Lösung der Einsteinschen Feldgleichungen mit leerem Raum und negativer kosmologischer Konstante. Der Anti-de-Sitter-Raum hat eine räumliche Grenzoberfläche im Unendlichen, die für die Implikationen, die sich entsprechend dem holographischen Prinzip deuten lassen, entscheidend ist.

[261] *Konform invariante Eichtheorien* (*CFTs*) zeichnen sich dadurch aus, dass ihre Kopplungskonstante, etwa im Gegensatz zu den Eichtheorien des quantenfeldtheoretischen Standardmodells, energieunabhängig ist.

Beide Theorien sind hinsichtlich ihrer phänomenologischen Konsequenzen äquivalent – arbeiten allerdings mit unterschiedlichen Freiheitsgraden[262] und grundsätzlich unterschiedlichen raumzeitlichen Voraussetzungen.

> *"[...] N = 4 U(N) Yang-Mills theory could be the same as ten dimensional superstring theory on AdS₅ x S⁵."* (Aharony et al. (1999) 6) – *"AdS/CFT Correspondence: String theory on spacetimes which asymptotically approach the product of anti de Sitter (AdS) and a compact space, is completely described by a conformal field theory 'living on the boundary at infinity'. [...] At first sight this conjecture seems unbelievable. How could an ordinary field theory describe all of string theory?."* (Horowitz (2005) 10)

Dies lässt die schon aus der Erörterung des *Holographischen Prinzips*[263] bekannte Frage aufkommen, was denn nun die tatsächlich wirksamen Freiheitsgrade sind, die der entsprechenden physikalischen Situation zugrundeliegen. Und dies wiederum zieht in grundsätzlicher Weise die realistische Interpretierbarkeit des jeweilig der Beschreibung zugrundeliegenden raumzeitlichen Hintergrundes nicht unerheblich in Zweifel.

## Probleme

Die Probleme des Stringansatzes sind vielfältiger Art.[264] Aber die wenigsten dieser Probleme sind vorrangig physikalischer Natur. Solche *externe Probleme* betreffen nicht zuletzt die immer noch unbewältigte Ankopplung an die bestehende Phänomenologie des quantenfeldtheoretischen Standardmodells der Elementarteilchenphysik (bzw. die Reproduktion seines empirischen Gehalts) sowie die ebenso ungelöste Frage nach den physikalischen Modalitäten des Zustandekommens der makroskopischen Raumzeit. Der Stringansatz trägt bisher nichts zum Verständnis der Raumzeit bei. Und er macht – wie im wesentlichen alle Ansätze zur Quantengravitation – keine konkreten, quantitativen Vorhersagen.

Die meisten der Probleme des Stringansatzes sind jedoch innertheoretischer Natur. Es handelt sich um *interne Probleme*, die aus der konzeptionellen Anlage wie aus dem modelltheoretischen Instrumentarium des Ansatzes resultieren, und die dann mit ihren Implikationen und daraus wiederum resultierenden Folgeproblemen, meist über mehrere Stufen hinweg, letztlich auch die Lösung der externen, physikalischen Probleme – insbesondere die Ableitung konkreter quantitativer Vorhersagen – unmöglich machen. Diese internen Probleme überwiegen beim Stringansatz hinsichtlich der zur Anwendung kommenden Bewältigungsstrategien bei weitem die externen, physikalischen Anforderungen, die an eine Theorie der Quantengravitation zu stellen sind.

Eines der elementarsten dieser internen Probleme besteht etwa darin, dass es – nach Auslotung der innertheoretischen Konsistenzanforderungen – immer noch (mindestens) fünf verschiedene (anoma-liefreie) perturbative Stringtheorien[265] gibt, und nicht etwa eine. Es wird zwar vermutet, dass sich diese fünf Stringtheorien als perturbative Entwicklungen bzw. Grenzfälle einer übergeordneten nicht-perturbativen, analytischen Theorie ansehen lassen, von der diese gerade unterschiedliche Aspekte erfassen. Es ist aber eben nicht so, dass hier (wie dies in vielen anderen Bereichen der Physik tatsächlich der Fall ist) perturbative Entwicklungen zu einer schon bekannten nicht-perturbativen Theorie zum Einsatz kommen, weil diese Theorie nicht oder nur sehr schwierig analytisch lösbar ist. Vielmehr ist diese Theorie im Falle des Stringansatzes gar nicht bekannt; es ist nicht einmal klar, ob es sie überhaupt gibt. Was bekannt ist, sind die fünf perturbativen Stringtheorien, von denen vermutet wird, dass sie gerade unterschiedliche Grenzfälle bzw. perturbative Entwicklungen zu einer solchen übergeordneten Theorie darstellen. Entsprechend nehmen die meisten Versuche, diese unbekannte nicht-perturbative Stringtheorie – schon im vorhinein als '*M-Theorie*'[266] apostrophiert – zu finden bzw. zu konstruieren, von den Dualitätsbeziehungen[267] ihren Ausgang, die sich zwischen den fünf perturbativen Stringtheorien (und der hier nun hinzukommenden, alten elfdimensionalen *Supergravity*-Theorie[268]) etablieren lassen.[269]

---

[262] Es sind inzwischen einige konkretere Beispiele für Korrespondenzen zwischen den Freiheitsgraden bekannt, die sich jeweils für die zueinander dualen Beschreibungen der gleichen physikalischen Situation ergeben:
  *"A string stretched across the bulk is represented by a point charge in the dual CFT."* (Bousso (2002) 40)
Einen systematischen Versuch der Erschliessung dieser Korrespondenzen unternimmt vor allem Freidel (2008).
[263] Siehe Kap. 3.2.
[264] Eine detailliertere Erörterung der Probleme des Stringansatzes findet sich in Hedrich (2002, 2002a, 2006, 2007, 2007a).

---

[265] Diese werden als Typ I, Typ IIA, Typ IIB, Heterotisch-SO(32) sowie Heterotisch-$E_8xE_8$ bezeichnet:

*"The Type IIA and Type IIB strings differ in that in the latter theory [the] clockwise / counterclockwise vibrations are identical, while in the former, they are exactly opposite in form.* Opposite *has a precise mathematical meaning in this context, but it's easiest to think about in terms of the spin of the resulting vibrational patterns in each theory. In the Type IIB theory, it turns out that all particles spin in the same direction (they have the same chirality), whereas in the Type IIA theory, they spin in both directions (they have both chiralities). [...] The heterotic theories differ in a similar but more dramatic way. Each of their clockwise string vibrations looks like those of the Type II string (when focusing on just the clockwise vibrations, the Type IIA and Type IIB theories are the same), but their counterclockwise vibrations are those of the original bosonic string theory. Although the bosonic string has insurmountable problems when chosen for both clockwise and counterclockwise string vibrations, in 1985 David Gross, Jeffrey Harvey, Emil Martinec, and Ryan Rhom (all then at Princeton University and dubbed the 'Princeton String Quartet') showed that a perfectly sensible theory emerges if it is used in combination with the Type II string. The really odd feature of this union is that it has been known [...] that the bosonic string requires a 26-dimensional spacetime, whereas the superstring [...] requires a 10-dimensional one. So the heterotic string constructions are a strange hybrid – a* heterosis *– in which counterclockwise vibrational patterns live in 26 dimensions and clockwise patterns live in 10 dimensions! Before you get caught up in trying to make sense of this perplexing union, Gross and his collaborators showed that the extra 16 dimensions on the bosonic side must be curled up into one of two very special higher-dimensional doughnutlike shapes, giving rise to the Heterotic-O and Heterotic-E theories. Since the extra 16 dimensions on the bosonic side are rigidly curled up, each of these theories behaves as though it really has 10 dimensions, just as in the Type II case. Again, both heterotic theories incorporate a version of supersymmetry. Finally, the Type I theory is a close cousin of the Type IIB string except that, in addition to the closed loops of string [...] it also has strings with unconnected ends – so called* open strings.*"* (Greene (1999) 405f)

[266] Siehe etwa Witten (1995), Duff (1996), Townsend (1995, 1996), Schwarz (1996, 1997), Green (1998), Sen (1998, 1998a), Mukhi (1997) sowie Lerche (2000).

[267] Vor allem die verschiedenen Ausprägungen der S-Dualität erweisen sich hier als entscheidend: Für eine flache sechsdimensionale Raumzeit etwa entspricht der Grenzwert der starken Kopplung der Typ-IIA-Theorie der schwachen Kopplung für die heterotische Theorie. Für eine flache zehndimensionale Raumzeit entspricht der Grenzwert der starken Kopplung der Typ-I-Theorie (unorientierte offene Strings) der schwachen Kopplung für die heterotische SO(32)-Theorie (orientierte geschlossene Strings). Damit ergibt sich eine Dualität zwischen einer Theorie mit unorientierten offenen Strings und einer Theorie mit orientierten geschlossenen Strings. Ebenfalls für eine flache zehndimensionale Raumzeit ergibt sich eine Äquivalenz zwischen dem Grenzwert der starken Kopplung der Typ-IIB-Theorie und der schwachen Kopplung der selben Theorie. Für die Typ-IIB-Theorie ist also für $g_s$ und für $1/ g_s$ das Zustandsspektrum gleich.

[268] Für eine flache zehndimensionale Raumzeit – und eine hier nun für den Fall starker Kopplung hinzutretende, kreisförmig aufgerollte, elfte Dimension – ist der Grenzwert der starken Kopplung sowohl der Typ-IIA-Theorie als auch der heterotischen $E_8xE_8$-Theorie identisch mit dem der schwachen Kopplung für die elfdimensionale *Supergravity*-Theorie

Zur *M-Theorie* gibt es die unterschiedlichsten Vorstellungen: Einerseits wird vermutet, dass es sich um eine elfdimensionale analytische Theorie handelt, deren unterschiedliche Reihenentwicklungen eben gerade den zehndimensionalen perturbativen Stringtheorien und der elfdimensionalen *Supergravity*-Theorie entsprechen. Dann wird – im Verbund wie auch unabhängig von der vorausgehenden Hypothese – vermutet, dass sich bestimmte Aspekte dieser *M-Theorie* als *Matrix-Theorie*[270] darstellen lassen, deren Variablen gerade die Dynamik von D-Branen erfassen. Andererseits gibt es aber auch die radikalere Vermutung, dass es sich bei der *M-Theorie* um eine vollends prägeometrische Theorie handeln muss, welche die Emergenz der Raumzeit – und in einer jeweils spezifischen Näherung auch die der zehndimensionalen Hintergrundraumzeiten der perturbativen Stringtheorien – in ihrem Zustandekommen erklären müsste.[271] Die Versuche, mittels der Dualitätsbeziehungen, die sich zwischen den fünf bekannten perturbativen Theorien etablieren lassen, eine konkrete und konsistente nicht-perturbative, analytische Formulierung des Stringansatzes zu entwickeln, sind jedoch bisher ohne Erfolg geblieben.[272] Auch fehlt insbesondere immer noch ein physikalisch motivierbares Grundprinzip, mit dem sich die konzeptionelle und modelltheoretische Basis des Stringansatzes rechtfertigen liesse. Dieser führt insofern ein ausschliesslich durch mathematisch-modell-

---

(siehe Cremmer / Julia / Scherk (1978).) Obwohl letztere, wie heute klar ist, nicht als fundamentale Theorie der Quantengravitation angesehen werden kann, kommt sie jedoch unter Umständen als effektive Niederenergienäherung einer fundamentaleren Theorie in Frage. Immerhin verfügt die Supergravity über keine dimensionslosen Parameter.

[269] Das Netz der Dualitäten deutet nach Ansicht der Stringtheoretiker gerade darauf hin, diese perturbativen Theorien als unterschiedliche Näherungen einer grundlegenderen nicht-perturbativen Theorie zu verstehen.

> *"For string theory the change in viewpoint is perhaps even wider and includes the discovery that there is only one theory. / For weak coupling the five string theories – and the wild card, eleven-dimensional supergravity – are all different. That is why they have been traditionally understood as different theories. Understanding them as different limits to one theory requires understanding what happens for strong coupling. / The novelty of the last couple of years, in a nutshell, is that we have learned that the strong-coupling behavior of supersymmetric string theories and field theories is governed by a web of dualities relating different theories. When one description breaks down because a coupling parameter becomes large, another description takes over. / [...] we learn that the different theories are all one. The different supertheories studied in different ways in the last generation are different manifestations of one underlying, and still mysterious, theory, sometimes called M-theory, where M stands for magic, mystery or membrane, according to taste. This theory is the candidate for superunification of the forces of nature. It has eleven-dimensional supergravity and all the traditionally studied string theories among its possible low-energy manifestations."* (Witten (1997) 32)

Alle perturbativen Stringtheorien entsprechen dieser Auffassung zufolge einer spezifischen 'Koordinatenwahl' im Parameterraum dieser grundlegenderen, aber noch unbekannten *M-Theorie*.

[270] Siehe etwa Banks (1998, 2001a), Banks et al. (1997), Bigatti / Susskind (1997), Bilal (1999), Taylor (2001).

> *"Matrix Theory is supposed to be the Discrete Light Cone Quantization (DLCQ) of M-theory. The spectrum of the DLCQ theory diverges more rapidly at large energy than that of the limiting, decompactified, theory, for 9 or fewer asymptotically flat dimensions. At D = 5 the DLCQ spectrum blows up faster than an exponential of light cone energy and we don't know how to define it. It is of greatest interest to work out the form of the decompactified quantum theory."* (Banks (2003) 29)

Aber die Matrix-Theorie ist ohnehin nur unter spezifischen Bedingungen einsetzbar:

> *"Not only is Matrix Theory awkward to use, but (more fundamentally) it seems to be applicable to only a limited class of quantum vacua. In particular, it does not seem able to describe realistic vacua in which all but four dimensions are compactified."* (Schwarz (1998) 6)

Insbesondere ist auch die Matrix-Theorie, wie schon die perturbativen Stringtheorien, keine hintergrundunabhängige Theorie.

[271] Sollte sich die *M-Theorie* tatsächlich formulieren lassen, könnte sich dies sehr leicht als unabwendbar herausstellen. Siehe weiter unten.

[272] Von einer solchen nicht-perturbativen Formulierung des Stringansatzes erhoffen sich manche zugleich eine Lösung für das Problem der Hintergrundabhängigkeit der perturbativen Stringtheorien. Siehe weiter unten.

theoretische Überlegungen und Konsistenzforderungen angetriebenes Eigenleben ohne weitergehende Verankerung in originär physikalischen Motivationen.

Die meisten der internen (und ebenso der externen) Probleme des Stringansatzes resultieren als Folgeprobleme aus konzeptionellen Komponenten des Stringansatzes, die aus Gründen der modelltheoretischen Kohärenz unumgänglich erscheinen, die phänomenologisch jedoch letztendlich völlig unmotiviert und physikalisch immerhin fragwürdig sind. Dies betrifft insbesondere die Supersymmetrie und die höhere Dimensionalität der Raumzeit. Nur eine supersymmetrische perturbative Stringtheorie, die auf einer zehndimensionalen Hintergrundraumzeit definiert ist, hat die spezifische Eigenschaft, mathematisch konsistent und anomaliefrei zu sein und das für eine nomologische Vereinheitlichung notwendige Spektrum an Oszillationszuständen des Strings aufzuweisen.

Aber nichts in der Natur, ausser der Konvergenz der (mutig über 16 bis 18 Grössenordnungen extrapolierten) Wechselwirkungsstärken auf der Planck-Ebene, spricht für die Supersymmetrie – schon gar nicht für eine ungebrochene Supersymmetrie, wie sie der Stringansatz erst einmal nahelegt. Es sind bisher keine supersymmetrischen Partner zu unseren bekannten Materieteilchen und Wechselwirkungsbosonen nachweisbar. Läge eine ungebrochene Supersymmetrie vor, so hätten diese Partnerteilchen die gleiche Masse (bzw. als String-Oszillationszustände die gleiche Energie) wie unsere bekannten Teilchen und wären längst in Erscheinung getreten. Ebenso spricht bisher nichts für eine zehndimensionale Raumzeit. Beide Komponenten – Supersymmetrie und höherdimensionale Raumzeit – sind ausschliesslich mathematisch-modelltheoretisch motiviert.

Der Stringansatz müsste also zumindest erklären können, wieso wir bisher keine supersymmetrischen Teilchen gesehen haben und wieso sich bisher keine Indizien für eine zehndimensionale Raumzeit aufzeigen liessen. Er müsste etwa die Mechanismen für eine gebrochene Supersymmetrie deutlich machen. Und er müsste, um die entsprechenden theoretischen Implikationen plausibel zu machen, nicht zuletzt quantitative Vorhersagen für die Massen der jeweiligen Superpartner liefern. Aber solche quantitative Vorhersagen scheitern letztendlich an gravierenden und folgenreichen Problemen, die wiederum gerade mit der Supersymmetrie und mit der höheren Dimensionalität der Raumzeit zusammenhängen. Versucht man etwa, aus dem Stringansatz, und damit auf der Grundlage einer aus Konsistenzgründen notwendigerweise zehndimensionalen perturbativen Stringtheorie mit ungebrochener Supersymmetrie, die entsprechenden Implikationen für unsere vierdimensionale phänomenologische Raumzeit (mit – entsprechend der vorliegenden Phänomenologie – gebrochener Supersymmetrie) abzuleiten, so führt dies nicht etwa zu einem eindeutigen Resultat. Vielmehr zeigt sich, dass die unterschiedlichen Möglichkeiten für den Übergang von der zehndimensionalen Theorie zum vierdimensionalen Szenario, etwa durch Kompaktifizierung[273] der überzähligen räum-

---

[273] Mit der raumzeitlichen Supersymmetrie sind nur Kompaktifizierungen vereinbar, bei der sich die zehndimensionale Raumzeit der (perturbativen) Stringtheorien als Produkt aus einer vierdimensionalen Minkowski-Raumzeit und einem kompakten sechsdimensionalen Calabi-Yau-Raum darstellt. Calabi-Yau-Räume sind Kähler-Mannigfaltigkeiten mit verschwindender erster Chern-Klasse. Kähler-Mannigfaltigkeiten wiederum sind komplexe Mannigfaltigkeiten mit spezieller Hermitscher Metrik. Komplexe Mannigfaltigkeiten bzw. komplexe n-Faltigkeiten lassen sich durch n komplexe Koordinaten charakterisieren und haben entsprechend d = 2n Dimensionen. Calabi-Yau-n-Faltigkeiten haben insofern n komplexe und somit 2n reale Koordinaten. Die für die Kompaktifizierung in den supersymmetrischen Stringtheorien infragekommenden sechsdimensionalen Calabi-Yau-Räume sind also CY-3-Faltigkeiten. Nach Yaus Theorem existiert für CY-n-Faltigkeiten eine Ricci-flache Metrik mit SU(n)-Holonomie. Dies ist gerade die Voraussetzung für die Supersymmetrie. Die Mathematik der Calabi-Yau-Räume ist bisher jedoch nur unzureichend erschlossen. Nach heutiger Einschätzung gibt es etwa zehntausend topologisch unterschiedliche Typen von sechsdimensionalen Calabi-Yau-Räumen, die sich noch einmal in ihren weiteren geometrischen Parametern unterscheiden können. Die

lichen Dimensionen, zu einem riesigen Spektrum[274] vierdimensionaler niederenergetischer Szenarien führen: der *String Landscape*.[275]

Einerseits enthält diese *String Landscape* Niederenergieszenarien[276] mit den unterschiedlichsten Eichsymmetrien und den unterschiedlichsten Oszillations- bzw. Teilchenspektren,[277] was die Aussicht auf eine tatsächliche Vorhersagekraft der Theorie, sollte sie einmal bestanden haben, vollends zunichte macht, da sich kein physikalisch motivierbares Selektionskriterum ausmachen lässt.[278] Andererseits liess sich, trotz intensiver Bemühungen, kein Szenario finden, welches das Standardmodell der Teilchenphysik, seine Phänomenologie oder wenigstens seine Eichsymmetrien reproduziert. Dies legt nahe, dass der Stringansatz vielleicht schlichtweg als empirisch inadäquat anzusehen ist, da er die tatsächlich vorliegenden physikalischen Verhältnisse einfach nicht beschreibt (was schon an ein Wunder grenzt angesichts des Übermasses an Szenarien, die er einschliesst). – Sollte eine solche Auffassung noch eine weitere Motivation benötigen, so findet sich diese vermutlich im zentralen konzeptionellen Problem des Stringansatzes, das dieser von den Quantenfeldtheorien und ihrer modelltheoretischen Grundlage geerbt hat, und das auch schon für die *Kovariante Quantisierung* des Gravitationsfeldes zum zentralen konzeptionellen Problem wurde: die Hintergrundabhängigkeit.

---

Geometrie und Topologie der kompakten Mannigfaltigkeit lässt sich dabei durch die sogenannten Moduli charakterisieren. Moduli sind geometrische Parameter, welche die Grösse und Form der kompakten Mannigfaltigkeit erfassen, vor allem aber ihre Topologie. Jeder Punkt im Raum der Moduli-Varianten steht für eine spezifische sechsdimensionale kompakte Mannigfaltigkeit. Die Kompaktifizierung selbst ist infolge der postulierten Ausdehnung der kompakten Dimensionen, die sich im Bereich der Planck-Länge bewegt, letztlich unbeobachtbar. Sie macht sich aber dadurch bemerkbar, dass die Geometrie des kompakten Raumes die möglichen Schwingungsmuster des String auf der resultierenden Vierer-Raumzeit bestimmt.

[274] Für die Zahl der Szenarien innerhalb dieses Spektrums kursieren Zahlen zwischen $10^{100}$ bis $10^{1000}$.

[275] Siehe etwa Banks (2003, 2004), Banks / Dine / Gorbatov (2004), Dine (2004), Douglas (2003, 2004, 2004a, 2004b, 2006), Susskind (2004, 2005, 2007), Freivogel / Susskind (2004).

[276] Die unterschiedlichen Möglichkeiten einer Kompaktifizierung der überzähligen Dimensionen treten innerhalb der Vierer-Raumzeit jeweils sehr unterschiedlich in Erscheinung. Unterschiedliche Kompaktifizierungsmodi führen zu unterschiedlicher Physik innerhalb der Vierer-Raumzeit. Und mit der Vielfalt an Optionen für die Geometrie des kompakten Raumes kommt es innerhalb des Stringansatzes zur Unbestimmtheit und Unvorhersagbarkeit hinsichtlich der physikalischen Implikationen für die Vierer-Raumzeit.

[277] Es gibt allerdings nicht die geringsten konkreten, quantitativen Vorhersagen, etwa für die zu erwartenden Massen innerhalb des jeweiligen Teilchenspektrums – nicht einmal für eines des vielen vierdimensionalen niederenergetischen Stringszenarien.

[278] In Ermangelung irgendeiner Aussicht auf einen physikalischen Selektionsprozess, etwa auf der Grundlage energetischer Minima, bringen einige Stringtheoretiker hier dann eine epistemische Selektion entsprechend dem schwachen anthropischen Prinzip ins Spiel: Die gesamte *String Landscape* wird in Form eines Multiversums als realisiert angesehen. Wir finden uns dann notwendigerweise in einem der wenigen Universen wieder, das eine Niederenergiephysik aufweist, die mit unserer Existenz vereinbar ist. – Der gedankliche Hintergrund hierfür ist der folgende: Dass eine geringe Abweichung hinsichtlich der fundamentalen Konstanten, die in unserem Universum vorliegen, die Unmöglichkeit unserer Existenz zur Folge hat, bestätigt (unabhängig von der Stringtheorie) die Besonderheit unseres Universums. Als Erklärung für diese Feinabstimmung kommt nur (i) ein Zufall wahrhaft kosmischen Ausmasses, (ii) ein absichtliches Design, das unserem Universum zugrundeliegt, oder eben (iii) ein Multiversum unterschiedlichster Szenarien in Frage, welches die Feinabstimmung unseres Universums als Ergebnis einer epistemischen Selektion im Sinne des schwachen anthropischen Prinzips plausibel macht. Schliesst man die ersten beiden Alternativen als letztendlich unwissenschaftlich aus, so kommen hier dann die Theorien ins Spiel, die ein solches Multiversum nahelegen. Siehe Hedrich (2006, 2007).

## *Die Hintergrundabhängigkeit*

Als unmittelbare konzeptionelle Erweiterung der Quantenfeldtheorien, bei der deren grundsätzliches modelltheoretisches Instrumentarium beibehalten wird, setzen die perturbativen Stringtheorien eine Hintergrundraumzeit mit fester Metrik voraus, die nicht in die auf ihr stattfindenden Prozesse dynamisch eingebunden ist. Für eine Quantentheorie der Gravitation – und mithin der Raumzeit –, für die noch dazu behauptet wird, dass sie die Allgemeine Relativitätstheorie als niederenergetischen Grenzfall reproduziert, ist diese Hintergrundabhängigkeit, für die der Stringansatz nicht den geringsten physikalisch motivierten Grund[279] liefert, als Rückschritt gegenüber der angeblich reproduzierten klassischen Theorie nicht nur unmotiviert, sondern letztlich konzeptionell äusserst problematisch.

Insbesondere ist es konzeptionell völlig unplausibel anzunehmen, dass sich die gravitative Wechselwirkung durch den Austausch von Gravitonen auf einer schon vorhandenen Raumzeit mit einer schon festgelegten Metrik beschreiben lässt – unabhängig davon, ob man diese Gravitonen als fundamentale Entitäten oder als Oszillationszustände des Strings versteht. Wenn der Austausch von Gravitonen die gravitative Wechselwirkung, und mithin das (quantisierte) Gravitationsfeld, repräsentieren soll, und wenn man das Gravitationsfeld – im Sinne der hier angeblich reproduzierten Allgemeinen Relativitätstheorie – als mit dem metrischen Feld identisch ansieht, so kann dieser Austausch nicht auf einer schon vorhandenen Metrik beschrieben werden. Die Metrik, auf der hier das quantisierte Gravitationsfeld beschrieben werden soll, entspricht schon dem Gravitationsfeld. Oder anders formuliert: Die Metrik, auf der sich die String-Dynamik angeblich abspielt, kann eigentlich erst auf der Grundlage der Konsequenzen der String-Dynamik zustandekommen. Sie kann nicht schon vorausgesetzt werden. Dann kann aber das modelltheoretische Instrumentarium der Quantenfeldtheorien, das der Stringansatz voraussetzt, eben nicht verwendet werden. Das modelltheoretische Instrumentarium des Stringansatzes steht somit in direktem Widerspruch mit seinen Implikationen, wenn diese tatsächlich die Reproduktion der Allgemeinen Relativitätstheorie einschliessen.

Dieses Problem tritt letztlich für jede Theorie auf, die mit einer Gravitonendynamik arbeitet, um die Dynamik des quantisierten Gravitationsfeldes – und damit die des quantisierten metrischen Feldes – zu erfassen. Da das Konzept des Gravitons das ausschliessliche Resultat des modelltheoretischen Instrumentariums der Quantenfeldtheorien ist, dieses Instrumentarium jedoch eine Hintergrundraumzeit mit vorgegebener fester Metrik voraussetzt, die gerade mit einer der Gravitonendynamik entsprechenden dynamischen Metrik unvereinbar ist, verliert die Idee, die gravitative Wechselwirkung – und mithin die Dynamik der Metrik – mittels des Austauschs von Gravitonen zu beschreiben, jede Berechtigung.[280] Sie erweist sich schlichtweg als konzeptionell widersprüchlich.[281]

---

[279] Es wäre im Rahmen eines solchen Ansatzes immerhin zu motivieren, wie ein hintergrundunabhängiger klassischer Grenzfall aus einer hintergrundabhängigen Quantentheorie resultieren kann.

[280] An dieser Sachlage ändert sich auch nichts, wenn man für die Hintergrundraumzeit gekrümmte Metriken zulässt (siehe etwa Sanchez (2003)), wie sie auch schon in den Quantenfeldtheorien einbeziehbar sind.

*"Does it really make sense to postulate gravitational exchange quanta in a theory that already is endowed with a curved spacetime?"* (Riggs (1996) 9)

[281] Die konzeptionelle Inkonsistenz von Ansätzen, die mit einer Gravitonenphysik operieren, ist also durchaus noch einmal zu unterscheiden von den Problemen, in die Theorien geraten, die auf eine fundamentale Beschreibung abzielen und dennoch mit einer (festen) Hintergrundraumzeit operieren, die aber eben nicht von einer Dynamik von Gravitonen auf dieser Hintergrundraumzeit ausgehen, sondern die Gravitation als emergentes, intrinsisch klassisches Phänomen

In den Quantenfeldtheorien des Standardmodells wird die Hintergrundabhängigkeit deshalb noch nicht zum konzeptionellen Problem, weil diese Theorien gerade nicht versuchen, die Gravitation – und damit eine dynamische Raumzeit – zu beschreiben. Wenn jedoch mit dem modelltheoretischen Instrumentarium der Quantenfeldtheorien genau dies erreicht werden soll, und wenn auf dieser Grundlage auch noch die Allgemeine Relativitätstheorie – wie dies im Stringansatz behauptet wird – reproduziert werden soll, erweist sich dieses modelltheoretische Instrumentarium, das notwendigerweise eine Hintergrundraumzeit mit fest vorgegebener Metrik voraussetzt, als nicht mehr anwendbar. Es ist gerade die allgemein-relativistische Identifizierung von Gravitation und dynamischer Metrik, die dazu führt, dass sich das modelltheoretische Instrumentarium der Quantenfeldtheorien nicht auf die Gravitation anwenden lässt.

## Perspektive: Emergenz der Raumzeit

Die Notwendigkeit einer hintergrundunabhängigen (und gleichzeitig nicht-perturbativen, analytischen) Formulierung des Stringansatzes wird inzwischen von nahezu allen Stringtheoretikern eingesehen.

> *"Finding the correct mathematical apparatus for formulating string theory without recourse to a preexisting notion of space and time is one of the most important issues facing string theorists. An understanding of how space and time emerge would take us a huge step closer to answering the crucial question of which geometrical form actually* does *emerge."* (Greene (1999) 380)

Sollte sich eine nicht-perturbative Stringtheorie – die oft bemühte *M-Theorie* – tatsächlich über das Netzwerk der Dualitäten, Korrespondenzen und Symmetrien entwickeln lassen, das sich zwischen den perturbativen Stringtheorien und der *Supergravity*-Theorie etablieren lässt, so müsste diese Theorie aufgrund der unterschiedlichen Hintergrundraumzeiten, mit denen die perturbativen Stringtheorien und die *Supergravity*-Theorie arbeiten, vermutlich ohnehin hintergrundunabhängig sein.

> *"If M theory is to be a unification of all the different background-dependent string theories, and hence treat them all on an equal footing, it cannot be formulated in terms of a single spacetime background. Hence, we expect that M theory must be a background-independent theory."* (Smolin (2006c) 222)

Aber eine solche *M-Theorie* gibt es noch nicht. – Dafür könnte es gute Gründe geben, die vielleicht sogar in der konzeptionellen Unmöglichkeit einer solchen hintergrundunabhängigen, nicht-perturbativen Fortschreibung der hintergrundabhängigen perturbativen Stringtheorien, die in grundlegender Weise auf der Idee einer symmetrieinduzierten Vereinheitlichung beruhen, zu suchen wären:

> *"[...] there is a built-in contradiction to the idea of M theory. [...] the idea that M theory is based on the largest possible symmetry is one that only makes sense in a background-dependent context. But as we have also just seen, M theory must be background-independent."* (Smolin (2006c) 222)

---

behandeln, zu dem es auf der Grundlage eines Substrats kommt, für das weder die Gravitation noch Gravitonen irgendeine Rolle spielen.

Die im Rahmen der Quantenfeldtheorien und schliesslich des Stringansatzes als grundlegend ange-sehen, immer umfassenderen Symmetrien und Symmetriegruppen sind ein typisches Zeichen hintergrundabhängiger Theorien. Es ist nicht zuletzt der raumzeitliche Hintergrund des dynami-schen Geschehens, der die in der jeweiligen Theorie nutzbaren Symmetrien überhaupt erst definier-bar werden lässt. Daher könnte es ohne weiteres sein, dass die *M-Theorie* als hintergrundunabhän-gige nicht-perturbative Fortschreibung des hintergrundabhängigen perturbativen Stringansatzes immer eine Chimäre bleiben wird.

Bisher jedenfalls gibt es im Hinblick auf eine hintergrundunabhängige Variante des Stringansatzes nicht viel mehr als eine halbwegs anschauliche Idee, die dann aber sogleich eine vollends prägeo-metrische Basis nahelegt: eine nicht-raumzeitliche fundamentale Dynamik, die erst die Raumzeit hervorbringt. – Mit den bildreichen Worten eines Stringtheoretikers, der allerdings für die Veran-schaulichung der grundlegenden Motivationen seines prägeometrischen Stringszenarios unglückli-cherweise weiterhin auf den Mythos des Gravitons setzt:

> *"But we can still ask whether the geometrical model of spacetime that plays such a pivotal role in ge-neral relativity and in string theory is solely a convenient shorthand for the spatial and temporal rela-tions between various locations, or whether we should view ourselves as truly being embedded in* something *when we refer to our immersion within the spacetime fabric. / Although we are heading into speculative territory, string theory does suggest an answer to this question. The graviton, the smallest bundle of gravitational force, is one particular pattern of string vibration. And just as an electromagnetic field such as visible light is composed of an enormous number of photons, a gravita-tional field is composed of an enormous number of gravitons – that is, an enormous number of strings executing the graviton vibrational pattern. Gravitational fields, in turn, are encoded in the warping of the spacetime fabric, and hence we are led to identify the fabric of spacetime itself with a colossal number of strings all undergoing the same, orderly, graviton pattern of vibration. In the language of the field, such an enormous, organized array of similarly vibrating strings is known as a* coherent state of string. *It's a rather poetic image – the strings of string theory as the threads of the spacetime fabric – but we should note that its rigorous meaning has yet to be worked out completely. / Nevertheless, describing the spacetime fabric in this string-stitched form does lead us to contemplate the following question. [...] we can ask ourselves whether there is a raw precursor to the fabric of spacetime – a configuration of the strings of the cosmic fabric in which they have not yet coalesced into the orga-nized form that we recognize as spacetime. Notice that it is somewhat inaccurate to picture this state as a jumbled mass of individual vibrating strings that have yet to stitch themselves together into an ordered whole because, in our usual way of thinking, this presupposes a notion of both space and time – the space in which a string vibrates and the progression of time that allows us to follow its changes in shape from one moment to the next. But in the raw state, before the strings that make up the cosmic fabric engage in the orderly, coherent vibrational dance we are discussing,* there *is no realization of space and time. [...] In a sense, it's as if individual strings are 'shards' of space and time. and only when they appropriately undergo sympathetic vibrations do the conventional notions of space and time emerge."* (Greene (1999) 377f)

Zusammengefasst: Wenn sich die Raumzeit auf Gravitationsfelder zurückführen bzw. mit diesen identifizieren lässt, und wenn Gravitationsfelder durch den Austausch von Gravitonen realisiert werden, und wenn Gravitonen Oszillationszustände von Strings sind, so muss sich die Raumzeit konsequenterweise auf Strings zurückführen lassen, die dann allerdings nicht schon die Raumzeit für ihre Dynamik voraussetzen können. – Es könnte durchaus sein, dass gerade der Umstand, dass sich der Begriff des 'Gravitons' spätestens vor dem Hintergrund der Annahme der konzeptionellen Angemessenheit der Allgemeinen Relativitätstheorie als konzeptionell widersprüchlich erweist, nicht unerheblich dazu beigetragen hat, dass dieses 'poetische Bild' bisher kein ausformuliertes theoretisches Gegenstück hervorgebracht hat.[282]

Eine gänzlich andere Strategie besteht dann schliesslich darin, zu behaupten, dass schon die perturbativen Stringtheorien letztendlich hintergrundunabhängig sind, die Hintergrundraumzeit schon hier letztendlich obsolet ist und durch eine zweidimensionale Feldtheorie ersetzt werden kann,[283] dass also eine tatsächliche Hintergrundabhängigkeit des Stringansatzes gar nicht vorliegt und insofern auch nicht zum konzeptionellen Problem werden kann, man also bestenfalls nach einer nicht-perturbativen Formulierung des Stringansatzes suchen müsse, um minder problematische Sachverhalte zu bessern.

> "[...] as Witten describes, it is simple to generalize from the assumption of a flat background by inserting your metric of choice into the Lagrangian for the string field. This observation, plus the fact that conformal invariance for the field on the string demands that the Einstein equation be satisfied by the spacetime metric, leads Witten to propose that we should not see spacetime as an absolute background in string theory after all." (Callender / Huggett (2001a) 16)

---

[282] Eine weitere Idee, die vor allem aus dem Versuch heraus motiviert ist, den Stringansatz und die *Loop Quantum Gravity* (siehe Kap. 4.4.) in irgendeiner Weise in Einklang zu bringen, geht in die Richtung, sowohl die Raumzeit als auch die Strings als Ergebnis einer diskreten (stringfreien) prägeometrischen Dynamik anzusehen:

> "In loop quantum gravity, the micro-state representing Minkowski space-time will have a highly non-trivial Planck-scale structure. The basic entities will be 1-dimensional and polymer-like. Even in absence of a detailed theory, one can tell that the fluctuations of these 1-dimensional entities will correspond not only to gravitons but also to other particles, including a spin-1 particle, a scalar and an anti-symmetric tensor. These 'emergent states' are likely to play an important role in Minkowskian physics derived from loop quantum gravity. A detailed study of these excitations may well lead to interesting dynamics that includes not only gravity but also a select family of non-gravitational fields. It may also serve as a bridge between loop quantum gravity and string theory. For, string theory has two apriori elements: unexcited strings which carry no quantum numbers and a background space-time. Loop quantum gravity suggests that both could arise from the quantum state of geometry, peaked at Minkowski (or, de Sitter) space. The polymer-like quantum threads which must be woven to create the classical ground state geometries could be interpreted as unexcited strings. Excitations of these strings, in turn, may provide interesting matter couplings for loop quantum gravity." (Ashtekar (2005) 32)

Konkrete Realisierungen dieser Idee existieren nicht und sind auch nicht absehbar. – Dies gilt in gleicher Weise für die durchaus interessante Möglichkeit, die AdS/CFT-Korrespondenz als Ausgangspunkt für eine hintergrundunabhängige Formulierung des Stringansatzes anzusehen und die Emergenz der Raumzeit im Sinne einer Festlegung des raumzeitlichen Geschehens 'von aussen', im Sinne des holographischen Prinzips, zu deuten. Auch hier sind die Strings abgeleitete Entitäten, aus denen die Raumzeit ganz sicher nicht hervorgeht:

> "If space and time are not fundamental, what replaces them? Here the answer is that there is an auxiliary spacetime metric which is fixed by the boundary conditions at infinity. The CFT uses this metric, but the physical spacetime metric is a derived quantity. It is important to emphasize that the spacetime is not emerging from the 'strings'. In this approach, the so-called fundamental strings of string theory are also derived quantities. Both the strings and spacetime are constructed from the CFT." (Horowitz (2005) 15)

Angesichts dieser Ideen stellt sich sofort die Frage: Wenn hier nun auch die Strings emergente Entitäten sein sollen, wozu benötigt man diese Strings dann überhaupt noch? Wozu benötigt man die Stringtheorien, wenn die Strings auf der fundamentalen Ebene keine Rolle spielen? Etwa zur nomologischen Vereinheitlichung aller Wechselwirkungen auf einer intermediären, emergenten Ebene? Dafür gibt es bessere Ideen im Rahmen prägeometrischer Ansätze, die gänzlich ohne Strings auskommen – auch ohne emergente, intermediäre Strings. Siehe Kap. 4.6. zur Kopplung von Geometrogenese und Materiegenese.

[283] Siehe Witten (1993, 1996), Seiberg (2006), Horowitz (2005). Siehe auch Smolin (1998, 1999).

Dieser Auffassung zufolge ist die Raumzeit schon im Kontext der perturbativen Stringtheorien als nichts mehr als ein für die makroskopische, phänomenologische Ebene relevantes klassisches Näherungsergebnis anzusehen.[284] Sie ist eine abgeleitete, emergente Grösse. – Leider spiegelt der Formalismus der perturbativen Stringtheorien dies nicht unbedingt wieder.

<center>*</center>

Ob die Überlegungen zu einer hintergrundunabhängigen (und nicht-perturbativen) Formulierung des Stringansatzes – nach vielen Jahrzehnten – noch zu einem Ergebnis führen werden, ist nicht absehbar. Da der Stringansatz konzeptionell und methodisch in den Instrumentarien der Quantenfeldtheorie verankert ist, diese aber in grundlegender Weise hintergrundabhängig sind, müsste er für eine hintergrundunabhängige Formulierung dieses Instrumentarium sehr wahrscheinlich hinter sich lassen – und nicht nur einfach im Sinne einer Sprachregelung in seiner Relevanz wegdiskutieren. Das Ergebnis einer solchen modelltheoretischen Transzendierung hätte, wenn es tatsächlich dazu kommen sollte, nur noch wenig mit den bekannten perturbativen Stringtheorien und ihren ansatzweise nicht-perturbativen Erweiterungen, wie sie bis zum heutigen Zeitpunkt erreicht werden konnten, zu tun.

Heute jedenfalls ist der Stringansatz – in seiner im wesentlichen perturbativen Form, in der er sich auch nach mehreren Jahrzehnten Entwicklung immer noch erschöpft – letztlich nichts anderes als ein von direkten inhaltlichen, physikalischen Motivationen weitestgehend unabhängiges, mathematisch-modelltheoretisches Konstrukt. Dieses Konstrukt beruht auf keinem physikalisch motivierbaren fundamentalen Prinzip. Auch nach Jahrzehnten liess sich kein solches finden oder benennen. Die einzige originär physikalisch zu nennende Motivation für den Stringansatz beruht auf der Tatsache, dass die quantisierte Stringdynamik Spin-2-Oszillationszustände beschreibt und sich auf dieser Grundlage die Einsteinschen Feldgleichungen formal reproduzieren lassen; erst dadurch ist der Stringansatz überhaupt als ernstzunehmender Theorieanwärter (wieder) ins Spiel gekommen. Vor dem Hintergrund der Tatsache, dass jede Theorie, die Spin-2-Bosonen enthält, unter sehr allgemeinen Bedingungen formal die Einsteinschen Feldgleichungen reproduziert, dass aber gleichzeitig auch jede Theorie, welche die Gravitation durch den Austausch von Gravitonen auf einer vorgegebenen Hintergrundraumzeit beschreibt, von vornherein mit der Allgemeinen Relativitätstheorie unverträglich und, so sie diese zu reproduzieren vorgibt, konzeptionell widersprüchlich ist, lässt diese einzige physikalische Motivation für den Stringansatz recht fragwürdig erscheinen. – Fragwürdig ist ohnehin jeder Ansatz zu einer Theorie der Quantengravitation, der, ohne dafür eine gute, physikalisch motivierte Begründung zu liefern, mit einem notwendigerweise hintergrundabhängigen mo-

---

[284] Die Idee, dass der vermeintliche Hintergrundraum schon in den perturbativen Stringtheorien als Teil der jeweiligen Lösung anzusehen ist, findet sich auch in Kaku (1999) 325f:

*"The novel feature of this approach is that nowhere have we made any mention of the background space-time metric. The background occurs only in the kinetic term, not in the interacting term. In fact, the choice of the background metric emerges when we expand about a classical solution to the equations of motion. This is why this approach is called the 'pregeometrical' theory. In principle, the geometric of space-time should emerge as one among many possible vacua. [...] So far, it can be shown that flat space is a consistent solution [...]. It remains to be seen, however, what other kinds of classical backgrounds can be found as solutions to the equation of motion."* (Kaku (1999) 326)

Aber auch hier bleibt es bei sehr kurzen, kursorischen Bemerkungen, die im Formalismus des Stringansatzes nicht nachvollzogen werden können und eher den Charakter einer Sprachregelung aufweisen, die offenbar dazu dient, das Problem der Hintergrundabhängigkeit wegzudiskutieren.

delltheoretischen Instrumentarium arbeitet und damit, ohne guten Grund, hinter die konzeptionellen Einsichten der Allgemeinen Relativitätstheorie zurückfällt.

Dass es sich beim Stringansatz um ein vor allem mathematisch-modelltheoretisch bestimmtes Konstrukt handelt, dessen minimale physikalische Motivationen nur wenig wirksam werden, wird vor allem deutlich, wenn man sich vor Augen führt, in welchem Ausmass seine explizite Form durch innertheoretische Erfordernisse bestimmt wird, die sich nahezu ausschliesslich auf die Versuche der Eliminierung von Anomalien und mathematischen Inkonsistenzen zurückführen lassen, nicht aber auf physikalische Anforderungen. Der Stringansatz bezieht nicht etwa deshalb die Supersymmetrie ein, weil wir offensichtlich in einer supersymmetrischen Welt leben – es gibt bisher keine empirischen Daten, die dafür sprächen –, sondern weil die Supersymmetrie aus rein innertheoretischen, mathematisch-modelltheoretischen Gründen unabdingbar erscheint. Der Stringansatz beschreibt die Dynamik der Strings nicht deshalb auf einer zehndimensionalen Hintergrundraumzeit, weil wir in einer offensichtlich zehndimensionalen Welt leben – auch dafür gibt es bisher nicht die geringsten empirischen Hinweise –, sondern weil die quantisierte Dynamik speziell-relativistischer Strings nur auf zehn raumzeitlichen Dimensionen anomaliefrei beschreibbar ist und auch nur unter diesen Bedingungen zur Reproduktion der Einsteinschen Feldgleichungen führt. Supersymmetrie und höherdimensionale Raumzeit beruhen auf keinen empirischen Indizien; sie sind nicht physikalisch motiviert; sie sind vielmehr für den Stringansatz schlichtweg aus Gründen innertheoretischer, mathematischer Konsistenz erforderlich. Wenn man sich nicht darauf festlegt, die Dynamik aller Wechselwirkungen in nomologisch vereinheitlichter Weise mittels der quantisierten Dynamik der Strings beschreiben zu wollen, spricht nichts für die Supersymmetrie oder die höherdimensionale Raumzeit. Supersymmetrie und höherdimensionale Raumzeit kommen nur ins Spiel, wenn man den Stringansatz als adäquate Theorie der Quantengravitation (und aller anderen Wechselwirkungen) zu etablieren versucht.[285]

Dass es dem Stringansatz auch mit diesen physikalisch erst einmal unmotivierten Postulaten der Supersymmetrie und der höherdimensionalen Raumzeit nicht gelingt, den empirischen Gehalt des quantenfeldtheoretischen Standardmodells zu reproduzieren, lässt dann schliesslich weitere Zweifel aufkommen, dass es sich vielleicht bei diesen Postulaten um unphysikalische Artefakte im Rahmen eines rein mathematisch-modelltheoretisch determinierten Konstruktes handeln könnte. Diese Zweifel verstärken sich noch, wenn man sieht, dass gerade die Supersymmetrie und die höherdimensionale Raumzeit einen wesentlichen Anteil daran haben, dass es zum Kontingenzproblem der *String Landscape* kommt und der Stringansatz damit eigentlich jegliche Aussicht auf konkrete quantitative Vorhersagen einbüsst.

Wenn man den Stringansatz nicht schon aufgrund seiner internen Probleme als ohnehin wenig aussichtsreichen Weg zu einer konsistenten Theorie der Quantengravitation betrachtet, so gibt es einen weiteren Faktor, der eine solche Sichtweise nahelegen könnte: Es ist die Tatsache, dass der Stringansatz auch nach vielen Jahrzehnten Entwicklung und trotz diverser spekulativer Ideen, die vermutlich z.T. über ihn hinausweisen und vielleicht in anderen konzeptionellen Kontexten besser verortet wären, letztendlich nichts (oder nur sehr wenig) zum Verständnis der Raumzeit beigetragen hat. Die sich mit ihm abzeichnende Diskretisierung, zu der es, quasi durch die Hintertür, auf der Basis eines Kontinuumsansatzes kommt, deutet – wie vieles andere auch – gerade über sein modelltheoretisches Instrumentarium hinaus.[286] Alle konkreteren Ideen zur Natur der Raumzeit sind im Kontext des Stringansatzes nicht nur hochgradig spekulativ; vielmehr sprengen sie dessen Kontext und seine modelltheoretischen und konzeptionellen Voraussetzungen.

---

[285] Das heisst noch nicht, dass sich nicht auch Theoretiker mit ganz anderen Ambitionen von der Idee der Supersymmetrie, die dann meist als 'mathematisch elegant' bezeichnet wird, anstecken lassen, und dann etwa versuchen, eine minimal-supersymmetrische Variante des Standardmodells zu etablieren. Wie könnte die Natur auch an einem so eleganten mathematischen Konstrukt wie der Supersymmetrie achtlos vorbeigehen? – Dem lässt sich mit gleicher Berechtigung entgegenhalten: Ginge es der Natur um Eleganz, so sähe die Welt vermutlich gänzlich anders aus – oder es gäbe sie vielleicht gar nicht.

## 4.3. Kanonische Quantisierung der Allgemeinen Relativitätstheorie I: Geometrodynamik

Der kanonische Quantisierungsansatz[287] besteht in einer direkten Quantisierung der Allgemeinen Relativitätstheorie auf der Grundlage ihrer Hamiltonschen Darstellung. Im Gegensatz zur *Kovarianten Quantisierung* ist die *Kanonische Quantengravitation* ein strikt nicht-perturbativer, hintergrundunabhängiger Ansatz.

Die Hamiltonsche Darstellung der Allgemeinen Relativitätstheorie, von der dieser Ansatz seinen Ausgang nimmt, impliziert – wie zuvor schon erläutert wurde[288] – insbesondere eine Zerlegung der Raumzeit in dreidimensionale raumartige Hyperflächen und einen Zeitparameter. Die formale Auszeichnung eines Zeitparameters ist nicht zuletzt zur Definition der kanonischen Impulse erforderlich, die Voraussetzung für die *Kanonische Quantisierung* sind.

> *"The usual starting point for developing the canonical formalism is the foliation of spacetime into three-dimensional spacelike hypersurfaces. A prerequisite for this is the global hyperbolicity of the spacetime. [...] In the canonical formalism, the three-metric will play the role of the configuration variable. [...] The classical equations are six evolution equations for the [configuration variables] $h_{ab}$ and their momenta $p^{ab}$ as well as four* constraints *for them. The momenta $p^{ab}$ are linear combinations of the extrinsic curvature of the three-dimensional space. The six evolution equations and four constraints are the canonical version of the ten Einstein field equations. Only after the classical equations have been solved, can one interpret spacetime as a 'trajectory of spaces'. / In the quantum theory, the trajectories will disappear as in ordinary quantum mechanics. There will thus be no spacetime at the most fundamental level; only the constraints for the three-dimensional space will remain."* (Kiefer (2008) 3f)

Die Diffeomorphismusinvarianz (bzw. Eichinvarianz) der Allgemeinen Relativitätstheorie wird in der Hamiltonschen Darstellung in Form der Zusatzbedingungen ('*constraints*') erfasst. Sie sind erforderlich, da die Wahl der kanonischen Variablen die Diffeomorphismusinvarianz noch nicht berücksichtigt. Der volle, kinematische Hamiltonsche Phasenraum enthält noch physikalisch unmögliche Zustände sowie insbesondere ungekennzeichnete Redundanzen (Eichfreiheiten), die erst mittels der Zusatzbedingungen (Generatoren von Eichtransformationen) erfasst werden.

---

[286] Gleiches wird sich dann, unter anderen modelltheoretischen Voraussetzungen, auch im Hinblick auf die *Loop Quantum Gravity* konstatieren lassen. Siehe Kap. 4.4.
[287] Siehe DeWitt (1967), Kuchar (1986, 1993), Friedrich / Ehlers (1994), Kiefer (2008a).
[288] Siehe Kap. 2.1.

Wie wir gesehen haben,[289] ergeben sich für die Allgemeine Relativitätstheorie nach der Zerlegung der Raumzeit in dreidimensionale raumartige Hyperflächen und einen Zeitparameter vier Zusatzbedingungen: die *Hamiltonsche Zusatzbedingung* und drei *Impuls- bzw. Diffeomorphismus-Zusatzbedingungen.*

> *"The invariance of the classical theory under coordinate transformation leads to four (local) constraints: the Hamiltonian constraint [...] and the three diffeomorphism (or momentum) constraints [...]."* (Kiefer (2005) 8)

Die Hamiltonsche Zusatzbedingung spiegelt die zeitliche Komponente (normal zur räumlichen Hyperfläche) der Diffeomorphismusinvarianz wider. Sie erfasst (zumindest für den Fall eines kompakten Raumes) nicht zuletzt die zeitliche Entwicklung des Systems. Die Impuls- bzw. Diffeomorphismus-Zusatzbedingungen erfassen die räumlichen Komponenten der Diffeomorphismusinvarianz (tangential zur räumlichen Hyperfläche).

> *"[...] the momentum and Hamiltonian constraints are believed to capture the invariance of general relativity under spacelike and timelike diffeomorphisms respectively."* (Callender / Huggett (2001a) 19)

Die Quantisierung des Hamiltonschen Systems mit Zusatzbedingungen erfolgt dann im kanonischen Quantisierungsansatz entsprechend der sogenannten Diracschen Methode.[290]

> *"[...] Dirac introduced a systematic quantization program. Here, one first ignores constraints and introduces a kinematic framework consisting of an algebra **a** of quantum operators and a representation thereof on a Hilbert space **H**$_{kin}$. This provides the arena for defining and solving the quantum constraints. When equipped with a suitable inner product, the space of solutions defines the physical Hilbert space **H**$_{phy}$."* (Ashtekar (2007a) 2)

Die *Dirac-Quantisierung* sieht drei sukzessive Schritte vor:

(1)     Zuerst wird das volle (nicht-reduzierte) Hamiltonsche System ohne Berücksichtigung der Zusatzbedingungen quantisiert. Man startet also mit dem vollen, kinematischen Phasenraum der Variablen des Hamiltonschen Systems. Diese Variablen erfassen im Rahmen der *geometrodynamischen* Variante des kanonischen Quantisierungsansatzes die metrischen Eigenschaften der Raumzeit. Zu diesen klassischen Variablen werden Quantenpendants (Operatoren) formuliert.[291] Es werden kanonische Vertauschungsrelationen für die Quantenvariablen und schliesslich eine Operator-Algebra definiert. Das Ergebnis ist ein kinematischer Hilbertraum, der (entsprechend dem klassischen kinematischen Phasenraum) immer noch unphysikalische Zustände und ungekennzeichnete Redundanzen (Eichfreiheiten) enthält, die erst im nachhinein (Schritt 3) mittels der Zusatzbedingungen bzw. ihrer quantenmechanischen Pendants erfasst werden müssen. (Dieser im Rahmen der Dirac-Quantisierung erst einmal zurückgestellte Schritt wird, sowohl im klassischen wie im quantenmechanischen Fall, als 'Lösung der Zusatzbedingungen' bezeichnet.)

---

[289] Siehe Kap. 2.1.
[290] Siehe Henneaux / Teitelboim (1992) sowie Belot (2003).
[291] Entscheidend ist hier, dass man schon für den klassischen Fall kanonische Variablen wählt, die nach der Quantisierung zu einer konsistenten und tragfähigen Quantentheorie führen. In der ursprünglichen (geometrodynamischen) kanonischen Quantisierung erfassen die kanonischen Variablen die metrischen Eigenschaften der Raumzeit. Dies führt für die Quantentheorie zu erheblichen Problemen, die sich, wie die *Loop Quantum Gravity* zeigt, durch die Wahl anderer klassischer Variablen zum Teil vermeiden lassen. Siehe Kap. 4.4.

(2) Erst nach der Quantisierung des vollen, kinematischen Hamiltonschen Systems werden, als Pendants zu den klassischen Zusatzbedingungen (erster Klasse), entsprechende Quanten-Zusatzbedingungen formuliert. Die Poisson-Algebra der klassischen Zusatzbedingungen wird durch die Kommutator-Algebra der entsprechenden Quanten-Zusatzbedingungen ersetzt. Für die zeitlichen Diffeomorphismen ergibt sich hier als quantisiertes Pendant zur klassischen Hamiltonschen Zusatzbedingung die *Wheeler-DeWitt-Gleichung*: eine zeitlose Null-Energie-Schrödingergleichung.

(3) Erst im letzten Schritt sollen dann die Zusatzbedingungen in ihrer quantisierten Form Berücksichtigung erfahren. D.h. die Quanten-Zusatzbedingungen sollen 'gelöst' werden, um dadurch den echten, physikalischen Hilbertraum zu erhalten, der die physikalisch möglichen Quantenzustände in eindeutiger, redundanzfreier Weise erfasst. Dazu ist es insbesondere erforderlich, ein inneres Produkt zu definieren. Als 'Observablen' kommen nur Grössen in Frage, die eichinvariant sind. Formal entspricht dies der Forderung, dass alle Observablen mit allen Quanten-Zusatzbedingungen kommutieren.

*"[...] Dirac invented a short-cut method referred to as constraint quantization, which consists in promoting the first class constraints to operators on a suitable Hilbert space and then identifying the physical sector of this space in terms of the state vectors that are annihilated by the operator constraints."* (Earman (2002a) 21)

Das Problem ist aber, dass Schritt (3) nicht so ohne weiteres funktioniert, insbesondere was die vollständige Formulierung des inneren Produktes betrifft, weil letztendlich schon im Schritt (2) unlösbare Probleme auftreten.

*"The mathematical problems of quantum geometrodynamics have to do with factor ordering, regularization, and Dirac consistency, which are themselves intertwined problems."* (Kiefer (2008) 8)

Die Probleme der geometrodynamischen Variante des kanonischen Quantisierungsansatzes lassen sich nicht zuletzt auf die *Wheeler-DeWitt-Gleichung* zurückführen. Diese stellt, wie schon erwähnt, das Quantenpendant zur klassischen Hamiltonschen (skalaren) Zusatzbedingung dar, die sich als Generator von Eichtransformationen verstehen lässt und schon im klassischen Fall zum *Problem der Zeit* führte.[292] Letzteres überträgt sich in vollem Umfang von der Allgemeinen Relativitätstheorie auf die *Kanonische Quantengravitation* und hat hier schliesslich noch wesentlich gravierendere Konsequenzen. Diese Konsequenzen des *Problems der Zeit* und mögliche Lösungsvorschläge sol-len allerdings erst im nächsten Abschnitt erörtert werden, wenn es um die *Loop Quantum Gravity* geht, die von diesem Problem in der gleichen Weise betroffen ist, jedoch zu deutlich weiterreichenden Resultaten als die geometrodynamische Variante des kanonischen Ansatzes führt. – Dass sie dies tut, liegt am entscheidenden Problem, hinsichtlich dessen sich *Loop Quantum Gravity* von der geometrodynamischen Variante der *Kanonischen Quantisierung* deutlich abhebt. Das vielleicht entscheidende Problem der letzteren manifestiert sich nämlich darin, dass sich die geometrodynamische *Wheeler-DeWitt-Gleichung* nicht nur als nicht ohne weiteres lösbar erweist,[293] sondern als mathematisch nicht einmal angemessen definiert.

---

[292] Da die Hamiltonsche Zusatzbedingung, die (zumindest im Fall eines kompakten Raumes) die zeitliche Entwicklung des Systems von einer Hyperfläche zur nächsten erfasst, als Generator einer Eichtransformation anzusehen ist, entspricht diese zeitliche Entwicklung des Systems von einer Hyperfläche zur nächsten nun einer Eichtransformation. Damit lässt sich die zeitliche Entwicklung, entsprechend der letztlich unumgänglichen (oder nur um den Preis einer metaphysischen Bezugnahme auf prinzipiell Unbeobachtbares vermeidbaren) eichinvarianten Interpretation, nicht als physikalisch real ansehen. Denn eichabhängige Grössen sind grundsätzlich nicht beobachtbar und messbar. Wenn die zeitliche Entwicklung aber einer unbeobachtbaren und damit unphysikalischen Eichtransformation entspricht, so sind alle Observablen des Systems notwendigerweise zeitunabhängig. Siehe Kap. 2.1.

> *"[...] a highly singular functional differential equation, which most likely cannot be made mathematically well defined [...]."* (Nicolai / Peeters / Zamaklar (2005) 10)

Der physikalische Gehalt der *Wheeler-DeWitt-Gleichung* ist in der geometrodynamischen Variante des kanonischen Quantisierungsansatzes letztlich völlig unklar. Die Ursache für diese Probleme liegt wohl bis zu einem gewissen Grade in einer letztendlich unglücklichen Wahl der klassischen kanonischen Ausgangsvariablen, auf denen die Quantisierung ansetzt.

> *"[...] geometrodynamics has so far not succeeded in constructing a suitable scalar product and an appropriate Hilbert space of wave functionals."* (Nicolai / Peeters / Zamaklar (2005) 10) – *"Most of the work in quantum geometrodynamics thus leaves the question of the inner product open [...]."* (Kiefer (2008) 9)

Die *Loop Quantum Gravity* zeigt, dass sich zumindest einige der Probleme der geometrodynamischen Variante des kanonischen Quantisierungsansatzes durch die Wahl geeigneterer klassischer Ausgangsvariablen lösen lassen. Darüberhinausgehend führt sie insbesondere zu überraschenden Ergebnissen, die sich in der älteren, geometrodynamischen Variante der *Kanonischen Quantisierung* nicht aufzeigen lassen.

## 4.4. Kanonische Quantisierung der Allgemeinen Relativitätstheorie II: Loop Quantum Gravity

Die *Loop Quantum Gravity*[294] ist eine Variante des kanonischen Quantisierungsansatzes, die von der Hamiltonschen Darstellung der Allgemeinen Relativitätstheorie in der *Ashtekar-Formulierung* ausgeht, die dann jedoch vor der Quantisierung noch einmal modifiziert wird.

---

[293] Als lösbar erweisen sich nur massiv reduzierte Versionen der *Wheeler-DeWitt-Gleichung*. Sie sind die Grundlage der Minisuperspace- und Midisuperspace-Modelle in der Quantenkosmologie.

[294] Siehe Ashtekar (2007, 2007a)), Ashtekar / Lewandowski (2004), Ashtekar et al. (1992), Rovelli (1991b, 1997, 1998, 2003, 2004), Smolin (1991a, 2000, 2001, 2004), Thiemann (2001, 2006), Nicolai / Peeters (2006), Nicolai / Peeters / Zamaklar (2005). Für eine Literaturübersicht siehe Hauser / Corichi (2005).

## *Der formale Ausgangspunkt*

Mitte der achtziger Jahre entwickelte Abhay Ashtekar[295] eine neue Variante der Hamiltonschen Dar-
stellung der Allgemeinen Relativitätstheorie, die nicht mehr mit den metrischen Grössen der Raum-
zeit arbeitet, sondern mit einem Satz neuer kanonischer Variablen, die schliesslich eine kanonische
Quantisierung der Allgemeinen Relativitätstheorie erheblich vereinfachen.

> *"The theory transforms the old Wheeler-DeWitt theory in a formalism that can be concretely used to*
> *compute physical quantities in quantum gravity."* (Rovelli (2004) 298)

Diese neuen kanonischen Variablen sind einerseits eine räumliche SU(2)-Konnektion, die als Kon-
figurationsgrösse die in der geometrodynamischen Variante verwendete räumliche Metrik ablöst,
andererseits eine orthonormale Triade (analog zur Struktur eines elektrischen Feldes), welche die
extrinsische Krümmung als Impulsgrösse ablöst. In der konnektionsdynamischen Ashtekar-Dar-
stellung vereinfachen sich dadurch gegenüber der alten geometrodynamischen Variante insbeson-
dere die Zusatzbedingungen: sowohl die skalare Hamiltonsche Zusatzbedingung als auch die Im-
puls- bzw. Diffeomorphismus-Zusatzbedingungen.

> *"On the relativity side, the third stage began with the following observation: the geometrodynamics*
> *program laid out by Dirac, Bergmann, Wheeler and others simplifies significantly if we regard a spa-*
> *tial connection – rather than the 3-metric – as the basic object. [...] these were 'spin-connections', re-*
> *quired to parallel propagate spinors, and they turn out to* simplify *Einstein's equations considerably."*
> (Ashtekar (2005) 8)

Mit diesem Übergang von der Geometrodynamik zur Konnektionsdynamik der Ashtekar-Variablen
stellt sich der kinematische Hamiltonsche Phasenraum der Allgemeinen Relativitätstheorie nun als
Raum von Konnektionen dar und gleicht damit formal dem Phasenraum einer klassischen Yang-
Mills-Theorie. In ihrer konnektionsdynamischen Darstellung lässt sich die Allgemeine Relativi-
tätstheorie nun auch formal leicht als Eichtheorie erkennen.

> *"Perhaps the most important advantage of the passage from metrics to connections is that the phase-*
> *space of general relativity is now the same as that of gauge theories."* (Ashtekar (2005) 8)

Dies hat insbesondere methodische Vorteile bei ihrer Quantisierung. Es ist jetzt möglich, in einem
deutlich umfassenderen Ausmass Prinzipien und Methoden aus den Quantenfeldtheorien aufzugrei-
fen, ohne dabei zugleich deren problematischere Implikationen übernehmen zu müssen.

> *"One could now import into general relativity techniques that have been highly successful in the*
> *quantization of gauge theories. [...] Since the canonical approach does not require the introduction of*
> *a background geometry or use of perturbative theory, and because one now has access to fresh, non-*
> *perturbative techniques from gauge theories, in relativity circles there is a hope that this approach*
> *may lead to well-defined,* non-perturbative *quantum general relativity [...]."* (Ashtekar (2005) 8f) –
> *"More precisely, in this approach one begins by formulating general relativity in the mathematical*
> *language of connections, the basic variables of gauge theories of electro-weak and strong interacti-*

---

[295] Siehe Ashtekar (1986, 1987).

*ons. [...] the emphasis is shifted from distances and geodesics to holonomies and Wilson loops. Consequently, the basic kinematical structures are the same as those used in gauge theories. A key difference, however, is that while a background space-time metric is available and crucially used in gauge theories, there are no background fields whatsoever now. Their absence is forced upon us by the requirement of diffeomorphism invariance (or 'general covariance')."* (Ashtekar (2005) 14)

Der moderate Preis, der für diese transparentere Darstellung als Yang-Mills-Eichtheorie zu zahlen ist, besteht darin, dass weitere Zusatzbedingungen erforderlich werden: die *Gaussschen Zusatzbedingungen*. Zu diesen kommt es, weil durch den Übergang zu den neuen kanonischen Variablen eine weitere Redundanz in die Theorie eingeführt wird, die bei der Auszeichnung der tatsächlichen, physikalisch unterscheidbaren Zustände des Systems, d.h. bei der Reduzierung auf den Phasenraum der echten Freiheitsgrade, berücksichtigt werden muss.

> *"In the connection and loop approaches, three additional (local) constraints emerge because of the freedom to choose the local triads upon which the formulation is based."* (Kiefer (2005) 9)

Es gibt also in der Ashtekar-Variante der Hamiltonschen Darstellung der Allgemeinen Relativitätstheorie nun drei Typen von Zusatzbedingungen:

(1) die *skalare* bzw. *Hamiltonsche Zusatzbedingung*: Sie generiert infinitesimale zeitliche Diffeomorphismen (normal zur räumlichen Hyperfläche).
(2) die *Impuls-* bzw. *Diffeomorphismus-Zusatzbedingungen*: Sie generieren infinitesimale räumliche Diffeomorphismen (tangential zur räumlichen Hyperfläche).
(3) die *Gaussschen Zusatzbedingungen*: Sie generieren SU(2)-Eichtransformationen und sind letztlich ein Artefakt, das mit der Einführung der neuen Variablen einhergeht.

### Diffeomorphismusinvarianz und Hintergrundunabhängigkeit

In der *Loop Quantum Gravity* wird die Darstellung der klassischen Theorie vor der Quantisierung schliesslich noch einmal modifiziert. Ausgangspunkt sind hier nicht die Ashtekar-Variablen selbst (Konnektionen und orthonormale Triaden), sondern die sich aus ihnen ergebenden Wilson-Loops. Ein Wilson-Loop ist nichts anderes als die Spur einer Holonomie, d.h. eines Integrals einer Konnektion auf einer geschlossenen Kurve.[296] Die so erhaltenen Loop-Variablen haben den Vorteil, dass sie von vornherein eichinvariant sind. – Die Wahl dieser kanonische Variablen ist von entscheidender Bedeutung im Hinblick auf das wichtigste Grundprinzip der Theorie: die aktive Diffeomorphismusinvarianz, die hier (entsprechend der eichinvarianten Deutung der Allgemeinen Relativitätstheorie) als Eichinvarianz verstanden wird.

> *"Because active diff[eomorphism] invariance is a gauge, the physical content of [general relativity] is expressed only by those quantities, derived from the basic dynamical variables, which are fully independent from the points of the manifold."* (Rovelli (2001) 108)

---

[296] Wie schon zuvor erwähnt, zeigt Lyre (2004) am Beispiel der Maxwellschen Elektrodynamik sehr überzeugend, dass es, wenn man sich im Sinne der *Leibniz-Äquivalenz* auf eichinvariante Grössen als ernstzunehmende physikalische Grössen beschränkt, im Kontext von Eichtheorien wohl vor allem die nichtlokalen Holonomien und Wilson-Loops sind, die als Repräsentanten der zugrundeliegenden echten, physikalischen Freiheitsgrade in Frage kommen.

Die aktive Diffeomorphismusinvarianz und ihre eichinvariante Deutung werden in der *Loop Quantum Gravity* nicht als zusätzlich zu verhandelnde interpretatorische Komponenten behandelt, wie dies vielleicht, wenn auch um einen hohen Preis, im Fall der Allgemeinen Relativitätstheorie noch möglich war. Sie werden vielmehr direkt in den formalen Apparat implementiert. Nur diffeomorphismusinvariante bzw. eichinvariante Grössen werden überhaupt als real anzusehende physikalische Grössen behandelt. – Ob die vollständige Erschliessung dieser als real anzusehenden physikalischen Grössen innerhalb der *Loop Quantum Gravity* jedoch letztlich gelingt, ist, wie noch zu erörtern sein wird, immer noch offen.

Die fest in den formalen Apparat der *Loop Quantum Gravity* eingeschriebene eichinvariante Interpretation der aktiven Diffeomorphismusinvarianz, die impliziert, dass nur eich- bzw. diffeomorphismusinvariante Grössen als physikalisch real anzusehen sind, soll vor allem die Hintergrundunabhängigkeit der Theorie garantieren. – Hintergrundunabhängig ist die *Loop Quantum Gravity* aber nur insofern, als sie keine Metrik voraussetzt. Ansonsten setzt die *Loop Quantum Gravity* weiterhin den gleichen festen Hintergrund voraus, den es auch in der Allgemeinen Relativitätstheorie schon bzw. noch gibt: eine kontinuierlich differenzierbare (pseudo-Riemannsche) Punktmannigfaltigkeit mit vorgegebener Dimensionalität, vorgegebener (3+1)-Signatur und in dynamischer Hinsicht unveränderlicher Topologie.

> *"To define the action principle one must assume that the topology, dimension and differential structure of spacelike surfaces [...] are fixed."* (Smolin (2006c) 212)

Die *Loop Quantum Gravity* ist formal so etwas wie eine Quantenfeldtheorie auf einer differenzierbaren Mannigfaltigkeit, ein nicht-perturbativer Ansatz ohne Voraussetzung einer Hintergrundmetrik. Die Unabhängigkeit von einer Hintergrundmetrik wird gerade durch die feste Implementierung der aktiven Diffeomorphismusinvarianz und ihrer eichinvarianter Deutung gewährleistet.

> *"[...] [Loop Quantum Gravity] is a straightforward quantization of [General Relativity] with its conventional matter couplings. [...] On the other hand, [Loop Quantum Gravity] has a radical and ambitious side: to merge the conceptual insight of [General Relativity] into [Quantum Mechanics]. In order to achieve this, we have to give up the familiar notions of space and time. The space continuum 'on which' things are located and the time 'along which' evolution happens are semiclassical approximate notions in the theory. [...] The price of taking seriously the conceptual novelty of [General Relativity] is that most of the traditional machinery of [Quantum Field Theory] becomes inadequate. This machinery is based on the existence of background spacetime, and [General Relativity] is the discovery that there is no background spacetime."* (Rovelli (2004) 10)

Die Diffeomorphismusinvarianz setzt notwendigerweise eine differenzierbare Mannigfaltigkeit voraus: zumindest als mathematisches Hilfskonstrukt. Dass diese nicht als physikalisch ernstzunehmende oder gar als substantiell zu deutende Entität angesehen werden muss, wird schon im Kontext der Allgemeinen Relativitätstheorie deutlich. Die (3+1)-Signatur der Raumzeit, wie sie schon in der Allgemeinen Relativitätstheorie vorliegt, ist infolge der als Ausgangspunkt gewählten Hamiltonschen Formulierung fest im Formalismus der *Loop Quantum Gravity* verankert. Die Dimensionalität der Raumzeit wird nicht einmal diskutiert. Und auch Topologiewechsel lassen sich mit dem modelltheoretischen Instrumentarium der *Loop Quantum Gravity* grundsätzlich nicht beschreiben. – Ob es jedoch für eine Theorie der Quantengravitation sehr motiviert erscheint, mit einer schon fest vorgegebenen Topologie[297], einer fest vorgegebenen Dimensionalität und einer fest vorgegebenen Signatur der Raumzeit zu starten, ist eine andere Frage. Vielleicht sollte man für eine fundamentale Theorie eher erwarten, dass sie diese hier fest vorgegebenen Randbedingungen in irgendeiner Weise erklärt oder zumindest motiviert. Die *Loop Quantum Gravity* tut dies nicht. Konsequenter sind in dieser Hinsicht, wenn man einmal von der aus Konsistenzforderungen resultierenden Dimensionalitätsfestlegung des Stringansatzes absieht, nur die prägeometrischen Ansätze.[298]

## Spinnetze: Diskretheit des Raumes als Ergebnis einer Kontinuumsmodellierung

Das Überraschende ist nun, dass sich nach der Quantisierung der klassischen Konnektionsdynamik eine diskrete räumliche Struktur ergibt.

> *"One finds that, at the Planck scale, geometry has a definite discrete structure. [...] space-time continuum arises only as a coarse-grained approximation."* (Ashtekar (2005) 9)

Erstaunlich ist dabei insbesondere auch, dass sich diese Diskretheit nicht erst im Rahmen einer Erfassung der echten physikalischen Freiheitsgrade einstellt, also nach der 'Lösung der Zusatzbedingungen', sondern schon auf der kinematischen Ebene nach der Quantisierung des vollen Phasenraums des Hamiltonschen Systems ohne Berücksichtigung der Zusatzbedingungen.[299] (Man verwendet also wiederum die Diracsche Quantisierungsmethode.[300])

> *"It is somewhat surprising that an important issue such as the fundamental discreteness of space emerges already at the kinematical level. One would have instead expected that it is a result that emerges from the treatment of the Hamiltonian constraint, which encodes the 'dynamical' features of Einstein's theory. The discreteness thus seems to hold for more general theories than quantum general relativity."* (Kiefer (2004 [²2007]) 194)

Die Raumzeit (genauer: der Raum und seine Metrik bzw. seine Konnektionen) stellt sich nach der Quantisierung als makroskopische Konsequenz einer diskreten Struktur auf der Planck-Ebene dar: der *Spinnetze* – diskreter Graphenstrukturen, die sich für die räumlichen Hyperflächen ergeben.

---

[297] Variable Topologien lassen sich für die Quantengravitation sicherlich nicht a priori ausschliessen; einiges spricht unter bestimmten konzeptionellen Bedingungen für ihr Auftreten. Und es ist immer noch unklar, ob die Allgemeine Relativitätstheorie (trotz der verwendeten Differentialgeometrie, die dies eigentlich ausschliesst) Topologiewechsel zumindest in sehr eingeschränkter Form zulässt.
> *"Of course, if topology change occurred classically, it would certainly be a part of any quantum theory of gravity."* (Horowitz (1991) 587)
Eine nur leicht erweiterte Variante der Allgemeinen Relativitätstheorie lässt sie offensichtlich zu; siehe Horowitz (1991).
> *"So these spacetimes provide a strong argument that topology change must be a part of any quantum theory of gravity."* (Horowitz (1991) 589)

[298] Siehe Kap. 4.6.

[299] Nur die Gaussschen Zusatzbedingungen, die ohnehin ein Artefakt der Einführung der neuen Variablen sind, sind an dieser Stelle schon gelöst.

[300] Siehe Kap. 4.3. Dass der hierbei erforderliche Übergang vom kinematischen zum echten physikalischen Hilbertraum durch die Lösung der Quanten-Zusatzbedingungen nicht ganz so einfach ist, stellt eines der grundsätzlichen, noch nicht vollends gelösten Probleme der *Loop Quantum Gravity* dar. Siehe weiter unten.

*"In the quantum theory, the fundamental excitations of geometry are most conveniently expressed in terms of holonomies. They are thus* one-dimensional, polymer-like *and in analogy with gauge theories, can be thought of as 'flux lines' of electric fields/triads. More precisely, they turn out to be* flux lines of area *[...]. [...] if quantum geometry were to be excited along just a few flux lines, most surfaces would have zero area and the quantum state would not at all resemble a classical geometry. [...] semiclassical geometries can result only if a huge number of these elementary excitations are superposed in suitable dense configurations. The state of quantum geometry around you, for example, must have so many elementary excitations that approximately $10^{68}$ of them intersect the sheet of paper you are reading. Even in such states, the geometry is still distributional, concentrated on the underlying elementary flux lines. But if suitably coarse-grained, it can be approximated by a smooth metric. Thus, the continuum picture is only an approximation that arises from coarse graining of semi-classical states."* (Ashtekar (2005) 16)

Diese diskrete Struktur der Spinnetze entspricht dem diskreten Eigenwertspektrum geometrischer Operatoren: des *Flächen-* und des *Volumen-Operators*, die sich auf der Grundlage der quantisierten Konnektionsdynamik in ihrer Loop-Variablen-Darstellung definieren lassen.

*"[...] a quantum spacetime can be decomposed in a basis of states that can be visualized as made by quanta of volume (the intersections) separated by quanta of area (the links). More precisely, we can view a spin network as sitting on the* dual *of a cellular decomposition of physical space. The nodes of the spin network sit in the center of the 3-cells, and their coloring determines the (quantized) 3-cell's volume. The links of the spin network cut the faces of the cellular decomposition, and their color j determine the (quantized) areas of these faces [...]."* (Rovelli (1998) 8)

Den Graphenlinien der Spinnetze lassen sich Spineigenwerte (*j*) zuordnen; sie entsprechen den Eigenwerten des Flächenoperators und stehen für elementare quantisierte Flächen. Den Knotenpunkten lassen sich die Eigenwerte des Volumenoperators zuordnen; sie stehen für elementare quantisierte Volumina. – Aber:

*"Spin networks are not primary concrete 'objects' like particles in classical mechanics: rather, they describe the way the gravitational field interacts, like the energy quanta of an oscillator do."* (Rovelli (2004) 269)

Interessanterweise stellt sich die Diskretheit der Raumzeit (bzw. der quantisierten räumlichen Hyperflächen) in der *Loop Quantum Gravity* als Ergebnis eines Ansatzes ein, der mit einer differenzierbaren Mannigfaltigkeit startet.[301] Die räumliche Diskretheit gehört also nicht zu den Voraussetzungen der Theorie. vielmehr ergibt sie sich als Implikation der Quantisierung: als Konsequenz einer nicht-perturbativen, hintergrundunabhängigen Quantenfeldtheorie auf einer differenzierbaren Mannigfaltigkeit. Die Diskretheit ist also (was die formale Seite anbelangt) das Ergebnis einer Kontinuumsmodellierung:

---

[301] Sogar die *Bekenstein-Hawking-Entropie* erweist sich als ableitbar, im Gegensatz zum Stringansatz sogar für realistische schwarze Löcher. Siehe etwa Meissner (2004), Carlip (2008a). Was sich nicht ableiten lässt, ist hingegen der 1/4-Faktor. Dieser wird vielmehr rückwirkend für die Festlegung des Barbero-Immirzi-Parameters verwendet, der einer parametrischen Wahlmöglichkeit innerhalb der *Loop Quantum Gravity* entspricht, die von der Theorie selbst nicht festgelegt wird.

> *"Space itself turns out to have a discrete and combinatorial character. Notice that this is not imposed on the theory, or assumed. It is the result of a completely conventional quantum mechanical calculation of the spectrum of the physical quantities that describe the geometry of space."* (Rovelli (2004) 14)

Diese resultierende Diskretheit auf der formalen Grundlage eines Kontinuumsansatzes wird von einigen Autoren als immenser Erfolg der *Loop Quantum Gravity* gesehen – insbesondere im Vergleich mit Ansätzen, die schon von einer diskreten Struktur ausgehen.

> *"Thus the theory defines a fundamental length scale, even though it takes a differentiable manifold structure as basic. This is much more satisfying than lattice approaches in which such a fundamental length is 'put in by hand'."* (Monk (1997) 17)

Die sich auf der Grundlage einer Kontinuumsmodellierung ergebende Diskretheit der *Loop Quantum Gravity* ist zweifellos ein Erfolg, zumindest im Vergleich mit Gitteransätzen, deren basale Länge tatsächlich nicht physikalisch motiviert ist, sondern eine artifizielle Komponente des modelltheoretischen Vehikels darstellt. – Dies spricht aber noch nicht unbedingt dafür, dass eine Theorie, für die sich ein solcher Erfolg einstellt, notwendigerweise eine realistische Beschreibung liefert oder als Instantiierung einer angemessenen Strategie angesehen werden muss. Und es heisst schon gar nicht, dass eine solche Theorie im Vergleich zu Theorien, die von Anfang an mit einer diskreten Struktur arbeiten, besser abschneidet; vorausgesetzt diese diskrete Struktur ab initio beruht auf physikalisch motivierbaren Grundprinzipien und ist nicht einfach nur ein Rechenkonstrukt.

Ein Indiz für die Begründetheit dieses Restzweifels hinsichtlich einer post-hoc-Diskretisierung manifestiert sich in der Tatsache, dass die Spinnetze für ihre Ableitung immer noch notwendigerweise das Kontinuum voraussetzen. Sie setzen insbesondere das Kontinuum der reellen Zahlen in der gleichen Weise voraus, wie dies etwa die diskreten Eigenzustände eines harmonischen Oszillators in der Quantenmechanik tun:

> *"This discreteness of the geometry, implied by the conjunction of [general relativity] and [quantum mechanics], is very different from the naive idea that the world is made by discrete bits of something. It is like the discreteness of the quanta of the excitations of a harmonic oscillator. A generic state of spacetime will be a continuous quantum superposition of states whose geometry has discrete features, not a collection of elementary discrete objects."* (Rovelli (2001) 110) – *"The discrete structure which [Loop Quantum Gravity] imposes is also entirely different from the discreteness of a lattice or naive discretisation of space (i.e. of a finite or countable set). Namely, it arises by 'polymerising' the continuum via an unusual scalar product."* (Nicolai / Peeters (2006) 4)

Und genau in dieser Hinsicht setzt die diskrete Struktur der Spinnetze tatsächlich immer noch die räumliche Mannigfaltigkeit voraus:

> *"Let us emphasize again that the 'discreteness' of the spin networks does not correspond to a naive discretisation of space. Rather, the underlying continuum, on which the spin networks 'float', the spatial manifold Σ, is still present."* (Nicolai / Peeters / Zamaklar (2005) 18)

Indem die *Loop Quantum Gravity* aber eine diskrete Struktur hervorbringt, die notwendigerweise ein Kontinuum der reellen Zahlen voraussetzt, steht sie, zumindest was ihren formalen Apparat und seine modelltheoretischen Voraussetzungen betrifft, in deutlichem Widerspruch zu den Argumenten für eine finite diskrete Struktur, wie sie sich in der Thermodynamik schwarzer Löcher mit der *Bekenstein-Hawking-Entropie* und der *holographischen Grenze* ergeben.[302] Die *Loop Quantum Gravity* arbeitet formal weiterhin mit dem Mythos der reellen Zahlen. Vor diesem Hintergrund lassen sich die Hinweise auf eine fundamentale Diskretheit, die sich im Kontext der *Loop Quantum Gravity* einstellen, – und insbesondere ihr Auftreten schon auf der kinematischen Beschreibungsebene, auf der die Hamiltonsche Zusatzbedingung, die erst die Dynamik erfasst, noch gar nicht berücksichtigt wurde – vielleicht auch als Anzeichen dafür lesen, dass es eine geeignetere modelltheoretische Ausgangsbasis geben könnte: eine solche, die ohne den Mythos der reellen Zahlen auskommt und gleich mit einem diskreten Substrat startet.[303] Die Ableitung der Diskretheit als Ergebnis einer Kontinuumsmodellierung ist eben noch nicht notwendigerweise überzeugender als ein Ansatz, der mit einer diskreten Struktur startet, wenn diese durch ausreichende konzeptionelle Motivationen begründet wird, insbesondere aber, wenn ein solcher Ansatz – im Gegensatz zu typenähnlichen wie typenunähnlichen Konkurrenten – schliesslich die richtigen, empirisch bestätigbaren, phänomenologischen Konsequenzen aufweist.

Andererseits ist die Entdeckung weiterer Indizien für die Diskretheit der Raumzeit, auch wenn sie vielleicht einem Kontext entstammen, der in seiner modelltheoretischen Ausrichtung von problematischen Voraussetzungen ausgeht, eine weitere wichtige Stützung für diese Idee. In dieser Hinsicht dient dann auch die direkte Quantisierung der Allgemeinen Relativitätstheorie als heuristisches Instrumentarium zur Stützung dieser Idee. Aber dieses heuristische Instrumentarium führt noch nicht notwendigerweise zu einer physikalisch und empirisch angemessenen Form der Erfassung der spezifischen Beschaffenheit einer diskreten Raumzeit. Eine erfolgreiche Erschliessung der tatsächlich vorliegenden diskreten Struktur, die eventuell der Raumzeit unterliegt, müsste womöglich auf andere Motivationen und auf eine andere konzeptionelle wie modelltheoretische Ausgangsbasis zurückgreifen. Dies lässt sich nicht unbedingt im Kontext einer direkten Quantisierung der Allgemeinen Relativitätstheorie erreichen. Dass sie überhaupt zu Anzeichen einer diskreten Raumzeitstruktur führt, ist vielleicht schon mehr als man von einem solchen Ansatz unbedingt erwarten konnte.

## *Von den Spinnetzen zu den echten physikalischen Freiheitsgraden*

Zurück zum kinematischen Bild der Raumzeit, wie es sich in der *Loop Quantum Gravity* abzeichnet: Aufgrund der Forderung nach Diffeomorphismusinvarianz bzw. Eichinvarianz sind es nicht die Spinnetze selbst, denen unmittelbare physikalische Relevanz zugesprochen werden kann. Denn die Spinnetze sind nicht diffeomorphismusinvariant. Die ihnen zugrundeliegenden geometrischen Operatoren, der Flächen- und der Volumenoperator, kommutieren nicht mit allen Quanten-Zusatzbedingungen.[304] Somit kommen sie als Observablen nicht in Frage; denn entsprechend der Dirac-Quantisierung eines Hamiltonschen Systems mit Zusatzbedingungen müssen alle Observablen mit allen Quanten-Zusatzbedingungen kommutieren. Dies entspricht gerade der Einforderung ihrer Eich- bzw. Diffeomorphismusinvarianz, die im ersten Schritt der Quantisierung mittels der Diracschen

---

[302] Siehe Kap.3.1.

[303] Weitere sich im Kontext der *Loop Quantum Gravity* ergebende Hinweise, die in diese Richtung zielen, werden weiter unten zu diskutieren sein. Hinzu kommen dann schliesslich Argumente für einen diskreten Ansatz, die völlig unabhängig von der *Loop Quantum Gravity* sind.

[304] Sie kommutieren ausschliesslich mit den artifiziellen Gaussschen Zusatzbedingungen.

Methode noch nicht berücksichtigt wurde; denn hier wird erst einmal der volle Phasenraum der kanonischen Variablen quantisiert, so dass die eichtheoretischen Redundanzen nach der Quantisierung zu berücksichtigen sind.

Es ist letztlich die in der *Loop Quantum Gravity* zentrale Forderung nach Hintergrundunabhängigkeit, die hier ausschlaggebend ist. Aus ihr resultiert die Notwendigkeit einer Berücksichtigung der Diffeomorphismusinvarianz, die bei Anwendung der Dirac-Methode erst nach der Quantisierung des klassischen Hamiltonschen Systems erfolgt. Die Diffeomorphismusinvarianz wird in der Hamiltonschen Darstellung mittels der Zusatzbedingungen eingefordert. Diese erhalten in der Dirac-Quantisierung entsprechende Quantenpendants – mit der Konsequenz, dass Grössen, die nicht mit allen Quanten-Zusatzbedingungen kommutieren, nicht diffeomorphismusinvariant und entsprechend grundsätzlich nicht beobachtbar sind. Sie kommen somit als Observablen nicht in Frage. Anders formuliert: Sie sind nicht eichinvariant und insofern nicht als physikalisch real anzusehen, wenn man die grundsätzliche Beobachtbarkeit als konstitutiv für physikalisch ernstzunehmende Grössen betrachtet. Der Hilbertraum der echten physikalischen Freiheitsgrade enthält dieser Auffassung zufolge nur Zustände, die von allen Quanten-Zusatzbedingungen annihiliert werden. Der Übergang zu diesem Hilbertraum der echten physikalischen Freiheitsgrade entspricht gerade der 'Lösung' der Quanten-Zusatzbedingungen. Diese 'Lösung' der Quanten-Zusatzbedingungen führt zur Auszeichnung der tatsächlichen physikalischen Grössen und gewährleistet deren Eich- und Diffeomorphismusinvarianz – und mithin die Hintergrundunabhängigkeit der Theorie.

Unter Festschreibung dieser Anforderungen durch die entsprechenden Prozeduren der Dirac-Quantisierung bleibt in der *Loop Quantum Gravity* kein Spielraum mehr für eine eventuelle andersgeartete Interpretation der Diffeomorphismusinvarianz. Vielmehr wird diese als Eichinvarianz direkt in die Theorie implementiert.

> *"[...] in the classical case, it must remain an optional* interpretative move *(albeit an obvious one) to regard equivalence classes of such points under 3-diffeomorphisms as what really correspond to instantaneous states. The corresponding move in loop quantum gravity is already hard-wired into the equations of the theory."* (Pooley (2006) 379)

Dennoch lässt sich dies nicht wirklich als Pioniertat der *Loop Quantum Gravity* ansehen: Der interpretatorische Spielraum, der diesbezüglich in der Allgemeinen Relativitätstheorie erst einmal noch bestehen mag, wird auch dort schon in der gleichen Weise aufgegeben, wenn man die Diffeomorphismusinvarianz der Theorie nicht nur als Eichinvarianz interpretiert, sondern im Übergang zu ihrer Hamiltonschen Darstellung in Form der Zusatzbedingungen, die auch formal als Generatoren von Eichtransformationen behandelt werden, festschreibt. Die *Loop Quantum Gravity* als konnektionsdynamische Variante der *Kanonischen Quantengravitation* übernimmt gerade diese formale Festschreibung durch die Wahl der Hamiltonschen Darstellung als Ausgangspunkt für die Quantisierung; die Behandlung der Zusatzbedingungen als Generatoren von Eichtransformationen findet nach der Quantisierung ihr Pendant in der Forderung, dass alle Observablen mit allen Zusatzbedingungen kommutieren müssen.

Für die geometrischen Operatoren bedeutet dies, dass sie als quantenmechanische Observablen nicht in Frage kommen, weil sie eben nicht mit allen Quanten-Zusatzbedingungen kommutieren; sie kommutieren erst einmal nur mit den Gaussschen Quanten-Zusatzbedingungen.[305]

> *"Note that the area operator is not invariant under three-dimensional diffeomorphisms. [...] It does also not commute with the Hamiltonian constraint. An area operator that is invariant should be defined intrinsically with respect to curvature invariants or matter fields. A concrete realization of such an operator remains elusive."* (Kiefer (2005) 11)

Die Bedeutung der geometrischen Operatoren stützt sich daher nicht zuletzt auf die Hoffnung, dass sich schliesslich diffeomorphismusinvariante, eichinvariante Pendants finden lassen. Dabei müssten nach Auffassung einiger Autoren vermutlich Materiefreiheitsgrade ins Spiel kommen:

> *"[...] it must be emphasized that the area and volume operators are* not *observables in the Dirac sense, as they do not commute with the Hamiltonian. To construct* physical *operators corresponding to area and volume is more difficult and would require the inclusion of matter (in the form of 'measuring rod fields')."* (Nicolai / Peeters (2006) 5)  –  *"Their supposed physical relevance is based on the general belief that proper Dirac observables corresponding to area and volume should exist, provided suitable matter degrees of freedom are included. The reason is that, for an operational definition of area and volume, one needs 'measuring rod fields' in the same way that the operational definition of 'time' in quantum gravity requires a 'clock field', and that the true observables are appropriate combinations of gravitational and matter fields (which then would commute with the constraints)."* (Nicolai / Peeters / Zamaklar (2005) 24)

Bisher beruht diese Hoffnung jedoch auf nichts mehr als einer Analogie zwischen klassischer und Quantentheorie.

> *"[...] we can gauge-fix the coordinates by choosing them to be determined by chosen physical rods and clocks. Then non-diff-invariant observables in the pure gravity theory corresponds precisely to diff-invariant observables in the matter+gravity theory. [...] The fact that this is possible in the classical theory suggests that the same could happen in the quantum theory."* (Rovelli (2007a) 5)

Ob sich diese Hoffnung aber tatsächlich erfüllen wird, bleibt vorerst unklar. Und für die materiefreie Formulierung der *Loop Quantum Gravity* hilft dies ohnehin nicht weiter. Diese setzt vielmehr erst einmal auf einen stufenweisen Prozess in der Berücksichtigung der Zusatzbedingungen.[306] Für die Spinnetze bedeutet dies: Berücksichtigt man erst einmal nur die räumlichen Diffeomorphismen,[307] – mit der naheliegenden Begründung, dass die Spinnetze auf den räumlichen Hyperflächen definiert sind –, so ergeben sich als relevante Grössen die entsprechenden Äquivalenzklassen von

---

[305] Man könnte sich hier nun fragen: Wie lassen sich eichabhängige Grössen, die entsprechend der eichinvarianten Sichtweise keine Observablen sein können, überhaupt als Quantenoperatoren verwenden? Wie lassen sich eichabhängige Grössen überhaupt quantisieren? – Solange für diese Grössen damit noch kein Observablenstatus und somit keine unmittelbare physikalische Relevanz postuliert wird, ist das erst einmal kein Problem; es handelt sich um nichts mehr als eine modelltheoretische Transposition im Übergang von den klassischen zur Quantentheorie.
   *"[...] non-physical things can be quantized."* (Rickles (2005a) 26)
[306] Zu diesem stufenweisen Lösungsansatz wird etwas weiter unten mehr zu sagen sein.
[307] Auch zu den zeitlichen Diffeomorphismen, die durch die Hamiltonsche Zusatzbedingung erfasst werden, wird bald mehr zu sagen sein.

Spinnetzen unter der Gruppe der räumlichen Diffeomorphismen. Diese werden als *S-Knoten* bezeichnet.

> *"Within the framework of loop quantum gravity, regarding s-knot states, rather than spin-network states, as the genuine physical states is not an optional move that one might be persuaded to take in response to some analogue of the hole argument. A quantum theory which countenances spin-network states as physical states is simply not a quantum version of general relativity."* (Pooley (2006) 378)

Die S-Knoten sind abstrakte topologische Gebilde, die nicht etwa auf einem Hintergrundraum existieren, sondern vielmehr den Raum selbst repräsentieren. Der quantisierte Raum lässt sich dann, indem er durch die S-Knoten repräsentiert wird, als vollständig relational bestimmtes Gefüge verstehen. Jegliche Lokalisierung erfolgt in Bezug auf die S-Knoten.[308]

> *"Spacetime itself is formed by loop-like states. Therefore the position of a loop state is relevant only* with respect to other loops*, and not with respect to the background."* (Rovelli (2004) 12) – *"The spin network represent relational quantum states: they are not located in a space. Localization must be defined in relation to them."* (Rovelli (2001) 110)

Die Zustände, die von der *Loop Quantum Gravity* beschrieben werden, sind also keine Anregungszustände eines Feldes, welches auf der Raumzeit definiert ist, sondern es sind Anregungszustände des Raumes (seiner Metrik bzw. seiner Konnektionen) selbst.

> *"A spin-network is a graph whose nodes represent 'chunks' of space and whose links represent surfaces separating these chunks. The spin-network (or rather the equivalence class of spin-networks under the group of diffeomorphisms of the spatial manifold) then represents a quantum state of the gravitational field, or of space."* (Rickles (2005a) 17)

Die als Eichinvarianz behandelte Diffeomorphismusinvarianz verbürgt auf diese Weise in ihrer direkten Implementierung in den Formalismus der *Loop Quantum Gravity* nicht nur deren Hintergrundunabhängigkeit, sondern hier nun auch die Relationalität der Raumzeit (bzw. des Raumes – denn bisher wurden nur die räumlichen Diffeomorphismen bzw. die entsprechenden Zusatzbedingungen berücksichtigt): einen Relationalismus, der im Rahmen der eichinvarianten Interpretation immer schon in die aktive Diffeomorphismusinvarianz hineingelesen wurde und hier nun fest implementiert wird.

<div align="center">*</div>

Dass es jedoch in der *Loop Quantum Gravity* nach der Quantisierung überhaupt noch Zusatzbedingungen gibt, die auf diese Weise, etwa im Übergang von den Spinnetzen zu den S-Knoten, (sukzessive) berücksichtigt werden müssen, liegt gerade an den schon mehrfach beschworenen Modalitäten der Diracschen Quantisierungsmethode, die in der *Loop Quantum Gravity* in der gleichen Weise wie schon in der geometrodynamischen Variante der *Kanonischen Quantisierung* zur Anwendung kommt. Bei dieser wird die Eich- bzw. Diffeomorphismusinvarianz eben erst nach der Quantisierung mittels der Quanten-Zusatzbedingungen eingefordert.

---

[308] Insbesondere wären alle Materiefelder und alle nicht-gravitativen Wechselwirkungsfelder in ihrer Lokalisierung in Bezug auf das System der S-Knoten zu bestimmen. Es gibt zudem die Idee, dass sich Materieteilchen vielleicht als Knotentypen in das System der S-Knoten integrieren lassen. Nicht-gravitative Felder, so wird dann vermutet, lassen sich vielleicht als Graphenattribute in die S-Knoten integrieren. – Vgl. auch die topologischen Eigenschaften der Materiekonstituenten in den prägeometrischen *Quantum Causal Histories*; siehe Kap. 4.6.

> *"[Loop Quantum Gravity] follows Dirac's constraint-quantization scheme. The idea is to turn the classical constraints into operators on a Hilbert space **H** and then to enforce gauge invariance by requiring that the physical sector $H_{phys}$ C **H** of the Hilbert space is the subspace corresponding to the kernel of the constraint operators."* (Earman (2006a) 19)

Angesichts der daraus resultierenden Problemlagen könnte man durchaus auf die Idee kommen, dass eine alternative Vorgehensweise, bei der überhaupt nur die echten physikalischen Freiheitsgrade – ohne die im vollen kinematischen Phasenraum enthaltenen Redundanzen und unphysikalischen Zustände – in die Quantisierung eingehen, einer adäquateren Implementierung der Diffeomorphismusinvarianz und ihrer eichinvarianten Deutung entspräche. Es gingen dann überhaupt nur eich- bzw. diffeomorphismusinvariante Grössen in die Quantisierung ein. Die Eich- bzw. Diffeomorphismusinvarianz müsste nicht erst nachträglich eingefordert bzw. berücksichtigt werden. – Von einer solchen, auf dem reduzierten Phasenraum und seiner Quantisierung beruhenden Strategie ist dann auch hin und wieder die Rede, wenn es um die formale Struktur der *Loop Quantum Gravity* und die aus ihr resultierenden Probleme geht:

> *"To pass to the quantum theory, one can use one of the two standard approaches: i) find the reduced phase space of the theory representing 'true degrees of freedom' thereby eliminating the constraints classically and then construct a quantum version of the resulting unconstrainted theory; or ii) first construct quantum kinematics for the full phase space ignoring the constraints, then find quantum operators corresponding to constraints and finally solve quantum constraints to obtain the physical states. Loop quantum gravity follows the second avenue [...]."* (Ashtekar / Lewandowski (2004) 51)

Grundsätzlich sind also zwei Strategien zur Quantisierung eines Hamiltonschen Systems mit Zusatzbedingungen denkbar:

(1) *Quantisierung des reduzierten Hamiltonschen Systems:*
   Man löse die (klassischen) Zusatzbedingungen und konstruiere auf diese Weise den reduzierten Phasenraum der eichinvarianten, echten, als physikalisch real anzusehenden Zustände des klassischen Systems. Erst dann quantisiere man die so erhaltene, auf dem reduzierten Phasenraum basierende klassische Theorie, die dann auch keine Redundanzen (Eichfreiheiten) und entsprechend keine Zusatzbedingungen, die diese erfassen, mehr aufweist, ebensowenig wie unphysikalische Zustände. Man quantisiert also nur die eichinvarianten, echten, physikalisch realen Zustände des klassischen Systems.

(2) *Diracsche Quantisierung:*
   Man quantisiere das auf dem vollen kinematischen Phasenraum definierte, noch mit Zusatzbedingungen ausgestattete Hamiltonsche System, ohne bei der Quantisierung die Zusatzbedingungen zu berücksichtigen. Man quantisiere also den vollen Phasenraum des klassischen Systems, der noch Redundanzen (Eichfreiheiten) und unphysikalische Zustände enthält.[309] Dann definiere man Quanten-Zusatzbedingungen als Pendants zu den klassischen Zusatzbedingungen. Schliesslich löse man (notfalls sukzessive) die Quanten-

---

[309] John Stachel notiert zu diesem Schritt in der Dirac-Quantisierung:
   *"Note that, in this approach, the commutation relations are simply postulated."* (Stachel (2006a) 73)

Zusatzbedingungen, um auf diese Weise den Hilbertraum der als physikalisch real anzu-
sehenden Quantenzustände zu erhalten. Letzteres geschieht, indem man für die als physi-
kalisch real anzusehenden Grössen – die Observablen – Eichinvarianz fordert, was nach
der Quantisierung gerade der Anforderung entspricht, dass die quantenmechanischen
Operatoren, die diese Grössen repräsentieren, mit allen Quanten-Zusatzbedingungen
kommutieren müssen. Die als physikalisch real anzusehenden Quantenzustände sind
dann genau die, die von allen Quanten-Zusatzbedingungen annihiliert werden.

Warum verwendet nun die *Loop Quantum Gravity* die Diracsche Quantisierungsmethode und nicht
die auf dem reduzierten Hamiltonschen Phasenraum basierende Methode, in der die Forderung nach
Eich- bzw. Diffeomorphismusinvarianz wesentlich direkter umgesetzt wird? – Dafür gibt es gute
Gründe: Die Alternative zur Dirac-Quantisierung steht nämlich eigentlich gar nicht zur Verfügung,
da es sich bisher als unmöglich erwiesen hat, den reduzierten Phasenraum der Allgemeinen Relati-
vitätstheorie, der deren 'echte' Freiheitsgrade erfassen soll, zu konstruieren.

> *"A distinct quantization method is the* reduced phase space quantization*, where the physical phase
> space is constructed classically by solving the constraints and factoring out gauge equivalence prior
> to quantization. But for a theory as complicated as general relativity it seems impossible to construct
> the reduced phase space."* (Gaul / Rovelli (2000) 9)

Über den reduzierten Phasenraum der Allgemeinen Relativitätstheorie ist immer noch sehr wenig
bekannt; das meiste davon ist Spekulation.[310]

> *"Relatively little is presently known about the structure of the reduced phase space of general relati-
> vity. It is known that this space has singularities corresponding to models of general relativity with
> symmetries, and is smooth elsewhere [...]."* (Belot / Earman (2001) 229)

Aber auch wenn dies anders sein sollte, blieben immer noch Restzweifel, ob die beiden alternativen
Quantisierungsmethoden überhaupt zum gleichen Ergebnis hinsichtlich des resultierenden Hilbert-
raums der als physikalisch real anzusehenden Quantenzustände führen würden.

> *"The two methods could lead to inequivalent quantum theories. Of course, it is possible, in principle,
> that more than one consistent quantum theory having general relativity as its classical limit might
> exist."* (Gaul / Rovelli (2000) 9)

Wenn jedoch die beiden Alternativen tatsächlich zu unterschiedlichen Ergebnissen hinsichtlich der
realen physikalischen Gegebenheiten kämen, würde dies den gesamten Ansatz der direkten kanoni-
schen Quantisierung in Frage stellen. – Aber wie gesagt, die Auffassung, dass es im Kontext der
kanonischen Quantisierung der Allgemeinen Relativitätstheorie überhaupt eine Alternative zur Di-
racschen Methode gibt, ist nichts mehr als eine Chimäre.

---

[310] Jenseits der Spekulation: Eine ansatzweise Erschliessung des reduzierten Phasenraumes unter spezifischen Bedin-
gungen findet sich in Fischer / Moncrief (1996). Siehe auch Giulini (2009), der insbesondere die Beziehungen zwischen
(i) dem reduzierten Phasenraum, (ii) dem *Superraum* (d.h. dem nur um die räumlichen, nicht aber um die zeitlichen
Diffeomorphismen reduzierten Phasenraum: *"Note that the cotangent bundle over Superspace is not the fully reduced
phase space for matter-free General Relativity. It only takes into account of the vector constraints and leaves the scalar
constraint unreduced."* (Giulini (2009) 25)) sowie (iii) dem *erweiterten Superraum* verdeutlicht und sich schliesslich
vor allem der Struktur des Superraums und der des erweiterten Superraums widmet.

*

Der *Loop Quantum Gravity* bleibt also erst einmal nichts anderes übrig, als auf die Diracsche Quantisierungsmethode zu setzen. Es ist aber immer noch unklar, ob sich alle damit verbundenen technischen, konzeptionellen und interpretativen Probleme lösen lassen.

> *"A common feature for all variables is that the quantum constraints are difficult if not impossible to solve."* (Kiefer (2004[²2007]) 243)

Die Lösung der Quanten-Zusatzbedingungen – der abschliessende Schritt der Diracschen Quantisierungsmethode – wird in der *Loop Quantum Gravity*, wie schon erläutert, sukzessive vollzogen: Zuerst werden die Gaussschen Zusatzbedingungen berücksichtigt, dann die Diffeomorphismus-Zusatzbedingungen, schliesslich die Hamiltonsche Zusatzbedingung – so lautet jedenfalls das Programm. Dazu müssen die Quanten-Zusatzbedingungen (als Pendants der klassischen Zusatzbedingungen) aber erst einmal adäquat definiert werden:

> *"To implement these constraints at the quantum level, one must first properly define them, i.e. express them in terms of the elementary variables, the holonomies and fluxes, and then investigate their properties."* (Nicolai / Peeters / Zamaklar (2005) 35)

Für die Gaussschen Zusatzbedingungen, die als Generatoren von SU(2)-Eichtransformationen anzusehen sind – sie resultieren aus der Formulierung der klassischen Theorie auf der Basis von Konnektionsvariablen –, ist dies noch unproblematisch. Nach der Lösung der Gaussschen Zusatzbedingungen lassen sich die geometrischen Operatoren für Fläche und Volumen formulieren. Ihr Eigenwertspektrum entspricht gerade den Spinnetzen. Diese, wie auch die geometrischen Operatoren selbst, sind jedoch noch nicht diffeomorphismusinvariant.

Die Einbeziehung der räumlichen Diffeomorphismusinvarianz führt dann zum Übergang von den Spinnetzen zu den S-Knoten. Die räumliche Lokalisierung der entsprechenden Operatoren geht hierbei verloren. Und es kommt schon hier zu einigen durchaus signifikanten Abweichungen vom vorgezeichneten Programm der Dirac-Quantisierung: So gibt es insbesondere für die Diffeomorphismus-Zusatzbedingungen keine quantenmechanischen Operatoren. Entsprechend können dann auch nicht irgendwelche Quantenpendants zu den räumlichen Diffeomorphismus-Zusatzbedingungen der klassischen Theorie gelöst werden, um die diffeomorphismusinvarianten Zustände zu ermitteln. Vielmehr geht man schlichtweg direkt von den Spinnetzen zu den entsprechenden Äquivalenzklassen unter der Gruppe der (klassischen) räumlichen Diffeomorphismen über, den S-Knoten.

Diese sind aber immer noch nicht invariant unter zeitlichen Diffeomorphismen. Die Hamiltonsche Zusatzbedingung ist an dieser Stelle immer noch unberücksichtigt. Die S-Knoten sind also noch nicht die Zustände des echten physikalischen Hilbertraums der *Loop Quantum Gravity*. – Aber hier liegen eigentlich schon die Grenzen der *Loop Quantum Gravity*; sie erschöpft sich letztlich in dem immer noch rein kinematischen Bild, das sich nach Berücksichtigung der Gaussschen und der Diffeomorphismus-Zusatzbedingungen einstellt.

> *"To summarize, the main results in the loop or connection representation have been found on the kinematical level. This holds in particular for the discrete spectra of geometric operators. The main* *open problem is the correct implementation (and solution) of the Hamiltonian constraint."* (Kiefer (2004 [²2007]) 198)

Mit dem letzten Teilschritt – der Berücksichtigung der Hamiltonschen Zusatzbedingung – beginnen definitiv die Probleme der *Loop Quantum Gravity*.

## Probleme

Eine konsistente Berücksichtigung der zeitlichen Diffeomorphismusinvarianz bzw. eine Lösung der Hamiltonschen Zusatzbedingung ist bisher noch nicht in eindeutiger und umfassender Weise gelungen.[311]

> *"[...] so far the problem of finding physical observables in quantum gravity is still very little explored territory [...]."* (Gaul / Rovelli (2000) 47)

Anscheinend wird eine solche Lösung von einigen Theoretikern der *Loop Quantum Gravity* auch gar nicht mehr in umfassender Hinsicht erwartet:

> *"The final step [...] remains to be done: the physical states of the theory should lie in the kernel of the quantum Hamiltonian constraint operator. Of course we do not expect to find a complete solution of the Hamiltonian constraint, which would correspond to a complete solution of the theory."* (Gaul / Rovelli (2000) 39)

Nicht einmal eine eindeutige Formulierung der Quantenversion der Hamiltonschen Zusatzbedingung liegt vor.[312][313]

> *"[...] there is still a large number of poorly controlled ambiguities in the definition of the Hamiltonian constraint."* (Ashtekar (2007a) 12)

Es gibt bisher ausschliesslich Ansätze mit reduzierten Ambitionen, die auf Einschränkungen (etwa auf den ausschliesslich euklidischen Fall) und auf vereinfachenden Plausibilitätsüberlegungen beruhen:

> *"A more complete treatment would include exponentially increasing efforts [...]."* (Gaul / Rovelli (2000) 41)

---

[311] Die bei der Lösung der Hamiltonschen Zusatzbedingung auftretenden Probleme legen in mancher Hinsicht einen Übergang von der *Loop Quantum Gravity* zu den kovarianten *Spin-Foam*-Modellen nahe. Aber auch hier lässt sich dann keine eindeutige Lösung für diese Probleme erreichen. Siehe weiter unten.

[312] So würde etwa ein imaginärer Immirzi-Parameter, zumindest in der Ashtekarschen Konnektionsdynamik, zu einer einfachen Hamiltonschen Zusatzbedingung führen, die sich polynomisch in den Ashtekar-Variablen darstellen lässt. Die Hamiltonsche und die Diffeomorphismus-Zusatzbedingungen liessen sich dann gemeinsam lösen. Jedoch führt ein solcher imaginärer Immirzi-Parameter, der im klassischen Fall unproblematisch wäre und einer Koordinatentransformation entspräche, zu grossen Problemen in der Formulierung der Quantentheorie.

[313] Es gibt schon Autoren, die diese Uneindeutigkeit mit der der *String-Landscape* vergleichen. Siehe etwa Nicolai / Peeters (2006).

Und fast alle weiteren Probleme der *Loop Quantum Gravity* hängen auf die eine oder andere Weise mit dem unvollendeten sukzessiven Abarbeiten in der Lösung der Quanten-Zusatzbedingungen und dem Scheitern an der Hamiltonschen Zusatzbedingung – nicht erst nur hinsichtlich ihrer Lösung, sondern schon in ihrer Formulierung als Quanten-Zusatzbedingung – zusammen.

> *"The core difficulties of canonical quantum gravity are all connected in one way or another to the Hamiltonian constraint – irrespective of which canonical variables are used."* (Nicolai / Peeters / Zamaklar (2005) 10)

Dies betrifft insbesondere das zentrale Problem der *Loop Quantum Gravity*: die Schwierigkeiten bei der Reproduktion der bekannten Niederenergiephysik und der Formulierung eines klassischen Grenzfalls.

> *"The main difficulties of loop quantum gravity lie in recovering low energy phenomenology. Quantum states corresponding to the Minkowski vacuum and its excitation have not yet been constructed, and particle scattering amplitudes have not been computed."* (Rovelli (2007) 1301)

Angesichts der Probleme mit der Hamiltonschen Zusatzbedingung, die (im allgemeinen Fall) gerade die Dynamik der Theorie erfasst, sind die Schwierigkeiten hinsichtlich einer (zumindest näherungsweisen) Reproduktion der Einsteinschen Feldgleichungen nicht verwunderlich.

> *"[...] there remain however, hard issues concerning whether and how classical general relativity dominates a suitably defined low energy limit. The fact that the theory is well defined and finite does not, so far as we know, guarantee that the low energy limit is acceptable."* (Smolin (2003) 27)

Vor diesem Hintergrund erweist es sich als grundsätzlich fragwürdig, ob die Dirac-Quantisierung oder zumindest ihre stufenweise Realisierung in der *Loop Quantum Gravity* zum gewünschten Ziel führen wird: eine Quantentheorie zu entwickeln, welche die Allgemeine Relativitätstheorie als klassischen Grenzfall (bzw. als makroskopische Niederenergienäherung) enthält, die bekannte niederenergetische Phänomenologie reproduziert und damit die wesentlichste Anforderung an jegliche Theorie der Quantengravitation erfüllt. – Sollte dieses Ziel sich auch in Zukunft nicht realisieren lassen, so wäre dies das Ende der *Loop Quantum Gravity*.

> *"Loop quantum gravity [...] will fail if it turns out that the low energy limit of quantum general relativity coupled to matter is not classical general relativity coupled to quantum matter fields."* (Smolin (2003) 32)

Sollte die *Loop Quantum Gravity* hingegen im Hinblick auf diese Zielsetzung schliesslich doch noch erfolgreich sein, so ist jetzt schon klar, dass die äusserst radikalen Konsequenzen, zu denen sie verglichen mit den etablierten Theorien führt, dann erst einmal als mögliche realistische Einschätzung des Naturgeschehens ernst zu nehmen wären:

> *"[...] the theory gives up unitarity, time evolution, Poincaré invariance at the fundamental level, and the very notion that physical objects are localized in space and evolve in time."* (Rovelli (2007) 1302)

Diese Konsequenzen würden dann jedoch in einiger Hinsicht ein Umdenken erforderlich machen. Die mangelnde Unitarität der *Loop Quantum Gravity* etwa wird oft als erhebliches Problem für eine

Quantentheorie angesehen. Die herkömmliche Quantenmechanik setzt die Unitarität der zeitlichen Entwicklung nämlich uneingeschränkt voraus. Mindestens eine konzeptionelle Erweiterung wäre erforderlich, um dieses Problem zu beheben.

*"In conventional [quantum mechanics] and [quantum field theory], unitarity is a consequence of the time translation symmetry of the dynamics. In [general relativity] there isn't, in general, an analogous notion of time translation symmetry. Therefore there is no sense in which conventional unitarity is necessary in the theory. One often hears that without unitarity a theory is inconsistent. This is a misunderstanding that follows from the erroneous assumption that all physical theories are symmetric under time translation."* (Rovelli (2004) 267)

Die Ursache dafür, dass es in der *Loop Quantum Gravity* keine unitären zeitlichen Entwicklungen gibt, hängt mit einer der radikalsten Konsequenzen dieser Theorie zusammen:

*"Absence of a fundamental notion of time evolution implies in particular that there is no unitary time evolution of the theory."* (Rovelli (2007) 1319) – *"There is no need to expect or to search for unitary time evolution in quantum gravity, because there is no time in which we could have unitary evolution."* (Rovelli (2001) 114)

Auch wenn eine Lösung der Hamiltonschen Zusatzbedingung noch nicht gelungen ist, so ist immerhin schon klar, dass diese zu einem quantenmechanischen Zustandsraum führen würde, in dem zeitliche Entwicklungen im herkömmlichen Sinne keinen Platz mehr finden. Wird dieser quantenmechanische Zustandsraum als Repräsentation der realen physikalischen Grössen verstanden, was im Rahmen der *Loop Quantum Gravity* mit ihrem eingeschriebenen eichinvarianten Deutung der Diffeomorphismusinvarianz unvermeidlich ist, so gibt es keine zeitlichen Entwicklungen mehr. Genau dies ist das vielbeschworene *Problem der Zeit*.

## Das Problem der Zeit

Das *Problem der Zeit* ist eine direkte Konsequenz der Hintergrundunabhängigkeit der *Loop Quantum Gravity*, die durch die als Eichinvarianz verstandene und als solche in den Formalismus eingeschriebene aktive Diffeomorphismusinvarianz gewährleistet wird. Oder anders gesagt: Es ist eine direkte Folge der relationalistischen Implikationen der als Eichinvarianz behandelten aktiven Diffeomorphismusinvarianz.

*"Dynamics is notoriously difficult to implement in most background independent approaches [...]."* (Markopoulou (2007) 16) – *"Such constraints result from any theory that is classically reparametrization invariant, that is, a theory without background structure."* (Kiefer (2005) 9)

Das *Problem der Zeit* ergibt sich unmittelbar aus der Hamiltonschen Formulierung der Allgemeinen Relativitätstheorie und ihrer mit dieser Formulierung einhergehenden Interpretation als Eichtheorie. Die schon in der klassischen Theorie letztlich unausweichlichen konzeptionellen Implikationen übertragen sich – im Rahmen ihrer *Kanonischen Quantisierung* fest in den Formalismus implemen-

tiert – auf die so entstehende Quantentheorie.[314] Das *Problem der Zeit* ist also nicht etwa das origi-
näre Ergebnis der *Kanonischen Quantisierung*. Denn, sobald man eine eichinvariante Deutung der
Diffeomorphismusinvarianz zur Anwendung bringt, tritt das *Problem der Zeit*, wie wir gesehen
haben, schon in der Allgemeinen Relativitätstheorie auf.[315] Schon hier ist es eine Folge der als
Eichinvarianz verstandenen aktiven Diffeomorphismusinvarianz, die dann in der Hamiltonschen
Darstellung in Form der Zusatzbedingungen implementiert wird.

> *"[...] progress has been difficult, precisely because of the background independence of the classical
> theory, a feature that distinguishes it from all other theories that we have successfully quantized. The
> equations of General Relativity are invariant under the diffeomorphism group of the manifold under
> investigation. A canonical analysis reveals that this means that the system is completely constrained:
> instead of generating time evolution, the Hamiltonian vanishes on solutions."* (Markopoulou (2007)
> 15)

Die Koordinatenzeit der Allgemeinen Relativitätstheorie (der Entwicklungsparameter der Feldglei-
chungen) erweist sich unter diesen Bedingungen als nicht eichinvariant und somit physikalisch ir-
relevant. Die Hamiltonsche Zusatzbedingung, die (zumindest im Falle eines kompakten Raumes)
die zeitliche Entwicklung des Systems von einer Hyperfläche zur nächsten repräsentiert, entspricht
gerade einer Eichtransformation.

> *"This means that each dynamical trajectory lies in a single gauge orbit: as the gravitational field
> evolves, it always stays in the same gauge orbit."* (Belot / Earman (2001) 225)

Die Allgemeine Relativitätstheorie beschreibt dann nicht etwa ein System, das sich in der Zeit ent-
wickelt: zumindest nicht in einer Zeit, die der Beobachtung zugänglich wäre und die unabhängig
von der in ihr ablaufenden Entwicklung verliefe.

> *"[General Relativity] does not describe evolution with respect to an external time, but only relative
> evolution of physical variables with respect to each other. In other words, temporal localization is
> relational like spatial localization. This is reflected in the fact that the theory has no hamiltonian (un-
> less particular structures are added), but only a 'hamiltonian' constraint."* (Rovelli (1998) 20)

Jede Auszeichnung eines Zeitparameters verletzt die (unter Vorgabe der *Leibniz-Äquivalenz* als
Identitätskriterium) als Eichinvarianz interpretierte aktive Diffeomorphismusinvarianz. Genau in
diesem Sinne gibt es schon in der Allgemeinen Relativitätstheorie keine physikalisch wirksame,
reale Zeit. – Die praktischen Auswirkungen des *Problems der Zeit* lassen sich in der Allgemeinen
Relativitätstheorie jedoch grundsätzlich abmildern, wenn man sich erst einmal nur auf einzelne Lö-
sungen der Theorie beschränkt:

> *"Such a weakening of the notion of time in classical [general relativity] is rarely emphasized, because,
> after all, in classical physics we may disregard the full dynamical structure of the dynamical theory
> and consider only a single solution of its equations of motion. [...] a single solution of the [general*

---

[314] Das *Problem der Zeit* tritt, wie im entsprechenden Kontext schon erwähnt, notwendigerweise auch in der alten geo-
metrodynamischen Variante der kanonischen Quantisierung der Allgemeinen Relativitätstheorie auf, in der die quanti-
sierte Version der Hamiltonschen Zusatzbedingung als zeitunabhängige *Wheeler-DeWitt-Gleichung* in Erscheinung tritt.
[315] Siehe Kap. 2.1.

*relativity] equations of motion determines a spacetime, where a notion of proper time is associated to each timelike worldline."* (Rovelli (2007) 1318)

Für den Bereich der Quantengravitation verliert diese Begrenzung auf einzelne Lösungen jedoch ihre Berechtigung. Hier hat das *Problem der Zeit* deutlich ernsthaftere und ungebremste Konsequenzen. Man hat es nicht mehr mit einzelnen klassischen Lösungen zu tun, sondern mit quantenmechanischen Superpositionen von Räumen bzw. Raumzeiten.

> *"In the quantum context, on the other hand, there is no single spacetime, as there is no trajectory for a quantum particle, and the very concept of time becomes fuzzy."* (Rovelli (2007) 1318) – *"[...] in quantum gravity the notion of spacetime disappears in the same manner in which the notion of trajectory disappears in the quantum theory of a particle."* (Rovelli (2004) 21)

Und durch die feste Implementierung der eichinvarianten Interpretation physikalischer Grössen gibt es im Rahmen der *Loop Quantum Gravity* nun auch keine interpretatorischen Ausweichoptionen mehr hinsichtlich des *Problems der Zeit*. Es bleibt in der *Loop Quantum Gravity* einfach kein Spielraum mehr für eine eventuell abweichende Interpretation der Diffeomorphismusinvarianz, da diese über die Quanten-Zusatzbedingungen als Eichinvarianz direkt in die Formulierung der Theorie eingeht.

Analog zur klassischen Theorie ist es hier nun die quantisierte Version der Hamiltonschen Zusatzbedingung, die aus der *Loop Quantum Gravity* eine Theorie ohne Zeit werden lässt. Da die zeitliche Entwicklung des Systems wiederum durch die Hamiltonsche Zusatzbedingung erfasst wird, diese aber einer Eichtransformation entspricht, lässt sich diese zeitliche Entwicklung nicht als physikalisch real ansehen – ebensowenig wie alle zeitabhängigen Grössen. Diese sind allesamt nicht eichinvariant und somit auch nicht beobachtbar und nicht messbar; sie kommen als Observablen nicht in Frage.

> *"Since the quantum Hamiltonian is zero, there is no evolution in time of the quantum states. This is the core of the* problem of time*: there appears to be no time or change in quantum gravity. This is not surprising: [...] embracing an interpretation of general relativity which is gauge invariant [...] involves, at least* prima facie*, renouncing the existence of time and change."* (Belot / Earman (1999) 176)

Alle Observablen der *Loop Quantum Gravity* sind notwendigerweise zeitunabhängig: aufgrund der Tatsache, dass die Hamiltonsche Zusatzbedingung, die die zeitliche Entwicklung des Systems erfasst, dem Generator einer Eichtransformation entspricht und die quantenmechanischen Operatoren, die möglichen Observablen entsprechen, insofern mit ihr kommutieren müssen. Operatoren, die nicht mit der Hamiltonschen Zusatzbedingung kommutieren – hierzu zählen etwa alle zeitabhängigen Operatoren –, entsprechen Grössen, die grundsätzlich unbeobachtbar sind und daher als Observable nicht in Frage kommen.

> *"The definition of 'observable' in the context of constrained systems is given as a variable that (weakly) commutes with all the first class constraints. However, since one of these is the generator of time evolution (the Hamiltonian constraint), the observables must be constants of motion."* (Rickles (2005a) 12)

Durch die feste Implementierung der aktiven Diffeomorphismusinvarianz als Eichinvarianz ist daran auch nichts zu ändern. Diese Implementierung besteht gerade darin, dass für quantenmechanische Observablen gefordert wird, dass sie mit allen Quanten-Zusatzbedingungen kommutieren. Eine solche Festschreibung der Eichinvarianz hat ihre guten Gründe: Würde man nämlich Quanten-Observablen zulassen, die nicht mit den Quanten-Zusatzbedingungen kommutieren, so würde die Theorie unvorhersagbar.[316] Nur, wenn die Quanten-Observablen mit allen Quanten-Zusatzbedingungen kommutieren, führt die Anwendung des entsprechenden quantenmechanischen Operators nicht zu Zuständen, die zwar im vollen kinematischen Hilbertraum enthalten sind, die aber letztlich unphysikalisch sind. Der reduzierte Hilbertraum der tatsächlichen physikalischen Zustände wird gerade durch die Quanten-Zusatzbedingungen definiert und strukturiert; er enthält nur Zustände, die von allen Quanten-Zusatzbedingungen annihiliert werden.

<div align="center">*</div>

Das *Problem der Zeit* – die Tatsache, dass sowohl die Allgemeine Relativitätstheorie als auch die Quantentheorien, die sich durch ihre *Kanonische Quantisierung* ergeben, eine Welt ohne realen physikalischen Zeitverlauf, eine Welt ohne wirklichen Wandel beschreiben – steht in krassem Kontrast zu der offensichtlichen Welt des Wandels und des Werdens, in der wir leben. Heisst dies, dass diese Theorien schlichtweg falsch sein müssen? Oder lässt sich dieser Konflikt zwischen Theorie und Erleben aufheben?[317] – Es gibt diesbezüglich inzwischen eine ganze Reihe von Deutungen und Lösungsansätzen:

<div align="center">*</div>

Julian Barbour[318] etwa geht davon aus, dass Zeit, zeitliche Entwicklung und zeitlicher Wandel nichts anderes als Resultate einer Illusion sind. Die Argumente für sein Szenario eines zeitlosen Universums leitet er direkt aus dem *Problem der Zeit* ab, wie es sich in der Allgemeinen Relativitätstheorie und dann wieder in ihrer *Kanonischen Quantisierung* ergibt. Hinsichtlich der letzteren bezieht er sich auf deren ursprüngliche geometrodynamische Variante; die Argumente lassen sich jedoch ohne weiteres auf den Kontext der *Loop Quantum Gravity* übertragen.

Nach Barbours Auffassung gibt es, wenn man die Implikationen der Allgemeinen Relativitätstheorie und ihrer kanonischen Quantisierung ernst nimmt (was er tut), nur eine scheinbare, aber keine reale Zeit. Realität kommt – im klassischen Fall – nur den Dreier-Räumen bzw. den Äquivalenzklassen von Dreier-Räumen unter räumlichen Diffeomorphismen zu, die sich in einem nur relational bestimmten Konfigurationsraum (*'Platonia'*) erfassen lassen. Im Rahmen der geometrodynamischen Variante der *Kanonischen Quantengravitation* ist es dann die zeitlose quantenmechanische Superposition von Dreier-Räumen (bzw. Äquivalenzklassen von Dreier-Räumen unter räumlichen Diffeomorphismen), der Realität zugesprochen wird.[319] Innerhalb dieser zeitlosen Superposition von

---

[316] Zudem würde etwa die Auszeichnung eines eichabhängigen Zeitparameters dazu führen, dass unterschiedliche Zeitflussraten innerhalb der Raumzeit auftreten können.

[317] Ein Lösungsvorschlag für diesen Konflikt im Kontext der Allgemeinen Relativitätstheorie findet sich in Healey (2003).

[318] Siehe etwa Barbour (1999). Vgl. auch Butterfield (2002). Für eine neuere, vom Nachfolgenden abweichende Auffassung Barbours, die sich der Auffassung Kuchars (s.u.) annähert, siehe Barbour / Foster (2008).

[319] Im Rahmen der *Loop Quantum Gravity* wären es – ohne dass Barbour hierauf explizit Bezug nimmt – anstelle der Dreier-Geometrien die Spinnetze bzw. S-Knoten.

Dreier-Räumen, die der Lösung der statischen *Wheeler-DeWitt-Gleichung* entspricht, ergibt sich nach Barbours Auffassung eine Wahrscheinlichkeitsverteilung. Den einzelnen Komponenten lassen sich demzufolge eindeutig Amplituden zuordnen, die durch die zeitlose *Wheeler-DeWitt-Gleichung* ein- für allemal festgelegt sind.

> *"The central structure in Barbour's vision is the space of Riemannian 3-metrics modulo the spatial diffeomorphism group (known as 'superspace') [...]. Choosing this space as the configuration space of the theory amounts to solving the diffeomorphism constraint; this is Barbour's relative configuration space that he labels 'Platonia'. The Hamiltonian constraint is then understood as giving (once solved, and 'once and for all' (Barbour 1994) 2875) a static probability distribution over Platonia that assigns amplitudes to 3-geometries [...]."* (Rickles (2006) 180)

Barbours Ansatz behandelt, wie hier offensichtlich wird, Raum und Zeit in ontologischer Hinsicht grundlegend unterschiedlich. Von der Allgemeinen Relativitätstheorie, die das *Problem der Zeit* überhaupt erst hervorgebracht hat, wird eine solche Sichtweise nicht unterstützt.[320]

> *"Barbour believes that space is fundamental, rather than spacetime. This emerges from his Machian analysis of general relativity."* (Rickles (2006) 180)

Die Illusion der Zeit entsteht, wie Barbour behauptet, nun dadurch, dass die hinsichtlich ihrer Wahrscheinlichkeitsamplituden dominanten Anteile innerhalb dieser zeitlosen quantenmechanischen Superposition von Dreier-Räumen gerade die Eigenschaft haben, vermeintliche Hinweise auf eine zeitliche Entwicklung (vermeintliche Entwicklungsspuren) zu enthalten; sie lassen sich in dieser Hinsicht als 'Zeitkapseln' deuten.

> *"Barbour* conjectures *that the relative probability distribution determined by the Wheeler-DeWitt equation is peaked on time capsules [...]."* (Rickles (2006) 180) – *"The space of Nows is given once and for all and does not alter, nor does the quantum state function defined over this space, and therefore the probability distribution is fixed too."* (Rickles (2006) 182)

Die Dominanz der 'Zeitkapseln' – ihre relative Häufigkeit innerhalb der Superposition und vor allem ihre hohen Wahrscheinlichkeitsamplituden – suggeriert das Vorliegen von Zeitlichkeit, die es aber nicht gibt. Und dass sich die 'Zeitkapseln' zudem in eine logische Ordnung bringen lassen, die eine zeitliche Ordnung suggeriert, führt schliesslich zur Illusion einer zeitlichen Abfolge.[321] – Es bleibt

---

[320] Zumindest Rickles ist der Auffassung, dass es aus diesem Problem vielleicht noch einen Ausweg gibt.
> *"Although Barbour's view is usually taken to imply a three-dimensionalist interpretation [...], I think it is also perfectly compatible with a kind of temporal parts type theory. However, rather than the structure of time being linear [...], it is, in a very rough sense, non-linear (modeled by relative configuration space) and the 'temporal evolution' is probabilistic (governed by a solution to the Hamiltonian constraint)."* (Rickles (2006) 181f)
Es bleibt jedoch sehr fraglich, ob Barbour einer solchen Auffassung zustimmen würde.

[321] Rickles suggeriert in seiner Erörterung des Barbourschen Szenarios, dass die Illusion einer zeitlichen Abfolge innerhalb des Zustandsraums der zeitlosen quantenmechanischen Superpositionen von Dreier-Räumen wiederum auf der Grundlage eines quantenmechanischen Prozesses (!) zustandekommt – also letztlich wiederum auf der Grundlage einer Dynamik. Ob er diese als reale Dynamik versteht, bleibt grundlegend unklar:
> *"[...] the quantum state 'jumps' around from Now to Now in accordance with the Hamiltonian constraint in such a way that the parts contain records that 'appear' to tell a story of linear evolution and persistence."* (Rickles (2006) 182) aber letztlich ein völliges Rätsel, wieso innerhalb des Zustandsraums der postulierten Superpositionen von Dreier-Geometrien gerade Zustände, die eine solche aussergewöhnliche Eigenschaft haben, wie sie den 'Zeitkapseln' zugeschrieben wird, in signifikanter Häufigkeit auftreten sollten und wieso ihnen hohe Wahrscheinlichkeitsamplituden zukommen sollten. Barbours Szenario kommt aber kaum ohne sie aus, wenn er die Illusion von Zeit erklären will. Die asymmetrische Behandlung von Raum und Zeit, die schon nicht mehr dem Geist der Allgemeinen Relativitätstheorie entspricht, ist da vermutlich das kleinere Problem.

*

Karel Kuchar[322] hingegen versucht, die radikalen Implikationen des *Problems der Zeit* zu überwinden. Sein Ziel ist es, innerhalb der *Kanonischen Quantisierung* der Allgemeinen Relativitätstheorie – er bezieht sich wiederum vor allem auf die geometrodynamische Variante – eine konzeptionelle Grundlage für echte zeitliche Entwicklungen zu etablieren. Sein Lösungsansatz besteht in gewisser Hinsicht im Versuch einer Übertragung der Idee einer lösungsspezifischen quasi-absoluten Zeit, wie sie sich im klassischen Fall im allgemeinen noch etablieren lässt, auf die Quantentheorie.

Kuchars Lösungsstrategie impliziert dabei eine Ablehnung der eichinvarianten Interpretation der Allgemeinen Relativitätstheorie in einigen, aber nicht in allen ihren Konsequenzen. Die (systemspezifische) absolute Zeit, die er zu etablieren versucht, kommutiert nämlich nicht mit der quantisierten Hamiltonschen Zusatzbedingung (also der *Wheeler-DeWitt-Gleichung*), ist also nach orthodoxer Auffassung nicht eichinvariant und somit nicht beobachtbar; aber Kuchar bestreitet gerade, dass die Hamiltonsche Zusatzbedingung als Generator einer Eichtransformation gedeutet werden kann. Und damit bestreitet er auch die Erfordernis, dass die Quantenobservablen mit der Hamiltonschen Zusatzbedingung kommutieren müssen. Die Diffeomorphismus-Zusatzbedingungen hingegen werden weiterhin als Generatoren von Eichtransformationen gesehen und behandelt. – Kuchars Begründung: Die Hamiltonsche Zusatzbedingung generiert gerade die zeitliche Entwicklung des Systems. Würde man sie als Generator einer Eichtransformation verstehen, so würde diese zeitliche Entwicklung nicht mehr als solche in Erscheinung treten können. Auch hier kommt also wieder eine asymmetrische Behandlung von Raum und Zeit zum tragen, die von der Allgemeinen Relativitätstheorie selbst nicht gestützt wird.

*"Viewed in this light, according to Belot and Earman's own taxonomy, Kuchar's position should more properly be seen as underwritten by a* relationist *interpretation of space coupled with a* substantivalist *interpretation of time!"* (Rickles (2006) 177)

Kuchars formale Strategie: Da die Diffeomorphismus-Zusatzbedingungen weiterhin als Generatoren von Eichtransformationen gesehen werden, können sie zu einer teilweisen Reduzierung des kinematischen Phasenraums genutzt werden. Für die Hamiltonsche Zusatzbedingung wird hingegen eine Eichfixierung verwendet – bzw. das, was einer Eichfixierung entspräche, wenn man die Hamiltonsche Zusatzbedingung als Generator einer Eichtransformation verstehen würde.

*"One frequently finds that the reduced space method is mixed with gauge-fixation methods, so that one has a partially reduced space, with the remaining gauge freedom frozen by imposing gauge conditi-*

---

Vielleicht sollte man hier besser von einer durch die Zeitkapselstruktur nahegelegten Illusion von Übergängen zwischen den Zeitkapseln ausgehen und das Reden über 'Quantensprünge' ausschliesslich als eloquente Metapher verstehen.
[322] Siehe Kuchar (1991, 1992).

*ons. Such an approach is used by a number of* internal time *responses to the problem of time. The idea is that one first solves the diffeomorphism constraint and then imposes gauge conditions on the gauge freedom generated by the Hamiltonian constraint. This is essentially the position of Kuchar [...]."*
(Rickles (2006) 166)

Das Besondere dabei ist, dass diese (uneigentliche) Eichfixierung nach Kuchars Auffassung vor der Quantisierung erfolgen sollte. Hierin kommt gerade die Idee zur Anwendung, eine lösungsspezifische quasi-absolute Zeit, wie sie sich im allgemeinen im klassischen Fall etablieren lässt, auf die entsprechende Quantentheorie zu übertragen – der dann allerdings keine universelle Gültigkeit zukommt, da sie dann eigentlich nur das Quantenpendant zu einer spezifischen klassischen Lösung darstellt.

*"The idea is to find a notion of time* before *quantization hidden amongst the phase space variables so that a time-dependent Schrödinger equation can be constructed [...]."* (Rickles (2006) 177)

Die (vermeintliche) Eichfixierung sollte Kuchars Strategie zufolge so vorgenommen werden, dass sie die echten geometrischen Freiheitsgrade, die den Übergang zwischen den räumlichen Hyperflächen bestimmen, identifizierbar werden lässt. Die Eichfixierung sollte also an den intrinsischen Freiheitsgraden der Theorie ansetzen. Dadurch soll eine an der eigentlichen Dynamik ausgerichtete und im Rahmen der tatsächlichen systemischen Gegebenheiten identifizierbare, 'echte' interne Zeit ausgezeichnet werden. – Die der Dynamik zugrundeliegenden intrinsischen Freiheitsgrade (soweit sie sich überhaupt ermitteln lassen) sind allerdings schon im klassischen Fall grundsätzlich abhängig von der jeweiligen Lösung der Theorie. Die Auszeichnung einer 'echten' internen Zeit kann immer nur durch die Auszeichnung einer spezifischen, für die entsprechende Lösung intrinsisch motivierbaren Form der Aufspaltung in räumliche Hyperflächen erfolgen. Die 'echte' interne Zeit ist also immer lösungsabhängig. Und es ist diese lösungsabhängige Zeit, die – wenn sie sich überhaupt identifizieren lässt – als Ausgangspunkt dienen soll für eine Etablierung zeitlicher Entwicklungen im Kontext der *Kanonischen Quantengravitation* (in der alten geometrodynamischen Variante). Diese Strategie geht einerseits, was den klassischen Ausgangspunkt betrifft, mit der Hoffnung einher, dass sich die Allgemeine Relativitätstheorie, zumindest unter jeweils spezifischen Bedingungen, doch noch als Theorie darstellt, welche die zeitliche Entwicklung räumlicher Dreier-Geometrien beschreibt. Andererseits ist es dann das Ziel, die im klassischen Fall erhofften Einsichten unmittelbar auf die entsprechende Quantentheorie zu übertragen, um schliesslich zu einer zeitabhängigen Schrödinger-Gleichung zu gelangen, welche die Quantendynamik räumlicher Dreier-Geometrien beschreibt.

Letztendlich impliziert eine solche Übertragung auf die entsprechende Quantentheorie jedoch, dass ein Zeitparameter eingeführt wird, der einer nicht diffeomorphismusinvarianten Grösse entspricht – mit allen schon beschriebenen Konsequenzen. Die Etablierung einer absoluten Zeit bricht sowohl im Kontext der klassischen als auch dem der Quantentheorie die substantielle Form der allgemeinen Kovarianz, die für die Allgemeine Relativitätstheorie gemeinhin als grundlegend angesehen wird.

*"Those who choose this route have to explain how they propose to implement general covariance and how they escape the indeterminism of the hole argument."* (Earman (2002b) 9)

Kuchars Lösungsansatz entspricht also der Etablierung einer absoluten Hintergrundzeit und verletzt somit die Anforderung der Hintergrundunabhängigkeit, ohne dass die Motivationen hierfür allzu überzeugend wären. – Denn bei Kuchars Begründung dafür, die Hamiltonsche Zusatzbedingung nicht als Generator einer Eichtransformation zu behandeln, da sie gerade die zeitliche Entwicklung des Systems erfasst, und es daher keine zeitliche Entwicklung mehr gäbe, wenn man sie als eine Eichtransformation generierend deuten würde, handelt es sich offensichtlich um nichts anderes als eine Petitio principii. Denn diese Begründung beruft sich ausschliesslich auf die Intuition, dass es so etwas wie eine zeitliche Entwicklung gibt, diese also um jeden Preis auch vom Formalismus der Theorie widergespiegelt werden muss, auch wenn es dafür keine anderweitigen Gründe geben mag.

Wenn die in stringenter Weise erschliessbaren Konsequenzen einer Theorie mit der Beobachtung nicht im Einklang stehen, so gibt es grundsätzlich verschiedene Möglichkeiten, mit diesem Problem umzugehen: Einerseits kann man sich überlegen, ob die entsprechende Beobachtung nicht auch anders gedeutet werden kann, so dass sie mit den vielleicht radikalen Konsequenzen der Theorie in Einklang gebracht werden kann. Dies führt etwa Barbour und, wie gleich zu erläutern sein wird, in abgeschwächter Weise Rovelli, die zu erklären versuchen, wie es zur Illusion einer zeitlichen Entwicklung in einer Welt ohne echte zeitliche Entwicklung im absoluten Sinne kommt.[323] Andererseits kann man sich überlegen, ob an der entsprechenden Theorie etwas Grundlegendes falsch ist. Vielleicht führt die Theorie zu einem konzeptionellen Artefakt, das für eine emergente Ebene vielleicht eine gewisse Berechtigung besitzt, nicht aber das fundamentale(re) Geschehen betrifft. Die Theorie wäre dann nur als effektive Theorie mit begrenzter Reichweite und begrenzter Durchschlagkraft hinsichtlich ihrer konzeptionellen Implikationen zu sehen. Diese Implikationen wären nicht in allen Fällen für eine fundamentalere Beschreibung konstitutiv. Eine solche Einstellung hinsichtlich bestimmter Implikationen der Allgemeinen Relativitätstheorie, die sich auch im Rahmen ihrer *Kanonischen Quantisierung* perpetuieren, findet sich etwa im Kontext mancher emergentistischer und prägeometrischer Szenarien.[324] Eine wenig motivierte Strategie ist es jedoch, ad hoc, punktuell, hochgradig selektiv und entsprechend ziemlich willkürlich etwas an den in stringenter Weise erschlossenen Konsequenzen einer Theorie zu drehen, um die aus der Beobachtung gewonnenen Intuitionen doch noch in irgendeiner Weise unterzubringen.

*

Ein wesentlich überzeugenderer Ansatz, der an der Interpretation der Diffeomorphismusinvarianz als Eichinvarianz uneingeschränkt festhält, Raum und Zeit in diesem Sinne – und mithin im Sinne der Allgemeinen Relativitätstheorie – auf gleichberechtigter Ebene behandelt, und damit die konzeptionellen Asymmetrien der Szenarien von Barbour und Kuchar vermeidet, stammt von Carlo Rovelli.[325] Seine Strategie im Umgang mit dem *Problem der Zeit* besteht in einem konsequenten Ausleuchten der Implikationen der eichinvarianten Deutung der Allgemeinen Relativitätstheorie, wie sie der *Loop Quantum Gravity* zugrundeliegt. Danach gibt es keinen externen, globalen, übergreifenden Zeitparameter; dieser ist aber nach Rovellis Auffassung auch gar nicht notwendig. Was sich vielmehr etablieren lässt, sind interne Zeitparameter, die sich entweder auf der Grundlage ausschliesslich eichinvarianter Grössen oder als Ergebnis der Einbeziehung von eichinvarianten, beobachtbaren Korrelationen zwischen nicht notwendigerweise eichinvarianten und damit auch nicht notwendigerweise beobachtbaren Variablen ergeben.

---

[323] Für den klassischen Fall siehe Healey (2003).
[324] Siehe Kap. 3.3. und 4.6.
[325] Siehe Rovelli (1991, 1991a, 1991c, 2001, 2002, 2004, 2007, 2009).

Rovellis Konzept einer internen, relationalen Auffassung von Zeit liegt in zwei verschiedenen Varianten vor. Die ältere Konzeption[326] nimmt ihren Ausgang von der Überzeugung, dass sich eine interne Zeit nur unmittelbar auf der Grundlage diffeomorphismusinvarianter bzw. eichinvarianter Grössen definieren lässt. Da aber sowohl in der Allgemeinen Relativitätstheorie als auch im Kontext ihrer *Kanonischen Quantisierung* nur zeitinvariante Grössen eichinvariant sind, kommen hierfür nur Konstanten in Frage. Rovellis Idee ist es nun, zeitliche Entwicklungen auf der Grundlage solcher Konstanten zu rekonstruieren, und zwar, indem man Konstanten auswählt, die sich in eine lineare Ordnung bringen lassen. Der Ordnungsparameter, der die Abfolge innerhalb der so entstehenden Ein-Parameter-Familien '*sich entwickelnder Konstanten*' erfasst, liesse sich dann als Zeitparameter interpretieren. Die formale Etablierung entsprechender Ein-Parameter-Familien 'sich entwickelnder Konstanten' ginge mit der Auszeichnung realer physikalischer Uhren einher. Die zugrundeliegende Hoffnung ist also die, dass sich – erst einmal für den klassischen Fall, und hier jeweils für eine spezifische Lösung der Einsteinschen Feldgleichungen – unter bestimmten Bedingungen tatsächlich solche Ein-Parameter-Familien 'sich entwickelnder Konstanten' finden lassen, die es dann ermöglichen eine interne Zeit auszuzeichnen. Ein solcher Ansatz bleibt aber wahrscheinlich grundsätzlich auf den klassischen Kontext beschränkt; denn eine solche interne Zeit wäre einerseits notwendigerweise lösungsabhängig und kaum auf quantenmechanische Superpositionen übertragbar, andererseits wäre sie auch für eine bestimmte klassische Lösung kaum auf der gesamten Raumzeit, sonder vermutlich nur unter spezifischen Bedingungen etablierbar. Sie beruht, wenn sie sich definieren lässt, ausschliesslich auf den Beziehungen, die sich unter möglicherweise sehr spezifischen Bedingungen zwischen den eichinvarianten (und damit beobachtbaren) Grössen aufzeigen lassen. Ob es also tatsächlich gelingt, zumindest im klassischen Kontext entsprechende Familien 'sich entwickelnder Konstanten' zu finden, bleibt selbst für Rovelli eine offene Frage.

> "*What does characterize the physical variables T that can be used as clocks? To our view, this is an open question.*" (Rovelli (1991) 453)

Noch fraglicher ist es aber eben, ob sich die Idee in irgendeiner Weise auf eine entsprechende Quantentheorie der Gravitation übertragen lässt. Beim Übergang von der klassischen zur Quantentheorie, für die es dann insbesondere keine lösungsspezifischen eichinvarianten Grössen und dementsprechend keine Eingrenzung wie im klassischen Fall mehr gibt, müssten übergreifend wirksame Pendants in Form (notwendigerweise zeitloser) quantenmechanischer Operatoren gefunden werden, die einerseits mit der Hamiltonschen Quanten-Zusatzbedingung kommutieren und sich andererseits, zumindest unter bestimmten Bedingungen, wiederum in eine lineare Ordnung bringen lassen, um so als Grundlage für die Etablierung eines internen Zeitbegriffs zu dienen.

> "*[...] technically, it is hard to construct such families of constants of motion as phase space functions on the phase space of general relativity. To the extent that they can be constructed at all, they result in rather complicated functions that are hard to represent at the quantum level (i.e. as quantum operators on a Hilbert space [...]), and face the full force of the factor ordering difficulties [...].*" (Rickles (2006) 184)

Konsequenterweise umgeht Rovelli in seiner neueren Konzeption[327] von vornherein diese Probleme, indem er versucht, die Etablierung einer internen, relationalen Zeit an Grössen festzumachen, die sich gleichermassen für den klassischen Fall wie im Kontext einer entsprechenden Quantentheorie finden lassen. Die zugrundeliegende Idee ist die, dass man zur Auszeichnung von internen Zeit-parametern nicht notwendigerweise ausschliesslich auf eichinvariante Grössen zurückgreifen muss, etwa auf Ein-Parameter-Familien 'sich entwickelnder Konstanten'. Es genügen dafür, wie Rovelli meint, beobachtbare eichinvariante (und damit zeitunabhängige) Korrelationen zwischen sich ver-ändernden, eichabhängigen und damit eigentlich unbeobachtbaren *partiellen Observablen*.[328] Ein zeitlicher Verlauf wird hier also in Form eichinvarianter relationaler Beziehungen zwischen *parti-ellen Observablen* rekonstruiert, die selbst nicht eichinvariant sind.

---

[326] Siehe Rovelli (1991).
[327] Siehe Rovelli (2002, 2004, 2009).

> *"The main novelty is that dynamics treats all physical variables (partial observables) on the same ground and predicts their correlations. It does not single out a special variable called 'time', to describe evolution with respect to it. Dynamics is not about time evolution, it is about relations between partial observables."* (Rovelli (2004) 265)

Entscheidend für die erfolgreiche Auszeichnung eines internen Zeitparameters sind wiederum vor allem die relationalen Beziehungen zwischen den Grössen, die physikalische Uhren im System er-fassen und somit in natürlicher Weise eine interne Zeit repräsentieren.[329]

> *"Internal times differ from a Hamiltonian time in many respects. First of all, the theory does not single out one or the other of these internal times. Second, none of the (proposed) internal times has all the features that characterize the t variable of Hamiltonian and quantum mechanics. [...] Third, by defini-tion an internal time refers to specific physical variables, unlike the t quantity, which is supposed to be universal. [...] The absolute quantity t has disappeared. In its place, there are different possible inter-nal times, related to specific physical variables."* (Rovelli (1991) 445)

Für die physikalischen Uhren, die jeweils als Grundlage für die Ausweisung interner Zeitparameter dienen, werden sehr wahrscheinlich – wie auch schon für die erhofften eichinvarianten Pendants zu den geometrischen Operatoren der *Loop Quantum Gravity* vermutet – Materiefreiheitsgrade eine wichtige Rolle spielen. Das Ergebnis ist aber in jedem Fall ein mittels der tatsächlich vorliegenden physikalischen Bedingungen bestimmter interner Zeitparameter, der nur soweit trägt, wie seine physikalischen Voraussetzungen gelten. Der jeweilige interne Zeitparameter ist genau soweit eta-blierbar, wie die Bedingungen reichen, die die interne physikalische Uhr definieren, auf der er be-ruht. Man hat es also ganz und gar nicht mit so etwas wie einer durchlaufenden, quasi-absoluten Zeit zu tun, sondern immer nur mit physikalisch motivierten Zeitparametern, die irgendwo beginnen und irgendwo enden können. Dem geometrisierten Bild der Zeit lässt sich hier nur stückchenweise Berechtigung zusprechen.[330] – Und die internen Zeitparameter sind als relationale, korrelative Grössen nie von vornherein in eindeutiger Weise festgelegt. Vielmehr kommen kompetitive Zeit-schemata ins Spiel, die auf unterschiedlichen physikalischen Uhren und entsprechend sehr unter-schiedlichen systemischen Bedingungen beruhen; diese werden im allgemeinen wiederum nur für sehr spezifische, jeweils unterschiedliche Parameterbereiche vorliegen.[331]

---

[328] Siehe Rovelli (2002). Rovelli behauptet, dass seine *partiellen Observablen* zwar nicht vorhersagbar, aber beobacht-bar und messbar sind, obwohl sie eichabhängig sind. Dies ist sicherlich nicht unproblematisch, soll hier aber nicht wei-terverfolgt werden, da die für die Auszeichnung der Zeitparameter entscheidende Grösse in den eichinvarianten (und damit sowohl beobachtbaren, als auch im klassischen Fall vorhersagbaren) Korrelationen zu sehen ist. Die Frage, ob *partiellen* Variablen selbst messbar sind, spielt dabei keine Rolle. – Kurios ist an der Problematik der *partiellen* Varia-blen nur, was – indem er sie ins Spiel gebracht hat – Rovelli, der sich selbst als einen der entschiedensten Vertreter des Relationalismus verstehen dürfte, aufgrund durchaus gut motivierbarer begrifflicher Kategorisierungen nachgesagt werden kann:
> *"Since Belot and Earman equate the view that there are physically real quantities that do not commute with the constraints with (straightforward) substantivalism, it appears that Rovelli would have to class as a substantiva-list, for his partial observables are just such quantities!"* (Rickles (2006) 187)
Alle Auffassungen, die von realen physikalischen Grössen einfordern, dass sie mit den Zusatzbedingungen kommutie-ren, gehören für Belot und Earman übrigens entweder zum Relationalismus oder zum 'sophisticated substantivalism', was wiederum nicht viel mehr als einer begrifflichen Differenzierung aufgrund einer unterschiedlichen Hintergrund-metaphysik entspricht. Siehe Kap. 2.1.
[329] Analoges gilt in dieser Konzeption für die Frage der räumlichen Zuordenbarkeit. Auch diese lässt sich nur mittels physikalischer Massstäbe erreichen, mit deren Hilfe sich physikalisch relevante 'Koordinatensysteme' auszeichnen las-sen.

> *"[...] it is difficult to imagine that in general relativity there could be a good clock that may run for-ever in* any *solution of Einstein equations [...]. [...] nontime behavior may appear at very short time intervals."* (Rovelli (1991) 454)

Was sich hiermit für Rovellis Ansatz abzeichnet, ist, dass interne, an physikalischen Systemen be-stimmte Zeitparameter (auch schon für den klassischen Fall) vermutlich überhaupt nur unter sehr spezifischen Bedingungen auszeichenbar sind.[332] Für den Planck-Bereich lassen sie sich, wie Ro-velli vermutet, wahrscheinlich gar nicht finden.

> *"It is possible that to make sense of the world at the Planck scale, and to find a consistent conceptual framework for [general relativity] and [quantum mechanics], we may have to give up the notion of time altogether, and learn ways to describe the world in atemporal terms."* (Rovelli (2007) 1290)  –
> *"At the fundamental level, we should, simply, forget time."* (Rovelli (2001) 114)

Zeit – und insbesondere die Idee einer quasi-globalen Zeit – ist nach Rovellis Auffassung damit auch im günstigsten Fall etwas, das nur im Kontext einer näherungsweise gültigen, effektiven Theo-rie, die makroskopische Dynamiken erfasst, eine Rolle spielt. Und die Etablierung einer solchen effektiven Theorie entspricht, zumindest vor dem konzeptionellen Hintergrund der in letzter Instanz zeitlosen Dynamik, wie sie die Allgemeine Relativitätstheorie und die *Loop Quantum Gravity* be-schreiben, dann schliesslich nicht viel mehr als der erfolgreichen theoretischen Erfassung der Illu-sion einer zeitlichen Entwicklung. Im fundamentalen Sinne gibt es nach Rovellis Einschätzung, wenn man die stringent erschliessbaren Konsequenzen der Allgemeinen Relativitätstheorie und ihre Auswirkungen im Rahmen ihrer *Kanonischen Quantisierung* ernst nimmt, keine Zeit und keine zeitliche Entwicklung. Das Äusserste, was sich erreichen lässt, ist eine Erklärung für das Zustande-kommen der Illusion einer zeitlichen Entwicklung auf der Grundlage relationaler Beziehungen zwi-schen grundlegend zeitlosen physikalischen Grössen. Diese Illusion sollte man keinesfalls mit einem absoluten Zeitverlauf verwechseln.

> *"We get the illusion of change by (falsely) taking the elements of these relative (correlation) obser-vables to be capable of being measured independently of the correlation."* (Rickles (2006) 183)

Rovellis Konzeption zum Verständnis der Implikationen des *Problems der Zeit* hat gegenüber den entsprechenden Ansätzen von Barbour und Kuchar deutliche Vorteile aufzuweisen: Im Gegensatz zu Kuchars Ansatz fügt Rovelli keine willkürlichen Ad-hoc-Ergänzungen in den Formalismus ein, wie etwa die, die Hamiltonsche Zusatzbedingung völlig anders als die Diffeomorphismus-Zusatz-bedingungen zu behandeln, um die konzeptionellen Implikationen der Theorie mit dem phänome-nologischen Anschein in Einklang zu bringen. Im Gegensatz zu Barbours wie zu Kuchars Ansatz behandelt Rovelli Raum und Zeit – im Geiste der Allgemeinen Relativitätstheorie – auf gleicher und gleichberechtigter Ebene.

---

[330] Die Geometrie der räumlichen Zuordnung reicht allerdings auch nicht weiter.
[331] Analoges gilt wiederum für die Frage der räumlichen Zuordenbarkeit. Auch diese lässt sich nur mittels kompetitiver und in ihrer Anwendbarkeit begrenzter physikalischer Massstäbe und entsprechender Koordinatensysteme erreichen.
[332] Vgl. Thiemann (2006a).

> *"Rovelli's approach has a certain appeal from a philosophical point of view. It bears similarities to four-dimensionalist views on time and persistence. The basic idea of both of these views is that a changing individual can be constructed from unchanging parts. Change over time is conceptually not different from variation over a region of space."* (Rickles (2006) 184)

Der grösste Vorzug von Rovellis Konzeption besteht gerade darin, dass seine *partiellen Obser-vablen* grundsätzlich in der gleichen Weise für die Etablierung zeitlicher wie räumlicher Relationen herangezogen werden können. Raum und Zeit werden im Sinne der Allgemeinen Relativitäts-theorie (und der mit ihr einhergehenden Geometrisierung der Zeit) von vornherein analog behan-delt. Dass hier das *Problem der Zeit* als konzeptioneller Ausgangspunkt dient – und nicht etwa ein entsprechendes 'Problem des Raumes' – liegt daran, dass ersteres als Implikation der Allgemeinen Relativitätstheorie wie ihrer *Kanonischen Quantisierung* unmittelbar mit der phänomenologisch motivierten Verlaufsintuition hinsichtlich der Zeit in Konflikt gerät, es aber keine vergleichbare Verlaufsintuition hinsichtlich des Raumes gibt. Obwohl also Raum und Zeit in der Allgemeinen Relativitätstheorie wie in ihrer *Kanonischen Quantisierung* auf gleicher Ebene und konzeptionell[333] analog behandelt werden, sind es schliesslich die Verräumlichung der Zeit und die sich aus ihr er-gebenden Konsequenzen, die zum Auslöser für die Einsicht werden, dass nur relationale Grössen als Observablen in Frage kommen – im klassischen Kontext wie in der entsprechenden Quanten-theorie. Und dies betrifft gleichermassen die räumlichen wie die zeitlichen Beziehungen.

Im Rahmen einer solchen Konzeption wird die raumzeitliche Welt zu einem Netzwerk relationaler Bezüge. Die hierin zum Ausdruck kommende, vermutlich allgemeinste Ausprägung von Hinter-grundunabhängigkeit liegt für den Fall ihrer eichinvarianten Interpretation, wenn man sich nicht auf einzelne ihrer Lösungen beschränkt, schon für die Allgemeine Relativitätstheorie vor. Dass man eine solche Sichtweise aber nicht notwendigerweise an den Kontext der Allgemeinen Relativitäts-theorie und ihrer *Kanonischen Quantisierung* binden und auf diesen beschränken sollte, wird vor allem vor dem Hintergrund der konzeptionellen Probleme letzterer deutlich. Ohnehin weist eine Sichtweise, welche die raumzeitliche Welt als vollständig hintergrundunabhängiges Netzwerk aus-schliesslich relationaler Bezüge versteht, über diesen Kontext hinaus und auf die abstrakten Netz-werke von Quantensystemen hin, die sich dann schliesslich im Kontext der prägeometrischen und insbesondere der computationalen Szenarien auf angemessenere, natürliche Weise modellieren lassen.[334]

---

[333] In modelltheoretischer Hinsicht werden sie jedoch im Kontext der *Loop Quantum Gravity* mit ihrer sukzessiven Implementierung der Zusatzbedingungen, die dann schliesslich unvollendet bleibt, unterschiedlich behandelt.

[334] Siehe Kap. 3.3. und insbesondere Kap. 4.6. Als expliziter Schritt im Übergang von der *Kanonischen Quantengravi-tation* zu diesen prägeometrischen Szenarien lässt sich in einiger Hinsicht der kovariante *Spin-Foam*-Ansatz sehen. Hier ergeben sich weitere Perspektiven, die für eine prägeometrische Modellierung im Bereich der Quantengravitation spre-chen. Siehe weiter unten.

## *Zwischenbilanz*

Die in dieser Hinsicht anstehende, grundsätzliche Frage lautet also vermutlich: In welchem Ausmass weisen die Probleme und die zum Teil radikalen Implikationen und Perspektiven, die sich für die *Kanonische Quantisierung* der Allgemeinen Relativitätstheorie ergeben, über diese hinaus? Inwieweit lassen sie sich als Anzeichen für eine zumindest partielle konzeptionelle oder modelltheoretische Unangemessenheit dieses kanonischen Quantisierungsansatzes sehen? – Zu den eindeutigen Problemen des Ansatzes zählt vor allem die immer noch fehlende Reproduktion eines klassischen Grenzfalles, der mit der Phänomenologie der klassischen Ausgangstheorie vereinbar wäre. Um vieles ambivalenter stellt sich die Situation dar, die sich im Kontext der *Kanonischen Quantengravitation* mit dem *Problem der Zeit* – und dem damit einhergehenden Konflikt mit der Phänomenologie einer Welt des Werdens – ergibt, vor allem aber mit den sich in diesem Kontext abzeichnenden Perspektiven, die eine relationalistische Lösung des *Problems der Zeit* nahelegen. Spricht die auf diese Weise motivierte relationalistische Sichtweise also insbesondere für eine umfassende Hintergrundunabhängigkeit, die über die in der Allgemeinen Relativitätstheorie wie in der *Loop Quantum Gravity* vorliegende vielleicht noch hinausgeht? Lassen sich die Probleme des kanonischen Quantisierungsansatzes, nicht zuletzt aber auch die Perspektiven, die sich auf seiner Grundlage ergeben, also vielleicht als Indizien dafür sehen, dass es gute Gründe gibt, nach Alternativen Ausschau zu halten, in denen diese Zusammenhänge in ein klareres Licht gerückt werden können? Stellt die direkte Quantisierung der Allgemeinen Relativitätstheorie also in strategischer und in modelltheoretischer Hinsicht vielleicht einen unnötigen Umweg dar?

Die sich an diese Problematik unmittelbar anschliessende Frage hinsichtlich der in der *Loop Quantum Gravity* auftretenden konzeptionellen Probleme ist dann, ob diese Probleme in allgemeinerer Weise als grundsätzliche Folge der Strategie einer direkten Quantisierung der Allgemeinen Relativitätstheorie anzusehen sind. Oder zeigen diese Probleme vielmehr in spezifischer Weise, dass die Verwendung der Hamiltonschen Darstellung der Allgemeinen Relativitätstheorie – infolge der mit ihr einhergehenden formalen Einschreibung der eichinvarianten Interpretation der klassischen Ausgangstheorie – als Ausgangspunkt für die Quantisierung in die falsche Richtung führt?

Für die erste Möglichkeit – dass also Probleme dieser Art für jede konzeptionell angemessene und konsistente Instantiierung der Strategie einer direkten Quantisierung der Allgemeinen Relativitätstheorie zu erwarten sind – spricht der Umstand, dass Konsequenzen der eichinvarianten Deutung der Allgemeinen Relativitätstheorie, auch wenn man diese nicht gleich fest implementiert, kaum zu umgehen sind. Und diese eichinvariante (und in vieler Hinsicht relationalistische) Deutung führt in der einen oder anderen Weise unabwendbar zu den Implikationen, die sich im Kontext des kanonischen Quantisierungsansatzes gerade in den Problemen manifestieren, die hier mit der Erfordernis der Lösung der Zusatzbedingungen verbunden sind – unabhängig davon, ob diese Lösung schon im klassischen Kontext oder erst in dem der entsprechenden Quantentheorie erfolgen soll, unabhängig davon also, ob man eine Quantisierung des schon reduzierten Phasenraums anstrebt oder ob man die Diracsche Quantisierungsmethode verwendet. Auch das *Problem der Zeit* stellt sich unter den Bedingungen, die sich mit der Deutung der Diffeomorphismusinvarianz als Eichinvarianz ergeben, notwendigerweise ein – und dies wiederum unabhängig davon, ob man die Implikationen einer eichinvarianten Deutung schon mit dem Charakterisierungsprofil relationalistischer Intuitionen gleichsetzt.

*"This is indeed true for any generally covariant theory, which means that a theory whose gauge group contains the diffeomorphism group of the underlying manifold has a weakly vanishing Hamiltonian. Here weakly vanishing refers to vanishing on physical configurations."* (Gaul / Rovelli (2000) 8)

Die Probleme einer direkten Quantisierung der Allgemeinen Relativitätstheorie lassen sich einer solchen Sichtweise zufolge vielleicht sogar hin auf eine grundlegende konzeptionelle Unverträglichkeit zwischen Allgemeiner Relativitätstheorie und Quantenmechanik bzw. Quantenfeldtheorien deuten:

*"In extending our knowledge of gravity from the macroscopic to the microscopic regime, we find that this goal cannot be achieved simply by applying quantum principles to [general relativity] (the active quantization approach). Instead, we have to construct our theory directly at the quantum level, and then to find an intelligible way to establish its connection with [general relativity]."* (Cao (2001) 200)

Einige Kritiker optieren dennoch für die zweite Möglichkeit. Sie machen die Ursachen für die Probleme, die in der *Loop Quantum Gravity* auftreten, vor allem an den Spezifika des kanonischen Quantisierungsansatzes und seiner Ausgangsbasis fest: der Hamiltonschen Darstellung der Allgemeinen Relativitätstheorie.

*"Although they carry out their calculations in a covariant way, in deriving their calculational rules they seem unable to wean themselves from canonical methods and Hamiltonians, which are holdovers from the nineteenth century, and are tied to the cumbersome (3+1)-dimensional baggage of conjugate momenta, bigger-than-physical Hilbert spaces and constraints."* (Stachel (2006a) 67, Bryce DeWitt paraphrasierend)

Hier werden also die Hamiltonsche Darstellung der Allgemeinen Relativitätstheorie, die in ihr formal festgeschriebene eichinvariante Deutung der Theorie und die mit ihr infolge der (3+1)-Aufspaltung der Raumzeit einhergehende (modelltheoretische) Einschränkung in der Erfassung der Allgemeinen Kovarianz als Ursachen für die Probleme des kanonischen Quantisierungsansatzes in den Vordergrund gerückt:

*"[...] identifying the causal structure, is skirted in canonical quantum gravity by positing a split of spacetime into space and time at the outset. [...] the diffeomorphism group [...] is represented in a distorted way in the canonical theory, so that it is unclear that one is actually quantizing general relativity at all."* (Weinstein (2001) 98)

Es wird insbesondere in Frage gestellt, dass die *Loop Quantum Gravity* tatsächlich die volle Diffeomorphismusinvarianz der Allgemeinen Relativitätstheorie berücksichtigt und dass sie tatsächlich in umfassender Weise als kovariant anzusehen ist:

*"[...] in any canonical approach there is the question whether one has succeeded in achieving (a quantum version of) full space-time covariance, rather than merely covariance under diffeomorphisms of the three-dimensional slices."* (Nicolai / Peeters (2006) 2)

Direkten Vorschub erhalten diese Zweifel durch das Faktum, dass die Quanten-Zusatzbedingungen in der *Loop Quantum Gravity* tatsächlich durchaus unterschiedlich behandelt werden und die sukzessive Implementierung dieser Zusatzbedingungen im letzten Schritt unvollendet bleibt, so dass die *Loop Quantum Gravity* eben noch nicht die echten physikalischen Freiheitsgrade beschreibt, die dem raumzeitlichen Geschehen zugrundeliegen:

> *"Spacetime covariance is a central property of Einstein's theory. Although the Hamiltonian formulation is not manifestly covariant, full covariance is still present in the classical theory, albeit in a hidden form, via the classical (Poisson or Dirac) algebra of constraints acting on phase space. However, this is not necessarily so for the quantised theory. [...] [Loop Quantum Gravity] treats the diffeomorphism constraint and the Hamiltonian constraint in a very different manner. Why and how then should one expect such a theory to recover full spacetime (as opposed to purely spatial) covariance? The crucial issue here is clearly what [Loop Quantum Gravity] has to say about the quantum algebra of constraints."* (Nicolai / Peeters (2006) 9)

Die Vertreter der *Loop Quantum Gravity* beharren jedoch darauf, dass ihr Ansatz die Diffeomorphismusinvarianz in vollem Umfang berücksichtigt, wenn auch in formaler Hinsicht vielleicht nicht ganz so offensichtlich:

> *"We insist on the fact that this breaking of covariance is not structurally needed in order to set up the canonical formalism; rather it is an artifact of the coordinatization we choose for the phase space."* (Gaul / Rovelli (2000) 7)

Oder sie argumentieren gar dafür, dass die volle raumzeitliche Kovarianz überhaupt nur für den (von der Theorie eben noch nicht reproduzierten) klassischen Grenzfall Bedeutung hat:

> *"[...] the notion of classical space-times and of space-time covariance is not likely to have a fundamental role in the full quantum theory. These notions have to be recovered in an appropriate semiclassical limit."* (Ashtekar (2005) 33)

Nachvollziehbar wird diese Haltung jedoch bestenfalls in einer kovarianten Erweiterung der *Loop Quantum Gravity:* dem *Spin-Foam*-Ansatz. Erst mit diesem ergibt sich – zumindest, was den formalen Apparat betrifft, nicht jedoch seine Erfolge – eine überzeugende Replik auf die letztgenannten Kritiken am kanonischen Quantisierungsansatz.

## Spin Foams

Bei den *Spin Foams*[335] handelt es sich, soweit es den ursprünglichen Ansatz betrifft, um das kovariante Lagrangesche Gegenstück zur kanonischen Hamiltonschen *Loop Quantum Gravity*. Die *Spin Foams* verhalten sich in dieser Hinsicht zur *Loop Quantum Gravity* wie die Lagrangesche Formulierung der Mechanik zu ihrer Hamiltonschen Formulierung bzw. wie die Pfadintegral-(*Sum-over-Histories*)-Formulierung der Quantenmechanik zu ihrer Hilbertraum-Formulierung. Sie lassen sich als so etwas wie eine Feynmansche Pfadintegral-Formulierung der *Loop Quantum Gravity* ansehen, die vor allem mit den Mitteln der topologischen Quantenfeldtheorien arbeitet.[336] – Das heisst jedoch

---

[335] Siehe Oriti (2001, 2003), Livine / Oriti (2003), Perez (2003, 2007), Baez (1998, 2000), Markopoulou / Smolin (1997).

[336] Es gibt auch deutliche Verbindungen zur Gruppenfeldtheorie:
> *"[...] spin foams are nothing else than Feynman graphs of [group field theory] [...]."* (Oriti (2006a) 16)

nicht, dass die genaue Beziehung zwischen *Spin-Foam*-Ansatz und *Loop Quantum Gravity* vollends geklärt wäre. Insbesondere ist eine stringente Ableitung des *Spin-Foam*-Ansatz aus der *Loop Quantum Gravity* – oder umgekehrt – bisher nicht möglich.

> *"It would be very good if we were able to translate directly between the two formalisms. [...] At present, neither of these two paths is under complete control. [...] Even at a more naive level, there are several gaps between the hamiltonian loop theory and the spinfoam models [...]."* (Rovelli (2004) 262)

Es gibt zudem unterschiedliche Versionen (bzw. Modelle) des *Spin-Foam*-Ansatzes. Nicht einmal deren Beziehung untereinander ist vollständig geklärt.

Das Entscheidende an den *Spin Foams* ist, dass sie von vornherein die gesamte Raumzeit in quantisierter Form erfassen – und nicht nur die räumlichen Hyperflächen, wie es die kanonischen Quantisierungsansätze erst einmal tun. Im Gegensatz zur *Kanonischen Quantengravitation* setzt der *Spin-Foam*-Ansatz – als auch in formaler Hinsicht vollständig kovarianter Ansatz – keine (3+1)-Aufspaltung der Raumzeit voraus.

> *"A spinfoam itself represents a spacetime, in the same sense in which a spin network represents a space."* (Rovelli (2004) 231)

Anfänglich waren die *Spin Foams* als Pfadintegrale über die zeitliche Entwicklung der Spinnetze (bzw. ihrer quantenmechanischen Superpositionen) gedacht, die sich im Kontext der *Loop Quantum Gravity* für die räumlichen Hyperflächen ableiten lassen.[337]

> *"The dynamics of the spin network states can be expressed also in a path integral formalism, called spin foams."* (Smolin (2003) 22)

Natürlich geht es eigentlich um die Superposition von S-Knoten und ihre Entwicklung – nicht um die Spinnetze:

> *"[...] sum over histories of evolutions of s-knot states."* (Gaul / Rovelli (2000) 44)

Insbesondere sollte es, wie man sich erhoffte, auf dieser Grundlage möglich sein, Übergangsamplituden zwischen raumzeitlichen Geometrien zu berechnen: als *Sum-over-Spin-Foams*.

> *"[...] we interpret the spinfoam model sum as a sum-over-paths definition of the transition amplitude between two quantum states of the gravitational field."* (Rovelli (2004) 238)

---

Die Gruppenfeldtheorie wird jedoch auch unabhängig von den *Spin Foams* als möglicher Ansatz zu einer Theorie der Quantengravitation gehandelt. Siehe etwa Oriti (2006a). Dieser Ansatz sieht eine lokale dritte Quantisierung des Gravitationsfeldes vor: Erzeugungs- und Vernichtungsoperatoren entsprechen, die lokalen Raumabschnitten ('Atomen' des Raumes) entsprechen. Die Topologie und die Geometrie des Raumes sollen sich dann aus der Dynamik dieser elementaren 'Atome' des Raumes ergeben: aus ihrer Erzeugung und Vernichtung auf der Grundlage eines abstrakten, prägeometrischen Quantenvakuums.

[337] Rovelli bezeichnet diese formal erfassbare Entwicklung als 'Histories of Spin Networks'.

Zudem gab es zeitweise die Hoffnung, eine mögliche Lösung der Hamiltonschen Zusatzbedingung, die im Kontext der *Loop Quantum Gravity* unerfüllt bleibt, nun im Rahmen des *Spin-Foam*-Ansatzes zu erreichen.

> *"The spinfoam formalism could suggest the correct form of the hamiltonian operator."* (Rovelli (2004) 262)

Die Berechnung der Übergangsamplituden zwischen raumzeitlichen Geometrien ist bisher noch nicht problemlos möglich, aber immerhin weiter gediehen als in der *Loop Quantum Gravity*. Die Aussichten auf eine Lösung der Hamiltonschen Zusatzbedingung sind im engeren, ursprünglichen *Spin-Foam*-Ansatz immer noch grundlegend unklar.

> *"Interesting as they are, however, these developments have so far not shed much new light on the problems with the Hamiltonian constraint, or the constraint algebra, in our opinion. A derivation of spin foam models from the Hamiltonian formulation remains incomplete due to the complexity of the Hamiltonian constraint. Hence, a decisive proof of the connection between spin foam models and the full Einstein theory and its canonical formulation still appears to be lacking."* (Nicolai / Peeters / Zamaklar (2005) 8)

Im Laufe der Zeit haben sich die *Spin Foams* stattdessen von ihrem Ausgangspunkt als kovariante Version der *Loop Quantum Gravity* weitgehend verabschiedet und sich entscheidend weiterentwickelt.

> *"Attempts to overcome the difficulties with the Hamiltonian constraint have led to another development,* spin foam models. *These were originally proposed as space-time versions of spin networks, to wit, evolutions of spin networks in 'time', but have since developed into a class of models of their own, disconnected from the canonical formalism."* (Nicolai / Peeters (2006) 11)

Es kommt inzwischen insbesondere zu einer deutlichen Annäherung an die prägeometrischen Ansätze, die nicht mehr in einer direkten Quantisierung der Allgemeinen Relativitätstheorie fussen.[338]

## *Prägeometrische Perspektiven*

Die Perspektiven, die sich mit den zentralen Problemen, auf die die *Loop Quantum Gravity* und der *Spin-Foam*-Ansatz stossen, und den entsprechenden Lösungsansätzen, die sich hier abzeichnen, ergeben, weisen deutlich über ihren Ursprung hinaus. Sie legen insbesondere Alternativen jenseits einer direkten Quantisierung der Allgemeinen Relativitätstheorie nahe.

So führt die *Loop Quantum Gravity* zwar zur Vorhersage einer diskreten Struktur, die dem Raumzeitkontinuum, von dem die Theorie erst einmal ihren Ausgang nimmt, zugrundeliegt – zumindest man den konzeptionellen Voraussetzungen und den formalen Ableitungen vertraut: Es sind die Spinnetze bzw. S-Knoten, die sich erstaunlicherweise schon für die kinematische Beschreibungsebene direkt nach der Quantisierung ergeben. Aber diese diskrete Struktur setzt, wie wir gesehen haben, immer noch notwendigerweise das Kontinuum der reellen Zahlen voraus. Damit steht die

---

[338] Siehe die nachfolgenden Ausführungen, insbesondere aber auch Kap. 4.6.

*Loop Quantum Gravity*, zumindest was ihren formalen Apparat und seine modelltheoretischen Voraussetzungen betrifft, immer noch im Widerspruch zu den Argumenten für eine finite diskrete Struktur und finite Informationsdichten, wie sie sich auf der Grundlage der *Bekenstein-Hawking-Entropie* und der holographischen Grenze motivieren lassen.[339] Die *Loop Quantum Gravity* arbeitet, zumindest in formaler Hinsicht, immer noch auf der Grundlage des Mythos der reellen Zahlen, der – wenn man den Implikationen der Thermodynamik schwarzer Löcher vertraut – letztendlich nicht aufrechtzuerhalten ist.

Die primäre Quelle für diese modelltheoretische Festlegung und die aus ihr resultierenden konzeptionellen Probleme der *Loop Quantum Gravity* ist in der Annahme einer differenzierbaren Mannigfaltigkeit zu sehen, die bei der direkten *Kanonischen Quantisierung* der Allgemeinen Relativitätstheorie notwendigerweise vorausgesetzt wird, auch wenn sie in der *Loop Quantum Gravity* dann schliesslich mit einigem argumentativen Aufwand wegdiskutiert werden soll. Nur auf ihrer Grundlage lässt sich überhaupt die Diffeomorphismusinvarianz erfassen und als elementare Anforderung in den Formalismus der Theorie implementieren. – Dass ein solcher Ansatz dann zu einer Diskretisierung führt, ist immerhin erstaunlich. Aber es handelt sich immer noch um eine Diskretisierung durch die Hintertür, auf der Basis eines formal mit dem Kontinuum arbeitenden Ansatzes.

Ohne dieses Kontinuum, ohne die differenzierbare Mannigfaltigkeit – und damit ohne die Diffeomorphismusinvarianz – käme es auch gar nicht zum *Problem der Zeit*, wie es in der *Loop Quantum Gravity* auftritt. Ein eventuelles Pendant zu diesem Problem, sollte es auch in einem originär diskreten Ansatz immer noch auftreten, würde vielleicht zu ganz anderen Implikationen oder Perspektiven führen.

Angesichts der Problemlagen und Implikationen der *Loop Quantum Gravity* könnte man sich also durchaus fragen: Warum sollte man bei der Entwicklung einer Quantentheorie der Gravitation und der Raumzeit unbedingt ein Konstrukt wie das einer differenzierbaren Mannigfaltigkeit voraussetzen, wenn dieses dann schliesslich (wie es zumindest die Vertreter der *Loop Quantum Gravity* für ihren Ansatz behaupten) im Rahmen der Konsequenzen der Theorie vollständig transzendiert wird, d.h. sich auf der Mikroebene als letztlich nicht haltbar erweist? Warum sollte man also im Rahmen einer Quantentheorie der Gravitation etwas postulieren, was sich letztlich nur im Kontext einer (im Falle der *Loop Quantum Gravity* leider immer noch nur) intendierten klassischen Näherung dieser Quantentheorie, nämlich der Allgemeinen Relativitätstheorie, als angemessen erweist, nicht aber in der Quantentheorie selbst? – Ein solcher Ansatz erscheint zumindest modelltheoretisch unvorteilhaft, wenn nicht gar konzeptionell unangemessen.

Warum sollte man also die in der *Loop Quantum Gravity* auftretenden Indizien für eine diskrete Substratstruktur – vor dem Hintergrund der konzeptionellen Probleme des Ansatzes – nicht als Motivation dafür sehen, gleich mit einem diskreten Ansatz zu beginnen, der eben nicht von einer Quantisierung der Allgemeinen Relativitätstheorie – und somit auch nicht von einer differenzierbaren Mannigfaltigkeit – ausgeht, um letztere dann im Rahmen dieser Quantisierung hinter sich zu lassen (oder wegzudiskutieren)? Warum sollte man diese naheliegende Möglichkeit nicht zumindest als Alternative zu den direkten Quantisierungsansätzen in Betracht ziehen?

---

[339] Siehe Kap. 3.1.

Aber letztlich geht es nicht nur um die Diskretheit des Substrats: Die über den *Spin-Foam*-Ansatz in seiner ursprünglichen Form hinausgehenden oder zumindest avisierten Erweiterungsmöglichkeiten der *Loop Quantum Gravity* weisen auf ein nicht nur diskretes, sondern vielmehr prägeometrisches Substrat hin, das sich besser im Rahmen eines Ansatzes erfassen lässt, der mit abstrakten, vollständig hintergrundunabhängigen Quantennetzwerkstrukturen arbeitet.[340]

> *"While loop quantum gravity [...] so far appears to be satisfactory as both a quantization of general relativity and a quantum theory of gravity, it may very well be that the quantization of general relativity does not in fact describe nature. The dimension of spacetime, physical degrees of freedom, and fundamental symmetries may be different from those which are presently observed. It turns out that there is a natural class of models which generalizes loop quantum gravity which addresses these possibilities."* (Smolin (2003) 27)

Ähnliches gilt, wie schon angedeutet, für die Perspektiven, die sich aus Rovellis Ansatz zur Lösung des *Problems der Zeit* ergeben. Einige der schon existierenden prägeometrischen Szenarien[341] verkörpern Rovellis Konzeption eines Netzwerkes von relationalen Bezügen, in denen sich das physikalische Weltgeschehen erschöpft, in wesentlich direkterer Weise als dies die *Loop Quantum Gravity* vermag. Sie realisieren die für eine Theorie der Quantengravitation eingeforderte Hintergrundunabhängigkeit in der denkbar radikalsten und umfassendsten Weise. – Warum sollte man also nicht die relationalistischen Implikationen der *Loop Quantum Gravity* ernst nehmen und vollständig hintergrundunabhängige, prägeometrische Ansätze in Betracht ziehen, natürlich wiederum unter Berücksichtigung der grundlegenden Anforderung, schliesslich die Allgemeine Relativitätstheorie bzw. ihren empirischen Gehalt im Rahmen eines klassischen Niederenergiegrenzfalles zu reproduzieren? Warum sollte man diese Alternative nicht zumindest weiterverfolgen? – In unübersichtlichen Problemsituationen ist es oft sinnvoll, nicht alles auf eine Karte zu setzen, sondern gleichzeitig alternative Lösungspfade in Betracht zu ziehen und zu verfolgen.

## 4.5. Quantisierung einer diskretisierten Variante der Allgemeinen Relativitätstheorie

Hinsichtlich ihrer raumzeitlichen Ausgangsbasis lassen sich die verschiedenen theoretischen Ansätze im Bereich der Quantengravitation in solche unterteilen, die schon eine Raumzeit voraussetzen und dann unter Umständen versuchen, deren Quanteneigenschaften zu erschliessen, und in solche, die völlig ohne Raumzeit starten, etwa auf der Grundlage diskreter relationaler Strukturen ohne raumzeitliche Einbettung, um schliesslich die Raumzeit als emergentes, makroskopisches Phänomen in seiner Entstehung zu erklären. Von diesem letzterem Typus, den *prägeometrischen* Ansätzen, soll erst im nachfolgenden Teilkapitel die Rede sein. Hinsichtlich des erstgenannten Falls ist es sinnvoll, noch einmal zu unterscheiden zwischen Ansätzen, die von einem raumzeitlichen Kontinuum ausgehen, um dessen vermutete Quanteneigenschaften zu erschliessen und dabei möglicherweise auf eine diskrete Struktur stossen (wie im Falle der *Loop Quantum Gravity*), und solchen, die von vornherein, sei es aus physikalischen oder aus rein modelltheoretischen Gründen, eine diskrete Struktur ansetzen, um im Grenzübergang das raumzeitliche Kontinuum als zumindest für den ma-kroskopischen Bereich gültiges, approximatives Konstrukt zu reproduzieren. Von diesen letztgenannten Ansätzen und ihrer nur begrenzten Durchschlagskraft[342] soll in diesem Teilkapitel die Rede sein.

---

[340] Siehe auch Markopoulou (2004).
[341] Siehe Kap. 4.6.

<p style="text-align:center">*</p>

Angesichts der Probleme, die sich für diejenigen Ansätze zu einer direkten Quantisierung der Allgemeinen Relativitätstheorie einstellen, die an der Kontinuumsannahme festhalten und eine differenzierbare Mannigfaltigkeit voraussetzen, könnte man sich fragen: Wie sieht es mit der Quantisierung einer diskretisierten Form der Allgemeinen Relativitätstheorie als Alternative diesbezüglich aus? Schliesslich lassen sich die Probleme der subtileren, nicht-perturbativen, kanonischen Ansätze zu einem erheblichen Teil auf die Diffeomorphismusinvarianz der klassischen Ausgangstheorie und ihre im Quantisierungsprozess wirksam werdenden Implikationen zurückführen. Die Diffeomorphismusinvarianz setzt aber selbst wiederum eine differenzierbare Mannigfaltigkeit voraus. Liefert also eine diskrete, aber immer noch raumzeitliche Ausgangsbasis vielleicht gerade einen Ausweg aus diesen Problemen? Und reproduziert ein solcher Ansatz vielleicht dann auch die klassische Raumzeit als Kontinuumsgrenzfall – was den kontinuumsgestützten kanonischen Quantisierungsansätzen gerade nicht gelungen ist?

Es gibt eine ganze Reihe von Ansätzen, die versuchen, entweder eine (schon im vorhinein) diskretisierte Form der Allgemeinen Relativitätstheorie zu quantisieren, oder aber, die Quantisierung der Allgemeinen Relativitätstheorie auf einer diskretisierten raumzeitlichen Grundlage zu vollziehen:

### Konsistente Diskretisierung

Zur ersteren Variante zählt die *Konsistente Diskretisierung*.[343] Hier wird im ersten Schritt der Versuch unternommen, die Allgemeine Relativitätstheorie, und zwar in ihrer Hamiltonschen Formulierung auf der Grundlage der traditionellen geometrodynamischen Variablen (Metrik und Krümmung), in eine diskrete Form zu bringen. – Nun führt aber eine Diskretisierung der klassischen Grundgleichungen nicht so ohne weiteres zu einem Satz miteinander verträglicher diskretisierter Gleichungen:

> *"Discretizing general relativity is more subtle than what one initially thinks. Consider a 3+1 decomposition of the Einstein equations. One has twelve variables to solve for (the six components of the spatial metric and the six components of the extrinsic curvature). Yet, there are* sixteen *equations to be solved, six evolution equations for the metric, six for the extrinsic curvature and four constraints. In the continuum, we know that these sixteen equations are* compatible*, i.e. one can find twelve functions that satisfy them. However, when one discretizes the equations, the resulting system of algebraic equations is in general incompatible."* (Gambini / Pullin (2004) 1)

Die 'konsistente Diskretisierung' besteht gerade darin, diskretisierte Pendants zu den klassischen Kontinuumsgleichungen zu finden, die miteinander in der Weise kompatibel sind, dass sich tatsächlich für alle Gleichungen gleichzeitig Lösungen finden lassen:

---

*"[...] consistently discretize the theory, in the sense that all the resulting discrete equations can be solved simultaneously."* (Gambini / Pullin (2004) 6)

Zudem muss das Ergebnis der *Konsistenten Diskretisierung* der klassischen Ausgangstheorie dann auch noch für eine Quantisierung geeignet sein; die auf diesem Wege zu entwickelnde Quantentheorie ist das eigentliche Ziel:

*"The situation is more involved if one is interested in discretizing the theory in order to quantize it. There, one needs to take into account all equations. In particular, in the continuum the constraints form an algebra. If one discretizes the theory the discrete version of the constraints will in many instances fail to close an algebra. Theories with constraints that do not form algebras imply the existence of more constraints which usually makes them inconsistent."* (Gambini / Pullin (2004) 2)

Rodolfo Gambini und Jorge Pullin haben schliesslich für die Realisierung dieses Ziels eine Diskretisierung der Allgemeinen Relativitätstheorie in ihrer Hamiltonschen Darstellung vorgeschlagen, die mit 16 Gleichungen und 16 Variablen arbeitet, aber keine Zusatzbedingungen mehr einschliesst:

*"The fact that we approximate the continuum theory (which has constraints) with a discrete theory that is constraint free allows us to bypass in the discrete theory many of the conceptual problems of canonical quantum gravity."* (Gambini / Pullin (2004) 3)

Dass diese konsistente diskretisierte Variante der Allgemeinen Relativitätstheorie über keine Zusatzbedingungen mehr verfügt, ist insofern nicht sonderlich verwunderlich, da nach der Diskretisierung keine differenzierbare Mannigfaltigkeit mehr vorausgesetzt wird, die Diffeomorphismusinvarianz in ihrer ursprünglichen Form also gar nicht mehr vorliegen kann, die Zusatzbedingungen aber gerade als die Bedingungen anzusehen sind, unter denen im Kontinuumsfall die Diffeomorphismusinvarianz in der Hamiltonschen Darstellung auf der Grundlage der gewählten kanonischen Variablen für den vollen kinematischen Phasenraum eingefordert wird.

Infolgedessen sind dann auch in der Quantisierung keine Zusatzbedingungen mehr zu beachten. Insbesondere gibt es in der *Konsistenten Diskretisierung* also die Hamiltonsche Zusatzbedingung nicht mehr, die in den kontinuumsbasierten Ansätzen der *Kanonischen Quantisierung* gerade zu den gravierendsten konzeptionellen Problemen führt.

*"[...] the Hamiltonian is a generator of infinitesimal time evolutions, and in a discrete theory, there is no concept of infinitesimal. What plays the role of a Hamiltonian is a canonical transformation that implements the finite time evolution from discrete instant n to n+1. [...] The theory is then quantized by implementing the canonical transformation as a unitary evolution operator."* (Gambini / Pullin (2004) 2)

Der Preis, der jedoch für eine solche Vorgehensweise zu zahlen ist, besteht dann gerade in der Brechung der Diffeomorphismusinvarianz der Allgemeinen Relativitätstheorie. Insofern wird hier eigentlich nicht mehr die Allgemeine Relativitätstheorie quantisiert, sondern eine andere, konzeptionell sehr unterschiedliche, nicht-diffeomorphismusinvariante Theorie.

*"[...] the discrete theory* is a different theory *which may approximate the continuum theory in some circumstances, but nevertheless is different and may have important differences even at the conceptual* *level. This is true of any discretization proposal, not only ours."* (Gambini / Pullin (2004) 2) – *"[...] in the discrete theory there will generally be several solutions that approximate a given solution of the continuum theory."* (Gambini / Pullin (2004) 3)

Die Quantisierung der diskreten, nicht-diffeomorphismusinvarianten Theorie geschieht dann in der Hoffnung, dass die *Konsistente Diskretisierung* (zufällig oder auf der Grundlage diesbezüglich hoffentlich relevanter Analogien zwischen der Kontinuumstheorie und der diskreten Theorie) zu einer Quantentheorie führen wird, die wiederum vor allem die Allgemeine Relativitätstheorie (und mithin ihre Diffeomorphismusinvarianz) als klassischen Kontinuumsgrenzfall enthält.

*"In the end the credibility of the whole approach will hinge upon us producing several examples of situations of interest where the discrete theories approximate continuum [general relativity] well."* (Gambini / Pullin (2004) 3)

Soweit es die tatsächliche Formulierung einer entsprechenden Quantentheorie betrifft, die auf der Grundlage der *Konsistenten Diskretisierung* der Allgemeinen Relativitätstheorie entwickelt werden soll, ist dies jedoch nicht mehr als ein Programm.

*"What is now needed is to demonstrate that the range of situations in which the discrete theory approximates general relativity well is convincingly large enough to consider its quantization as a route for the quantization of general relativity."* (Gambini / Pullin (2004) 6)

Ein auch nur ansatzweiser Erfolg der *Konsistenten Diskretisierung* ist noch nicht abzusehen.

## Regge Calculus

Der *Regge Calculus*[344] beschreitet einen konzeptionell gänzlich anderen Weg zur Diskretisierung der Allgemeinen Relativitätstheorie. Hier wird nicht die klassische Theorie durch ein konsistentes System diskreter Gleichungen ersetzt; vielmehr wird die Raumzeit selbst schlichtweg durch Triangulation (mit variablen Längen) diskretisiert. Die modelltheoretische Ausgangsbasis besteht aus einer kombinatorischen Struktur von Simplices (beliebig-dimensionalen Verallgemeinerungen von Dreiecken). Diese Triangulationsstruktur und insbesondere die Zahl der Simplices ist fest vorgegeben; nur die Längen der Verbindungslinien sind variabel. Die Längen innerhalb des Triangulationskonstruktes werden als Gravitationsvariablen interpretiert. Die Krümmung der Raumzeit wird durch die Winkel innerhalb der Triangulationsstruktur erfasst.

In einer anschliessenden Quantisierung soll dann über diese Längen integriert werden. Danach soll ein Grenzübergang zu einer unendlichen Zahl von Simplices vollzogen werden. In dieser Hinsicht ist der Ansatz formal vergleichbar mit den Gittereichtheorien. – Die immensen Schwierigkeiten, zu denen es hierbei insbesondere im Hinblick auf die Rekonstruktion eines quasi-kontinuierlichen Niederenergiegrenzfalles kommt, sind bisher völlig ungelöst.

---

[344] Siehe Regge / Williams (2000), Williams / Tuckey (1992), Gentle (2002), Barrett (1987).

## Dynamische Triangulation

Als wesentlich weiter ausgearbeitet und als wesentlich aussichtsreicher als etwa der *Regge Calculus* stellen sich die Ansätze zur *Dynamischen Triangulation* [345] dar. Hier erfolgt die Diskretisierung der Raumzeit wiederum auf der Basis einer kombinatorischen Struktur von Simplices, nun jedoch mit festen Längen. Die Dimensionalität der Simplices ist von vornherein fest vorgegeben. Der einzige Freiraum, der zugelassen wird, betrifft die Verbindungsstruktur selbst. Für die mikroskopische Textur der Raumzeit wird im Rahmen der *Dynamischen Triangulation* also eine polymerartige Struktur angenommen, die den Spinnetzen der *Loop Quantum Gravity* nicht unähnlich ist, nur dass die *Dynamische Triangulation* diese Struktur schon im Ansatz und insbesondere vor der Quantisierung voraussetzt, diese also nicht, wie in der *Loop Quantum Gravity*, aus der Quantisierung einer klassischen Kontinuumstheorie auf einer differenzierbaren Mannigfaltigkeit gewonnen wird.

Der Übergang zur Quantentheorie erfolgt im Rahmen der *Dynamischen Triangulation* dann mittels eines diskreten *Sum-over-Histories*-Ansatzes: über die Zuweisung von quantenmechanischen Wahrscheinlichkeitsamplituden für die einzelnen diskreten Strukturen entsprechend ihrem (aus dem klassischen Kontext heraus motivierbaren) Beitrag zur Wirkung. Bisher sieht alles danach aus, als ob sich im anschliessenden Grenzübergang zu einer unendlichen Zahl von Simplices tatsächlich die klassische Raumzeit reproduzieren lässt. Dies gilt allerdings nur für die *Kausale Dynamische Triangulation*: für Triangulationsmodelle also, die schon eine Lorentzsche Raumzeitsignatur voraussetzen. Für die ursprünglichen Euklidischen Triangulationsmodelle ist dies nachgewiesenermassen nicht möglich. Damit ist die anfängliche Hoffnung, etwa die Signatur der Raumzeit aus einer fundamentaleren Dynamik abzuleiten, mit der die dynamischen Triangulationsansätze ursprünglich, im Kontext ihrer frühen Euklidischen Modelle, spielten, wohl definitiv dahin.

> *"It has been suggested that the signature of spacetime may be explained from a dynamical principle. Being somewhat less ambitious, we will assume it has Lorentzian signature [...] motivated by the uncontroversial fact that the universe has three space and one time dimension."* (Ambjorn / Jurkiewicz / Loll (2006) 5)

Bei den frühen Euklidischen Ansätzen zur *Dynamischen Triangulation* wurde nicht einmal eine fest vorgegebene Topologie vorausgesetzt. Geometrie und Topologie sollten erst durch die dynamische Verbindungsstruktur des Triangulationskonstruktes festgelegt werden. Die Idee war, dass sich eine d-dimensionale Raumzeit (quasi automatisch) auf der Grundlage der Dynamik von (d-1)-dimensionalen Simplices ergibt. Die Modelle der Lorentzschen *Kausalen Dynamischen Triangulation*, mit denen sich im Gegensatz zu ihren Euklidischen Pendants die klassische Raumzeit erfolgreich als Kontinuumsgrenzfall reproduzieren lässt, arbeiten nun von vornherein mit einer kombinatorischen Struktur von Simplices, für die Dimensionalität, Topologie und Signatur der Raumzeit fest vorgeben sind.

> *"We will in the following take a conservative point of view and only sum over geometries (with Lorentzian signature) which permit a foliation in (proper) time and are causally well-behaved in the sense that no topology changes are allowed as a function of time."* (Ambjorn / Jurkiewicz / Loll (2006) 6)

---

[345] Siehe Ambjorn (1995) Ambjorn / Jurkiewicz / Loll (2000, 2001, 2001a, 2004, 2005, 2005a, 2005b, 2006, 2009), Ambjorn / Loll (1998), Loll (1998, 2001, 2003, 2007), Loll / Ambjorn / Jurkiewicz (2005).

Und es gibt nun ausgeprägte Parallelen zum kanonischen Quantisierungsansatz: Im diskretisierten (3+1)-Szenario der *Kausalen Dynamischen Triangulation* werden die 'Schichten' der kombinatorischen Grundstruktur als diskrete (einen Simplex dicke) Pendants zu den dreidimensionalen raumartigen Hyperflächen der Hamiltonschen Kontinuumsdarstellung angesehen.

> *"The discretized analogue of an infinitesimal proper-time 'sandwich' in the continuum will be a finite sandwich [...] consisting of a single layer of four simplices."* (Ambjorn / Jurkiewicz / Loll (2006) 6)

Diese Pendants zu den raumartigen Hyperflächen werden dann 'kausal' miteinander verbunden.

> *"To obtain extended spacetimes, one glues together sandwiches pairwise along their matching three-dimensional boundary geometries."* (Ambjorn / Jurkiewicz / Loll (2006) 6)

Die miteinander verbundenen diskreten Pendants zu den raumartigen Hyperflächen entsprechen dann den diskreten Schritten einer zeitlichen Entwicklung. Nach der Pfadintegration über diese Lorentzschen (3+1)-Geometrien und dem anschliessenden Grenzübergang zu einer unendlichen Zahl von Simplices lassen sich dann unter spezifischen Bedingungen – und im Rahmen aufwendigster numerischer Berechnungen – überzeugende Anzeichen für einen Grenzfall mit phänomenologisch adäquater kontinuierlicher Raumzeit finden.

> *"For certain values of the bare gravitational and cosmological coupling constants we have found evidence that a continuum limit exists."* (Ambjorn / Jurkiewicz / Loll (2006) 18)

Dieser durchaus spektakuläre Erfolg der *Kausalen Dynamischen Triangulation* sollte jedoch nicht darüber hinwegtäuschen, dass es zunächst keinen guten Grund gibt davon auszugehen, dass es sich bei diesem Ansatz um eine fundamentale Theorie handelt, die ein Bild des mikroskopischen Substrats liefert, auf dem die klassische makroskopische Raumzeit tatsächlich zustandekommt. Vielmehr handelt es sich bei den Modellen der *Kausalen Dynamischen Triangulation* bestenfalls um effektive Theorien, vergleichbar den Gittereichtheorien, die unter bestimmten Bedingungen formal den makroskopischen Kontinuumsgrenzfall reproduzieren: nämlich durch eine Quantisierung auf einer aus modelltheoretischen – und nicht etwa aus physikalischen – Gründen angesetzten diskreten Struktur (im Fall der Gittereichtheorien: einem Gitter; im Fall der *Kausalen Dynamischen Triangulation*: der subtileren Triangulationsstruktur der Simplices) und einen anschliessenden Grenzübergang zum Kontinuum (Gitterparameter gegen Null bzw. Zahl der Simplices gegen Unendlich).[346] – In dieser Hinsicht unterscheidet sich die *Kausale Dynamische Triangulation* massgeblich insbesondere von den *Spin-Foam*-Ansätzen:

> *"The spinfoam formalism has formal similarities with lattice gauge theory. The interpretation of the two formalisms, however, is quite different. In the case of lattice theories, the discretized action depends on a parameter, the lattice spacing* a. *The physical theory is recovered as* a *is taken to zero. In this limit, the discretization introduced by the lattice is removed. [...] In gravity, on the other hand, there is not lattice spacing parameter* a *in the discretized action. Therefore there is no sense in the* a → 0 *limit. The discrete structure of the spinfoams must reflect actual features of the physical theory."* (Rovelli (2004) 261f)

---

[346] Die Ergebnisse sind zudem, wie schon angedeutet, nur im Rahmen sehr aufwendiger numerischer Verfahren erreichbar.

Für die *Kausale Dynamische Triangulation* gilt also erst einmal das gleiche wie für jeden Ansatz, der mit einer rein modelltheoretisch (und nicht etwa physikalisch) motivierten Diskretisierung der Raumzeit bzw. der Allgemeinen Relativitätstheorie startet, um auf diese Weise bestimmte Probleme kontinuumsbasierter Ansätze wie etwa der *Loop Quantum Gravity* oder des *Spin-Foam*-Ansatzes traktibel zu machen oder gar gänzlich zu umgehen: Es gibt nicht den geringsten Grund sie realistisch zu interpretieren. Die *Kausale Dynamische Triangulation* ist nichts anderes als ein mathematisches Instrumentarium: ein effektiver Theorieansatz mit mehr oder weniger Erfolg, was die Reproduktion des klassischen Kontinuumsgrenzfalles betrifft, ohne Aussagekraft jedoch, was das tatsächliche Substrat betrifft, auf dessen Grundlage es zum raumzeitlichen Geschehen kommt.

> "The current approaches to non-perturbatively construct in detail a mathematically well defined theory of quantum gravity both at the canonical level and at the path integral level resort to discretizations to regularize the theory." (Gambini / Pullin (2004) 1)

Geht man davon aus, dass für den Bereich der Quantengravitation ohnehin nicht mehr als eine effektive Modellierung in diesem Sinne erreicht werden kann, oder, dass es in der Physik ohnehin nur effektive Theorien gibt und vermeintlich fundamentale Theorien nichts anderes als das Ergebnis einer grundsätzlich unzulässigen Extrapolation metaphysischer Ambitionen darstellen, so mag man sich mit den Erfolgen der *Kausalen Dynamischen Triangulation* durchaus zufriedengeben. Hält man metaphysische Ambitionen jedoch nicht von vornherein für physikalisch nichtig und erwartet man von einer Theorie der Quantengravitation eine aus einleuchtenden physikalischen Prinzipien resultierende und über entsprechende konzeptionelle Querverbindungen motivierte (und vor allem schliesslich auch empirisch bestätigbare) Beschreibung des Substrates, auf der das raumzeitliche Geschehen tatsächlich beruht, so muss man offensichtlich nach Alternativen Ausschau halten.

## 4.6.   Prägeometrische Ansätze

### Die Motivation prägeometrischer Theorieansätze

Die vorrangige Motivation für die Beschäftigung mit prägeometrischen Ansätzen zur 'Quantengravitation' ergibt sich aus den massiveren Problemen der auf den ersten Blick konzeptionell naheliegenderen und, was nicht notwendigerweise zusammentreffen muss, der etablierteren Theorieansätze. Gemeint sind hierbei insbesondere einerseits die Probleme, die sich im Rahmen der *Loop Quantum Gravity* als des wohl konzeptionell überzeugendsten und zudem schon am weitesten ausgearbeiteten Versuchs einer direkten Quantisierung der Allgemeinen Relativitätstheorie ergeben. Andererseits sind es die Probleme, auf die der Stringansatz als der (gemessen an der hinter ihm stehenden Personalstärke und nicht etwa aufgrund der ihm eigenen konzeptionellen Kohärenz) populärste Ansatz zu einer Theorie der Quantengravitation stösst, der nebenbei auch gleich noch das Problem einer nomologischen Vereinigung aller Wechselwirkungen lösen möchte. – Hinzu kommen die sich in diesen Ansätzen abzeichnenden Perspektiven, die in zum Teil deutlicher Weise auf eine prägeometrische Modellierung als den vielleicht adäquateren Weg hin zu einer Theorie der 'Quantengravitation' hinweisen.

Dabei sollte die Beschäftigung mit prägeometrischen Theorien zur Quantengravitation, die sich auf diese Weise auf der Grundlage der massiveren Probleme der etablierteren Ansätze als auch der über sie hinausweisenden Perspektiven motivieren lässt, nicht im Sinne einer Ablösung dieser etablierteren Ansätze verstanden werden. Vielmehr geht es um einen Pluralismus der Strategien angesichts einer unübersichtlichen Problemlage. Haben die etablierteren und z.T. konzeptionell naheliegenderen Theorieansätze massivere Probleme, für die sich momentan (und vielleicht schon über längere Zeit) keine vielversprechenden Lösungen abzeichnen, so heisst dies, dass man sich vielleicht nicht weiterhin *allein* auf diese Ansätze verlassen sollte, sondern eben *auch* Alternativen erwägen und ausarbeiten sollte, die auf den ersten Blick vielleicht fernliegender erscheinen könnten. Es könnte immerhin sein, dass die Probleme der etablierteren Ansätze gerade darauf zurückgehen, dass diese in ihren Grundannahmen und in der Einbeziehung von konzeptionellen Elementen aus den Vorgängertheorien viel zu konservativ sind, um hinsichtlich der Umsetzung der Zielvorgaben für eine Theorie der Quantengravitation erfolgreich sein zu können. Es könnte also sein, dass sie dafür einfach nicht radikal genug sind. – Schaut man sich die Probleme der etablierteren Theorieansätze etwas genauer an, so kann man zudem vielleicht noch etwas über geeignete Ansatzpunkte für prägeometrische Theorien lernen. Manche der eher offensichtlichen konzeptionellen Unangemessenheiten, modelltheoretischen Ungeschicklichkeiten und vielleicht sogar Fehler der etablierteren Theorieansätze lassen sich dann vielleicht vermeiden.

Die Probleme etwa, auf die der Stringansatz stösst, sind vielfältiger Art. Sie betreffen sowohl die konzeptionellen und modelltheoretischen Voraussetzungen, von denen der Ansatz seinen Ausgang nimmt, als auch die Folgeprobleme und Implikationen, die sich daraus schliesslich ergeben. Letztendlich ist der Stringansatz ein mathematisch-modelltheoretisches Konstrukt, das auf keinen unmittelbaren physikalischen Motivationen, insbesondere auf keinem physikalisch motivierbaren, fundamentalen Prinzip, beruht und nach mehreren Jahrzehnten immer noch nahezu ausschliesslich in einer störungstheoretischen Formulierung vorliegt. Und es sind vor allem die phänomenologisch völlig unmotivierten und physikalisch immerhin fragwürdigen, aber aus Gründen der modelltheoretischen Kohärenz unumgänglichen Zutaten wie die Supersymmetrie und die höhere Dimensionalität der Raumzeit, die zu erheblichen Folgeproblemen für den Ansatz führen. Diese internen Probleme überwiegen hinsichtlich der erforderlichen Bewältigungsstrategien bei weitem die externen, physikalischen Anforderungen, die an eine Theorie der Quantengravitation zu stellen sind. Insbesondere ergibt sich auf der Grundlage der Anforderung der Supersymmetrie und der Notwendigkeit, aus der höherdimensionalen perturbativen Beschreibung, die der Stringansatz bereitstellt, niederenergetische Implikationen für die vierdimensionale phänomenologische Raumzeit abzuleiten, das Kontingenzproblem der *String-Landscape*: eine extrem hohe Zahl möglicher Stringszenarien und dennoch keine – und erst recht keine eindeutige – Chance zu einer Ableitbarkeit des quantenfeldtheoretischen Standardmodells.

Das zentrale konzeptionelle Problem des Stringansatzes ist aber vermutlich seine Hintergrundabhängigkeit. Als unmittelbare konzeptionelle Erweiterung der Quantenfeldtheorien, bei der deren modelltheoretisches Instrumentarium beibehalten wird, setzen die perturbativen Stringtheorien eine Hintergrundraumzeit mit fest vorgegebener Metrik voraus, die nicht in die auf ihr stattfindenden Prozesse dynamisch eingebunden ist. Für eine Theorie der Raumzeit, die noch dazu von sich behauptet, die Allgemeine Relativitätstheorie als niederenergetischen Grenzfall zu reproduzieren, ist diese Hintergrundunabhängigkeit, für die der Stringansatz keinen guten Grund liefert, weil sie schlichtweg aus dem ihm zugrundeliegenden modelltheoretischen Instrumentarium resultiert, als Rückschritt gegenüber der angeblich reproduzierten klassischen Theorie nicht nur unmotiviert, son-dern letztlich nicht hinnehmbar. Spätestens dies macht die Suche nach konzeptionell weniger fragwürdigen Alternativen unabdinglich.

Hier spätestens kommt dann als eine der in konzeptioneller Hinsicht plausibelsten Alternativen die *Loop Quantum Gravity* ins Spiel. Sie verfolgt mit ihrer direkten, nicht-perturbativen Quantisierung der Allgemeinen Relativitätstheorie im Gegensatz zum Stringansatz eine konzeptionell ausgewogene und ausgereifte Strategie zur Formulierung einer Theorie der Quantengravitation. Dabei ist sie extrem konservativ in ihren Grundannahmen. Sie übernimmt alle wesentlichen konzeptionellen Elemente ihrer Vorgängertheorien, soweit diese nicht zu unmittelbaren Inkompatibilitäten führen. Insbesondere hält sie an der Geometrisierbarkeit der Gravitation fest, verankert die aktive Diffeomorphismusinvarianz der Allgemeinen Relativitätstheorie fest in ihrem Ansatz und übernimmt damit deren Hintergrundunabhängigkeit in uneingeschränkter Weise. Die Quantenmechanik wird weiterhin als universell gültig angesehen und konzeptionell nur in minimalster Weise verallgemeinert, um den relationalen Implikationen der Allgemeinen Relativitätstheorie Rechnung tragen zu können. Man kann also durchaus sagen: Die *Loop Quantum Gravity* macht alles richtig, was man in konzeptioneller Hinsicht richtig machen kann – unter der Voraussetzung, dass die ihr zugrundeliegenden Annahmen tatsächlich zutreffen; diese sind:

(a) Die Gravitation ist eine fundamentale Wechselwirkung.
(b) Die Quantenmechanik ist universell gültig.

Dennoch führt die *Loop Quantum Gravity* zu erheblichen Problemen. Diese sind nicht unbedingt in der Tatsache zu sehen, das sie zu äusserst radikalen Konsequenzen – insbesondere der Nichtlokalität der Observablen, der fehlenden Unitarität und dem *Problem der Zeit* mit den sich aus ihm ergebenden relationalistischen Implikationen – führt; diese könnten eben gerade Teil der neuen Einsichten sein, die sich im Rahmen einer Theorie der Quantengravitation notwendigerweise einstellen werden. Auch, dass die *Loop Quantum Gravity* bisher noch zu keinen empirisch direkt überprüfbaren Vorhersagen führt, kann man ihr schwerlich als entscheidendes Negativum anrechnen; alle anderen Ansätze zur Quantengravitation erzielen diesbezüglich keine besseren Ergebnisse. Schwerwiegender ist schon die Tatsache, dass es in der Theorie noch eine ganze Reihe konzeptioneller Ambiguitäten gibt, vor allem in der Formulierung des Hamilton-Operators und in der Festlegung der Repräsentation der Operator-Algebra. Das entscheidende Problem besteht aber wohl darin, dass es bisher nicht möglich war, eine der zentralen Anforderungen an jede Theorie der Quantengravitation zu erfüllen: die Reproduktion des empirischen Gehalts ihrer unmittelbaren Vorgängertheorie bzw. der ihr zugrundeliegenden Phänomenologie. Die *Loop Quantum Gravity* hat erhebliche Schwierigkeiten mit der Ableitung niederenergetischer Implikationen. Und es ist ihr insbesondere nicht möglich, den klassischen Ausgangspunkt der Quantisierung, die Allgemeine Relativitätstheorie, mittels der so gewonnenen Quantentheorie wieder als klassischen Grenzfall zu reproduzieren.

Wenn also die *Loop Quantum Gravity* unter den gegebenen Voraussetzungen tatsächlich alles richtig macht, was man in konzeptioneller Hinsicht richtig machen kann, dann könnte man vermuten, dass es eigentlich nur an den ihr zugrundeliegenden Annahmen liegen kann, dass es immer noch zu diesen Problemen kommt. Die naheliegendste Vermutung, die sich dann einstellt: Vielleicht ist die Gravitation eben doch keine fundamentale Wechselwirkung. Dies würde am einfachsten erklären, wieso eine konzeptionell ausgereifte und überzeugende Quantisierung der Allgemeinen Relativitätstheorie bisher nicht in der Lage war, diese wiederum als klassischen Grenzfall der so gewonnenen Quantentheorie zu reproduzieren.

Eine differenziertere, mit einer nicht fundamentalen, aber dennoch geometrisierbaren Gravitation durchaus verträgliche Erklärungsvariante, die sich dann hinsichtlich des Auftretens der konzeptionellen Probleme in der *Loop Quantum Gravity* einstellen könnte und schliesslich zu nicht weniger radikalen Einsichten als diese führt, sieht folgendermassen aus: Könnte es sein, dass die konzeptionellen Probleme der *Loop Quantum Gravity* darauf zurückgehen, dass diese trotz allem noch nicht hintergrundunabhängig genug ist, um zu einer funktionierenden Theorie der Quantengravitation zu führen? Im formalen Ansatz der *Loop Quantum Gravity* wird nämlich (ebenso wie schon in der Allgemeinen Relativitätstheorie) ein rudimentärer Hintergrund vorausgesetzt: Die Dimensionalität, die Signatur und die Topologie der Raumzeit sind festgelegt; zudem wird für die Implementierung der Diffeomorphismusinvarianz eine differenzierbare Mannigfaltigkeit angesetzt, auch wenn diese dann im Rahmen der weiteren konzeptionellen Entwicklung wegdiskutiert werden soll. Nur die Metrik wird (in Form einer konnektionsdynamischen Darstellung) als dynamische Grösse behandelt und entsprechend quantisiert. Sollte man diese rudimentäre Hintergrundabhängigkeit der *Loop Quantum Gravity* als Ursache für ihre Probleme ansehen, so wäre dies gleichbedeutend mit der Annahme, dass die phänomenologische Raumzeit mit ihrer Dimensionalität, Signatur und Topologie aus etwas hervorgebracht wird, das gänzlich anders geartet ist und noch nicht über diese spezifischen Eigenschaften der Raumzeit verfügt; dies käme der Annahme der Emergenz der Raumzeit gleich.

In eine ähnliche Richtung liesse sich ausgehend von der Kontinuumsannahme argumentieren, mit der die *Loop Quantum Gravity*, was ihre modelltheoretische Grundlagen betrifft, immer noch arbeitet. Diese Kontinuumsannahme hat sich mit den Entwicklungen im Rahmen der Thermodynamik schwarzer Löcher[347] als zumindest konzeptionell fragwürdig erwiesen: als ein modelltheoretischer Mythos, der in seiner Anwendung gleichermassen zu Artefakten wie zu unlösbaren konzeptionellen Problemen führen kann. Die *Loop Quantum Gravity* benötigt jedoch die differenzierbare raumzeitliche Mannigfaltigkeit für die Implementierung der Diffeomorphismusinvarianz im Rahmen ihrer kanonischen Quantisierungsprozedur. Und auch die diskrete Struktur der Spinnetze setzt immer noch einen kontinuierlichen Hintergrund voraus. Mit den Argumenten, die gegen die konzeptionelle Adäquatheit der Kontinuumsannahme sprechen, wird jedoch die Implementierung der Diffeomorphismusinvarianz, wie sie in der *Loop Quantum Gravity* vorgenommen wird, zu einer sehr fragwürdigen Angelegenheit. – Die gleiche Einsicht geht letztlich schon mit der Vermutung einher, dass es sich bei der Raumzeit um eine emergente Grösse handelt, und mit den entsprechenden Argumenten, die sich hierfür anführen lassen:

> *"[...] if spacetime is a phenomenological concept of limited applicability, then so will be the diffeomorphisms of the manifold that models spacetime in this limited sense."* (Isham (1997) 23)

## Hintergrundunabhängigkeit, Prägeometrie und Emergenz der Raumzeit

Eine der entscheidendsten Fragen, die sich im Rahmen der Entwicklung einer Theorie der Quantengravitation stellen, ist die folgende: Auf welcher Strukturebene sollte eine solche Theorie ansetzen? Was sollte also als schon gegeben vorausgesetzt werden? – Die aus der Allgemeinen Relativitätstheorie heraus motivierbaren Argumente, die erst einmal für eine Hintergrundunabhängigkeit als elementare Anforderung für eine Theorie der Quantengravitation sprechen, lassen – im Verbund mit den sich im Kontext der Thermodynamik schwarzer Löcher abzeichnenden Argumenten, die nahelegen, dass die Kontinuumsannahme letztendlich physikalisch unangemessen ist, so dass ihre modelltheoretische Verwendung vielleicht gerade zu unnötigen Problemen führt – vermutlich am ehesten den Schluss zu, dass man mit so wenig wie möglich an Hintergrundstrukturen und vor allem ohne Kontinuum auskommen sollte. Man sollte also vielleicht erst einmal versuchen herauszufinden, wie weit man kommt, wenn man auf jeglichen raumzeitlichen Hintergrund verzichtet und eine diskrete, basale Struktur annimmt, die keine der bekannten geometrischen Freiheitsgrade einschliesst, insbesondere solche nicht, die auf der Ebene der phänomenologischen, anscheinend kontinuierlichen Raumzeit, wie sie von der Allgemeinen Relativitätstheorie beschrieben wird, eine Rolle spielen:

> *"It is reasonable, therefore, to think of a future quantum theory of gravitation as a theory that is not a spacetime theory in the ordinary sense. Rather, the usual spatiotemporal concepts should emerge from the theory, presumably in the non-relativistic classical limit."* (Dieks (2001) 153)

---

[347] Siehe Kap. 3.1. und 3.2.

Ein solcher Ansatz wird dann allerdings, da er über den bisherigen Erfahrungskontext physikalischer Theoriebildung deutlich hinausgeht und die bekannten und erprobten mathematischen Instrumentarien weitestgehend nicht mehr anwendbar sind, mit erheblichen modelltheoretischen Schwierigkeiten zu kämpfen haben:

> *"The usual tools of mathematical physics depend so strongly on the real-number continuum and its generalizations (from elementary calculus 'upwards' to manifolds and beyond), that it is probably even harder to guess what non-continuum structure is needed by such radical approaches than to guess what novel structures of dimension, metric and so on are needed by the more conservative approaches that retain manifolds. Indeed, there is a more general point: space and time are such crucial categories for thinking about, and describing, the empirical world, that it is bound to be ferociously difficult to understand their emerging, or even some aspects of them emerging, from 'something else'."* (Butterfield / Isham (1999) 132)

Die Helden, die sich diesen Schwierigkeiten stellen, sind gerade die prägeometrischen Ansätze. Dabei ist der Begriff 'prägeometrisch' noch nicht unbedingt mit 'ungeometrisch' gleichzusetzen; erst einmal bedeutet er nichts anderes als 'prä-raumzeitlich'.[348]

> *"[...] pregeometry is synonymous to pre-Riemannian-manifold physics [...]."* (Meschini / Lehto / Piilonen (2004) 13)

Bei den prägeometrischen Theorien handelt es sich um Ansätze, die eine Dynamik diskreter Strukturen bzw. Entitäten ohne die Voraussetzung einer Raumzeit beschreiben, auch nicht einer diskreten Raumzeit, wie sie sich etwa in den Spinnetzen der *Loop Quantum Gravity* abzeichnet. Sie gehen von einem diskret strukturierten Substrat aus, das selbst über keine raumzeitlichen Freiheitsgrade

---

[348] Die meisten prägeometrischen Ansätze arbeiten mit irgendwelchen rudimentären geometrischen Eigenschaften, allerdings nicht mit geometrischen Freiheitsgraden, die der (makroskopischen) Raumzeit zugeschrieben werden könnten.
> *"Varied though pregeometric approaches can be, they all share a feature in common: their all-pervading geometric understanding of the world. Such a kind of understanding is by no means particular to pregeometry; it is also present in all of quantum gravity and, moreover, in all of physics."* (Meschini / Lehto / Piilonen (2004) 3) verfügt. Dies hat nicht zuletzt den Vorteil, dass diese Theorien hinsichtlich der makroskopischen Raumzeit notwendigerweise hintergrundunabhängig sind. Die zugrundeliegende Idee ist die, dass sich Raumzeit und Gravitation als emergente Phänomene ergeben: als Konsequenzen einer Dynamik auf der Substratebene, auf der es keine Raumzeit gibt; als makroskopischer Ausdruck anderer nicht-raumzeitlicher und nicht-gravitativer Freiheitsgrade.

> *"[...] spacetime geometry resting on entities ontologically prior to it and of an essentially new character."* (Meschini / Lehto / Piilonen (2004) 13)

Prägeometrische Theorien sind also immer Theorien, die von einer emergenten Raumzeit ausgehen. Sie widersprechen insbesondere der weit verbreiteten Auffassung, dass die makroskopische Raumzeit eine fundamentale Entität ist, die spezifische Quantenfreiheitsgrade aufweist, die in der Allgemeinen Relativitätstheorie als klassischer Theorie der Raumzeit noch nicht erfasst werden, und damit Gegenstand einer Theorieerweiterung im Sinne einer Quantisierung sein sollten.

> *"It has been a common assumption that the quantum geometry describes a 'bumpy' classical geometry [...]."* (Markopoulou (2006) 25)

Den prägeometrischen Theorieansätzen liegt damit die Auffassung zugrunde, dass die konzeptionellen Voraussetzungen, die sich für eine Theorie der Quantengravitation vielleicht aus den etablierten Theorien ableiten lassen, nur bedingt verlässlich sind. Insbesondere liefern sie keine Grundlage für eine direkte Amalgamierung der Vorläufertheorien. In den prägeometrischen Theorieansätzen geht es nicht mehr einfach um Quantenkorrekturen zu einer klassischen Raumzeitauffassung, also etwa um Quantenfluktuationen der Metrik,[349] um quantenmechanische Unschärfen der Raumzeit, um Superpositionen von Raumzeiten oder gar um die Ausformulierung einer Gravitonenphysik.[350] Es geht in ihnen vielmehr um die Eigenschaften – möglicherweise die Quanteneigenschaften – des nicht-raumzeitlichen Substrats, auf dessen Grundlage sich die klassische Raumzeit ergibt.

Prägeometrische Theorieansätze lassen sich insbesondere nicht über die Quantisierung einer klassischen Raumzeittheorie gewinnen; vielmehr starten sie mit einer Substrathypothese, also einer direkten Hypothese hinsichtlich der Mikrodynamik, mit dem Ziel, auf dieser Grundlage schliesslich in kohärenter und konzeptionell angemessener Weise die bekannte Phänomenologie von Gravitation und Raumzeit zu reproduzieren. In die Hypothesenbildung hinsichtlich des Substrats fliessen natürlich wiederum Elemente aus den Vorgängertheorien ein, allerdings in wesentlich geringerem Ausmass als in den etablierteren Theorieansätzen zur Quantengravitation. Entscheidend ist hierbei wiederum, welche konzeptionellen Elemente der Vorgängertheorien als basal bzw. als essentiell angesehen werden. Ob die entsprechende Einschätzung physikalisch angemessen ist, lässt sich dann erst über den Erfolg des entsprechenden Ansatzes feststellen. Dieser Erfolg muss einerseits die Reproduzierbarkeit der Phänomenologie und des empirischen Gehalts der Vorgängertheorien einschlies-

---

[349] Zu den methodologischen Problemen, die sich aus der Annahme von Quantenfluktuationen der Metrik für eine Quantisierung ergeben, siehe Kap. 2.2.

[350] Zu der Einsicht, dass eine Gravitonenphysik keine angemessene Lösung für die Probleme der Quantengravitation darstellt, kann man jedoch offensichtlich auch kommen, ohne gleich auf prägeometrische Theorien zu setzen:

> *"The failure of the perturbative approach to quantum gravity in terms of linear fluctuations around a fixed background metric implies that the fundamental dynamical degrees of freedom of quantum gravity at the Planck scale are definitively not gravitons. At this stage, we do not yet know what they are."* (Loll (2007) 2) sen, andererseits muss er in spezifischen Vorhersagen zum Ausdruck kommen, die über die Vorgängertheorien hinausgehen und als Grundlage für differentielle Tests dienen können, in denen sich der Ansatz schliesslich zu bewähren hat.

Die meisten der prägeometrischen Ansätze gehen einerseits davon aus, dass die Quantenmechanik schon für das postulierte Substrat gültig ist, und andererseits, dass für die emergente makroskopische Ebene die Geometrisierbarkeit der Gravitation, wie sie in der Allgemeinen Relativitätstheorie vorliegt, weiterhin als gegeben angesehen werden kann. Beide Annahmen sind jedoch nicht notwendigerweise konstitutiv für eine prägeometrische Theorie der 'Quantengravitation', deren Ziel – entsprechend der in der Einleitung gegebenen Minimaldefinition[351] – darin besteht, die konzeptionelle Unverträglichkeit zwischen Allgemeiner Relativitätstheorie und Quantenmechanik bzw. Quantenfeldtheorie zu überwinden bzw. als nur scheinbare Unverträglichkeit in ihrer Wirksamkeit aufzuheben, indem sie etwa die Gravitation in ihrem Zustandekommen als emergentes Phänomen erklärt. Dieses Ziel schliesst pro forma auch solche Ansätze ein, die nicht von einer fundamentalen Gültigkeit der Quantenmechanik ausgehen,[352] ebenso wie solche, die noch nicht unbedingt die in der Allgemeinen Relativitätstheorie vorliegende Identifizierbarkeit der (nun emergenten) Gravitation mit der raumzeitlichen Metrik reproduzieren. Entscheidend ist, dass der empirische Gehalt der Vorläufertheorien in angemessener Weise reproduziert wird. Die grundlegende Anforderung an die prägeometrischen Ansätze besteht also weiterhin vor allem in der Notwendigkeit der Reproduktion der in der Allgemeinen Relativitätstheorie beschriebenen Phänomenologie als Näherung bzw. als Kontinuumsgrenzfall. Dies wird sich im Rahmen der prägeometrischen Theorieansätze vermutlich nur bewerkstelligen lassen, wenn es gelingt, die Modalitäten des Zustandekommens der Raumzeit (und der Gravitation) auf der Grundlage des dafür angenommenen prägeometrischen Substrats explizit zu klären.

## *Konzeptionelle Alternativen*

Die entscheidende Frage, die sich für alle prägeometrischen Theorieansätze stellt, ist die nach der Struktur, der Dynamik und den Konstituenten auf der Substratebene. Wenn man auf kontinuierliche Freiheitsgrade und eine raumzeitliche Einbettung verzichtet, bleiben für das Substrat erst einmal nur noch relationale Beziehungen und/oder Wechselwirkungen zwischen diskreten elementaren Konstituenten.[353]

> *"Suppose we want to construct a completely background independent quantum field theory. Such a theory must be independent of any of the ingredients of a classical field theory, including manifolds, coordinates, metrics, connections and fields. What is left of quantum theory when we remove all references to these structures? The answer is just algebra, representation theory and combinatorics."*
> (Smolin (2003) 28)

---

[351] Siehe Kap. 1.2.

[352] Ansätze, die Raumzeit, Gravitation und Quantenmechanik als emergent behandeln, wurden schon im Rahmen der Erörterung der computationalen Szenarien in Kap. 3.3. erwähnt. Siehe etwa 't Hooft (1999, 2000a, 2001, 2001a, 2007), Adler (2002), Elze (2009), Requardt (1996, 1996a, 2000), Cahill (2002, 2005), Cahill / Klinger (1996, 1997, 1998, 2005).

[353] Perez Bergliaffa / Romero / Vucetich (1997) schlagen etwa einen axiomatischen Ansatz hinsichtlich der prägeometrischen Entitäten und Strukturen vor.

Es wäre also erst einmal zu klären, welche elementaren Konstituenten und welche diskreten relationalen Strukturen zwischen diesen Konstituenten für eine solche Substratebene überhaupt in Frage kommen könnten.

*"The first problem is to identify the fundamental constituents of spacetime."* (Mäkelä (2007) 2)

Fast alle prägeometrischen Ansätze verwenden als Basiselemente auf der Substratebene elementare (Quanten-)Ereignisse ohne raumzeitliche Einbettung. Die strukturelle Einbettung erfolgt durch basale relationale Strukturen zwischen den elementaren Konstituenten; diese relationalen Strukturen lassen sich etwa in Form von Graphen[354] modellieren: als Netzwerke aus (gerichteten oder ungerichteten) äquivalenten, zweistelligen Relationen. Im allgemeinen werden sowohl den elementaren Ereignissen (Vertices) wie auch den Relationen (Linien) Quanteneigenschaften zugesprochen.

Entscheidend ist innerhalb der prägeometrischen Modelle dann aber vor allem die Frage, wofür die basalen Relationen stehen. – In den meisten Fällen werden sie als kausale Relationen zwischen elementaren Ereignissen (wie etwa bei den *Causal Sets*) oder als elementare dynamische Quanteninformationskanäle zwischen elementaren Quantenereignissen (wie etwa bei den *Quantum Causal Histories*)[355] gedeutet.

Schliesslich kann man sich dann noch die Frage stellen, in welcher Weise sich eventuelle dynamische Änderungen auf der Substratebene vollziehen. Setzen sie etwa eine Begrifflichkeit von Zeit voraus, die im Konflikt mit der Allgemeinen Relativitätstheorie steht? Wie liesse sich ein solcher möglicher Konflikt auflösen? – Eine Möglichkeit, die sich diesbezüglich insbesondere mit den *Quantum Causal Histories* abzeichnet, besteht in der vollständigen dynamischen Entkopplung der emergenten Ebene und ihrer zeitlichen Entwicklung von der Substratebene und der dort stattfindenden zeitlichen Entwicklung.

Werfen wir also einen Blick auf zwei konkrete Ansätze. Der zweite – die Modellierung eines prägeometrischen Substrats im Rahmen der *Quantum Causal Histories* – lässt sich dabei in mancher Hinsicht als paradigmatisch ansehen; er stellt eine Synthese vieler Denkansätze zu einer prägeometrischen Theorie der 'Quantengravitation' dar; und er schliesst nicht zuletzt auch die Grundideen des ersten, des *Causal-Set*-Ansatzes, ein und löst insbesondere dessen konzeptionelle Probleme. Die Vorteile der Modellierung mittels der *Quantum Causal Histories* lassen sich daher, ebenso wie die Perspektiven, die sich hier einstellen, viel besser einschätzen, wenn man schon die konzeptionellen Probleme kennt, auf die der *Causal-Set*-Ansatz stösst.

## Causal Sets

Dem vor allem von Rafael D. Sorkin und seinen Mitarbeitern entwickelten *Causal-Set*-Ansatz[356] liegt die Idee zugrunde, dass die kausale Ordnung fundamentaler als die Metrik oder die Topologie einer Raumzeit ist.[357] Ausgangspunkt ist die Annahme, dass das diskret strukturierte, nicht-raum-zeitliche Substrat, aus dem die Raumzeit schliesslich als emergentes Phänomen hervorgeht, aus elementarsten kausalen Vernetzungsstrukturen besteht: den *Causal Sets*.

---

[354] Diese lassen sich als die einfachsten diskreten mathematischen Strukturen überhaupt ansehen. Gitterstrukturen, ob regulär oder irregulär, könnten schon zu voraussetzungsreich sein und eine weitergehende dynamische Erklärung oder gar eine tieferliegende, die Struktur verursachende Schicht erforderlich machen.

[355] Siehe weiter unten.

[356] Siehe Bombelli / Lee / Meyer / Sorkin (1987), Sorkin (1997, 2003, 2007), Rideout / Sorkin (2000, 2001), Rideout (2002), Henson (2006), Surya (2007), Rideout / Wallden (2009).

> *"For the purposes of quantum gravity, a causal set is, of course, meant to be the deep structure of spacetime. Or to say this another way, the basic hypothesis is that spacetime ceases to exist on suffi-ciently small scales and is superseded by an ordered discrete structure to which the continuum is only a coarse-grained, macroscopic approximation."* (Sorkin (2003) 7)

Zu den Grundprinzipien, deren Gültigkeit schon für die Substratebene angenommen wird, gehört neben der Kausalität die Allgemeine Kovarianz, die hier nun in entsprechend transponierter Form auf eine basale diskrete Vernetzungsstruktur angewandt wird, für die dann jegliche Kennzeichnung oder Benennung ihrer Elemente ohne irgendeine physikalische Bedeutung ist:

> *"Discrete general covariance is simply the requirement that this labeling be 'pure gauge' [...]."* (Sor-kin (2003) 13)

Die so transponierte Allgemeine Kovarianz impliziert, dass es für die kausalen Vernetzungsstruktu-ren auf der Substratebene keinen festgelegten strukturellen Hintergrund in irgendeiner Form mehr gibt. Die Substratstruktur setzt insbesondere keine raumzeitliche Einordenbarkeit irgendeiner Form voraus. Sie erschöpft sich vielmehr in der umfassenden Relationalität der *Causal Sets*. – Modelliert werden diese in Form einer diskreten, ausschliesslich relationalen Struktur zwischen elementaren ('punktförmigen') Ereignissen. Für die diskrete relationale Struktur wird lokale Finitheit vorausge-setzt:

> *"[...] a finite number of elements causally between any two elements in the structure [...]."* (Henson (2006) 3)

Die physikalische Deutbarkeit der so modellierten Substratstruktur als Kausalstruktur und die ihrer elementaren Relationen als kausale Relationen setzt die Gerichtetheit dieser elementaren Relationen voraus. Die partielle kausale Ordnung von Ereignissen ist damit transitiv und irreflexiv. Zwei Er-eignisse sind entweder kausal miteinander verbunden oder sie werden als nominell (in einem dis-kreten prägeometrischen Sinne) 'räumlich' zueinander aufgefasst.

---

[357] Dies ist auch die Grundidee des *Twistor*-Ansatzes von Roger Penrose. Siehe etwa Penrose (1967, 2004), Penrose / MacCallum (1973), Mason (1991). Im *Twistor*-Ansatz werden kausale Relationen in topologische Relationen übersetzt. Das Ergebnis ist eine diskrete kombinatorische Basis, deren elementare Entitäten ausschliesslich über Spineigenschaf-ten verfügen: Spinnetze ohne raumzeitlichen Hintergrund, die gewisse Ähnlichkeiten mit den Spinnetzen bzw. den S-Knoten der *Loop Quantum Gravity* aufweisen. Die Idee ist dabei, dass sich die Raumzeit als effizientes Instrumenta-rium zur makroskopischen Beschreibung der dynamischen relationalen Struktur der *Twistor*-Netze ergibt und sich die Allgemeine Relativitätstheorie als makroskopische Implikation dieser basalen Dynamik erweist. Es zeichnet sich im *Twistor*-Ansatz schliesslich sogar eine Erklärung für das Auftreten von Nichtseparierbarkeiten bzw. Nichtlokalitäten in der Quantenmechanik wie in der Allgemeinen Relativitätstheorie ab:

> *"In twistor terms, massless particles are described by null geodesics; the emergence of space-time points is con-nected with the intersection of these geodesics. [...] Penrose suggests that the non-localisability inherent in any theory which assumes both quantum mechanics and gravity [...] could arise from the failure of null geodesics to intersect."* (Monk (1997) 19)

Bisher gibt es jedoch, obwohl der Ansatz schon lange Zeit existiert, noch keine konsistente Quantentheorie für die *Twistor*-Netze.

*"[...] these observations provide the kinematical starting point for a theory of discrete quantum gravity based on causal sets. The dynamics must then be obtained in the form of a 'quantum law of motion' for the causet."* (Sorkin (2003) 7)

Das Ziel des *Causal-Set*-Ansatzes ist es dann, im Rahmen einer mit der Quantenmechanik konformen Konkretisierung dieses elementaren modelltheoretischen Szenarios – also im Rahmen einer Quantentheorie elementarer kausaler Vernetzungsstrukturen – die Bedingungen für die Emergenz der Raumzeit aufzuzeigen. Sollte dies gelingen, dann sollte sich die Raumzeit, wie die Vertreter des *Causal-Set*-Ansatzes hoffen, als relationales, kausal bestimmtes Gefüge ihrer mikroskopischen (prägeometrischen) Konstituenten ('Punktereignisse') und der (ausschliesslich lokal bestimmten) Entwicklung dieses kausalen Gefüges rekonstruieren lassen – bzw. als das Ergebnis der Dynamik quantenmechanischer Superpositionen solcher prägeometrischer, relationaler Strukturen. Die kausale Struktur der makroskopischen Raumzeit wäre dann eine direkte oder indirekte Folge einer diskreten prägeometrischen kausalen Mikrostruktur. Insbesondere entsprächen makroskopische raumzeitliche Volumina, der Auffassung des *Causal-Set*-Ansatzes zufolge, naheliegenderweise der Zahl der mikroskopischen Elemente (Punktereignisse), aus denen heraus sie konstituiert würden. Geometrie reduzierte sich auf die Abzählbarkeit von Mengen:

*"If [...] we postulate that a finite volume of space-time contains only a (large but) finite number of elements, then we can – as Riemann suggested – measure its size by* counting.*"* (Bombelli / Lee / Meyer / Sorkin (1987) 522)

Auch die Dimensionalität der Raumzeit wäre ein dynamisches Ergebnis, das erst auf der emergenten makroskopischen Ebene eine Rolle spielt; auf der Ebene der *Causal Sets* gäbe es sie noch nicht.[358] Die effektive Dimensionalität der Raumzeit könnte sich dabei unter Umständen als abhängig von der Längenskala erweisen. Auch die Möglichkeit des Einschlusses von unterschiedlichen und insbesondere dynamischen Topologien ist im *Causal-Set*-Ansatz grundsätzlich gegeben.[359] – Im Gegensatz dazu ist jedoch die Signatur der makroskopischen Raumzeit in indirekter Weise schon konstitutiver Teil des *Causal-Set*-Ansatzes; kausale Ordnungen setzen, auch wenn auf der fundamentalen Ebene keine raumzeitliche Einbettung vorliegt, dort zumindest lokal schon eine eindeutige zeitliche Gerichtetheit voraus.

*"[...] why the spacetime metric is Lorentzian. The point is that no other metric signature, Riemannian or (++--) or whatever, can lend a partial ordering to the events of spacetime."* (Sorkin (1997) 18)

Damit sagt der *Causal-Set*-Ansatz eine notwendigerweise ungebrochene lokale Lorentz-Invarianz voraus, was immerhin zu einer konkreten differentiellen empirischen Testinstanz führen könnte, da viele der etablierten Ansätze zu einer Theorie der Quantengravitation eine zumindest minimale Brechung der Lorentz-Invarianz voraussagen.

---

[358] Es gäbe also keine grundsätzliche Festlegung hinsichtlich einer bestimmten Dimensionalität der Mannigfaltigkeiten, in die sich *Causal Sets* einbetten lassen.

[359] Es ist immer noch unklar, ob die Allgemeine Relativitätstheorie Topologiewechsel zumindest in eingeschränkter Form zulässt; eine nur leicht erweiterte Variante der Allgemeinen Relativitätstheorie lässt sie offensichtlich zu. Siehe etwa Horowitz (1991).

> *"[...] the Lorentzian signature [...] is singled out as the only one compatible with a consistent distinction between past and future [...]."* (Sorkin (2003) 5)

Interessanterweise führt der *Causal-Set*-Ansatz zudem zu einer Vorhersage einer kosmologischen Konstante, die bei $10^{-120}$ in natürlichen Einheiten liegt.[360] Er ist der einzige Ansatz der diesbezüglich eine konkrete und quantitative Vorhersage macht.

> *"[...] Sorkin's [quantum gravity] theory predicted a small value of the constant with the correct order of magnitude, before its observation."* (Rovelli (2007) 1322)

Das entscheidende Problem des Ansatzes besteht jedoch darin, dass sich noch keine konsistente quantenmechanische Formulierung für die Dynamik der *Causal Sets* abzeichnet, geschweige denn eine Rekonstruktion der makroskopischen Raumzeit oder die Ableitung der Einsteinschen Feldgleichungen. – Hinsichtlich der Dynamik werden zwei unterschiedliche Strategien verfolgt:

> *"A priori, one can imagine at least two routes to a 'quantum causet dynamics'."* (Sorkin (2003) 12)

Die erste zielt auf einen *Sum-over-histories*-Ansatz ab: auf eine Superposition bzw. Interferenz von *Causal Sets*, die als *Histories* aufgefasst werden und denen dabei entsprechende Wahrscheinlichkeitsamplituden zugeschrieben werden müssen.

> *"[...] the most pressing question is how to construct a dynamics, with causal sets as the histories, that would be a satisfying theory of quantum gravity. The question is, perhaps unsurprisingly, a difficult one."* (Henson (2006) 9)

Eine solche Netzwerksuperposition führt unabdinglich zu einer nicht-unitären Dynamik. Die konkrete Formulierung des *Causal-Set*-Ansatzes in Form eines solchen quantenmechanischen *Sum-over-histories*-Ansatzes steht immer noch aus:

> *"While progress is being made on the dynamics, a final theory is still not available."* (Henson (2006) 13) – *"The chief defect of the causal set approach is that so far it is not really a quantum theory; that is, it has not been able to take the step from transition probabilities to transition probability amplitudes, which would allow a Feynman formulation of the theory, leading to a 'sum over histories' quantum theory of causets [...]."* (Stachel (2006a) 76)

Nicht besser steht es um die zweite Strategie: Das klassische stochastische Wachstum der Netzwerkstruktur soll hier als Ausgangspunkt für eine anschliessende Quantisierung dienen.

> *"[...] one could try to identify certain general principles or rules powerful enough to lead, more or less uniquely, to a family of dynamical laws sufficiently constrained that one could then pick out those members of the family that reproduced the Einstein equations in an appropriate limit or approximation. [...] Expressed more technically, the idea is to seek a quantum causet dynamics by first formulating the causet's growth as a classical stochastic process and then generalizing the formulation to the case of a 'quantum measure' [...].[...] A dynamics or 'law of growth' is then simply an assignment of probabilities to each possible sequence of transitions."* (Sorkin (2003) 12f)

---

[360] Siehe Sorkin (1997, 2007).

Auch in Bezug auf diese zweite Strategie bleiben bisher die Möglichkeiten für eine Konstruktion einer zumindest approximativ ableitbaren phänomenologischen Raumzeit völlig im Unklaren. Der *Causal-Set*-Ansatz führt im Versuch, eine makroskopische Raumzeit in ihrem Entstehen nachvoll­ziehbar werden zu lassen, immer nur zu einer differenzierbaren Struktur mit konformer Metrik ohne irgendein Längenmass. Und schon die Frage nach der entsprechenden Mannigfaltigkeit, in die sich ein *Causal Set* einbetten lässt, führt nicht unbedingt zu einer eindeutigen Antwort.

> *"Can such a structure really contain enough information to provide a good manifold approximation? We do not want one causal set to be well-approximated by two spacetimes that are not similar on large scales. The conjecture that this cannot happen (sometimes called the 'causal set hauptvermu­tung', meaning 'fundamental conjecture') is central to the program. [...] One of the chief difficulties has been the lack of a notion of similarity between Lorentzian manifolds, or more properly, a distance measure on the space of such manifolds."* (Henson (2006) 5)

Vielleicht sind kausale Relationen allein für eine Ableitung der phänomenologischen Raumzeit mit allen ihren geometrischen Strukturen einfach noch nicht hinreichend.[361] Die prägeometrischen *Quantum Causal Histories* legen nahe, dass man dazu noch ein wenig mehr benötigt:

## Quantum Causal Histories

Die prägeometrischen *Quantum Causal Histories*[362], die vor allem von Fotini Markopoulou als kon­zeptionelle Idee mit möglicher Relevanz für den Bereich der 'Quantengravitation' propagiert wer­den, können, auch wenn sie noch nicht in Form einer vollständig ausformulierten Theorie vorliegen, dennoch in gewisser Weise gleichermassen als paradigmatisches Beispiel eines prägeometrischen Ansatzes wie als Konvergenzpunkt und Synthese der unterschiedlichsten Denkansätze zu einer prä­geometrischen, kausal bestimmten Quantenstruktur angesehen werden. So lassen sich die *Quantum Causal Histories* einerseits, indem sie Wheelers Idee des *It from bit*[363], Elemente aus Lloyds Kon­zeption eines *Computationalen Universums*[364] sowie die Idee der *Holographischen Schirme*[365] ein­beziehen, als konzeptionelle Erweiterung von Sorkins *Causal-Set*-Ansatz[366] verstehen; andererseits lassen sie sich ebenso als Verallgemeinerung kausaler Spinnetze und des *Spin-Foam*-Ansatzes[367] ansehen, angereichert um Elemente aus den *algebraischen Quantenfeldtheorien*.

---

[361] Ebenso bleibt zunächst die Frage unbeantwortet, wie sich in einem solchen Ansatz Materie- und Wechselwirkungs­felder unterbringen liessen. Diese müssten grundsätzlich ebenso aus der Menge der basalen Ereignisse und ihren kau­salen Relationen ableitbar bzw. mit ihren Mitteln implementierbar sein. Der *Causal-Set*-Ansatz schickt sich aber erst einmal nur an, die Raumzeit auf diskreter Grundlage zu reproduzieren. Materie- und Wechselwirkungsfelder bleiben dabei aussen vor. Auch hier bieten die prägeometrischen *Quantum Causal Histories* eine überzeugendere Lösung.

[362] Siehe Markopoulou (2000, 2000a, 2000b, 2004, 2006, 2007), Dreyer (2004, 2006, 2007) (Dreyer nennt seinen An­satz *Internal Gravity*), Kribs / Markopoulou (2005) (Emergente Teilchenphysik), Konopka / Markopoulou / Smolin (2006) (*Quantum Graphity*), Konopka / Markopoulou / Severini (2008), Hawkins / Markopoulou / Sahlmann (2003).

[363] Siehe Kap. 3.3. sowie Wheeler (1989).

[364] Siehe Kap. 3.3. sowie Lloyd (1999, 2005, 2005a, 2007).

[365] Siehe Kap. 3.3. sowie Markopoulou / Smolin (1999).

[366] Siehe oben, sowie Bombelli / Lee / Meyer / Sorkin (1987), Sorkin (2003), Rideout / Sorkin (2000, 2001), Rideout (2002), Henson (2006), Surya (2007), Rideout / Wallden (2009).

[367] Siehe Kap. 4.4. sowie Oriti (2001, 2003), Livine / Oriti (2003), Perez (2003, 2006), Baez (1998, 2000), Markopoulou / Smolin (1997).

Die prägeometrischen[368] *Quantum Causal Histories* sind relationale, sich lokal verändernde Netzwerke von Quantensystemen ohne jeglichen raumzeitlichen Hintergrund. Sie lassen sich ohne weiteres als eine konsistente Konkretisierung von Rovellis Konzeption der Welt als Netzwerk relationaler Bezüge ansehen,[369] zu der es hier aber eben gerade ohne eine vorausgehende Quantisierung der Allgemeinen Relativitätstheorie kommt, denn diese wird im Kontext der *Quantum Causal Histories* als effektive Theorie verstanden, die ausschliesslich für eine emergente Ebene Relevanz besitzt. Man könnte in dieser Hinsicht durchaus behaupten, dass die *Quantum Causal Histories* – indem sie versuchen, Rovellis Idee einer fundamentalen Relationalität auf einer prägeometrischen Ebene zu erfassen – gerade die Leerstellen ausfüllen, die Rovellis Konzeption noch offen lässt, indem sie insbesondere die Idee einer völligen Hintergrundunabhängigkeit soweit umsetzen, wie dies überhaupt möglich ist, ohne sich mit ihrer konkreten Substratkonstruktion die Perspektive auf eine Rekonstruktion der makroskopischen raumzeitlichen Welt zu verschliessen:

> *"The approaches of Rovelli, however, left open the precise structure that is to tie together the network of Hilbert spaces and algebras necessary to describe a whole universe."* (Smolin (2006c) 231)

Zu den konzeptionellen Ausgangspunkten in der Theoriebildung zählt dementsprechend – über die Einschätzung hinaus, dass eine Theorie der Quantengravitation notwendigerweise eine hintergrundunabhängige Theorie sein muss – vor allem die Annahme, dass es auf der fundamentalen Ebene keine raumzeitlichen Freiheitsgrade gibt, dass also die Raumzeit ein emergentes Phänomen ist:

> *"It is peculiar that the approaches that advocate that gravity is only an effective theory (string theory, condensed matter) are based explicitly on a spacetime being present while approaches that are background independent consider gravity to be fundamental. / Here, we will advocate an approach orthogonal to the quantum field theory-like approaches above (we are background independent) but also orthogonal to the usual background independent approaches (there will be no fundamental degrees of freedom for the gravitational field)."* (Markopoulou (2006) 2) – *"[...] the fundamental theory is free of geometric notions. [...] This should be viewed as an advantage of our proposal since it has been extremely difficult to construct Hilbert spaces of spacetime geometries or to make sense of superpositions of spacetimes [...]."* (Dreyer (2007) 10)

Konstitutiv für den Ansatz ist zudem die Annahme, dass keine Form von Kontinuum – nicht nur nicht das raumzeitliche – auf der fundamentalen Ebene irgendeine Rolle spielt, dass also die sich im Kontext der Thermodynamik schwarzer Löcher abzeichnende Einsicht,[370] dass einer finiten raumzeitlichen Region immer nur eine finite Zahl von Freiheitsgraden zugeschrieben werden kann, aus einer fundamentalen (lokalen) Finitheit auf der Substratebene resultiert.

---

[368] Die Konstruktion einer prägeometrischen Theorie der Quantengravitation ist nur eine der Möglichkeiten, die sich mit den *Quantum Causal Histories* ergibt. Diese sind darüberhinausgehend dafür geeignet, die unterschiedlichsten Systeme zu modellieren: Einerseits kann man sie dazu verwenden, ein diskretes Analogon zur algebraischen Quantenfeldtheorie, definiert auf einem *Causal Set*, zu formulieren. Andererseits sind sie dazu geeignet, die Quantengeometrie hintergrundunabhängiger Systeme im Sinne eines *Causal Spin Foam* in Form von Quantensuperposition von Geometrien zu beschreiben. Und schliesslich lässt sich mit ihnen eben nicht zuletzt die Prägeometrie hintergrundunabhängiger Systeme modellieren. Um diese geht es hier im Folgenden; der Begriff der *Quantum Causal Histories* wird im Folgenden ohne weiteren Zusatz in diesem Sinne verstanden.

[369] Siehe Kap. 4.4.

[370] Siehe Kap. 3.1.

> *"In the picture of the universe as a [Quantum Causal History], it can be thought of as a very large collection of quantum-mechanical building blocks. Everything is (locally) finite, both the dimensionality of these systems as well as the causal set. We regard this as a simple way to model the idea that there is a finite number of fundamental degrees of freedom in a finite region of spacetime."* (Hawkins / Markopoulou / Sahlmann (2003) 17)

Hinzu kommt, wie schon im *Causal-Set*-Ansatz, die Annahme, dass die kausale Ordnung fundamentaler ist als die metrischen oder topologischen Eigenschaften der Raumzeit, dass kausale Relationen auf der Substratebene in Form von elementarsten kausalen Vernetzungsstrukturen vorliegen, und schliesslich, dass die Quantentheorie auch auf der fundamentalen Ebene gültig ist. – Der entscheidende Unterschied zum *Causal-Set*-Ansatz besteht darin, dass die als fundamental angesehene kausale Ordnung in den *Quantum Causal Histories* modelltheoretisch völlig anders erfasst und implementiert wird; insbesondere wird sie nicht zuerst klassisch rekonstruiert, um dann einer Quantisierung unterzogen zu werden, zu der es im *Causal-Set*-Ansatz aufgrund konzeptioneller Schwierigkeiten letztendlich dann gar nicht kommt.

<div align="center">*</div>

Hinsichtlich ihrer konkreten Substratkonstruktion gehen die *Quantum Causal Histories* von diskreten, gerichteten, lokal finiten, zirkelfreien Graphen aus, die als relationale, sich lokal verändernde Netzwerke von Quantensystemen ohne raumzeitlichen Hintergrund stehen.[371] Die Vertices der Graphen stehen für elementare Ereignisse, die gerichteten Linien für kausale Relationen zwischen jeweils zwei elementaren Ereignissen. Für jeden Vertex innerhalb des gerichteten Graphen wird ein endlich-dimensionaler Hilbertraum definiert (bzw. eine Matrixalgebra von Operatoren).

> *"The idea is to use a causal set to describe the causal structure while a quantum theory being introduced through the assignment of finite-dimensional Hilbert spaces to the elementary events."* (Livine / Terno (2006) 1)

Jeder Vertex, jedes elementare Ereignis, wird also als elementares Quantensystem aufgefasst. Jede (gerichtete) Linie, also jede kausale Verbindung zwischen zwei elementaren Ereignissen, wird mit einem Quantenkanal ('*completely positive map*') identifiziert, der einen Übergang von einem Hilbertraum zu einem anderen vermittelt und innerhalb der relationalen Graphenstruktur dem Analogon einer unitären Abbildung entspricht.[372] Dieser Quantenübergang repräsentiert bzw. instantiiert die elementarste Form einer kausalen Verbindung zwischen zwei elementaren Ereignissen. Die Graphenstruktur wird somit zu einem Netzwerk von Informationsflüssen zwischen den als Quantensystemen verstandenen elementaren Ereignissen. Prägeometrische *Quantum Causal Histories* sind damit formal identisch mit informationsverarbeitenden Quantensystemen bzw. Quantencomputern.

> *"[...] this suggests that the universe modeled as a [Quantum Causal History] is nothing but a background-independent quantum computer."* (Hawkins / Markopoulou / Sahlmann (2003) 17f)

---

[371] Insbesondere wird keine Metrik vorausgesetzt; es wird also mit den Relationen zwischen den Elementen des Netzwerkes kein Abstand zwischen diesen Elementen definiert.

[372] Dies sind die entscheidenden Modifikationen gegenüber dem *Causal-Set*-Ansatz. Die so definierte Graphenstruktur lässt sich gerade als die gesuchte quantenmechanische Erweiterung der *Causal Set* interpretieren.

*

Der erste Schritt auf dem Weg zur Reproduktion der phänomenologischen Implikationen der Allgemeinen Relativitätstheorie, also zur Einlösung der basalsten Anforderung an jede Theorie der Quantengravitation, besteht nun in einer Rekonstruktion der Geometrogenese. – Die Idee, die diesbezüglich bei den *Quantum Causal Histories* zur Anwendung kommt, ist die folgende: Klassische Raumzeit und Gravitation ergeben sich nicht in Folge einer Mittelung quantengeometrischer Freiheitsgrade – solche gibt es im Kontext der *Quantum Causal Histories* nicht –, sondern auf der Grundlage von langreichweitig (innerhalb der Graphenstruktur) propagierenden kohärenten Anregungszuständen (sogenannten *rauschfreien Subsystemen*), die von der Mikrodynamik der Substratebene bestimmt werden und durch sie erhalten bleiben.[373]

> *"[...] instead of looking for ways to coarse-grain the quantum geometry directly, one can first look for long-range propagating degrees of freedom and reconstruct the geometry from these."* (Markopoulou (2006) 15) – *"[...] we will take up the idea that the effective description of a background independent theory can be characterized by the dynamics of coherent excitations in the fundamental theory and implement it by importing the method of noiseless subsystems from quantum information theory."* (Markopoulou (2006) 3)

Damit es überhaupt zur Emergenz der Raumzeit und der Gravitation kommen kann, muss die Dynamik dieser kohärenten Anregungszustände, auf deren Grundlage Raumzeit und Gravitation konstituiert werden soll, von der Dynamik auf der nicht-raumzeitlichen Substratebene entkoppelt sein;[374] wäre sie es nicht, gäbe es auch auf der Makroebene keine Gravitation und keine Raumzeit, so wie es diese auf der Mikroebene nicht gibt. Die Kausalität, die in den Wechselwirkungen auf der emergenten Makroebene zur Geltung kommt, ist dann keinesfalls mit den kausalen Relationen auf der Substratebene zu identifizieren; sie ist ebenfalls eine Implikation der Dynamik der kohärenten Anregungszustände, auf deren Grundlage die Makroebene überhaupt erst zustandekommt. Ebenso ist im Kontext der prägeometrischen *Quantum Causal Histories* die raumzeitliche Lokalität (bzw. Nichtlokalität) der Dynamik auf der Makroebene von den lokalen relationalen Beziehungen auf der Substratebene zu unterscheiden.[375]

> *"In a given graph (the fundamental theory) there will be a notion of locality: in a graph two nodes are neighbors if they are connected by a link. We call this* microlocality. *In the known background independent theories, the dynamics is generated by moves that are local in the microscopic sense. But if this is to be a good theory, there should be a notion of classical spacetime geometry that emerges from the quantum geometry. This will give rise to another notion of locality, which we may characterize as* macrolocality. *[...] they do not coincide. [...] the notion of macrolocality should be defined directly from the interactions of the noiseless subsystems that we identify with the emergent degrees of freedom [...] It is the fundamental evolution that is non-local with respect to our spacetime."* (Markopoulou (2006) 24f)

Analoges gilt für die zeitliche Entwicklung auf der Makroebene, also der Ebene der emergenten Raumzeit, die durch die Dynamik der kohärenten Anregungszustände zustandekommt; diese ist notwendigerweise von der zeitlichen Abfolge auf der Substratebene entkoppelt. Letztere liegt nur lokal in Form elementarer kausaler Beziehungen vor. Auf der Makroebene hingegen könnten sich sogar für spezifische Umgebungen quasi-globale Zeitentwicklungen ergeben.

---

[373] Siehe insbesondere Kribs / Markopoulou (2005), die deutlich machen, dass eine solche Annahme nicht zuletzt auch mit dem *Spin-Foam*-Ansatz kompatibel ist.
[374] Siehe hierzu insbesondere auch Jannes (2008).
[375] Die kohärenten Anregungszustände, welche die Makroebene konstituieren, sind nicht notwendigerweise lokalisiert.

> *"[...] truly effective spacetime means effective locality and effective time direction that are not simply Planck scale quantum corrections on the classical ones."* (Markopoulou (2006) 29) – *"[...] it may be possible to have an underlying ('micro') time without running into the observationally excluded preferred frame."* (Markopoulou (2006) 26) – *"This viewpoint is orthogonal to the much discussed relational approach where the introduction of a background time is seen as something that should be avoided. "* (Markopoulou (2006) 28)

Die Allgemeine Relativitätstheorie ist unter diesen Bedingungen als eine effektive Theorie anzusehen, welche die emergente Raumzeit bzw. Gravitation (zumindest unter bestimmten Voraussetzungen) beschreibt. Auch die Diffeomorphismusinvarianz der Allgemeinen Relativitätstheorie ist dann das emergente Resultat einer relationalen 'geometriefreien' Dynamik auf der prägeometrischen Ebene – ein Resultat, zu dem es wiederum vielleicht nur unter spezifischen Bedingungen im Rahmen der Emergenz der Makrodynamik kommt.

> *" Here, symmetries such as gauge invariance, diffeomorphism invariance, and time re-parameterization invariance are not fundamental features of the full system comprising the various environments. [...] the situation is different in quantum gravity approaches [...] in which general relativity is expected to be an effective theory of the excitations of a system with no fundamental gravitational degrees of freedom."* (Markopoulou (2006) 28)

Es ist nicht zu erwarten, dass alle prägeometrischen *Quantum Causal Histories* notwendigerweise eine geometrische Phase aufweisen. Wenn es aber zu einer makroskopischen Raumzeit kommt, so ist diese notwendigerweise dynamisch, weil sie überhaupt erst auf der Grundlage einer (hintergrundunabhängigen) prägeometrischen Dynamik zustandekommt.

<div align="center">*</div>

Welcher Art sind nun aber die kohärenten Anregungszustände, die von der relationalen Substratdynamik hervorgebracht werden, um dann selbst wieder diese Wunder hervorzubringen: eine emergente, dynamisch vom Substrat entkoppelte raumzeitliche Makrowelt, die über eine eigenständige Form von Kausalität und über gravitative Freiheitsgrade verfügt? – Die Antwort, die im Kontext der *Quantum Causal Histories* auf diese Frage gegeben wird, besteht in einer Kopplung der Geometrogenese an die Materiegenese. Beide sind hier untrennbar miteinander verbunden. Die Materiegenese ist die Voraussetzung zur Emergenz der Raumzeit. Die Emergenz der Raumzeit ist eine Implikation der Materiegenese. Letztlich gibt es keine Trennung in geometrogenetische und materiegenetische Freiheitsgrade.[376]

> *"In our view, matter and geometry have a more dual role. One can not have one without the other. Both emerge from the fundamental theory simultaneously."* (Dreyer (2007) 4)

---

[376] Für eine prä-raumzeitliche Substratebene gäbe es auch kaum noch eine Rechtfertigung für die Unterscheidung in geometrogenetische und materiegenetische Freiheitsgrade. Siehe weiter unten.

Das, was auf der makroskopischen Ebene als Materieteilchen in Erscheinung tritt, sind gerade die langreichweitig propagierenden kohärenten Anregungszustände, die sich auf der Grundlage der Substratdynamik ergeben. Die gesuchten kohärenten Anregungszustände sind also mit den Materieteilchen zu identifizieren. – Und diese langreichweitig propagierenden kohärenten Anregungszustände konstituieren mit ihrer spezifischen, entkoppelten Dynamik gerade deshalb die Raumzeit, weil sie sich so verhalten, als würden sie sich in einer Raumzeit bewegen. Die Raumzeit ist also nicht anderes als das Ergebnis – oder genauer: eine Implikation – des Verhaltens der Materie.

> *"We propose that it is properties of the interactions of these excitations that we understand as space-time."* (Markopoulou (2006) 2) – *"[...] we only look at the available matter degrees of freedom and ask what geometry we obtain if we only use these degrees of freedom. It is important here that we do not include information, like an absolute time, that is only available to an observer external to the system."* (Dreyer (2007) 2)

Es gibt also gerade deswegen eine emergente Raumzeit, weil die kohärenten Anregungszustände (Materieteilchen) sich so verhalten, als würden sie sich in einer Raumzeit bewegen, nicht zuletzt indem ihr dynamisches Verhalten bestimmte Symmetrien aufweist.

> *"[...] to have a good low energy limit it must be the case that the emergent symmetries that act on the space of noiseless subsystems includes an approximate translation invariance in space and time. In this case the conserved quantities will include momenta and energy. [...] these translation symmetries should emerge as additional symmetries which protect the degrees of freedom we have identified as elementary particles."* (Bilson-Thompson / Markopoulou / Smolin (2006) 9) – *"[...] all we can mean by a Minkowski spacetime is that all coherent degrees of freedom and their interactions are Poincaré invariant at the relevant scale."* (Markopoulou (2006) 18)

Raumzeitliche Invarianzen wie etwa die Poincaré-Invarianz werden von der prägeometrischen Theorie also nicht vorausgesetzt, sondern ergeben sich erst aus dem Verhalten der kohärenten Freiheitsgrade.

> *"In our approach the relationship between particles and symmetry group is exactly reversed. It is the particles that determine structures like the light cone and the symmetry group. We are thus proposing not the use the Poincaré group and its representation theory in the basic setup of the theory."* (Dreyer (2007) 10)

Wenn die Raumzeit und ihre Struktur jedoch in nichts anderem als im Verhalten der Materie zum Ausdruck kommt, so ist sie nichts anderes als ein rein relationales Konstrukt, das sich aus diesem Verhalten ergibt.

Unter diesen Voraussetzungen stellt sich nicht zuletzt auch das Problem der kosmologischen Konstante als Artefakt einer hintergrundabhängigen Beschreibungsweise dar, die letztlich eine unabhängige raumzeitliche Substanz unterstellt – ein Problem, das unter Annahme einer vollständig hintergrundunabhängigen, relationalen Dynamik seine natürliche Auslösung findet:

> *"If matter is viewed as propagating on geometry then the zero mode energy of matter fields should contribute to the curvature of geometry. This view leads to one of the worst predictions of theoretical physics. The cosmological constant obtained in this was off by more than 120 orders of magnitude.*

*[...] In internal gravity matter is to be viewed as giving rise to geometry and so the above line of argument is seen to be fallacious. It is the excitations that make the geometry. Zero mode energies only appear in the effective description of the matter on a given spacetime. Fundamentally they should not be viewed as energies residing on the spacetime."* (Dreyer (2007) 9)

Dass es durch das kollektive Verhalten der materiellen Freiheitsgrade nicht nur zur Emergenz der Raumzeit kommt, sondern – mit dieser verbunden – auch zur Emergenz der Gravitation, lässt sich im Kontext der *Quantum Causal Histories* verstehen, wenn man annimmt, dass Materiekonstellationen ihr Umfeld (bzw. den Grundzustand dieses Umfeldes) verändern.

*"[...] the order parameter near a [bound state of excitations] will deviate from [the ground state]."* (Dreyer (2007) 6)

Die Ausbreitung einer solchen Veränderung erfolgt schrittweise, was die endliche Wirkungsausbreitungsgeschwindigkeit der Gravitation im emergenten raumzeitlichen Umfeld erklärt:

*"Once the velocities of the bound objects are no longer small we have to take into account that the change of [the state of the order parameters] is not instantaneous. Gravity here has a finite propagation speed."* (Dreyer (2007) 8)

Eines der zentralen Ziele im Kontext der *Quantum Causal Histories* betrifft das Verständnis der dynamischen Implementierung geometrisierter gravitativer Freiheitsgrade in die Struktur der emergenten Raumzeit, die sich im Verhalten der kohärenten Materiefreiheitsgrade manifestiert. Inzwischen gibt es immerhin Indizien für eine gekrümmte raumzeitliche Mannigfaltigkeit mit Lorentz-Signatur:

*"What then is the geometry that we will find? [...] generically the geometry is that of a curved manifold with a Lorentzian signature."* (Dreyer (2007) 2)

Die zentrale Vermutung hinsichtlich der *Quantum Causal Histories* ist jedoch bisher noch unbewiesen. Sie besteht in der Annahme, dass die Einsteinschen Feldgleichungen für die emergente raumzeitliche Struktur und Dynamik, die sich gerade in nichts anderem als im Verhalten der kohärenten Materiefreiheitsgrade manifestiert, notwendigerweise erfüllt sind:

*"[...] we use the excitations and their interactions to define* both *the geometry and the energy-momentum tensor $T_{\mu\nu}$. This leads to the following Conjecture on the role of General Relativity [...]: If the assignment of geometry and $T_{\mu\nu}$ from the same excitations and interactions is done consistently, the geometry and $T_{\mu\nu}$ will not be independent but will satisfy Einstein's equations as identities."* (Markopoulou (2006) 29) – *"[...] the same excitations of the underlying system (characterizing the geometrogenesis phase transition) and their interactions will be used to define* both *the geometry and the energy-momentum tensor $T_{\mu\nu}$. This leads to the following Conjecture on the role of General Relativity: / If the assignment of geometry and $T_{\mu\nu}$ from the same excitations and interactions is done consistently, the geometry and $T_{\mu\nu}$ will not be independent but will satisfy Einstein's equations as identities. / What is being questioned here is the separation of physical degrees of freedom into matter and gravitational ones. In theories with a fixed background, such as quantum field theory, the separation is unproblematic, since the gravitational degrees of freedom are not really free and do not interact with the matter. In the classical background independent theory, general relativity, we are left with an in-* *tricate non-linear relation between the two sets: the Einstein equations. As the practitioners of cano-
nical quantum gravity know well, cleanly extracting dynamical gravitational degrees of freedom from
the matter is fraught with difficulties. If such a clean separation could be achieved, canonical quantum
gravity would have succeeded at least two decades ago. / The new direction unifies matter and gravity
in the pre-geometric phase and provides a path towards* explaining *gravity rather than just quantizing
it."* (Markopoulou (2007) 19)

<div align="center">*</div>

Welcher Art ist jedoch die im Kontext der *Quantum Causal Histories* beschriebene Materie? Inwie-
fern lässt sich die Beschaffenheit der Materie tatsächlich aus der Substratdynamik erschliessen?
Wodurch zeichnen sich die kohärenten Anregungszustände aus, die sich gerade so verhalten, als
befänden sie sich in einer Raumzeit? Was ist es, was dazu führt, dass sie im Rahmen der Substrat-
dynamik erhalten bleiben? – Eine mögliche direkte Antwort auf die letzte Frage (und mithin eine
zumindest indirekte Antwort auf die vorausgehenden Fragen) lautet: Topologie. Die Idee ist die,
dass sich die kohärenten Anregungszustände mit stabilen topologischen Strukturen identifizieren
lassen, genauer: mit topologischen Knotenstrukturen mit Überkreuzungen und Verdrillungen.[377]

> *"Given a particular theory of dynamical quantum geometry, however, it is not immediately apparent
> whether it has any such [noiseless subsystems]. We show that in a large class of theories, it does.
> These are theories in which the microscopic quantum states are defined in terms of the embedding, up
> to diffeomorphisms, of a framed, or ribbon, graph in a three manifold and in which the allowed evolu-
> tion moves are the standard local exchange and expansion moves."* (Bilson-Thompson / Markopoulou
> / Smolin (2006) 1)

Es sind offensichtlich gerade solche topologischen Knotenstrukturen, die im Rahmen der Substrat-
dynamik stabil erhalten bleiben.

> *"We have shown that braidings of graph edges are unaffected by the usual evolution moves. Any phy-
> sical information contained in the braids will propagate coherently [...]."* (Markopoulou (2006) 19)

Grundlage der Erhaltung dieser topologischen Strukturen sind entsprechende topologische Symme-
trien bzw. Erhaltungssätze.

> *"The states are bound here, not by fields, but by quantum topology. [...] the states are bound because
> there are conserved quantum numbers that measure topological properties of the states."* (Bilson-
> Thompson / Markopoulou / Smolin (2006) 2)

Was dem Ansatz zusätzliche Plausibilität verleiht, ist dann schliesslich die Tatsache, dass sich die
topologischen Eigenschaften dieser Strukturen mit den typischen Eigenschaften unserer bekannten
Elementarteilchen identifizieren lassen: Ladung entspricht etwa der Verdrillung eines Bandes. Ent-
sprechend gibt es topologische Pendants zur Ladungskonjugation, zur Parität sowie zu den Farb-
eigenschaften von Quarks.

---

[377] Siehe Bilson-Thompson / Markopoulou / Smolin (2006), Bilson-Thompson (2005), Bilson-Thompson / Hackett /
Kauffman (2009).

*"It is then possible that all the quantum numbers, including the geometric labels used in loop quantum gravity, can be regarded as composites of fundamentally topological properties."* (Bilson-Thompson / Markopoulou / Smolin (2006) 11) – *"Twist is interpreted as U(1) charge, so that a ± 2 π twist in a ribbon represents charge ± e/3."* (Bilson-Thompson / Markopoulou / Smolin (2006) 4) – *"[...] parity inversion [...] for a braid is equivalent to a left-right inversion, while not affecting the handedness of any twists on the strands."* (Bilson-Thompson / Markopoulou / Smolin (2006) 6)

Inzwischen ist es sogar gelungen, die Teilchen des quantenfeldtheoretischen Standardmodells jeweils mit spezifischen topologischen Strukturen zu identifizieren.[378] Gravitonen treten naheliegenderweise nicht in Erscheinung, da die Gravitation im Kontext der *Quantum Causal Histories* ein emergentes makroskopisches Phänomen und nicht etwa eine fundamentale Wechselwirkung ist.

*"The lightest states are the simplest non-trivial braids made of ribbons with no twists (no charge) or one full twist (positive or negative charge) / Quantum numbers are assigned only to braids with no positive and negative charge mixing."* (Bilson-Thompson / Markopoulou / Smolin (2006) 7) – *"The simplest non-trivial braids can be made with three ribbons and two crossings [...]. It is remarkable that with a single condition, these map to the first generation of the standard model."* (Bilson-Thompson / Markopoulou / Smolin (2006) 4) – *"It is natural to hypothesize then that the second generation standard model fermions come from the next most complicated states, which have three crossings."* (Bilson-Thompson / Markopoulou / Smolin (2006) 8) – *"[...] it is also proposed that the gauge vector bosons of the standard model are composite, and are represented by triplets of ribbons with no crossings. Braids with three ribbons and no crossings are mapped to the bosons of the electroweak interaction. The electroweak interactions between the fermions and the photon and vector bosons are then described by cutting and joining operations on 3-ribbon braids. These preserve the relevant quantum numbers."* (Bilson-Thompson / Markopoulou / Smolin (2006) 8f)

Was allerdings noch fehlt, ist eine *dynamische* Erklärung für das Verhalten und für die phänomenologischen Eigenschaften der stabilen topologischen Strukturen – ebenso wie für ihre Stabilität überhaupt:

*"Ultimately such rules have to arise from the dynamics."* (Bilson-Thompson / Markopoulou / Smolin (2006) 7)

Insbesondere sollten nicht zuletzt auch die Massen der Teilchen auf der Grundlage der Dynamik der rauschfreien Subsysteme verstehbar werden. Damit es dazu aber kommen kann, ist offensichtlich eine Modifikation des Ansatzes hin auf eine nur approximativ gültige Rauschfreiheit der kohärenten Anregungszustände erforderlich; erst dies ermöglicht die Einführung von Wechselwirkungen zwischen den basalen topologischen Strukturen:

*"In order to describe interactions we will have to weaken the notion of a noiseless subsystem to a more realistic notion of an approximate noiseless subsystem."* (Bilson-Thompson / Markopoulou / Smolin (2006) 10)

\*

---

[378] Siehe Bilson-Thompson / Markopoulou / Smolin (2006), Bilson-Thompson (2005), Bilson-Thompson / Hackett / Kauffman / Smolin (2008), Bilson-Thompson / Hackett / Kauffman (2009).

Auch wenn hier sicherlich noch einiges an Entwicklung aussteht, so zeichnet sich mit den prägeometrischen *Quantum Causal Histories* immerhin schon ein kohärentes Bild eines konzeptionell überzeugenden Emergenzszenarios ab. – Die Vereinheitlichung von Materie und raumzeitlicher Geometrie, deren emergentes Zustandekommen zumindest konzeptionell schon nachvollziehbar wird, ergibt sich im Kontext der *Quantum Causal Histories* auf der Grundlage emergenter topologischer Freiheitsgrade, die von einer prägeometrischen, rein relationalen Substratdynamik hervorgebracht und stabilisiert werden.

> *"[...] one may expect that the emergent spacetime is the collection of events that are the interactions of the excitations of an underlying pre-spacetime quantum theory, with matter being also emergent as these same excitations."* (Bilson-Thompson / Markopoulou / Smolin (2006) 10)

Materie und raumzeitliche Geometrie sind hier das Ergebnis einer umfassenden, wenngleich emergenten Topologisierung, quasi einer topologischen Variante umfassender, prä-raumzeitlicher Geometrisierung, die im Gegensatz zum alten Einsteinschen und Wheelerschen Traum einer umfassenden raumzeitlichen Geometrisierung[379] auf einem rein relationalen Substrat aus elementaren Quantenereignissen und elementaren Quanteninformationsflüssen fusst.

> *"Ever since the notion that geometry is dynamical was advanced in the 19th Century and, even more since that idea was realized in general relativity, there has been a dream: to unify matter with geometry and gravity by demonstrating that matter arises from singularities of topological defects in geometry."* (Bilson-Thompson / Markopoulou / Smolin (2006) 1)

Der traditionelle Geometrisierungsgedanke zielte in seiner radikalsten Form darauf ab, die Raumzeit als einzige Substanz zu erweisen. Die Materie und alle Wechselwirkungen sollten als Ausprägungen der Dynamik komplexerer raumzeitlicher Strukturen und komplexerer raumzeitlicher Topologien rekonstruiert werden. Am Ende sollte nichts anderes mehr übrigbleiben als dynamische raumzeitliche Geometrie. – In der neuen topologisierten, emergenztheoretischen Variante dieser Idee einer umfassenden Geometrisierung bleibt nun aber nicht nur kein Platz mehr für eine raumzeitliche Substanz. Nicht nur die Raumzeit stellt sich als rein relationales Konstrukt dar, das sich aus der Dynamik emergenter kollektiver Anregungszustände ergibt. Auch auf der Substratebene, die diese Anregungszustände und die entsprechende makroskopische Dynamik hervorbringt, gibt es nur mehr ein relationales Gefüge aus Informationsflüssen, dem sich schwerlich irgendeine Form von Substantialität zusprechen lässt. Es sind diese Informationsflüsse, die – über dynamische Entkopplungen – die raumzeitliche Geometrie hervorbringen.

Die grundsätzliche konzeptionelle Plausibilität eines solchen Szenarios wird, soweit es die emergente raumzeitliche Geometrie betrifft, letztlich schon im Konzept der *Holographischen Schirme* deutlich.[380] Um aber schliesslich nachvollziehen zu können, wie Informationsflüsse gleichzeitig die raumzeitliche Geometrie und die Materie hervorbringen könnten, benötigt man mindestens einen strukturellen Kontext, wie er mit den *Quantum Causal Histories* vorliegt. Der tatsächliche Zusammenhang zwischen Substrat und raumzeitlicher Makroebene könnte sich jedoch als noch deutlich komplizierter herausstellen, als es diese nahelegen. Dennoch liefern die *Quantum Causal Histories* zumindest einen nachvollziehbaren konzeptionellen Rahmen, der die möglichen Beziehungen zwi-schen einer raumzeitlichen Makroebene und einem prägeometrischem Substrat grundsätzlich verstehbar macht.

---

[379] Siehe Kap. 1.3.
[380] Siehe Kap. 3.3.

## Geometrogenese und Materiegenese

Eines der besonderen Kennzeichen der *Quantum Causal Histories* ist die intrinsisch in ihrer Konzeption festgeschriebene Kopplung von Materie- und Geometrogenese. Angesicht der Strategien, die von den etablierten Theorieansätzen im Bereich der Quantengravitation verfolgt werden, mag die konzeptionelle Unabdinglichkeit dieser Kopplung bei den *Quantum Causal Histories* vielleicht erst einmal überraschend erscheinen. Eigentlich ist sie dies aber ganz und gar nicht. Bei genauerer Betrachtung erweist sich vielmehr eine separate Behandlung von Materie und Raumzeit spätestens dann als konzeptionell völlig unplausibel, wenn es um die Erfassung einer fundamentalen prägeometrischen Dynamik- und Strukturebene geht. Sie lässt sich dann bestenfalls noch als das Erbe diesbezüglich unzulänglicher Traditionen im naturphilosophischen wie im physikalischen Denken ansehen,[381] die sich im Kontext der Bemühungen um eine Theorie der Quantengravitation als sehr hinderlich herausstellen könnten. Spätestens im Kontext prägeometrischer Szenarien – so diese beanspruchen, eine fundamentale physikalische Beschreibung zu leisten, sich also auf die elementarste Substratebene physikalischen Geschehens zu beziehen – ist es unumgänglich, eine gemeinsame Geometro- und Materiegenese anzunehmen. Ein Szenario mit einer prägeometrischen Basis, aus der die makroskopische Raumzeit hervorgeht, und einer weiteren, separaten Basis, welche die Emergenz der Materie und der nicht-gravitativen Wechselwirkungen gewährleisten soll,[382] wäre nicht nur sehr künstlich. Das Vorhandensein zweier Basen wäre vielmehr ein eindeutiges Zeichen dafür, dass es sich um keine fundamentale Beschreibung handelt; denn die Dynamik zwischen beiden Komponenten bliebe hierbei erst einmal unberücksichtigt. Sie müsste noch hinzukommen – im Rahmen einer Modellierung, die dann auf eine wiederum fundamentalere Ebene bezöge. Für prägeometrische Ansätze zu einer Theorie der Quantengravitation ist die Kopplung von Geometro- und Materiegenese also konzeptionell zwingend, wenn diese den Anspruch erheben, eine fundamentale Beschreibung des physikalischen Geschehens zu liefern. Für alle anderen Ansätze zu einer Theorie der Quantengravitation wäre eine solche Kopplung zumindest alles andere als unplausibel.[383]

---

[381] Schon die Allgemeine Relativitätstheorie weist mit ihrer dynamischen Kopplung von Materie (inklusive aller nicht-gravitativen Wechselwirkungsfelder) und geometrisierter Gravitation in einer Hinsicht über diese alte Tradition einer ontologischen wie physikalisch-konzeptionellen Trennung von Materie und Raumzeit hinaus.

[382] Man könnte hier vielleicht auf die Idee kommen, dass sich hinsichtlich der Argumente für eine notwendige Kopplung von Geometro- und Materiegenese ein Ausweg bietet, indem man annimmt, dass die Materie schon auf der Substratebene existiert, also nicht erst einer emergenten Ebene angehört. Ein solcher Ausweg existiert im Kontext der prägeometrischen Szenarien aber eigentlich nicht mehr. Wie sollte Materie, wie wir sie aus dem Kontext ihres raumzeitlichen Verhaltens kennen, auf einer nicht-raumzeitlichen Substratebene schon existieren? Wie sollte diese Materie, die sich nicht zuletzt gerade durch ihr raumzeitliches Verhalten auszeichnen, dort implementiert werden? Wie sollte sie mit irgendwelchen Grössen auf dieser Substratebene identifizierbar sein?

[383] In einigen Ansätzen, wie etwa in der *Loop Quantum Gravity*, zeichnet sich schon ab, dass sich bestimmte konzeptionelle Probleme wahrscheinlich nur unter Berücksichtigung von Materiefreiheitsgraden lösen lassen werden. Siehe die diesbezüglichen Hinweise in Kap. 4.4.

# 5.    Resümee

## *Die Sequenz der Strategien in der Theoriebildung*

Die entscheidende Motivation für die Entwicklung einer Theorie der Quantengravitation ergibt sich mit der (tatsächlichen oder auch nur augenscheinlichen) konzeptionellen Unverträglichkeit der Allgemeinen Relativitätstheorie mit der Quantenmechanik bzw. den Quantenfeldtheorien.[384] Für eine Theorieentwicklung, in deren Rahmen diese konzeptionelle Unverträglichkeit überwunden, aufgehoben oder als Missverständnis erwiesen werden soll, kommen grundsätzlich sehr unterschiedliche Strategien in Frage. In ihrer Gesamtheit stellen sie sich in mancher Hinsicht als eine Sequenz von Alternativen dar, die von naheliegenden, konservativen Strategien zu immer radikaleren reicht. Naheliegend sind die ersteren dabei vor allem vor dem Hintergrund bisheriger Erfahrungen mit der Theorienbildung in der Physik und der in ihnen zur Anwendung kommenden modelltheoretischen Vorgehensweisen; radikal sind die letzteren, indem sie die expliziten wie impliziten konzeptionellen Voraussetzungen der konservativeren Lösungswege ebenso wie die Angemessenheit der dort zum Einsatz kommenden modelltheoretischen Prozeduren zunehmend in Frage stellen.[385] Je mehr gravierende ungelöste Probleme sich für die naheliegenden, konservativen Strategien einstellen, desto plausibler erscheint es, sich mit den radikaleren Lösungsansätzen zu beschäftigen – die dann aber noch nicht notwendigerweise die konservativeren Strategien ersetzen, sondern vielmehr erst einmal als parallellaufende Alternativen verstanden werden können.

Die naheliegendste, konservativste Herangehensweise an eine Lösung des Problems der konzeptionellen Konflikte der Allgemeinen Relativitätstheorie mit der Quantenmechanik und den Quantenfeldtheorien besteht vermutlich darin, die Gravitation als fundamentale Wechselwirkung anzusehen und zu versuchen, das Gravitationsfeld zu quantisieren. In seiner einfachsten Form bedeutet dies, mit dem Gravitationsfeld so umzugehen wie mit dem klassischen elektromagnetischen Feld, dessen Quantisierung zur Quantenelektrodynamik führte. Das Ziel wäre demzufolge die Formulierung einer Quantenfeldtheorie der Gravitation. Infolge der mit der Allgemeinen Relativitätstheorie einhergehenden Geometrisierung der Gravitation entspricht eine Quantisierung des Gravitationsfeldes jedoch einer Quantisierung raumzeitlicher Geometrie. Und letztlich führt genau dies dazu, dass die grundlegend perturbative *Kovariante Quantisierung*, die gerade versucht, eine solche herkömmliche Quantenfeldtheorie der Gravitation zu formulieren, scheitert.[386] Sie scheitert an der völligen Unverträglichkeit des hier zur Anwendung kommenden modelltheoretischen Instrumentariums der Quantenfeldtheorien, die notwendigerweise einen Hintergrundraum mit fest vorgegebener Metrik voraussetzen, mit der sich aus der Allgemeinen Relativitätstheorie ergebenden Erfordernis einer hintergrundunabhängigen Beschreibung der Dynamik von Gravitation und Raumzeit. Will man die Allgemeine Relativitätstheorie quantisieren, kann man deren grundlegendste konzeptionelle Implikationen nicht einfach ausser acht lassen, nur weil sie mit dem gewählten modelltheoretischen Instrumentarium einfach nicht zu vereinbaren sind. Die *Kovariante Quantisierung* ist eine hintergrundabhängige Quantisierung einer hintergrundunabhängigen Theorie – eine Quantisierung der Raumzeit auf der Raumzeit. Noch dazu, oder gerade deswegen, ist sie nicht-renormierbar.

---

[384] Siehe Kap. 1.1.
[385] Siehe Kap. 1.3.
[386] Siehe Kap. 4.1.

Wenn sich aber das herkömmliche quantenfeldtheoretische Instrumentarium als ungeeignet für eine Quantisierung des Gravitationsfeldes bzw. der Raumzeit erweist, so ist die nächstliegende Alternative die, ein Instrumentarium zu entwickeln und zur Anwendung zu bringen, das die Formulierung einer hintergrundunabhängigen Quantentheorie des Gravitationsfeldes bzw. seiner raumzeitlichen Pendants möglich macht. Ein solcher Ansatz liegt mit der *Kanonischen Quantisierung* vor,[387] einer nicht-perturbativen, hintergrundunabhängigen Quantisierung der Allgemeinen Relativitätstheorie, ausgehend von ihrer Hamiltonschen Formulierung. Viele der formalen Probleme, die sich in der geometrodynamischen Variante dieses Ansatzes einstellen,[388] erweisen sich in der *Loop Quantum Gravity*,[389] einer konnektionsdynamischen Variante der *Kanonischen Quantisierung*, als besser lösbar. Hier ergeben sich dann nicht zuletzt Hinweise auf eine diskrete Struktur des Raumes (Spinnetze und S-Knoten) sowie auf ein basales Netzwerk relationaler bzw. korrelativer Bezüge zwischen den physikalisch relevanten bzw. als physikalisch real deutbaren Grössen. Dieses relationale Netzwerk setzt selbst keinen raumzeitlichen Hintergrund mehr voraus, sondern konstituiert vielmehr erst die Raumzeit. Die Indizien für ein solches relationales Netzwerk, das dem raumzeitlichen Geschehen zugrundeliegt, lassen sich, über die *Loop Quantum Gravity* hinausgehend, unter Umständen sogar hin auf eine abstrakte, vielleicht computationale Substruktur deuten, etwa ein prägeometrisches, nur relational bestimmtes Substrat, auf dessen Grundlage es überhaupt erst zum raumzeitlichen Geschehen kommt.

Das grundlegendste Problem, das die *Loop Quantum Gravity* (neben diversen formalen Ungeklärtheiten und Uneindeutigkeiten) weiterhin belastet, besteht darin, dass sich eine der essentiellen Anforderungen an eine Theorie der Quantengravitation bisher nicht einlösen liess: die Rekonstruktion der Allgemeinen Relativitätstheorie (oder immerhin einer phänomenologisch adäquaten makroskopischen Raumzeit) als klassischer Grenzfall, hier aus einer Quantentheorie, die gerade aus der Quantisierung dieser klassischen Theorie resultiert, die sich dann nicht wieder ableiten lässt. – Dies legt wiederum nahe, sich auch mit Alternativen zu beschäftigen.

Eine solche Alternative, die weiterhin an der Auffassung festhält, dass die Gravitation eine fundamentale Wechselwirkung ist, dem Gravitationsfeld also Quanteneigenschaften zuzusprechen wären, die aber gerade nicht versucht, die Beschreibung des quantisierten Gravitationsfeldes über eine direkte Quantisierung der Allgemeinen Relativitätstheorie zu gewinnen, sondern vielmehr über die einer anderen klassischen Dynamik, ist der *Stringansatz*.[390] Dieser verbindet die Erfassung des quantisierten Gravitationsfeldes – wiederum in Form einer (hier allerdings aus der Stringdynamik abgeleiteten) perturbativen Gravitonenphysik, wie sie schon in der *Kovarianten Quantisierung* vorlag – mit einer nomologischen Vereinigung aller Wechselwirkungen. Im Gegensatz zur *Kovarianten Quantisierung* entgeht der *Stringansatz* wohl in der Tat dem Problem der Nichtrenormierbarkeit – und dies gerade infolge der formalen nomologischen Vereinigung aller Wechselwirkungen in der Stringdynamik, die offenbar zu einer wechselseitigen Aufhebung der Divergenzen führt. Es bleibt aber bei der modelltheoretisch bedingten Hintergrundabhängigkeit, die schon die *Kovariante Quantisierung* auszeichnete; sie ist auch im Kontext des *Stringansatzes* unvereinbar mit den elementaren konzeptionellen Implikationen der Allgemeinen Relativitätstheorie, die auch hier als klassischer Grenzfall reproduziert werden muss (und in formaler Hinsicht auch wird). Eine hinter-grundunabhängige, nicht-perturbative Variante des Stringansatzes, deren Notwendigkeit von seinen Proponenten weitgehend eingesehen wird, liess sich trotz intensiver Bemühungen bisher nicht formulieren. Hinzu kommen eine ganze Reihe weiterer konzeptioneller Probleme, die zum grössten Teil auf die nur intern, mathematisch und modelltheoretisch motivierten Implikationen des Stringansatzes (höhere Dimensionalität der Raumzeit, Supersymmetrie etc.) zurückzuführen sind und diesen als nahezu vollständig durch interne Erfordernisse bestimmtes mathematisches Konstrukt erweisen, für das sich immer noch, auch nach mehreren Jahrzehnten Entwicklung, kein physikalisch motivierbares Grundprinzip formulieren lässt. Insbesondere war es bisher trotz der formalen nomologischen Vereinigung der Wechselwirkungen nicht möglich, das Standardmodell der Quantenfeldtheorien zu reproduzieren. Und die Aussicht auf eine konkrete Vorhersagekraft des Stringansatzes versinkt zusehends in der überbordenden Vielfalt der niederenergetischen Szenarien innerhalb der *String-Landscape*.

---

[387] Siehe Kap. 4.3. und 4.4.
[388] Siehe Kap. 4.3.
[389] Siehe Kap. 4.4.
[390] Siehe Kap. 4.2.

Die gravierenden Probleme, die sich bisher für alle die Ansätze eingestellt haben, die von der Gravitation als einer fundamentalen Wechselwirkung ausgehen, legen in ihrer Gesamtheit schliesslich nahe, auch die Option in Betracht zu ziehen, dass diese Annahme vielleicht einfach falsch ist.[391] Es könnte sich bei der Gravitation durchaus um ein residuales bzw. induziertes und vielleicht sogar intrinsisch klassisches Phänomen handeln, zu dem es auf der Grundlage einer fundamentaleren Dynamik kommt, für die gravitative Freiheitsgrade keine Rolle spielen. Eine intrinsisch klassische Gravitation würde, wenn sie ein emergentes Phänomen sein sollte, insbesondere auch zu keinem Konflikt mit den Argumenten gegen semi-klassische Theorien der Gravitation führen; auf der fundamentalen Ebene eines Quantensubstrats käme sie gar nicht vor. Diverse thermodynamische, festkörperphysikalische, hydrodynamische und computationale Ansätze, die aber allesamt noch nicht den Status einer ausformulierten Theorie erreicht haben, versuchen diese Möglichkeit auszuloten.[392] Die Allgemeine Relativitätstheorie wäre dann nichts mehr als eine (effektive) Theorie, die gerade dieses emergente, intrinsisch klassische Phänomen Gravitation beschreibt. Ihre Quantisierung würde unter diesen Umständen keineswegs zu einer angemessenen Beschreibung des der Gravitation zugrundeliegenden Substrats führen.

Sollte die Allgemeine Relativitätstheorie dennoch hinsichtlich der Geometrisierbarkeit der Gravitation in physikalischer Hinsicht richtig liegen und einen essentiellen Sachverhalt erfassen, so wäre vielleicht auch die Raumzeit als emergentes, intrinsisch klassisches Phänomen anzusehen, das auf der Substratebene nicht vorliegt. Man hätte es also mit einem originär prägeometrischen Substrat zu tun, das zwar die Raumzeit hervorbringt, aber selbst über keine raumzeitlichen Freiheitsgrade verfügt. Es ginge dann in einer Theorie der 'Quantengravitation' nicht einfach um Quantenkorrekturen zu einer klassischen Raumzeitauffassung (im naiven Bild: Quantenfluktuationen der Metrik, quantenmechanische Unschärfen der Raumzeit, Superpositionen von Raumzeiten oder gar eine Form von Gravitonenphysik –in einer subtileren Variante: eine diskrete raumzeitliche Struktur, wie sie etwa die Spinnetze der *Loop Quantum Gravity* nachzeichnen), sondern vielmehr um die prägeometrische Substratstruktur und -dynamik selbst.[393] Die konzeptionellen Voraussetzungen, die sich für eine Theorie der Quantengravitation vielleicht aus den etablierten Theorien, insbesondere aus der Allgemeinen Relativitätstheorie, ableiten lassen, erweisen sich unter diesen Bedingungen nur bedingt als relevant und verlässlich – zumindest soweit sie über die nun auf den emergenten klassischen Kontext beschränkte allgemein-relativistische Idee einer Geometrisierbarkeit der Gravitation hinausgehen. Die unterschiedlichen Indizien, die darauf hindeuten, dass die Allgemeine Relativitätstheorie und die Quantenmechanik offenbar viel zu verschieden sind, um eine direkte Amalgamierung zuzulassen, lassen sich durchaus als Erscheinungsformen eines fundamentaleren Problems sehen, welches sich vielleicht erst im Rahmen prägeometrischer Ansätze zu einer Theorie der 'Quantengravitation' lösen lässt. Hierzu gibt es inzwischen eine ganze Reihe von Theorie- und Denkansätzen: *Causal Sets*[394], *Holographische Schirme*[395], Lloyds *Computational Universe*[396], prägeometrische *Quantum Causal Histories*[397].

---

[391] Siehe Kap. 3.3. und 4.6.
[392] Siehe Kap. 3.3.
[393] Siehe Kap. 3.3. und 4.6.

Erst wenn all dies zu keinem Ergebnis führen sollte, kommen dann noch radikalere Ansätze ins Spiel, die etwa die universelle Gültigkeit der Quantenmechanik oder die grundsätzliche Nomologizität der Natur in Frage stellen.[398] – Eine Auffassung jedoch, derzufolge physikalische Theorien ausschliesslich als Instrumentarien zur Erfassung der Phänomene und, entsprechend, die Bemühungen der Physik um die Erschliessung fundamentaler Strukturen als unangemessene Extrapolation nomologischer Ambitionen anzusehen sind – eine Extrapolation, die dem Pluralismus eigenständiger Gegenstandsbereiche physikalischer Theorien nie gerecht werden kann –,[399] hätte die prinzipielle Unmöglichkeit der Etablierung einer physikalisch ernstzunehmenden, realistisch interpretierbaren Theorie der Quantengravitation zur Folge. Eine solche Auffassung wäre erst in Erwägung zu ziehen, wenn alle Alternativen für die Formulierung einer solchen Theorie als definitiv gescheitert angesehen werden müssten: also nie!

Die weiterhin bestehende Grundfrage ist also vor allem: Wie weit müssen die Ansätze zur Quantengravitation in konzeptioneller und auch in modelltheoretischer Hinsicht gehen, um erfolgreich zu sein? Angesichts der Probleme, auf die alle bisherigen Ansätze zur Quantengravitation gestossen sind, vor allem auch die konzeptionell naheliegendsten und konservativsten, ist es kaum empfehlenswert ausschliesslich auf einen spezifischen Ansatz und dessen erhofften Erfolg zu setzen. Vielmehr empfiehlt sich auf der Grundlage dieser Problemlage eine pluralistisch ausgerichtete Strategie der Theorieentwicklung. Alle konzeptionell kohärenten Möglichkeiten sollten erst einmal ernstgenommen werden.

<div align="center">*</div>

Trotz der Offenheit der gegenwärtigen Situation im Kontext der Bemühungen um eine Theorie der Quantengravitation und trotz der Tatsache, dass es bisher keine eindeutigen empirischen Daten gibt, die über die Vorgängertheorien hinausweisen, so dass sich die gesamte Entwicklung weiterhin ausschliesslich im Bereich konzeptioneller Spekulationen bewegt, sollen nun im folgenden dennoch einige hochgradig tentative Schlüsse aus der bisherigen Entwicklung und ihrem heutigen Stand gezogen werden. Diese stützen sich vor allem auf die schon ausgeloteten konzeptionellen Unverträglichkeiten, Unvereinbarkeiten und Sackgassen einerseits und auf die sich abzeichnenden konzeptionellen Konvergenzen andererseits, die in ihrer Gesamtheit inzwischen den Konfigurationsraum der konzeptionellen Möglichkeiten immerhin schon erheblich einschränken und zudem bestimmte Lösungen wesentlich plausibler machen als andere:

---

[394] Siehe Kap. 4.6.
[395] Siehe Kap. 3.3.
[396] Siehe Kap. 3.3.
[397] Siehe Kap. 4.6. Siehe auch weiter unten den Abschnitt "Emergente vs. fundamentale Raumzeit".
[398] Siehe Kap. 1.3.
[399] Siehe Kap. 1.3.

## *Höhere Dimensionalität der Raumzeit ?*

Der einzige aktuelle Ansatz zu einer Theorie der Quantengravitation, der zwingend – infolge interner, konzeptioneller, modelltheoretischer Anforderungen – von einer höheren Dimensionalität der Raumzeit ausgeht, ist der *Stringansatz*.[400] Als physikalische Plausibilisierung für die konzeptionellen Anforderungen, die hier eine höhere Dimensionalität der Raumzeit erforderlich machen, wird manchmal die Idee angeführt, dass die Einbindung aller Wechselwirkungen im Sinne einer nomologischen Vereinheitlichung gerade eine solche höhere Dimensionalität erforderlich macht – so wie schon im Kaluza-Klein-Ansatz zur Einbindung des Elektromagnetismus eine weitere räumliche Dimension erforderlich war, mittels derer gerade die elektromagnetischen Freiheitsgrade erfasst wurden.[401] Hier klingt dann letztlich wieder so etwas wie eine Geometrisierung aller Wechselwirkungen an, wenn auch nicht gleich im radikalen monistischen Sinne Wheelers.

> *"It is of course possible that, as has been proposed for decades, the other interactions also come from the dynamics of spacetime geometry, such as the curvature of extra dimensions."* (Smolin (2003) 8)

Die Annahme einer höheren raumzeitlichen Dimensionalität steht und fällt jedoch mit dem konzeptionellen, aber insbesondere auch mit dem empirisch nachweislichen Erfolg des *Stringansatzes* – die sich beide auch nach Jahrzehnten noch nicht eingestellt haben und mit der *String-Landscape* immer unwahrscheinlicher werden. Damit erscheint die Annahme einer höheren Dimensionalität der Raumzeit bis auf Weiteres als sehr wenig plausibel.

Hinsichtlich der Erfassung nicht-gravitativer Freiheitsgrade bieten sich zudem – als Alternative zur Erweiterung der raumzeitlichen Dimensionalität – topologische Strukturen und Eigenschaften an, die entweder der Raumzeit oder aber Entitäten, die überhaupt erst die Raumzeit hervorbringen, zugesprochen werden. Hier wird dann die Idee einer (mehr oder weniger umfassenden) Geometrisierung aller Wechselwirkungen (und vielleicht auch der Materie) durch die ihrer (eventuell prägeometrischen) Topologisierung abgelöst.[402]

## *Diskrete vs. kontinuierliche Raumzeit*

Das entscheidendste Argument gegen die Annahme kontinuierlicher Freiheitsgrade auf der fundamentalsten Ebene des physikalischen Geschehens ergibt sich im Kontext der Thermodynamik schwarzer Löcher mit der auf der *Bekenstein-Hawking-Entropie* beruhenden *holographischen* bzw. *kovarianten Entropiegrenze*.[403] Dieses Argument lässt sich einerseits hin auf eine letztendlich diskret strukturierte Raumzeit deuten – für den Fall, dass die Raumzeit eine fundamentale Entität und die Gravitation eine fundamentale Wechselwirkung sein sollten, denen im Rahmen einer Theorie der Quantengravitation Quanteneigenschaften zuzuschreiben wären, so die Quantenmechanik universell gültig sein sollte und die Allgemeine Relativitätstheorie mit der Geometrisierbarkeit der Gravitation recht behalten sollte. Gerade in den Quanteneigenschaften der Raumzeit könnte ihre letztendlich diskrete Struktur zum Ausdruck kommen. – Andererseits ist das Argument für eine diskrete fundamentale Strukturebene aber ebenso mit einer emergenten Raumzeit und einer emergenten Gravitation vereinbar. Dann wäre es womöglich ein diskret strukturiertes prä-raumzeitliches Substrat, aus dem Raumzeit und Gravitation hervorgingen.

---

[400] Siehe Kap. 4.2.
[401] Siehe Kap. 1.3.
[402] Siehe Kap. 4.6.
[403] Siehe Kap. 3.1.

Sogar in einigen der Ansätze zu einer Theorie der Quantengravitation, die – zumindest soweit es ihre modelltheoretischen Grundlagen betrifft – das Kontinuum voraussetzen, finden sich (in ihrer Ableitung unabhängig von der Thermodynamik schwarzer Löcher) Hinweise auf eine diskrete Substratstruktur. So führt etwa die *Loop Quantum Gravity*, die eine differenzierbare Mannigfaltigkeit voraussetzt (nicht zuletzt, um die Diffeomorphismusinvarianz der Allgemeinen Relativitätstheorie implementieren zu können), nach der Quantisierung des Hamilton-Systems (schon auf der kinematischen Beschreibungsebene) zu einer diskreten Spinnetzstruktur.[404] Im *Stringansatz*, der mit den in modelltheoretischer Hinsicht nur wenig erweiterten Methoden der Quantenfeldtheorien arbeitet und insbesondere ebenso wie diese ein raumzeitliches Kontinuum voraussetzt, kommt es zu eindeutigen Anzeichen für eine dynamisch begründete minimale Länge.[405] Beide Ansätze reproduzieren zudem unter bestimmten Bedingungen die *Bekenstein-Hawking-Entropie*. – Weitere, wenn auch indirektere Indizien für die Unangemessenheit der Kontinuumsannahme lassen sich womöglich aus der Vorhersage von Singularitäten im Kontext der Allgemeinen Relativitätstheorie, dem Auftreten von Divergenzen in den Quantenfeldtheorien und der Nichtrenormierbarkeit der *Kovarianten Quantisierung* der Allgemeinen Relativitätstheorie herauslesen.

Das Kontinuum ist in letzter Instanz sehr wahrscheinlich nichts anderes als eine modelltheoretische Fiktion. Und das Problempotential dieser Fiktion, das sich schon im Rahmen der etablierten Theorien andeutet, wird im Bereich der Quantengravitation wohl endgültig virulent. Will man hier das ständige Auftreten modelltheoretischer Artefakte vermeiden, die sehr schnell unkontrollierbar werden können, so heisst dies letztlich, das gesamte modelltheoretische Instrumentarium der Kontinuumsphysik ausser Kraft zu setzen. Aber vielleicht gibt es hierzu, wenn alle Relevanzen und wechselseitigen konzeptionellen Implikationen abgeklärt sind, ohnehin keine Alternative mehr. Vielleicht muss spätestens dann, wenn es um die fundamentale Ebene des physikalischen Geschehens geht, eine konzeptionell widerspruchsfreie Beschreibung notwendigerweise ohne die Fiktion des Kontinuums auskommen.

## *Emergente vs. fundamentale Raumzeit*

Die Frage, ob die Raumzeit konstitutiver Teil des physikalischen Geschehens auf allen seinen Ebenen, auch der fundamentalsten, ist – was noch nicht unbedingt heisst, dass die Raumzeit eine Substanz oder eine unabhängig existierende Entität ist – oder ob es sich bei ihr vielmehr um ein emergentes Phänomen handelt, das auf der fundamentalsten Ebene des physikalischen Geschehens noch keine Rolle spielt, ist zum heutigen Stand der Dinge noch grundlegend unentschieden.

Die Allgemeine Relativitätstheorie selbst, wie auch alle Ansätze, die deren konzeptionelle Unverträglichkeit mit der Quantenmechanik durch ihre direkte Quantisierung zu beheben versuchen, gehen davon aus, dass die Raumzeit schon auf der fundamentalsten Ebene des physikalischen Geschehens eine Rolle spielt, wenngleich nicht unbedingt als eigenständige substantielle Entität, son-dern vielmehr – wie es spätestens im Rahmen der *Loop Quantum Gravity* deutlich wird – als relational bestimmtes raumzeitliches Gefüge. Im Kontext der Quantengravitation geht es dann nicht zuletzt darum zu klären, welche Quanteneigenschaften einer solchen fundamentalen Raumzeit zuzuschreiben sind. In den naiveren, quantenfeldtheoretischen Ansätzen kann diese Zuschreibung etwa in einer Form einer Gravitonenphysik erfolgen; in den subtileren Ansätzen kann sie etwa in einer aus der Quantisierung resultierenden diskreten Strukturierung ihren Ausdruck finden.

---

[404] Siehe Kap. 4.4.
[405] Siehe Kap. 4.2.

Es sind vor allem die gravierenden konzeptionellen und formalen Probleme, in die diese direkten Quantisierungsansätze geraten, die als Argument dafür gesehen werden können, auch die Alternative in Betracht zu ziehen, dass die Raumzeit vielleicht auf der fundamentalsten Ebene des physikalischen Geschehens keine Rolle spielt, sondern vielmehr als emergentes Phänomen von einem Substrat hervorgebracht wird, das selbst über keine raumzeitlichen Freiheitsgrade verfügt.[406] Die Emergenz der Raumzeit wäre dann das Ergebnis der Dynamik basalerer, nicht-raumzeitlicher Entitäten bzw. Prozesse und der relationalen Beziehungen, in denen diese zueinander stehen. Die Raumzeit wäre sekundär gegenüber diesem prä-raumzeitlichen Substrat. Und es ginge in einer Theorie der 'Quantengravitation' dann eben nicht mehr einfach um Quantenkorrekturen zu einer klassischen Raumzeitauffassung, sondern vielmehr um die Beschreibung dieses prägeometrischen Substrats. Die konzeptionellen Voraussetzungen, die sich für eine Theorie der Quantengravitation vielleicht aus den etablierten Theorien, insbesondere aus der Allgemeinen Relativitätstheorie, ableiten liessen, würden sich unter diesen Bedingungen als wenig verlässlich erweisen.

Von den diversen Ansätzen, die versuchen, die Raumzeit als emergentes Phänomen verstehbar werden zu lassen und zu diesem Zweck von einer direkten, möglichst einfachen und möglichst physikalisch plausiblen Substratkonstruktion ihren Ausgang nehmen, lassen sich vor allem die prägeometrischen *Quantum Causal Histories* als in konzeptioneller Hinsicht paradigmatisch ansehen.[407] Auch wenn sie sehr wahrscheinlich noch kein realistisches Bild nachzeichnen, machen sie immerhin plausibel, wie man sich überhaupt die Hervorbringung von Raumzeit und Gravitation aus einem originär prägeometrischen Substrat vorstellen könnte, das in diesem Fall aus nichts mehr als einer relationalen Struktur von Quanteninformationsflüssen besteht. Und sie zeigen vor allem, dass die Hervorbringung der Raumzeit aus einem prägeometrischen Substrat sehr wahrscheinlich eine Kopplung von Materie- und Geometrogenese impliziert.

Ein weiteres Argument für eine emergente Raumzeit, das noch dazu den Vorteil hat, sich nicht auf irgendeinen spezifischen Theorieansatz im Kontext der Quantengravitation berufen zu müssen, resultiert aus dem *Holographischen Prinzip*.

---

[406] Insbesondere die im Kontext der direkten Quantisierungsansätze naheliegende Annahme von Quantenfluktuationen der Metrik – und damit der raumzeitlichen Kausalstruktur – erweist sich als vollständig unverträglich mit dem Instrumentarium der Quantenmechanik. Siehe Kap. 2.2. – Weitere indirekte Hinweise für eine emergente Raumzeit und eine basalere prä-raumzeitliche Dynamik lassen sich aus dem Kontext der *Loop Quantum Gravity* heraus motivieren:
> "If [Loop Quantum Gravity] leads to what comes to be regarded as the correct quantum theory of gravity, then classical *general relativistic spacetime can no longer be taken as a fundamental entity because none of the states of the physical Hilbert space of [Loop Quantum Gravity] describes a classical general relativistic spacetime at the sub-Planck scale, and some of the states fail to describe a classical general relativistic spacetime at the macroscopic scale."* (Earman (2006a) 20)

[407] Siehe Kap. 4.6.

## Holographische Determinierung ?

Das *Holographische Prinzip*[408] lässt sich auf der Grundlage der Oberflächenproportionalität der *Bekenstein-Hawking-Entropie* und der sich mit ihr ergebenden *holographischen* bzw. *kovarianten Entropiegrenze* motivieren.[409] Gemäss der holographischen Grenze ist der maximale Informationsgehalt, der einem räumlichen Volumen zugeschrieben werden kann, proportional zur Fläche, die dieses Volumen begrenzt. Die maximale, finite Zahl von echten, physikalisch wirksamen Freiheitsgraden, auf deren Grundlage sich das Geschehen in einem Raumvolumen vollständig erfassen lässt, erschöpft sich offensichtlich gerade in der Zahl der Freiheitsgrade, die sich – wie ihre numerische Proportionalität deutlich macht – grundsätzlich der einschliessenden Oberfläche zuschreiben liessen.

Was, wenn dies nicht erst für die holographische Grenze gilt, also für das erreichbare Maximum an Information bzw. an echten, physikalisch wirksamen Freiheitsgraden: wenn sich also die Freiheitsgrade, die einem raumzeitlichen Volumen zugeschrieben werden, letztlich immer schon – vor allem auf der fundamentalsten (raumzeitlichen) Ebene des Geschehens – verlustfrei der dieses Volumen begrenzenden Oberfläche zuschreiben liessen, wenn sie also Freiheitsgrade wären, die eigentlich dem dynamischen Geschehen auf der Oberfläche zukommen? Die holographische Grenze legt in dieser ihrer Verallgemeinerung als *Holographisches Prinzip* immerhin nahe anzunehmen, dass spätestens auf der fundamentalsten Ebene des physikalischen Geschehens Bedingungen vorliegen könnten, die dazu führen, dass die Freiheitsgrade, die sich einem räumlichen Volumen zuschreiben lassen, grundsätzlich und vollständig (und nicht nur im Hinblick auf die maximal einschliessbare Information) durch die Freiheitsgrade auf der einschliessenden Fläche determiniert sind bzw. sich in diesen erschöpfen, zumindest soweit es echte, physikalisch wirksame Freiheitsgrade betrifft. – Und diese Verallgemeinerung der Implikationen der holographischen Grenze in Form des *Holographischen Prinzips* könnte sich letztlich als unabhängig von der Frage herausstellen, ob die Raumzeit fundamentalen Status besitzt oder emergent ist. Träfe sie zu, so gäbe es keine das gesamte Raumvolumen betreffende Information, die über die auf der Oberfläche kodierte finite Information hinausginge, unabhängig davon, ob diese schon der fundamentalsten Ebene des physikalischen Geschehens angehören. Die Gesamtheit der physikalisch wirksamen Freiheitsgrade, die sich einem räumlichen Volumen zuschreiben lassen, erschöpfte sich in den auf seiner Oberfläche vorliegenden Freiheitsgraden. Diese entsprächen den physikalisch wirksamen Freiheitsgraden des raumzeitlichen Geschehens.

Das *Holographische Prinzip* lässt sich mit seiner Hypothese der vollständigen Determinierung der raumzeitlichen Gegebenheiten durch die dynamischen Strukturen auf der Oberfläche der entsprechenden Raumzeit (bzw. des räumlichen Geschehens durch das Geschehen auf der Oberfläche des entsprechenden Raumes) geradezu als Vorstufe zu einer emergenten Auffassung hinsichtlich der Raumzeit interpretieren. Es legt mit seiner These der vollständigen Kodierung und Determinierung der Volumeninformation durch die Flächeninformation darüber hinaus ein computationales, informationstheoretisches Substrat nahe: etwa einen Quantencomputer, der infolge einer geschickten Informationsdekompression ein raumzeitliches Geschehen vorgaukelt, das mehr zu sein scheint, als es ist – das mehr an Information bzw. an physikalischen Freiheitsgraden zu enthalten scheint, als ihm tatsächlich zugrundeliegen.

---

[408] Siehe Kap. 3.2.
[409] Siehe Kap. 3.1.

Sollte das *Holographische Prinzip* – als durchaus plausible Extrapolation der holographischen Informationsgrenze und ihrer Oberflächenproportionalität – für die Gegebenheiten auf der fundamentalsten raumzeitlichen Ebene des physikalischen Geschehens tatsächlich zutreffen, so hiesse dies, dass der Informationsgehalt, der dem raumzeitlichen Geschehen anscheinend zukommt, mit sehr effizienten Mitteln kodiert ist: mit Mitteln, die es insbesondere nicht erforderlich machen, Raumzeitpunkten (oder diskreten raumzeitlichen Vertices oder Zellen) jeweils singulär Informationen bzw. physikalische Freiheitsgrade direkt zuzuschreiben.

Vielleicht – und dies wäre dann eine Extrapolation, die noch über das *Holographische Prinzip* hinausgeht – kommt man letztlich sogar mit noch deutlich weniger aus als einer diskretisierten Fläche, auf der die basale Information kodiert ist; vielleicht genügt ein vollständig nicht-raumzeitliches, prägeometrisches Substrat. Die Emergenz der Raumzeit, die eben auch ohne ein weiteres umfassender sein könnte, als dies im *Holographischen Prinzip* deutlich wird, käme dann – wenn man im informationstheoretischen Bild bleiben möchte – so etwas wie einer Dekompression von Information gleich, die auf einer vollständig nicht-raumzeitlichen Substrateebene höchst effizient kodiert ist. Vielleicht wird also eben nicht nur *eine* räumliche Dimension mehr vorgegaukelt als (soweit es die physikalisch wirksamen Freiheitsgrade betrifft) tatsächlich vorhanden ist, sondern vielmehr die Raumzeit und das scheinbare raumzeitliche Geschehen überhaupt. Die natürliche Basis für eine solche virtuelle Raumzeit wäre ein computationales (Quanten-)Substrat.

Erstaunlicherweise ergibt sich sogar im *Stringansatz* mit der *AdS/CFT-Dualität* eine Instantiierung des *Holographischen Prinzips*.[410] Und eine sich über die *AdS/CFT-Dualität* hinausgehend abzeichnende Universalität[411] von grundlegend verschiedenen Gravitations- und Eichtheorien – Theorien, die unterschiedliche raumzeitliche Dimensionalitäten voraussetzen, unterschiedliche raumzeitliche und interne Symmetrien aufweisen, von gänzlich verschiedenen basalen Freiheitsgraden ausgehen, aber dennoch die gleiche niederenergetische Phänomenologie reproduzieren – liefert schliesslich weitere Argumente für eine emergente Raumzeit. Spätestens die offensichtliche Austauschbarkeit von raumzeitlichen und internen Freiheitsgraden im Rahmen der entsprechenden phänomenologischen Äquivalenzen lässt sich dann als Argument für ein rein computationales Substrat deuten.

Die, soweit es unsere tiefsitzenden raumzeitlichen Intuitionen betrifft, wohl entscheidende Frage, wie man sich überhaupt die Hervorbringung einer Raumzeit aus reinen (Quanten-) Informationsflüssen, vorstellen kann, lässt sich schliesslich mittels der Idee der *Holographischen Schirme* beantworten – einer konzeptionellen Umkehrung des *Holographischen Prinzips* hinsichtlich der Beziehung zwischen Fläche und Information.[412] Diese Idee verdeutlicht, wie sich geometrische Grössen, erst einmal eben Flächen, grundsätzlich aus Informationsflüssen gewinnen bzw. auf ihrer Grundlage definieren lassen und wie sich insofern geometrische Verhältnisse und Grössen innerhalb einer Dynamik ergeben können, die sich ausschliesslich auf der Grundlage abstrakter relationaler Bezüge vollzieht, die also letztlich nur informationstheoretisch interpretiert werden kann. Ohne die Idee der *Holographischen Schirme* bliebe das Postulat der Möglichkeit einer prägeometrischen Erzeugung von raumzeitlicher Geometrie – der Möglichkeit der Erzeugung von Geometrie aus einem Substrat, für das diese raumzeitliche Geometrie noch nicht vorliegt – vielleicht erst einmal im wesentlichen unklar.

---

[410] Siehe Kap. 4.2.
[411] Siehe Kap. 3.2.
[412] Siehe Kap. 3.3.

## *Raumzeitliche Lokalität ?*

Das *Holographische Prinzip* – die Annahme, dass die echten, physikalisch wirksamen Freiheits­grade, die einem räumlichen Volumen zugeschrieben werden können, sich in den Freiheitsgraden auf der umgebenden Oberfläche erschöpfen – impliziert eine entschiedene Durchbrechung der An­nahme raumzeitlicher Lokalität. Aber letztlich muss man nicht erst das *Holographische Prinzip* bemühen, um die Gültigkeit der Lokalitätsannahme hinsichtlich des raumzeitlichen Geschehens in Frage zu stellen. Sie wird ohnehin in allen hintergrundunabhängigen Theorien sehr fragwürdig. Diese sind für ihre ausgeprägte Virulenz hinsichtlich der Lokalitäts- bzw. Nichtlokalitätsproblema­tik bekannt:[413]

> *"Locality is a tricky issue in background independent quantum theories of gravity because there is no background metric with which to measure distances or intervals. And it is non-trivial to construct dif­feomorphism invariant observables that measure local properties of fields."* (Bilson-Thompson / Mar­kopoulou / Smolin (2006) 9)

Schon in der Allgemeinen Relativitätstheorie gibt es keine lokalen (oder auch nur quasi-lokalen) diffeomorphismusinvarianten und damit beobachtbaren Grössen.[414] Damit gibt es im Rahmen ihrer eichinvarianten Interpretation keine physikalisch ernstzunehmenden und als real anzusehenden lo­kalen (oder auch nur quasi-lokalen) Grössen.

> *"[...] in [general relativity] there are no local or even quasi-local observables."* (Earman (2006a) 13)

Die Ursache dafür ist gerade die aktive Diffeomorphismusinvarianz der Allgemeinen Relativitäts­theorie.

> *"[...] a diffeomorphism maps one spacetime point to another, and therefore one obvious way of con­structing a diffeomorphism-invariant object is to take a scalar function of spacetime fields and inte­grate it over the whole of spacetime, which gives something that is very* non-*local. The idea that 'phy­sical observables' are naturally non-local is an important ingredient in some approaches to quantum gravity."* (Butterfield / Isham (2001) 61)

Die Implikationen der aktiven Diffeomorphismusinvarianz, insbesondere die Nichtlokalität der Ob­servablen – und damit im Rahmen einer eichinvarianten Deutung: die Nichtlokalität aller ernstzu­nehmenden physikalischen Grössen –, übertragen sich auf jeden Ansatz zu einer Theorie der Quan­tengravitation, der die substantiell interpretierte allgemeine Kovarianz und entsprechend die Hinter­grundunabhängigkeit der Allgemeinen Relativitätstheorie als konstitutives Element betrachtet. Dies betrifft insbesondere die Ansätze zu einer *Kanonischen Quantisierung* der Allgemeinen Relativi­tätstheorie.

Völlig unabhängig von dieser und in völlig anderer Ausprägung zeigen aber auch schon die Quan­tenmechanik und insbesondere die quantenmechanischen Nichtseparierbarkeiten, dass die Idee einer durchgängigen raumzeitlichen Lokalität des physikalischen Geschehens kaum aufrechtzuerhalten sein wird.[415] – Berücksichtigt man die verschiedenen Einschränkungen, Relativierungen und Transzendierungen, welche die Annahme einer durchgängigen lokalen, raumzeitlichen Geordnetheit des Weltgeschehens aus den verschiedensten theoretischen Kontexten erfährt, so ist es eher schon erstaunlich, wie weit man in der Physik mit Beschreibungen kommt, die von der Gültigkeit der Lokalitätsannahme ausgehen, wie weit also die Illusion einer durchgängigen lokalen Geordnetheit des Weltgeschehens trägt.

---

[413] Siehe Kap. 2.1. und 4.4.
[414] Siehe Kap. 2.1.

Letztlich lassen sich die aus verschiedenen, insbesondere auch empirisch bestätigten Theoriekontexten stammenden Anzeichen für eine Durchbrechung der Lokalitätsannahme hinsichtlich des raumzeitlichen Geschehens mit nur wenig Anstrengung als Indizien für die Emergenz dieses raumzeitlichen Geschehens lesen. Wollte man diese Lesart konkretisieren, so hiesse dies: Die Illusion einer raumzeitlichen Geordnetheit und einer im raumzeitlichen Geschehen ausschliesslich lokal wirksamen Kausalität lässt sich wohl nur deshalb bis zu einem bestimmten Grade aufrechterhalten, weil ihr eine andere Art von Strukturiertheit zugrundeliegt, auf deren Grundlage die nur unvollkommene und nur unter einschränkenden Bedingungen gültige raumzeitliche Geordnetheit zustandekommt. Die makroskopische raumzeitliche Geordnetheit, insbesondere die makroskopische raumzeitliche Lokalität und Kausalität, würde einem solchen Szenario zufolge von der Dynamik auf einer Strukturebene hervorgebracht, deren kausale Ordnung und deren Lokalität – so diese vorliegen sollten – nicht mit ihren makroskopischen Pendants zu verwechseln wären. Sollte die Substratebene eine kausale Ordnung aufweisen und sollte auf ihr ein Lokalitätsprinzip gelten, so läge zwischen diesen und ihren makroskopischen Pendants mindestens eine dynamische Entkopplung. Die makroskopische Lokalität gilt dann – in Übereinstimmung mit den tatsächlich feststellbaren Gegebenheiten – bestenfalls in einer Näherung, nicht aber strikt und umfassend. Für das Substrat, auf dessen Grundlage es dazu kommen könnte, würde man diese Striktheit jedoch erwarten. Eine Beschreibungsebene, welche eine weitgehende Aufrechterhaltung der Lokalitätsannahme zuliesse, die dann jedoch in ganz spezifischer Weise durch nichtlokale Aspekte oder Komponenten durchbrochen würde, erscheint als Substrat wenig plausibel. Solche kuriosen hybriden Eigenschaften sind vielmehr gerade für eine emergente Ebene zu erwarten und lassen sich mithin als Indiz für die Emergenz der entsprechenden Beschreibungsebene verstehen.

## *Relationale vs. substantielle Raumzeit*

Für den Fall einer emergenten Raumzeit werden die Begrifflichkeiten des Relationalismus und des Substantialismus, so sie im Sinne ihrer traditionellen Begriffsfassung weiterhin auf die Raumzeit angewandt werden sollen, weitestgehend hinfällig. Die Frage der Substantialität liesse sich hier in ernsthafter Weise bestenfalls noch in Bezug auf das Substrat stellen. Und insbesondere für ein rein computationales Substrat, etwa ein relationales Gefüge von Quanteninformationsflüssen, müsste sie ziemlich sicher verneint werden.

Etwas diffiziler sieht es immer noch für den Fall einer fundamentalen Raumzeit aus. Schon im Kontext der Allgemeinen Relativitätstheorie spricht vieles für eine relationalistische Sichtweise hinsichtlich der Raumzeit.[416] Die Argumente für den Relationalismus beruhen hier vor allem auf der aktiven Diffeomorphismusinvarianz, die gerade eine eichinvariante Interpretation nahelegt.[417] Letz-tere lässt sich am ehesten dem Relationalismus subsumieren und wird von manchen Autoren sogar als konstitutives Element des Relationalismus definiert.

---

[415] Siehe Kap. 2.2.
[416] Siehe Kap. 2.1.
[417] Siehe Kap. 2.1.

Negiert man, etwa um einem Substantialismus Vorschub zu leisten, die eichinvariante Interpretation der Allgemeinen Relativitätstheorie, so impliziert dies auch eine Negierung der *Leibniz-Äquivalenz* als Kriterium für physikalische Identität, und damit einen Bezug auf unbeobachtbare Entitäten. Der zusätzliche Preis, der hierfür im Falle der Allgemeinen Relativitätstheorie zu zahlen ist, besteht in der letztendlich völlig unmotivierten (bzw. nur metaphysisch motivierten) Annahme eines Krypto-indeterminismus, der wiederum zu keinen beobachtbaren Konsequenzen führt.

Die subtileren Formen des Substantialismus, die dieses Problem gerade vermeiden können, sind hingegen mit der eichinvarianten Interpretation der Allgemeinen Relativitätstheorie durchaus vereinbar.[418] Wenn man diese als konstitutiv für eine relationalistische Sichtweise hinsichtlich der Raumzeit ansieht, so sind sie schliesslich sogar formal dem Relationalismus subsumierbar. Sie unterscheiden sich vom Relationalismus ausschliesslich durch ihre Hintergrundmetaphysik.

Und wenn man, insbesondere im Hinblick auf die Perspektiven, die sich für den Bereich der Quantengravitation einstellen, nicht die Frage, ob eine bestimmte Sichtweise nominell als relationalistisch oder als substantialistisch eingestuft werden sollte, für entscheidend hält, sondern vielmehr die inhaltlichen Konkreta der jeweiligen Sichtweise, verliert diese Hintergrundmetaphysik ihre Bedeutung. Entscheidend ist dann vielmehr, wie man – vor allem wiederum im Übergang zu einer Theorie der Quantengravitation – mit der in der Allgemeinen Relativitätstheorie vorliegenden substantiellen Form der allgemeinen Kovarianz bzw. ihrer aktiven Diffeomorphismusinvarianz umgeht, die gerade die eichinvariante Interpretation nahelegen. Vor allem im Falle einer Strategie, die auf eine direkte Quantisierung der Allgemeinen Relativitätstheorie setzt, sollten diese entscheidenden konzeptionellen Komponenten der hier zu quantisierenden klassischen Vorläufertheorie Berücksichtigung finden, zumindest soweit dem keine gravierenden Argumente entgegenstehen – die dann vermutlich ohnehin den Rahmen einer direkten Quantisierungsstrategie und der Annahme, dass diese aussichtsreich sein könnte, sprengen würden.

Konsequenterweise sind die aktive Diffeomorphismusinvarianz und ihre eichinvariante Interpretation fest in den Formalismus der *Loop Quantum Gravity* eingeschrieben.[419] Über die unmittelbaren Implikationen der eichinvarianten Interpretation der aktiven Diffeomorphismusinvarianz hinausgehend liefert die *Loop Quantum Gravity* dann schliesslich zusätzliche Indizien für ein basales Netzwerk relationaler Bezüge zwischen physikalisch relevanten Grössen, die selbst ohne fest vorgegebenen Bezug auf einen raumzeitlichen Kontext auskommen. Mit diesem Bild einer Welt, die sich als hintergrundunabhängiges Netzwerk relationaler Bezüge darstellt, stimmt die *Loop Quantum Gravity* in ihren Implikationen letztendlich mit einigen der prägeometrischen Ansätze überein. Vor allem Rovellis Lösungsansatz zum *Problem der Zeit*[420] und seine darüber hinausgehenden, allgemeineren Implikationen für das Verständnis physikalischer Grössen in einer hintergrundunabhängigen Beschreibung des Naturgeschehens lassen sich als direkte Hinweise auf ein letztlich prägeometrisches, nur relational bestimmtes Substrat lesen, auf dessen Grundlage es erst zum raumzeitlichen Geschehen kommt. Die ausschliesslich relationale Bestimmtheit legt dabei eine abstrakte, vielleicht ausschliesslich informationstheoretisch erfassbare, computationale Substratstruktur nahe.

---

[418] Siehe Kap. 2.1.
[419] Siehe Kap. 4.4.
[420] Siehe Kap. 4.4.

Eine solche Perspektive transzendiert dann schliesslich zwei der grundlegenden Intuitionen, die am Ausgangspunkt der *Loop Quantum Gravity* stehen: einerseits die, dass es sich bei der Raumzeit – wenn auch vielleicht nicht beim Raumzeitkontinuum, sondern in einer diskretisierten Form – um eine fundamentale Entität handelt, andererseits die, dass eine direkte Quantisierung der Allgemeinen Relativitätstheorie die angemessenste Strategie zur Entwicklung einer Theorie der Quantengravitation darstellt. Wenn letztere schliesslich ohnehin auf die Erfassung und Beschreibung einer prägeometrischen Basis hinausläuft, die sich im Kontext der Strategie einer direkten Quantisierung jedoch nur über massive Umwege andeutet und in mancher Hinsicht verschleiert darstellt, so liesse sich mutmassen, dass sich diese prägeometrische Basis vielleicht ohne den gewaltigen Umweg über die direkte Quantisierung der Allgemeinen Relativitätstheorie direkter und wesentlich konkreter modellieren lassen könnte: im Kontext eines originär prägeometrischen Ansatzes wie etwa dem der prägeometrischen *Quantum Causal Histories*.[421]

Und die sich hier aus dem Kontext der *Loop Quantum Gravity* heraus ergebende Perspektive transzendiert schliesslich nicht zuletzt auch die hinsichtlich der Raumzeit geführte Substantialismus-Relationalismus-Debatte. Entweder man hält diese Debatte ohnehin für irrelevant, wenn es um eine emergente Raumzeit geht, oder man deklariert eine solche emergente Raumzeit, die auf der Basis einer völlig anders gearteten nicht-raumzeitlichen Substratstruktur zustandekommt, trivialerweise als relationales Konstrukt. Ein substantieller Charakter im ursprünglichen Sinne kann ihr wohl kaum noch zugesprochen werden.

## Zeit ?

Die Frage nach dem Status der Zeit, die in vielfältige und zum Teil unüberwindliche Probleme zu führen scheint, ist eng mit den relationalistischen Implikationen und Intuitionen verbunden, die sich sowohl in der Allgemeinen Relativitätstheorie als auch insbesondere in den Ansätzen zu ihrer *Kanonischen Quantisierung* ergeben. Die diesbezüglichen Probleme sind vor allem wiederum mit der letztlich unumgänglichen und – soweit es die *Kanonische Quantisierung* betrifft – fest implementierten eichinvarianten Interpretation der aktiven Diffeomorphismusinvarianz verbunden.

Die Tatsache, dass es in der Allgemeinen Relativitätstheorie wie auch in den Ansätzen zu ihrer *Kanonischen Quantisierung* keinen physikalisch wirksamen externen Zeitparameter gibt,[422] führt einerseits zu Konflikten mit der Quantenmechanik, die schon am Ausgangspunkt der Motivationen für eine zu entwickelnde Theorie der Quantengravitation eine Rolle spielen.[423] Andererseits führt diese Tatsache dazu, dass Theorieansätze zur Quantengravitation, die von einem solchen physikalisch wirksamen externen Zeitparameter ausgehen, der noch dazu nicht in das dynamische Geschehen eingebunden ist, und die dennoch die Allgemeine Relativitätstheorie zu reproduzieren vorgeben, in einige Erklärungsnot geraten. Dies betrifft erst einmal alle hintergrundabhängigen Ansätze: den *Stringansatz*[424], die *Kovariante Quantisierung*[425] sowie die hydrodynamischen und die festkörper-physikalischen Modelle[426] für das Zustandekommen der Gravitation. Die einzige Möglichkeit, an einem externen Zeitparameter festzuhalten, dem noch dazu physikalische Relevanz zugesprochen wird, besteht dann letztlich darin, dafür zu argumentieren, dass die Allgemeine Relativitätstheorie eine effektive Theorie mit nur begrenzter Gültigkeit ist – eine Theorie, deren konzeptionelle Grundlagen in einiger Hinsicht als Artefakte anzusehen sind und sich insofern nicht beliebig extrapolieren lassen. Bisher ist eine solche Argumentation jedoch noch nicht in überzeugender Weise gelungen.

---

[421] Siehe Kap. 4.6.
[422] Siehe Kap. 2.1.
[423] Siehe Kap. 1.1. und 1.2.
[424] Siehe Kap. 4.2.
[425] Siehe Kap. 4.1.

Es ist aber unzweifelhaft, dass es im Kontext einer solchen Argumentation dann nicht zuletzt auch um das aus der aktiven Diffeomorphismusinvarianz der Allgemeinen Relativitätstheorie und ihrer eichinvarianten Interpretation resultierende *Problem der Zeit*[427] gehen müsste, das mit noch gravierenderen Folgen in den Ansätzen zu ihrer *Kanonischen Quantisierung* wiederkehrt. Schon in der Allgemeinen Relativitätstheorie[428] und erst recht in den kanonischen Quantisierungsansätzen wie der *Loop Quantum Gravity*[429] sieht es so aus, als gäbe es überhaupt keine physikalisch reale Zeit. Diese ist entsprechend der eichinvarianten Interpretation der Diffeomorphismusinvarianz vielmehr eine unphysikalische Eichvariable.

Dieses Problem nimmt im Kontext der Hamiltonschen Darstellung der Allgemeinen Relativitätstheorie, die insbesondere als Ausgangspunkt für ihre *Kanonische Quantisierung* dient, noch konturiertere Züge an: Alle zeitlichen, dynamischen Entwicklungen werden hier durch die Hamiltonsche Zusatzbedingung erfasst, die im Sinne der eichinvarianten Deutung als Generator einer Eichtransformation verstanden wird. Dies hat zur Folge, dass alle Veränderungen und alle zeitlichen Entwicklungen grundsätzlich unbeobachtbaren Eichtransformationen entsprechen; sie sind damit nicht als physikalisch real anzusehen. Es gibt hier also überhaupt keine physikalisch relevanten zeitlichen Entwicklungen und Veränderungen mehr. Alle Observablen sind notwendigerweise zeitunabhängig. Zeit und zeitliche Veränderung sind unter diesen Voraussetzungen nichts anderes als eine Illusion.

Die Allgemeine Relativitätstheorie beschreibt, wie spätestens im Kontext ihrer Hamiltonschen Darstellung deutlich wird, keine Welt des Wandels, sondern ein statisches Blockuniversum. Im Kontext ihrer *Kanonischen Quantisierung* wird daraus dann ein statischer (quantenmechanischer) Zustandsraum; es bleibt also nicht einmal mehr ein raumzeitliches oder auch nur räumliches Gebilde zurück.[430]

Damit steht sowohl die Allgemeine Relativitätstheorie als auch ihre *Kanonische Quantisierung* letztlich in eklatanter Spannung zur Phänomenologie der Zeit, zu einer offensichtlichen Welt des Werdens und der Veränderung. – Die plausibelste Auflösung dieses Konflikts theoretischer Implikationen mit einer offensichtlich im Wandel befindlichen Welt besteht nun aber nicht unbedingt

---

[426] Siehe Kap. 3.3.

[427] Siehe Kap. 2.1. und 4.4.

[428] Siehe Kap. 2.1.

[429] Siehe Kap. 4.4.

[430] Die zeitliche Gerichtetheit der Welt sowie die Auszeichnung von (wenigstens lokal wirksamen) Jetztzeitpunkten wird in diesen Theorien ohnehin nicht erfasst, genauso wenig wie in allen anderen physikalischen Theorien, die von einer zeitumkehrbaren Dynamik ausgehen:
*"So, where does temporality, with all its peculiar features ('flow' of time, whatever this means, irreversibility, memory, awareness ...) come from? I think that all this has nothing to do with mechanics. It has to do with statistical mechanics, thermodynamics, perhaps psychology or biology. [...] temporality is an artifact of our largely incomplete knowledge of the state of the world, not an ultimate property of reality."* (Rovelli (2006) 35)

darin, die entsprechende Theorie einfach *deshalb* aufzugeben, weil es zum Konflikt mit der Phäno-menologie der Zeit kommt. Dafür müssten schon bessere Gründe ins Feld geführt werden. Denn es gibt durchaus die Möglichkeit, den Konflikt mit der Phänomenologie der Zeit auf andere Weise zu beheben: Zeit lässt sich nämlich, soweit ihr physikalische Relevanz zukommt, als interne, relatio-nale Grösse rekonstruieren, ohne dass ihr ein fundamentales externalisierbares Pendant zugespro-chen werden müsste.[431]

Zeitliche Entwicklungen und 'Dynamiken' sind dann, ebenso wie räumliche Verhältnisse, nichts anderes als der Ausdruck eines nur intern bestimmten relationalen Gefüges. Und in der Form, in der wir sie kennen, spielen Raum und Zeit erst im Makroskopischen, auf einer möglicherweise emer-genten Ebene, eine Rolle.[432] Die Welt, wie wir sie in unserer Erfahrung erleben (und wie wir sie auch in unseren bisherigen physikalischen Experimenten kennengelernt haben), ist dann vielleicht nichts anderes als ein vollständig selbstbezügliches Spiel über der Leere bzw. Substanzlosigkeit – so wie es Nagarjuna schon im 2. Jahrhundert im Rahmen seiner *Madhyamika*-Lehre beschrieben hat.

*     *     *

---

[431] Eine Zeit, die über ihre ausschliesslich relationale Bestimmtheit hinausgehend als quasi-globale, externalisierbare physikalische Grösse angesehen werden kann, wäre nur um den Preis der Aufgabe der wesentlichsten Einsichten der Allgemeinen Relativitätstheorie zu haben. Das in ihrem Kontext etablierte Bild einer quasi geometrisierten Zeit, die auf gleicher Ebene mit den räumlichen Freiheitsgraden behandelt wird, müsste sich hierbei als theoretisches Artefakt erwei-sen. Nicht nur die tatsächlichen Verhältnisse, sondern vor allem auch die Frage, wie es zu diesem Artefakt und seinem nicht nur konzeptionellen, sondern eben auch empirischen Erfolg kommen konnte, wären zu klären.

[432] Dies gilt letztlich auch für den Fall, dass es auf der Substratebene so etwas wie einen zumindest lokal wirksamen Zeitverlauf geben sollte, der vielleicht in den dort vorliegenden lokalen Instantiierungen von Kausalitätsbeziehungen zum Ausdruck kommt oder für eventuelle Informationsflüsse eine Rolle spielt. Denn die zeitlichen Relationen und die auf diesen beruhenden (vielleicht nur scheinbaren) dynamischen Verläufe auf der Makroebene wären ohnehin von die-ser Substratdynamik entkoppelt. Es spielt für die makroskopische Zeit als interne, ausschliesslich relational bestimm-bare Grösse erst einmal keine Rolle, ob es auf der Substratebene so etwas wie lokale Zeitverläufe gibt oder nicht.

# Nachwort

Schliesslich gibt es noch eine Besonderheit des vorliegenden Textes, die wenigstens an dieser Stelle einmal thematisiert werden sollte. Sie ist – wie ich mir zumindest gerne einreden möchte – eine mehr oder weniger direkte Folge seines inhaltlichen Gegenstands, also der zugrundeliegenden Themenwahl – und damit letztlich der am Beginn der Untersuchung stehenden persönlichen Interessenlage. Hätte ich noch zwei oder drei Jahrzehnte mit der Fertigstellung der vorliegenden, endgültigen Version meiner Ausführungen gewartet, hätte sich womöglich die Zahl der im Text vorfindlichen wilden Spekulationen um ein Viertel oder gar ein Drittel senken lassen; vielleicht wäre die Verwendung solcher Worte wie 'womöglich', 'vielleicht', 'mutmasslich', 'wahrscheinlich', 'scheinbar', 'anscheinend', 'offensichtlich' etc. sogar um die Hälfte reduzierbar gewesen. Die mangelnde Bereitschaft, bis zu dem Zeitpunkt zu warten, zu dem aus dem 'offensichtlich' das Offensichtliche wird, die mangelnde Bereitschaft also, den Impuls des Wissenwollens aufzuschieben, bis er abgehangen und ausgereift oder hinsichtlich seiner intendierten Ziele vielleicht sogar irrelevant geworden ist – sie ist wahrscheinlich nur um den Preis einer eklatanten Sicherheits- und Gewissheitsreduzierung zu haben. Und diese mutmassliche Meta-Gewissheit verfügt offensichtlich im Kontext der vielfältigen und vielgestaltigen Bemühungen um eine Theorie der Quantengravitation und der mit diesen Bemühungen einhergehenden konzeptionellen und philosophischen Implikationen über eine noch grössere Brisanz als im scheinbar normalen Leben – wenn es so etwas denn überhaupt geben sollte.

R. H., 2010

# 6.    Literaturverzeichnis

Aalok, 2007: Background Independent Quantum Mechanics, Metric of Quantum States, and Gravity: A Comprehensive Perspective, arXiv: quant-ph/0701189

Adler, S., 2002: *Statistical Dynamics of Global Unitary Invariant Matrix Models as Pre-Quantum Mechanics*, arXiv: hep-th/0206120

Adler, S., 2004: *Quantum Theory as an Emergent Phenomenon*, New York

Ahmed, M. / Dodelson, S. / Greene, P.B. / Sorkin, R., 2004: Everpresent Lambda, *Physical Review* **D 69**, 103523, arXiv: astro-ph/0209274

Akama, K.-I. / Terazawa, H., 1983: Pregeometric Origin of the Big Bang, *General Relativity and Gravitation* **15**, 201-207

Akhmedov, E.T. / Singleton, D., 2007: On the Physical Meaning of the Unruh Effect, arXiv: 0705.2525 [hep-th]

Aldrovandi, R. / Pereira, J.G. / Vu, K.H., 2007: The Nonlinear Essence of Gravitational Waves, arXiv: 0709.1603 [gr-qc]

Alexander, S., 2007: Isogravity: Toward an Electroweak and Gravitational Unification, arXiv: 0706.4481 [hep-th]

Alexander, S.H.S. / Calcagni, G., 2008: Quantum Gravity as a Fermi Liquid, arXiv: 0807.0225 [hep-th]

Alexander, S. / Crane, L. / Sheppeard, M.D., 2003: The Geometrization of Matter Proposal in the Barrett-Crane Model and Resolution of Cosmological Problems, arXiv: gr-qc/0306079

Alfonso-Faus, A., 2008: Three Findings to model a Quantum-Gravitational Theory, arXiv: 0807.1711 [physics]

Alvarez-Gaumé, L. / Vazquez-Mozo, M.A., 1995: Topics in String Theory and Quantum Gravity, in: J. Zinn-Justin / B. Julia (Eds.): *Les Houches Summer School on Gravitation and Quantizations, 1992*, Amsterdam, 481-636, arXiv: hep-th/9212006

Amati, D., 2006: The Information Paradox, arXiv: hep-th/0612061

Amati, D. / Ciafaloni, M. / Veneziano, G., 1989: Can spacetime be probed below the string size?, *Physics Letters* **B 216**, 41-47

Ambjorn, J., 1995: Quantum Gravity represented as Dynamical Triangulation, *Classical and Quantum Gravity* **12**, 2079-2134

Ambjorn, J. / Gorlich, A. / Jurkiewicz, J. / Loll, R., 2007: Planckian Birth of the Quantum de Sitter Universe, arXiv: 0712.2485 [hep-th]

Ambjorn, J. / Jurkiewicz, J. / Loll, R., 2000: A Non-perturbative Lorentzian Path Integral for Gravity, *Physical Review Letters* **85**, 924-927, arXiv: hep-th/0002050

Ambjorn, J. / Jurkiewicz, J. / Loll, R., 2001: Non-perturbative 3D Lorentzian Quantum Gravity, *Physical Review* **D 64**, 044011, arXiv: hep-th/0011276

Ambjorn, J. / Jurkiewicz, J. / Loll, R., 2001a: Dynamically Triangulating Lorentzian Quantum Gravity, *Nuclear Physics* **B 610**, 347-382, arXiv: hep-th/0105267

Ambjorn, J. / Jurkiewicz, J. / Loll, R., 2004: Emergence of a 4D World from Causal Quantum Gravity, *Physical Review Letters* **93**, 131301, arXiv: hep-th/0404156

Ambjorn, J. / Jurkiewicz, J. / Loll, R., 2005: Semiclassical Universe from First Principles, *Physics Letters* **B 607**, 205-213, arXiv: hep-th/0411152

Ambjorn, J. / Jurkiewicz, J. / Loll, R., 2005a: Spectral Dimension of the Universe, arXiv: hep-th/0505113

Ambjorn, J. / Jurkiewicz, J. / Loll, R., 2005b: Reconstructing the Universe, arXiv: hep-th/0505154

Ambjorn, J. / Jurkiewicz, J. / Loll, R., 2006: Quantum Gravity, or The Art of Building Spacetime, arXiv: hep-th/0604212

Ambjorn, J. / Jurkiewicz, J. / Loll, R., 2009: Quantum Gravity as Sum over Spacetimes, arXiv: 0906.3947 [gr-qc]

Ambjorn, J. / Loll, R., 1998: Non-perturbative Lorentzian Quantum Gravity, Causality and Topology Change, *Nuclear Physics* **B 536**, 407-434, arXiv: hep-th/9805108

Ambjorn, J. / Loll, R. / Watabiki, Y. / Westra, W. / Zohren, S., 2008: A String Field Theory based on Causal Dynamical Triangulation, arXiv: 0802.0719 [hep-th]

Ambjorn, J. / Loll, R. / Watabiki, Y. / Westra, W. / Zohren, S., 2008a: Topology Change in Causal Quantum Gravity, arXiv: 0802.0896 [hep-th]

Ambjorn, J. / Loll, R. / Westra, W. / Zohren, S., 2009: Stochastic Quantization and the Role of Time in Quantum Gravity, arXiv: 0908.4224 [hep-th]

Amelino-Camelia, G., 2008: Quantum Gravity Phenomenology, arXiv: 0806.0339 [gr-qc]

Amelino-Camelia, G. / Arzano, M. / Procaccini, A., 2004: A glimpse at the flat-spacetime limit of quantum gravity using the Bekenstein argument in reverse, *International Journal of Modern Physics* **D 13**, 2337-2343, arXiv: hep-th/0506182

Amelino-Camelia, G. / Smolin, L., 2009: Prospects for Constraining Quantum Gravity Dispersion with Near Term Observations, arXiv: 0906.3731 [astro-ph]

Anderson, A. / DeWitt, B., 1986: Does the Topology of Space Fluctuate?, *Foundation of Physics* **16**, 91-105

Anderson, E., 2007: Foundations of Relational Dynamics, arXiv: 0706.3934 [gr-qc]

Anderson, E., 2007: Records Theory, arXiv: 0709.1892 [gr-qc]

Arcos, H.I. / Pereira, J.G., 2007: Spacetime: Arena or Reality, arXiv: 0710.0301 [gr-qc]

Artrinaux, D., 2009: *Dijon – Samsara*, Frankfurt am Main

Artrinaux, D., 2010: *The Dynamics of Dirt*, Singapore

Arvanitaki, A. / Dimopoulos, S. / Dubovsky, S. / Graham, P.W. / Harnik, R. / Rajendran, S., 2008: Astrophysical Probes of Unification, arXiv: 0812.2075 [hep-ph]

Ashtekar, A., 1984: *Asymptotic Quantization*, Naples; auch: http://cgpg.gravity.psu.edu/research/asymquant-book.pdf

Ashtekar, A., 1986: New Variables for Classical and Quantum Gravity, *Physical Review Letters* **57**, 2244-2247

Ashtekar, A., 1987: New Hamiltonian Formulation for General Relativity, *Physical Review* **D 36**, 1587-1602

Ashtekar, A., 1991: *Lectures on Non-Perturbative Canonical Gravity*, Singapore

Ashtekar, A., 1991a: Introduction: The Winding Road to Quantum Gravity, in: Ashtekar / Stachel (1991), 1-9

Ashtekar, A., 1991b: Old Problems in the Light of New Variables, in: Ashtekar / Stachel (1991), 401-426

Ashtekar, A., 2000: Quantum Mechanics of Geometry, in: N. Dadhich / A. Kembhavi (Eds.): *The Universe – Visions and Perspectives*, Dordrecht, arXiv: gr-qc/9901023

Ashtekar, A., 2002: Quantum Geometry in Action: Big Bang and Black Holes, arXiv: math-ph/0202008

Ashtekar, A., 2005: Gravity and the Quantum, *New Journal of Physics* **7**, 198, arXiv: gr-qc/0410054

Ashtekar, A., 2006: Gravity, Geometry and the Quantum, arXiv: gr-qc/0605011

Ashtekar, A., 2007: An Introduction to Loop Quantum Gravity through Cosmology, arXiv: gr-qc/0702030

Ashtekar, A., 2007a: Loop Quantum Gravity: Four Recent Advances and a Dozen Frequently Asked Questions, arXiv: 0705.2222 [gr-qc]

Ashtekar, A., 2009: Some surprising implications of background independence in canonical quantum gravity, arXiv: 0904.0184 [gr-qc]

Ashtekar, A. / Corichi, A. / Singh, P., 2007: On the Robustness of Key Features of Loop Quantum Cosmology, arXiv: 0710.3565 [gr-qc]

Ashtekar, A. / Geroch, R., 1974: Quantum Theory of Gravitation, *Reports on Progress in Physics* **37**, 1211-1256

Ashtekar, A. / Lewandowski, J., 1999: Quantum Field Theory of Geometry, in: T.Y. Cao (Ed.): *Conceptual Foundations of Quantum Field Theory*, Cambridge, 187-206

Ashtekar, A. / Lewandowski, J., 2004: Background Independent Quantum Gravity – A Status Report, *Classical and Quantum Gravity* **21**, R53, arXiv: gr-qc/0404018

Ashtekar, A. / Pawlowski, T. / Singh, P., 2006: Quantum Nature of the Big Bang, *Physical Review Letters* **96**, 141301, arXiv: gr-qc/0602086

Ashtekar, A. / Rovelli, C. / Smolin, L., 1992: Weaving a Classical Metric with Quantum Threads, *Physical Review Letters* **69**, 237-240

Ashtekar, A. / Stachel, J. (Eds.), 1991: *Conceptual Problems of Quantum Gravity*, Boston

Audretsch, J. / Lämmerzahl, C., 1991: Reasons for a Physical Field to Obey Partial Differential Equations, *Journal of Mathematical Physics* **32**, 1354-1359

Auyang, S.Y., 2001: Spacetime as a Fundamental and Inalienable Structure of Fields, *Studies in History and Philosophy of Modern Physics* **32**, 205-215

Bachas, C., 1996: D-Brane dynamics, *Physics Letters* **B 374**, 37, arXiv: hep-th/9511043

Bachas, C., 1997: (Half) A Lecture on D-Branes, arXiv: hep-th/9701019

Bachmat, E., 2007: Discrete Spacetime and its Applications, arXiv: gr-qc/0702140

Baez, J. (Ed.), 1994: *Knots and Quantum Gravity*, Oxford

Baez, J., 1994a: Strings, Loops, Knots and Gauge Fields, in: Baez (1994), arXiv: gr-qc/9309067

Baez, J., 1998: Spin Foam Models, *Classical and Quantum Gravity* **15**, 1827-1858, arXiv: gr-qc/9709052

Baez, J., 2000: An Introduction to Spin Foam Models of Quantum Gravity and BF Theory, *Lecture Notes in Physics* **543**, 25-94, arXiv: gr-qc/9905087

Baez, J., 2006: Quantum Quandaries: A Category-Theoretic Perspective, in: Rickles / French / Saatsi (2006) 240-265

Bahr, B. / Dittrich, B., 2009: Breaking and Restoring of Diffeomorphism Symmetry in Discrete Gravity, arXiv: 0909.5688 [gr-qc]

Baierlein, R.F. / Sharp, D.H. / Wheeler, J.A., 1962: Three-Dimensional Geometry as a Carrier of Information about Time, *Physical Review* **126**, 1864-1865

Bain, J., 2001: Spacetime Structuralism, in: Dieks (2006), 37-65

Bak, D. / Rey, S.-J., 2000: Holographic Principle and String Cosmology, *Classical and Quantum Gravity* **17**, L1, arXiv: hep-th/9811008

Bak, D. / Rey, S.-J., 2000a: Cosmic Holography, *Classical and Quantum Gravity* **17**, L83, arXiv: hep-th/9902173

Balasubramanian, V. / Jejjala, V. / Simón, J., 2005: The Library of Babel, *International Journal of Modern Physics* **D 14**, 2181-2186, arXiv: hep-th/0505123

Balasubramanian, V. et al., 2007: Quantum Geometry and Gravitational Entropy, arXiv: 0705.4431 [hep-th]

Bañados, M., 2007: The Ground-State of General Relativity, Topological Theories and Dark Matter, arXiv: hep-th/0701169

Banks, T., 1998: Matrix Theory, *Nuclear Physics Proceedings Supplement* **B 67**, 180-224, arXiv: hep-th/9710231

Banks, T., 2000: On Isolated Vacua and Background Independence, arXiv: hep-th/0011255

Banks, T., 2001: Cosmological Breaking of Supersymmetry or Little Lambda goes back to the Future II, *International Journal of Modern Physics* **A 16**, 910, arXiv: hep-th/0007146

Banks, T., 2001a: TASI Lectures on Matrix Theory, in: Harvey / Kachru / Silverstein (2001), arXiv: hep-th/9911068

Banks, T., 2002: Heretics of the False Vacuum – Gravitational Effects On and Of Vacuum Decay 2., arXiv: hep-th/0211160

Banks, T., 2003: A Critique of Pure String Theory: Heterodox Opinions of Diverse Dimensions, arXiv: hep-th/0306074

Banks, T., 2004: Landscepticism or Why Effective Potentials Don't Count String Models, arXiv: hep-th/0412129

Banks, T., 2009: Deriving Particle Physics from Quantum Gravity: A Plan, arXiv: 0909.3223 [hep-th]

Banks, T. / Dine, M. / Gorbatov, E., 2004: Is there a String Theory Landscape?, *Journal of High Energy Physics* **0408**, 058, arXiv: hep-th/0309170

Banks, T. / Fischler, W., 2001: A Holographic Cosmology, arXiv: hep-th/0111142

Banks, T. / Fischler, W., 2005: Holographic Cosmology 3.0, *Physica Scripta* **T 117**, 56-63, arXiv: hep-th/0310288

Banks, T. / Fischler, W. / Shenker, S.H. / Susskind, L., 1997: M Theory as a Matrix Model: A Conjecture, *Physical Review* **D 55**, 5112-5128, arXiv: hep-th/9610043

Barbour, J., 1982: Relational Concepts of Space and Time, *British Journal for the Philosophy of Science* **33**, 251-274

Barbour, J., 1994: The Timelessness of Quantum Gravity I – The Evidence from the Classical Theory, *Classical and Quantum Gravity* **11**, 2853-2873

Barbour, J., 1994a: The Timelessness of Quantum Gravity II – The Appearance of Dynamics in Static Configurations, *Classical and Quantum Gravity* **11**, 2875-2897

Barbour, J., 1995: General Relativity as a Perfectly Machian Theory, in: Barbour / Pfister (1995) 214-236

Barbour, J., 1999: *The End of Time. The Next Revolution in Physics,* London

Barbour, J., 1999a: The Development of Machian Themes in the Twentieth Century, in: Butterfield (1999) 83-109

Barbour, J., 2001: On general covariance and best matching, in: Callender / Huggett (2001)

Barbour, J., 2003: Dynamics of Pure Shape, Relativity, and the Problem of Time, *Lecture Notes in Physics* **633**, 15ff, arXiv: gr-qc/0309089

Barbour, J., 2009: The Nature of Time, arXiv: 0903.3489 [gr-qc]

Barbour, J. / Foster, B.Z., 2008: Constraints and Gauge Transformations: Dirac's Theorem is not always valid, arXiv: 0808.1223 [gr-qc]

Barbour, J. / Pfister, H. (Eds.), 1995: *Mach's Principle: From Newton's Bucket to Quantum Gravity*, Boston

Bardeen, J.M. / Carter, B. / Hawking, S.W., 1973: The Four Laws of Black Hole Mechanics, *Communications in Mathematical Physics* **31**, 161-170

Barcelo, C. / Liberati, S. / Visser, M., 2005: Analogue Gravity, *Living Reviews in Relativity* (Electronic Journal) **8/12**, www.livingreviews.org, auch: arXiv: gr-qc/0505065

Barrett, J.W., 1987: The Geometry of Classical Regge Calculus, *Classical and Quantum Gravity* **4**, 1565-1576

Barrett, J.W., 1995: Quantum Gravity as Topological Quantum Field Theory, *Journal of Mathematical Physics* **36**, 6161-6179, arXiv: gr-qc/9506070

Barrow, J.D., 2007: *New Theories of Everything*, Oxford

Barrow, J.D. / Davies, P.C.W. / Harper, C.L. (Eds.), 2004: *Science and Ultimate Reality: Quantum Theory, Cosmology, and Complexity*, Cambridge

Bartels, A., 1994: Von Einstein zu Aristoteles – Raumzeit-Philosophie und Substanz-Metaphysik, *Philosophia Naturalis* **31**, 294-308

Bartels, A., 1996: Modern Essentialism and the Problem of Individuation of Spacetime Points, *Erkenntnis* **45**, 25-43

Bartels, A., 1999: Objects or Events?: Towards an Ontology for Quantum Field Theory, *Philosophy of Science* **66**, Supplement (Proceedings), S170-S184

Batlle, C. / Gomis, J. / Pons, J.M. / Roman-Roy, N., 1986: Equivalence between the Lagrangian and the Hamiltonian Formalism for Constrained Systems, *Journal of Mathematical Physics* **27**, 2953-2962

Bekenstein, J.D., 1973: Black Holes and Entropy, *Physical Review* **D 7**, 2333-2346

Bekenstein, J.D., 1974: Generalized Second Law of Thermodynamics in Black Hole Physics, *Physical Review* **D 9**, 3292-3300

Bekenstein, J.D., 1981: Universal Upper Bound on the Entropy-to-Energy Ratio for Bounded Systems, *Physical Review* **D 23**, 287-298

Bekenstein, J.D., 2000: Holographic Bound from Second Law, arXiv: gr-qc/0007062

Bekenstein, J.D., 2001: The Limits of Information, *Studies in History and Philosophy of Modern Physics* **32**, 511-524, arXiv: gr-qc/0009019

Bekenstein, J.D., 2003: Das holografische Universum, *Spektrum der Wissenschaft* (11/2003), 34-41

Belot, G., 1996: *Whatever is Never and Nowhere is Not: Space, Time, and Ontology in Classical and Quantum Gravity*, Dissertation, University of Pittsburgh, Pittsburgh, www.pitt.edu/~gbelot/Papers/dissertation.pdf

Belot, G., 1996a: Why General Relativity Does Need an Interpretation, *Philosophy of Science* **63** (Supplement), S80-S88

Belot, G., 1999: Rehabilitating Relationalism, *International Studies in Philosophy of Science* **13**, 35-52

Belot, G., 2000: Geometry and Motion, *British Journal for the Philosophy of Science* **51**, 561-595

Belot, G., 2003: Symmetry and Gauge Freedom, *Studies in the History and Philosophy of Modern Physics* **34**, 189-225

Belot, G., 2007: An Elementary Notion of Gauge Equivalence, http://philsci-archive.pitt.edu

Belot, G. / Earman, J., 1999: From metaphysics to physics, in: Butterfield / Pagonis (1999)

Belot, G. / Earman, J., 2001: Pre-socratic quantum gravity, in: Callender / Huggett (2001)

Belot, G. / Earman, J., / Ruetsche, L., 1999: The Hawking Information Loss Paradox: The Anatomy of a Controversy, *British Journal for the Philosophy of Science* **50**, 189-229

Benedetti, D., 2008: Fractal Properties of Quantum Spacetime, arXiv: 0811.1396 [hep-th]

Benioff, P., 2003: Resource Limited Theories and their Extensions, arXiv: quant-ph/0303086

Bennett, D.L. / Brene, N. / Nielsen, H., 1987: Random Dynamics, *Physica Scripta* **T 15**, 158-163

Bergmann, P., 1961: Observables in General Relativity, *Reviews of Modern Physics* **33**, 510-514

Bergmann, P., 1977: Geometry and Observables, in: J. Earman / C. Glymour / J. Stachel (Eds.): *Foundations of Space-time Theories*, University of Minnesota Press, 275-280

Bernal, A.N. / Sánchez, M. / Soler Gil, F.J., 2008: Physics from Scratch – Letter on M. Tegmark's "The Mathematical Universe", arXiv: 0803.0944 [gr-qc]

Bertolami, O. / Paramos, J. / Turyshev, S.G., 2006: General Theory of Relativity: Will it survive the next decade?, arXiv: gr-qc/0602016

Bianchi, E., 2008: The Length Operator in Loop Quantum Gravity, arXiv: 0806.4710 [gr-qc]

Bigatti, D. / Susskind, L., 1997: Review of Matrix Theory, arXiv: hep-th/9712072

Bilal, A., 1999: M(atrix) Theory – A Pedagogical Introduction, *Fortschritte der Physik* **47**, 5-28, arXiv: hep-th/9710136

Bilson-Thompson, S.O., 2005: A Topological Model of Composite Preons, arXiv: hep-ph/0503213

Bilson-Thompson, S. / Hackett, J. / Kauffman, L. / Smolin, L., 2008: Particle Identifications from Symmetries of Braided Ribbon Network Invariants, arXiv: 0804.0037 [hep-th]

Bilson-Thompson, S. / Hackett, J. / Kauffman, L.H., 2009: Particle Topology, Braids, and Braided Belts, arXiv: 0903.1376 [math.AT]

Bilson-Thompson, S.O. / Markopoulou, F. / Smolin, L., 2006: Quantum Gravity and the Standard Model, arXiv: hep-th/0603022

Bleyer, U. / Liebscher, D.-E., 1995: Mach's Principle and Local Causal Structure, in: Barbour / Pfister (1995) 293-307

Blute, R. / Ivanov, I.T. / Panangaden, P., 2001: Decoherent Histories on Graphs, arXiv: gr-qc/0111020

Blute, R. / Ivanov, I.T. / Panangaden, P., 2003: Discrete Quantum Causal Dynamics, *International Journal of Theoretical Physics* **42**, 2025-2041, arXiv: gr-qc/0109053

Bojowald, M., 2001: *Quantum Geometry and Symmetry*, Aachen

Bojowald, M., 2004: Loop Quantum Cosmology – Recent Progress, arXiv: gr-qc/0402053

Bojowald, M., 2005: Nonsingular Black Holes and Degrees of Freedom in Quantum Gravity, *Physical Review Letters* **95**, 061301, arXiv: gr-qc/0506128

Bojowald, M., 2006: Loop Quantum Cosmology, *Living Reviews in Relativity* (Electronic Journal), www.livingreviews.org, auch: arXiv: gr-qc/0601085

Bojowald, M., 2006a: Quantum Riemannian Geometry and Black Holes, arXiv: gr-qc/0602100

Bojowald, M., 2006b: Quantum Cosmology, arXiv: gr-qc/0603110

Bojowald, M., 2007: Quantum Gravity and Cosmological Observations, arXiv: gr-qc/0701142

Bojowald, M., 2007a: Singularities and Quantum Gravity, arXiv: gr-qc/0702144

Bojowald, M., 2008: Canonical Relativity and the Dimensionality of the World, arXiv: 0807.4874 [gr-qc]

Bojowald, M. / Morales-Tecotl, H.A., 2004: Cosmological Applications of Loop Quantum Gravity, in: *Lecture Notes in Physics* **646**, 421-462, arXiv: gr-qc/0306008

Bokulich, P., 2001: Black Hole Remnants and Classical vs. Quantum Gravity, *Philosophy of Science* (Proceedings), **68**, S407-S423

Bombelli, L. / Corichi, A. / Winkler, O., 2009: Semiclassical Quantum Gravity: Obtaining Manifolds from Graphs, arXiv: 0905.3492 [gr-qc]

Bombelli, L. / Lee, J. / Meyer, D. / Sorkin, R.D., 1987: Space-Time as a Causal Set, *Physical Review Letters* **59**, 521-524

Booss-Bavnbek, B. / Esposito, G. / Lesch, M., 2007: Quantum Gravity: Unification of Principles and Interactions, and Promises of Spectral Geometry, arXiv: 0708.1705 [hep-th]

Boughn, S., 2008: Nonquantum Gravity, arXiv: 0809.4218 [gr-qc]

Bousso, R., 2002: The Holographic Principle, *Reviews of Modern Physics* **74**, 825-874, arXiv: hep-th/0203101

Bousso, R., 2006: Holographic Probabilities in Eternal Inflation, arXiv: hep-th/0605263

Boyarsky, A. / Gora, P., 2007: The Measurement of Time, arXiv: 0705.1000 [hep-th]

Brading, K. / Brown, H.R., 2004: Are Gauge Symmetry Transformations Observable?, *British Journal for the Philosophy of Science* **55**, 645-665, auch: http://philsci-archive.pitt.edu, Dokument: 1436

Brading, K. / Castellani, E. (Eds.), 2003: *Symmetries in Physics: Philosophical Reflections,* Cambridge

Brading, K. / Castellani, E., 2003a: Symmetries in Physics: Philosophical Reflections – Introduction, in: Brading / Castellani (2003), 1-18, auch: arXiv: quant-ph/0301097

Brandenberger, R.H., 2007: String Theory, Space-Time Non-Commutativity and Structure Formation, arXiv: hep-th/0703173

Braunstein, S.L. / Pati, A.K., 2006: Quantum Information cannot be completely hidden in correlations: Implications for the Black-Hole Information Paradox, arXiv: gr-qc/0603046

Brax, P. / Bruck, C. van der, 2003: Cosmology and Brane Worlds – A Review, *Classical and Quantum Gravity* **20**, R201-R232, arXiv: hep-th/0303095

Breuer, T., 1997: Universell und unvollständig: Theorien über alles?, *Philosophia Naturalis* **34**, 1-20

Breuer, T., 1997a: What Theories of Everything Don't Tell, *Studies in History and Philosophy of Modern Physics* **28**, 137-143

Brighouse, C., 1994: Spacetime and Holes, in: D. Hull / M. Forbes / R.M. Burian (Eds.): *PSA 1994*, Vol. 1, East Lansing, 117-125

Brightwell, G. / Dowker, F. / García, H. / Henson, J. / Sorkin, R.D., 2002: General Covariance and the 'Problem of Time' in a Discrete Cosmology, arXiv: gr-qc/0202097

Brown, H.R., 2005: *Physical Relativity: Space-Time Structure from a Dynamical Perspective*, Oxford

Brown, H.R., 2009: The Behaviour of Rods and Clocks in General Relativity, and the Meaning of the Metric Field, arXiv: 0911.4440 [gr-qc]

Brown, H. / Pooley, O., 1999: The Origin of the Spacetime Metric: Bell's 'Lorentzian Pedagogy' and its Significance in General Relativity, arXiv: gr-qc/9908048

Brunetti, R. / Fredenhagen, K., 2006: Towards a Background Independent Formulation of Perturbative Quantum Gravity, arXiv: gr-qc/0603079

Brunetti, R. / Fredenhagen, K. / Hoge, M., 2009: Time in Quantum Physics: From an External Parameter to an Intrinsic Observable, arXiv: 0909.1899 [math-ph]

Brustein, R. / Hadad, M., 2009: The Einstein Equations for Generalized Theories of Gravity and the Thermodynamic Relation δQ = T δS are Equivalent, arXiv: 0903.0823 [hep-th]

Buchholz, D., 2008: Quantenfeldtheorie ohne Quantenfelder, *Physik Journal* **7**, 8/9: 45-50

Burgess, C.P., 2004: Quantum Gravity in Everyday Life – General Relativity as an Effective Field Theory, *Living Reviews in Relativity* (Electronic Journal), www.livingreviews.org/lrr-2004-5

Burgess, C.P., 2009: Effective Theories and Modifications of Gravity, arXiv: 0912.4295 [gr-qc]

Busch, J., 2003: What Structures could not be, *International Studies in Philosophy of Science* **17**, 211-223

Butterfield, J., 1989: The Hole Truth, *British Journal for the Philosophy of Science* **40**, 1-28

Butterfield, J. (Ed.), 1999: *The Arguments of Time*, Oxford

Butterfield, J., 2002: Julian Barbour: The End of Time – Critical Notice, *British Journal for the Philosophy of Science* **53**, 289-330

Butterfield, J. / Hogarth, M. / Belot, G. (Eds.), 1996: *Spacetime*, Aldershot

Butterfield, J. / Isham, C., 1999: On the Emergence of Time in Quantum Gravity, in: Butterfield (1999) 111-168; auch: arXiv: gr-qc/9901024

Butterfield, J. / Isham, C., 2000: Some Possible Roles for Topos Theory in Quantum Theory and Quantum Gravity, *Foundations of Physics* **30**, 1707-1735

Butterfield, J. / Isham, C., 2001: Spacetime and the Philosophical Challenge of Quantum Gravity, in: Callender / Huggett (2001), auch: arXiv: gr-qc/9903072

Butterfield, J. / Pagonis, C. (Eds.), 1999: *From Physics to Philosophy*, Cambridge

Cahill, R.T., 2002: Process Physics – From Quantum Foam to General Relativity, arXiv: gr-qc/0203015

Cahill, R.T., 2005: *Process Physics – From Information Theory to Quantum Space and Matter*, New York, auch: www.scieng.flinders.edu.au/cpes/people/cahill_r/HPS13.pdf

Cahill, R.T., o.J.: *Process Physics Website*, www.scieng.flinders.edu.au/cpes/people/cahill_r/processphysics.html

Cahill, R.T. / Klinger, C.M., 1996: Pregeometric Modelling of the Spacetime Phenomenology, *Physics Letters* **A 223**, 313-319, arXiv: gr-qc/9605018

Cahill, R.T. / Klinger, C.M., 1997: Bootstrap Universe from Self-Referential Noise, arXiv: gr-qc/9708013

Cahill, R.T. / Klinger, C.M., 1998: Self-Referential Noise and the Synthesis of Three-Dimensional Space, arXiv: gr-qc/9812083

Cahill, R.T. / Klinger, C.M., 2005: Bootstrap Universe from Self-Referential Noise, *Progress in Physics*, **2**, 108-112, auch: www.scieng.flinders.edu.au/cpes/people/cahill_r/CahillBoot.pdf

Callender, C., 2000: Shedding Light on Time, *Philosophy of Science* **67**, S587-S599

Callender, C., (Ed.), 2002: *Time, Reality and Experience*, Cambridge

Callender, C., 2002a: Philosophy of Space-Time Physics, in: P. Machamer / M. Silberstein (Eds.): *The Blackwell Guide to the Philosophy of Science*, Cambridge, 173-198

Callender, C. / Huggett, N. (Eds.), 2001: *Physics meets Philosophy at the Planck Scale. Contemporary Theories of Quantum Gravity*, Cambridge

Callender, C. / Huggett, N., 2001a: Introduction, in: Callender / Huggett (2001)

Callender, C. / Huggett, N., 2001b: Why quantize Gravity (or any other field for the matter)?, *Philosophy of Science* (Proceedings), **68**, S382-S394

Callender, C. / Weingard, R., 2000: Topology Change and the Unity of Space, *Studies in History and Philosophy of Modern Physics* **31**, 227-246

Calmet, X. / Gong, W. / Hsu, S.D.H., 2008: Colorful Quantum Black Holes at the LHC, arXiv: 0806.4605 [hep-ph]

Calmet, X. / Graesser, M. / Hsu, S.D.H., 2005: Minimum Length from First Principles, *International Journal of Modern Physics* **D 14**, 2195-2200, arXiv: hep-th/0505144

Cao, T.Y., 2001: Prerequisites for a Consistent Framework of Quantum Gravity, *Studies in History and Philosophy of Modern Physics* **32**, 181-204

Cao, T.Y., 2003: Can We Dissolve Physical Entities into Mathematical Structures?, *Synthese* **136/1**, 57-71

Cao, Y.T., 2006: Structural Realism and Quantum Gravity, in: Rickles / French / Saatsi (2006) 40-52

Carlip, S., 2001: Quantum Gravity: A Progress Report, *Reports on Progress in Physics* **64**, 885, arXiv: gr-qc/0108040

Carlip, S., 2007: Black Hole Entropy and the Problem of Universality, arXiv: gr-qc/0702094

Carlip, S., 2008: Is Quantum Gravity Necessary?, arXiv: 0803.3456 [gr-qc]

Carlip, S., 2008a: Black Hole Thermodynamics and Statistical Mechanics, arXiv: 0807.4520 [gr-qc]

Carlip, S., 2008b: Black Hole Entropy and the Problem of Universality, arXiv: 0807.4192 [gr-qc]

Carlip, S., 2009: Spontaneous Dimensional Reduction in Short-Distance Quantum Gravity?, arXiv: 0909.3329 [gr-qc]

Carr, B. (Ed.), 2007: *Universe or Multiverse*, Cambridge

Carroll, S.M., 2008: What if Time Really Exists?, arXiv: 0811.3772 [gr-qc]

Cartwright, N., 1983: *How the Laws of Physics lie*, Oxford

Cartwright, N., 1989: *Nature's Capacities and Their Measurement*, Oxford

Cartwright, N., 1994: Fundamentalism vs the Patchwork of Laws, *Proceedings of the Aristotelian Society* **93**, 279-292

Cartwright, N., 1995: Precis of *Nature's Capacities and Their Measurement*, *Philosophy and Phenomenological Research* **LV/1**, 153-161

Cartwright, N., 1999: *The Dappled World*, Cambridge

Caticha, A., 2005: The Information Geometry of Space and Time, arXiv: gr-qc/0508108

Caticha, A., 2007: Information and Entropy, arXiv: 0710.1068 [physics]

Chaves, M., 2007: The Quantization of Gravity and the Vacuum Energy of Quantum Fields, arXiv: 0707.1066 [gr-qc]

Chew, G.F., 1963: The Dubious Role of the Space-Time Continuum in Microscopic Physics, *Science Progress* **51**, 529-539

Chiao, R.Y., 2004: Conceptual Tensions between Quantum Mechanics and General Relativity: Are there Experimental Consequences?, in: Barrow / Davies / Harper (2004), 254-279

Chu, Chong-Sun / Lechtenfeld, O., 2006: Emergence of Time from Dimensional Reduction in Noncommutative Geometry, *Modern Physics Letters* **A 21**, 639-648, arXiv: hep-th/0508055

Ciufolini, I. / Wheeler, J.A., 1995: *Gravitation and Inertia*, Princeton

Cline, J.M., 2006: String Cosmology, arXiv: hep-th/0612129

Cole, E.A.B., 1970: Transition from a Continuous to a Discrete Space-Time Scheme, *Il Nuovo Cimento* **66A**, 645-656

Connes, A., 1994: *Noncommutative Geometry*, San Diego

Connes, A., 1998: Noncommutative Differential Geometry and the Structure of Space-Time, in: S.A. Huggett et al. (Eds.): *The Geometric Universe*, Oxford (1998)

Connes, A., 2000: Noncommutative Geometry Year 2000, arXiv: math.QA/0011193

Connes, A. / Marcolli, M., 2006: A Walk in the Noncommutative Garden, arXiv: math.QA/0601054

Corichi, A., 1997: *Interplay between Topology, Gauge Fields and Gravity*, Ph.D. Thesis, Pennsylvania State University, http://cgpg.gravity.psu.edu/archives/thesis/1997/corichi.pdf

Corichi, A., 2008: On the Geometry of Quantum Constrained Systems, arXiv: 0801.1119 [gr-qc]

Corichi, A. / Ryan, M. / Sudarsky, D., 2002: Quantum Geometry as a Relational Construct, *Modern Physics Letters* **A 17**, 555-567, arXiv: gr-qc/0203072

Corichi, A. / Zapata, J.A., 2007: Quantum Structure of Geometry: Loopy and Fuzzy?, arXiv: 0705.2440 [gr-qc]

Crane, L., 1995: Clocks and Categories: Is Quantum Gravity Algebraic?, *Journal of Mathematical Physics* **36**, 6180-6193, arXiv: gr-qc/9504038

Crane, L., 2001: A New Approach to the Geometrization of Matter, arXiv: gr-qc/0110060

Crane, L., 2007: What is the Mathematical Structure of Quantum Spacetime?, arXiv: 0706.4452 [gr-qc]

Crane, L., 2008: A Pointless Model for the Continuum as the Foundation for Quantum Gravity, arXiv: 0804.0030 [gr-qc]

Crane, L., 2008a: Model Categories and Quantum Gravity, arXiv: 0810.4492 [gr-qc]

Cremmer, E. / Julia, B. / Scherk, J., 1978: Supergravity Theory in Eleven Dimensions, *Physics Letters* **B 76**, 409-412

Curiel, E., 2001: Against the Excess of Quantum Gravity: A Plea for Modesty, *Philosophy of Science* (Proceedings), **68**, S424-S441

Curiel, E., 2009: General Relativity Needs no Interpretation, http://philsci-archive.pitt.edu, Dokument: 4567

Dadhich, N. / Maeda, H., 2007: Origin of Matter out of Pure Curvature, arXiv: 0705.2490 [hep-th]

Dambmann, H., 1990: Die Bedeutung des Machschen Prinzips in der Kosmologie, *Philosophia Naturalis* **27**, 234-271

Damour, T. / Nordtvedt, K., 1993: General Relativity as a Cosmological Attractor of Tensor-Scalar Theories, *Physical Review Letters* **70**, 2217-2219

Danielsson, U.H. / Johansson, N. / Larfors, M., 2006: The World Next Door – Results in Landscape Topography, arXiv: hep-th/0612222

Das, S.R. / Mathur, S.D., 2001: The Quantum Physics of Black Holes – Results from String Theory, *Annual Reviews of Nuclear Particle Science* **50**, 153-206, arXiv: gr-qc/0105063

Das, S. / Shankaranarayanan, S., 2007: Where are the Black Hole Entropy Degrees of Freedom, arXiv: gr-qc/0703082

Das, S. / Shankaranarayanan, S. / Sur, S., 2008: Black Hole Entropy from Entanglement – A Review, arXiv: 0806.0402 [gr-qc]

David, J.R. / Mandal, G. / Wadia, S.R., 2002, Microscopic Formulation of Black Holes in String Theory, *Physics Reports* **369**, 549-686, arXiv: hep-th/0203048

Dawid, R., 1999: A Speculative Remark on Holography, *Physics Letters* **B 451**, 19, arXiv: hep-th/9811243

Dehnen, H., 1995: The Higgs Field and Mach's Principle of Relativity of Inertia, in: Barbour / Pfister (1995) 479-490

Dehnen, H. / Frommert, H., 1990: Scalar Gravity and Higgs Potential, *International Journal of Theoretical Physics* **29**, 361-370

Dehnen, H. / Frommert, H., 1991: Higgs-Field Gravity Within the Standard Model, *International Journal of Theoretical Physics* **30**, 985-998

Deser, S. / Woodard, R.P., 2007: Nonlocal Cosmology, arXiv: 0706.2151 [astro-ph]

Deutsch, D., 1999: Quantum Theory of Probability and Decisions, arXiv: quant-ph/9906015

DeWitt, B.S., 1967: Quantum Theory of Gravity. I. The Canonical Theory, *Physical Review* **160**, 1113-1148

DeWitt, B.S., 1967a: Quantum Theory of Gravity. II. The Manifestly Covariant Theory, *Physical Review* **162**, 1195-1239

DeWitt, B.S., 1967b: Quantum Theory of Gravity. III. Applications of the Covariant Theory, *Physical Review* **162**, 1239-1256

DeWitt, B.S., 1972: Covariant Quantum Geometrodynamics, in: J.R. Klauder (Ed.): *Magic without Magic – John Archibald Wheeler*, San Francisco

DeWitt, B.S., 1999: Quantum Field Theory and Space-Time – Formalism and Reality, in: T.Y. Cao (Ed.): *Conceptual Foundations of Quantum Field Theory*, Cambridge, 176-186

DeWitt, B.S., 2007: An Introduction to Quantum Gravity, arXiv: 0711.2445 [hep-th]

DeWitt, B.S., 2008: Quantum Gravity Yesterday and Today, arXiv: 0805.2935 [physics]

Dieks, D., 2001: Introduction – Special Issue: Spacetime, Fields and Understanding: Perspectives on Quantum Field Theory, *Studies in History and Philosophy of Modern Physics* **32**, 151-156

Dieks, D., 2001a: Space and Time in Particle and Field Physics, *Studies in History and Philosophy of Modern Physics* **32**, 217-241

Dieks, D., 2001b: Space-Time Relationism in Newtonian and Relativistic Physics, *International Studies in the Philosophy of Science* **15**, 5-17, Manuskript auch: http://philsci-archive.pitt.edu

Dieks, D. (Ed.), 2006: *The Ontology of Spacetime*, Amsterdam

Dieks, D., 2006a: Introduction, in: Dieks (2006), ix-xviii

Dieks, D., 2006b: Becoming, Relativity and Locality, in: Dieks (2006), 157-176

Dine, M., 2004: Is there a String Theory Landscape: Some Cautionary Remarks, arXiv: hep-th/0402101

Dipert, R.R., 1997: The Mathematical Structure of the World: The World as a Graph, *Journal of Philosophy* **94**, 329-358

DiSalle, R., 1994: On Dynamics, Indiscernibility, and Spacetime Ontology, *British Journal for the Philosophy of Science* **45**, 265-287

DiSalle, R., 1995: Spacetime Theory as Physical Geometry, *Erkenntnis* **42**, 317-337

Dittrich, B., 2004: Partial and Complete Observables for Hamiltonian Systems, arXiv: gr-qc/0411013

Dittrich, B., 2006: Partial and Complete Observables for Canonical General Relativity, *Classical and Quantum Gravity* **23**, 6155, arXiv: gr-qc/0507106

Dittrich, B., 2008: Diffeomorphism Symmetry in Quantum Gravity Models, arXiv: 0810.3594 [gr-qc]

Dittrich, B. / Hoehn, P.A., 2009: From Covariant to Canonical Formulations of Discrete Gravity, arXiv: 0912.1817 [gr-qc]

Dittrich, B. / Thiemann, T., 2007: Are the Spectra of Geometrical Operators in Loop Quantum Gravity really discrete, arXiv: 0708.1721 [gr-qc]

Doering, A. / Isham, C., 2008: What is a Thing? – Topos Theory in the Foundations of Physics, arXiv: 0803.0417 [quant-ph]

Donoghue, J.F. / Pais, P., 2009: Gauge Federation as an Alternative to Unification, arXiv: 0903.3929 [hep-ph]

Doplicher, S., 2006: Quantum Field Theory on Quantum Spacetime, arXiv: hep-th/0609124

Doplicher, S., 2009: The Principle of Locality: Effectiveness, Fate and Challenges, arXiv: 0911.5136 [math-ph]

Doplicher, S. / Fredenhagen, K. / Roberts, J.E., 1995: The Quantum Structure of Spacetime at the Planck Scale and Quantum Fields, *Communications in Mathematical Physics* **172**, 187-220, arXiv: hep-th/0303037

Dorato, M., 2000: Substantivalism, Relationism, and Structural Spacetime Realism, *Foundations of Physics* **30**, 1605-1628

Dorato, M., 2002: On Becoming, Cosmic Time and Rotating Universes, in: Callender (2002) 253-276

Dorato, M. / Pauri, M., 2006: Holism and Structuralism in Classical and Quantum General Relativity, in: Rickles / French / Saatsi (2006) 121-151

Dorofeev, V.Y., 2008: Non-Associativity as Gravity, arXiv: 0807.2151 [gr-qc]

Douglas, M.R., 1996: Superstring dualities, Dirichlet branes and the small scale structure of space, arXiv: hep-th/9610041

Douglas, M.R., 1998: Gauge Fields and D-Branes, *Journal of Geometry and Physics* **28**, 255-262, arXiv: hep-th/9604198

Douglas, M.R., 2003: The Statistics of String/M Theory Vacua, *Journal for High Energy Physics* **0305**:046, arXiv: hep-th/0303194

Douglas, M.R., 2004: Statistical Analysis of the Supersymmetry Breaking Scale, arXiv: hep-th/0405279

Douglas, M.R., 2004a: Statistics of String Vacua, arXiv: hep-ph/0401004

Douglas, M.R., 2004b: Basic Results in Vacuum Statistics, *Comptes Rendues Physique* **5**, 965-977, arXiv: hep-th/0409207

Douglas, M.R., 2006: Understanding the Landscape, arXiv: hep-th/0602266

Dreyer, O., 2001: *Isolated Horizons and Black Hole Entropy*, Ph.D. Thesis, Pennsylvania State University, http://cgpg.gravity.psu.edu/archives/thesis/2001/dreyer.pdf

Dreyer, O., 2004: Background Independent Quantum Field Theory and the Cosmological Constant Problem, arXiv: hep-th/0409048

Dreyer, O., 2006: Emergent General Relativity, arXiv: gr-qc/0604075

Dreyer, O., 2007: Why Things Fall, arXiv: 0710.4350 [gr-qc]

Dreyer, O., 2009: Time is not the problem, arXiv: 0904.3520 [gr-qc]

Duff, M.J., 1996: M-Theory (the Theory Formerly known as Strings), *International Journal of Modern Physics* **A 11**, 5623-5642, arXiv: hep-th/9608117

Dzhunushaliev, V. / Serikbayev, N. / Myrzakulov, R., 2009: Topology Change in Quantum Gravity and Ricci Flows, arXiv: 0912.5326 [gr-qc]

Earman, J., 1986: Why Space is not a Substance (at least not to the first degree), *Pacific Philosophical Quarterly* **67**, 225-244

Earman, J., 1989: *World Enough and Spacetime – Absolute Versus Relational Theories of Space and Time*, Cambridge, Ma.

Earman, J., 2002: Thoroughly Modern McTaggart. Or what McTaggart would have said if he had learned General Relativity Theory, *Philosopher's Imprint* **2**, 1-28, http://www.philosophersimprint.org

Earman, J., 2002a: Response to Maudlin, in: Maudlin (2002), 19-23

Earman, J., 2002b: Gauge Matters, *Philosophy of Science* **69**, S209-S220, http://philsci-archive.pitt.edu, Dokument: 0070

Earman, J., 2003: Tracking Down Gauge: An Ode to the Constrained Hamiltonian Formalism, in: Brading / Castellani (2003), 140-162

Earman, J., 2003a: The Cosmological Constant, the Fate of the Universe, Unimodular Gravity, and all that, *Studies in History and Philosophy of Modern Physics* **34**, 559-577

Earman, J., 2006: Two Challenges to the Requirement of Substantive General Covariance, *Synthese* **148**, 443-468

Earman, J., 2006a: The Implications of General Covariance for the Ontology and Ideology of Spacetime, in: Dieks (2006), 3-23

Earman, J. / Butterfield, J. (Eds.), 2007: *Handbook of the Philosophy of Science, Vol. 2: The Philosophy of Physics*, Amsterdam

Earman, J. / Butterfield, J., 2007a: Introduction, in: Earman / Butterfield (2007)

Earman, J. / Norton, J.D., 1987: What Price Spacetime Substantivalism? – The Hole Story, *British Journal for the Philosophy of Science* **38**, 515-525

Egan, G., 2003: *Schild's Ladder*, London

Ehlers, J., 1995: Machian Ideas an General Relativity, in: Barbour / Pfister (1995) 458-473

Ehlers, J. / Friedrich, H. (Eds.), 1994: *Canonical Gravity – From Classical to Quantum*, Berlin

Eling, C., 2008: Hydrodynamics of Spacetime and Vacuum Viscosity, arXiv: 0806.3165 [hep-th]

Eling, C. / Bekenstein, J.D., 2008: Is the Black Hole Area Theorem always classically valid?, arXiv: 0810.5255 [gr-qc]

Eling, C. / Guedens, R. / Jacobson, T., 2006: Non-Equilibrium Thermodynamics of Spacetime, *Physical Review Letters* **96**, 121301, arXiv: gr-qc/0602001

Ellis, G.F.R., 2006: Physics in the Real Universe: Time and Spacetime, *General Relativity and Gravitation* **38**, 1797-1824, arXiv: gr-qc/0605049

Ellis, G.F.R., 2008: On the Flow of Time, arXiv: 0812.0240 [gr-qc]

Ellis, G.F.R. / Rothman, T., 2009: Time and Spacetime: The Crystallizing Block Universe, arXiv: 0912.0808 [quant-ph]

Ellis, G.F.R. / Smolin, L., 2009: The Weak Anthropic Principle and the Landscape of String Theory, arXiv: 0901.2414 [hep-th]

Elze, H.-T., 2009: Does Quantum Mechanics tell an Atomistic Spacetime?, arXiv: 0906.1101 [quant-ph]

Emam, M.H., 2008: So what will you do if String Theory is wrong?, arXiv: 0805.0543 [physics]

Engle, J., 2007: On the Physical Intepretation of States in Loop Quantum Cosmology, arXiv: gr-qc/0701132

Erdmenger, J. / Meyer, R. / Park, J.-H., 2007: Spacetime Emergence in the Robertson-Walker-Universe from a Matrix Model, arXiv: 0705.1586 [hep-th]

Esfeld, M., 2008: Die Metaphysik dispositionaler Eigenschaften, *Zeitschrift für philosophische Forschung* **62**, 323-342

Esfeld, M. / Lam, V., 2007: Moderate Structural Realism about Space-Time, *Synthese* **160**, 27-46

Esposito-Farese, G., 2007: Summary of Session A4 at the GRG18 Conference: Alternative Theories of Gravity, arXiv: 0711.0332 [gr-qc]

Esposito-Farese, G., 2009: Motion in Alternative Theories of Gravity, arXiv: 0905.2575 [gr-qc]

Espriu, D. / Puigdomenech, D., 2009: Gravity as an Effective Theory, arXiv: 0910.4110 [hep-th]

Ezawa, K., 1997: *Nonperturbative Solutions for Canonical Quantum Gravity – An Overview*, Ph.D. Thesis, Osaka University, *Physics Reports* **286**, 271-348, arXiv: gr-qc/9601050

Fabbri, L., 2008: Higher-Order Theories of Gravitation, Ph.D. Thesis, arXiv: 0806.2610 [hep-th]

Fadeev, L.N., 2009: New Variables for the Einstein Theory of Gravitation, arXiv: 0911.0282 [hep-th]

Fairbairn, W. / Rovelli, C., 2004: Separable Hilbert Space in Loop Quantum Gravity, *Journal of Mathematical Physics* **45**, 2802-2814

Falkenburg, B., 1993: The Concept of Spatial Structure in Microphysics, *Philosophia Naturalis* **30**, 208-228

Falkenburg, B., 1995: *Teilchenmetaphysik – Zur Realitätsauffassung in Wissenschaftsphilosophie und Mikrophysik*, 2. Aufl., Heidelberg

Falkenburg, B., 1997: Modelle, Korrespondenz und Vereinheitlichung in der Physik, *Dialektik* **1**, 27-42

Falkenburg, B., 1998: Korrespondenz, Vereinheitlichung und die Grenzen physikalischer Erkenntnis, *Logos (Neue Folge)* **5**, 215-234

Falkenburg, B., 2002: Correspondence and the Non-Reductive Unity of Physics, in: C. Mataix / A. Rivadulla (Eds.): *Física Cuántica y Realidad – Quantum Physics and Reality*, Madrid

Falkenburg, B., 2002a: Measurement and ontology: What kind of evidence can we have for quantum fields? In: Kuhlmann / Lyre / Wayne (2002), 235-254

Falkenburg, B., 2002b: Symbol and Intuition in Modern Physics, in: M. Ferrari / I.-O. Stamatescu (Eds.): *Symbol and Physical Knowledge,* Heidelberg, 149-176

Falkenburg, B., 2004: Experience and Completeness in Physical Knowledge: Variations on a Kantian Theme, *Philosophiegeschichte und logische Analyse* **7**

Falkenburg, B., 2007: *Particle Metaphysics – A Critical Account of Subatomic Reality*, Berlin

Finelli, F. / Tronconi, A. / Venturi, G., 2007: Dark Energy, Induced Gravity and Broken Scale Invariance, arXiv: 0710.2741 [astro-ph]

Finkelstein, D.R., 1996: *Quantum Relativity: A Synthesis of the Ideas of Einstein and Heisenberg*, Berlin

Finkelstein, D.R., 2004, *Finite Quantum Relativity*, Work in Progress

Finkelstein, D.R., o.J., *Cosmputation*, Manuskript

Finkelstein, D.R. / Gibbs, J.M., 1993, Quantum Relativity, *International Journal of Theoretical Physics* **32**, 1801-1813

Fischer, A.E. / Moncrief, V., 1996: A Method of Reduction of Einstein's Equations of Evolution and a Natural Symplectic Structure on the Space of Gravitational Degrees of Freedom, *General Relativity and Gravitation* **28**, 207-219

Fischler, W. / Susskind, L., 1998: Holography and Cosmology, arXiv: hep-th/9806039

Fliessbach, T., 1995: *Allgemeine Relativitätstheorie*, Heidelberg

Flori, C., 2009: Approaches to Quantum Gravity, arXiv: 0911.2135 [gr-qc]

Fouxon, I., 2008: Page-Unruh-Wald Radiation Entropy Bound from the Second Law of Thermodynamics, arXiv: 0806.2388 [hep-th]

Frampton, P.H., 2008: Innocuous Implications of a Minimum Length in Quantum Gravity, arXiv: 0808.0283 [hep-th]

Frampton, P. / Hsu, S.D.H. / Kephart, T.W. / Reeb, D., 2008: What is the Entropy of the Universe, arXiv: 0801.1847 [hep-th]

Freidel, L., 2008: Reconstructing AdS/CFT, arXiv: 0804.0632 [hep-th]

Freivogel, B. / Susskind, L., 2004: A Framework for the Landscape, *Physical Review* **D 70**, 126007, arXiv: hep-th/0408133

Friedman, J.L. / Jack, I., 1991: Formal Commutators of the Gravitational Constraints are not well-defined, in: Ashtekar / Stachel (1991), 490-498

Friedman, J.L., 1991: Space-Time Topology and Quantum Gravity, in: Ashtekar / Stachel (1991), 539-572

Friedman, M., 1983: *Foundations of Space-Time Theories: Relativistic Physics and Philosophy of Science*, Princeton

Friedman, M., 2007: Understanding Space-Time, *Studies in History and Philosophy of Modern Physics* **38**, 216-225

Frogatt, C.D. / Nielsen, H.B., 1991: *Origin of Symmetries*, Singapore

Galison, P. / Stump, D. (Eds.), 1996: *The Disunity of Science: Boundaries, Contexts and Power*, Stanford

Gallo, E. / Marolf, D., 2008: Resource Letter BH-2: Black Holes, arXiv: 0806.2316 [astro-ph]

Gambini, R. / Porto, R.A. / Pullin, J., 2003: Consistent Discrete Gravity Solution of the Problem of Time: A Model, in: K. Kokkotas / N. Stergioulas: *Recent Developments in Gravity – Proceedings of the 10th Hellenic Relativity Conference*, Singapore, arXiv: gr-qc/0302064

Gambini, R. / Porto, R.A. / Pullin, J., 2004: A Relational Solution to the Problem of Time in Quantum Mechanics and Quantum Gravity induces a Fundamental Mechanism for Quantum Decoherence, *New Journal of Physics* **6**, 45, arXiv: gr-qc/0402118

Gambini, R. / Porto, R.A. / Pullin, J., 2005: Fundamental Gravitational Limitations to Quantum Computing, arXiv: quant-ph/0507262

Gambini, R. / Porto, R.A. / Pullin, J., 2006: Fundamental Decoherence from Quantum Gravity: A Pedagogical Review, arXiv: gr-qc/0603090

Gambini, R. / Pullin, J., 1996: *Loops, Knots, Gauge Theories and Quantum Gravity*, Cambridge

Gambini, R. / Pullin, J., 1999: Nonstandard Optics from Quantum Spacetime, *Physical Review* **D59**, 124021, arXiv: gr-qc/9809038

Gambini, R. / Pullin, J., 2003: Discrete Quantum Gravity: A Mechanism for Selecting the Value of Fundamental Constants, *International Journal of Modern Physics* **D 12**, 1775-1782, arXiv: gr-qc/0306095

Gambini, R. / Pullin, J., 2004: Consistent Discretization and Quantum Gravity, arXiv: gr-qc/0408025

Gambini, R. / Pullin, J., 2005: Classical and Quantum General Relativity: A New Paradigm, arXiv: gr-qc/0505052

Gambini, R. / Pullin, J., 2005a: Consistent Discretizations as a Road to Quantum Gravity, arXiv: gr-qc/0512065

Gambini, R. / Pullin, J., 2005b: Discrete Space-Time, arXiv: gr-qc/0505023

Gambini, R. / Pullin, J., 2008: Modern Space-Time and Undecidability, arXiv: 0801.2564 [gr-qc]

Gambini, R. / Pullin, J., 2008a: Emergent Diffeomorphism Invariance in a Discrete Loop Quantum Gravity Model, arXiv: 0807.2808 [gr-qc]

Gambini, R. / Pullin, J., 2009: Free Will, Undecidability, and the Problem of Time in Quantum Gravity, arXiv: 0903.1859 [quant-ph]

Gambini, R. / Pullin, J., 2009a: Quantum Cosmic Censor – Gravitation makes Reality Undecidable, arXiv: 0903.2438 [gr-qc]

Gambini, R. / Pullin, J., 2009b: The Montevideo Interpretation of Quantum Mechanics: Frequently Asked Questions, arXiv: 0905.4402 [quant-ph]

Gaul, M. / Rovelli, C., 2000: Loop Quantum Gravity and the Meaning of Diffeomorphism Invariance, *Lecture Notes in Physics* **541**, 277-324, arXiv: gr-qc/9910079

Gentle, A.P., 2002: Regge Calculus – A Unique Tool for Numerical Relativity, *General Relativity and Gravitation* **34**, 1701-1718, arXiv: gr-qc/0408006

Gérard, J.-M., 2007: The Strong Equivalence Principle from a Gravitational Gauge Structure, *Classical and Quantum Gravity* **24**, 1867-1877, arXiv: gr-qc/0607019

Gibbons, G.W. / Shellard, E.P.S. / Rankin, S.J. (Eds.), 2003: *The Future of Theoretical Physics and Cosmology*, Cambridge

Gibbs, P., 1995: The Small-Scale Structure of Space-Time: A Bibliographic Review, arXiv: hep-th/9506171

Giddings, S.B., 2003: The Fate of Four Dimensions, *Physical Review* **D 68**, 026006, arXiv: hep-th/0303031

Giddings, S.B., 2005: Gravity and Strings, arXiv: hep-th/0501080

Giddings, S.B., 2006: Black Hole Information, Unitarity, and Nonlocality, *Physical Review* **D 74**, 106005, arXiv: hep-th/0605196

Giddings, S.B., 2006a: Locality in Quantum Gravity and String Theory, *Physical Review* **D 74**, 106006, arXiv: hep-th/0604072

Giddings, S.B., 2007: Black Holes, Information, and Locality, arXiv: 0705.2197 [hep-th]

Giddings, S.B., 2007a: Universal Quantum Mechanics, arXiv: 0711.0757 [quant-ph]

Giddings, S.B., 2009: Nonlocality vs. Complementarity: A Conservative Approach to the Information Problem, arXiv: 0911.3395 [hep-th]

Giddings, S.B. / Mangano, M.M., 2008: Astrophysical Implications of Hypothetical Stable TeV-Scale Black Holes, arXiv: 0806.3381 [hep-ph]

Giesel, K. / Thiemann, T. 2007: Algebraic Quantum Gravity (AQG) I. Conceptual Setup, *Classical and Quantum Gravity* **24**, 2465-2497, arXiv: gr-qc/0607099

Girelli, F. / Liberati, S. / Sindoni, L., 2008: On the Emergence of Time and Gravity, arXiv: 0806.4239 [gr-qc]

Girelli, F. / Liberati, S. / Sindoni, L., 2009: Is the Notion of Time Really Fundamental?, arXiv: 0903.4876 [gr-qc]

Giulini, D., 1995: On the Configurations Space Topology in General Relativity, *Helvetica Physica Acta* **68**, 86-111, arXiv: gr-qc/9301020

Giulini, D., 1995a: What is the Geometry of Superspace, *Physical Review* **D 51**, 5630-5635, arXiv: gr-qc/9311017

Giulini, D., 2003: That Strange Procedure Called Quantization, in: D. Giulini / C. Kiefer / C. Lämmerzahl (Eds.): *Quantum Gravity – From Theory to Experimental Search*, Berlin, 17-40, arXiv: quant-ph/0304202

Giulini, D., 2006: What is (not) wrong with Scalar Gravity?, arXiv: gr-qc/0611100

Giulini, D., 2007: Some Remarks on the Notions of General Covariance and Background Independence, in: I.O. Stamatescu (Ed.): *Approaches to Fundamental Physics – An Assessment of Current Theoretical Ideas, Lecture Notes in Physics* **721**, 105ff, arXiv: gr-qc/0603087

Giulini, D., 2008: Concepts of Symmetry in the Work of Wolfgang Pauli, arXiv: 0802.4341 [physics]

Giulini, D., 2009: The Superspace of Geometrodynamics, arXiv: 0902.3923 [gr-qc]

Giulini, D., 2009a: Matter from Space, arXiv: 0910.2574 [physics]

Giulini, D. / Kiefer, C. / Lämmerzahl, C. (Eds.), 2003: *Quantum Gravity – From Theory to Experimental Search*, Berlin

Giveon, A. / Porati, M. / Rabinovici, E, 1994: Target Space Duality in String Theory, *Physics Reports* **244**, 77, arXiv: hep-th/9401139

Glinka, L.A., 2008: Quantum Gravity as the Way from Spacetime to Space Quantum States Thermodynamics, arXiv: 0803.1533 [gr-qc]

Goenner, H.F.M., 1995: Mach's Principle and Theories of Gravitation, in: Barbour / Pfister (1995) 442-457

Goenner, H.F.M., 2008: On the History of Geometrization of Space-time: From Minkowski to Finsler Geometry, arXiv: 0811.4529 [gr-qc]

Goheer, N. / Leach, J.A. / Dunsby, P.K.S., 2007: Compactifying the State Space for Alternative Theories of Gravity, arXiv: 0710.0819 [gr-qc]

Gopakumar, R., 2001: Geometry and String Theory, *Current Science* **81(12)**, 1568-1575

Gotay, M., 1986: Constraints, Reduction, and Quantization, *Journal of Mathematical Physics* **27**, 2051-2066

Goursac, A. de, 2009: Noncommutative Geometry, Gauge Theory and Renormalization, arXiv: 0910.5158 [math-ph]

Govaerts, J., 2002: The Quantum Geometer's Universe: Particles, Interactions and Topology, arXiv: hep-th/0207276

Gray, J. / He, Y.-H. / Jejjala, V. / Nelson, B.D., 2005: The Geometry of Particle Physics, arXiv: hep-th/0511062

Green, M.B., 1998: Connections between M-Theory and Superstrings, *Nuclear Physics Proceedings Supplement* **68**, 242-251, arXiv: hep-th/9712195

Green, M.B. / Schwarz, J.H. / Witten, E., 1987: *Superstring Theory*, 2 Vols., Cambridge

Greene, B., 1999: *The Elegant Universe: Superstrings, Hidden Dimensions, and the Quest for the Ultimate Theory*, New York (dt.: *Das elegante Universum: Superstrings, verborgene Dimensionen und die Suche nach der Weltformel*, Berlin, 2000)

Greene, B., 2004: *The Fabric of the Cosmos: Space, Time, and the Texture of Reality*, New York (dt.: *Der Stoff, aus dem der Kosmos ist – Raum, Zeit und die Beschaffenheit der Wirklichkeit*, München, 2004)

Grinbaum, A., 2008: On the Eve of the LHC: Conceptual Questions in High-Energy Physics, arXiv: 0806.4268 [physics]

Grishchuk, L.P., 2009: Some Uncomfortable Thoughts on the Nature of Gravity, Cosmology and the Early Universe, arXiv: 0903.4395 [gr-qc]

Grünbaum, A., 1973: *Philosophical Problems of Space and Time*, 2nd enl. Ed., Dordrecht / Boston

Guth, A.H., 2000, Inflation and Eternal Inflation, *Physics Reports* **333**, 555-574, arXiv: astro-ph/0002156

Hackett, J. / Wan, Y., 2008: Conserved Quantities for Interacting Four Valent Braids in Quantum Gravity, arXiv: 0803.3203 [hep-th]

Hackett, J. / Wan, Y., 2008a: Infinite Degeneracy of States in Quantum Gravity, arXiv: 0811.2161 [hep-th]

Hajicek, P., 1996: Time Evolution and Observables in Constrained Systems, *Classical and Quantum Gravity* **13**, 1353-1375

Hajicek, P., 1996a: Time Evolution of Observable Properties of Reparametrization-Invariant Systems, arXiv: gr-qc/9612051

Hajicek, P. / Kijowski, J., 1999: Covariant Gauge Fixing and Kuchar Decomposition, arXiv: gr-qc/9908051

Halvorson, H., 2004: On Information-Theoretic Characterizations of Physical Theories, *Studies in History and Philosophy of Modern Physics* **35**, 277-293

Hamber, H.W., 2007: Discrete and Continuum Quantum Gravity, arXiv: 0704.2895 [hep-th]

Hardy, L., 2006: Towards Quantum Gravity: A Framework for Probabilistic Theories with Non-Fixed Causal Structure, arXiv: gr-qc/0608043

Hardy, L., 2007: Quantum Gravity Computers: On the Theory of Computation with Indefinite Causal Structure, arXiv: quant-ph/0701019

Hardy, L., 2008: Formalism Locality in Quantum Theory and Quantum Gravity, arXiv: 0804.0054 [gr-qc]

Harvey, J. / Kachru, S. / Silverstein, E. (Eds.), 2001: *Strings, Branes, and Gravity – TASI 99*, Singapore

Harnad, J., 2007: Trouble with Physics?, arXiv: 0709.1728 [physics]

Hartle, J.B., 1995: Spacetime Quantum Mechanics and the Quantum Mechanics of Spacetime, in: B. Julia / J. Zinn-Justin (Eds.): *Gravitation and Quantizations, Les Houches Summer School Proceedings* **LVII**, Amsterdam, arXiv: gr-qc/9508023

Hartle, J.B., 1996: Scientific Knowledge from the Perspective of Quantum Cosmology, in: J.L. Casti / A. Karlqvist (Eds.): *Boundaries and Barriers – On the Limits to Scientific Knowledge*, Reading, Ma., arXiv: gr-qc/9601046

Hartle, J.B., 2006: Generalizing Quantum Mechanics for Quantum Spacetime, arXiv: gr-qc/0602013

Hauser, A. / Corichi, A., 2005: Bibliography of Publications related to Classical Self-Dual Variables and Loop Quantum Gravity, arXiv: gr-qc/0509039

Hawking, S.W., 1974: Black Hole Explosions, *Nature* **248**, 30ff

Hawking, S.W., 1975: Particle Creation by Black Holes, *Communications in Mathematical Physics* **43**, 199-220

Hawking, S.W., 1976: The Breakdown of Predictability in Gravitational Collapse, *Physical Review* **D 14**, 2460-2473

Hawking, S.W., 1980: *Is the End in Sight for Theoretical Physics?*, Cambridge

Hawking, S.W., 1982: The Unpredictability of Quantum Gravity, *Communications in Mathematical Physics* **87**, 395-415

Hawking, S.W., 1993: *A Brief History of Time*, New York

Hawking, S.W., 2001: Future Science, *Current Science* **81(12)**, 1614-1616

Hawking, S.W., 2005: Information Loss in Black Holes, *Physical Review* **D 72**, 084013, arXiv: hep-th/0507171

Hawking, S.W. / Hertog, T., 2002: Why does Inflation start at the Top of the Hill?, *Physical Review* **D 66**, 123509, arXiv: hep-th/0204212

Hawking, S.W. / Penrose, R., 1998: *Raum und Zeit*, Reinbek

Hawking, S.W. / Thorne, K. / Novikov, I. / Ferris, T. / Lightman, A., 2002: *The Future of Spacetime*, New York

Hawkins, E. / Markopoulou, F. / Sahlmann, H., 2003: Evolution in Quantum Causal Histories, arXiv: hep-th/0302111

Healey, R., 1995: Substance, Modality and Spacetime, *Erkenntnis* **42**, 287-316

Healey, R., 2002: Can Physics Coherently Deny the Reality of Time, in: Callender (2002) 293-316

Healey, R., 2003: Change without Change, and How to Observe it in General Relativity, http://philsci-archive.pitt.edu, Dokument: 1162

Healey, R., 2004: Gauge Theories and Holism, http://philsci-archive.pitt.edu, Dokument: 1163

Healey, R., 2007: Gauge Symmetry and the Theta Vacuum, http://philsci-archive.pitt.edu, Dokument: 3419

Hedrich, R., 1990: *Komplexe und fundamentale Strukturen – Grenzen des Reduktionismus*, Mannheim / Wien / Zürich

Hedrich, R., 1993: Die nicht ganz so unglaubliche Effizienz der Mathematik in den Naturwissenschaften, *Philosophia Naturalis* **30/1**, 106-125

Hedrich, R., 1994: *Die Entdeckung der Komplexität – Skizzen einer strukturwissenschaftlichen Revolution*, Frankfurt am Main / Thun

Hedrich, R., 1995: Unsere epistemische Situation, ihre Grenzen und ihre neuronalen Determinanten – Der Objektbegriff, *Philosophia Naturalis* **32/1**, 117-139

Hedrich, R., 1995a: Was ist eine physikalische Theorie ?, *Praxis der Naturwissenschaften – Physik* **1/44**, 10-16

Hedrich, R., 1996: Artikel 'Naturphilosophie', in: P. Prechtl / F.-P. Burkard (Hrsg.): *Metzler-Philosophie-Lexikon*, Stuttgart, 2., erw. Aufl. (1999)

Hedrich, R., 1996a: Artikel 'Physik', in: P. Prechtl / F.-P. Burkard (Hrsg.): *Metzler-Philosophie-Lexikon*, Stuttgart, 2., erw. Aufl. (1999)

Hedrich, R., 1998: *Erkenntnis und Gehirn – Realität und phänomenale Welten innerhalb einer naturalistisch-synthetischen Erkenntnistheorie*, Paderborn

Hedrich, R., 1998a: Las bases materiales de nuestras capacidades epistémicas – Ensayo de una revisión sintética de la epistemología, *Argumentos de Razón Técnica* **1**, 91-109

Hedrich, R., 1999: Die materialen Randbedingungen epistemischer Leistungen, *Philosophia Naturalis* **36/2**, 237-262

Hedrich, R., 1999a: Artikel 'Naturwissenschaft', in: H.J. Sandkühler (Hrsg.): *Enzyklopädie Philosophie*, Hamburg, 930-933

Hedrich, R., 2001: The Naturalization of Epistemology and the Neurosciences, *Epistemologia – Rivista Italiana di Filosofia della Scienza / An Italian Journal for the Philosophy of Science* **24/2**, 271-300

Hedrich, R., 2002: Anforderungen an eine physikalische Fundamentaltheorie, *Zeitschrift für Allgemeine Wissenschaftstheorie / Journal for General Philosophy of Science* **33/1**, 23-60

Hedrich, R., 2002a: Superstring Theory and Empirical Testability, http://philsci-archive.pitt.edu, Dokument: 0608

Hedrich, R., 2002b: Epistemische Grenzen, in: W. Hogrebe (Hrsg.): *Grenzen und Grenzüberschreitungen – XIX. Deutscher Kongress für Philosophie – Bonn 2002*, Bonn, 317-327

Hedrich, R., 2002c: Zelluläre Automaten – Diskrete Modelle der Welt?, *Philosophia Naturalis* **39/1**, 1-24

Hedrich, R., 2004: Beschleunigte Expansion und neue Kosmologie, *Praxis der Naturwissenschaften – Physik* **2/53**, 25-29

Hedrich, R., 2005: In welcher Welt leben wir? – Superstrings, Kontingenz und Selektion, in: G. Abel (Hrsg.): *Kreativität – XX. Deutscher Kongress für Philosophie – Sektionsbeiträge*, Band 1, Universitätsverlag der TU Berlin, Berlin, S. 867-879

Hedrich, R., 2005a: In welcher Welt leben wir? – Physikalische Vereinheitlichung, Kontingenz und Selektion im Superstring-Ansatz, in: W. Lütterfelds (Hrsg.): *Vom Sinn und Unsinn des menschlichen Lebens*, Universität Passau, Passau, S. 5-26

Hedrich, R., 2006: String Theory – From Physics to Metaphysics, (i) *Physics and Philosophy* (Online-Zeitschrift) 2006, 005; (ii) http://philsci-archive.pitt.edu, Dokument: 2709; (iii) arXiv: physics/0604171

Hedrich, R., 2007: *Von der Physik zur Metaphysik – Physikalische Vereinheitlichung und Stringansatz*, Frankfurt am Main / Paris / Ebikon / Lancaster / New Brunswick

Hedrich, R., 2007a: The Internal and External Problems of String Theory – A Philosophical View, (i) *Zeitschrift für Allgemeine Wissenschaftstheorie / Journal for General Philosophy of Science* **38** (2007) 261-278; (ii) http://philsci-archive.pitt.edu, Dokument: 3012; (iii) arXiv: physics/0610168

Hedrich, R., 2007b: Kohärenz und Kontingenz – Grundlagen der Superstring-Theorie, in: B. Falkenburg (Hrsg.): *Natur – Technik – Kultur: Philosophie im interdisziplinären Dialog*, Paderborn (2007), 91-107

Hedrich, R., 2008: Motivationen, konzeptionelle Randbedingungen und Alternativen für eine Theorie der Quantengravitation, in: C.F. Gethmann (Hrsg.): *Lebenswelt und Wissenschaft – XXI Deutscher Kongress für Philosophie – Sektionsbeiträge*, Essen (2008) (CD-ROM-Veröffentlichung)

Hedrich, R., 2009: Quantum Gravity: Has Spacetime Quantum Properties?, (i) arXiv:0902.0190 [gr-qc]; (ii) http://philsci-archive.pitt.edu, Dokument: 4445

Hedrich, R., 2009a: Quantum Gravity: Motivations and Alternatives, (i) arXiv:0908.0355 [gr-qc]; (ii) http://philsci-archive.pitt.edu, Dokument: 4820

Hedrich, R. / Kuhn, W., 1989: Der drohende infinite Regress materieller Strukturen, *Praxis der Naturwissenschaften – Physik* **7/38**, 31-36; wiederabgedruckt in: W. Kuhn: *Physik: Erleben – Lehren – Lernen*, Aulis-Verlag Deubner, Köln (1993) S. 188-193

Hehl, F.W. / Mashhoon, B., 2009: A Formal Framework for a Nonlocal Generalization of Einstein's Theory of Gravitation, arXiv: 0902.0560 [gr-qc]

Helfer, A., 2008: The Produktion of Time, arXiv: 0812.0605 [gr-qc]

Henneaux, M. / Teitelboim, C., 1992: *Quantization of Gauge Systems*, Princeton

Henson, J., 2005: Comparing Causality Principles, *Studies in History and Philosophy of Modern Physics* **36**, 519-543, arXiv: quant-ph/0410051

Henson, J., 2006: The Causal Set Approach to Quantum Gravity, arXiv: gr-qc/0601121

Henson, J., 2009: Quantum Histories and Quantum Gravity, arXiv: 0901.4009 [gr-qc]

Hilgevoord, J. (Ed.), 1995: *Physics and Our View of the World*, Cambridge

Hoefer, C., 1996: The Metaphysics of Space-Time Substantialism, *Journal of Philosophy* **93**, 5-27

Hoefer, C., 1998: Absolute Versus Relational Spacetime: For Better or Worse, the Debate goes on, *British Journal for the Philosophy of Science* **49**, 451-467

Hogan, C.J., 2007: Holographic Indeterminacy, Uncertainty and Noise, arXiv: 0709.0611 [astro-ph]

Hogan, C.J., 2007a: Quantum Indeterminacy of Emergent Spacetime, arXiv: 0710.4153 [gr-qc]

Hogan, C.J., 2007b: Spacetime Indeterminacy and Holographic Noise, arXiv: 0706.1999 [gr-qc]

Hogan, C.J., 2007c: The New Science of Gravitational Waves, arXiv: 0709.0608 [astro-ph]

Hong, S.E. / Yeom, D.-H. / Zoe, H., 2009: Critical Reviews on Holographic Measure over the Multiverse, arXiv: 0903.2370 [gr-qc]

Horava, P., 1999: M-Theory as a Holographic Field Theory, *Physical Review* **D 59**, 046004, arXiv: hep-th/9712130

Horowitz, G.T., 1991: String Theory without Space-Time, in: Ashtekar / Stachel (1991), 299-325

Horowitz, G.T., 1991a: Topology Change in Classical and Quantum Gravity, *Classical and Quantum Gravity* **8**, 587-601

Horowitz, G.T., 1997: Quantum States of Black Holes, arXiv: gr-qc/9704072

Horowitz, G.T., 2005: Spacetime in String Theory, in: J. Pullin / R. Price (Eds.): *Spacetime 100 Years Later*, arXiv: gr-qc/0410049

Horowitz, G.T., 2007: Black Holes, Entropy, and Information, arXiv: 0708.3680 [astro-ph]

Horowitz, G.T. / Maldacena, J., 2004: The Black-Hole Final State, *Journal of High Energy Physics* **0402**, 008, arXiv: hep-th/0310281

Horwich, P., 1987: *Asymmetries in Time*, Cambridge, Mass.

Hossenfelder, S., 2006: Phenomenological Quantum Gravity, arXiv: hep-th/0611017

Hossenfelder, S. / Smolin, L., 2009: Phenomenological Quantum Gravity, arXiv: 0911.2761 [physics]

Hsu, S.D.H., 2007: Information, Information Processing and Gravity, arXiv: 0704.1154 [hep-th]

Hsu, S.D.H. / Reeb, D., 2007: Black Hole Entropy, Curved Space and Monsters, arXiv: 0706.3239 [hep-th]

Hsu, S.D.H. / Reeb, D., 2009: Black Holes, Information and Decoherence, arXiv: 0903.2258 [gr-qc]

Hsu, S.D.H. / Reeb, D., 2009a: Monsters, Black Holes and the Statistical Mechanics of Gravity, arXiv: 0908.1265 [gr-qc]

Hu, B.L., 2005: Can Spacetime be a Condensate, *International Journal of Theoretical Physics* **44**, 1785-1806, arXiv: gr-qc/0503067

Hu, B.L., 2009: Emergent Quantum Gravity: Macro/Micro Structures of Spacetime, arXiv: 0903.0878 [gr-qc]

Hu, B.L. / Verdaguer, E., 2003: Stochastic Gravity: A Primer with Applications, *Classical and Quantum Gravity* **20**, R1-R42, arXiv: gr-qc/0211090

Hu, B.L. / Verdaguer, E., 2004: Stochastic Gravity: Theory and Applications, *Living Reviews in Relativity* (Electronic Journal) **7/3**, www.livingreviews.org; auch: arXiv: gr-qc/0307032

Hu, B.L. / Verdaguer, E., 2008: Stochastic Gravity: Theory and Applications, arXiv: 0802.0658 [gr-qc]

Huggett, N., 2006: The Regularity Account of Relational Spacetime, *Mind* **115**, 41-73

Huggett, N., 2007: Why the Parts of Absolute Space are Immobile, Manuskript

Huggett, N. / Hoefer, C., 2006: Absolute and Relational Theories of Space and Motion, in: E.N. Zalta (Ed.): *Stanford Encyclopedia of Philosophy*, http://plato.stanford.edu

Huggett, N. / Weingard, R., 1999: Gauge Fields, Gravity and Bohm's Theory, in: T.Y. Cao (Ed.): *Conceptual Foundations of Quantum Field Theory*, Cambridge, 287-297

Huggett, S.A. / Tod, K.P., 1985: *An Introduction to Twistor Theory*, Cambridge

Huggett, S.A. et al. (Eds.), 1998: *The Geometric Universe*, Oxford

Hughes, S.A., 2005: Trust but verify: The case for astrophysical black holes, arXiv: hep-ph/0511217

Hull, C.M. / Townsend, P.K., 1995: Unity of Superstring Dualities, *Nuclear Physics* **B 438**, 109, arXiv: hep-th/9410167

Husain, V., 2009: Time, Vacuum Energy, and the Cosmological Constant, arXiv: 0906.5562 [gr-qc]

Hut, P. / Alford, M. / Tegmark, M., 2005: On Math, Matter and Mind, arXiv: physics/0510188

Iliopoulos, J., 2008: Physics Beyond the Standard Model, arXiv: 0807.4841 [hep-ph]

Isham, C.J., 1981: Quantum Gravity – An Overview, in: Isham / Penrose / Sciama (1981) 1-62

Isham, C.J., 1991: Canonical Groups and the Quantization of Geometry and Topology, in: Ashtekar / Stachel (1991), 351-400

Isham, C.J., 1993: Canonical Quantum Gravity and the Problem of Time, in: *Integrable Systems, Quantum Groups, and Quantum Field Theory*, Dordrecht, 157-288, arXiv: gr-qc/9210011

Isham, C.J., 1994: Prima Facie Questions in Quantum Gravity, in: Ehlers / Friedrich (1994), 1-21, arXiv: gr-qc/9310031

Isham, C.J., 1997: Structural Issues in Quantum Gravity, in: M. Francaviglia / G. Longhi / L. Lusanna / E. Sorace (Eds.): *Florence 1995, General Relativity and Gravitation*, Singapore, arXiv: gr-qc/9510063

Isham, C.J. / Penrose, R. / Sciama, D.W. (Eds.), 1975: *Quantum Gravity. An Oxford Symposium*, Oxford

Isham, C.J. / Penrose, R. / Sciama, D.W. (Eds.), 1981: *Quantum Gravity 2. A Second Oxford Symposium*, Oxford

Ismael, J., 2002: Rememberances, Mementos, and Time Capsules, in: Callender (2002) 317-328

Jackson, M.G. / Hogan, C.J., 2007: A New Spin on Quantum Gravity, arXiv: hep-th/0703133

Jacobson, T., 1991: Unitarity, Causality and Quantum Gravity, in: Ashtekar / Stachel (1991), 212-216

Jacobson, T., 1991a: Black-Hole Thermodynamics and the Space-Time Continuum, in: Ashtekar / Stachel (1991), 597-600

Jacobson, E., 1995: Thermodynamics of Spacetime: The Einstein Equation of State, *Physical Review Letters* **75**, 1260-1263, arXiv: gr-qc/9504004

Jacobson, T., 1999: On the Nature of Black Hole Entropy, arXiv: gr-qc/9908031

Jacobson, T., 1999a: Trans-Planckian Redshifts and the Substance of the Space-Time River, *Progress in Theoretical Physics (Supplement)* **136**, 1-17, arXiv: hep-th/0001085

Jacobson, T., 2007: Renormalization and Black Hole Entropy in Loop Quantum Gravity, arXiv: 0707.4026 [gr-qc]

Jacobson, T., 2007a: When is $g_{tt} g_{rr} = -1$ ?, arXiv: 0707.3222 [gr-qc]

Jacobson, T., 2008: Einstein-Aether Gravity: A Status Report, arXiv: 0801.1547 [gr-qc]

Jacobson, T. / Liberati, S. / Mattingly, D., 2006: Lorentz Violation at High Energy: Concepts, Phenomena and Astrophysical Constraints, *Annals of Physics* **321**, 150-196, arXiv: astro-ph/0505267

Jacobson, T. / Marolf, D. / Rovelli, C., 2005: Black Hole Entropy: inside or out?, *International Journal of Theoretical Physics* **44**, 1807-1837, arXiv: hep-th/0501103

Jacobson, T. / Parentani, R., 2003: Horizon Entropy, *Foundations of Physics* **33**, 323

Jaekel, M.-T. / Reynaud, S., 2008: Mass, Inertia and Gravitation, arXiv: 0812.3936 [gr-qc]

Jammer, M., 1993: *Concepts of Space: The History of Theories of Space in Physics* (3rd Ed.), New York

Jannes, G., 2008: On the Condensed Matter Scheme for Emergent Gravity and Interferometry, arXiv: 0810.0613 [gr-qc]

Jannes, G., 2009: Emergent Gravity: The BEC Paradigm, arXiv: 0907.2839 [gr-qc]

Jannes, G., 2009a: Some comments on "The Mathematical Universe, arXiv: 0904.0867 [gr-qc]

Jejjala, V. / Kavic, M. / Minic, D., 2007: Time and M-Theory, arXiv: 0706.2252 [hep-th]

Jejjala, V. / Kavic, M. / Minic, D. / Tze, C.-H., 2008: On the Origin of Time and the Universe, arXiv: 0804.3598 [hep-th]

Jenkins, A., 2009: Constraints on Emergent Gravity, arXiv: 0904.0453 [gr-qc]

Johansson, L.-G. / Matsubara, K., 2009: String Theory and General Methodology; a Reciprocal Evaluation, arXiv: 0912.3160 [physics]

Johnson, C., 2002: *D Branes*, Cambridge

Johnston, S., 2008: Particle Propagators on Discrete Spacetime, arXiv: 0806.3083 [hep-th]

Jones, N., 2009: General Relativity and the Standard Model: Why Evidence for one does not Disconfirm the other, *Studies in the History and Philosophy of Modern Physics* **40**, 124-132

Jones, R., 1991: Realism about What?, *Philosophy of Science* **58**, 185-202

Joos, E., 1986: Why Do We Observe a Classical Spacetime?, *Physics Letters* **A 116**, 6-8

Kaiser, G., 2009: Quantum Physics, Relativity, and Complex Spacetime: Towards a New Synthesis, arXiv: 0910.0352 [math-ph]

Kaku, M., 1999: *Introduction to Superstrings and M-Theory*, 2nd Ed., New York

Kaluza, T., 1921: Zum Unitätsproblem der Physik, *Sitzungsberichte der Preussischen Akademie der Wissenschaften Berlin 1921*, 967ff

Kanitscheider, B., 1971: *Geometrie und Wirklichkeit*, Berlin

Kanitscheider, B., 1976: *Vom absoluten Raum zur dynamischen Geometrie*, Mannheim

Kaplunovsky, V. / Weinstein, M., 1985: Space-Time: Arena or Illusion?, *Physical Review* **D 31**, 1879-1898

Katz, E. / Okui, T., 2007: The 't Hooft Model as a Hologram, arXiv: 0710.3402 [hep-th]

Kauffman, S. / Smolin, L., 2000: Combinatorial Dynamics in Quantum Gravity, *Lecture Notes in Physics* **541**, 101-129, arXiv: hep-th/9809161

Kempf, A. / Martin, R., 2007: On Information Theory, Spectral Geometry and Quantum Gravity, arXiv: 0708.0062 [gr-qc]

Kent, A., 2009: One World versus Many: The Inadequacy of Everettian Accounts of Evolution, Probability, and Scientific Confirmation, arXiv: 0905.0624 [quant-ph]

Kiefer, C., 1990: Der Zeitbegriff in der Quantengravitation, *Philosophia Naturalis* **27**, 43ff

Kiefer, C., 1994: Probleme der Quantengravitation, *Philosophia Naturalis* **31**, 309-327

Kiefer, C., 2004: *Quantum Gravity*, Oxford ($^2$2007)

Kiefer, C., 2005: Quantum Gravity: General Introduction and Recent Developments, *Annalen der Physik* **15**, 129-148, arXiv: gr-qc/0508120

Kiefer, C., 2008: Quantum Geometrodynamics: Whence, Whither?, arXiv: 0812.0295 [gr-qc]

Kiefer, C., 2009: Does Time Exist in Quantum Gravity?, arXiv: 0909.3767 [gr-qc]

Kilmister, C., 1996: Time: And Space?, *Studies in History and Philosophy of Modern Physics* **27**, 525-531

Kiriushcheva, N. / Kuzmin, S.V., 2008: The Hamiltonian Formulation of General Relativity: Myths and Reality, arXiv: 0809.0097 [gr-qc]

Kiriushcheva, N. / Kuzmin, S.V., 2009: The Hamiltonian of Einstein Affine-Metric Formulation of General Relativity, arXiv: 0912.3396 [gr-qc]

Klammer, D. / Steinacker, H., 2009: Cosmological Solutions of Emergent Noncommutative Gravity, arXiv: 0903.0986 [gr-qc]

Klauder, J.R., 2006: Fundamentals of Quantum Gravity, arXiv: gr-qc/0612168

Klebanov, I.R., 2001: TASI Lectures – Introduction to the AdS/CFT Correspondence, in: Harvey / Kachru / Silverstein (2001), arXiv: hep-th/0009139

Klein, O., 1926: Quantentheorie und fünfdimensionale Relativitätstheorie, *Zeitschrift für Physik* **37**, 895-906

Klinkhamer, F.R., 2007: Fundamental Length Scale of Quantum Spacetime Foam, arXiv: gr-qc/0703009

Kober, M., 2009: Quantum Theory of Ur Objects and General Relativity, arXiv: 0905.3828 [hep-th]

Kochan, D., 2008: Does Path Integral Really Need a Lagrangian/Hamiltonian?, arXiv: 0812.0678 [quant-ph]

Kocharyan, A.A., 2009: Is Nonrelativistic Gravity Possible?, arXiv: 0905.4204 [hep-th]

Konopka, T. / Markopoulou, F., 2006: Constrained Mechanics and Noiseless Subsystems, arXiv: gr-qc/0601028

Konopka, T. / Markopoulou, F. / Severini, S., 2008: Quantum Graphity: A Model of Emergent Locality, *Physical Review* **D 77**, 104029, arXiv: 0801.0861 [hep-th]

Konopka, T. / Markopoulou, F. / Smolin, L., 2006: Quantum Graphity, arXiv: hep-th/0611197

Kowalski-Glikman, J., 2005: Introduction to Doubly Special Relativity, *Lecture Notes in Physics* **669**, 131-159, arXiv: hep-th/0405273

Kragh, H. / Carazza, B., 1994: From Time Atoms to Space-Time Quantization: The Idea of Discrete Time, *Studies in History and Philosophy of Science* **25**, 437-462

Krasnov, K., 1999: *Spin Foam Models*, Ph.D. Thesis, Pennsylvania State University, http://cgpg.gravity.psu.edu/archives/thesis/1999/kirill.pdf

Krasnov, K., 2007: Non-Metric Gravity: A Status Report, arXiv: 0711.0697 []

Kreimer, D., 2008: Not so Non-Renormalizable Gravity, arXiv: 0805.4545 [hep-th]

Kribs, D.W. / Markopoulou, F., 2005: Geometry from Quantum Particles, arXiv: gr-qc/0510052

Kuchar, K., 1986: Canonical Geometrodynamics and General Covariance, *Foundations of Physics* **16**, 193-208

Kuchar, K., 1991: The Problem of Time in Canonical Quantization of Relativistic Systems, in: Ashtekar / Stachel (1991), 141-171

Kuchar, K., 1992: Time and Interpretation of Quantum Gravity, in: G. Kunstatter et al. (Eds.): *Proceedings of the 4th Canadian Conference on General Relativity and Relativistic Astrophysics*, Singapore, auch: www.phys.lsu.edu/faculty/pullin/kvk.pdf

Kuchar, K., 1993: Canonical Quantum Gravity, arXiv: gr-qc/9304012

Kuchar, K., 1993a: Matter Time in Canonical Quantum Gravity, in: B. Hu / M. Ryan / C. Vishveshvara (Eds.): *Directions in General Relativity*, Cambridge, 201-221

Kuchar, K., 1999: The Problem of Time in Quantum Geometrodynamics, in: Butterfield (1999) 169-195

Kuchar, K. / Torre, C.G., 1991: Strings as Poor Relatives of General Relativity, in: Ashtekar / Stachel (1991), 326-348

Kuhlmann, M. / Lyre, H. / Wayne, A. (Eds.), 2002: *Ontological Aspects of Quantum Field Theory*, Singapore

Kuhlmann, M. / Lyre, H. / Wayne, A., 2002a: Introduction, in: Kuhlmann / Lyre / Wayne (2002)

Kurita, Y. / Kobayashi, M. / Morinari, T. / Tsubota, M. / Ishihara, H., 2008: Spacetime Analogue of Bose-Einstein Condensates: Bogoliubov-de Gennes Formulation, arXiv: 0810.3088 [cond-mat]

Kutasov, D. / Lunin, O. / McOrist, J. / Royston, A.B., 2009: Dynamical Vacuum Selection in String Theory, arXiv: 0909.3319 [hep-th]

Lam, V., 2007: The Singular Nature of Space-Time, http://philsci-archive.pitt.edu

Lämmerzahl, C., 2004: General Relativity in Space and Sensitive Tests of the Equivalence Principle, arXiv: gr-qc/0402122

Laughlin, R.B., 2003: Emergent Relativity, *International Journal of Modern Physics* **A18**, 831-854, arXiv: gr-qc/0302028

Leeds, S., 1995: Holes and Determinism: Another Look, *Philosophy of Science* **62**, 425-437

Lehnert, R., 2007: Quantum Gravity and Spacetime Symmetries, arXiv: gr-qc/0701006

Lehnert, R., 2007a: Violations of Spacetime Symmetries, arXiv: 0711.4851 [hep-th]

Lemos, J.P.S., 2005: Black Holes and Fundamental Physics, arXiv: gr-qc/0507101

Lerche, W., 2000: *Recent Developments in String Theory*, Wiesbaden; auch: arXiv: hep-th/9710246 (ältere Fassung)

Lev, F., 1993: Finiteness of Physics and its Possible Consequences, *Journal of Mathematical Physics* **34**, 490-527

Lev, F.M., 2009: Is Gravity and Interaction?, arXiv: 0905.0767 [physics]

Liberati, S. / Girelli, F. / Sindoni, L., 2009: Analogue Models for Emergent Gravity, arXiv: 0909.3834 [gr-qc]

Liebscher, D.-E., 1985: Classical and Quantum Pregeometry, in: M.A. Markov et al. (Eds.): *Proceedings of the Third Seminar on Quantum Gravity*, Singapore, 223-227

Lin, C.-Y., 2009: Emergence of General Relativity from Loop Quantum Gravity, arXiv: 0912.0554 [gr-qc]

Linde, A., 2004: Inflation, Quantum Cosmology, and the Anthropic Principle, in: Barrow / Davies / Harper (2004), 426-458

Liu, C., 1996: Realism and Spacetime: Of Arguments against Metaphysical Realism and Manifold Realism, *Philosophia Naturalis* **33**, 243-263

Livine, E.R. / Oriti, D., 2003: Implementing Causality in the Spin Foam Quantum Geometry, *Nuclear Physics* **B 663**, 231-279, arXiv: gr-qc/0210064

Livine, E.R. / Terno, D.R., 2007: Quantum Causal Histories in the Light of Quantum Information, *Physical Review* **D75**, 084001, arXiv: gr-qc/0611135

Lizzi, F., 2008: The Structure of Spacetime and Noncommutative Geometry, arXiv: 0811.0268 [hep-th]

Lloyd, S., 1999: Universe as Quantum Computer, arXiv: quant-ph/9912088

Lloyd, S., 2005: The Computational Universe – Quantum Gravity from Quantum Computation, arXiv: quant-ph/0501135

Lloyd, S., 2005a: A Theory of Quantum Gravity based on Quantum Computation, arXiv: quant-ph/0501135

Lloyd, S., 2006: Almost Certain Escape from Black Holes, *Physical Review Letters* **96**, 061302, arXiv: quant-ph/0406205

Lloyd, S., 2007: *Programming the Universe: A Quantum Computer Scientist takes on the Cosmos*, New York

Lobo, F.S.N., 2007: Nature of Time and Causality in Physics, arXiv: 0710.0428 [gr-qc]

Lobo, F.S.N., 2008: The Dark Side of Gravity: Modified Theories of Gravity, arXiv: 0807.1640 [gr-qc]

Loll, R., 1998: Discrete Approaches to Quantum Gravity in Four Dimensions, *Living Reviews in Relativity* (Electronic Journal), www.livingreviews.org

Loll, R., 2001: Discrete Lorentzian Quantum Gravity, *Nuclear Physics* **B 94** (Proc. Suppl.), 96-107, arXiv: hep-th/0011194

Loll, R., 2003: A Discrete History of the Lorentzian Path Integral, in: D. Giulini / C. Kiefer / C. Lämmerzahl (Eds.): *Quantum Gravity – From Theory to Experimental Search*, Berlin, auch: arXiv: hep-th/0212340

Loll, R., 2007: The Emergence of Spacetime, or, Quantum Gravity on Your Desktop, arXiv: 0711.0273 [gr-qc]

Loll, R. / Ambjorn, J. / Jurkiewicz, J., 2005: The Universe from Scratch, arXiv: hep-th/0509010

Lopis, F. /Tegmark, M., 2008: Relativity Revisited, arXiv: 0804.0016 [astro-ph]

Lorente, M., 2007: Causal Spin Foams and the Ontology of Spacetime, arXiv: 0712.1600 [gr-qc]

Lusanna, L., 2009: Post-Minkowskian Gravity: Dark Matter as a Relativistic Inertial Effect, arXiv: 0912.2935 [gr-qc]

Lusanna, L. / Pauri, M., 2004: The Physical Role of Gravitational and Gauge Degrees of Freedom in General Relativity – I: Dynamical Synchronization and Generalized Inertial Effects, *General Relativity and Gravitation* **38**, 187-227, arXiv: gr-qc/0403081

Lusanna, L. / Pauri, M., 2004a: The Physical Role of Gravitational and Gauge Degrees of Freedom in General Relativity – II: Dirac versus Bergmann Observables and the Objectivity of Spacetime, *General Relativity and Gravitation* **38**, 229-267, arXiv: gr-qc/0407007

Lusanna, L. / Pauri, M., 2005: General Covariance and the Objectivity of Space-Time Point-Events, arXiv: gr-qc/0503069

Lusanna, L. / Pauri, M., 2006: Explaining Leibniz Equivalence as Difference of Non-Inertial Appearances: Dis-Solution of the Hole Argument and Physical Individuation of Point-Events, http://philsci-archive.pitt.edu, Dokument: 2714

Lusanna, L. / Pauri, M., 2006a: Dynamical Emergence of Instantaneous 3-Spaces in a Class of Models of General Relativity, http://philsci-archive.pitt.edu, Dokument: 3032

Lüst, D. / Stieberger, S. / Taylor, T.R., 2008: The LHC String Hunter's Companion, arXiv: 0807.3333 [hep-th]

Lynden-Bell, D., 1967: On the Origins of Spacetime and Inertia, *Monthly Notices of the Royal Astronomical Society* **135**, 453-467

Lyre, H., 1999: Gauges, Holes, and their 'Connections', arXiv: gr-qc/9904036

Lyre, H., 2001: The Principles of Gauging, *Philosophy of Science* **68**, S371-S381, auch: http://philsci-archive.pitt.edu, Dokument: 0113

Lyre, H., 2004: *Lokale Symmetrien und Wirklichkeit. Eine naturphilosophische Studie über Eichtheorien und Strukturenrealismus,* Paderborn

Lyre, H., 2005: Metaphysik im 'Handumdrehen' – Kant und Earman, Parität und moderne Raumauffassung, *Philosophia Naturalis* **42**, 49-76

Lyre, H., 2007: Philosophische Probleme von Raumzeit-Theorien, in: A. Bartels / M. Stöckler (Hg.): *Wissenschaftstheorie – Ein Studienbuch,* Paderborn

Lyre, H., 2008: Time in Philosophy of Physics – The Central Issues, *Physics and Philosophy* (Online-Zeitschrift) 2008, 012

Lyre, H. / Eynck, T.O., 2003: Curve it, Gauge it or Leave it? – Practical Underdetermination in Gravitational Theories, *Journal for General Philosophy of Science* **34**, 277-303; auch: http://philsci-archive.pitt.edu, Dokument: 0514

Macias, A. / Quevedo, H., 2006: Time Paradox in Quantum Gravity, arXiv: gr-qc/0610057

Mack, G., 1994: Gauge Theory of Things Alive and Universal Dynamics, arXiv: hep-lat/9411059

Mack, G., 1997: Pushing Einstein's Principles to the Extreme, arXiv: gr-qc/9704034

MacKenzie, R., 2008: Is Faith the Enemy of Science, arXiv: 0807.3670 [physics]

Maidens, A., 1998: Symmetry Groups, Absolute Objects and Action Principles in General Relativity, *Studies in the History and Philosophy of Modern Physics* **29**, 245-272

Majid, S., 2006: Algebraic Approach to Quantum Gravity II: Noncommutative Spacetime, arXiv: hep-th/0604130

Majid, S., 2007: Algebraic Approach to Quantum Gravity I: Relative Realism, http://philsci-archive.pitt.edu

Major, S., 1997: *q-Quantum Gravity*, Ph.D. Thesis, Pennsylvania State University, http://cgpg.gravity.psu.edu/archives/thesis/1997/seth.pdf

Majumdar, P., 2009: Holography, Gauge-Gravity Connection and Black Hole Entropy, arXiv: 0903.5080 [gr-qc]

Mäkelä, J., 2006: Area and Entropy – A New Perspective, arXiv: gr-qc/0605098

Mäkelä, J., 2007: Quantum-Mechanical Model of Spacetime, arXiv: gr-qc/0701128

Mäkelä, J., 2008: Quantum-Mechanical Model of Spacetime I: Microscopic Properties of Spacetime, arXiv: 0805.3952 [gr-qc]

Mäkelä, J., 2008a: Quantum-Mechanical Model of Spacetime II: Thermodynamics of Spacetime, arXiv: 0805.3955 [gr-qc]

Mäkelä, J., 2008b: Partition Function of Spacetime, arXiv: 0810.4910 [gr-qc]

Mäkelä, J. / Peltola, A., 2006: Gravitation and Thermodynamics: The Einstein Equation of State Revisited, arXiv: gr-qc/0612078

Maldacena, J.M., 1996: *Black Holes in String Theory*, Ph.D. Thesis, arXiv: hep-th/9607235

Maldacena, J.M., 1999: The Large N Limit of Superconformal Field Theories and Supergravity, *International Journal of Theoretical Physics* **38**, 1113-1133

Maldacena, J.M., 2003: TASI 2003 Lectures on AdS/CFT, arXiv: hep-th/0309246

Maldacena, J.M., 2004: Quantum Gravity as an Ordinary Gauge Theory, in: Barrow / Davies / Harper (2004), 153-166

Manders, K., 1982: On the Space-Time Ontology of Physical Theories, *Philosophy of Science* **49**, 575-590

Mannheim, P.D., 2007: Conformal Gravity Challenges String Theory, arXiv: 0707.2283 [hep-th]

Manrique, E. / Oeckl, R. / Weber, A. / Zapata, J., 2006: Loop Quantization as a Continuum Limit, *Classical and Quantum Gravity* **23**, 3393-3404, arXiv: hep-th/0511222

Markopoulou, F., 2000: The Internal Description of a Causal Set: What the Universe looks like from the inside, *Communications in Mathematical Physics* **211**, 559, arXiv: gr-qc/9811053

Markopoulou, F., 2000a: Quantum Causal Histories, *Classical and Quantum Gravity* **17**, 2059, arXiv: hep-th/9904009

Markopoulou, F., 2000b: An Insider's Guide to Quantum Causal Histories, *Nuclear Physics* **88** (Proc. Suppl.), 308-313, arXiv: hep-th/9912137

Markopoulou, F., 2004: Planck-Scale Models of the Universe, in: Barrow / Davies / Harper (2004) 550-563, arXiv: gr-qc/0210086

Markopoulou, F., 2006: Towards Gravity from the Quantum, arXiv: hep-th/0604120

Markopoulou, F., 2007: New Directions in Background Independent Quantum Gravity, arXiv: gr-qc/0703097

Markopoulou, F., 2009: Space does not exist, so time can, arXiv: 0909.1861 [gr-qc]

Markopoulou, F. / Smolin, L., 1997: Causal Evolution of Spin Networks, *Nuclear Physics* **B 508**, 409, arXiv: gr-qc/9702025

Markopoulou, F. / Smolin, L., 1999: Holography in a Quantum Spacetime, arXiv: hep-th/9910146

Markopoulou, F. / Smolin, L., 2004: Quantum Theory from Quantum Gravity, *Physical Review* **D 70**, 124029, arXiv: gr-qc/0311059

Markosian, N., 2002: Time, in: E.N. Zalta (Ed.): *Stanford Encyclopedia of Philosophy*, http://plato.stanford.edu

Marlow, A.R., 1986: Quantum Theoretical Origin of Spacetime Structure, *International Journal of Theoretical Physics* **25**, 561-571

Marolf, D., 2004: Resource Letter: The Nature and Status of String Theory, *American Journal of Physics* **72**, 730, arXiv: hep-th/0311044

Marolf, D., 2008: Unitarity and Holography in Gravitational Physics, arXiv: 0808.2842 [gr-qc]

Marolf, D., 2008a: Holographic Thought Experiments, arXiv: 0808.2845 [gr-qc]

Marsh, G.E., 2007: Charge, Geometry, and Effective Mass, arXiv: 0708.1958 [gr-qc]

Martin, C., 2002: Gauge Principles, Gauge Arguments and the Logic of Nature, *Philosophy of Science* **69**, S221-S234

Martin, C., 2003: On Continuous Symmetries and the Foundations of Modern Physics, in: Brading / Castellani (2003), 29-60

Martin, T.D., 2004: Comments on Cahill's Quantum Foam Inflow Theory of Gravity, arXiv: gr-qc/0407059

Mason, L.J., 1991: Insights from Twistor Theory, in: Ashtekar / Stachel (1991), 499-511

Mathur, S.D., 2007: Falling into a Black Hole, arXiv: 0705.3828 [hep-th]

Mathur, S.D., 2009: The Information Paradox: A Pedagogical Introduction, arXiv: 0909.1038 [hep-th]

Matsas, G.E.A. / Pleitez, V. / Saa, A. / Vanzella, D.A.T., 2007: The Number of Dimensional Fundamental Constants, arXiv: 0711.4276 [physics]

Mattingly, D., 2005: Is Quantum Gravity Necessary?, in: A. Kox / J. Eisenstaedt (Eds.): *The Universe of General Relativity*, Boston, 327-338

Maudlin, T., 1988: The Essence of Space-Time, in: A. Fine / J. Leplin (Eds.): *PSA 1988*, Vol. 2, East Lansing, 82-91

Maudlin, T., 1990: Substances and Space-Time: What Aristotle would have said to Einstein, *Studies in History and Philosophy of Science* **21**, 531-561

Maudlin, T., 1993: Buckets of Water and Waves of Space: Why Spacetime is Probably a Substance, *Philosophy of Science* **60**, 183-203

Maudlin, T., 2002: Thoroughly Muddled McTaggart. Or How to Abuse Gauge Freedom to Generate Metaphysical Monstrosities, *Philosopher's Imprint* **2/4**, 1-23, http://www.philosophersimprint.org

Mavromatos, N.E., 2007: CPT and Decoherence in Quantum Gravity, arXiv: 0707.3422 [hep-ph]

Maxwell, N., 2001: Special Relativity, Time, Probabilism and Ultimate Reality, in: Dieks (2006), 229-245

Maziashvili, M., 2007: Space-Time Uncertainty Relation and Operational Definition of Dimension, arXiv: 0709.0898 [gr-qc]

McCall, S., 2001: Philosophical Consequences of the Twin Paradox, in: Dieks (2006), 191-204

McInnes, B., 2006: Arrow of Time in String Theory, arXiv: hep-th/0611088

McInnes, B., 2007: Bad Babies or Vacuum Selection and The Arrow of Time, arXiv: 0705.4141 [hep-th]

McInnes, B., 2008: Black Hole Final State Conspiracies, arXiv: 0806.3818 [hep-th]

Meissner, K., 2004: Black Hole Entropy in Loop Quantum Gravity, *Classical and Quantum Gravity* **21**, 5245-5252, arXiv: gr-qc/0407052

Mellor, D.H., 1998: *Real Time II*, London / New York

Mersini-Houghton, L., 2006: Do We Have Evidence for New Physics in the Sky?, *Modern Physics Letters* **A 21**, 1-22, arXiv: hep-th/0510101

Mersini-Houghton, L., 2008: Thoughts on Defining the Multiverse, arXiv: 0804.4280 [gr-qc]

Mersini-Houghton, L., 2009: Notes on Time's Enigma, arXiv: 0909.2330 [gr-qc]

Mersini-Houghton, L. et al., 2007: Nontrivial Geometries: Bounds on the Curvature of the Universe, arXiv: 0705.0332 [astro-ph]

Meschini, D., 2006: Planck-Scale Physics – Facts and Beliefs, arXiv: gr-qc/0601097

Meschini, D., 2008: *A Metageometric Enquiry Concerning Time, Space, and Quantum Physics*, Dissertation, Universität Jyväskylä, arXiv: 0804.3742 [gr-qc]

Meschini, D. / Lehto, M., 2005: Is Empty Spacetime a Physical Thing?, arXiv: gr-qc/0506068

Meschini, D. / Lehto, M. / Piilonen, J., 2005: Geometry, Pregeometry and Beyond, *Studies in History and Philosophy of Modern Physics* **36**, 435-464, Langversion: arXiv: gr-qc/0411053

Mikovic, A., 2009: Temporal Platonic Metaphysics, arXiv: 0903.1800 [physics.hist-ph]

Misner, C.W. / Thorne, K.S. / Wheeler, J.A., 1973: *Gravitation*; San Francisco

Misner, C.W. / Wheeler, J.A., 1957: Classical Physics as Geometry, *Annals of Physics* **2**, 525-603

Mittelstaedt, P., 1976: *Der Zeitbegriff in der Physik*, Mannheim

Moffat, J.W., 1993: Do Black Holes Exist?, arXiv: gr-qc/9302032

Moffat, J.W. / Toth, V.T., 2007: Modified Gravity and the Origin of Inertia, arXiv: 0710.3415 [gr-qc]

Moffat, J.W. / Toth, V.T., 2007a: Modified Gravity: Cosmology without Dark Matter of a Cosmological Constant, arXiv: 0710.0364 [astro-ph]

Mohaupt, T., 2003: Introduction to String Theory, in: D. Giulini / C. Kiefer / C. Lämmerzahl (Eds.): *Quantum Gravity – From Theory to Experimental Search*, Berlin, auch: arXiv: hep-th/0207249

Monk, N.A.M., 1997: Conceptions of Space-Time: Problems and Possible Solutions, *Studies in History and Philosophy of Modern Physics* **28**, 1-34

Montani, G., 2007: Evolutionary Reformulation of Quantum Gravity, arXiv: gr-qc/0702023

Montani, G. / Cianfrani, F., 2008: General Relativity as Classical Limit of Evolutionary Quantum Gravity, arXiv: 0802.0942 [gr-qc]

Monton, B., 2001: Presentism and Quantum Gravity, in: Dieks (2006), 263-280

Morrison, M., 2000: *Unifying Scientific Theories. Physical Concepts and Mathematical Structures.* Cambridge

Mukhi, S., 1997: Recent Developments in String Theory: A Brief Review for Particle Physicists, arXiv: hep-ph/9710470

Müller, T. / Grave, F., 2009: Catalogue of Spacetimes, arXiv: 0904.4184 [gr-qc]

Muller-Hoissen, F., 2007: Noncommutative Geometries and Gravity, arXiv: 0710.4418 [gr-qc]

Myers, R.C. / Pospelov, M., 2003: Experimental Challenges for Quantum Gravity, arXiv: gr-qc/0402028

Nakanishi, N., 2006: Spacetime in the Ultimate Theory, arXiv: hep-th/0610090

Nelson, W. / Sakellariadou, M., 2007: Dark Energy from Corrections to the Wheeler-DeWitt Equation, arXiv: 0709.1625 [gr-qc]

Nerlich, G., 2003: Space-Time Substantivalism, in: M.J. Loux (Ed.): *The Oxford Handbook of Metaphysics*, Oxford, 281-314

Newton-Smith, W.H., 1980: *The Structure of Time*, London

Nicolai, H. / Peeters, K., 2006: Loop and Spin Foam Quantum Gravity, arXiv: hep-th/0601129

Nicolai, H. / Peeters, K. / Zamaklar, M., 2005: Loop Quantum Gravity: An Outside View, *Classical and Quantum Gravity* **22**, R193, arXiv: hep-th/0501114

Nikolic, H., 2008: Time in Relativistic and Nonrelativistic Quantum Mechanics, arXiv: 0811.1905 [quant-ph]

Nielsen, H.B., 1983: Field Theories without Fundamental Gauge Symmetries, *Philosophical Transactions of the Royal Society London* **A 310**, 261-272

Nielsen, H.B., 1989: Random Dynamics and Relations Between the Number of Fermion Generations and the Fine Structure Constants, *Acta Physica Polonica* **B 20**, 427ff

Nielsen, H.B. / Ninomiya, M., 2007: Search for Future Influence from L.H.C., arXiv: 0707.1919 [hep-ph]

Nielsen, H.B. / Ninomiya, M., 2008: Test of Influence from Future in Large Hadron Collider – A Proposal, arXiv: 0802.2991 [physics]

Nielsen, H.B. / Rugh, S.E., 1994: Why do we live in 3+1 dimensions?, arXiv: hep-th/9407011

Nielsen, H.B. / Rugh, S.E. / Surlykke, C., 1994: Seeking Inspiration from the Standard Model in order to go beyond it, arXiv: hep-th/9407012

Niewenhuizen, T.M., 2007: Einstein vs Maxwell: Is Gravitation a Curvature of Space, a Field in Flat Space, or both, arXiv: 0704.0228 [gr-qc]

North, J., 2009: The "Structure" of Physics: A Case Study, *Journal of Philosophy* **106**, 57-88

Norton, J.D., 1987: Einstein, the Hole Argument and the Reality of Space, in: J. Forge (Ed.): *Measurement, Realism and Objectivity*, Dordrecht, 153-188

Norton, J.D., 1988: The Hole Argument, in: A. Fine / J. Leplin (Eds.): *PSA 1988*, Vol. 2, East Lansing, 56-64

Norton, J.D., 1989: Coordinates and Covariance: Einstein's View of Space-Time and the Modern View, *Foundations of Physics* **19**, 1215-1263

Norton, J.D., 1992: Philosophy of Space and Time, in: M.H. Salmon et al. (Eds.): *Introduction to the Philosophy of Science*, Englewood Cliffs, NJ, 179-231

Norton, J.D., 1992a: The Physical Content of General Covariance, in: J. Eisenstaedt / A. Kox (Eds.): *Studies in the History of General Relativity* (Einstein Studies 3), Boston, 281-315

Norton, J.D., 1993: General Covariance and the Foundations of General Relativity, *Reports on Progress in Physics* **56**, 791-858

Norton, J.D., 1994: Science and Certainty, *Synthese* **99**, 3-22

Norton, J.D., 2000: What Can We Learn about the Ontology of Space and Time from the Theory of Relativity?, http://philsci-archive.pitt.edu, Dokument: 0138

Norton, J.D., 2001: General Covariance, Gauge Theories and the Kretschmann Objection, http://philsci-archive.pitt.edu, Dokument: 0380

Norton, J.D., 2003: General Covariance, Gauge Theories, and the Kretschmann Objection, in: Brading / Castellani (2003), 110-123

Norton, J.D., 2004: The Hole Argument, in: E.N. Zalta (Ed.): *Stanford Encyclopedia of Philosophy*, http://plato.stanford.edu

Oaklander, N. / Smith, Q. (Eds.), 1994: *The New Theory of Time*, London / New Haven

Olson, S.J. / Dowling, J.P., 2007: Probability, Unitarity, and Realism from Generally Covariant Quantum Information, arXiv: 0708.3535 [quant-ph]

Oppenheim, J., 2002: Area Scaling Entropies for Gravitating Systems, *Physical Review* **D 65**, 024020, arXiv: gr-qc/0105101

Oriti, D., 2001: Spacetime Geometry from Algebra: Spin Foam Models for non-perturbative Quantum Gravity, *Reports on Progress in Physics* **64**, 1489-?, arXiv: gr-qc/0106091

Oriti, D., 2003: *Spin Foam Models of Quantum Spacetime*, Ph.D. Thesis, arXiv: gr-qc/0311066

Oriti, D., 2006: A Quantum Field Theory of Simplicial Geometry and the Emergence of Spacetime, arXiv: hep-th/0612301

Oriti, D., 2006a: The Group Field Theory Approach to Quantum Gravity, arXiv: gr-qc/0607032

Oriti, D., 2007: Quantum Gravity as a Quantum Field Theory of Simplicial Geometry, in: B. Fauser / J. Tolksdorf / E. Zeidler (Eds.): Mathematical and Physical Aspects of Quantum Gravity, Basel, arXiv: gr-qc/0512103

Oriti, D., 2007a: Group Field Theory as the Micrsocopic Description of the Quantum Spacetime Fluid: A New Perspective on the Continuum in Quantum Gravity, arXiv: 0710.3276 [gr-qc]

Oriti, D. (Ed.), 2009: *Approaches to Quantum Gravity: Toward a New Understanding of Space, Time and Matter*, Cambridge

Oriti, D., 2009a: The Group Field Theory Approach to Quantum Gravity: Some Recent Results, arXiv: 0912.2441 [hep-th]

Padmanabhan, T., 2002: Is Gravity an Intrinsically Quantum Phenomenon?, *Modern Physics Letters* **A 17**, 1147-1158, arXiv: hep-th/0205278

Padmanabhan, T., 2002a: Gravity from Spacetime Thermodynamics, arXiv: gr-qc/0209088

Padmanabhan, T., 2003: Cosmological Constant – The Weight of the Vacuum, *Physics Reports* **380**, 235-320, arXiv: hep-th/0212290

Padmanabhan, T., 2004: Gravity as Elasticity of Spacetime: A Paradigm to understand Horizon Thermodynamics and Cosmological Constant, *International Journal of Modern Physics* **D 13**, 2293-2298, arXiv: gr-qc/0408051

Padmanabhan, T., 2007: Dark Energy and Gravity, arXiv: 0705.2533 [gr-qc]

Padmanabhan, T., 2007a: Gravity as an Emergent Phenomenon: A Conceptual Description, arXiv: 0706.1654 [gr-qc]

Padmanabhan, T., 2008: Emergent Gravity and Dark Energy, arXiv: 0802.1798 [gr-qc]

Padmanabhan, T., 2009: Entropy Density of Spacetime and Thermodynamic Interpretation of Field Equations of Gravity in any Diffeomorphism Invariant Theory, arXiv: 0903.1254 [hep-th]

Padmanabhan, T., 2009a: Thermodynamical Aspects of Gravity: New Insights, arXiv: 0911.5004 [gr-qc]

Padmanabhan, T., 2009b: A Dialogue on the Nature of Gravity, arXiv: 0910.0839 [gr-qc]

Padmanabhan, T., 2009c: A Physical Interpretation of Gravitational Field Equations, arXiv: 0911.1403 [gr-qc]

Padmanabhan, T. / Paranjape, A., 2007: Entropy of Null Surfaces and Dynamics of Spacetime, arXiv: gr-qc/0701003

Parikh, M.K. / Sarkar, S., 2009: Beyond the Einstein Equation of State: Wald Entropy and Thermodynamical Gravity, arXiv: 0903.1176 [hep-th]

Pauri, M. / Vallisneri, M., 2002: Ephemeral Point-Events: Is there a Last Remnant of Physical Objectivity, *Dialogos* **79**, 263-303, auch: http://philsci-archive.pitt.edu

Pavsic, M., 2007: Towards a New Paradigm – Relativity in Configuration Space, arXiv: 0712.3660 [gr-qc]

Pavsic, M., 2009: On the Relativity in Configuration Space: A Renewed Physics in Sight, arXiv: 0912.3669 [gr-qc]

Pavsic, M., 2009a: Towards the Unification of Gravity and other Interactions: What has been Missed?, arXiv: 0912.4836 [gr-qc]

Peacock, K.A., 2001: Temporal Presentness and the Dynamics of Spacetime, in: Dieks (2006), 247-261

Peet, A.W., 1998: The Bekenstein Formula and String Theory, *Classical and Quantum Gravity* **15**, 3291-3338, arXiv: hep-th/9712253

Peet, A.W., 2001: TASI Lectures on Black Holes in String Theory, in: Harvey / Kachru / Silverstein (2001), arXiv: hep-th/0008241

Penrose, R., 1967: Twistor Theory, *Journal of Mathematical Physics* **8**, 345-366

Penrose, R., 2004: *The Road to Reality. A Complete Guide to the Laws of the Universe*, London

Penrose, R. / Isham, C.J. (Eds.), 1986: *Quantum Concepts in Space and Time*, Oxford

Penrose, R. / MacCallum, M.A.H., 1973: Twistor Theory: An Approach to the Quantization of Fields and Space-Time, *Physics Reports* **6**, 241-316

Per, M.A. / Segui, A., 2005: Holographic Cosmology and Uncertainty Relation, arXiv: gr-qc/0502025

Percacci, R., 2009: Gravity from a Particle Physicists' Perspective, arXiv: 0910.5167 [hep-th]

Perez Bergliaffa, S.E. / Romero, G.E. / Vucetich, H., 1998: Steps Towards an Axiomatic Pregeometry of Space-Time, *International Journal of Theoretical Physics* **37**, 2281-2298, arXiv: gr-qc/9710064

Perez, A., 2003: Spin Foam Models for Quantum Gravity, *Classical and Quantum Gravity* **20**, R43-R104, arXiv: gr-qc/0301113

Perez, A., 2006: The Spin-Foam-Representation of Loop Quantum Gravity, arXiv: gr-qc/0601095

Peres, A. / Terno, D.R., 2001: Hybrid Classical-Quantum Dynamics, *Physical Review* **A 63**, 022101

Pesci, A., 2007: From Unruh Temperature to Generalized Bousso Bound, arXiv: 0708.3729 [gr-qc]

Pesci, A., 2008: On the Statistical-mechanical Meaning of Bousso Bound, arXiv: 0803.2642 [gr-qc]

Petkov, V., 2001: Is there an Alternative to the Block Universe View?, in: Dieks (2006), 207-228

Phillips, P.R., 2007: Is the Mass Scale for Elementary Particles Classically Determined?, arXiv: 0707.4323 [gr-qc]

Philpott, L., 2009: Particle Simulations in Causal Set Theory, arXiv: 0911.5595 [gr-qc]

Philpott, L. / Dowker, F. / Sorkin, R., 2008: Energy-Momentum Diffusion from Spacetime Discreteness, arXiv: 0810.5591 [gr-qc]

Piazza, F., 2005: Quantum Degrees of Freedom in a Region of Spacetime, arXiv: hep-th/0511285

Pirogov, Y.F., 2005: Space-time: emerging vs. existing, arXiv: gr-qc/0503091

Pitts, J.B., 2008: *General Covariance, Artificial Gauge Freedom and Empirical Equivalence*, Dissertation, University of Notre Dame, Notre Dame, In.

Pitts, J.B., 2009: Empirical Equivalence, Artificial Gauge Freedom and a Generalized Kretschmann Obejction, arXiv: 0911.5400 [physics]

Plyuschchay, M.S. / Razumov, A.V., 1996: Dirac versus Reduced Phase Space Quantization, *International Journal of Modern Physics* **A 11**, 1427-1462, arXiv: hep-th/9306017

Polawski, N.J., 2009: Spacetime and Fields, arXiv: 0911.0334 [gr-qc]

Polchinski, J.G., 1995: Dirichlet-Branes and Ramond-Ramond-Charges, *Physical Review Letters* **74**, 4724-4727, arXiv: hep-th/9510017

Polchinski, J.G., 1996: String Duality, *Reviews of Modern Physics* **68**, 1245-1258, arXiv: hep-th/9607050

Polchinski, J.G., 1996a: TASI Lectures on D-Branes, TASI-96 Summer School Lectures, arXiv: hep-th/9611050

Polchinski, J.G., 1999: Quantum Gravity at the Planck Length, *International Journal of Modern Physics* **A14**, 2633-2658, arXiv: hep-th/9812104

Polchinski, J.G., 2000: *String Theory. Vol. 1: An Introduction to the Bosonic String*, Cambridge

Polchinski, J.G., 2000a: *String Theory. Vol. 2: Superstring Theory and Beyond*, Cambridge

Polchinski, J.G. / Chaudhuri, S. / Johnson, C.V., 1996: Notes on D-Branes, arXiv: hep-th/9602052

Polhemus, G. / Hamilton, A.J.S. / Wallace, C.S., 2009: Entropy Creation Inside Black Holes Points to Observer Complementarity, arXiv: 0903.2290 [gr-qc]

Polyakov, A.M., 2006: Beyond Space-Time, arXiv: hep-th/0602011

Pons, J.M. / Salisbury, D.C., 2005: The Issue of Time in Generally Covariant Theories and the Komar-Bergmann Approach to Observables in General Relativity, *Physical Review* **D 71**, 124012, arXiv: gr-qc/0503013

Pons, J.M. / Salisbury, D.C. / Sundermeyer, K.A., 2009: Gravitational Observables, Intrinsic Coordinates, and Canonical Maps, arXiv: 0902.0401 [gr-qc]

Pooley, O., 2001: Relationism Rehabilited? II: Relativity, http://philsci-archive.pitt.edu, Dokument: 0221

Pooley, O., 2006: A Hole Revolution, or are we back where we started, *Studies in History and Philosophy of Modern Physics* **37**, 372-380

Pooley, O., 2006a: Points, Particles and Structural Realism, in: Rickles / French / Saatsi (2006) 83-120, auch: http://philsci-archive.pitt.edu

Poundstone, W., 1985: *The Recursive Universe – Cosmic Complexity and the Limits of Scientific Knowledge*, Chicago

Preskill, J., 2000: Quantum Information and Physics: Some Future Directions, *Journal of Modern Optics* **47**, 127-137, arXiv: quant-ph/9904022

Price, H., 1996: *Time's Arrow and Archimedes' Point: New Directions of the Physics of Time*, Oxford

Quevedo, H. / Vazquez, A., 2007: The Geometry of Thermodynamics, arXiv: 0712.0868 [math-ph]

Quiros, I., 2007: Time-like vs. Space-like Extra Dimensions, arXiv: 0707.0714 [gr-qc]

Ramond, P., 2007: Memoirs of an Early String Theorist, arXiv: 0708.3656 [hep-th]

Redhead, M., 2002: The Interpretation of Gauge Symmetry, in: Kuhlmann / Lyre / Wayne (2002)

Redhead, M., 2003: The Interpretation of Gauge Symmetry, in: Brading / Castellani (2003), 124-139

Regge, T. / Williams, R.M., 2000: Discrete Structures in Quantum Gravity, *Journal of Mathematical Physics* **41**, 3964, arXiv: gr-qc/0012035

Requardt, M., 1995: Discrete Mathematics and Physics on the Planck-Scale, arXiv: hep-th/9504118

Requardt, M., 1996: Discrete Mathematics and Physics on the Planck-Scale exemplified by means of a Class of 'Cellular Network Models' and their Dynamics, arXiv: hep-th/9605103

Requardt, M., 1996a: Emergence of Space-Time on the Planck-Scale Described as an Unfolding Phase Transition within the Scheme of Dynamical Cellular Networks and Random Graphs, arXiv: hep-th/9610055

Requardt, M., 1996b: Emergence of Space-Time on the Planck-Scale within the Scheme of Dynamical Cellular Network Models and Random Graphs, arXiv: hep-th/9612185

Requardt, M., 2000: Let's Call it Nonlocal Quantum Physics, arXiv: gr-qc/0006063

Requardt, M., 2001: The Continuum Limit of Discrete Geometries, *Int. J. Geom. Meth. Mod. Phys.* **3**, 285-314, arXiv: math-ph/0507017

Requardt, M., 2009: Wormhole Spaces: the Common Cause for the Black Hole Entropy-Area Law, the Holographic Principle and Quantum Entanglement, arXiv: 0910.4017 [hep-th]

Requardt, M. / Roy, S., 2001: (Quantum) Space-Time as a Statistical Geometry of Fuzzy Lumps and the Connection with Random Metric Spaces, *Classical and Quantum Gravity* **18**, 3039-3058, arXiv: gr-qc/0011076

Rescher, N., 2000: The Price of an Ultimate Theory, *Philosophia Naturalis* **37**, 1-20

Rickles, D., 2005: A New Spin on the Hole Argument, *Studies in History and Philosophy of Modern Physics* **36**, 415-434

Rickles, D., 2005a: Interpreting Quantum Gravity, http://philsci-archive.pitt.edu, Document: 2407

Rickles, D., 2006: Time and Structure in Canonical Gravity, in: Rickles / French / Saatsi (2006) 152-195, auch: http://philsci-archive.pitt.edu

Rickles, D., 2006a: Bringing the Hole Argument back in the Loop: A Response to Pooley, *Studies in History and Philosophy of Modern Physics* **37**, 381-387

Rickles, D., 2008: *Symmetry, Structure and Spacetime*, Amsterdam

Rickles, D., 2008a: Who´s Afraid of Background Independence, http://philsci-archive.pitt.edu, Dokument: 4223

Rickles, D. / French, S., 2006: Quantum Gravity Meets Structuralism: Interweaving Relations in the Foundations of Physics, in: Rickles / French / Saatsi (2006) 1-39

Rickles, D. / French, S. / Saatsi, J. (Eds.), 2006: *The Structural Foundations of Quantum Gravity*, Oxford

Rideout, D.P., 2002: Dynamics of Causal Sets, arXiv: gr-qc/0212064

Rideout, D.P. / Sorkin, R.D., 2000: A Classical Sequential Growth Dynamics for Causal Sets, *Physical Review* **D 61**, 024002, arXiv: gr-qc/9904062

Rideout, D.P. / Sorkin, R.D., 2001: Evidence for a Continuum Limit in Causal Set Dynamics, *Physical Review* **D 63**, 104011, arXiv: gr-qc/0003117

Rideout, D. / Wallden, P., 2008: Spacelike Distance from Discrete Causal Order, arXiv: 0810.1768 [gr-qc]

Rideout, D. / Wallden, P., 2008a: Emergent Continuum Spacetime from a Random, Discrete, Partial Order, arXiv: 0811.1178 [gr-qc]

Rideout, D. / Wallden, P., 2009: Emergence of Spatial Structure from Causal Sets, arXiv: 0905.0017 [gr-qc]

Riggs, P.J., 1996: Spacetime or Quantum Particles: The Ontology of Quantum Gravity, in: ders. (Ed.): *Natural Kinds, Laws of Nature and Scientific Methodology*, Dordrecht

Romano, J.D., 1993: *Geometrodynamics vs. Connection Dynamics (in the Context of (2+1)- and (3+1)-Gravity)*, Ph.D. Thesis, Syracuse University, *General Relativity and Gravitation* **25**, 759-854, arXiv: gr-qc/9303032

Rosenbaum, M. / Vergara, J.D. / Juarez, L.R., 2008: Space-Time Diffeomorphisms in Noncommutative Gauge Theories, arXiv: 0807.2508 [hep-th]

Rovelli, C., 1991: Time in Quantum Gravity: An Hypothesis, *Physical Review* **D 43**, 442-456

Rovelli, C., 1991a: Is there Incompatibility between the Way Time is treated in General Relativity and in Standard Quantum Mechanics, in: Ashtekar / Stachel (1991), 126-140

Rovelli, C., 1991b: Loop Representation in Quantum Gravity: The Transform Approach, in: Ashtekar / Stachel (1991), 427-439

Rovelli, C., 1991c: What is Observable in Classical and Quantum Gravity, *Classical and Quantum Gravity* **8**, 297-316

Rovelli, C., 1991d: Quantum Reference Systems, *Classical and Quantum Gravity* **8**, 317-331

Rovelli, C., 1994: On Quantum Mechanics, arXiv: hep-th/9403015

Rovelli, C., 1996: Relational Quantum Mechanics, *International Journal of Theoretical Physics* **35**, 1637, arXiv: quant-ph/9609002

Rovelli, C., 1997: Loop Quantum Gravity, *Living Reviews in Relativity* (Electronic Journal), www.livingreviews.org; auch: arXiv: gr-qc/9710008

Rovelli, C., 1998: Strings, Loops, and the Others: A Critical Survey on the Present Approaches to Quantum Gravity, in: N. Dadhich / J. Narlikar (Eds.): *Gravitation and Relativity: At the Turn of the Millenium*, Poona, arXiv: gr-qc/9803024

Rovelli, C., 1999: 'Localization' in Quantum Field Theory: How much of QFT is Compatible with What we Know about Space-Time?, in: T.Y. Cao (Ed.): *Conceptual Foundations of Quantum Field Theory*, Cambridge, 207-232

Rovelli, C., 2000: Notes for a Brief History of Quantum Gravity, arXiv: gr-qc/0006061

Rovelli, C., 2001: Quantum Spacetime: What do we know?, in: Callender / Huggett (2001); auch: arXiv: gr-qc/9903045 (ältere Fassung)

Rovelli, C., 2002: Partial Observables, *Physical Review* **D 65**, 124013, arXiv: gr-qc/0110035

Rovelli, C., 2003: A Dialog on Quantum Gravity, *International Journal of Modern Physics* **12**, 1, arXiv: hep-th/0310077

Rovelli, C., 2004: *Quantum Gravity*, Cambridge; auch: www.cpt.univ-mrs.fr/~rovelli

Rovelli, C., 2005: *What is Time? – What is Space?* (Übers. a. d. Ital.: J.C. van den Berg), www.cpt.univ-mrs.fr/~rovelli

Rovelli, C., 2006: Unfinished Revolution, arXiv: gr-qc/0604045

Rovelli, C., 2006a: The Disappearance of Space and Time, in: Dieks (2006), 25-36

Rovelli, C., 2007: Quantum Gravity, in: Earman / Butterfield (2007)

Rovelli, C., 2007a: Comment on 'Are the Spectra of Geometrical Operators in Loop Quantum Gravity Really Discrete' by B. Dittrich and T. Thiemann, arXiv: 0708.2481 [gr-qc]

Rovelli, C., 2009: Forget Time, arXiv: 0903.3832 [gr-qc]

Rovelli, C. / Upadhya, P., 1998: Loop Quantum Gravity and Quanta of Space – A Primer, arXiv: gr-qc/9806079

Rozali, M., 2008: Comments on Background Independence and Gauge Redundancies, arXiv: 0809.3962 [gr-qc]

Ryckman, T.A., 2006: Early Philosophical Interpretations of General Relativity, in: E.N. Zalta (Ed.): *Stanford Encyclopedia of Philosophy*, http://plato.stanford.edu

Rynasiewicz, R., 1994: The Lessons of the Hole Argument, *British Journal for the Philosophy of Science* **45**, 407-436

Rynasiewicz, R., 1996: Absolute Versus Relational Space-Time: An Outmoded Debate?, *Journal of Philosophy* **93**, 279-306

Rynasiewicz, R., 2000: On the Distinction between Absolute and Relative Motion, *Philosophy of Science* **67**, 70-93

Sakellariadou, M., 2007: The Origin of Spacetime Dimensionality, arXiv: hep-th/0701024

Sakharov, A.D., 2000: Vacuum Quantum Fluctuations in Curved Space and the Theory of Gravitation, *General Relativity and Gravitation* **32**, 365-367 (Reprint; Original: *Doklady Akademii Nauk SSSR* **177** (1967) 70-71 / *Soviet Physics Doklady* **12** (1968) 1040-1041)

Salisbury, D., 2008: Book Review: The Genesis of General Relativity, arXiv: 0807.3706 [physics]

Samuel, J., 2000: Canonical Gravity, Diffeomorphisms and Objective Histories, *Classical and Quantum Gravity* **17**, 4645-4654, arXiv: gr-qc/0005094

Sánchez, M., 2008: On the Foundations and Necessity of Classical Gauge Invariance, arXiv: 0803.1290 [math-ph]

Sánchez-Guillen, J. / Vazquez, R.A., 2009: On the Objective Existence of Physical Processes: A Physicist's Philosophical View, arXiv: 0903.4250 [quant-ph]

Sánchez-Rodriguez, I., 2008: Geometrical Structures of Space-Time in General Relativity, arXiv: 0803.1929 [gr-qc]

Sarfatti, J. / Levit, C., 2009: The Emergence of Gravity as a retro-causal post-inflation macro-quantum-coherent holographic vacuum Higgs-Goldstone Field, arXiv: 0902.0032 [physics]

Satheesh Kumar, V.H. / Suresh, P.K., 2005: Are we Living in a Higher Dimensional Universe?, arXiv: gr-qc/0506125

Saunders, S., 2001: Indiscernibles, General Covariance, and Other Symmetries – The Case for Non-Eliminativist Relationalism, in: A. Ashtekar / D. Howard / J. Renn / S. Sarkar / A. Shimony (Eds.): *Revisiting the Foundations of Relativistic Physics: Festschrift in Honour of John Stachel*, Dordrecht, http://philsci-archive.pitt.edu, Dokument: 0459

Saunders, S., 2002: How Relativity Contradicts Presentism, in: Callender (2002) 277-292

Savitt, S. (Ed.), 1995: *Time's Arrow Today: Recent Physical and Philosophical Work on the Direction of Time*, Cambridge

Sawayama, S., 2007: Time Operator and Static Restriction in the Quantum Gravity, arXiv: 0705.2916 [gr-qc]

Schellekens, A.N., 2008: The Emperor's Last Clothes?, arXiv: 0807.3249 [physics]

Schilling, T., 1996: *Geometry of Quantum Mechanics*, Ph.D. Thesis, Pennsylvania State University, http://cgpg.gravity.psu.edu/archives/thesis/1996/

Schimmrigk, R., 2008: Applied String Theory, arXiv: 0810.1743 [hep-th]

Schimmrigk, R., 2008a: Emergent Spacetime from Modular Motives, arXiv: 0812.4450 [hep-th]

Schmid, C., 2008: Mach's Principle: Exact Frame-Dragging by Energy Currents in the Universe, arXiv: 0806.2085 [astro-ph]

Schmidhuber, J., 1997: A Computer Scientist's View of Life, the Universe, and Everything, in: C. Freksa (Ed.): *Foundations of Computer Science: Potential – Theory – Cognition, Lecture Notes in Computer Science*, Berlin, 201-208, arXiv: quant-ph/9904050

Schmidhuber, J., 2000: Strings from Logic, arXiv: hep-th/0011065

Schmidhuber, J., 2000a: Algorithmic Theories of Everything, arXiv: quant-ph/0011122

Schroer, B., 2006: String Theory deconstructed, arXiv: hep-th/0611132

Schroer, B., 2007: Localization and the Interface between Quantum Mechanics, Quantum Field Theory and Quantum Gravity, arXiv: 0711.4600 [hep-th]

Schroer, B., 2008: String Theory and the Crisis of Particle Physics II, arXiv: 0805.1911 [hep-th]

Schücker, T., 2009: Ashtekar's Variables without Spin, arXiv: 0906.4918 [gr-qc]

Schucking, E.L., 2009: Einstein's Apple and Relativity's Gravitational Field, arXiv: 0903.3768 [physics.hist-ph]

Schutz, B.F. / Centrella, J. / Cutler, C. / Hughes, S.A., 2009: Will Einstein Have the Last Word on Gravity, arXiv: 0903.0100 [gr-qc]

Schwarz, A., 2006: Space and Time from Translation Symmetry, arXiv: hep-th/0601035

Schwarz, J.H., 1996: The Power of M-theory, *Physics Letters* **B 367**, 97-103, arXiv: hep-th/9510086

Schwarz, J.H., 1997: Lectures on Superstring and M Theory Dualities, in: Efthimiou / Greene (1997), 359-418, arXiv: hep-th/9607201

Schwarz, J.H., 1998: Beyond Gauge Theories, arXiv: hep-th/9807195

Schwarz, J.H., 2007: String Theory: Progress and Problems, arXiv: hep-th/0702219

Schwarz, J.H., 2007a: The Early Years of String Theory: A Personal Perspektive, arXiv: 0708.1917 [hep-th]

Segal, G., 1998: Space from the Point of View of Loop Groops, in: S.A. Huggett et al. (Eds.): *The Geometric Universe*, Oxford (1998)

Seiberg, N., 2006: Emergent Spacetime, arXiv: hep-th/0601234

Sen, A., 1982: Gravity as a Spin System, *Physics Letters* **B 11**, 89ff

Sen, A., 1998: An Introduction to Non-perturbative String Theory, arXiv: hep-ph/9802051

Sen, A., 1998a: String Network, *Journal of High Energy Physics* **9803**, 005, arXiv: hep-th/9711130

Shenker, S.H., 1995: Another Length Scale in String Theory, arXiv: hep-th/9509132

Shestakova, T.P., 2008: The Wheeler-DeWitt Quantum Geometrodynamics: Its Fundamental Problems and Tendencies of their Resolution, arXiv: 0801.4854 [gr-qc]

Shestakova, T.P., 2008a: The "Extended Phase Space" Approach to Quantum Geometrodynamics: What can it give for the Development of Quantum Gravity, arXiv: 0810.4031 [gr-qc]

Shestakova, T.P., 2009: The Formulation of General Relativity in Extended Phase Space as a Way to its Quantization, arXiv: 0911.4852 [gr-qc]

Shestakova, T.P., 2009a: Hamiltonian Formulation of General Relativity 50 Years after the Dirac celebrated paper: do unsolved Problems still exist?, arXiv: 0911.5252 [gr-qc]

Shimony, A, 1999: Can the fundamental laws of nature be the results of evolution?, in: Butterfield / Pagonis (1999)

Silvestri, A. / Trodden, M., 2009: Approaches to Understanding Cosmic Acceleration, arXiv: 0904.0024 [astro-ph]

Sindoni, L. / Girelli, F. / Liberati, S., 2009: Emergent Gravitational Dynamics in Bose-Einstein Condensates, arXiv: 0909.5391 [gr-qc]

Singh, T.P., 2007: Quantum Measurement and Quantum Gravity: Many-Worlds or Collapse of the Wave-Function?, arXiv: 0711.3773 [gr-qc]

Slowik, E., 2004: Spacetime, Ontology, and Structural Realism, *International Studies in Philosophy of Science* **19**, 147-166, http://philsci-archive.pitt.edu, Dokument: 2872

Slowik, E., 2006: Spacetime, Structural Realism, and the Substantival/Relational Debate: An Ontological Investigation from the Perspective of Structural Realism in the Philosophy of Mathematics, http://philsci-archive.pitt.edu, Dokument: 2873

Smolin, J.A. / Oppenheim, J., 2006: Information Locking in Black Holes, *Physical Review Letters* **96**, 081302, arXiv: hep-th/0507287

Smolin, L., 1991: Space and Time in the Quantum Universe, in: Ashtekar / Stachel (1991), 228-291

Smolin, L., 1991a: Nonperturbative Quantum Gravity via the Loop Representation, in: Ashtekar / Stachel (1991), 440-489

Smolin, L., 1995: Linking Topological Quantum Field Theory and Nonperturbative Quantum Gravity, *Journal of Mathematical Physics* **36**, 6417-6455, arXiv: gr-qc/9505028

Smolin, L., 1995a: The Bekenstein Bound, Topological Quantum Field Theory and Pluralistic Quantum Cosmology, arXiv: gr-qc/9508064

Smolin, L., 1998: Towards a background independent approach to M theory, arXiv: hep-th/9808192

Smolin, L., 2000: *Three Roads to Quantum Gravity*, London

Smolin, L., 2000a: A Holographic Formulation of Quantum General Relativity, *Physical Review* **D 61**, 084007, arXiv: hep-th/9808191

Smolin, L., 2000b: A Candidate for a Background Independent Formulation of M Theory, *Physical Review* **D 62**, 086001, arXiv: hep-th/9903166

Smolin, L., 2001: The Strong and the Weak Holographic Principles, *Nuclear Physics* **B 601**, 209, arXiv: hep-th/0003056

Smolin, L., 2002: Matrix Models as Non-Local Hidden Variables Theories, arXiv: hep-th/0201031

Smolin, L., 2003: How far are we from the quantum theory of gravity?, arXiv: hep-th/0303185

Smolin, L., 2004: An Invitation to Loop Quantum Gravity, arXiv: hep-th/0408048

Smolin, L., 2004a: Quantum Theories of Gravity: Results and Prospects, in: Barrow / Davies / Harper (2004) 492-527

Smolin, L., 2006: *The Trouble with Physics*, Boston

Smolin, L., 2006a: Generic Predictions of Quantum Theories of Gravity, arXiv: hep-th/0605052

Smolin, L., 2006b: Could Quantum Mechanics be an Approximation to another Theory?, arXiv: quant-ph/0609109

Smolin, L., 2006c: The Case for Background Independence, in: Rickles / French / Saatsi (2006) 196-239, auch: arXiv: hep-th/0507235

Smolin, L., 2008: Matrix Universality of Gauge and Gravitational Dynamics, arXiv: 0803.2926 [hep-th]

Sorkin, R.D., 1989: A Specimen of Theory Construction from Quantum Gravity, arXiv: gr-qc/9511063

Sorkin, R.D., 1994: Quantum Mechanics as Quantum Measure Theory, *Modern Physics Letters* **A 9**, 3119-3128, arXiv: gr-qc/9401003

Sorkin, R.D., 1997: Forks on the Road on the Way to Quantum Gravity, *International Journal of Theoretical Physics* **36**, 2757-2781, arXiv: gr-qc/9706002

Sorkin, R.D., 1997a: The Statistical Mechanics of Black Hole Thermodynamics, arXiv: gr-qc/9705006

Sorkin, R.D., 2003: Causal Sets: Discrete Gravity, arXiv: gr-qc/0309009

Sorkin, R.D., 2005: Ten Theses on Black Hole Entropy, *Studies in History and Philosophy of Modern Physics* **36**, 291-301, arXiv: hep-th/0504037

Sorkin, R.D., 2007: Is the Cosmological 'Constant' a Nonlocal Quantum Residue of Discreteness of the Causal Set Type, arXiv: 0710.1675 [gr-qc]

Sorkin, R.D., 2007a: Relativity Theory does not imply that the Future already exists: A Counterexample, arXiv: gr-qc/0703098

Sorkin, R.D., 2007b: Does Locality Fail at Intermediate Length Scales, arXiv: gr-qc/0703099

Sotiriou, T.P. / Faraoni, V. / Liberati, S., 2007: Theory of Gravitation Theories: A No-Progress Report, arXiv: 0707.2748 [gr-qc]

Spaans, M., 2009: The Role of Information in Gravity, arXiv: 0903.4315 [gr-qc]

Sparling, G.A.J., 2006: Spacetime is Spinorial: New Dimensions are Timelike, arXiv: gr-qc/0610068

Sparling, G.A.J., 2006a: Esquisse d'une Synthese, arXiv: gr-qc/0610069

Spencer, J.A. / Wheeler, J.T., 2008: The Emergence of Time, arXiv: 0811.0112 [gr-qc]

Stachel, J., 1986: What can a Physicist Learn from the Discovery of General Relativity?, in: R. Ruffini (Ed.): *Proceedings of the Fourth Marcel Grossmann Meeting on Recent Developments in General Relativity*, Amsterdam

Stachel, J., 1991: Einstein and Quantum Mechanics, in: Ashtekar / Stachel (1991), 13-42

Stachel, J., 1993: The Meaning of General Covariance, in: J. Earman, / I. Janis / G.J. Massey / N. Rescher (Eds.): *Philosophical Problems of the Internal and the External World*, Pittsburgh, 129-160

Stachel, J., 2002: 'The Relations between Things' versus 'The Things between Relations': The Deeper Meaning of the Hole Argument, in: D.B. Malament (Ed.): *Reading Natural Philosophy. Essays in the History and Philosophy of Science and Mathematics*, Chicago, 231-266

Stachel, J., 2006: Prolegomena to any future Quantum Gravity, arXiv: gr-qc/0609108

Stachel, J., 2006a: Structure, Individuality and Quantum Gravity, in: Rickles / French / Saatsi (2006) 53-82, auch: arXiv: gr-qc/0507078

Stachel, J. / Iftime, M., 2005: Fibered Manifolds, Natural Bundles, Structured Sets, G-Spaces and all that: The Hole Story from Space-Time to Elementary Particles, in: J. Stachel (Ed.): *Going Critical*, Vol. 2, *The Practice of Relativity*, Dordrecht, arXiv: gr-qc/0505138

Stapp, H.P., 1988: Spacetime and Future Quantum Theory, *Foundations of Physics* **18**, 833-849

Steinacker, H., 2009: Matrix Models, Emergent Gravity, and Gauge Theory, arXiv: 0903.1015 [hep-th]

Stephani, H., 1977: *Allgemeine Relativitätstheorie*, Berlin

Stöckler, M., 1990: Materie in Raum und Zeit? – Philosophische Aspekte der Raum-Zeit-Interpretation der Quantenfeldtheorie, *Philosophia Naturalis* **27**, 111-135

Stoica, O.C., 2008: World Theory, http://philsci-archive.pitt.edu, Dokument: 4355

Stoyanovsky, A.V., 2009: Mathematical Definition of Quantum Field Theory on a Manifold, arXiv: 0910.2296 [math-ph]

Strominger, A., 1996: Open p-Branes, *Physics Letters* **B 383**, 44, arXiv: hep-th/9512059

Strominger, A., 2009: Five Problems in Quantum Gravity, arXiv: 0906.1313 [hep-th]

Strominger, A. / Vafa, C., 1996: Microscopic Origin of the Bekenstein-Hawking Entropy, *Physics Letters* **B 379**, 99-104, arXiv: hep-th/9601029

Stuckey, W.M., 2001: Metric Structure and Dimensionality over a Borel Set via Uniform Spaces, arXiv: gr-qc/0109030

Stuckey, W.M. / Silberstein, M., 2000: Uniform Spaces in the Pregeometric Modelling of Quantum Non-Separability, arXiv: gr-qc/0003104

Suarez, A., 2007: Classical Demons and Quantum Angels: On 't Hooft's Deterministic Quantum Mechanics, arXiv: 0705.3974 [quant-ph]

Sudarsky, D., 2007: Unspeakables and the Epistemological Path towards Quantum Gravity, arXiv: 0712.3242 [gr-qc]

Surya, S., 2007: Causal Set Topology, arXiv: 0712.1648 [gr-qc]

Susskind, L., 1995: The World as a Hologram, *Journal of Mathematical Physics* **36**, 6377-6396, arXiv: hep-th/9409089

Susskind, L., 2004: Supersymmetry Breaking in the Anthropic Landscape, arXiv: hep-th/0405189

Susskind, L., 2005: *The Cosmic Landscape – String Theory and the Illusion of Intelligent Design*, New York

Susskind, L., 2007: The Anthropic Landscape of String Theory, in: Carr (2007), arXiv: hep-th/0302219

Susskind, L., 2007a: The Census Taker's Hat, arXiv: 0710.1129 [hep-th]

Susskind, L. / Arcioni, G. / Haro, S. de, 2006: Holographics Views of the World, arXiv: physics/0611143

Szabo, R.J., 2009: Quantum Gravity, Field Theory and Signatures of Noncommutative Spacetime, arXiv: 0906.2913 [hep-th]

Tahim, M.O. et al., 2007: Spacetime as a deformable Solid, arXiv: 0705.4120 [gr-qc]

Tartaglia, A. / Radicella, N., 2009: A Tensor Theory of Space-Time as a Strained Material Continuum, arXiv: 0903.4096 [gr-qc]

Tate, R.S., 1992: *An Algebraic Approach to the Quantization of Constrained Systems: Finite Dimensional Examples*, Ph.D. Thesis, Syracuse University, arXiv: gr-qc/9304043

Tavakol, R. / Ellis, G., 1999: On Holography and Cosmology, *Physics Letters* **B 469**, 37-45, arXiv: hep-th/9908093

Taylor, W., 2001: M(atrix) Theory – Matrix Quantum Mechanics as a Fundamental Theory, *Reviews of Modern Physics* **73**, 419-462, arXiv: hep-th/0101126

Tegmark, M., 1998: Is 'the Theory of Everything' Merely the Ultimate Ensemble Theory?, *Annals of Physics* **270**, 1-51, arXiv: gr-qc/9704009

Tegmark, M., 2007: Shut Up and Calculate, arXiv: 0709.4024 [physics.pop-ph]

Tegmark, M., 2009: The Second Law and Cosmology, arXiv: 0904.3931 [physics]

Teitelboim, C., 1973: How Commutators of Constraints Reflect the Spacetime Structure, *Annals of Physics (New York)* **79**, 542-557

Teller, P., 1991: Substance, Relations, and Arguments about the Nature of Space-Time, *Philosophical Review*, 363-397

Teller, P., 1991a: Relativity, Relational Holism, and the Bell Inequalities, in: J.T. Cushing / E. McMullin (Eds.): *Philosophical Consequences of Quantum Theory – Reflections on Bell's Theorem*, Notre Dame, Ind., 208-223

Terno, D.R., 2005: From Qubits to Black Holes: Entropy, Entanglement and all that, *International Journal of Modern Physics* **D 14**, 2307-2314, arXiv: gr-qc/0505068

Terno, D.R., 2006: Inconsistency of Quantum-Classical Dynamics, and What it Implies, *Foundations of Physics* **36**, 102-111, arXiv: quant-ph/0402092

Terno, D.R., 2009: Black Hole Information Problem and Quantum Gravity, arXiv: 0909.4143 [gr-qc]

Tetteh-Lartey, E., 2007: Toward a Quantum Theory of Gravity and a Resolution of the Time Paradox, arXiv: 0709.0494 [hep-th]

Thiemann, T., 2001: *Introduction to Modern Canonical Quantum General Relativity*, arXiv: gr-qc/0110034

Thiemann, T., 2002: Lectures on Loop Quantum Gravity, arXiv: gr-qc/0210094

Thiemann, T., 2006: Loop Quantum Gravity: An Inside View, arXiv: hep-th/0608210

Thiemann, T., 2006a: Solving the Problem of Time in General Relativity and Cosmology with Phantoms and k-Essence, arXiv: astro-ph/0607380

Thiemann, T. 2006b: Reduced Phase Space Quantization and Dirac Observables, *Classical and Quantum Gravity* **23**, 1163, arXiv: gr-qc/0411031

Thiemann, T., 2007: *Modern Canonical Quantum General Relativity*, Cambridge

't Hooft, G., 1993: Dimensional Reduction in Quantum Gravity, arXiv: gr-qc/9310026

't Hooft, G., 1995: Questioning the Answers or Stumbling upon Good and Bad Theories of Everything, in: Hilgevoord (1995)

't Hooft, G., 1999: Quantum Gravity as a Dissipative Deterministic System, *Classical and Quantum Gravity* **16**, 3263-3279, arXiv: gr-qc/9903084

't Hooft, G., 2000: The Holographic Principle, arXiv: hep-th/0003004

't Hooft, G., 2000a: Determinism and Dissipation in Quantum Gravity, arXiv: hep-th/0003005

't Hooft, G., 2001: How does God play dice? – (Pre-)Determinism at the Planck-Scale, arXiv: hep-th/0104219

't Hooft, G., 2001a: Obstacles on the Way Towards the Quantisation of Space, Time and Matter – and Possible Resolutions, *Studies in History and Philosophy of Modern Physics* **32**, 157-180

't Hooft, G., 2007: Emergent Quantum Mechanics and Emergent Symmetries, arXiv: 0707.4568 [hep-th]

't Hooft, G., 2007a: The Grand View of Physics, arXiv: 0707.4572 [hep-th]

't Hooft, G., 2008: A Locally Finite Model for Gravity, arXiv: 0804.0328 [gr-qc]

't Hooft, G., 2009: Quantum Gravity without Space-Time Singularities or Horizons, arXiv: 0909.3426 [gr-qc]

Tipler, F.J., 2005: The Structure of the World from Pure Numbers, *Reports on Progress in Physics* **68**, 897-964

Tiwari, T.C., 2006: Thermodynamics of Spacetime and Unimodular Relativity, arXiv: gr-qc/0612099

Tooley, M., 1997: *Time, Tense, and Causation*, Oxford

Toretti, R., 2000: Spacetime Models for the World, *Studies in History and Philosophy of Modern Physics* **31B**, 171-186

Torre, C.G., 1993: Gravitational Observables and Local Symmetries, *Physical Review* **D 48**, R2373-R2376

Townsend, P.K., 1995: The eleven-dimensional supermembrane revisited, *Physics Letters* **B 350**, 184-187, arXiv: hep-th/9501068

Townsend, P.K., 1995a: P-Brane Democracy, arXiv: hep-th/9507048

Townsend, P.K., 1996: Four Lectures on M-Theory, in: E. Gava et al. (Eds.): *High-Energy Physics and Cosmology, Trieste 1996*, Singapore, 385-438, arXiv: hep-th/9612121

Treder, H.-J., 1974: *Philosophische Probleme des Physikalischen Raumes*, Berlin

Trivedi, S.P., 2001: Holography, Black Holes and String Theory, *Current Science* **81(12)**, 1582-1590

Tsujikawa, S. / Singh, P. / Maartens, R., 2004: Loop Quantum Gravity Effects on Inflation and the CMB, *Classical and Quantum Gravity* **21**, 5767-5775, arXiv: astro-ph/0311015

Turner, M.S. / Huterer, D., 2007: Cosmic Acceleration, Dark Energy and Fundamental Physics, arXiv: 0706.2186 [astro-ph]

Turyshev, S.G., 2008: Experimental Tests of General Relativity: Recent Progress and Future Directions, arXiv: 0809.3730 [gr-qc]

Tye, S.-H. H., 2006: A New View of the Cosmic Landscape, arXiv: hep-th/0611148

Unruh, W., 1989: Unimodular Theory of Canonical Quantum Gravity, *Physical Review* **D 40**, 1048-1052

Unruh, W., 1991: No Time and Quantum Gravity, in: Mann, R. / Wesson, P. (Eds.): *Gravitation*, Singapore, 260-275

Unruh, W., 1995: Time, Gravity, and Quantum Mechanics, in: Savitt (1995) 23-65

Unruh, W.G., 2001: Black Holes, Dumb Holes, and Entropy, in: Callender / Huggett (2001)

Unruh, W. / Wald, R., 1989: Time and the Interpretation of Canonical Quantum Gravity, *Physical Review* **D 40**, 2598-2614

Unzicker, A., 2007: Why do we still believe in Newton's Law? – Facts, Myths and Methods in Gravitational Physics, arXiv: gr-qc/0702009

Uzan, J.-P., 2009: Fundamental Constants and Tests of General Relativity – Theoretical and Cosmological Consideration, arXiv: 0907.3081 [gr-qc]

Uzan, J.-P., 2009a: Dark Energy, Gravitation and the Copernican Principle, arXiv: 0912.5452 [gr-qc]

Vafa, C., 1997: Lectures on Strings and Dualities, 1996 ICTP Summer School Lectures, arXiv: hep-th/9702201

Vafa, C., 2009: Geometry of Grand Unification, arXiv: 0911.3008 [math-ph]

Vanchurin, V., 2007: Numerical Search for Fundamental Theory, arXiv: hep-th/0701147

Verozub, L.V., 2009: On some Fundamental Problems of the Theory of Gravitation, arXiv: 0911.5512 [gr-qc]

Visser, M., 2002: Sakharov's Induced Gravity: A Modern Perspective, *Modern Physics Letters* **A17**, 977-992, arXiv: gr-qc/0204062

Visser, M., 2007: Emergent Rainbow Spacetimes – Two Pedagogical Examples, arXiv: 0712.0810 [gr-qc]

Visser, M., 2009: Black Holes in General Relativity, arXiv: 0901.4365 [gr-qc]

Visser, M. / Barcelo, C. / Liberati, S. / Sonego, S., 2009: Small, Dark, and Heavy: But is it a Black Hole, arXiv: 0902.0346 [gr-qc]

Visser, M. / Weinfurtner, S., 2007: Analogue Spacetimes – Toy Models for "Quantum Gravity", arXiv: 0712.0427 [gr-qc]

Volovik, G.E., 2000: Links between Gravity and Dynamics of Quantum Liquids, arXiv: gr-qc/0004049

Volovik, G.E., 2001: Superfluid Analogies of Cosmological Phenomena, *Physics Reports* **352**, 195-348, arXiv: gr-qc/0005091

Volovik, G.E., 2003: *The Universe in a Helium Droplet*, Oxford, auch: http://boojum.hut.fi/personnel/THEORY/volovik.html

Volovik, G.E., 2006: From Quantum Hydrodynamics to Quantum Gravity, arXiv: gr-qc/0612134

Volovik, G.E., 2006a: Vacuum Energy: Myths and Reality, *International Journal of Modern Physics* **D 15**, 1987-2010, arXiv: gr-qc/0604062

Volovik, G.E., 2007: Fermi-Point Scenario for Emergent Gravity, arXiv: 0709.1258 [gr-qc]

Volovik, G.E., 2008: Emergent Physics: Fermi Point Scenario, arXiv: 0801.0724 [gr-qc]

Volovik, G.E., 2009: Osmotic Pressure of Matter and Vacuum Energy, arXiv: 0909.1044 [physics]

Vongehr, S., 2009: Supporting Abstract Relational Space-Time as Fundamental without Doctrinism against Emergence, arXiv: 0912.3069 [physics]

Wadia, S.R., 2001: A Microscopic Theory of Black Holes in String Theory – Thermodynamics and Hawking Radiation, *Current Science* **81(12)**, 1591-1597

Wald, R.M., 1994: *Quantum Field Theory on Curved Spacetime and Black Hole Thermodynamics*, Chicago

Wald, R.M., 2001: The Thermodynamics of Black Holes, *Living Reviews in Relativity* (Electronic Journal) **4/6**, www.livingreviews.org; auch: arXiv: gr-qc/9912119

Wald, R.M., 2005: The Arrow of Time and the Initial Conditions of the Universe, arXiv: gr-qc/0507094

Wald, R.M., 2006: The History and Present Status of Quantum Field Theory in Curved Spacetime, arXiv: gr-qc/0608018

Wald, R.M., 2009: The Formulation of Quantum Field Theory in Curved Spacetime, arXiv: 0907.0416 [gr-qc]

Wallace, D., 2003: Time-Dependent Symmetries: The Link between Gauge Symmetries and Indeterminism, in: Brading / Castellani (2003), 163-173

Wan, Y., 2007: On Braid Excitations in Quantum Gravity, arXiv: 0710.1312 [hep-th]

Wang, C.H.-T., 2006: New "Phase" of Quantum Gravity, arXiv: gr-qc/0605124

Ward, R.S., 1998: Twistors, Geometry, and Integrable Systems, in: S.A. Huggett et al. (Eds.): *The Geometric Universe*, Oxford (1998)

Weinberg, S., 1992: *Dreams of a Final Theory*, New York (dt.: *Der Traum von der Einheit des Universums*, München, 1993)

Weinberg, S., 1997: What is Quantum Field Theory, and What Did We Think It Is?, arXiv: hep-th/9702027

Weinberg, S., 2009: Living with Infinities, arXiv: 0903.0568 [hep-th]

Weinfurtner, S.C.E., 2007: *Emergent Spacetimes*, Ph.D. Thesis, arXiv: 0711.4416 [gr-qc]

Weinfurtner, S. / Visser, M. / Jain, P. / Gardiner, C.W., 2008: On the Phenomenon of Emergent Spacetimes: An Instruction Guide for Experimental Cosmology, arXiv: 0804.1346 [gr-qc]

Weinstein, S., 1998: *Conceptual and Foundational Issues in the Quantization of Gravity*, Dissertation, Northwestern University, Evanston, Il.

Weinstein, S., 1999: Gravity and Gauge Theory, *Philosophy of Science* **66**, Supplement (Proceedings), S146-S155

Weinstein, S., 2001: Naive Quantum Gravity, in: Callender / Huggett (2001)

Weinstein, S., 2005: Quantum Gravity, in: E.N. Zalta (Ed.): *Stanford Encyclopedia of Philosophy*, http://plato.stanford.edu

Wess, J., 2009: From Symmetry to Supersymmetry, arXiv: 0902.2201 [hep-th]

Wesson, P.S., 1998: *Space – Time – Matter: Modern Kaluza-Klein Theory*, Singapore

Wesson, P.S., 2007: The Meaning of Dimension, arXiv: 0712.1315 [gr-qc]

Wesson, P.S., 2009: Time as an Illusion, arXiv: 0905.0119 [physics]

Westman, H. / Sonego, S., 2007: Events and Observables in Generally Invariant Spacetime Theories, arXiv: 0708.1825 [gr-qc]

Westman, H. / Sonego, S., 2007a: Coordinates, Observables and Symmetry in Relativity, arXiv: 0711.2651 [gr-qc]

Wetterich, C., 2008: Emergence of Quantum Mechanics from Classical Statistics, arXiv: 0811.0927 [quant-ph]

Wetterich, C., 2009: Quantum Mechanics from Classical Statistics, arXiv: 0906.4919 [quant-ph]

Wheeler, J.A., 1957: On the Nature of Quantum Geometrodynamics, *Annals of Physics* **2**, 604-614

Wheeler, J.A., 1962: *Geometrodynamics*, New York

Wheeler, J.A., 1973: Beyond the End of Time, in: C.W. Misner / K.S. Thorne / J.A. Wheeler: *Gravitation*; San Francisco, 1196-1217

Wheeler, J.A., 1979: Frontiers of Time, in: N. Toraldo di Francia (Ed.): *Problems in the Foundations of Physics. Proceedings of the International School of Physics 'Enrico Fermi'*, Course 72, Amsterdam

Wheeler, J.A., 1983: Law without Law, in: J.A. Wheeler / W.H. Zurek (Eds.): *Quantum Theory and Measurement*, Princeton, N.J.

Wheeler, J.A., 1989: Information, Physics, Quantum: the Search for Links, in: *Proceedings 3rd International Symposium on the Foundation of Quantum Mechanics*, Tokyo, 354-368; auch in: W.H. Zurek (Ed.): *Complexity, Entropy and the Physics of Information*, New York (1990), 3-28

Wheeler, J.A., 1991: *Gravitation und Raumzeit – Die vierdimensionale Ereigniswelt der Relativitätstheorie*, Heidelberg

Wheeler, J.A. / Ford, K., 1998: *Black Holes and Quantum Foam: A Life in Physics*, New York

Will, C.M., 2005: The Confrontation between General Relativity and Experiment, *Living Reviews in Relativity* (Electronic Journal) **9/3**, www.livingreviews.org, auch: arXiv: gr-qc/0510072

Williams, R.M. / Tuckey, P.A., 1992: Regge Calculus – A Brief Review and Bibliography, *Classical and Quantum Gravity* **9**, 1409-1422

Willis, J., 2004: *On the Low-Energy Ramifications and a Mathematical Extension of Loop Quantum Gravity*, Ph.D. Thesis, Pennsylvania State University, http://cgpg.gravity.psu.edu/archives/thesis/2004/willis_thesis.pdf

Wiltshire, D.L., 2009: From Time to Timescape – Einstein's Unfinished Revolution, arXiv: 0912.4563 [gr-qc]

Witten, E., 1993: Quantum background independence in string theory, arXiv: hep-th/9306122

Witten, E., 1995: String Theory Dynamics in Various Dimensions, *Nuclear Physics* **B 443**, 85-126, arXiv: hep-th/9503124

Witten, E., 1996: Reflections on the Fate of Spacetime, *Physics Today* **49/4**, 24-30; auch in: Callender / Huggett (2001)

Witten, E., 1996a: Bound States of Strings and p-Branes, *Nuclear Physics* **B4 60**, 335, arXiv: hep-th/9510135

Witten, E., 1997: Duality, Spacetime and Quantum Mechanics, *Physics Today* **50/5**, 28-33

Witten, E., 2001: Black Holes and Quark Confinement, *Current Science* **81(12)**, 1576-1581

Witten, E., 2008: The Problem of Gauge Theory, arXiv: 0812.4512 [math.DG]

Woit, P., 2006: *Not Even Wrong: The Failure of String Theory and the Search for Unity in Physical Law*, New York

Wolfram, S., 2002: *A New Kind of Science*, Champaign, Ill.

Woodard, R.P., 2009: How Far Are We from the Quantum Theory of Gravity?, arXiv: 0907.4238 [gr-qc]

Wüthrich, C., 2005: To Quantize or Not to Quantize: Fact and Folklore in Quantum Gravity, *Philosophy of Science* **72**, 777-788

Wüthrich, C., 2006: *Approaching the Planck Scale from a General Relativistic Point of View: A Philosophical Appraisal of Loop Quantum Gravity*, Ph.D. Thesis, University of Pittsburgh

Yang, H.S., 2007: Noncommutative Spacetime and Emergent Gravity, arXiv: 0711.0234 [hep-th]

Yang, H.S., 2007a: Emergent Gravity and the Cosmological Constant Problem, arXiv: 0711.2797 [hep-th]

Yang, H. S., 2009: Dark Energy and Emergent Spacetime, arXiv: 0902.0035 [hep-th]

Zapata, J.A., 1998: *A Combinatorial Approach to Quantum Gauge Theories and Quantum Gravity*, Ph.D. Thesis, Pennsylvania State University, http://cgpg.gravity.psu.edu/archives/thesis/1998/zapata.pdf

Zee, A., 2008: Gravity and its Mysteries: Some Thoughts and Speculations, arXiv: 0805.2183 [hep-th]

Zhang, S.-C., 2002: To see a World in a Grain of Sand, arXiv: hep-th/0210162

Ziaeepour, H., 2009: And what if Gravity is intrinsically quantic?, arXiv: 0901.4634 [gr-qc]

Zimmermann, R.E., 2000: Loops and Knots as Topoi of Substance – Spinoza Revisited, http://philsci-archive.pitt.edu

Zimmermann, R.E., 2001: Recent Conceptual Consequences of Loop Quantum Gravity – Part I: Foundational Aspects, http://philsci-archive.pitt.edu

Zimmermann, R.E., 2001a: Recent Conceptual Consequences of Loop Quantum Gravity – Part II: Holistic Aspects, http://philsci-archive.pitt.edu

Zimmermann, R.E., 2001b: Recent Conceptual Consequences of Loop Quantum Gravity – Part III: A Postscript on Time, http://philsci-archive.pitt.edu

Zinkernagel, H., 2006: The Philosophy behind Quantum Gravity, *Theoria – An International Journal for Theory, History and Foundations of Science* **21/3**, 295-312, http://philsci-archive.pitt.edu, Dokument: 4050

Zizzi, P.A., 2001: Quantum Computation toward Quantum Gravity, *General Relativity and Gravitation* **33**, 1305-1318, arXiv: gr-qc/0008049

Zizzi, P.A., 2004: Computability at the Planck Scale, arXiv: gr-qc/0412076

Zizzi, P.A., 2005: A Minimal Model for Quantum Gravity, *Modern Physics Letters* **A 20**, 645-653, arXiv: gr-qc/0409069

Zumino, B., 1979: Supersymmetry – A Way to the Unitary Field Theory, in: H. Nelkowski et al. (Eds.): Einstein Symposium Berlin, Berlin (1979)

Zwiebach, B., 2004: *A First Course in String Theory*, Cambridge